\documentclass{aastex62}

\turnoffedit

\received{ddmmyy}
\revised{ddmmyy}
\accepted{ddmmyy}

\submitjournal{ApJ}

\shorttitle{}
\shortauthors{Duan \& Wang}

\begin{document}
\title{A SPECTRAL ANALYSIS OF $FERMI$-LLE GRBs}

\correspondingauthor{Xiang-Gao, Wang}
\email{wangxg@gxu.edu.cn}

\author[0000-0001-5487-4537]{Ming-Ya Duan}
\affil{GXU-NAOC Center for Astrophysics and Space Sciences, School of Physical Science and Technology, Guangxi University, Nanning 530004, People's Republic of China \\}
\affiliation{Guangxi Key Laboratory for the Relativistic Astrophysics, Nanning 530004, People's Republic of China}

\author[0000-0001-8411-8011]{Xiang-Gao Wang}
\affil{GXU-NAOC Center for Astrophysics and Space Sciences, School of Physical Science and Technology, Guangxi University, Nanning 530004, People's Republic of China \\}
\affiliation{Guangxi Key Laboratory for the Relativistic Astrophysics, Nanning 530004, People's Republic of China}

\begin{abstract}
The prompt emission of gamma-ray bursts remains mysterious since the mechanism is difficult to understand even though there are much more observations with the development of detection technology. But most of the gamma-ray bursts spectra show the Band shape, which consists of the low energy spectral index $\alpha$, the high energy spectral index $\beta$, the peak energy $E_{p}$ and the normalization of the spectrum. We present a systematic analysis of the spectral properties of 36 GRBs, which were detected by the Gamma-ray Burst Monitor (GBM), simultaneously, were also observed by the Large Area Telescope (LAT) and the LAT Low Energy (LLE) detector on the $Fermi$ satellite. We performed the detailed time-resolved spectral analysis for all of the bursts in our sample. We found that the time-resolved spectrum at peak flux can be well fitted by the empirical Band function for each burst in our sample. Moreover, the evolution patterns of $\alpha$ and $E_{p}$ have been carried for statistical analysis, and the parameter correlations have been obtained such as $E_{p}-F$, $\alpha-F$, and $E_{p}- \alpha$, all of them are presented by performing the detailed time-resolved spectral analysis. We also demonstrated that the two strong positive correlations $\alpha-F$ and $E_{p}-\alpha$ for some bursts originate from a non-physical selection effects through simulation.

\end{abstract}


\section{Introduction} \label{sec:intro}
As we all know, gamma-ray bursts (GRBs) are the brightest explosions in the universe. \edit1{\replaced{It's}{It is}} generally believed that they are from the magnetars or black holes \edit1{\replaced{since}{resulting from}} the mergers of compact binaries (NS-NS or BH-NS) or  the death of massive stars  \citep{1974ApJ...187..333C,1986ApJ...308L..43P,1989Natur.340..126E,1992ApJ...395L..83N,1993ApJ...405..273W,1999ApJ...524..262M,2006ARA&A..44..507W,2015PhR...561....1K}. The Band function \citep{1993ApJ...413..281B} can fit the gamma-ray burst spectra such as the time-integrated spectra and the time-resolved spectra, which is contained four parameters, the \edit1{\replaced{low-energy powerlaw}{low energy power-law}} index $\alpha$, the high energy \edit1{\replaced{powerlaw}{power-law}} index $\beta$, the peak energy $E_{p}$ and the normalized constant. It is proved that these parameters \edit1{\replaced{are evolves}{evolve}} with time instead of remaining constant. Many \edit1{\replaced{literatures}{references}}, such as \citet{1983Natur.306..451G}, \citet{1986ApJ...301..213N}, \citet{1994ApJ...422..260K}, \citet{1994ApJ...426..604B}, \citet{1995ApJ...439..307F}, \citet{1997ApJ...479L..39C}, \citet{2006ApJS..166..298K}, \citet{2009ApJ...698..417P} in \edit1{\added{the}} pre-$Fermi$ era and \citet{2012ApJ...756..112L}, \citet{2016A&A...588A.135Y}, \citet{2018MNRAS.475.1708A}, \citet{2019ApJS..242...16L}, \citet{2019ApJ...886...20Y} in \edit1{\added{the}} $Fermi$ era have shown the \edit1{\replaced{evolution}{evolutional}} characteristics of $\alpha$ and $E_{p}$ in Band function \citep{1993ApJ...413..281B}. There are three types for the evolution patterns of peak energy $E_{p}$, (i) \edit1{\deleted{it is named }}`hard-to-soft' trend, the value of $E_{p}$ is decreasing monotonically\citep{1986ApJ...301..213N,1994ApJ...426..604B,1997ApJ...486..928B}; (ii) \edit1{\replaced{it varies}{those varing}} with flux, i.e., $E_{p}$ will be increasing/decreasing since the flux is increasing/decreasing, named `flux-tracking' trend \citep{1983Natur.306..451G,1999ApJ...512..693R}; (iii) `soft-to-hard' trend or chaotic evolutions \citep{1985ApJ...290..728L,1994ApJ...422..260K}. Recently, \citet{2012ApJ...756..112L} and \citet{2019ApJ...886...20Y} pointed out that the first two patterns are dominated. \edit1{\replaced{And the evolution for}{For the evolution of}} the low energy photon index $\alpha$, it does not show \edit1{\added{a}} strong general trend compared with $E_{p}$ although it evolves with time instead of remaining constant. However, \edit1{\replaced{it is not very clear for the physical origin of the evolution patterns in $E_{p}$ and $\alpha$}{the physical origin of the evolution patterns in $E_{p}$ and $\alpha$ is not very clear}}. On the other hand, \edit1{\replaced{there is no}{the}} analysis of a large sample of \edit1{\added{LLE}} GRBs for the parameters evolution and the parameter correlations \edit1{\added{is lacking}}, except for the single burst analysis, such as GRB 131231A in \citet{2019ApJ...884..109L} which is a single-pulse burst, and GRB 180720B \edit1{\replaced{(to be submitted)}{in \citet{2019ApJ...884...61D}}} which is a multi-peaked burst in \edit1{\added{the}} prompt light curve.

\edit1{\replaced{On the other hand,}{Furthermore,}} the launch of \edit1{\added{the}} Fermi Space Gamma-ray Telescope \edit1{\added{($Fermi$)}} in 2008 \citep{2009ApJ...697.1071A} \edit1{\replaced{make}{makes}} it possible to detect GRBs in \edit1{\added{a}} broad energy \edit1{\replaced{range}{band}} both in \edit1{\added{the}} prompt emission and \edit1{\added{the}} afterglow phase. $Fermi$ satellite consists of the Gamma-ray Burst Monitor (GBM) and \edit1{\added{the}} Large Area Telescope (LAT) with the LAT Low Energy (LLE) detector. The GBM consists of 12 NaI detectors (8 \edit1{\deleted{keV }}to 900 keV) and 2 BGO (200 keV to 40 MeV) detectors. \edit1{\replaced{That is to say,}{Obviously,}} the energy range in GBM detection is from 8 keV to 40 MeV. The LAT can detect the photons with \edit1{\added{the}} energy range from 100 MeV to 300 GeV. Moreover, the LLE can collect those lower energy gamma-ray photons down to 10 MeV. \edit1{\replaced{There are about}{About}} 2000 GRBs detected by $Fermi$ in the last ten years while the fewer of them were detected by $Fermi$-LAT, which is the number with \edit1{\added{a}} value of more than one hundred. \edit1{\replaced{And}{In addition,}} the GRBs with the detection of LLE are less than one hundred according to the available data at the Fermi Science Support Center (FSSC).\footnote{\url{https://fermi.gsfc.nasa.gov/ssc/data/access/}} \edit1{\added{\citet{2019ApJ...878...52A} gives that only 74 GRBs co-detected by the GBM and LAT (include LAT-LLE). We called them LLE GRBs.}} The photons \edit1{\replaced{possess the energy range across}{cover}} 8 orders of magnitude \edit1{\added{in the energy range}} for LLE bursts. \edit1{\deleted{Then, what is the power to produce these photons? It naturally leads to a question that whether the mechanism is the same, synchrotron origin or photosphere model?}}

In this work, after performing the detailed time-resolved spectral analysis of the bright gamma-ray bursts with the detection of $Fermi$-LLE in \edit1{\added{the}} prompt phase, \edit1{\replaced{we cluster these bursts to synchrotron origin or photosphere model by using Gaussian Mixture Models according to the parameters from the Band function.}{we present the time-resolved spectra around their peak flux, which they all can be fitted well by the Band function.}} Then we \edit1{\deleted{will }}give the evolution patterns \edit1{\replaced{for}{of}} the peak energy $E_{p}$ and \edit1{\replaced{low-energy}{low energy}} spectral index $\alpha$. \edit1{\replaced{And the}{The}} parameter correlations \edit1{\replaced{also will}{will also}} be presented in our analysis such as $E_{p}-F$, $\alpha-F$, and $E_{p}-\alpha$. \edit1{\replaced{At last,}{Besides,}} we \edit1{\replaced{also}{will}} make \edit1{\deleted{a}} statistical analysis for whether the \edit1{\replaced{low-energy powerlaw}{low energy power-law}} indices $\alpha$ exceed the synchrotron \edit1{\replaced{limits}{limit ($\alpha=-\frac{2}{3}$)}} given by \citet{1998ApJ...506L..23P} in these slices. \edit1{\added{We will perform a simulation to identify whether the two strong positive correlations $\alpha-F$ and $E_{p}-\alpha$ for some bursts are intrinsic or artificial.}}

\section{sample selection and method} \label{sec:sec2}
Up to now, more than one hundred bursts have been co-detected by the $Fermi$/GBM and LAT. \edit1{\replaced{And there are 62 GRBs}{But only 74 GRBs \citep{2019ApJ...878...52A}}} were also detected by LLE which can collect those lower energy gamma-ray photons down to 10 MeV in all of these bursts if there is no omission in our collection. This work makes use of all available LLE bursts observed until 20 July 2018. We remove \edit1{\deleted{all of 9 short bursts that have been identified as photosphere origin,}} a pure black body burst GRB 090902B, three extremely bright bursts (\edit1{\replaced{include GRB 080916C, GRB 130427A, GRB 160625B}{GRBs 080916C, 130427A and 160625B}}) and 2 long bursts that have been studied in \citet{2019ApJ...884..109L} (GRB 131231A) and \edit1{\replaced{Duan \& Wang (2019)}{\citet{2019ApJ...884...61D}}} (GRB 180720B) in detail. \edit1{\replaced{For these two long bursts, they}{These two long bursts}} are originated from synchrotron emission in \edit1{\added{the}} prompt phase.

We download data from the FSSC described as above. To complete this study, we take RMFIT as the tool of making \edit1{\added{the}} time-resolved spectral analysis. We perform the detailed time-resolved spectral analysis by using the TTE event data files of two NaI detectors and the corresponding BGO detector(s) on $Fermi$/GBM, 
but \edit1{\replaced{we gave up the use of LLE data}{the use of LAT and LLE data was abandoned}} because of \edit1{\replaced{its}{their}} lower impact for peak energy $E_{p}$ and low energy spectral index $\alpha$. \edit1{\replaced{And the}{The}} background photon counts were estimated by fitting the light curve before and after the operated burst with a one-order background polynomial model. We selected all of the prompt phase as the source. We take the signal-to-noise ratio (S/N) as 40 in all of the slices for each burst and they all can be well fitted by \edit1{\added{the}} Band function \citep{1993ApJ...413..281B}. \edit1{\replaced{In order to}{To}} show the spectral evolution, the sample in our analysis includes only those bursts which at least five time-resolved spectra can be produced from the data. Based on this, \edit1{\replaced{18}{32}} GRBs have been excluded due to the insufficiency of the number of time-resolved spectra. \edit1{\replaced{And}{Finally,}} \edit1{\replaced{there are 29 samples of bursts}{we get a sample of 36 GRBs}} by filtering described as above. The reduced $\chi^{2}$ has been taken into measuring the goodness-of-fit. The $\chi^{2}/dof$ is typically in the range of 0.75-1.5 in each slice. 

\edit1{\replaced{It's necessary to evoke a more suitable new method to identify the prompt origin in gamma-ray bursts. In our work, we will identify the prompt mechanism for these LLE bursts by using the Gaussian Mixture Models (GMM) to cluster these bursts according to the parameters from the Band function such as low energy spectral index $\alpha$, high energy spectral index $\beta$ and peak energy $E_{p}$.}{In our work, we present the Band-fitting spectra for all of the bursts around their peak flux firstly.}} For the evolution patterns of $\alpha$ and $E_{p}$, \edit1{\added{then,}} we will identify them as `hard-to-soft' (h.t.s.), `soft-to-hard' (s.t.h.), `intensity-tracking' (i.t.), `rough-tracking' (r.t.), `anti-tracking' (a.t.)\edit1{\added{,}} and `no' which means that it evolves without rule. \edit1{\replaced{And}{It is notable that}} all `-tracking' patterns based on the evolution of energy flux. \edit1{\replaced{Then,}{Finally,}} the statistical analysis of  the linear dependence in the parameter correlations such as $E_{p}-F$, $\alpha-F$, and $E_{p}-\alpha$ will be made by using \edit1{\deleted{the}} Pearson's correlation coefficient r. \edit1{\added{We also address whether the two observed correlations $\alpha-F$ and $E_{p}-\alpha$ are intrinsic or artificial by simulation.}}

\section{data analysis and results} \label{sec:sec3}
\begin{figure}[ht!]
\plotone{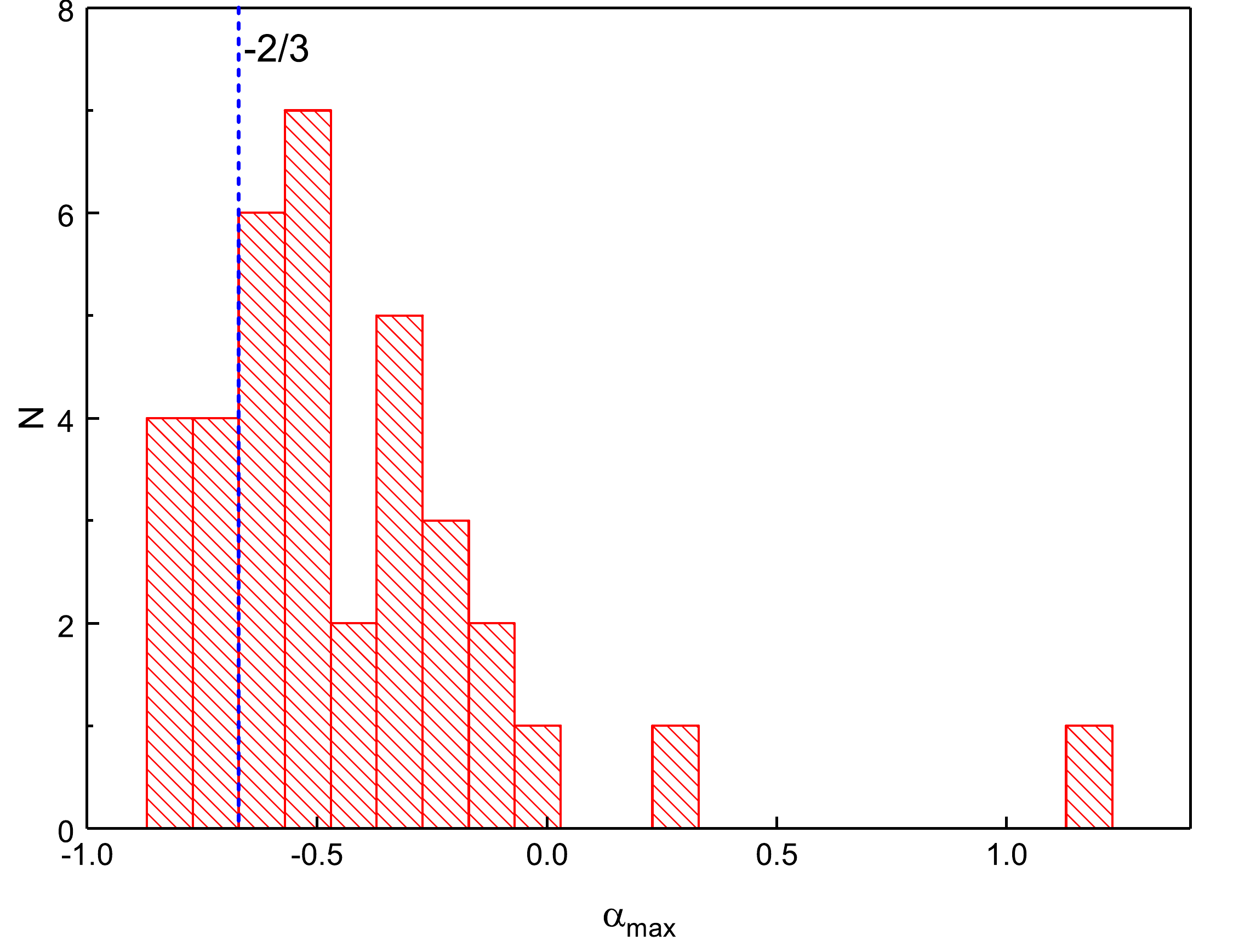} 
\caption{The histogram of the maximal value of $\alpha$ in the detailed time-resolved spectra for each burst. The blue short dash line indicates the synchrotron limit ($-\frac{2}{3}$). \edit1{\replaced{$79.3\%$}{$77.8\%$}} of the bursts have an $\alpha_{max}$ which is larger than the synchrotron limit in our sample of bursts.\label{fig:a_max}}
\end{figure}

\edit1{\added{
The data analysis results have been presented in Tables \ref{tab:peak_flux_results}, \ref{tab:integrated_results}, \ref{tab:resolved_results}, Figures \ref{fig:a_max}, \ref{fig:peak_flux}, \ref{fig:comparison_peak_flux}, \ref{fig:a_integrated}, \ref{fig:comparison_integrated}, \ref{fig:spectral evolutions}, \ref{fig:resolved}, \ref{fig:parameter correlations}, \ref{fig:statistical}, \ref{fig:simulation}, \ref{fig:simulation1}. Table \ref{tab:peak_flux_results} shows the results of the time-resolved spectral fits at peak flux for all samples. Table \ref{tab:integrated_results} shows the results of the time-integrated spectral fits for all samples. The fitting results of the parameter correlations and the spectral evolution patterns of $\alpha$ and $E_{p}$ have been shown in Table \ref{tab:resolved_results}, simultaneously, we also present the linear-fitting results from simulation for those bursts (23 GRBs) which exhibit a strong positive correlation in $\alpha-F$ and $E_{p}-\alpha$ correlations in this table. Figure \ref{fig:a_max} is the histogram of the maximal value of $\alpha$ in the detailed time-resolved spectra for each burst. Figure \ref{fig:peak_flux} presents those spectra with the best Band-fitting results around the peak-flux for all of our bursts. Figure \ref{fig:comparison_peak_flux} shows the comparison between our fitting results and the results of the GBM catalog \citep{2014ApJS..211...12G,2016ApJS..223...28N} at peak flux. Figure \ref{fig:a_integrated} is the comparison between the histogram of $\alpha$ in the time-integrated spectra in our energy range and the BATSE energy range. Figure \ref{fig:comparison_integrated} shows the comparison between our time-integrated spectral analysis results and the corresponding results of GBM catalog \citep{2014ApJS..211...12G}. Figure \ref{fig:spectral evolutions} represents the temporal characteristics of energy flux for all bursts in our sample (the left-hand, y-axis), along with time evolution of the $E_{p}$ and $\alpha$, both are marked with red stars in the right-hand y-axis. That is to say, Figure \ref{fig:spectral evolutions} shows the spectral evolutions for all of the bursts in our sample. The histograms of $E_{p}$ and $\alpha$ obtained by performing the detailed time-resolved spectral analysis have been shown in Figure \ref{fig:resolved}. The correlations such as $E_{p}-F$, $\alpha-F$, and $E_{p}- \alpha$ obtained from the time-resolved spectra are shown in Figure \ref{fig:parameter correlations}. Figure \ref{fig:statistical}, the histograms of Pearson's correlation coefficient from the fitting results of parameter correlations such as $E_{p}-F$, $\alpha-F$, and $E_{p}- \alpha$ have been shown on it. The last two figures, Figures \ref{fig:simulation}, \ref{fig:simulation1}, are the linear-fitting results in $\alpha-F$ and $E_{p}-\alpha$ correlations from simulation for 23 GRBs.}}

\subsection{Band-fitting Results at Peak Flux for All of the Bursts} \label{subsec:subsec3.1}

\begin{figure}
\centering
\resizebox{8cm}{!}{\includegraphics{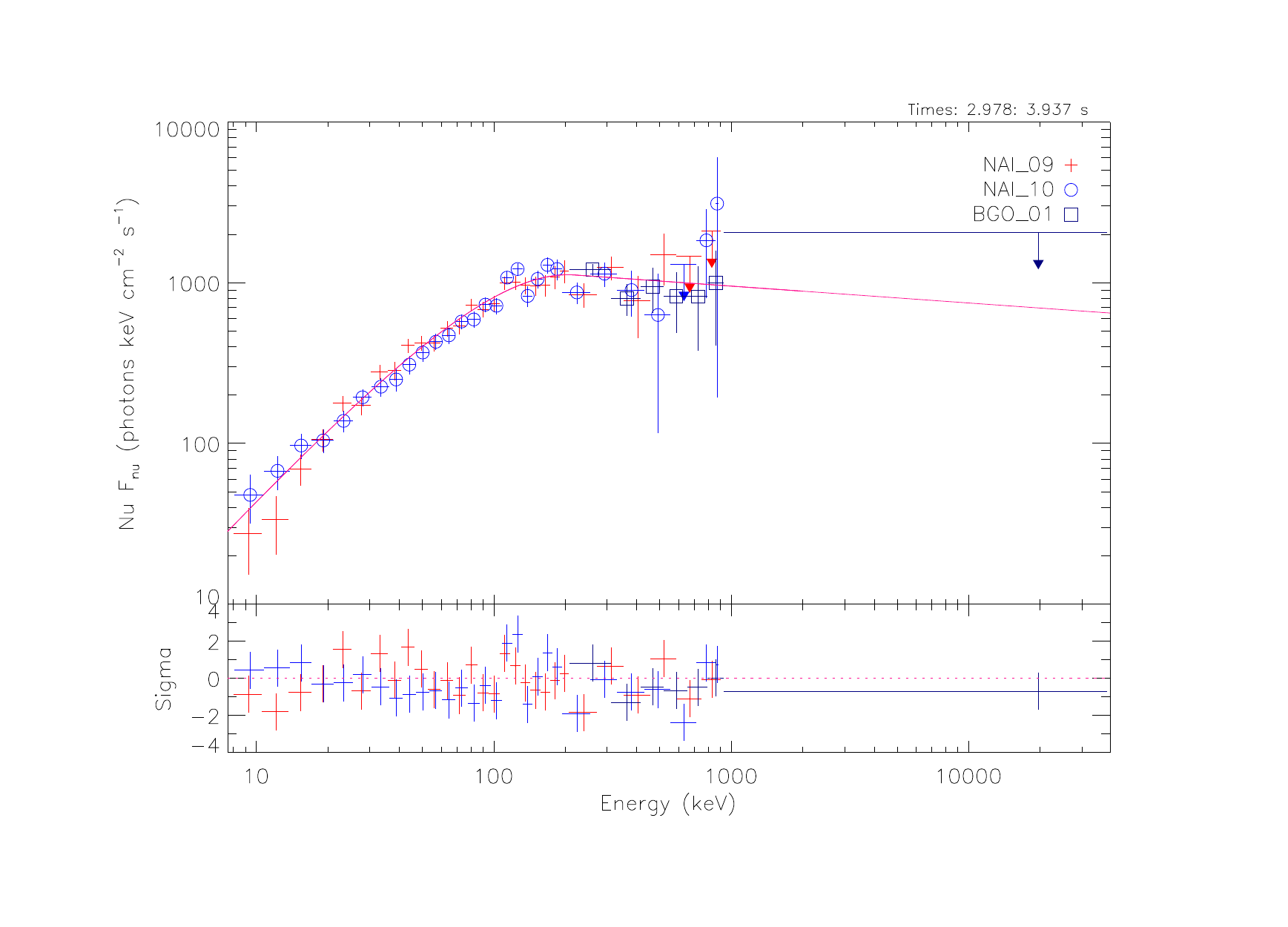}}
\resizebox{8cm}{!}{\includegraphics{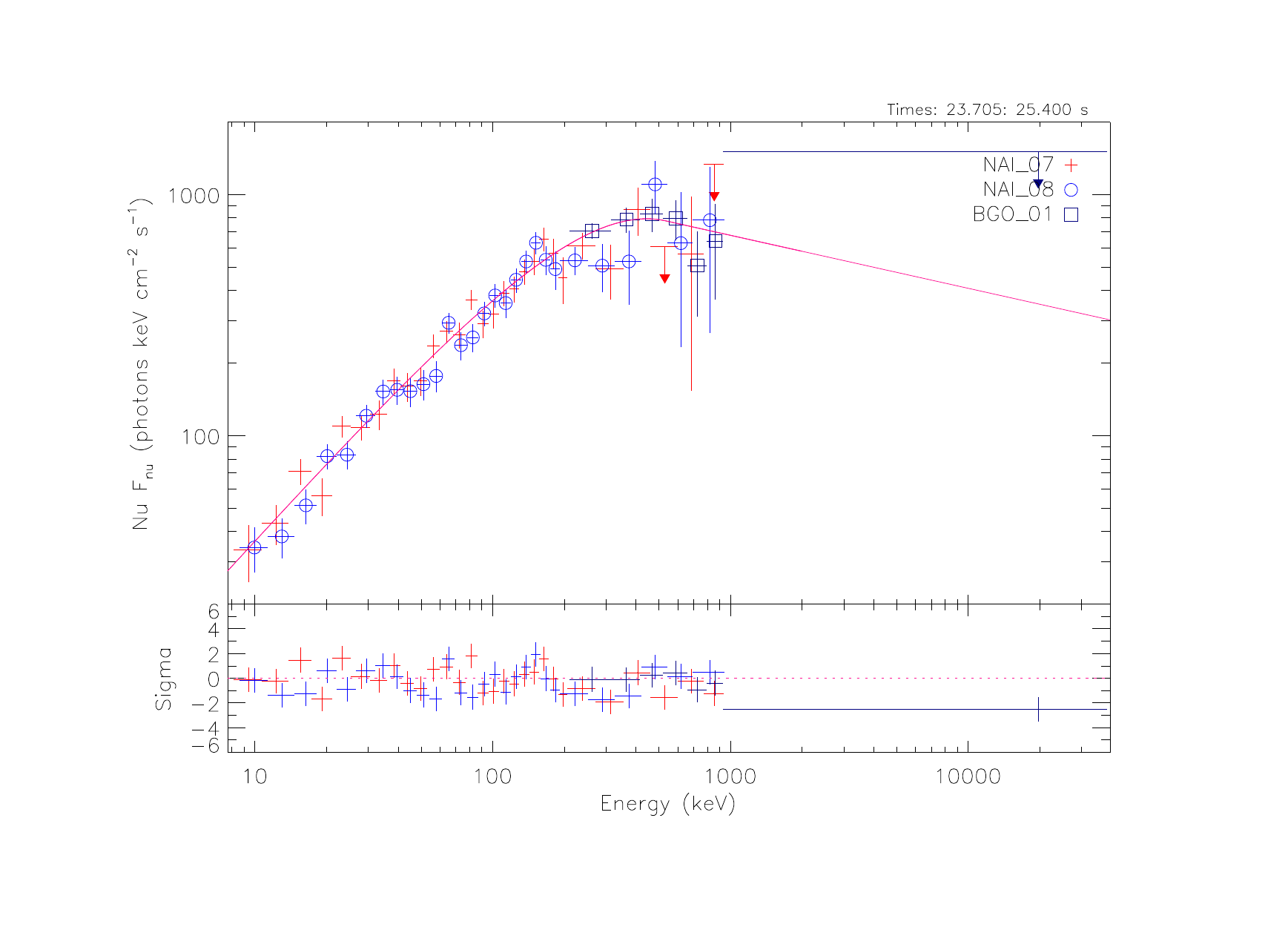}}
\resizebox{8cm}{!}{\includegraphics{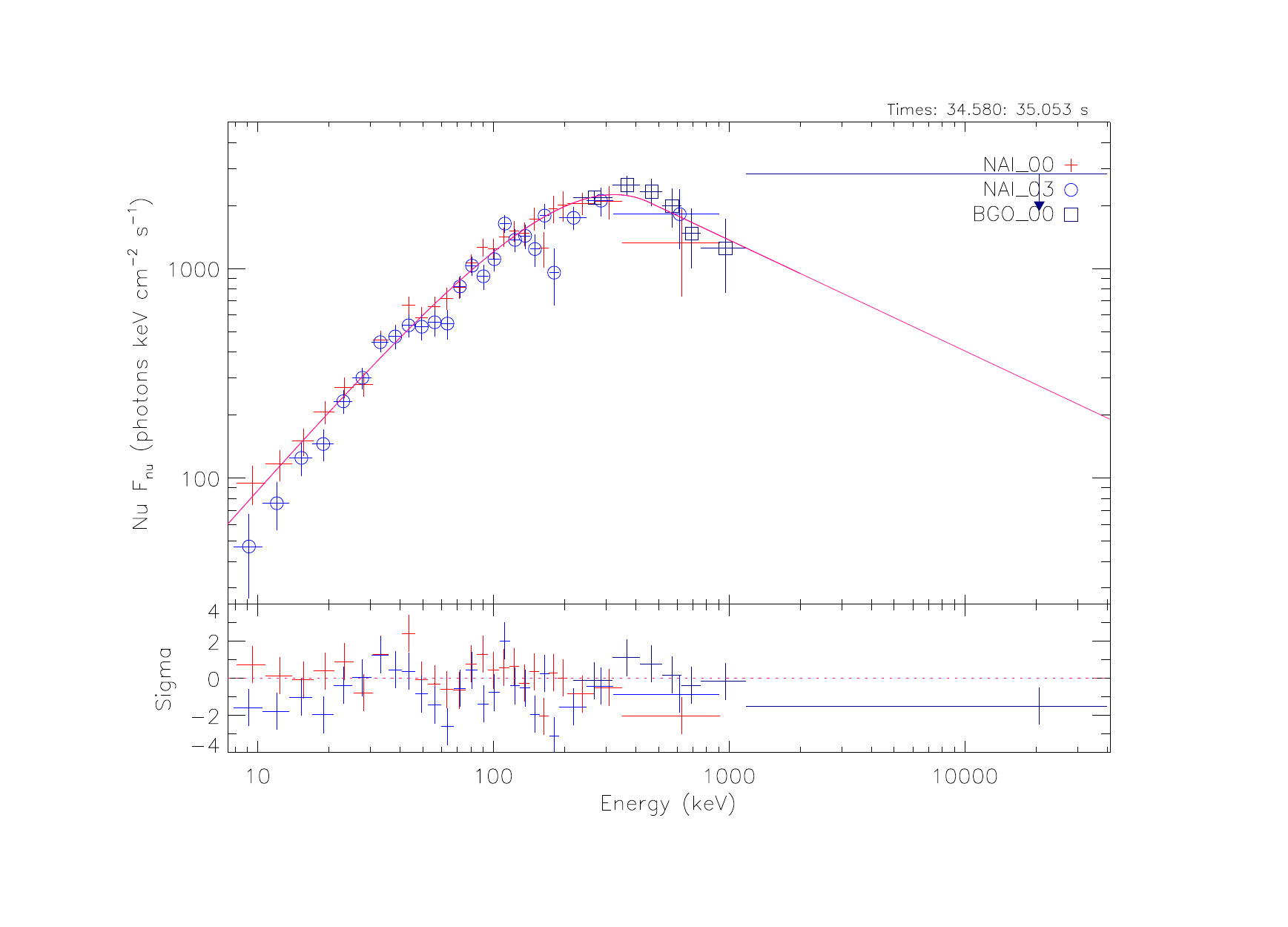}}
\resizebox{8cm}{!}{\includegraphics{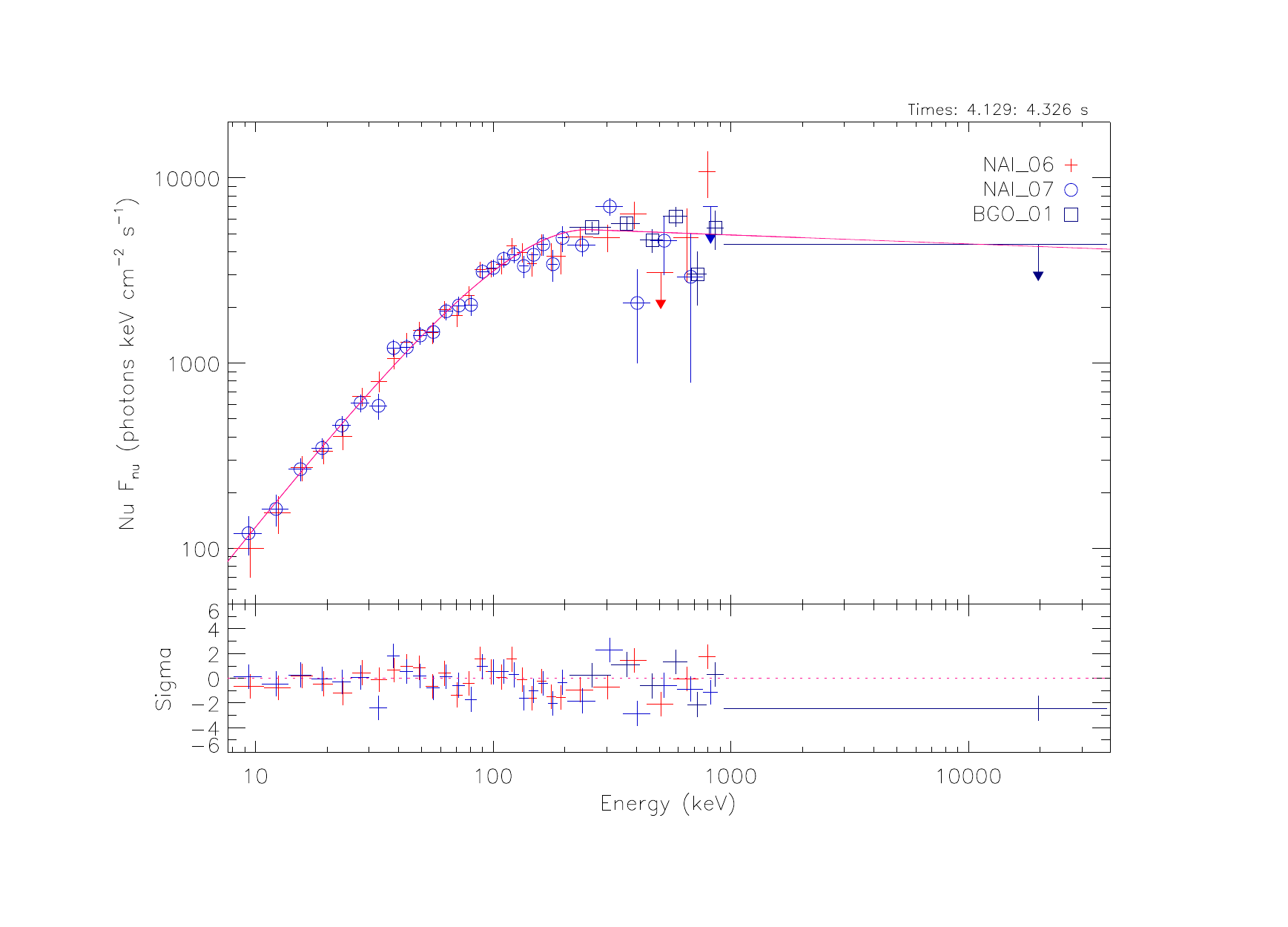}}  
\resizebox{8cm}{!}{\includegraphics{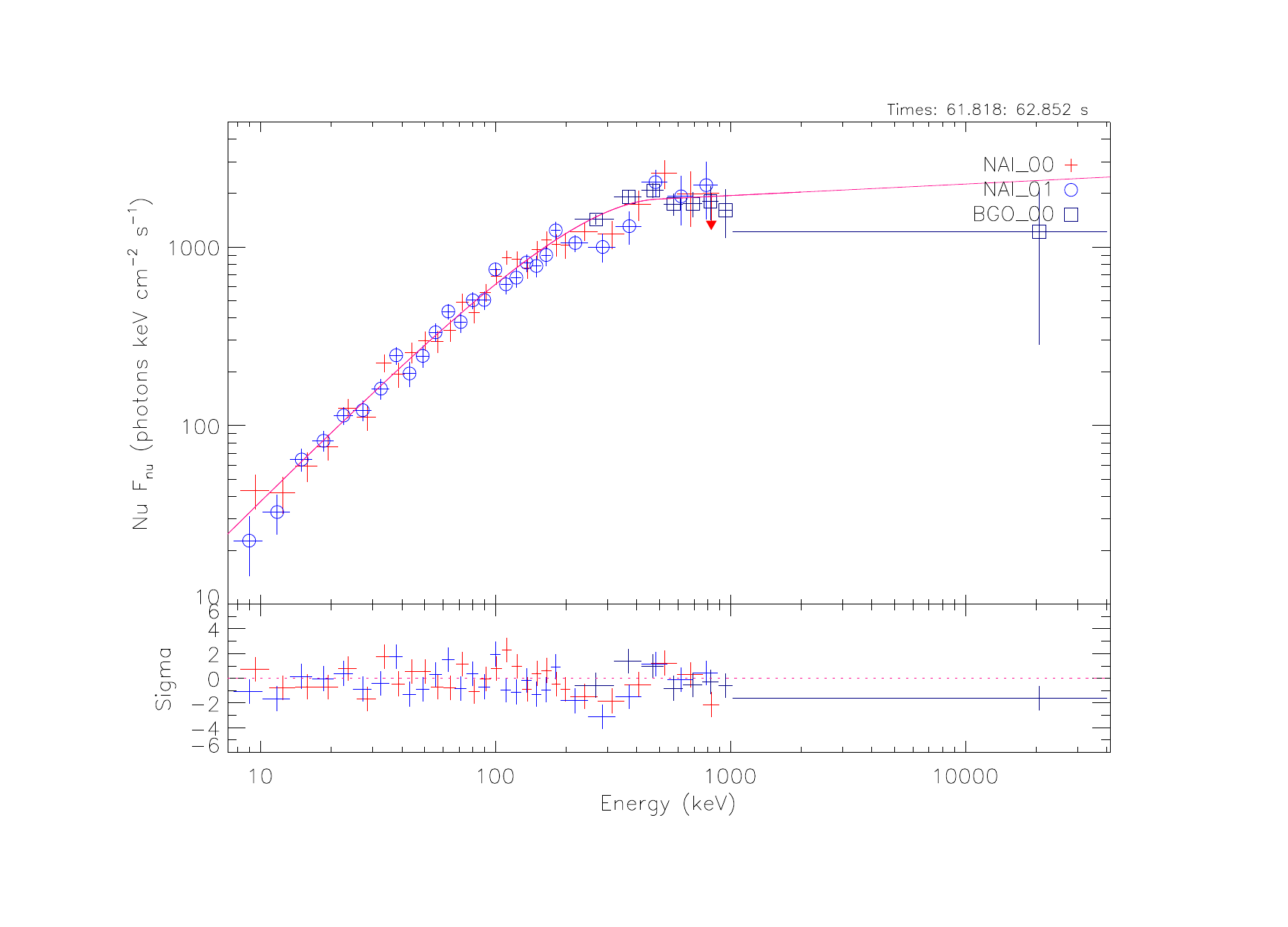}}
\resizebox{8cm}{!}{\includegraphics{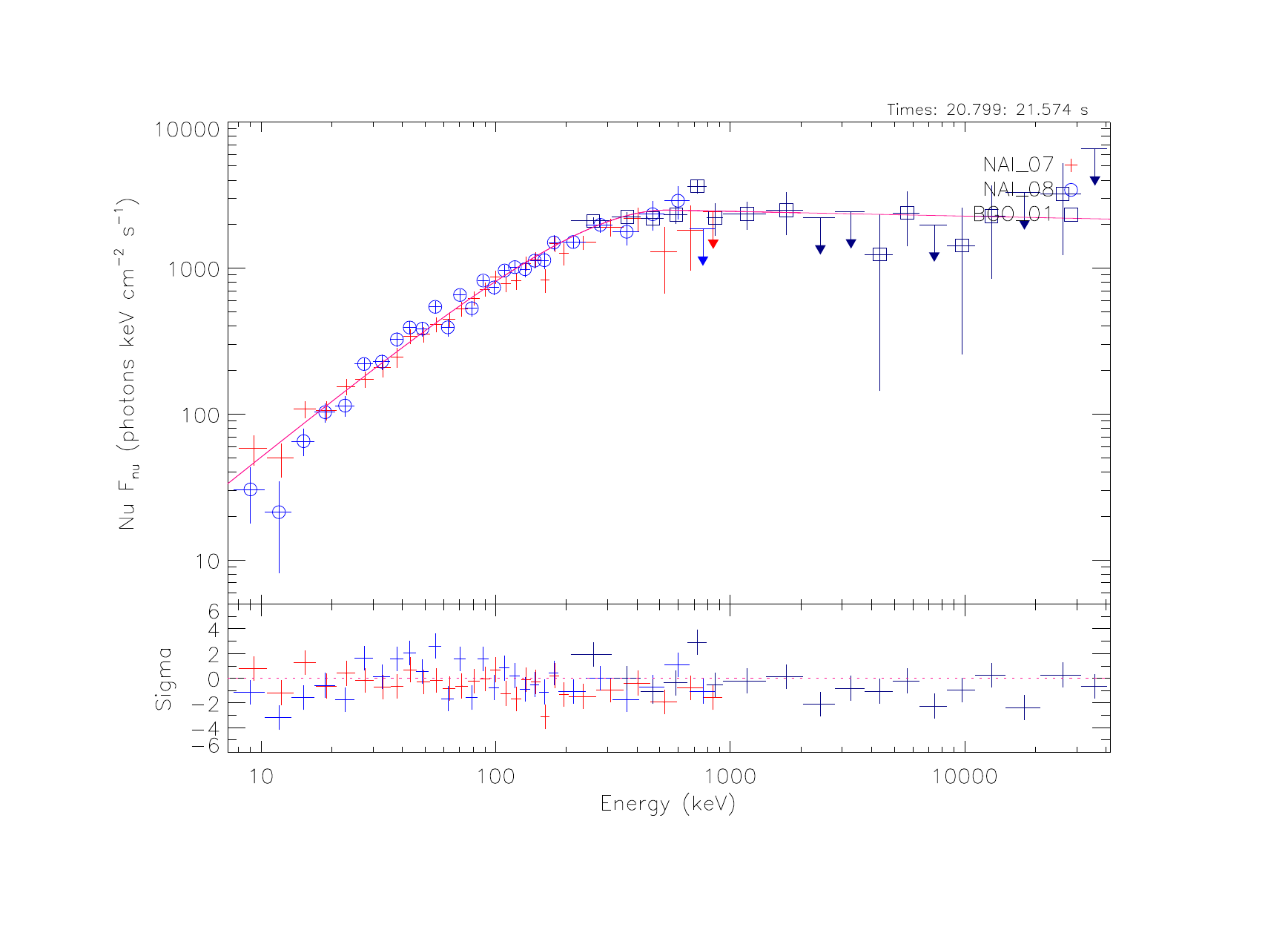}}
\resizebox{8cm}{!}{\includegraphics{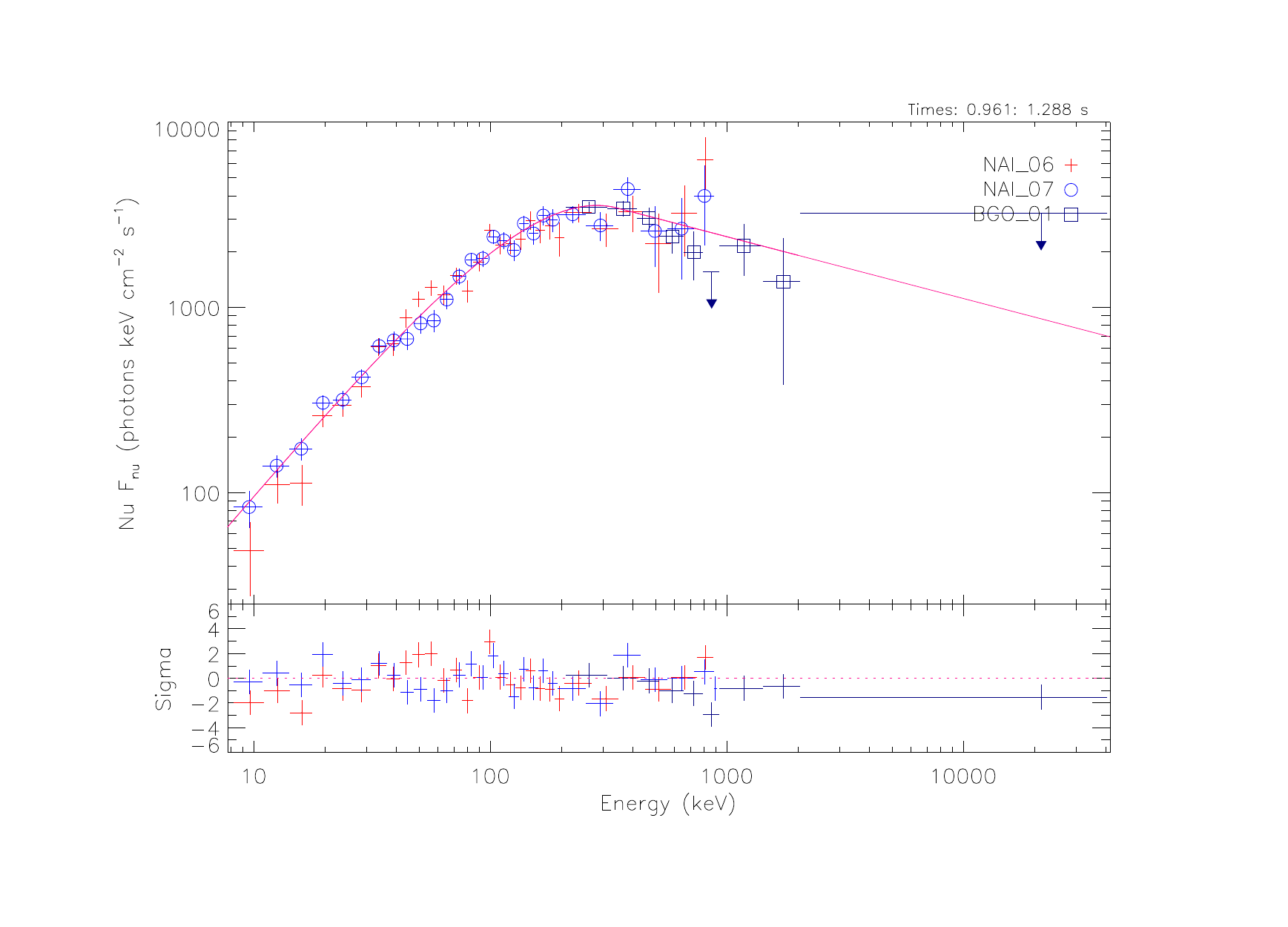}}
\resizebox{8cm}{!}{\includegraphics{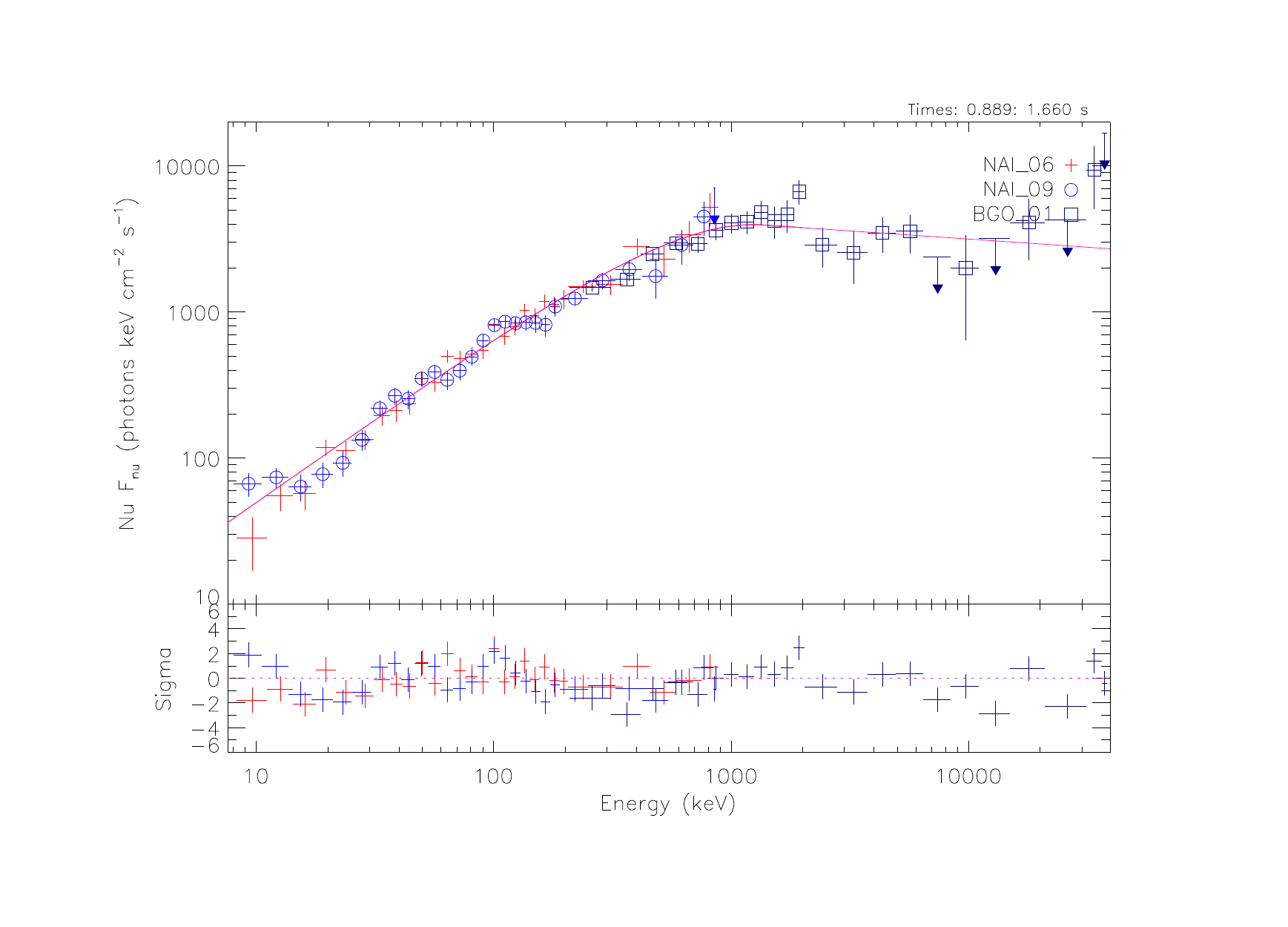}}    
\caption{\edit1{\added{The spectra with the best Band-fitting results around the peak-flux for all of the bursts in our sample. The first one is consistent with GRB 080825C, the last one is consistent with GRB 180305A. All of them are consistent with the results in Table \ref{tab:peak_flux_results} from GRB 080825C to GRB 180305A.}}}\label{fig:peak_flux}
\end{figure}

\addtocounter{figure}{-1}
\begin{figure}
\centering 
\resizebox{8cm}{!}{\includegraphics{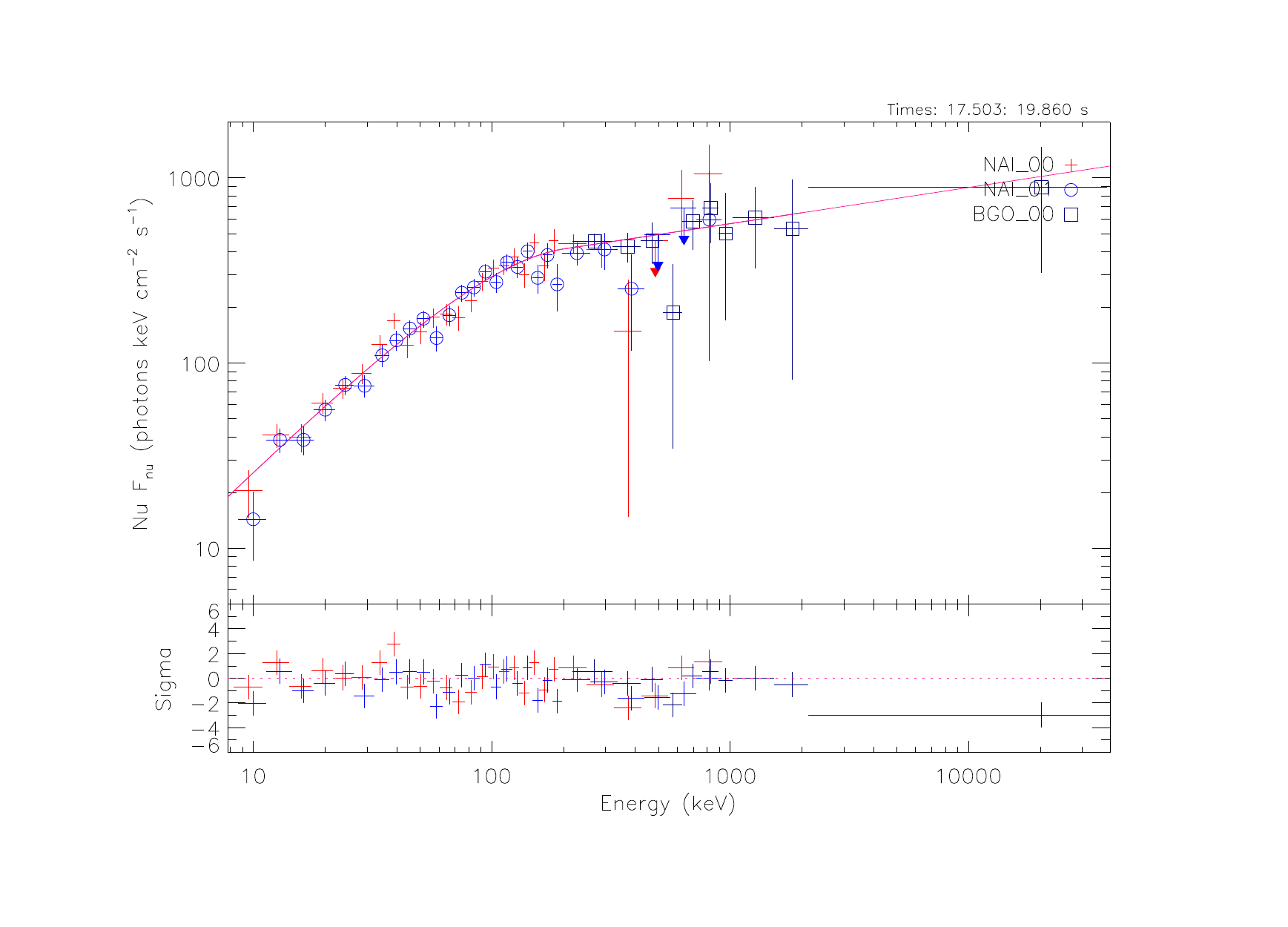}}
\resizebox{8cm}{!}{\includegraphics{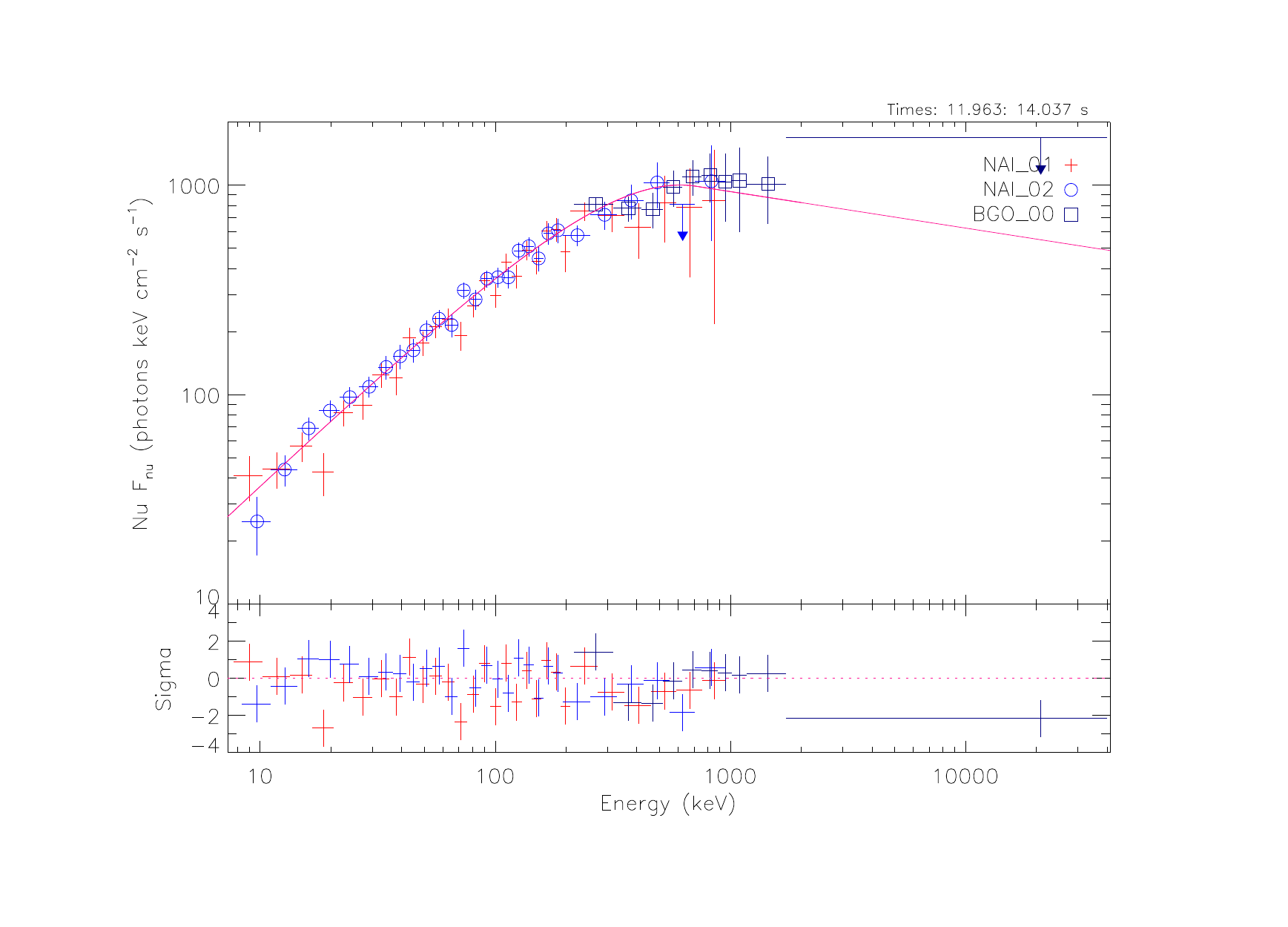}}
\resizebox{8cm}{!}{\includegraphics{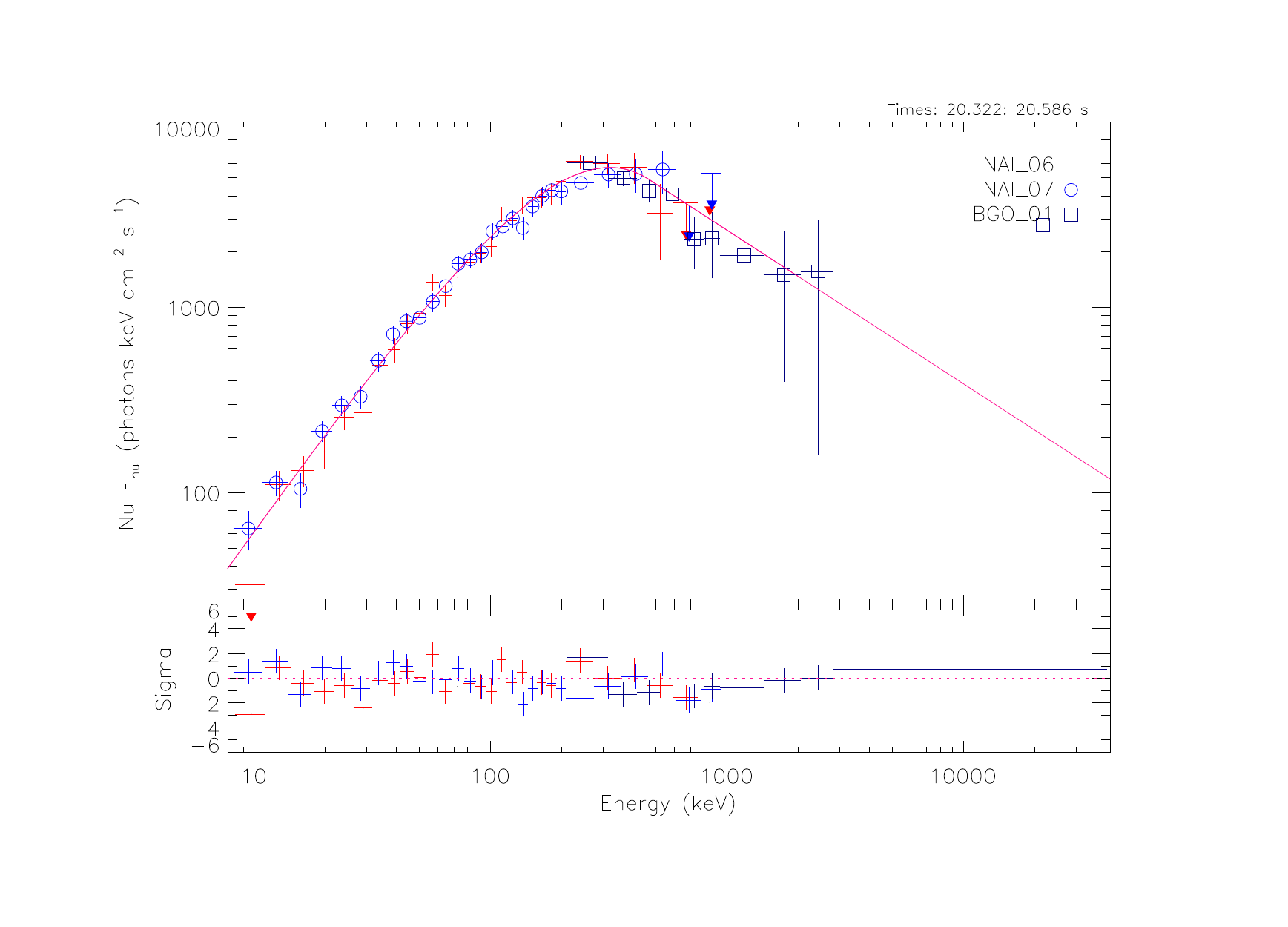}}
\resizebox{8cm}{!}{\includegraphics{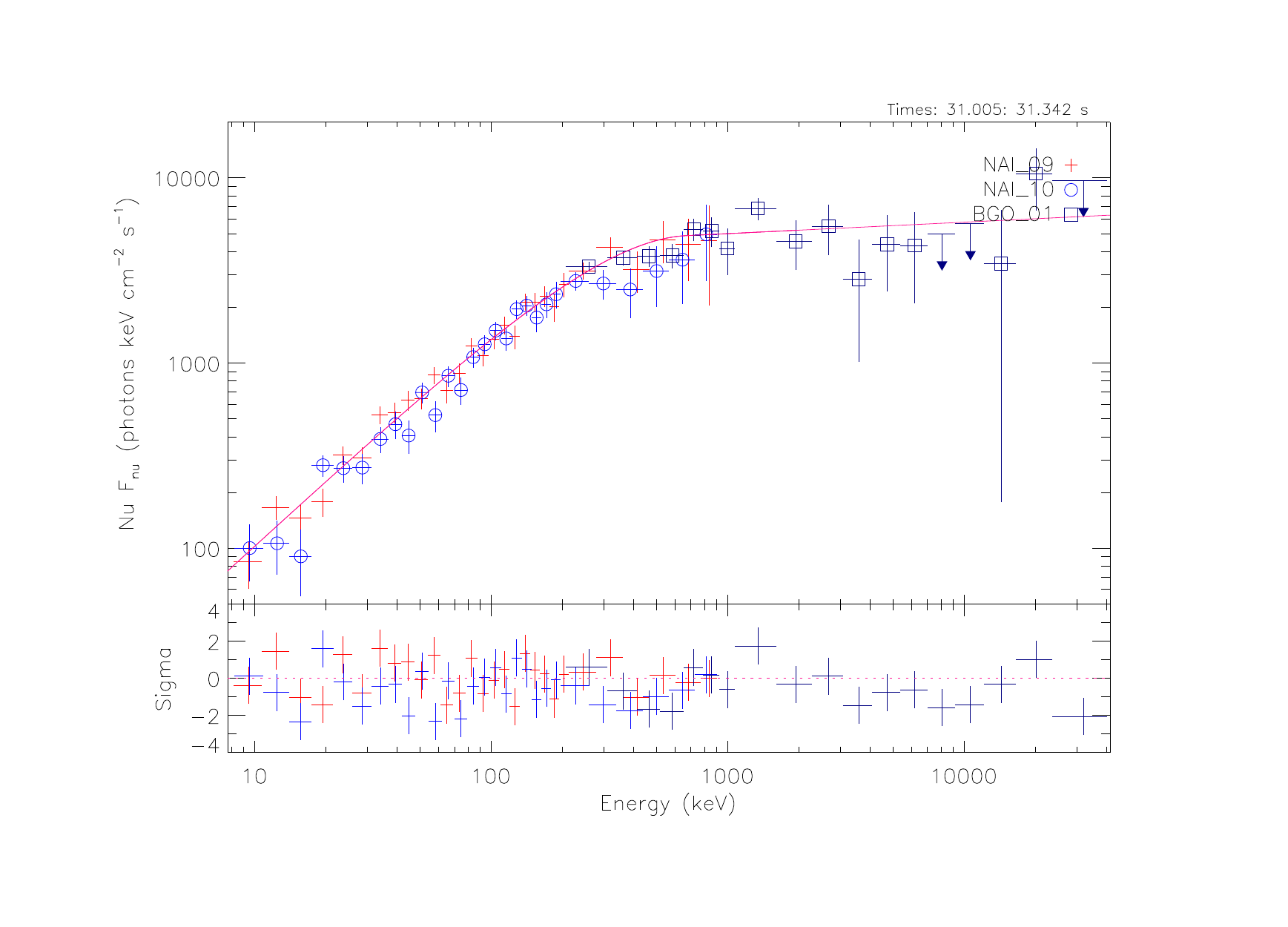}} 
\resizebox{8cm}{!}{\includegraphics{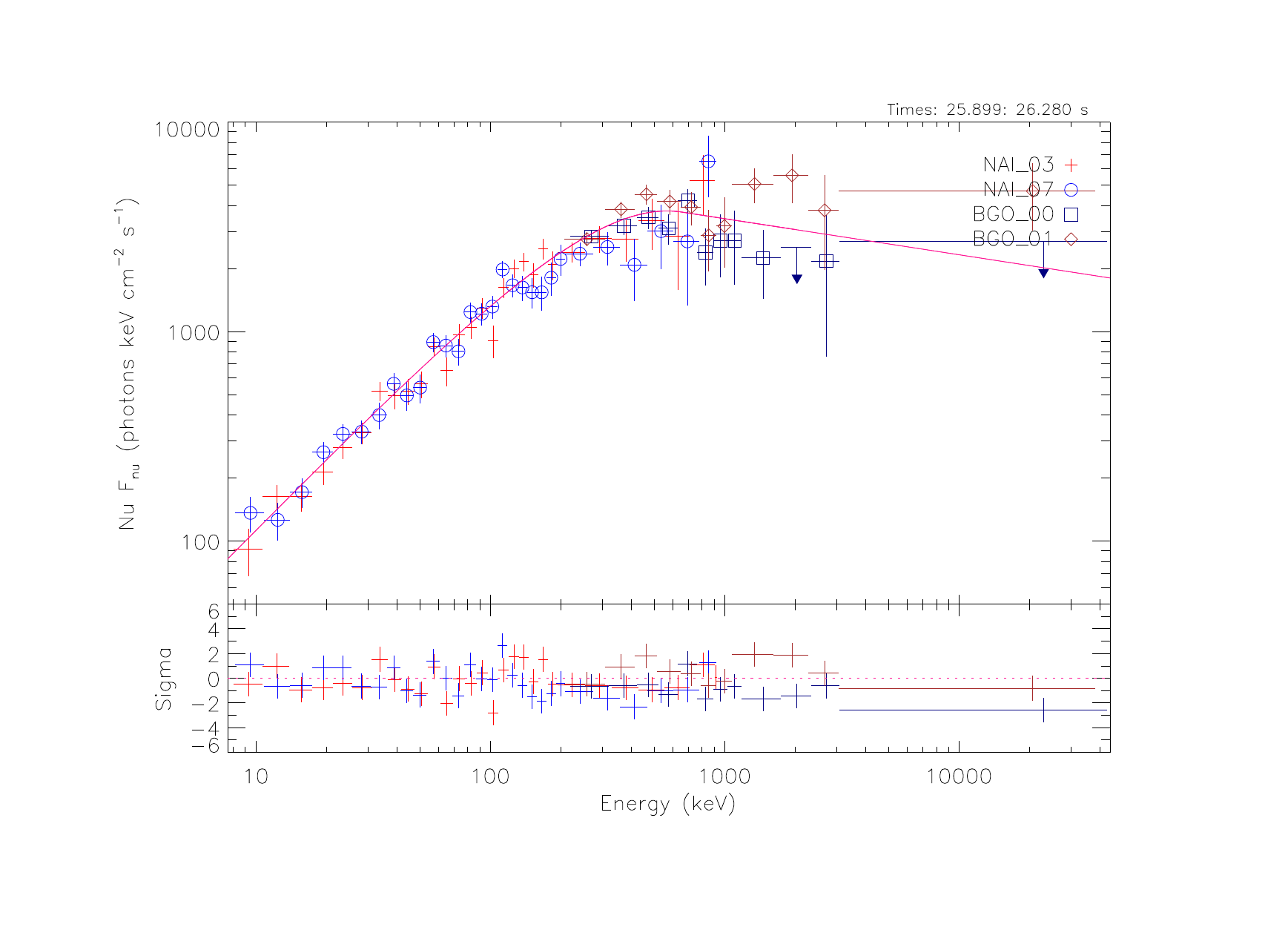}}
\resizebox{8cm}{!}{\includegraphics{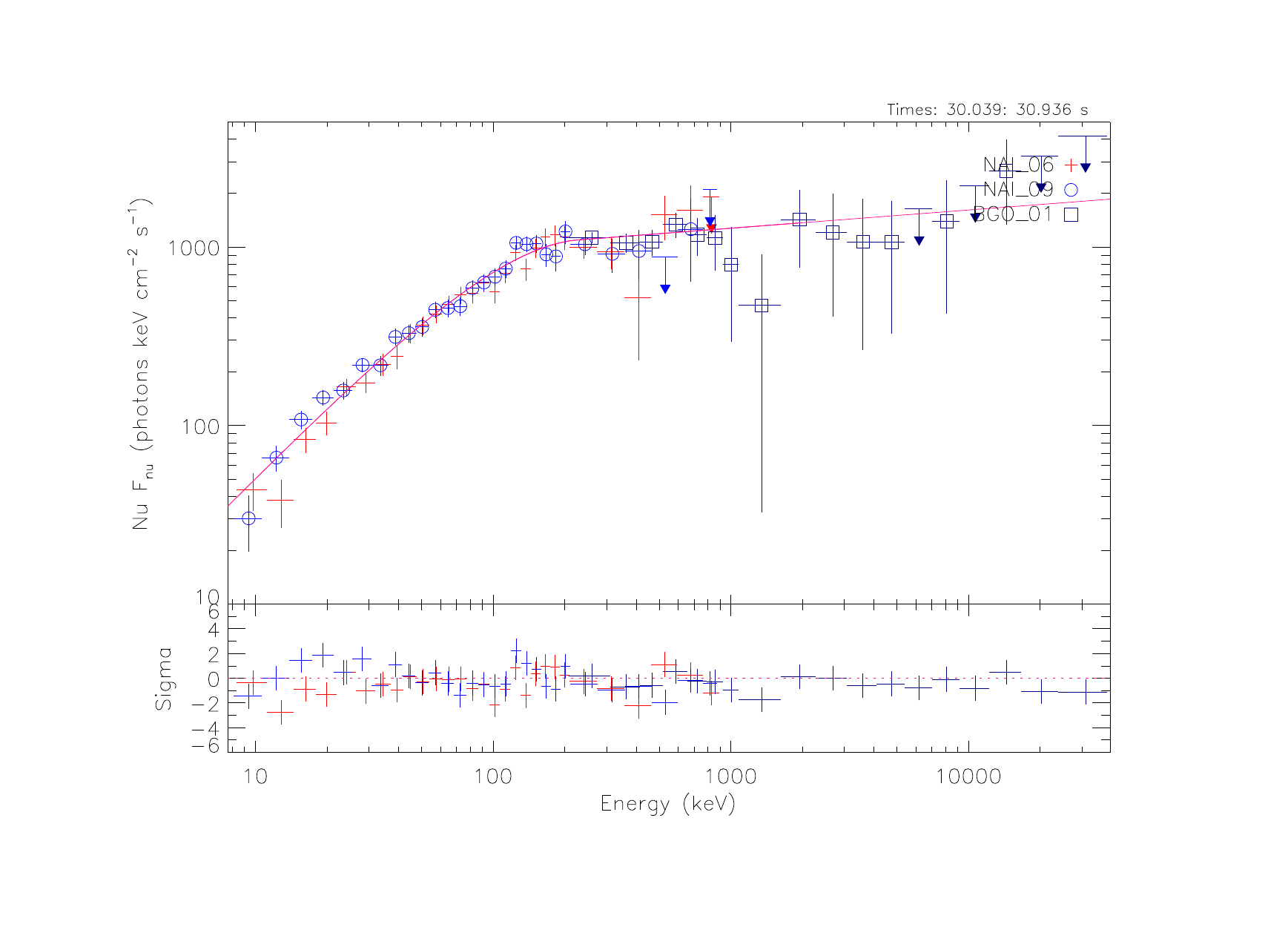}}
\resizebox{8cm}{!}{\includegraphics{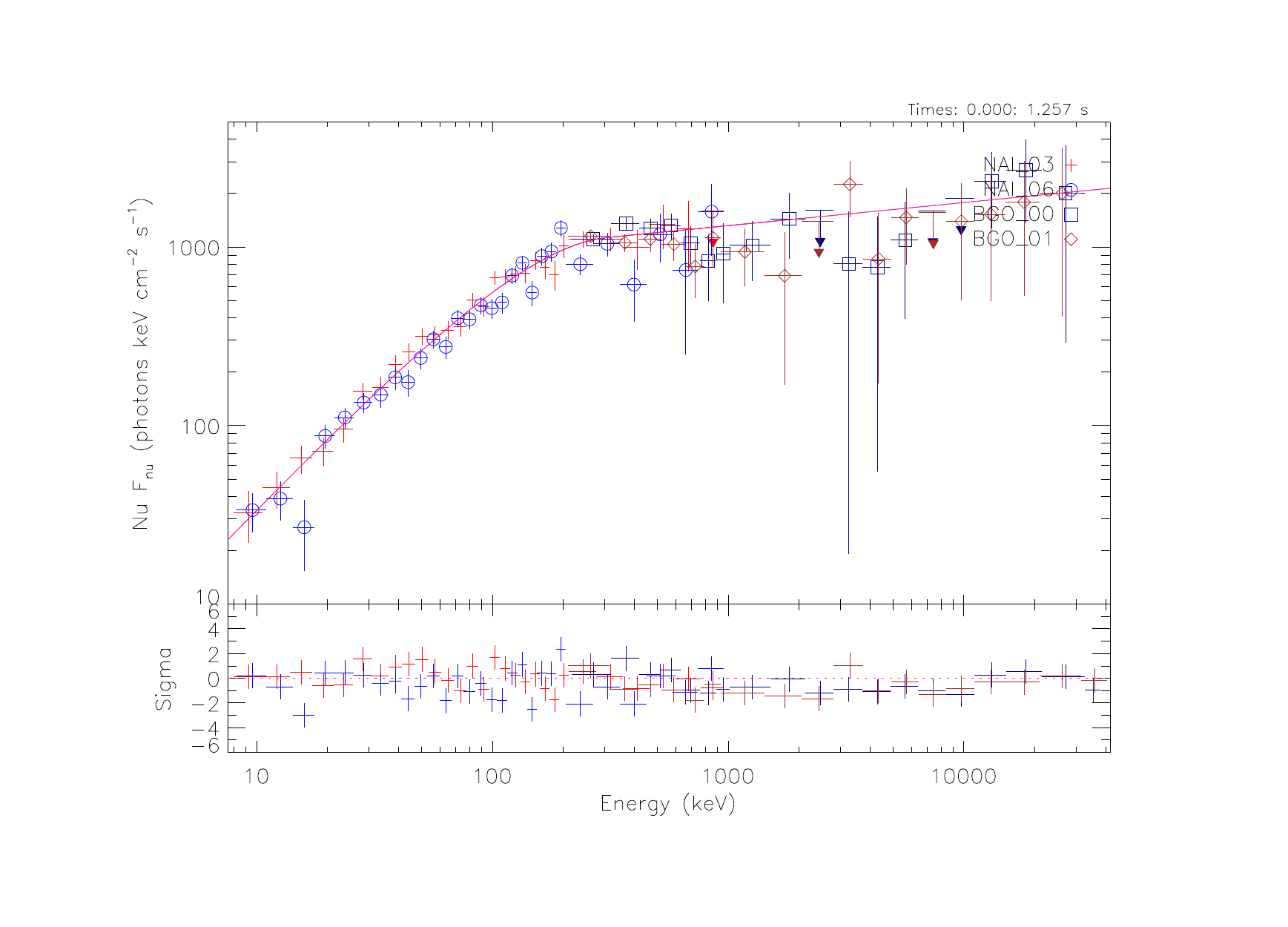}}
\resizebox{8cm}{!}{\includegraphics{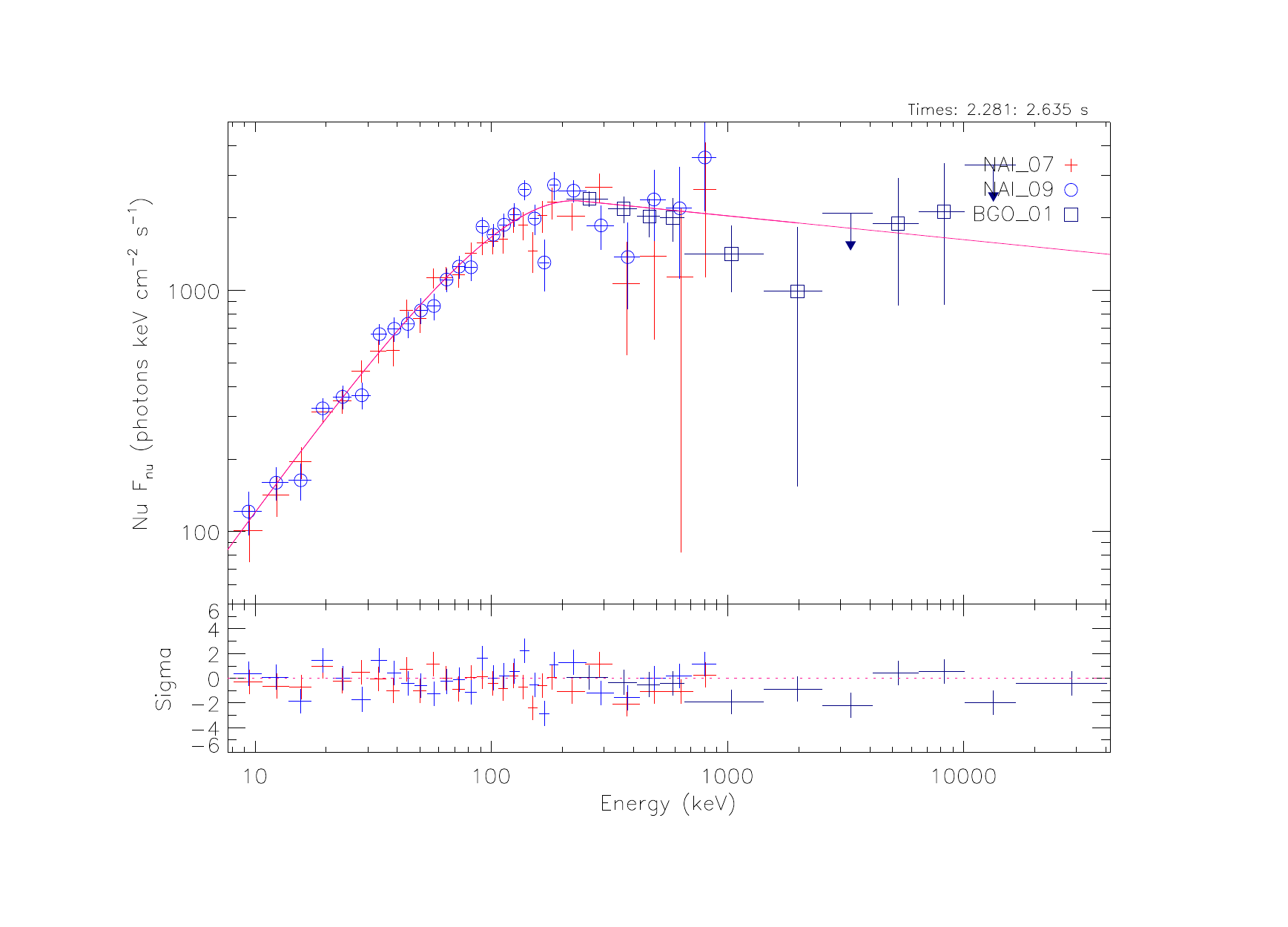}}
\caption{\it-continued}
\end{figure}

\addtocounter{figure}{-1}
\begin{figure}
\centering 
\resizebox{8cm}{!}{\includegraphics{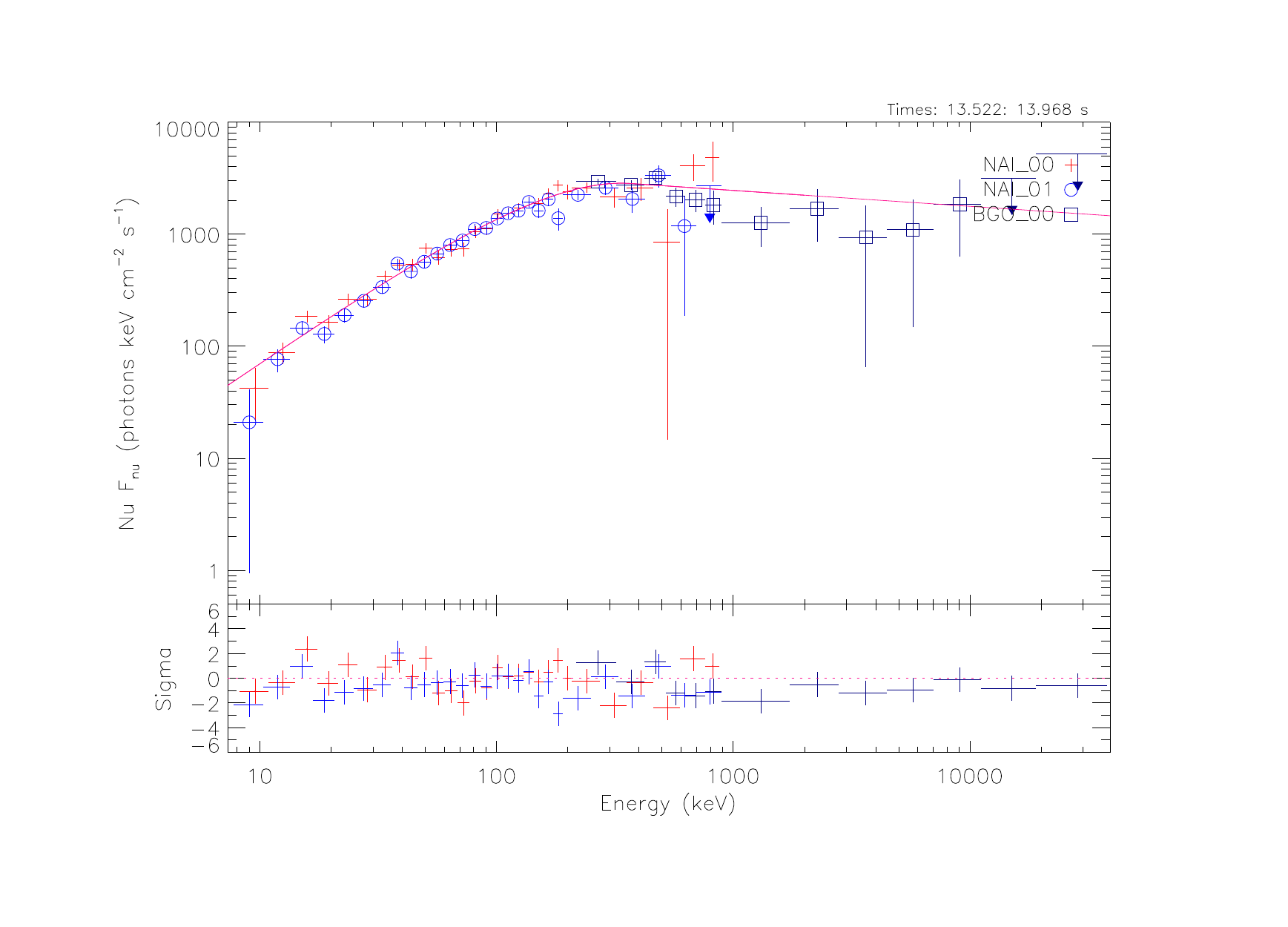}}
\resizebox{8cm}{!}{\includegraphics{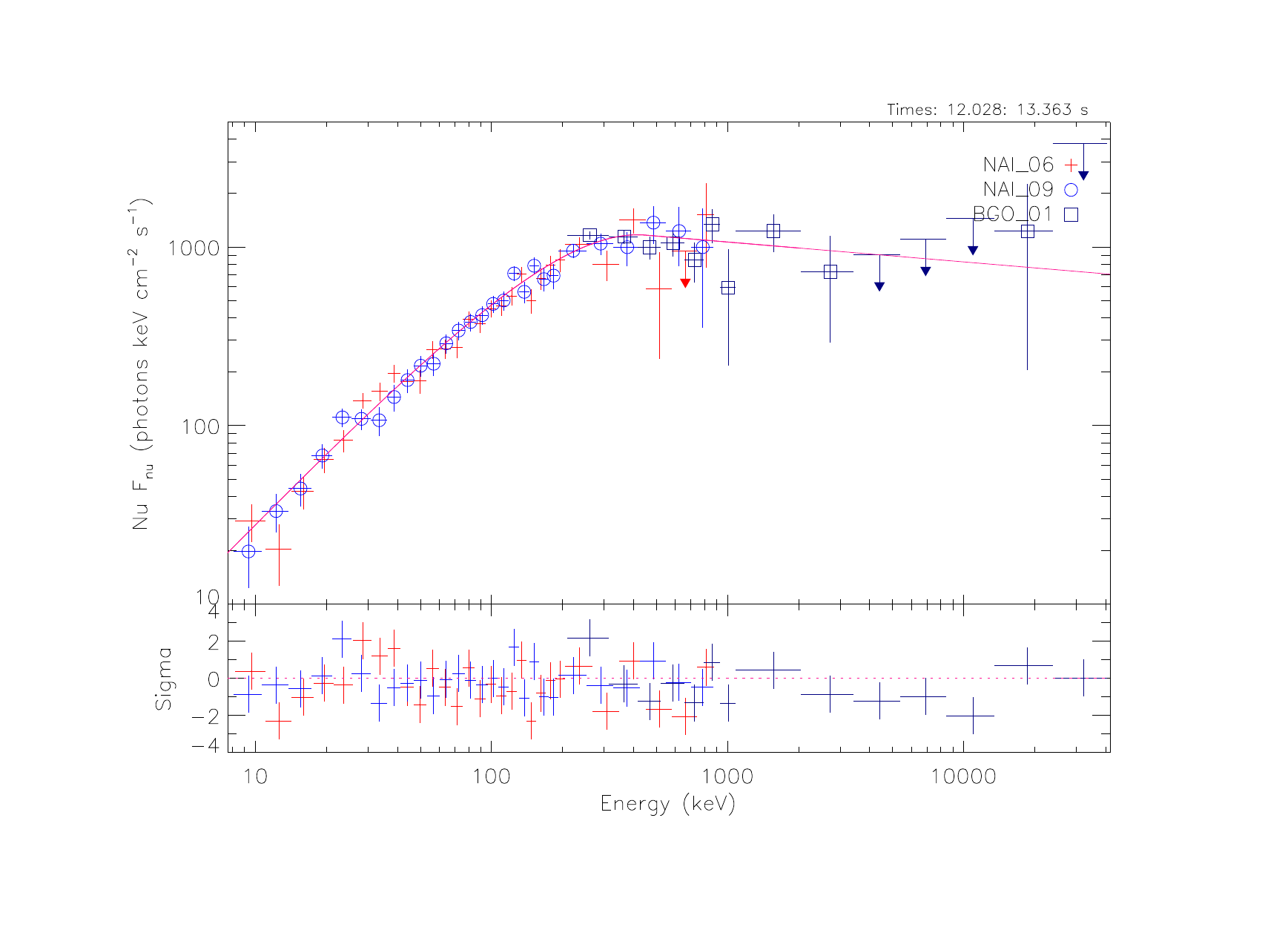}}
\resizebox{8cm}{!}{\includegraphics{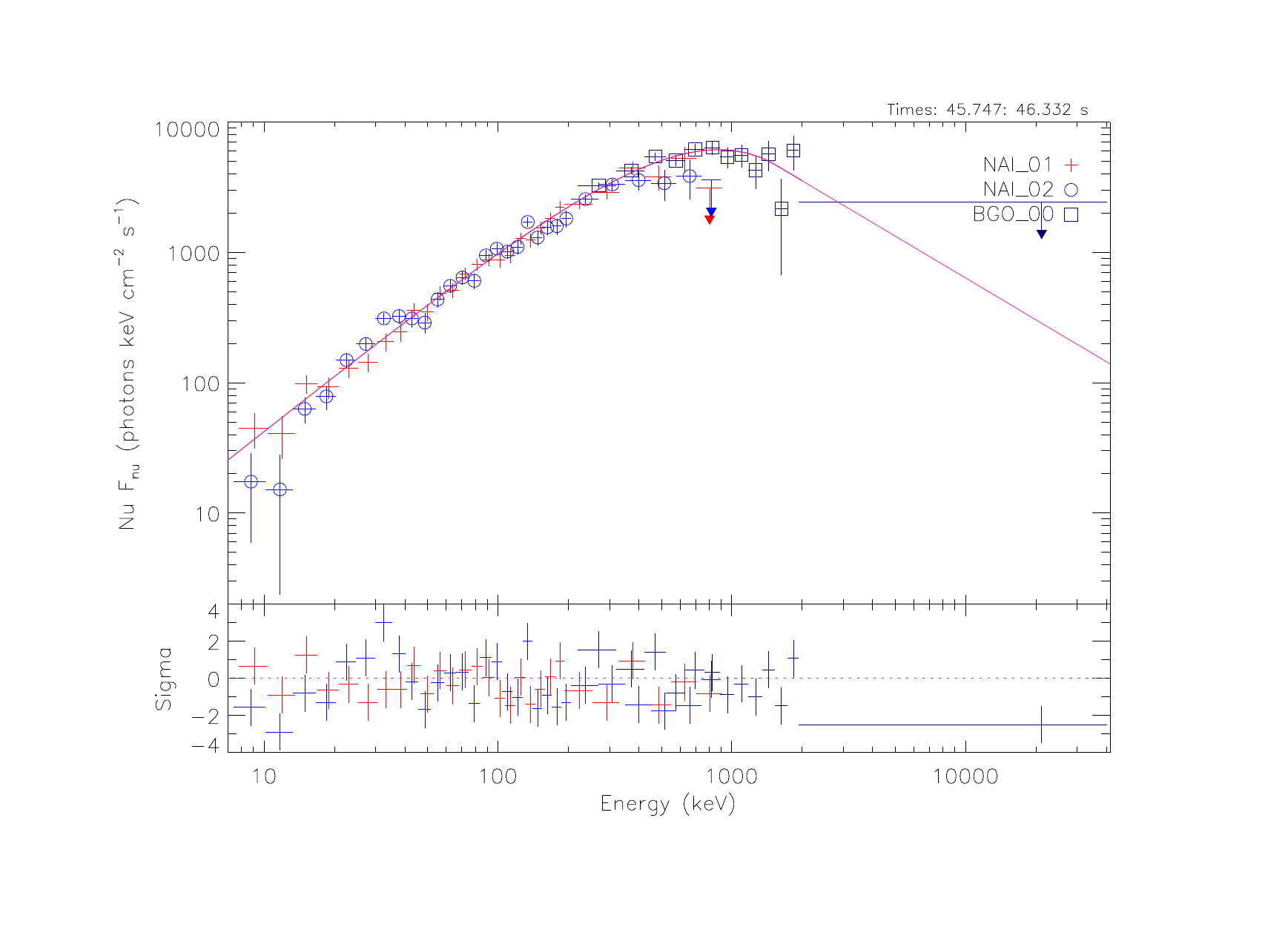}}
\resizebox{8cm}{!}{\includegraphics{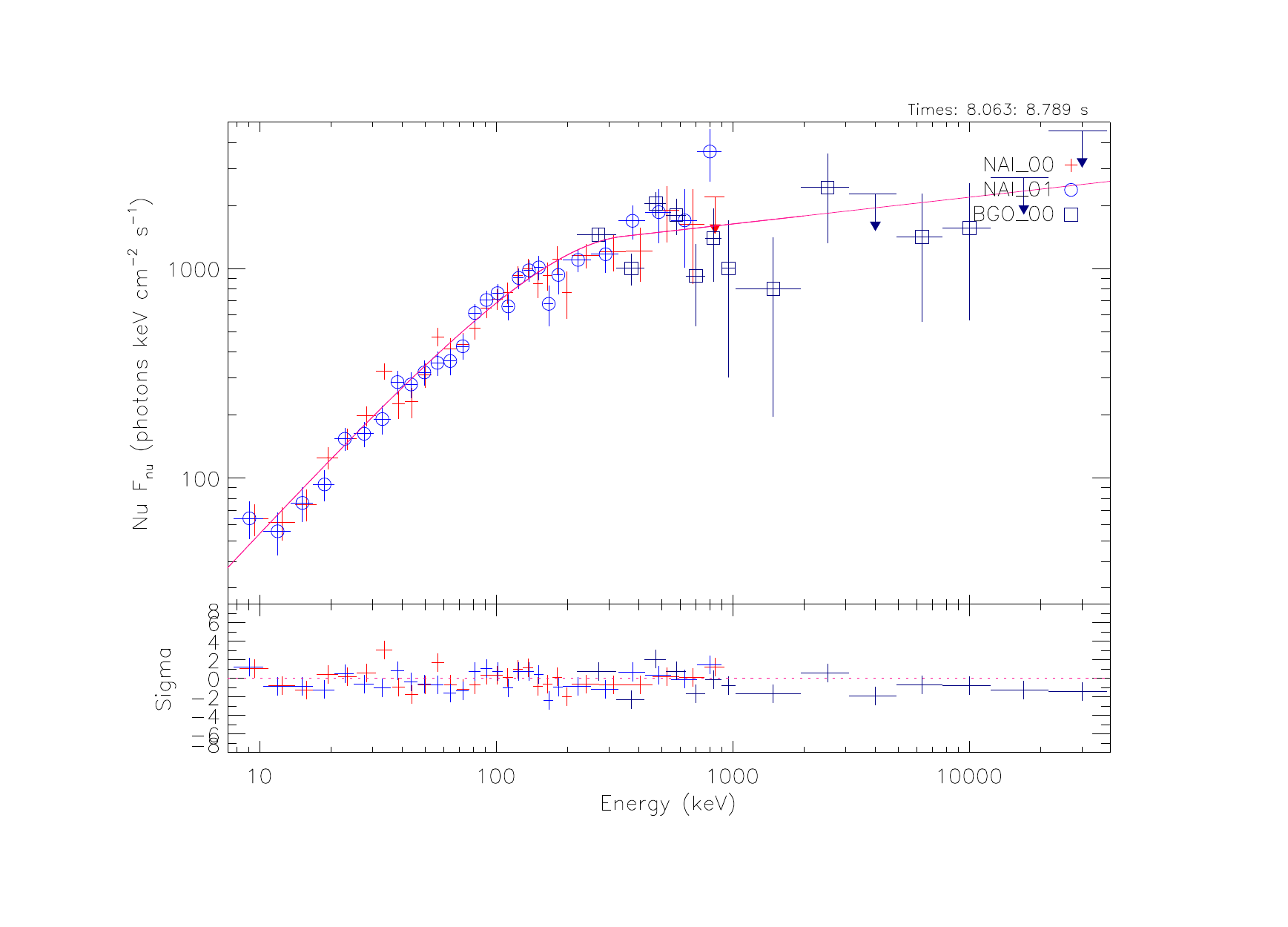}} 
\resizebox{8cm}{!}{\includegraphics{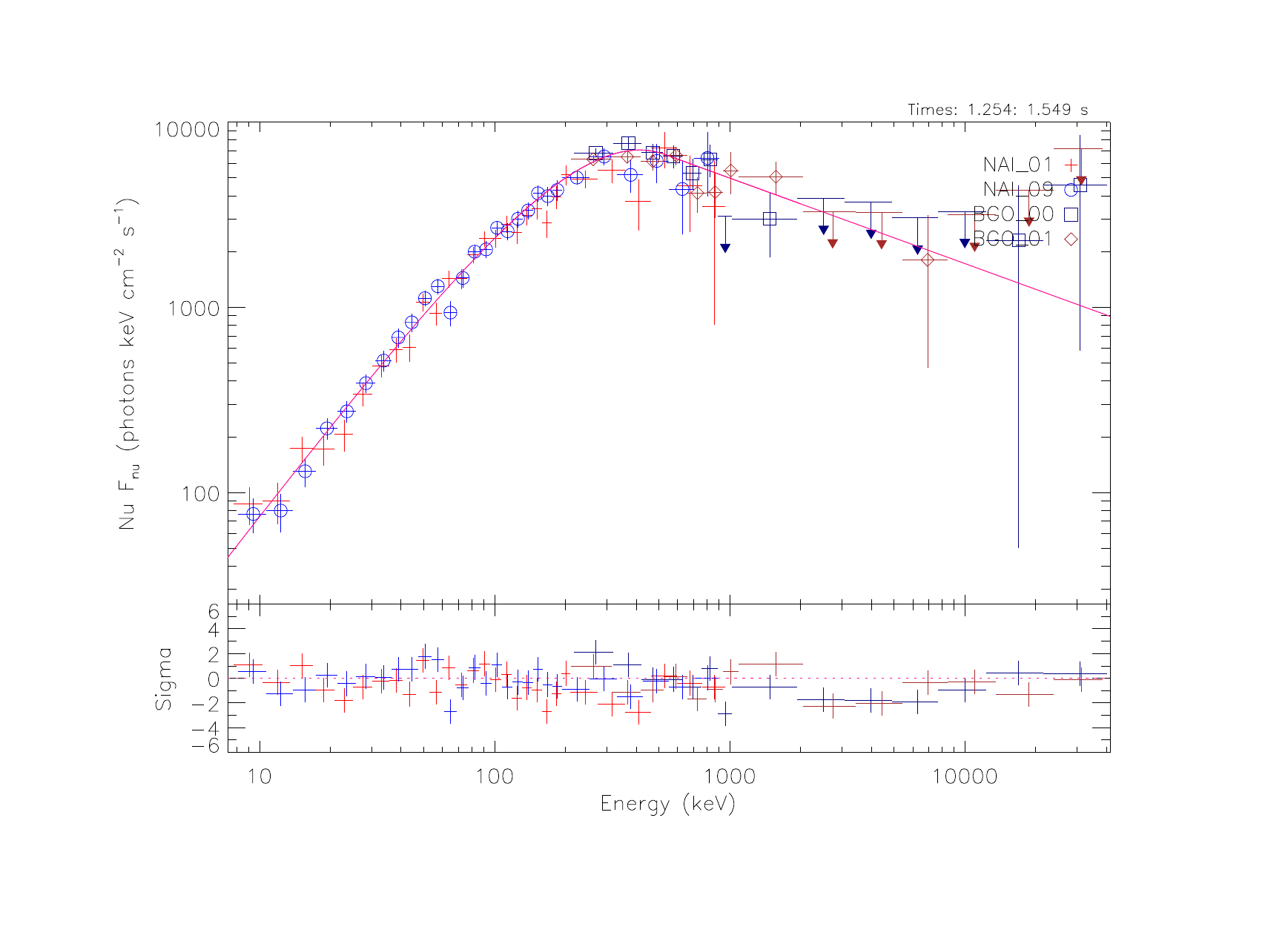}}
\resizebox{8cm}{!}{\includegraphics{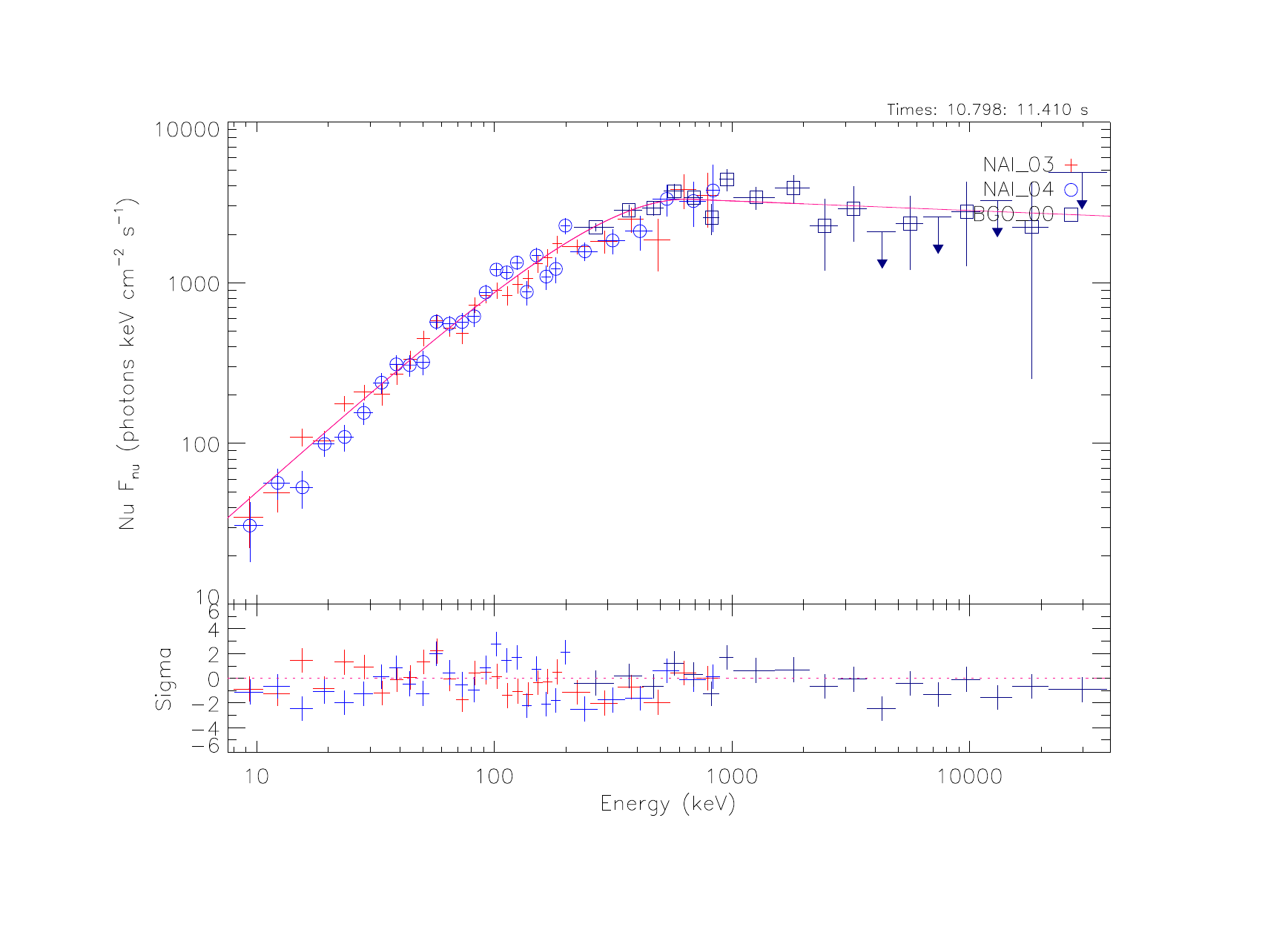}}
\resizebox{8cm}{!}{\includegraphics{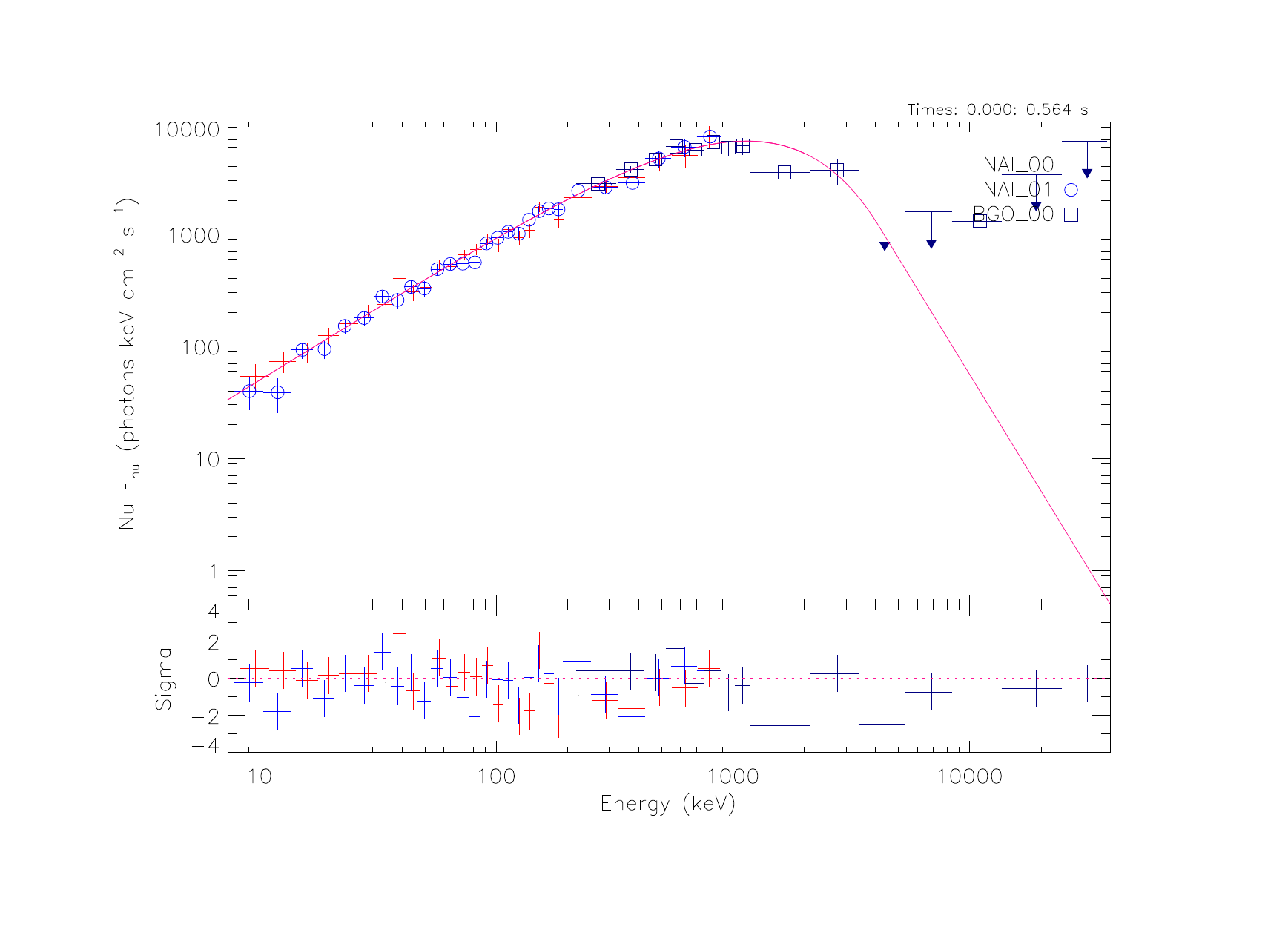}}
\resizebox{8cm}{!}{\includegraphics{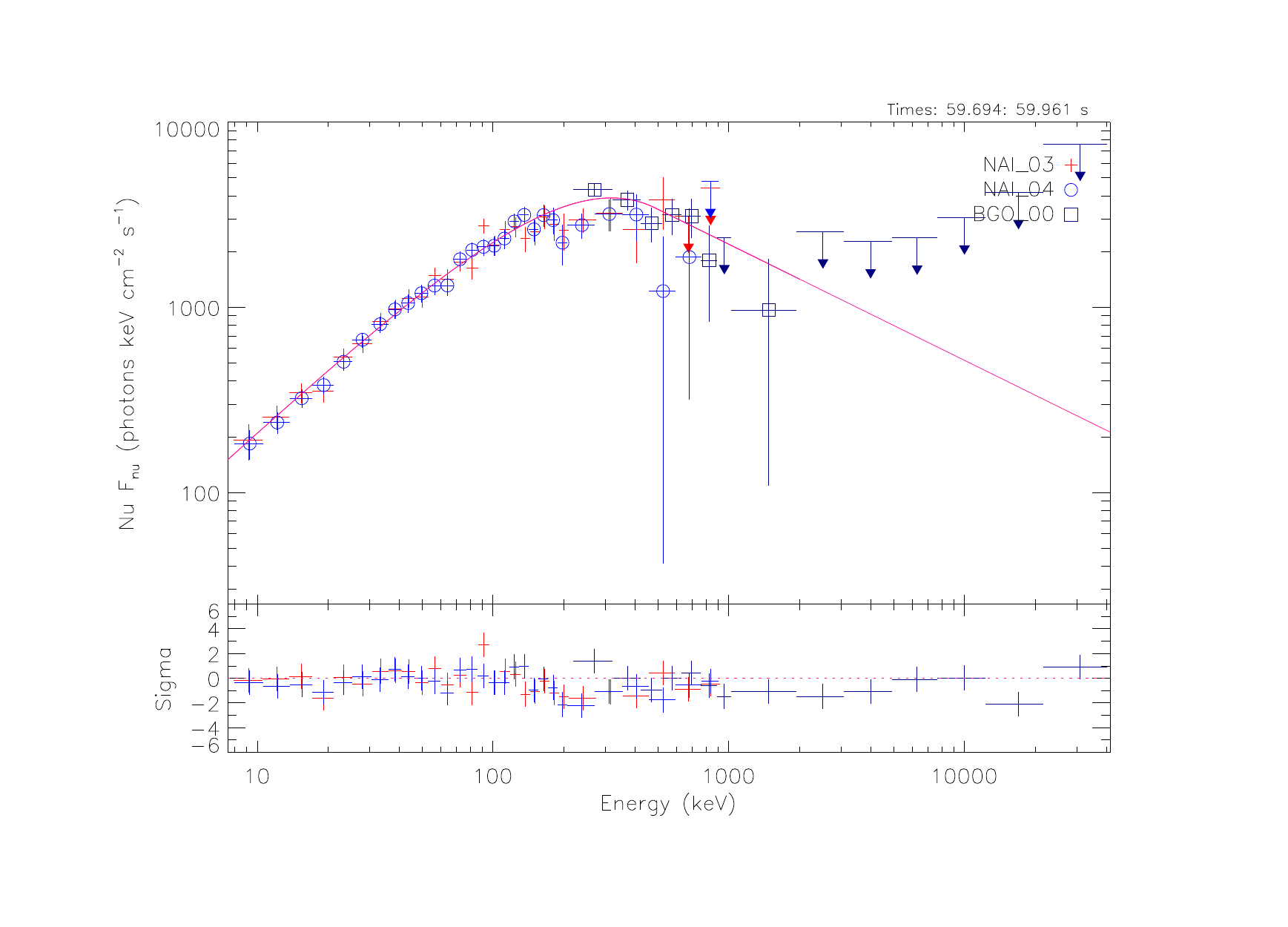}}
\caption{\it-continued}
\end{figure}

\addtocounter{figure}{-1}
\begin{figure}
\centering 
\resizebox{8cm}{!}{\includegraphics{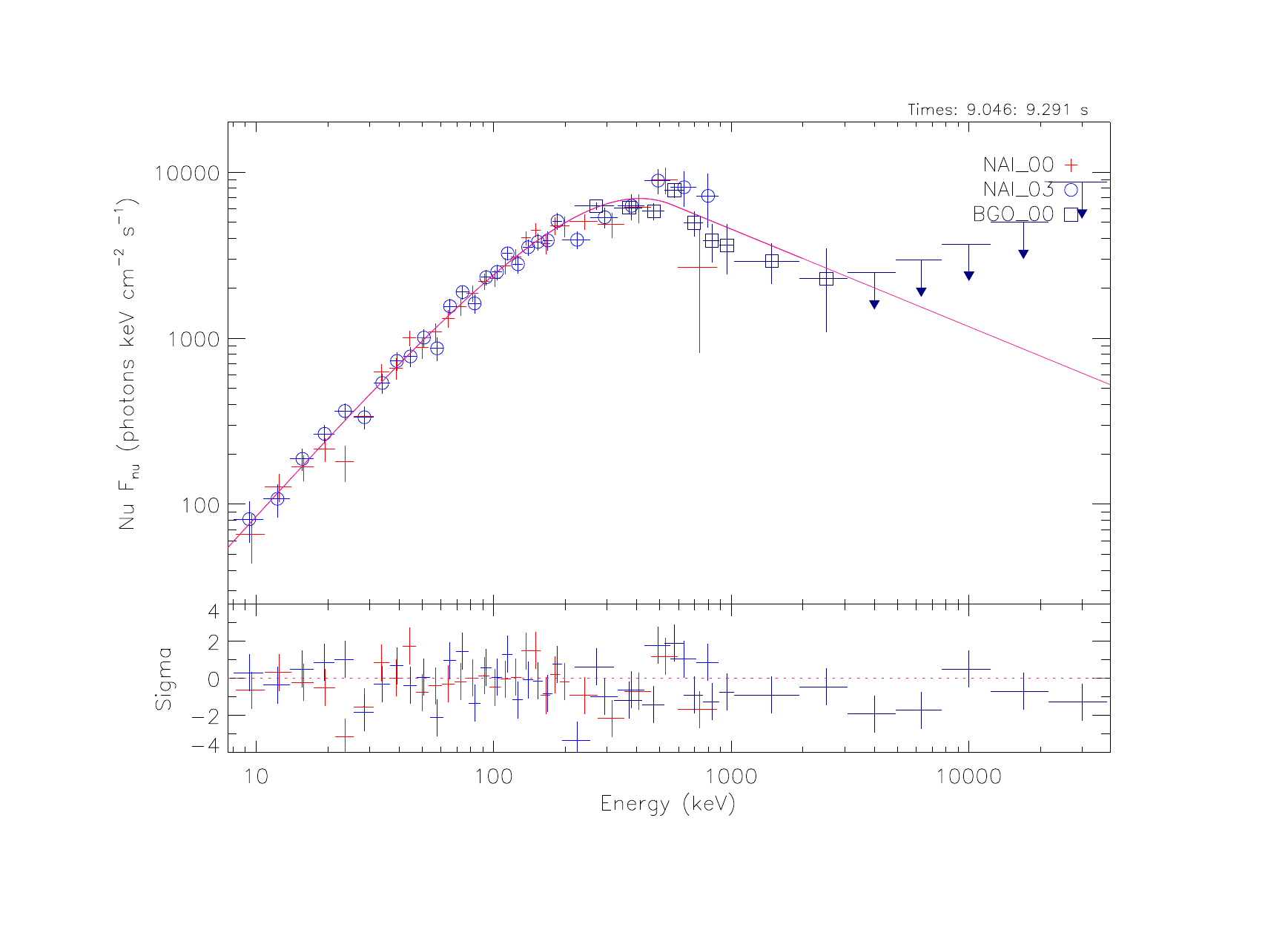}}
\resizebox{8cm}{!}{\includegraphics{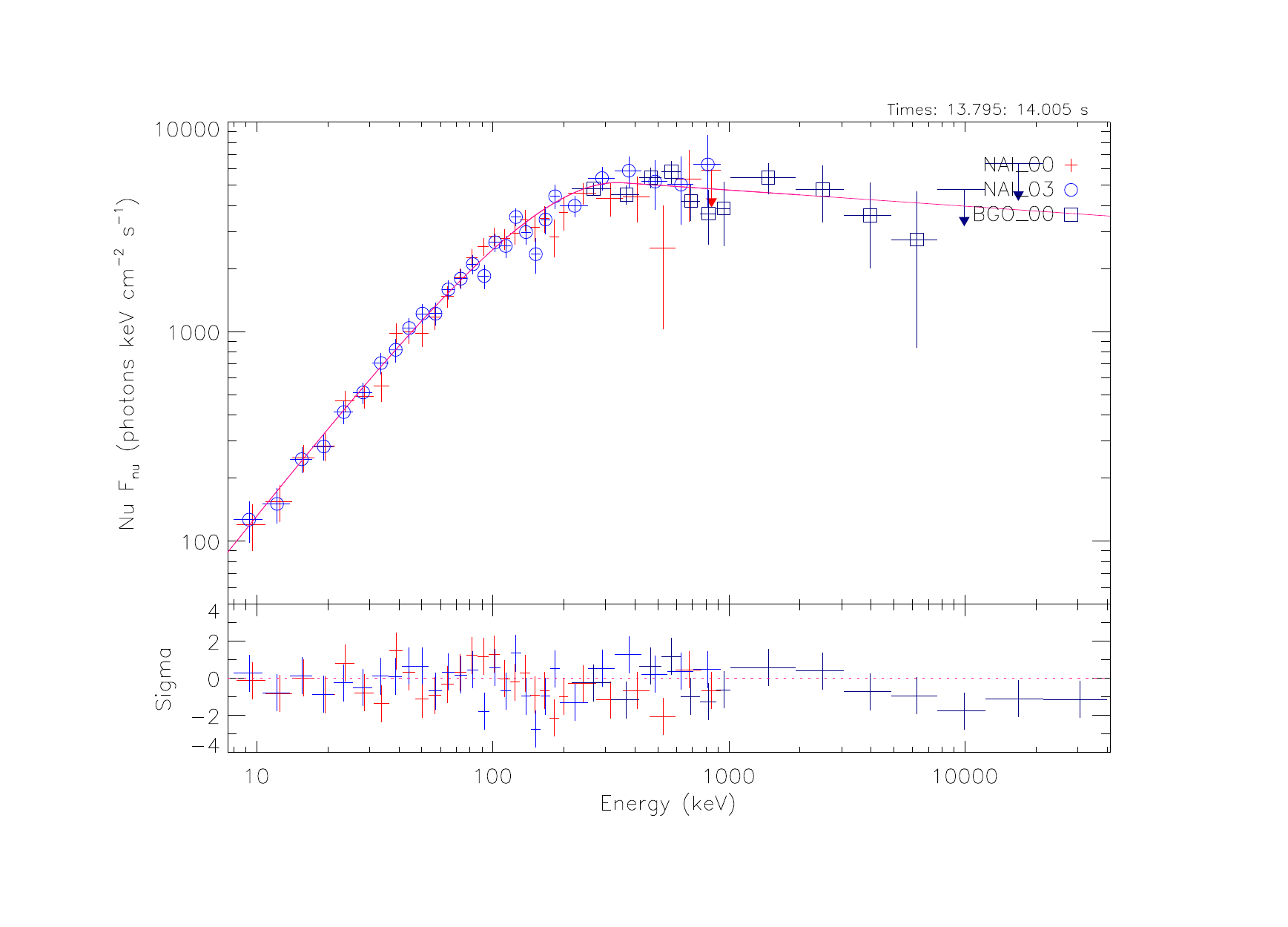}}
\resizebox{8cm}{!}{\includegraphics{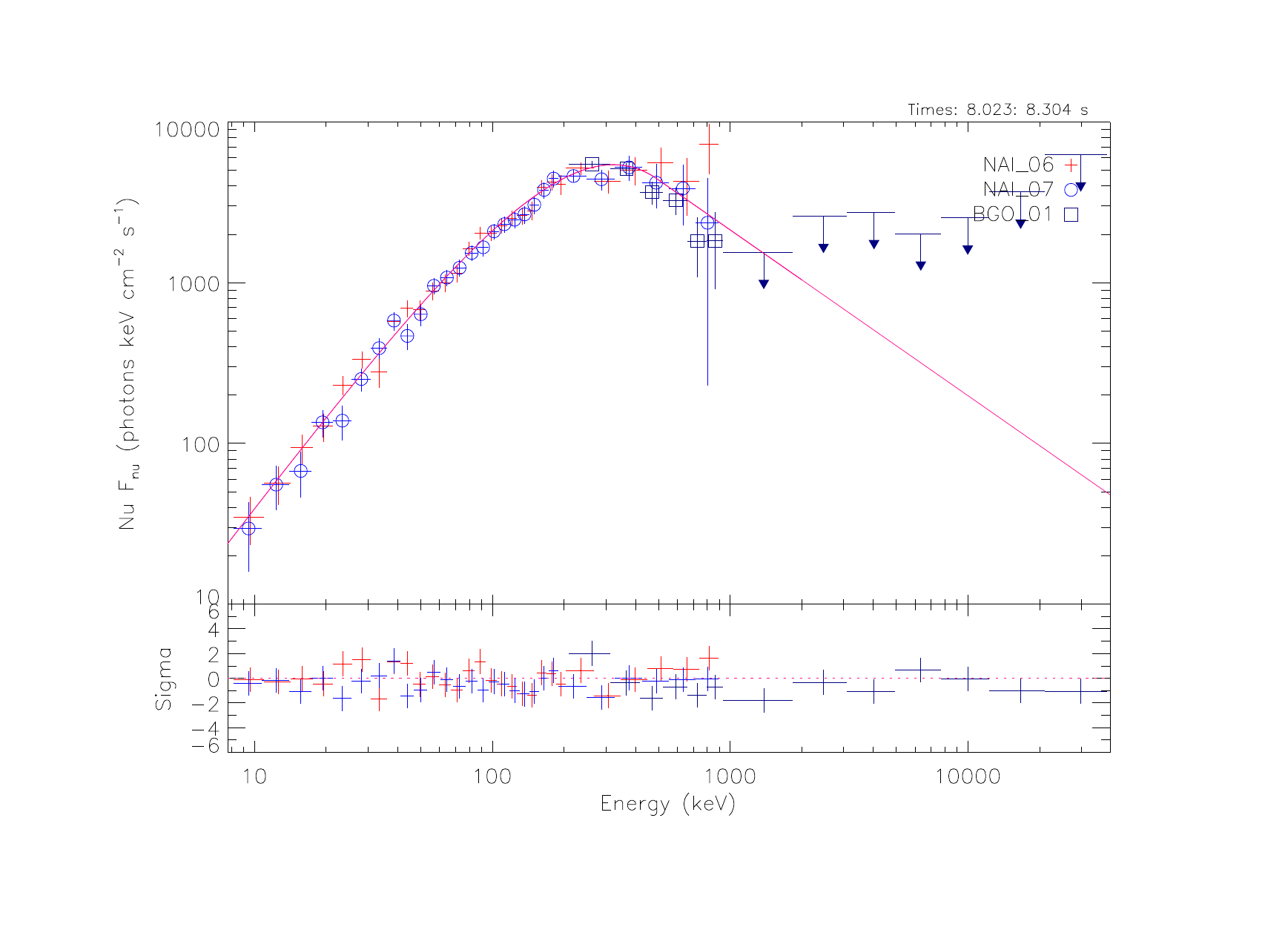}}
\resizebox{8cm}{!}{\includegraphics{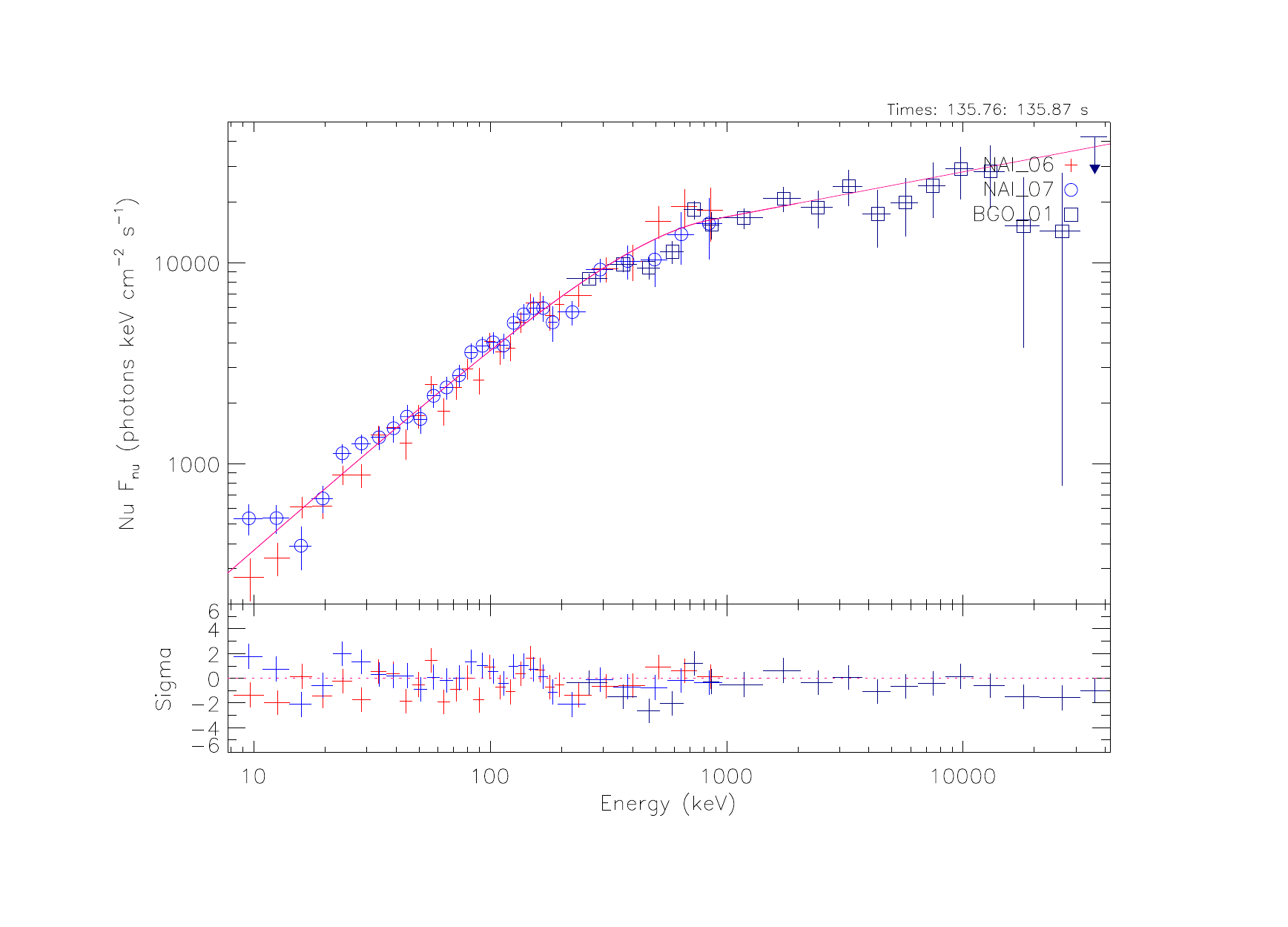}} 
\resizebox{8cm}{!}{\includegraphics{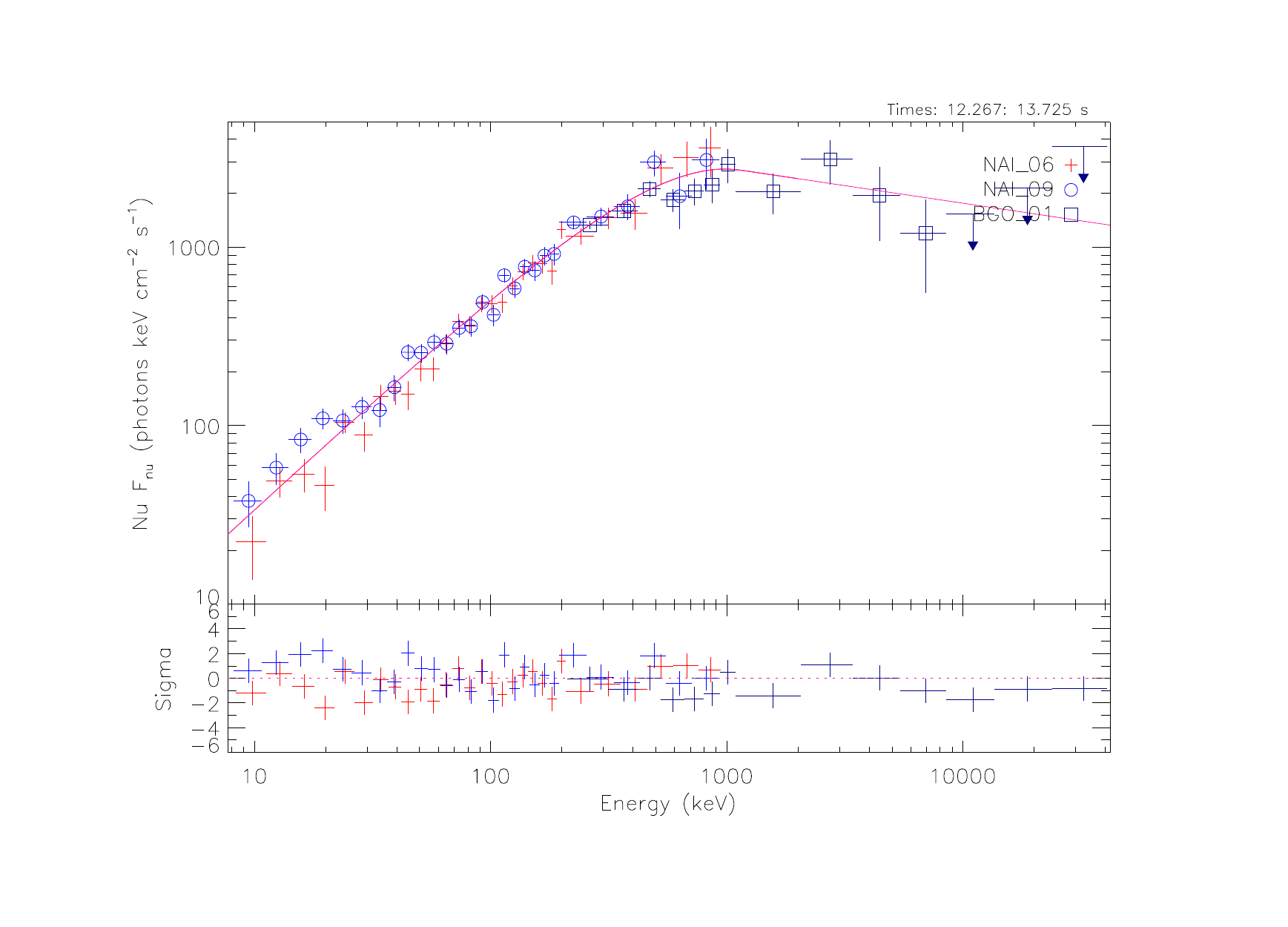}}
\resizebox{8cm}{!}{\includegraphics{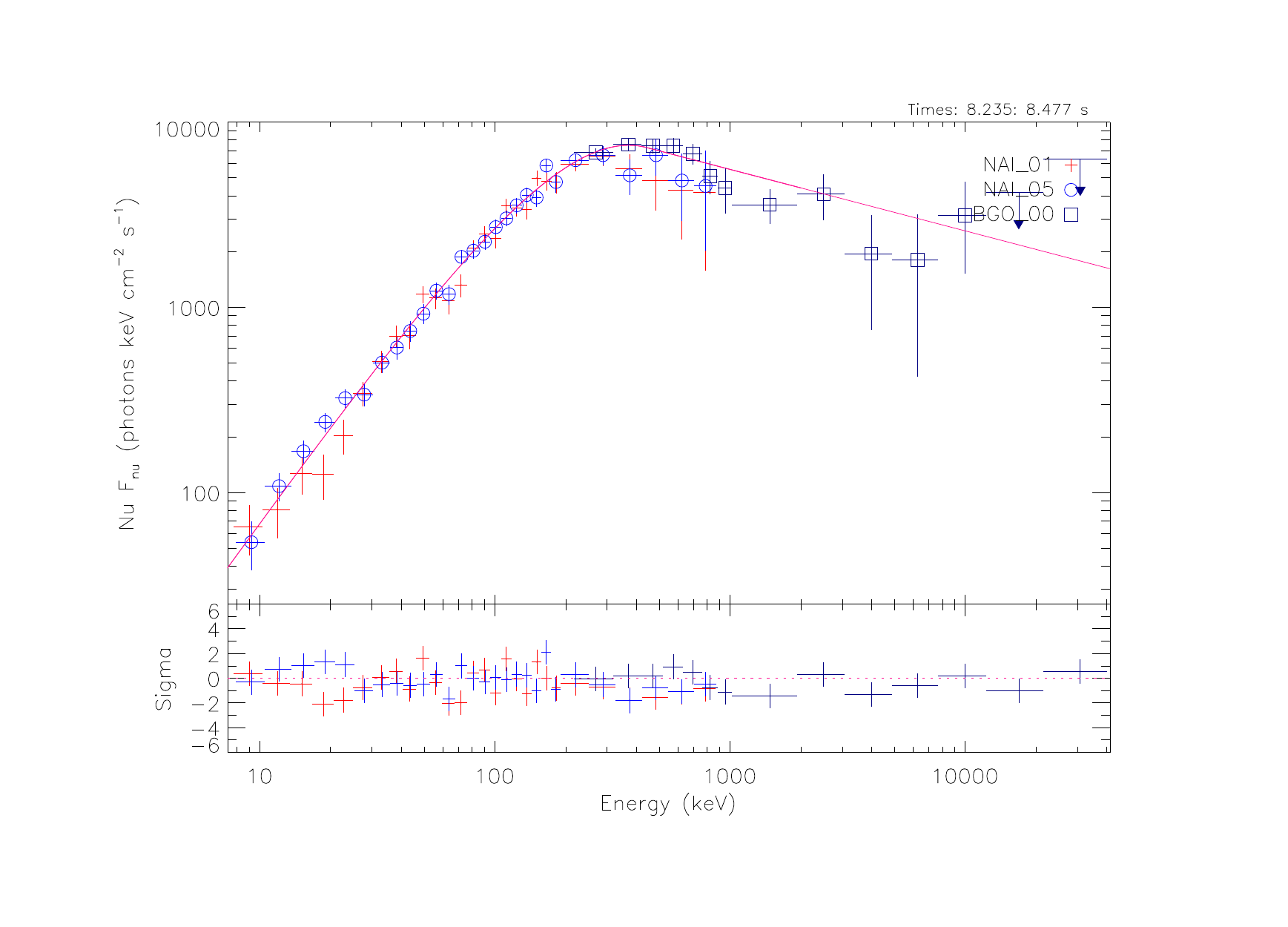}}
\resizebox{8cm}{!}{\includegraphics{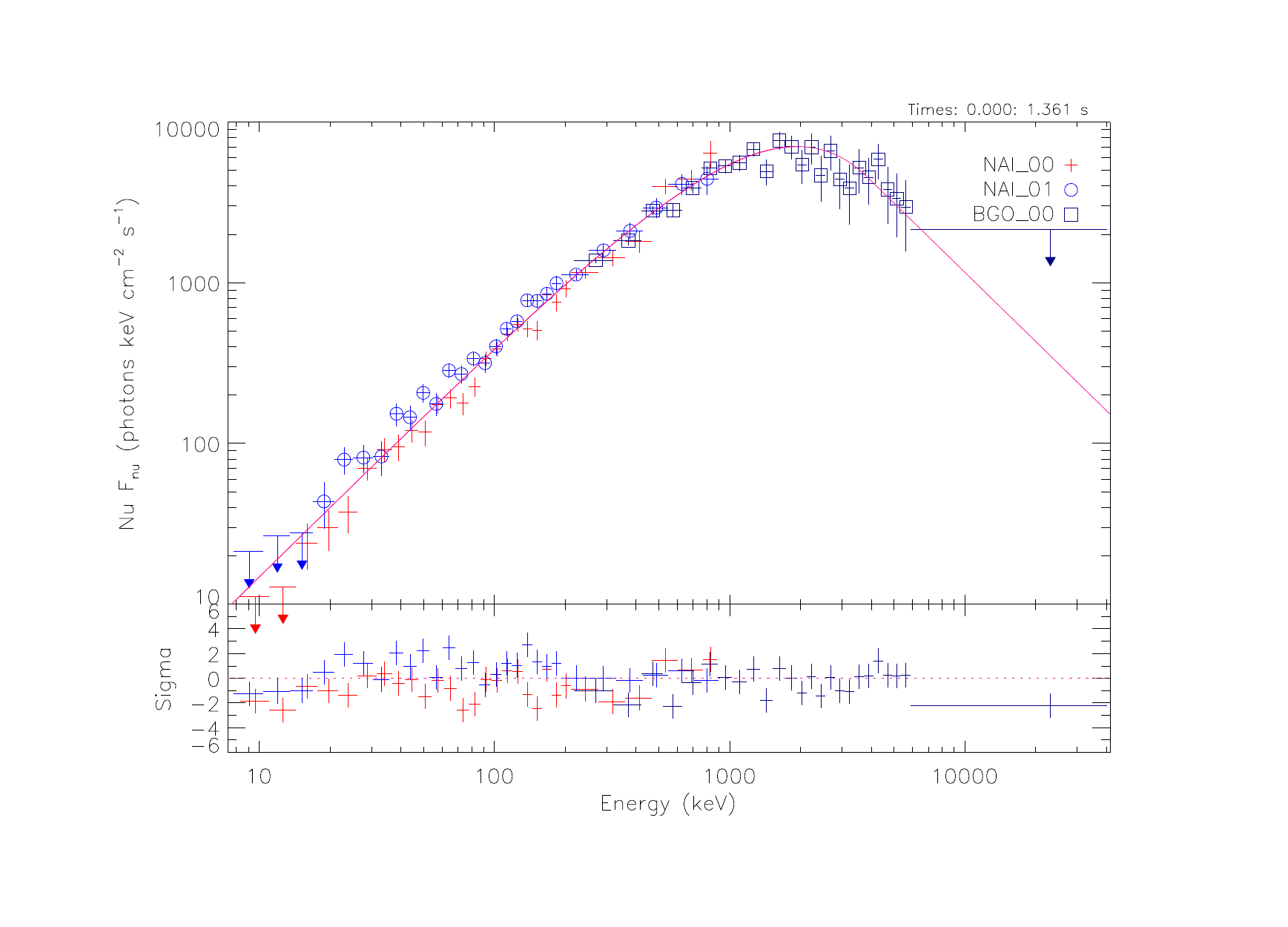}}
\resizebox{8cm}{!}{\includegraphics{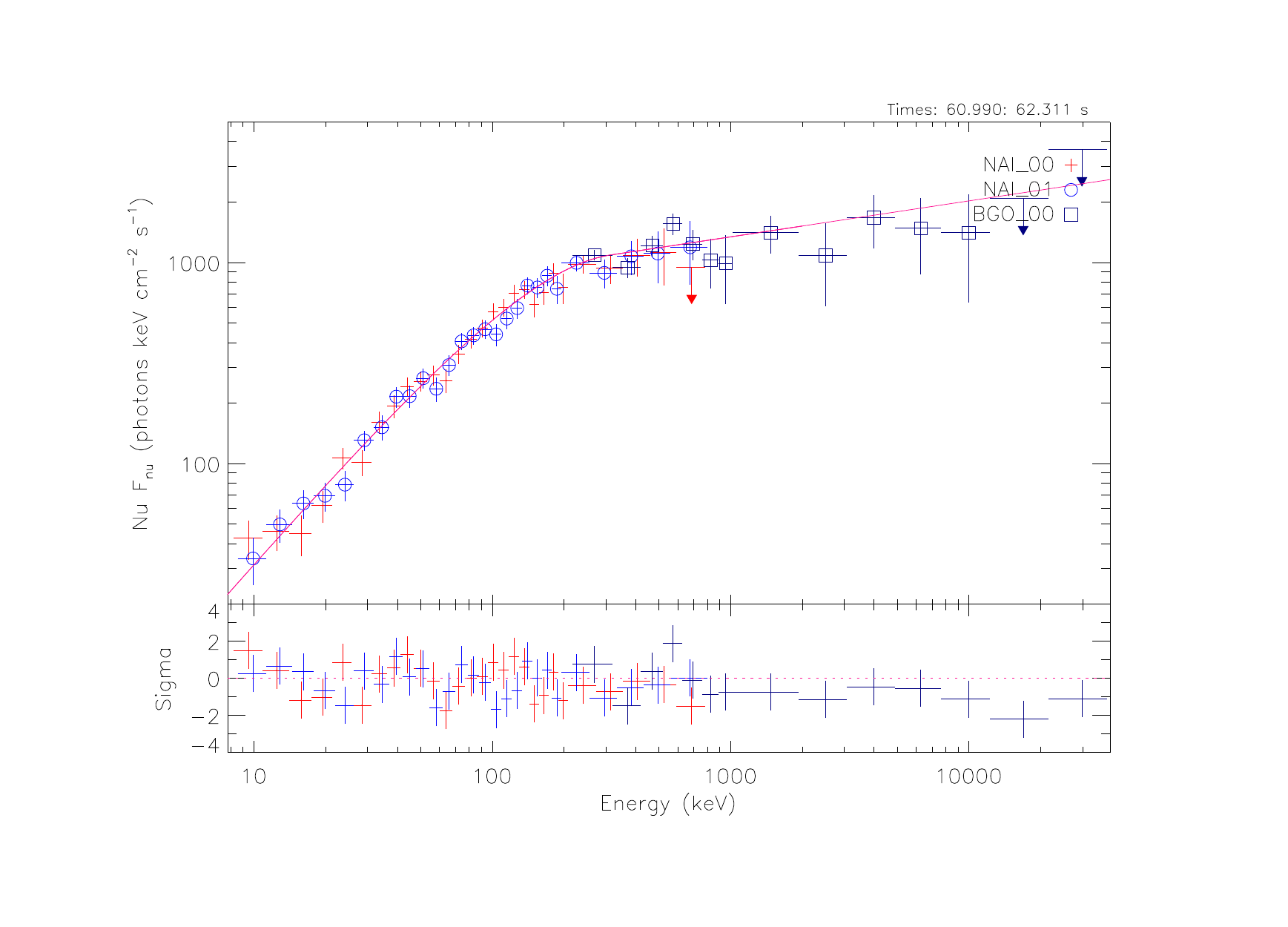}}
\caption{\it-continued}
\end{figure}

\addtocounter{figure}{-1}
\begin{figure}
\centering 
\resizebox{8cm}{!}{\includegraphics{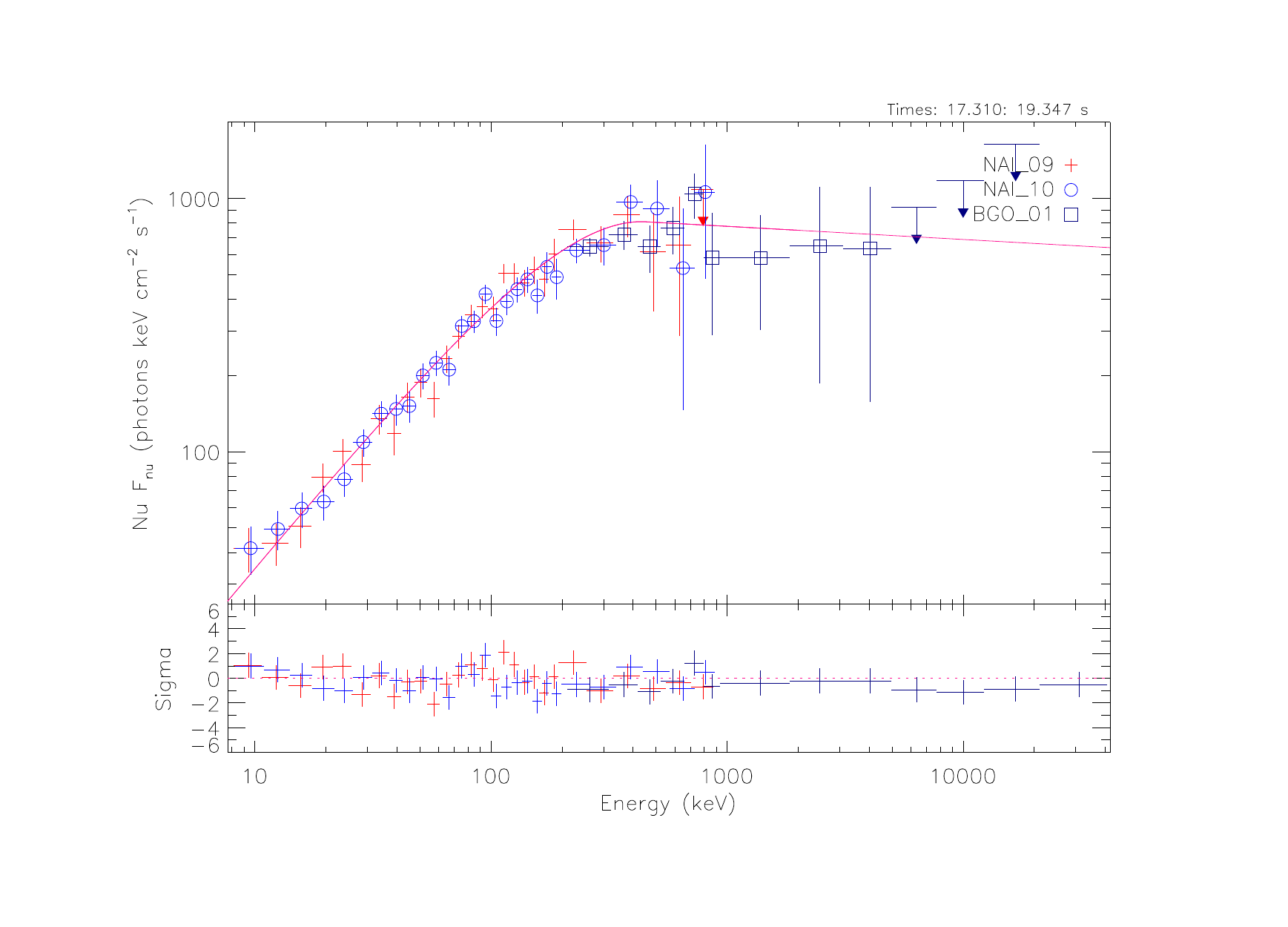}}
\resizebox{8cm}{!}{\includegraphics{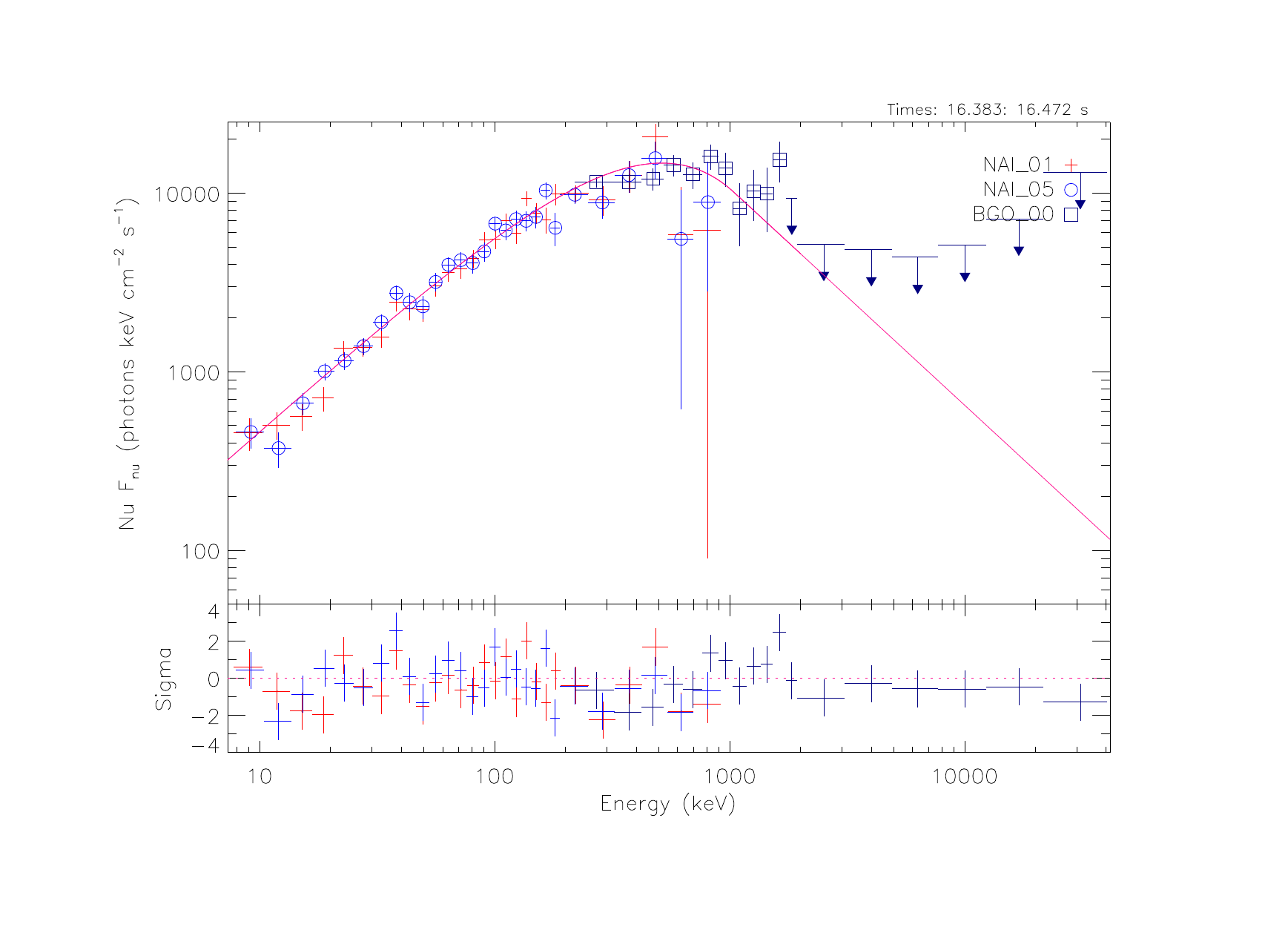}}
\resizebox{8cm}{!}{\includegraphics{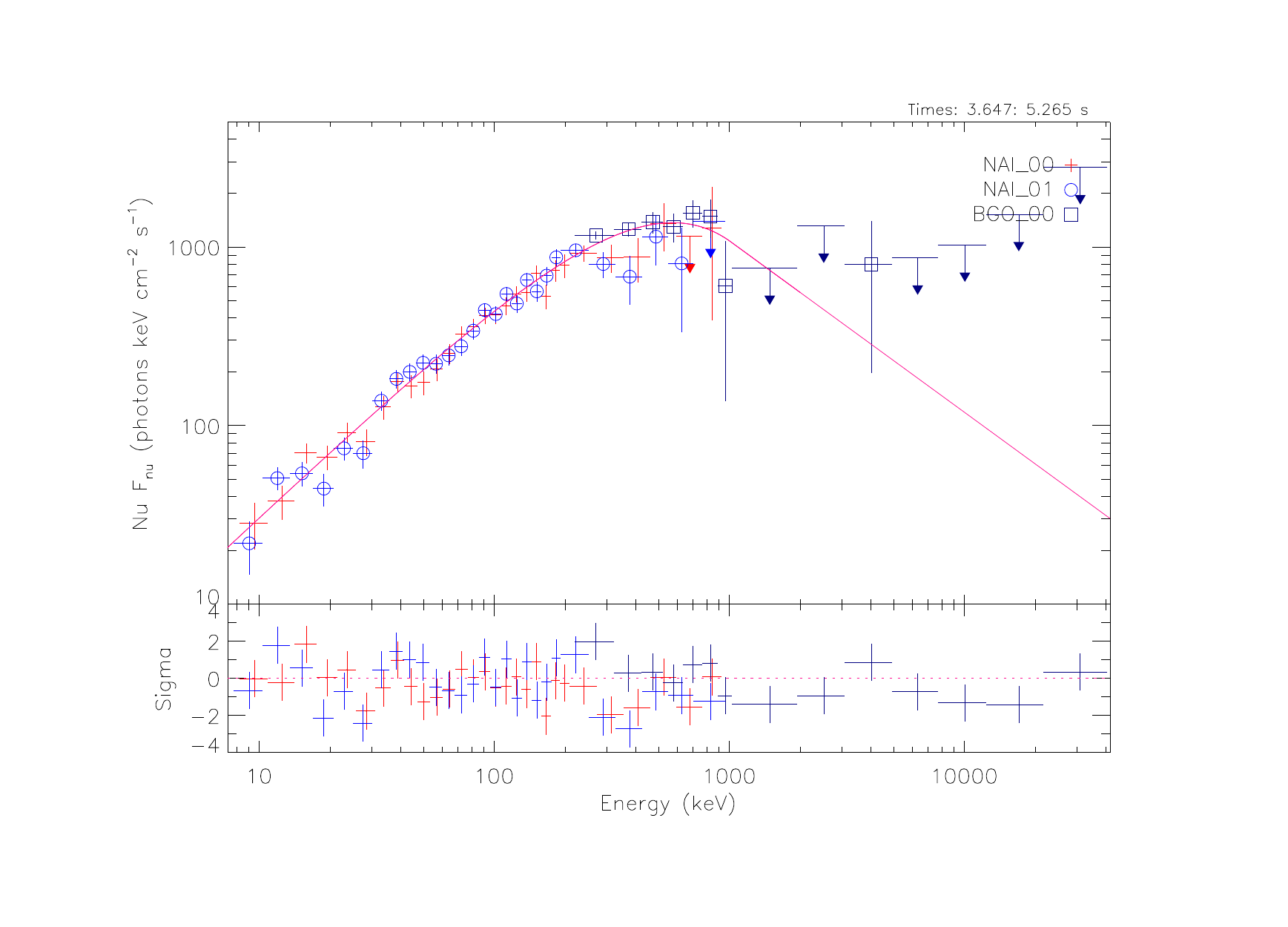}}
\resizebox{8cm}{!}{\includegraphics{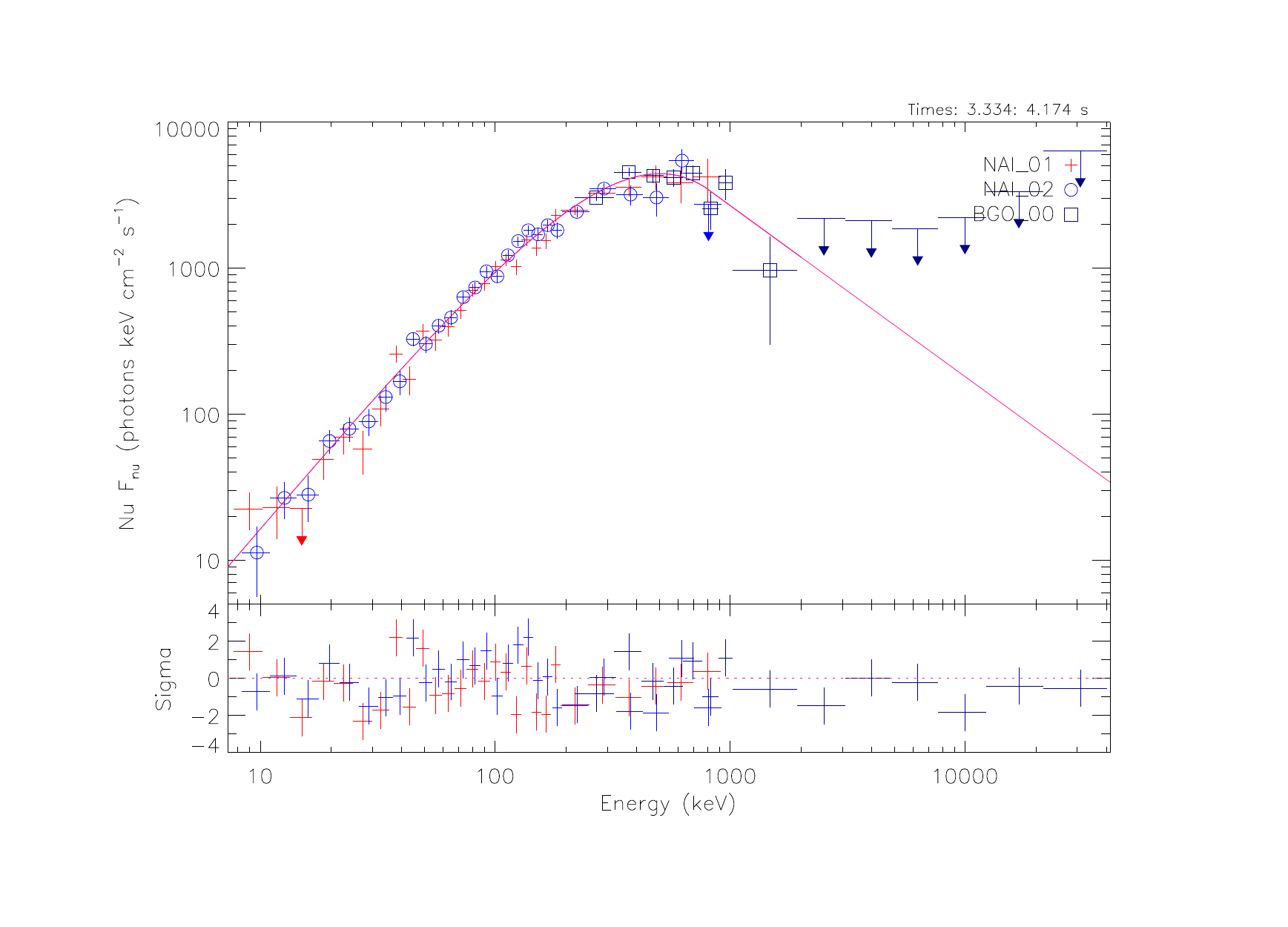}} 
\caption{\it-continued}
\end{figure}

\begin{deluxetable}{cccccc}
\renewcommand\tabcolsep{12.0pt}
\tablecaption{Results of the Time-resolved Spectral Fits at Peak Flux for All Samples \label{tab:peak_flux_results}}
\tablehead{
\colhead{GRB} 
&\colhead{$t_{1} \sim t_{2}$}
& \colhead{$\alpha$}
& \colhead{$\beta$}
& \colhead{$E_{p}$}
& \colhead{Red.$\chi^{2}$}\\
\colhead{}& \colhead{(s)}& \colhead{}& \colhead{} & 
\colhead{(keV)} & \colhead{}
}
\colnumbers
\startdata
080825C&2.978$\sim$3.937&-0.4269$\pm$0.0924&-2.105$\pm$0.102&205.1$\pm$19.5&0.96\\
090328A&23.705$\sim$25.400&-0.9062$\pm$0.0500&-2.220$\pm$0.192&444.0$\pm$57.2&1.15\\
090626A&34.580$\sim$35.053&-0.7057$\pm$0.0541&-2.530$\pm$0.239&324.7$\pm$27.3&0.92\\
090926A&4.129$\sim$4.326&-0.3629$\pm$0.0699&-2.048$\pm$0.055&249.5$\pm$18.7&0.97\\
100724B&61.818$\sim$62.852&-0.6834$\pm$0.0469&-1.936$\pm$0.060&517.2$\pm$52.6&1.09\\
100826A&20.799$\sim$21.574&-0.7023$\pm$0.0461&-2.033$\pm$0.072&536.0$\pm$53.3&1.12\\
101014A&0.961$\sim$1.288&-0.4757$\pm$0.0542&-2.334$\pm$0.101&281.6$\pm$18.0&1.02\\
110721A&0.889$\sim$1.660&-0.8542$\pm$0.0321&-2.111$\pm$0.095&1236.0$\pm$145&1.18\\
120226A&17.503$\sim$19.860&-0.7359$\pm$0.0857&-1.805$\pm$0.063&238.4$\pm$37.3&1.04\\
120624B&11.963$\sim$14.037&-0.9411$\pm$0.0443&-2.174$\pm$0.172&611.3$\pm$85.0&0.88\\
130502B&20.322$\sim$20.586&-0.1871$\pm$0.0530&-2.829$\pm$0.199&320.3$\pm$15.0&0.95\\
130504C&31.005$\sim$31.342&-0.8189$\pm$0.0500&-1.938$\pm$0.070&705.9$\pm$97.9&1.00\\
130518A&25.899$\sim$26.280&-0.8515$\pm$0.0394&-2.172$\pm$0.075&567.6$\pm$51.3&0.97\\
130821A&30.039$\sim$30.936&-0.6272$\pm$0.0733&-1.898$\pm$0.055&246.9$\pm$27.3&0.99\\
131108A&0.000$\sim$1.257&-0.6219$\pm$0.0672&-1.871$\pm$0.040&341.0$\pm$34.6&0.98\\
140102A&2.281$\sim$2.635&-0.6150$\pm$0.0710&-2.099$\pm$0.075&223.4$\pm$20.5&0.93\\
140206B&13.522$\sim$13.968&-0.5438$\pm$0.0569&-2.142$\pm$0.079&336.6$\pm$26.7&1.02\\
141028A&12.028$\sim$13.363&-0.6414$\pm$0.0555&-2.111$\pm$0.103&416.2$\pm$40.0&0.97\\
150118B&45.747$\sim$46.332&-0.5728$\pm$0.0329&-3.067$\pm$0.316&881.3$\pm$53.4&0.96\\
150202B&8.063$\sim$8.789&-0.7736$\pm$0.0612&-1.872$\pm$0.070&383.2$\pm$53.8&1.07\\
150314A&1.254$\sim$1.549&-0.3399$\pm$0.0448&-2.462$\pm$0.088&413.5$\pm$19.8&1.03\\
150403A&10.798$\sim$11.410&-0.6775$\pm$0.0418&-2.059$\pm$0.074&639.9$\pm$59.6&1.11\\
150510A&0.000$\sim$0.564&-0.6889$\pm$0.0275&unconstrained&1141.0$\pm$65.9&0.97\\
150627A&59.694$\sim$59.961&-0.8258$\pm$0.0441&-2.627$\pm$0.228&317.8$\pm$24.3&0.87\\
150902A&9.046$\sim$9.291&-0.3920$\pm$0.0471&-2.587$\pm$0.142&411.5$\pm$22.7&0.98\\
160509A&13.795$\sim$14.005&-0.5605$\pm$0.0573&-2.077$\pm$0.069&336.7$\pm$28.7&0.91\\
160816A&8.023$\sim$8.304&-0.0321$\pm$0.0625&-3.032$\pm$0.287&322.8$\pm$15.0&0.91\\
160821A&135.76$\sim$135.87&-0.9698$\pm$0.0376&-1.776$\pm$0.054&1093.0$\pm$192.0&1.12\\
160905A&12.267$\sim$13.725&-0.7799$\pm$0.0423&-2.197$\pm$0.113&987.2$\pm$120.0&1.15\\
160910A&8.235$\sim$8.477&-0.2183$\pm$0.0540&-2.332$\pm$0.072&370.8$\pm$19.8&0.94\\
170115B&0.000$\sim$1.361&-0.5548$\pm$0.0284&-3.430$\pm$0.423&1931.0$\pm$102.0&1.04\\
170214A&60.990$\sim$62.311&-0.6362$\pm$0.0650&-1.821$\pm$0.050&360.1$\pm$41.8&0.96\\
170510A&17.310$\sim$19.347&-0.8697$\pm$0.0543&-2.052$\pm$0.121&433.2$\pm$57.5&0.91\\
170808B&16.383$\sim$16.472&-0.8287$\pm$0.0341&-3.215$\pm$0.447&514.0$\pm$33.0&0.91\\
171210A&3.647$\sim$5.265&-0.7582$\pm$0.0415&-2.960$\pm$0.658&572.5$\pm$49.8&0.96\\
180305A&3.334$\sim$4.174&-0.0916$\pm$0.0525&-3.172$\pm$0.461&502.8$\pm$24.1&1.00\\
\enddata
\end{deluxetable}

\edit1{\added{
We have extracted the maximal value of $\alpha$ after performing the detailed time-resolved spectral analysis for each burst (Figure \ref{fig:a_max}). The fact that most of them ($77.8\%$) in our sample are larger than the synchrotron limit which is the value of $-\frac{2}{3}$ is amazing. Historically, one thought that the fitted spectrum can't be produced by synchrotron emission when the spectral slope $\alpha\geq -\frac{2}{3}$. However, the recent study in \citet{2019NatAs.tmp..471B} showed that the synchrotron model can fit most of the bursts with Band $\alpha$ parameter harder than the line-of-death limit. Additionally, \citet{2013MNRAS.428.2430L} pointed out that some structured jet photosphere models can also account for slopes softer than $-\frac{2}{3}$ even though in the majority of the cases it is not easy to do so \citep{2014ApJ...785..112D}. \citet{2014ApJ...784...17B} illustrated that the Band function can't be representative of a non-thermal synchrotron emission component because of the blackbody component will be more significant when a physical synchrotron model was used to perform the spectral fitting analysis instead of the Band function. Based on the above, it seems difficult to identify whether they originated from the synchrotron emission or photosphere model. As well as, it is difficult to address the question whether the thermal component was detected in each burst. Maybe, the spectral information at peak flux is representative among all the time-resolved spectra. In this section, we present the spectra with the best Band-fitting results at peak flux for all of our bursts in Figure \ref{fig:peak_flux}. Correspondingly, the GRB name, the fitting interval, as well as, the fitting results such as $\alpha$, $\beta$, $E_{p}$, and the reduced $\chi^{2}$ were listed in Table \ref{tab:peak_flux_results}. Undoubtedly, a single Band function is enough to perform the spectral fitting for every burst from those fitting lines in Figure \ref{fig:peak_flux} even though there are papers argued that the blackbody component was detected in some bursts such as GRB 100724B \citep{2011ApJ...727L..33G}, GRB 110721A \citep{2012ApJ...757L..31A, 2012ApJ...758L..34Z} and so on. Additionally, we found that the maximal value of the low energy spectral index $\alpha_{max}$ in the time-resolved spectra is equal to the value of $\alpha$ around the peak flux for 7 GRBs (GRBs 080825C, 101014A, 130821A, 131108A, 140102A, 150510A, 160816A) due to the value of $\alpha$ is maximal while the peak flux is emerging. For the rest of the bursts, the maximal value of $\alpha$ is larger than the value of $\alpha$ at peak flux. Especially, the two values are greatly different for 7 GRBs (GRBs 090626A, 100826A, 141028A, 150627A, 170115B, 170808B, 171210A), the $\alpha_{max}$ is much larger than the value of $\alpha$ at peak flux for them.}}

\begin{figure}
\gridline{\fig{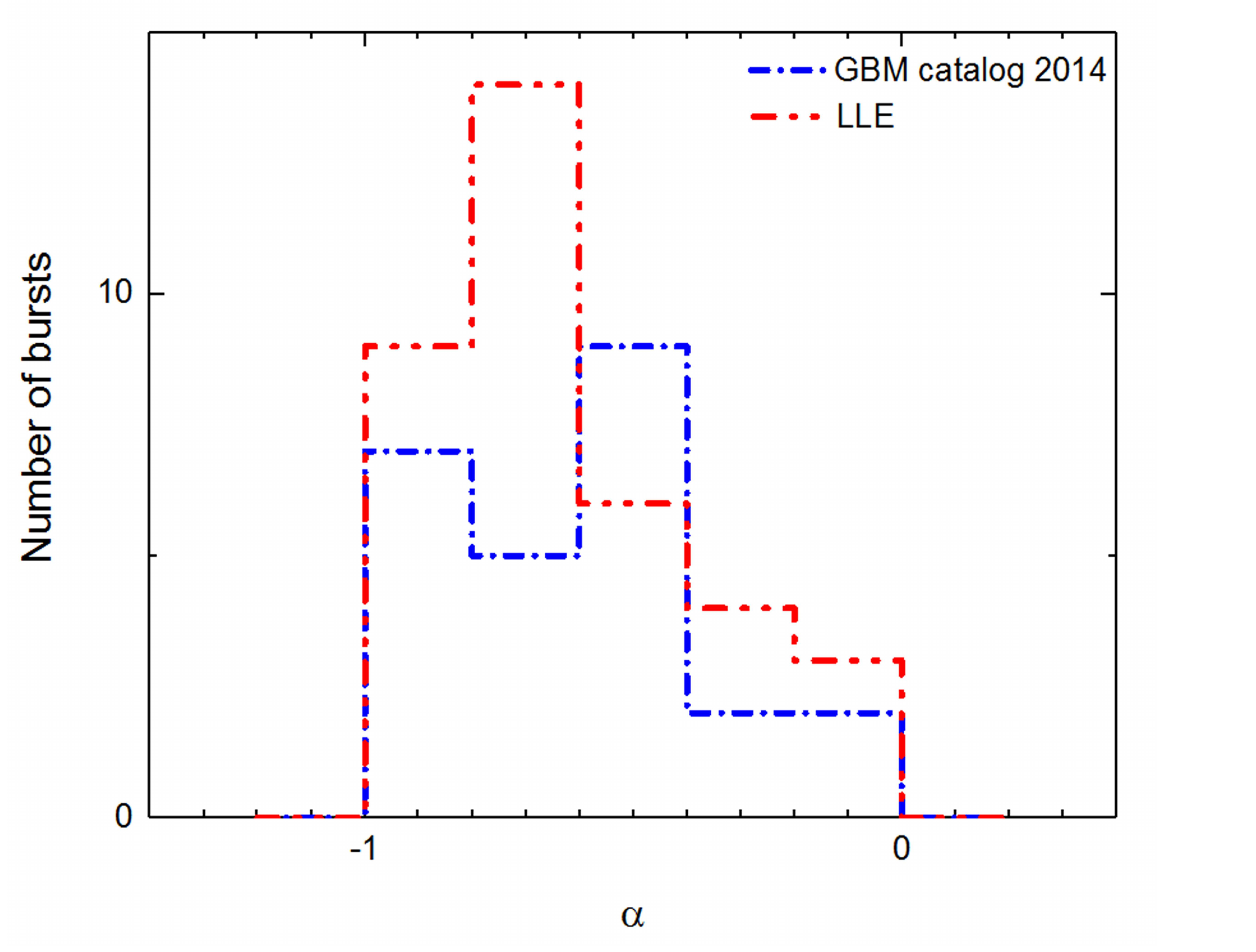}{0.5\textwidth}{}
          \fig{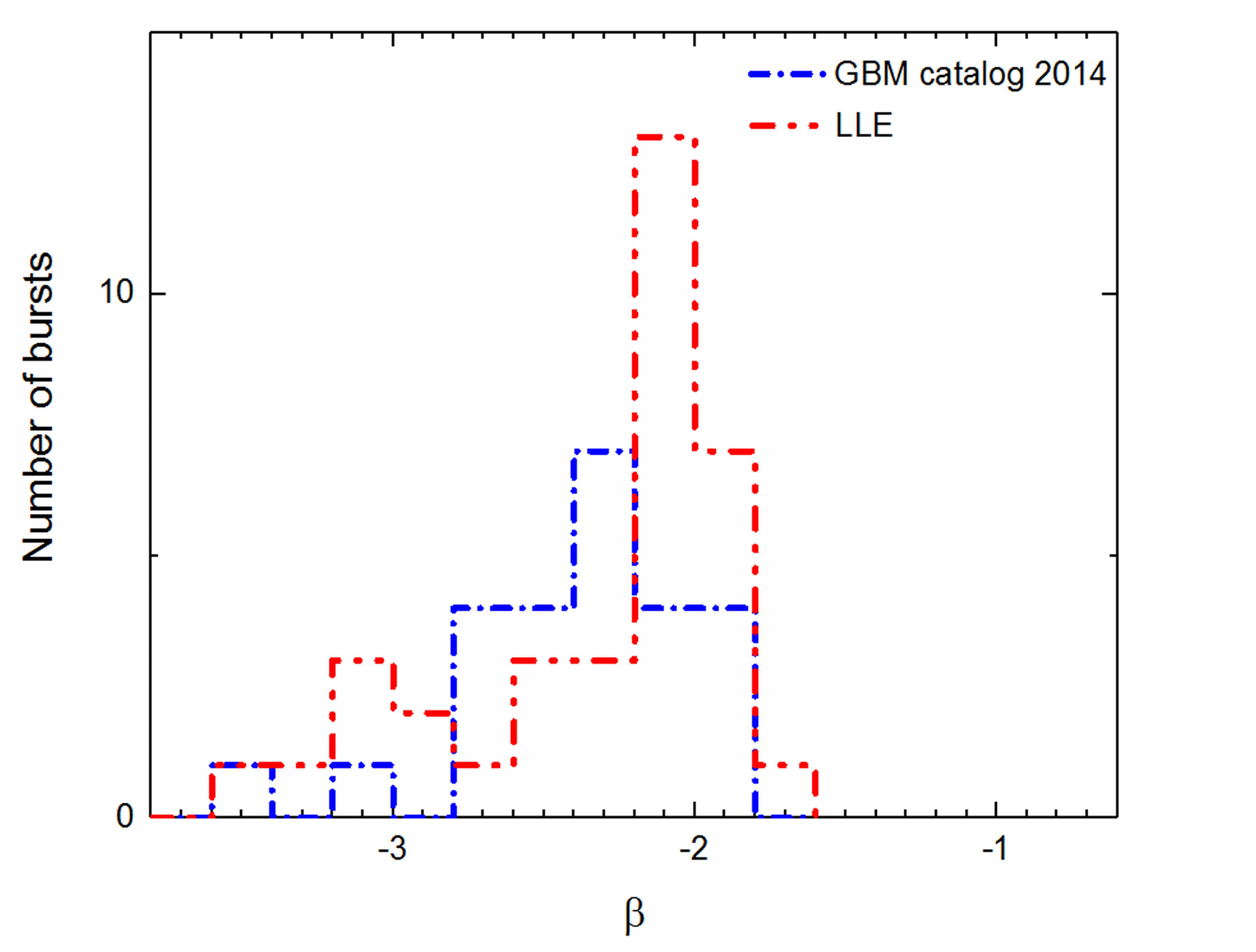}{0.5\textwidth}{}
          }
\gridline{
          \fig{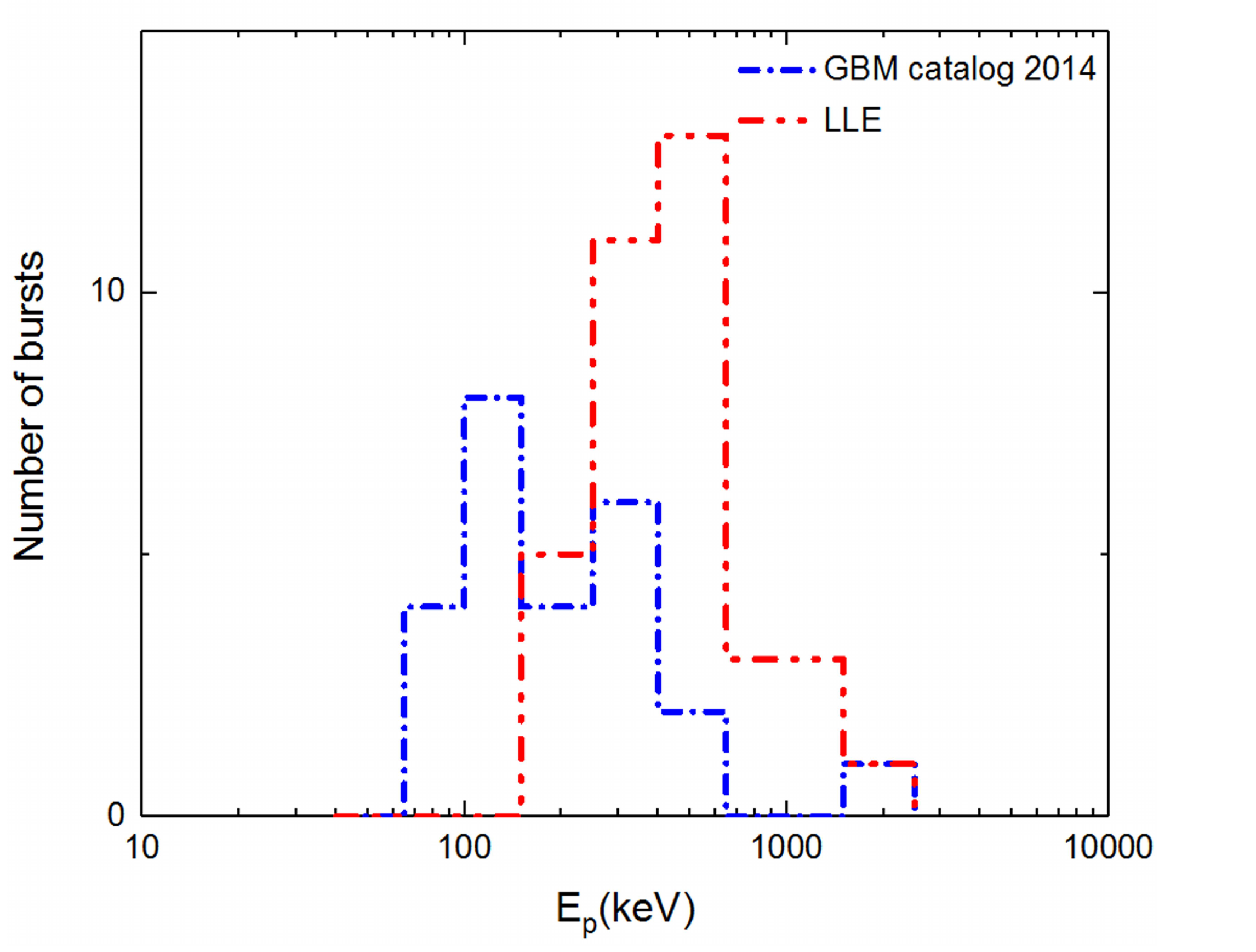}{0.5\textwidth}{}
          \fig{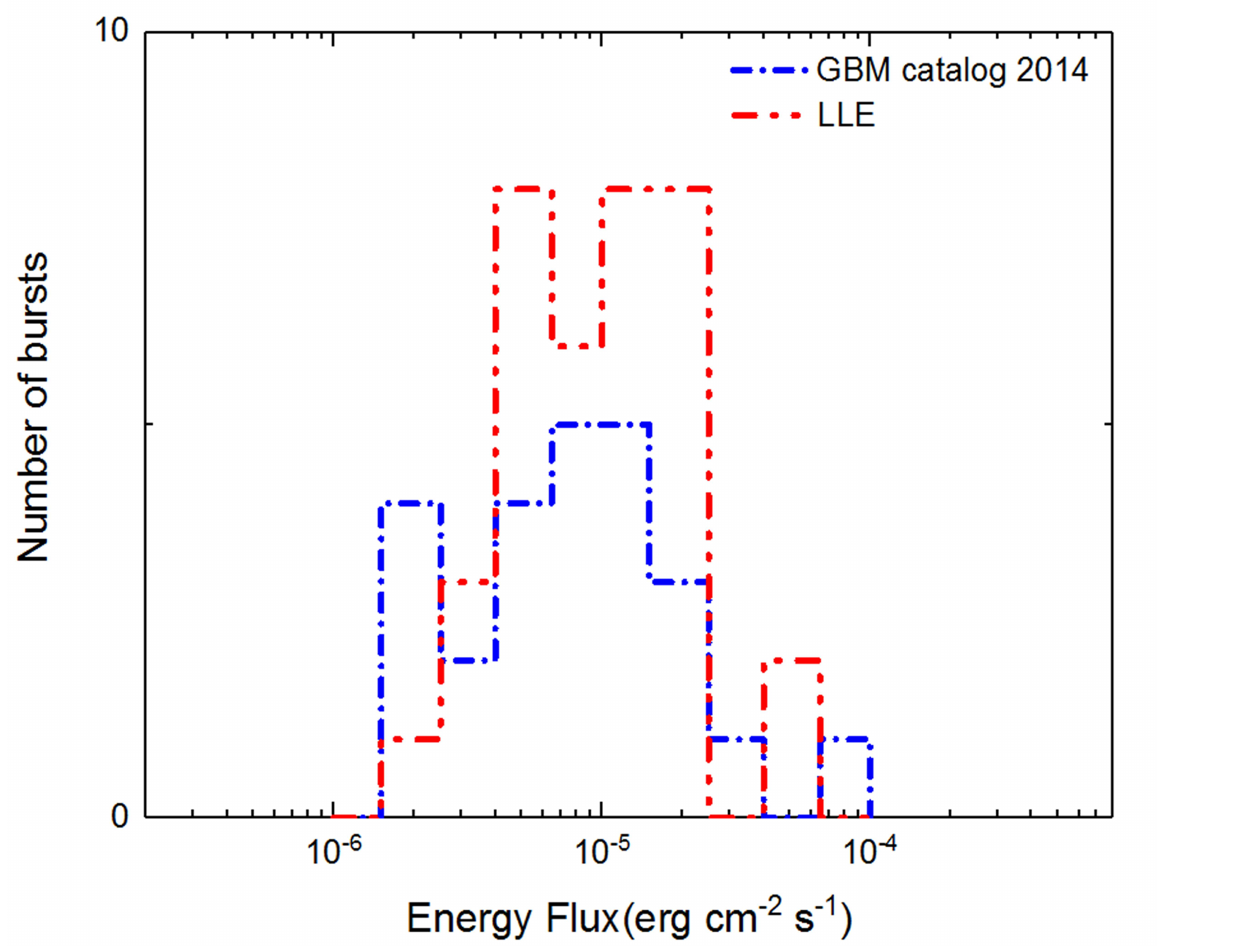}{0.5\textwidth}{}
          }
\gridline{
          \fig{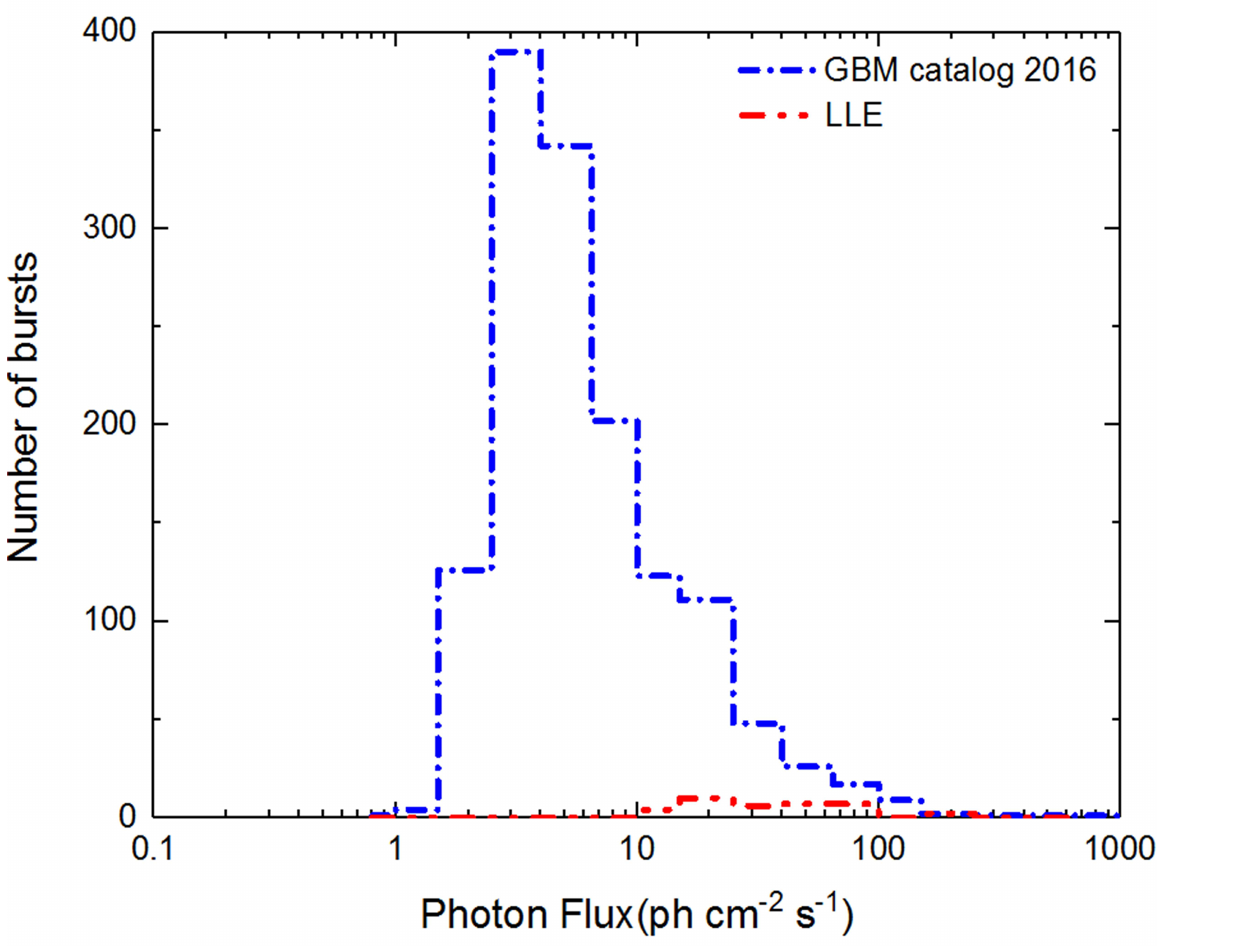}{0.5\textwidth}{}
          \fig{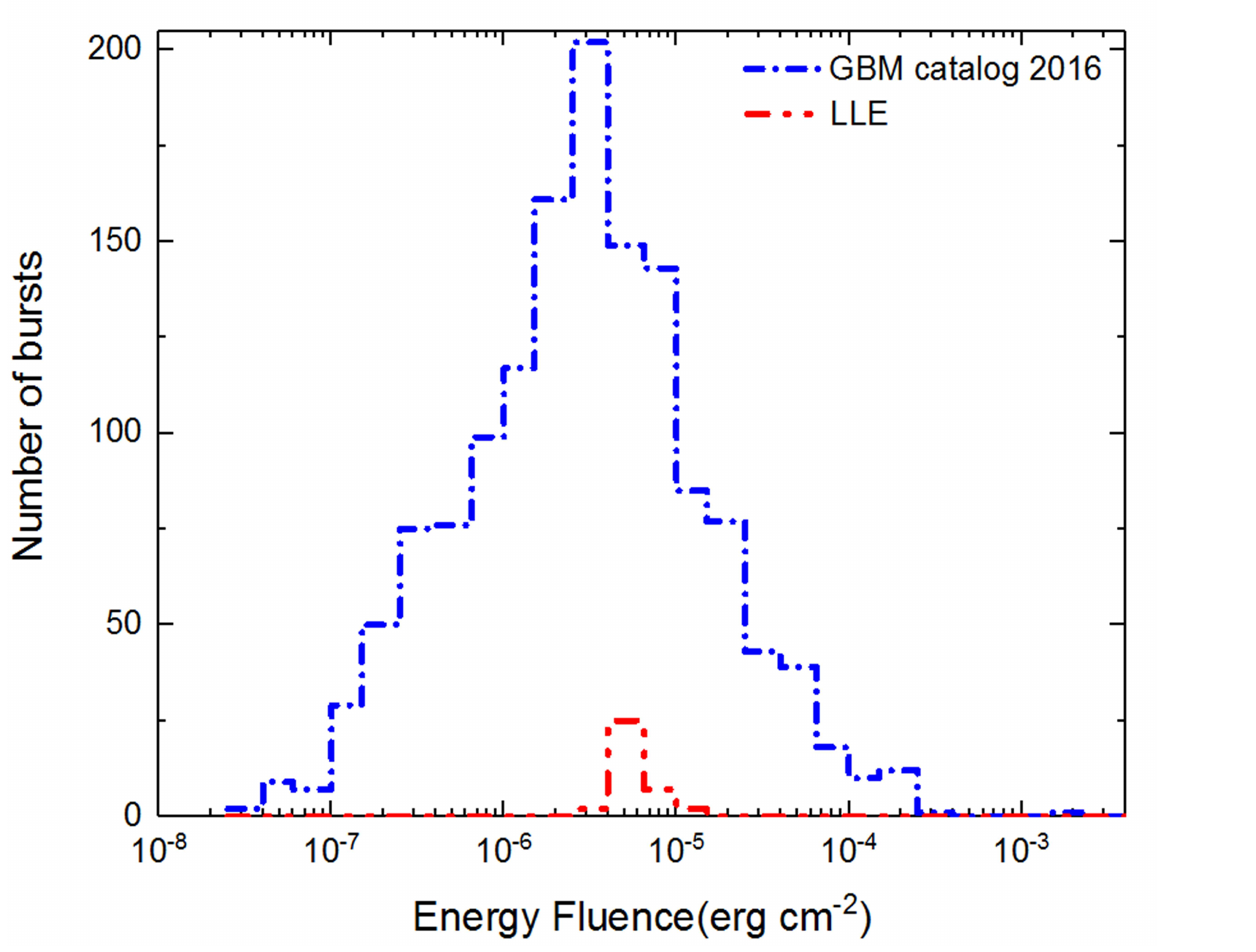}{0.5\textwidth}{}
          }
\caption{\edit1{\added{Distributions of the low energy spectral indices, high energy spectral indices, peak energy $E_{p}$, energy flux, photon flux, and energy fluence obtained from our time-resolved spectral fits around the peak flux (red dash-dot-dot lines). The blue short dash-dot lines show the corresponding distributions in \citet{2014ApJS..211...12G} or \citet{2016ApJS..223...28N}.}}}
\end{figure}\label{fig:comparison_peak_flux}

\edit1{\added{Since we used RMFIT to fit the GRB spectra, we also compared the results in our sample with those published in the $Fermi$ GRB spectral catalogs such as \citet{2014ApJS..211...12G} and \citet{2016ApJS..223...28N}. In Figure \ref{fig:comparison_peak_flux}, the distributions of the low energy spectral indices, high energy spectral indices, peak energy $E_{p}$, energy flux, photon flux, and energy fluence obtained from our time-resolved spectral fits at peak flux are shown in red dash-dot-dot lines. Meanwhile, the blue short dash-dot lines show the corresponding distributions in \citet{2014ApJS..211...12G} or \citet{2016ApJS..223...28N}. The BEST sample that was fitted by the Band function (in short, the BEST-Band sample) in \citet{2014ApJS..211...12G} was used for comparison. The energy flux, photon flux, and energy fluence are in the energy range from 10 keV to 1 MeV. The values of $\alpha$ are in the interval from -1 to 0 both for the two distributions although they have different distribution structures and peaks, which peak around $-0.7\pm0.1$ (LLE bursts) and $-0.5\pm0.1$ (BEST-Band sample), respectively. For the $\beta$ distribution, from -2.8 to -1.8, they are in $75\%$ (LLE bursts) and $92\%$ (BEST-Band sample), respectively. It is obvious that the peak energies have a median value of around 500 keV (LLE sample) and 200 keV (BEST-Band sample), respectively. Especially, $55.6\%$ of the LLE bursts have an $E_{p}$ value which is larger than 400 keV, and only $12\%$ of the BEST-Band bursts have an $E_{p}$ with the value of $>400$ keV. The energy flux values are larger than $1\times10^{-6}$ erg cm$^{-2}$ s$^{-1}$ both for the LLE sample and BEST-Band sample. $94.4\%$ of the LLE bursts and $92\%$ of the BEST-Band bursts are in the interval from $1\times10^{-6}$ erg cm$^{-2}$ s$^{-1}$ to $2.5\times10^{-5}$ erg cm$^{-2}$ s$^{-1}$. For the distributions of photon flux and energy fluence, all of the bursts in \citet{2016ApJS..223...28N} (1405 GRBs) have been selected (see the two bottom panels in Figure \ref{fig:comparison_peak_flux}). The distribution of photon flux covers an interval from 0.8 to 1000 photons cm$^{-2}$ s$^{-1}$ based on these 1405 GRBs. However, our sample only covers the interval from 10 to 100 photons cm$^{-2}$ s$^{-1}$. Similarly, our bursts cover just two orders of magnitude although these 1405 GRBs cover six orders of magnitude in the distributions of the energy fluence.}}

\subsection{Evolution Patterns of $E_{p}$ and $\alpha$} \label{subsec:subsec3.2}
In this section, we give the spectral analysis results which include the time-integrated spectral results and the time-resolved spectral results. Table \ref{tab:integrated_results} shows the results of the time-integrated spectral fits for all samples. \edit1{\added{Table \ref{tab:resolved_results} shows all pieces of information in the time-resolved spectral analysis. Figure \ref{fig:a_integrated} is the comparison between the histogram of $\alpha$ in the time-integrated spectra in our energy range and the BATSE energy range. Figure \ref{fig:comparison_integrated} shows the comparison between our results and the results of GBM catalog.}} \edit1{\deleted{Figure \ref{fig:spectral evolutions} represents the temporal characteristics of energy flux for all bursts in our sample (the left-hand, y-axis), along with time evolution of the $E_{p}$ and $\alpha$, both are marked with red stars in the right-hand y-axis. That is to say,}}Figure \ref{fig:spectral evolutions} shows the spectral evolutions for all of the bursts in our sample. The histograms of $E_{p}$ and $\alpha$ obtained by performing the detailed time-resolved spectral analysis have been shown in Figure \ref{fig:resolved}.

\subsubsection{The Time-integrated Spectral Results} \label{subsubsec:subsubsec3.2.1}

\begin{deluxetable}{cccccccc}
\tablecaption{Results of the Time-integrated Spectral Fits for All Samples \label{tab:integrated_results}}
\tablehead{
\colhead{GRB}
&\colhead{z} 
&\colhead{$T_{90}$}
&\colhead{$t_{1} \sim t_{2}$ \tablenotemark{a}}
& \colhead{$\alpha$}
& \colhead{$\beta$}
& \colhead{$E_{p}$}
& \colhead{Red.$\chi^{2}$}\\
\colhead{}& \colhead{}& \colhead{(s)}& \colhead{(s)}& \colhead{}& \colhead{} & 
\colhead{(keV)} & \colhead{}
}
\colnumbers
\startdata
080825C&...&22&0$\sim$30.016&-0.6197$\pm$0.0595&-2.243$\pm$0.119&174.7$\pm$11.6&1.14\\
090328A&0.736&80&0$\sim$80.064&-1.1790$\pm$0.0294&-2.352$\pm$0.366&756.0$\pm$121.0&1.19\\
090626A&...&70&0$\sim$70.016&-1.1920$\pm$0.0448&-2.061$\pm$0.074&152.2$\pm$15.8&1.10\\
090926A&2.106&20&0$\sim$25.024&-0.7967$\pm$0.0108&-2.428$\pm$0.054&312.4$\pm$6.1&1.97\\
100724B&...&111.6&0$\sim$100.031&-0.7046$\pm$0.0251&-1.904$\pm$0.035&384.6$\pm$19.3&1.38\\
100826A&...&100&0$\sim$100.032&-0.8828$\pm$0.0224&-1.897$\pm$0.029&289.4$\pm$14.4&2.03\\
101014A&...&450&0$\sim$50.047&-1.1690$\pm$0.0190&-2.470$\pm$0.128&186.7$\pm$8.1&1.46\\
110721A&0.382&24.45&0$\sim$30.015&-1.0790$\pm$0.0343&-1.742$\pm$0.035&411.1$\pm$56.3&1.10\\
120226A&...&57&0$\sim$60.032&-0.9439$\pm$0.0390&-2.008$\pm$0.090&266.1$\pm$25.1&1.27\\
120624B&2.20&271&0$\sim$30.016&-0.9902$\pm$0.0328&-2.505$\pm$0.383&685.4$\pm$78.3&1.13\\
130502B&...&24&0$\sim$35.006&-0.6279$\pm$0.0129&-2.404$\pm$0.051&303.8$\pm$5.9&1.83\\
130504C&...&74&0$\sim$80.064&-1.2830$\pm$0.0114&-2.250$\pm$0.110&858.8$\pm$66.4&1.45\\
130518A&2.49&48&0$\sim$50.045&-0.8689$\pm$0.0157&-2.288$\pm$0.055&408.5$\pm$13.5&1.38\\
130821A&...&84&0$\sim$100.031&-1.1860$\pm$0.0226&-2.044$\pm$0.073&317.3$\pm$26.4&1.78\\
131108A&2.4&19&0$\sim$25.024&-0.9453$\pm$0.0253&-2.337$\pm$0.104&381.0$\pm$20.6&1.07\\
140102A&...&65&0$\sim$30.015&-1.2550$\pm$0.0300&unconstrained&211.2$\pm$13.2&1.21\\
140206B&...&120&0$\sim$55.039&-1.0260$\pm$0.0158&-2.041$\pm$0.032&271.9$\pm$10.6&2.11\\
141028A&2.332&31.5&0$\sim$35.008&-0.6429$\pm$0.0415&-1.884$\pm$0.037&254.9$\pm$16.0&1.16\\
150118B&...&40&0$\sim$50.048&-0.8896$\pm$0.0098&-3.435$\pm$0.439&743.1$\pm$20.5&1.42\\
150202B&...&167&0$\sim$50.048&-0.7537$\pm$0.0440&-2.260$\pm$0.166&235.0$\pm$17.7&1.23\\
150314A&1.758&14.79&0$\sim$20.032&-0.8268$\pm$0.0104&-2.897$\pm$0.136&404.7$\pm$7.9&1.55\\
150403A&2.06&40.9&0$\sim$50.046&-0.7383$\pm$0.0266&-1.986$\pm$0.044&312.8$\pm$15.6&1.18\\
150510A&...&52&0$\sim$60.032&-1.0530$\pm$0.0104&unconstrained&1640.0$\pm$82.4&1.27\\
150627A&...&65&0$\sim$80.063&-1.0660$\pm$0.0104&-2.154$\pm$0.030&239.4$\pm$6.1&2.49\\
150902A&...&14&0$\sim$20.032&-0.7066$\pm$0.0125&-2.480$\pm$0.063&431.9$\pm$9.5&1.62\\
160509A&1.17&371&0$\sim$50.047&-0.8953$\pm$0.0107&-2.041$\pm$0.024&373.2$\pm$9.8&1.92\\
160816A&...&14&0$\sim$20.032&-0.7409$\pm$0.0215&-3.350$\pm$0.492&235.8$\pm$6.7&1.14\\
160821A&...&120&109.952$\sim$170.048&-1.0680$\pm$0.0034&-2.299$\pm$0.021&966.3$\pm$14.9&...\\
160905A&...&64&0$\sim$80.064&-1.0950$\pm$0.0174&-2.844$\pm$0.359&1392.0$\pm$143.0&1.82\\
160910A&...&24.3&0$\sim$30.016&-0.9891$\pm$0.0126&-1.776$\pm$0.012&506.9$\pm$22.2&3.86\\
170115B&...&44&0$\sim$50.048&-0.8061$\pm$0.0239&-2.504$\pm$0.156&997.4$\pm$65.6&2.39\\
170214A&2.53&123&0$\sim$150.016&-0.9511$\pm$0.0133&-2.519$\pm$0.137&465.7$\pm$16.1&2.03\\
170510A&...&128&0$\sim$135.040&-1.2760$\pm$0.0315&unconstrained&563.2$\pm$84.9&1.47\\
170808B&...&17.7&0$\sim$25.024&-0.9949$\pm$0.0101&-2.297$\pm$0.035&249.1$\pm$5.2&2.28\\
171210A&...&143&0$\sim$145.024&-0.7107$\pm$0.0383&-2.244$\pm$0.063&136.3$\pm$5.6&1.30\\
180305A&...&12.5&0$\sim$15.040&-0.3126$\pm$0.0266&-2.490$\pm$0.098&329.5$\pm$9.6&1.27\\
\enddata
\tablenotetext{a}{Time intervals.}
\end{deluxetable}

\edit1{\added{The time-integrated spectra reflect the overall emission properties but do not exhibit any spectral evolution.}} Table \ref{tab:integrated_results} shows the results of the time-integrated spectral fits for all samples. Listed in this Table are the \edit1{\replaced{29}{36}} GRBs in our sample which satisfy our criteria in this study (Col.1), the redshift of them (Col.2), the duration interval of $T_{90}$ (Col.3), the integrated range in our analysis (Col.4), the low energy photon index $\alpha$ in \edit1{\added{the}} time-integrated analysis (Col.5), the high energy photon index $\beta$ in \edit1{\added{the}} time-integrated analysis (Col.6), the peak energy in \edit1{\added{the}} time-integrated analysis (Col.7) and \edit1{\added{the}} reduced $\chi^{2}$ (Col.8). 

There are \edit1{\replaced{10}{11}} GRBs with known redshift. The duration values of $T_{90}$ for most of them in our sample seem to be from 20 s to 100 s. \edit1{\replaced{And as we all know,}{As we all know,}} \edit1{\replaced{the typical value of low energy photon index $\alpha$ is $\sim -1.0$ and the peak energy $E_{p} \sim 300$ keV}{the typical values of the low energy photon index $\alpha$ and peak energy $E_{p}$ are $\sim -1.0$ and $\sim 300$ keV, respectively,}} for the time-integrated \edit1{\replaced{spectrum}{spectra}} based on the statistical study such as \citet{2000ApJS..126...19P}, \citet{2006ApJS..166..298K}, \citet{2011ApJ...730..141Z}, \citet{2012ApJS..199...19G}, and \citet{2013ApJ...764...75G}. While the typical value of $\alpha$ in our sample is \edit1{\replaced{$\sim-0.8$}{$\sim-0.9$}} obtained from Table \ref{tab:integrated_results}, which is larger than the statistical study of \edit1{\added{a}} large sample of GRBs\edit1{\deleted{and it is close to the synchrotron limit}}, \edit1{\replaced{when}{but}} the $E_{p}$ is similar to the previous statistics. \edit1{\added{It is curious that the typical $\alpha$ value for the LLE bright bursts in our sample is different from the BATSE bright bursts \citep{2000ApJS..126...19P}. To explore the possible cause of the discrepancy, we limit the $Fermi$ spectral fitting only to the BATSE energy range, but we do not get a similar typical $\alpha$ value as \citet{2000ApJS..126...19P}. Whereas, we found that this typical value would be smaller than the situation when we select fewer bursts as the sample in our study. So, we guess that the two typical $\alpha$ values for LLE bright bursts and BATSE bright bursts would be similar if we have enough bursts in the study.}} \edit1{\replaced{And}{Besides,}} \edit1{\deleted{there are }}\edit1{\replaced{three}{four}} time-integrated values of $\alpha$, in GRB 080825C ($\sim -0.6197$), \edit1{\replaced{GRB 130305A ($\sim -0.5633$)}{GRB 130502B ($\sim -0.6279$)}}, GRB 141028A ($\sim -0.6429$), \edit1{\added{and GRB 180305A ($\sim -0.3126$),}} \edit1{\deleted{which }}violate the synchrotron limit.

\begin{figure}
\gridline{\fig{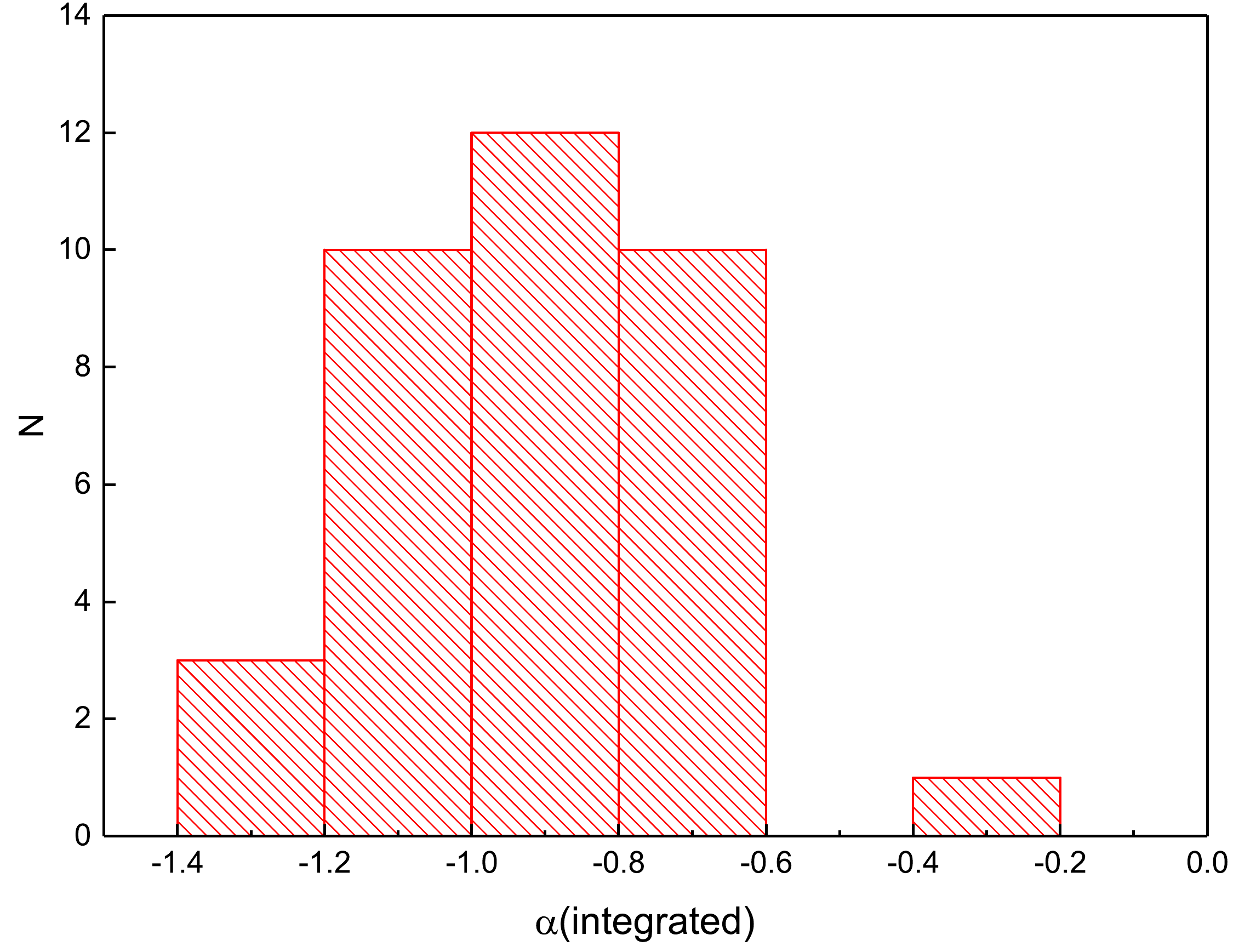}{0.5\textwidth}{}
          \fig{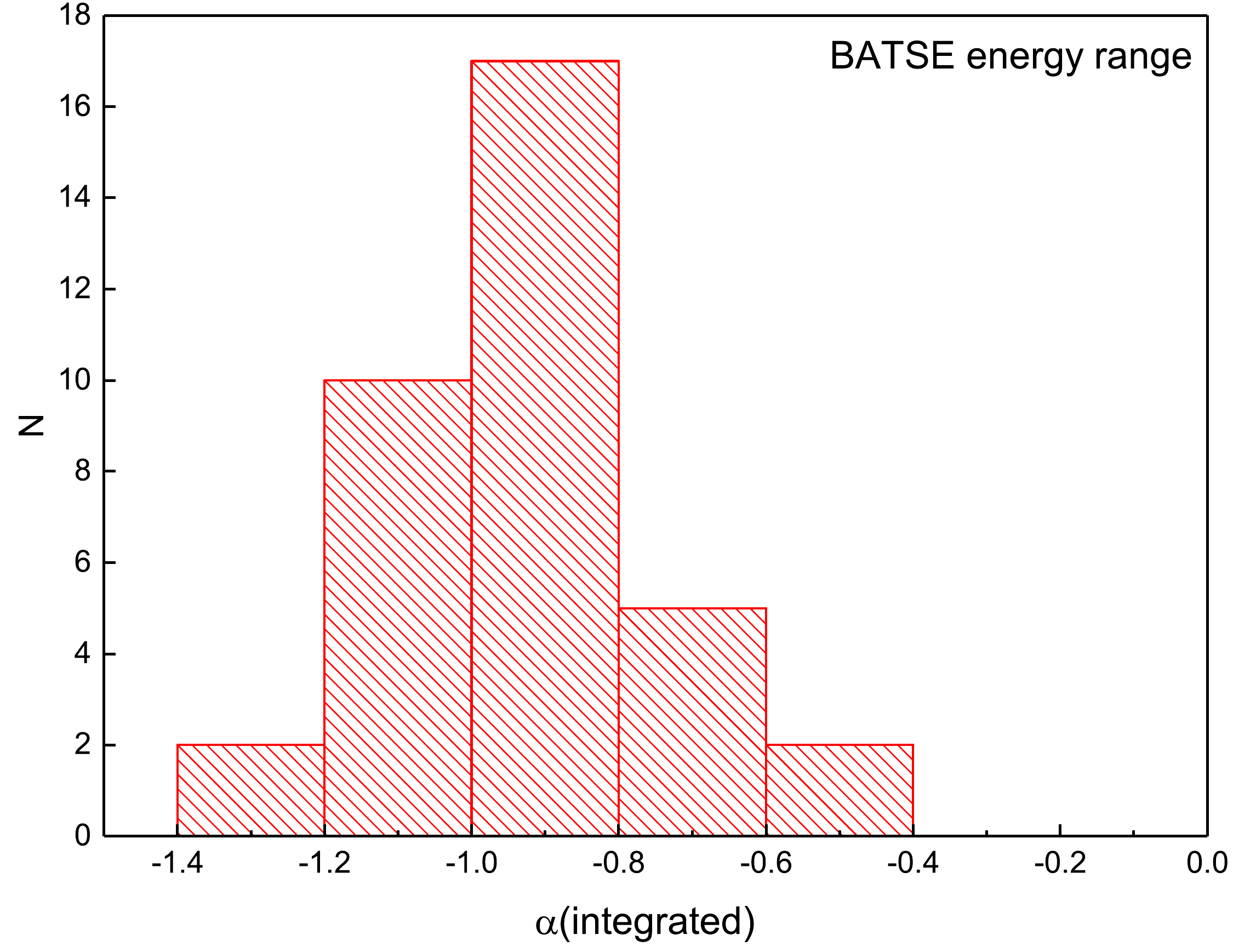}{0.5\textwidth}{}
          }
\caption{\edit1{\added{The comparison between the histogram of $\alpha$ in the time-integrated spectra in our energy range and the BATSE energy range. The left panel represents the histogram of $\alpha$ in the time-integrated spectra in the $Fermi$-GBM energy range (from 8 keV to 40 MeV). The other one is in the BATSE energy range (from 28 keV to 1800 keV). }}\label{fig:a_integrated}}
\end{figure}

\begin{figure}
\gridline{\fig{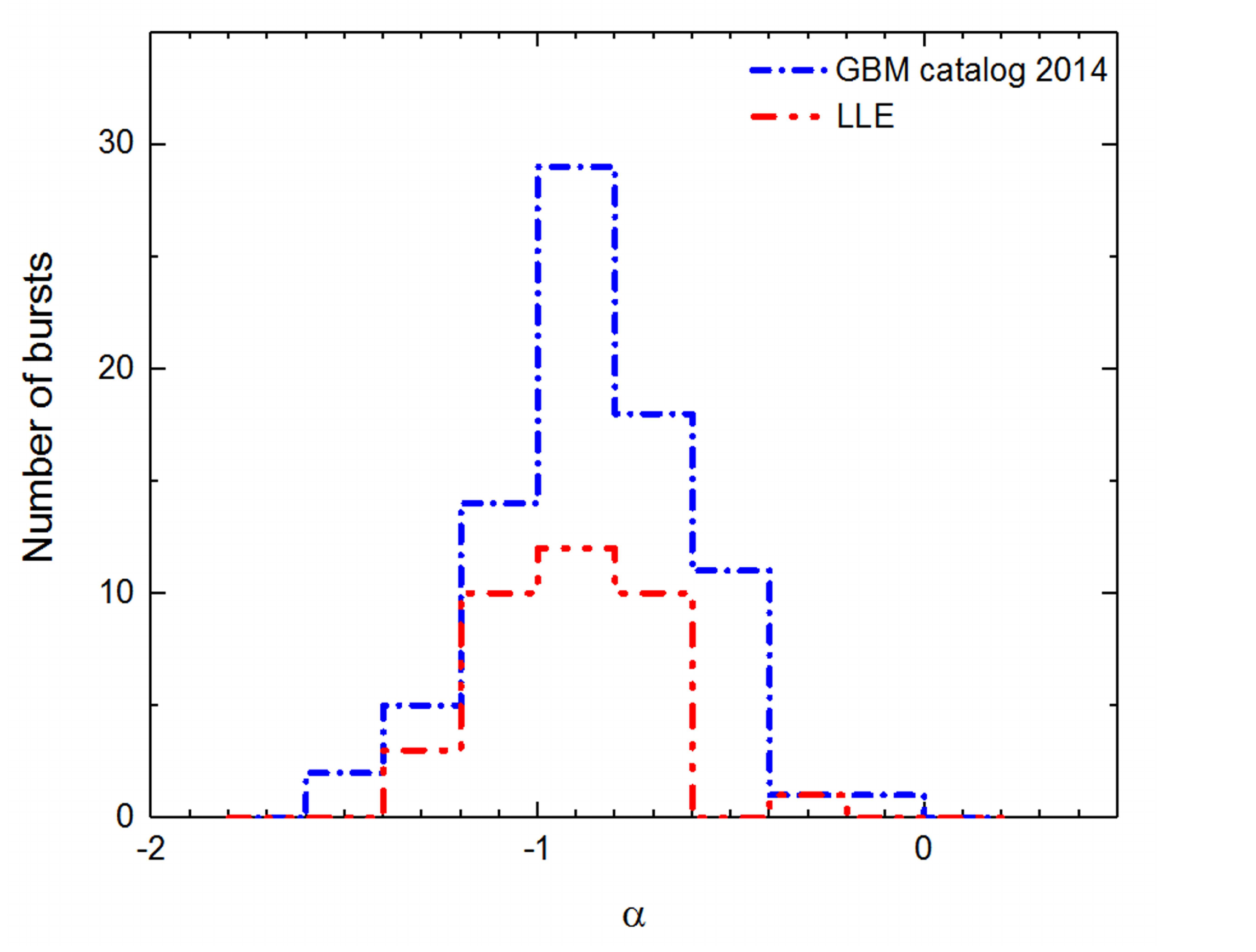}{0.5\textwidth}{}
          \fig{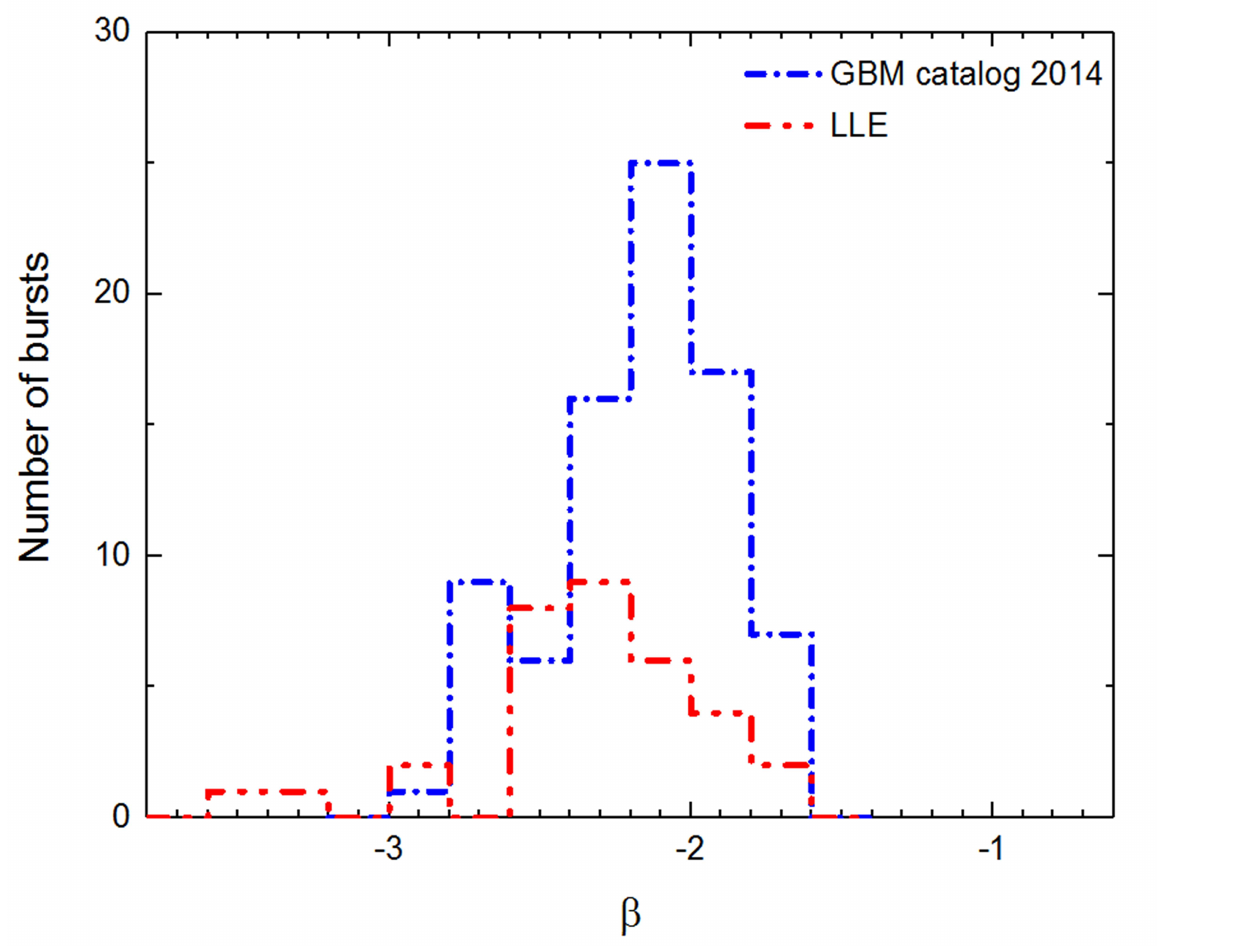}{0.5\textwidth}{}
          }
\gridline{
          \fig{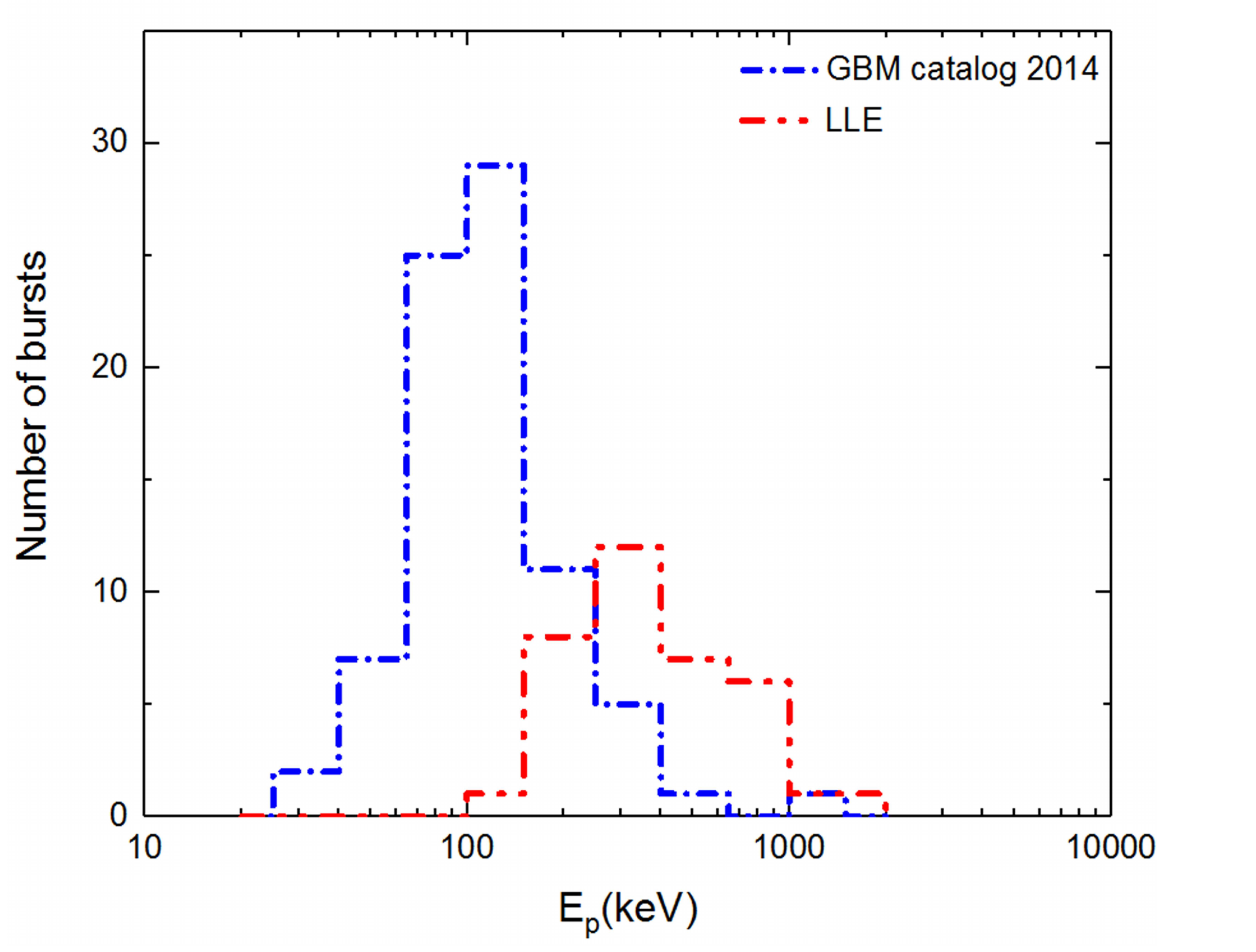}{0.5\textwidth}{}
          \fig{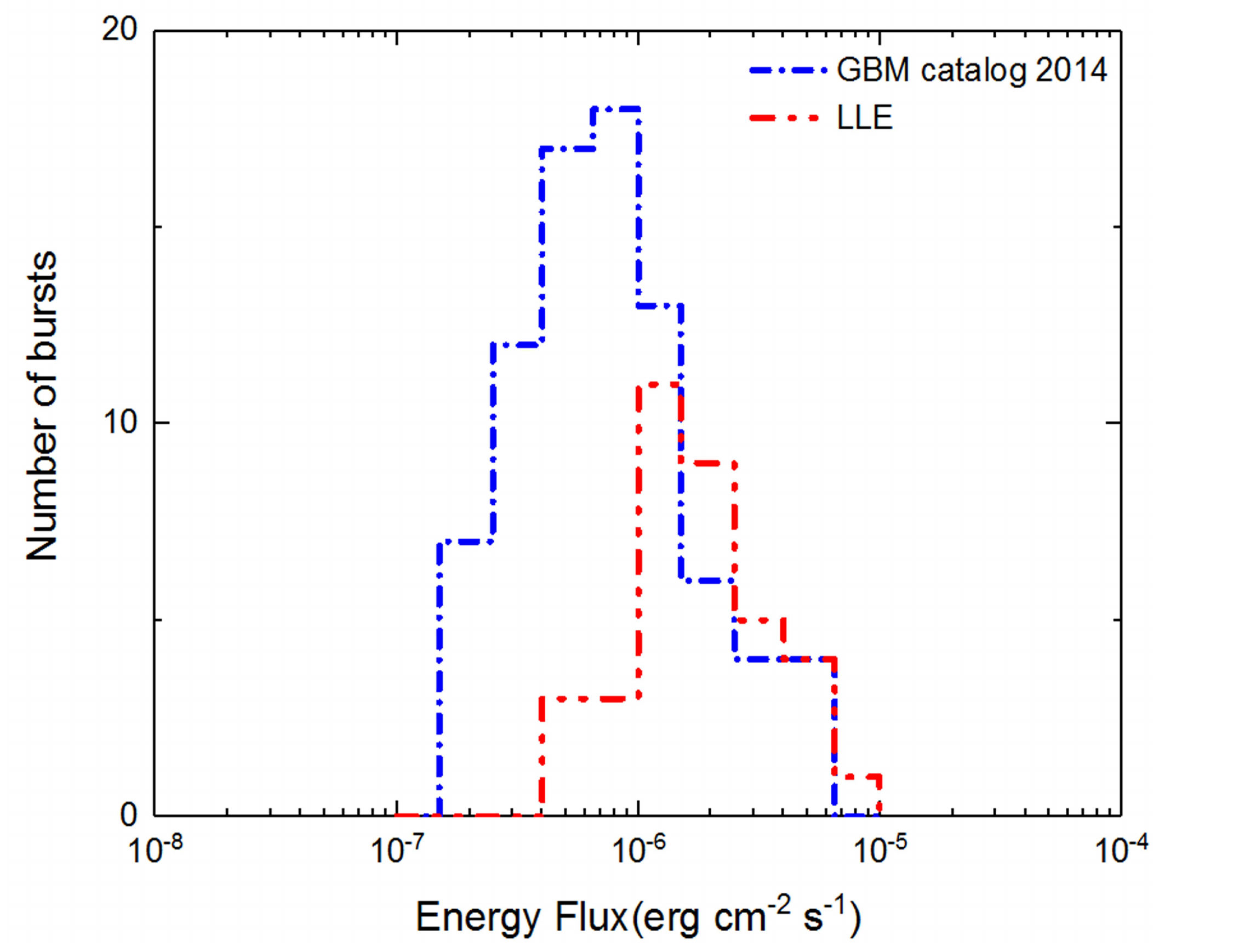}{0.5\textwidth}{}
          }
\gridline{
          \fig{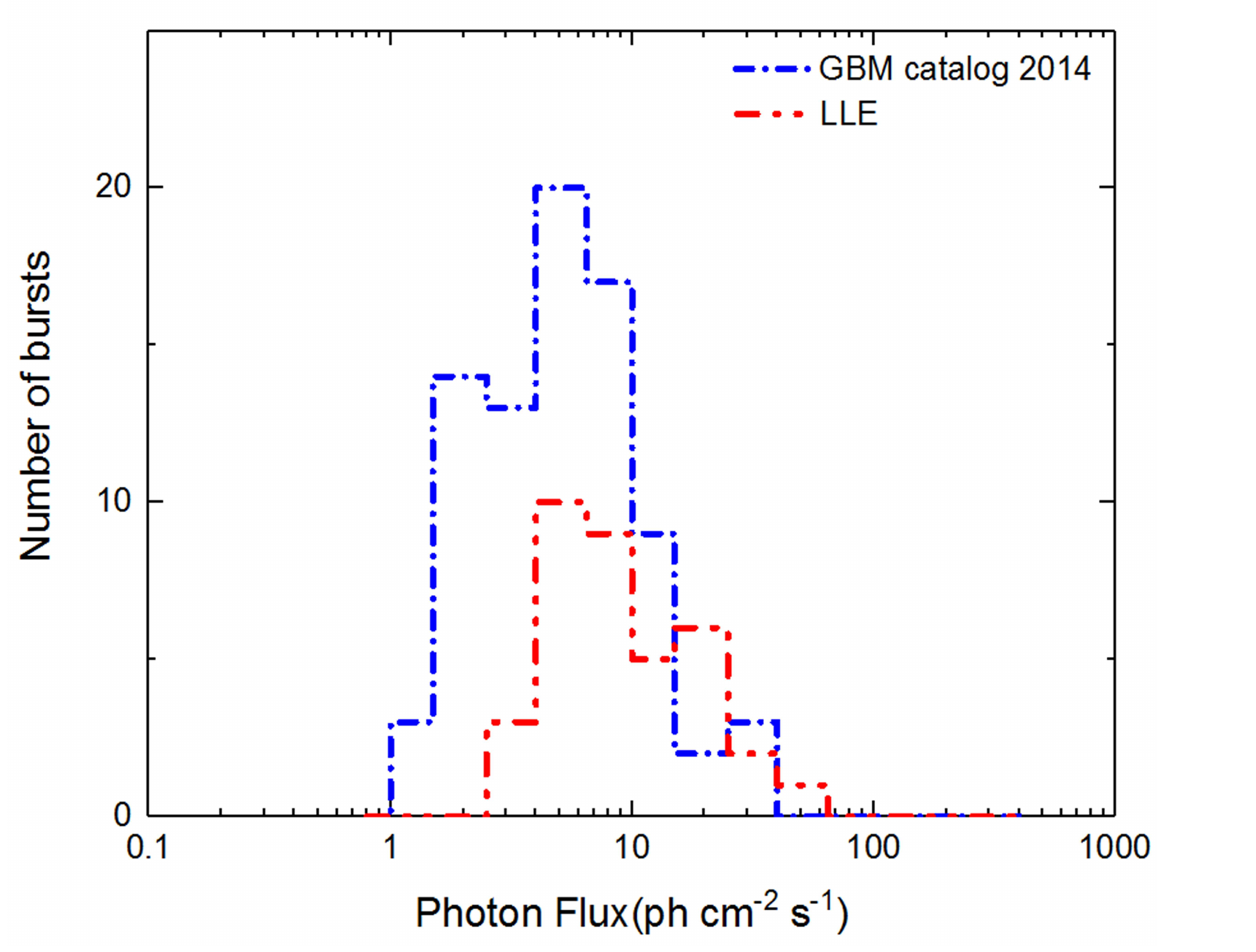}{0.5\textwidth}{}
          \fig{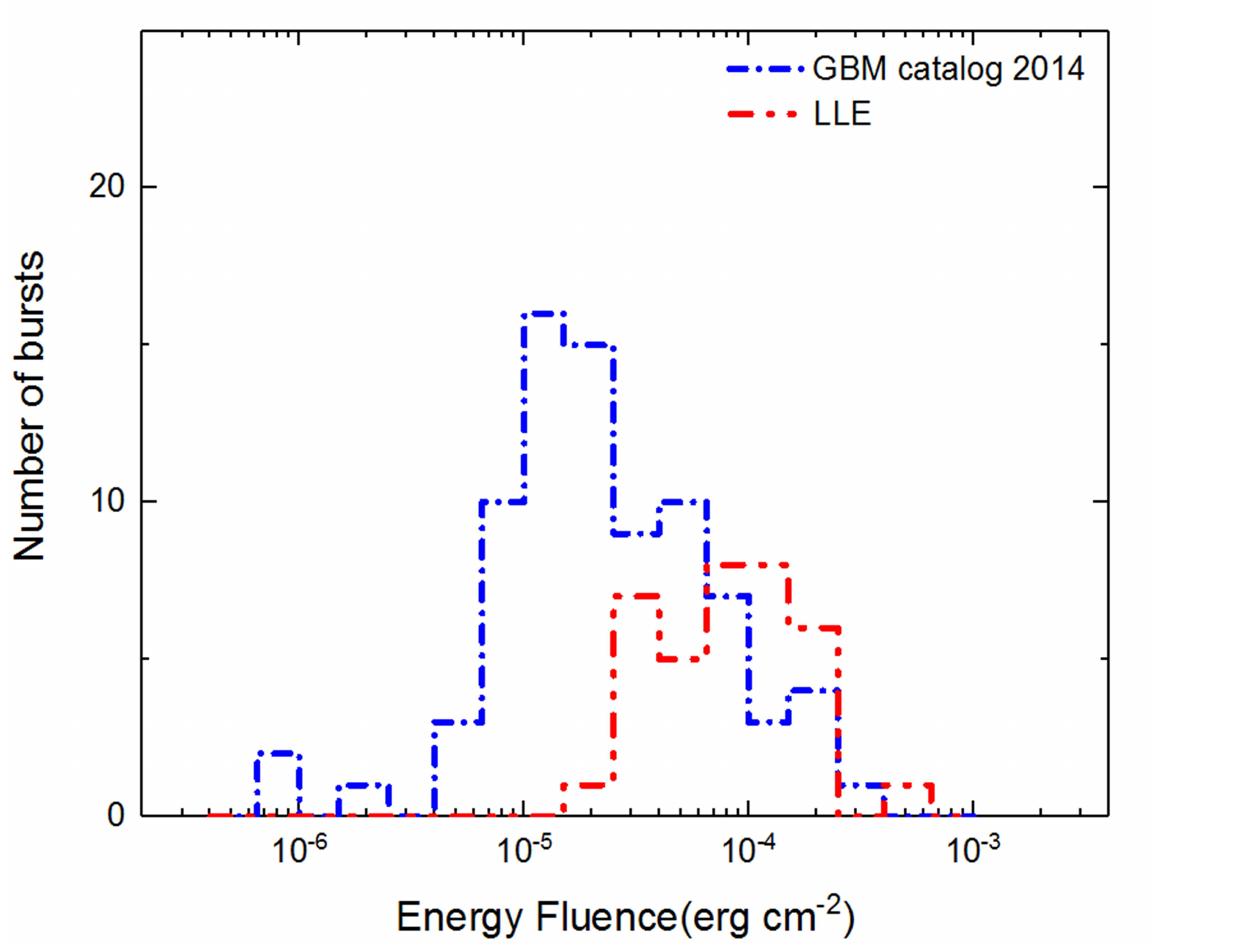}{0.5\textwidth}{}
          }
\caption{\edit1{\added{Distributions of the low energy spectral indices, high energy spectral indices, peak energy $E_{p}$, energy flux, photon flux, and energy fluence obtained from our time-integrated spectral fits (red dash-dot-dot lines). The blue short dash-dot lines show the corresponding distributions in \citet{2014ApJS..211...12G}.}}}
\end{figure}\label{fig:comparison_integrated}

\edit1{\added{Similarly, we also compared our results with \citet{2014ApJS..211...12G}.  In Figure \ref{fig:comparison_integrated}, the distributions of the low energy spectral indices, high energy spectral indices, peak energy $E_{p}$, energy flux, photon flux, and energy fluence obtained from our time-integrated spectral fits during the whole interval are shown in red dash-dot-dot lines. Meanwhile, the blue short dash-dot lines show the corresponding distributions for the BEST-Band sample in \citet{2014ApJS..211...12G}. The energy flux, photon flux, and energy fluence are in the interval from 10 keV to 1 MeV. The overall distribution of $\alpha$ is similar to that found in the BEST-Band sample, in which the typical value is $\sim -0.9$ both for them. In the distribution of $\beta$, they are different because of their different distribution structures and peaks. However, they are both concentrated in the interval from -2.6 to -1.6 although the $\beta$ values in our bursts are generally smaller. \citet{2012ApJ...754..121F} pointed out that the inclusion of $Fermi$/LAT upper limits in the fitting process can make $\beta$ steeper. Maybe the reason why our $\beta$ values are generally smaller is that the LAT detector observed these bursts. On the contrary, the rest of 4 parameters, peak energy, energy flux, photon flux, and energy fluence, are generally larger than the BEST-Band bursts. For most of the LLE bursts, the $E_{p}$ is larger than 150 keV. But, it is smaller than 150 keV for most of the BEST-Band sample. $66.7\%$ of the BEST-Band bursts have an energy flux value which is smaller than $1\times10^{-6}$ erg cm$^{-2}$ s$^{-1}$. While $83.3\%$ of our bursts have a value which is larger than $1\times10^{-6}$ erg cm$^{-2}$ s$^{-1}$. The two distributions of the photon flux both generally peak around 4-6.5 photon cm$^{-2}$ s$^{-1}$. Besides, $61.7\%$ of the BEST-Band bursts have a photon flux value which is smaller than 6.5 photon cm$^{-2}$ s$^{-1}$ while $63.9\%$ of the LLE bursts have a value which is larger than 6.5 photon cm$^{-2}$ s$^{-1}$. More than half of the BEST-Band bursts have an energy fluence with the value of $<2.5\times10^{-5}$ erg cm$^{-2}$, but all of the LLE bursts have an energy fluence with the value of $>2.5\times10^{-5}$ erg cm$^{-2}$ except for GRB 140102A. Meanwhile, 15 GRBs show an energy fluence with the value of $>1\times10^{-4}$ erg cm$^{-2}$ for the LLE sample, but only 8 GRBs show this value for the BEST-Band sample.}}

\subsubsection{The Time-resolved Spectral Results} \label{subsubsec:subsubsec3.2.2}
 
\begin{deluxetable}{cccccccccc}
\tablecaption{Fitting Results of the Parameter Correlations and the Spectral Evolutions of $E_{p}$ and $\alpha$ \label{tab:resolved_results}}
\tablehead{
\colhead{GRB}
&\colhead{Detectors} 
&\colhead{N}
& \colhead{$E_{p}-F$}
& \colhead{$\alpha-F$}
& \colhead{$E_{p}-\alpha$}
& \colhead{Spectral Evolutions}
&\colhead{$\alpha>-\frac{2}{3}$}
&\colhead{$\alpha-F$} 
&\colhead{$E_{p}-\alpha$} \\
\colhead{}& \colhead{}& \colhead{}& \colhead{r}& \colhead{r} & 
\colhead{r} & \colhead{$E_{p}/\alpha$} & \colhead{} &\colhead{r(S)} &\colhead{r(S)}
}
\colnumbers
\startdata
080825C&n9,na,b1&8&0.94&0.70&0.54&h.t.s./r.t.& yes & -0.38 & -0.96\\
090328A&n7,n8,b1&8&0.70&0.93&0.83&h.t.s./i.t.& no & -0.20 & -0.86\\
090626A&n0,n3,b0&20&0.61&0.69&0.01&r.t./r.t.& not all & -0.56 & -0.88\\
090926A&n6,n7,b1&37&0.61&0.67&0.35&r.t./r.t.& not all & -0.36 & -0.86\\
100724B&n0,n1,b0&30&0.59&0.35&-0.08&r.t./r.t.& not all & ... &...\\
100826A&n7,n8,b1&24&0.93&0.08&-0.01&r.t./r.t.& not all & ... & ...\\
101014A&n6,n7,b1&21&0.86&0.83&0.62&r.t./r.t.& not all & 0.28 & -0.54\\
110721A&n6,n9,b1&7&0.62&0.76&0.07&h.t.s./s.t.h. to h.t.s.& no & -0.71 & -0.88\\
120226A&n0,n1,b0&12&0.47&0.73&-0.11&r.t./r.t.& no & -0.57 & -0.87\\
120624B&n1,n2,b0&5&0.52&0.61&0.94&h.t.s./h.t.s.& no & -0.40 & -0.80\\
130502B&n6,n7,b1&25&0.64&0.75&0.24&r.t./r.t.&not all & -0.13 & -0.67\\
130504C&n9,na,b1&29&0.54&0.45&-0.18&r.t./r.t.& no & ... & ...\\
130518A&n3,n7,b0,b1&19&0.61&0.69&0.32&r.t./r.t.& no & -0.71 & -0.81\\
130821A&n6,n9,b1&11&0.67&0.71&-0.06&r.t./r.t.& not all & -0.002 & -0.95\\
131108A&n3,n6,b0,b1&6&0.84&0.77&0.44&s.t.h. to h.t.s./r.t.& not all & -0.14 & -0.32\\
140102A&n7,n9,b1&6&0.89&0.84&0.71&i.t./i.t.& not all & -0.002 & -0.93\\
140206B&n0,n1,b0&23&0.67&0.58&0.38&r.t./r.t.& not all & ... & ...\\
141028A&n6,n9,b1&5&0.91&-0.07&0.18&i.t./h.t.s.& yes & ... & ...\\
150118B&n1,n2,b0&20&0.86&0.50&0.26&r.t./r.t.& not all & ... & ...\\
150202B&n0,n1,b0&7&0.72&-0.48&-0.69&r.t./a.t.& not all & ... & ...\\
150314A&n1,n9,b0,b1&17&0.05&0.95&0.05&no/r.t.& not all & -0.64 & -0.89\\
150403A&n3,n4,b0&9&0.83&0.39&0.01&r.t./r.t.& not all & ... & ...\\
150510A&n0,n1,b0&11&0.56&0.95&0.55&s.t.h. to h.t.s./r.t.+h.t.s. & not all & 0.27 & -0.86\\
150627A&n3,n4,b0&39&0.66&0.75&0.59&r.t./r.t.& not all & -0.45 & -0.79\\
150902A&n0,n3,b0&17&0.58&0.85&0.29&r.t./r.t.& not all & -0.68 & -0.91\\
160509A&n0,n3,b0&39&0.46&0.83&0.39&r.t./r.t.& not all & -0.18 & -0.96\\
160816A&n6,n7,b1&10&0.76&0.70&0.27&i.t./r.t.& not all & -0.08 & -0.64\\
160821A&n6,n7,b1&130&0.43&0.81&0.08&r.t./r.t.& no & 0.07 & -0.72\\
160905A&n6,n9,b1&12&0.71&0.97&0.73&r.t./r.t.& no & 0.65 & -0.76\\
160910A&n1,n5,b0&13&0.83&0.17&-0.06&h.t.s./no & not all & ... & ...\\
170115B&n0,n1,b0&5&0.99&-0.95&-0.97&i.t./a.t.& yes & ... & ...\\
170214A&n0,n1,b0&24&0.30&0.73&-0.18&r.t./r.t.& not all & -0.64 & -0.90\\
170510A&n9,na,b1&7&0.16&0.82&-0.02&no/r.t.& no & -0.43 & -0.84\\
170808B&n1,n5,b0&31&0.81&0.33&0.27&r.t./r.t.& not all & ... & ...\\
171210A&n0,n1,b0&17&0.90&-0.50&-0.56&r.t.+h.t.s./no& not all & ... & ...\\
180305A&n1,n2,b0&8&0.83&-0.31&0.02&i.t./no& yes & ... & ...\\
\enddata
\end{deluxetable}

\begin{figure}
\centering
\resizebox{4cm}{!}{\includegraphics{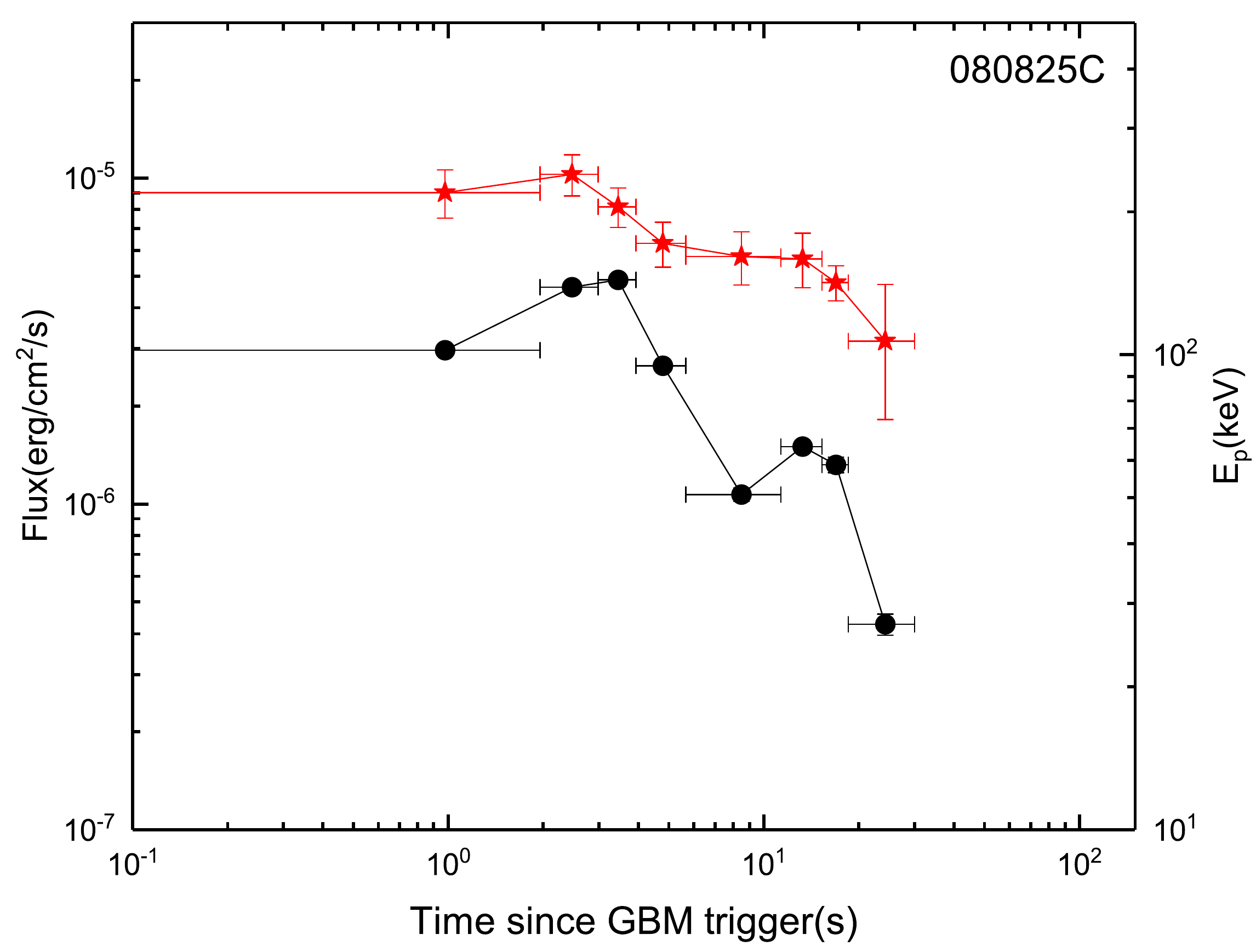}}
\resizebox{4cm}{!}{\includegraphics{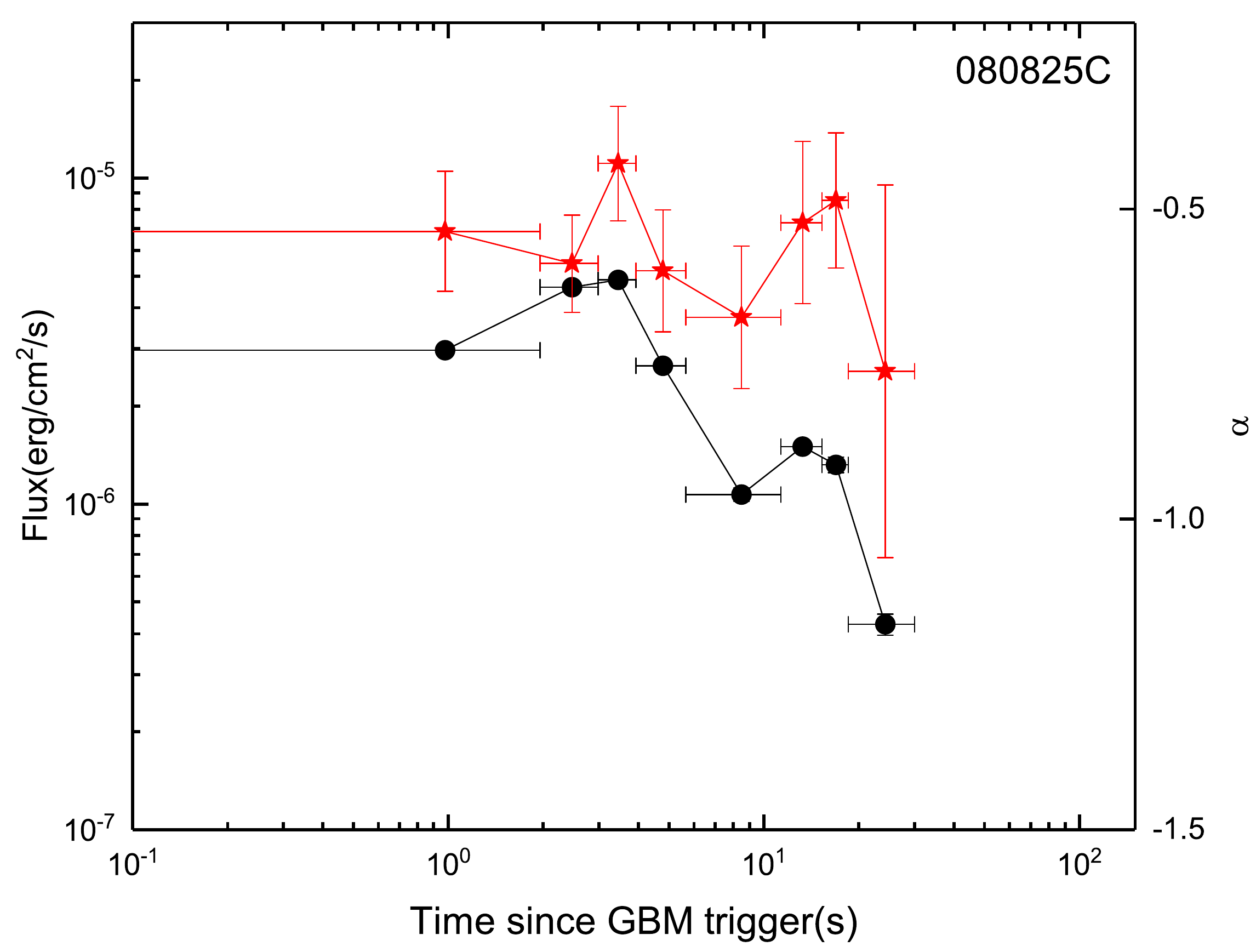}}
\resizebox{4cm}{!}{\includegraphics{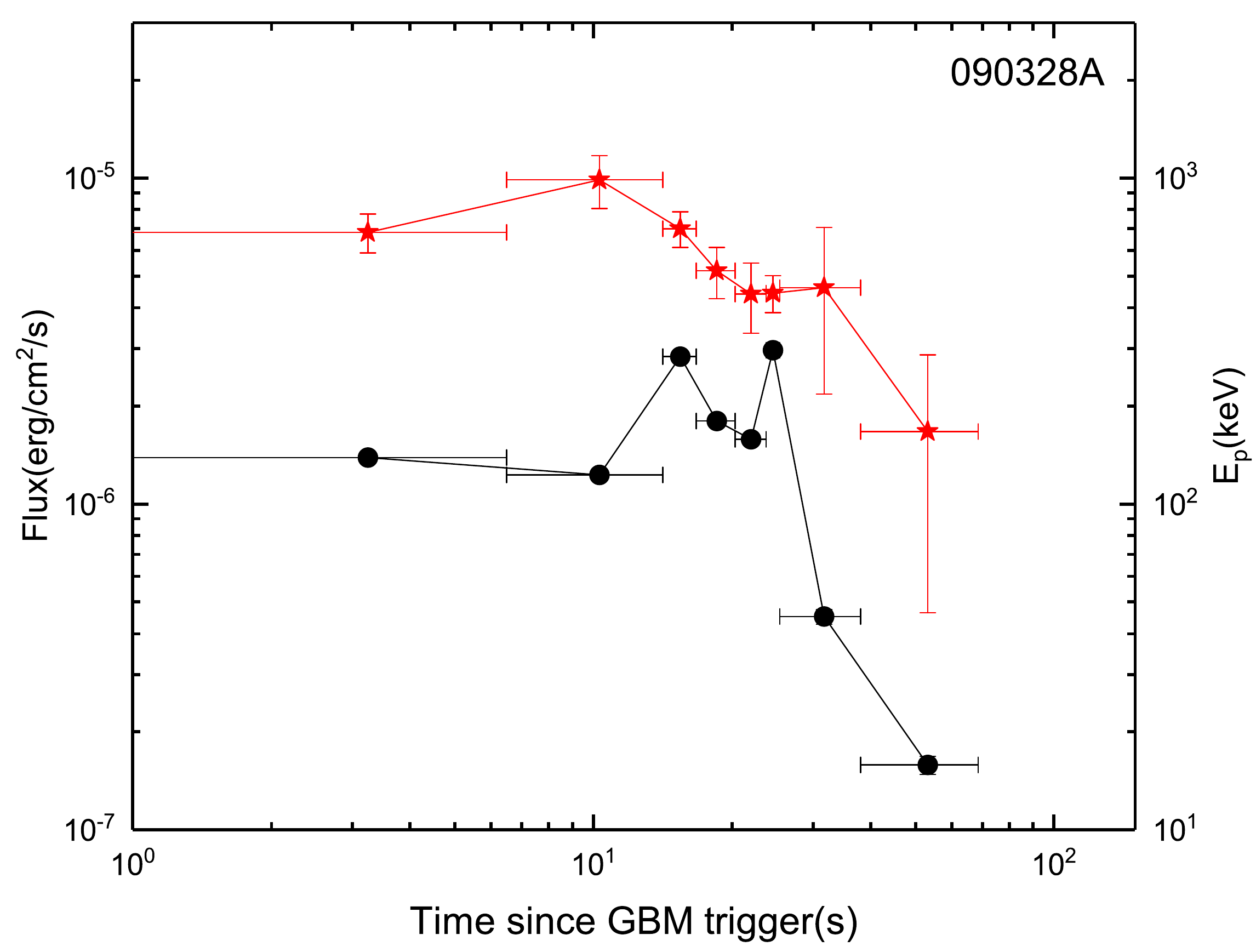}}
\resizebox{4cm}{!}{\includegraphics{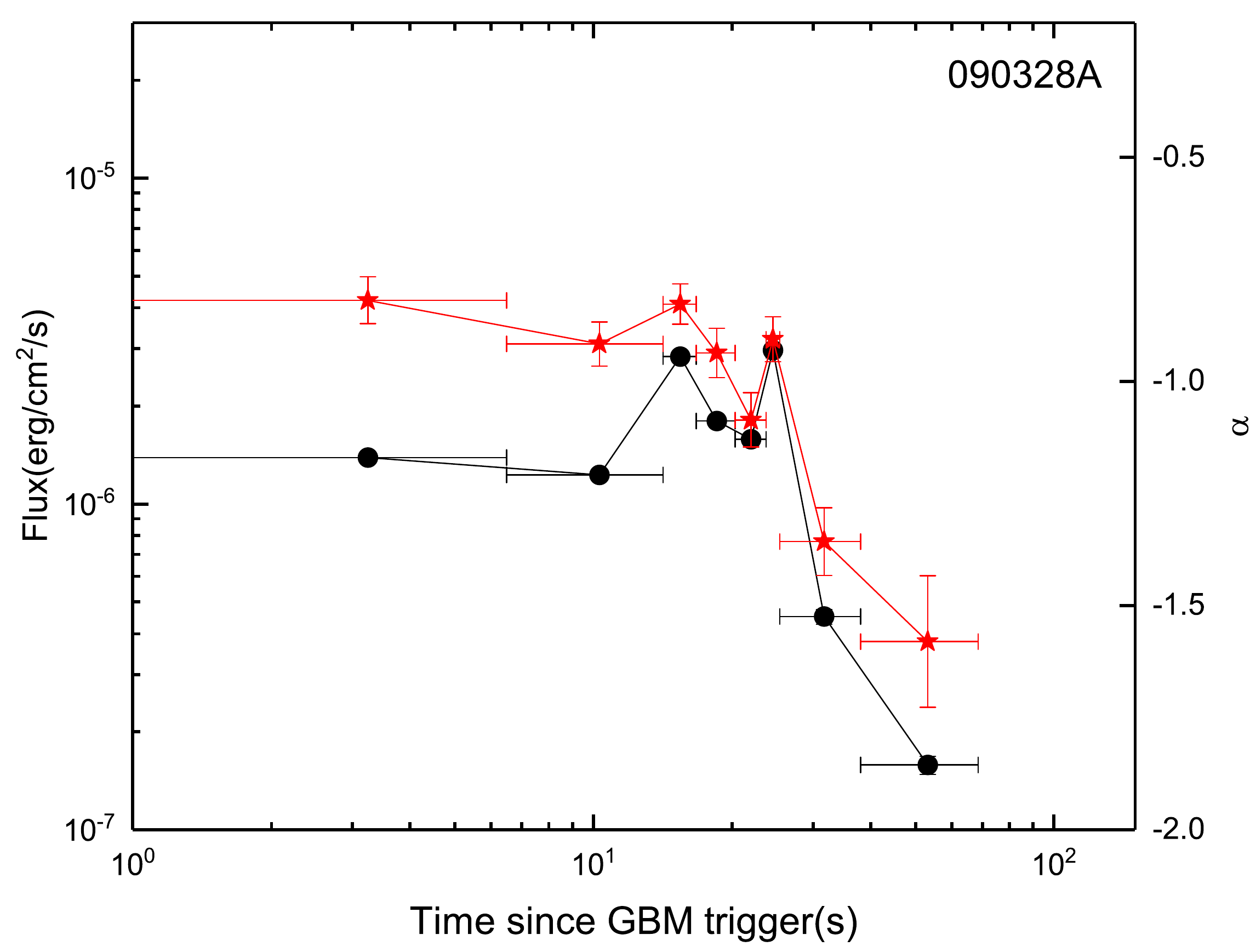}}
\resizebox{4cm}{!}{\includegraphics{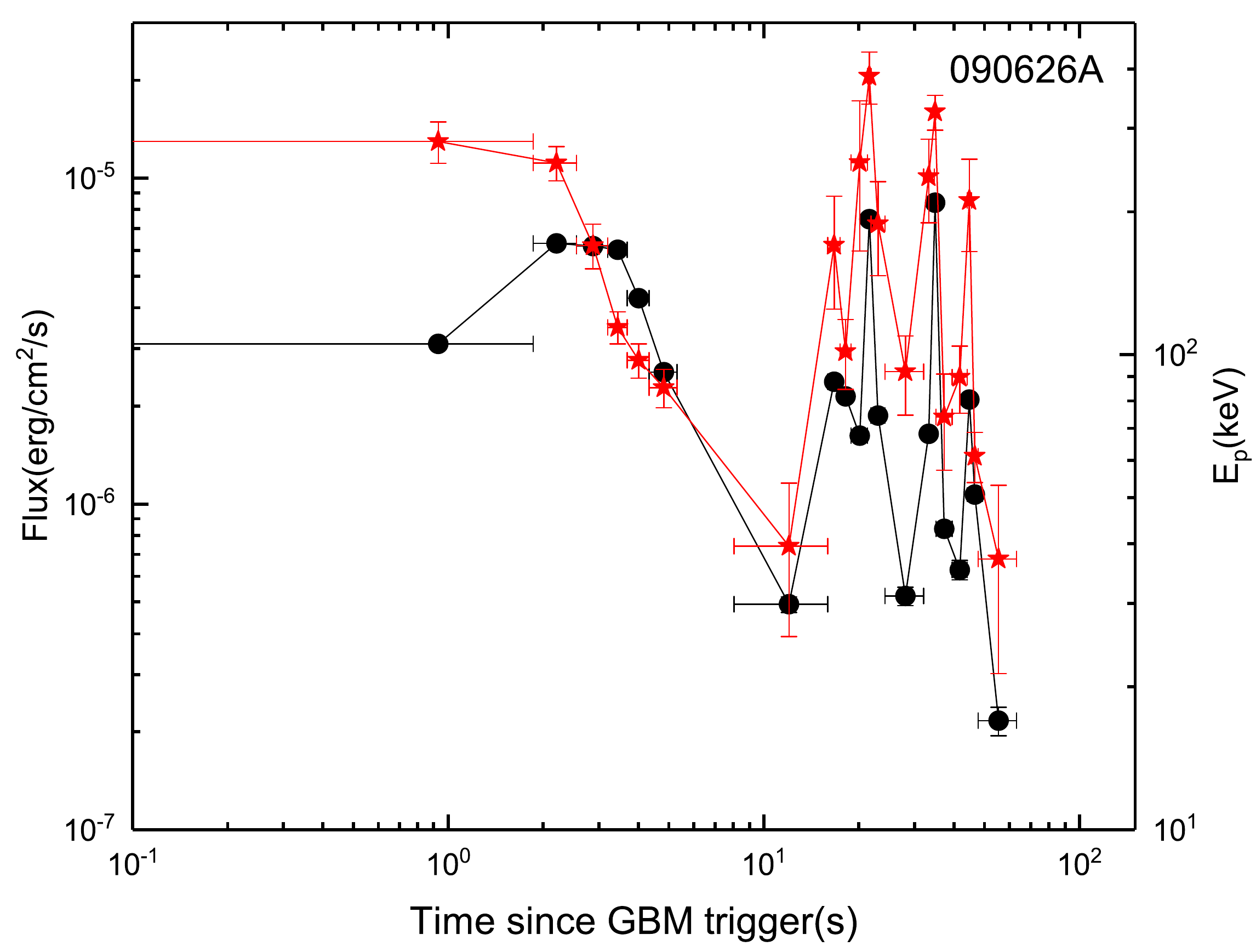}}
\resizebox{4cm}{!}{\includegraphics{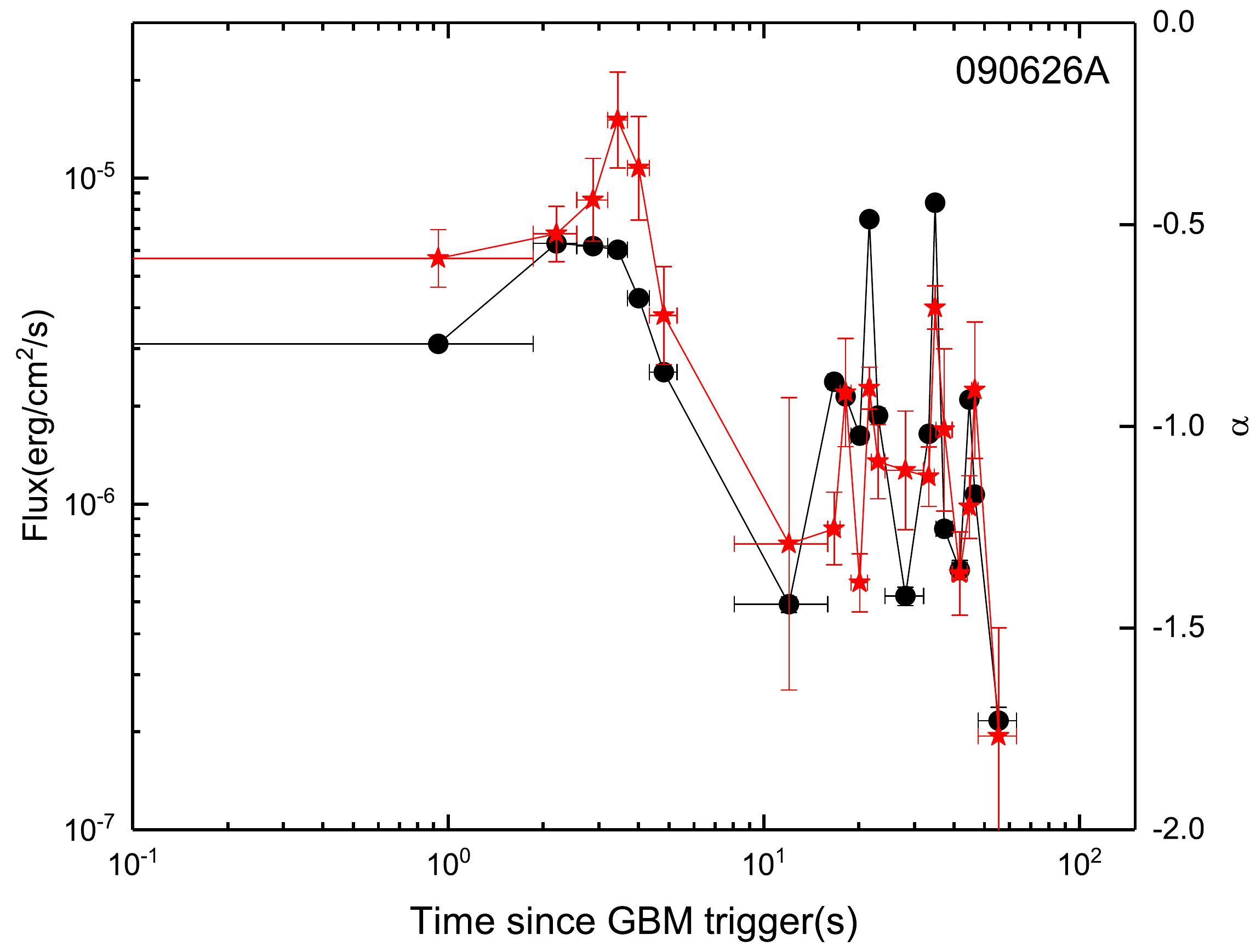}}
\resizebox{4cm}{!}{\includegraphics{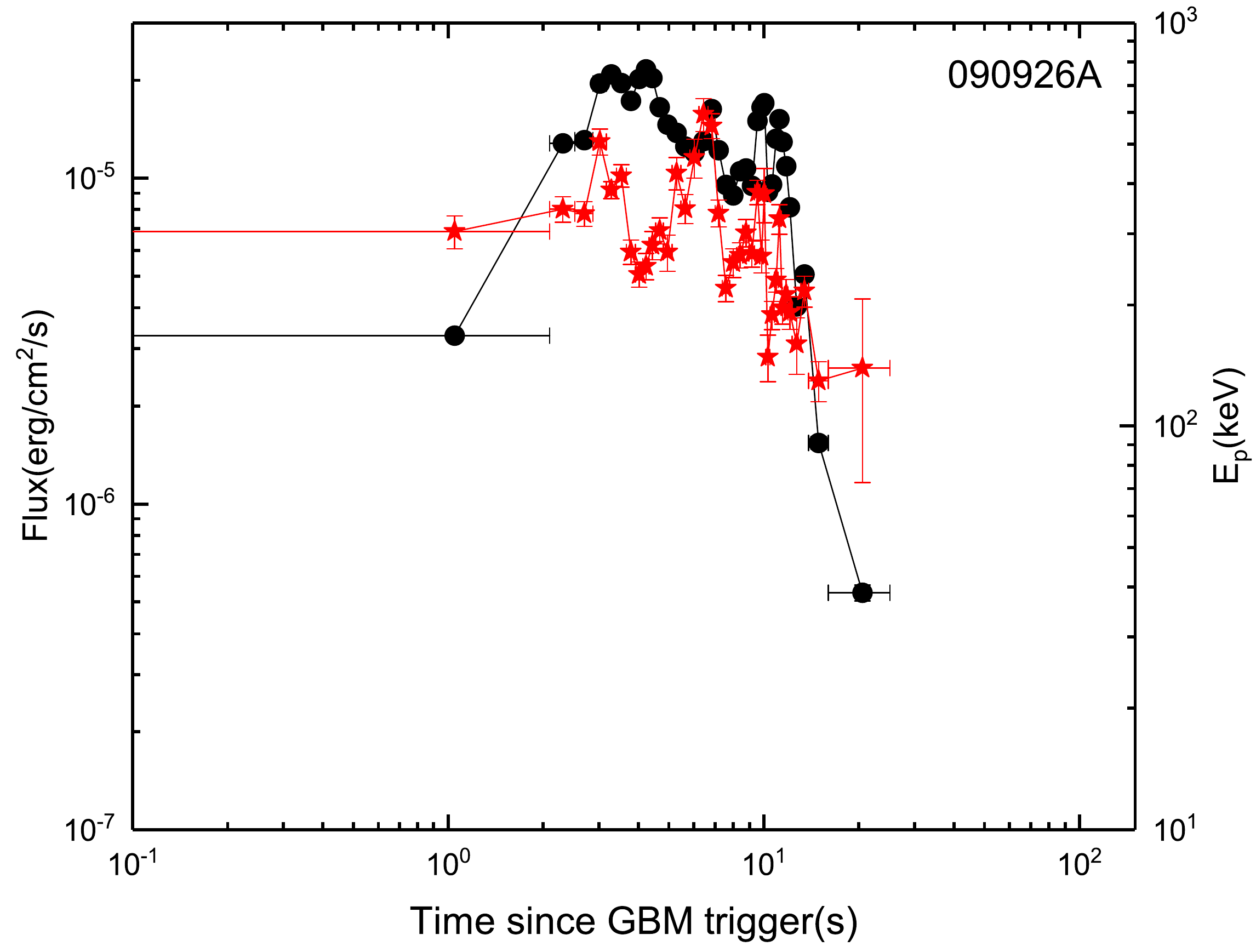}}
\resizebox{4cm}{!}{\includegraphics{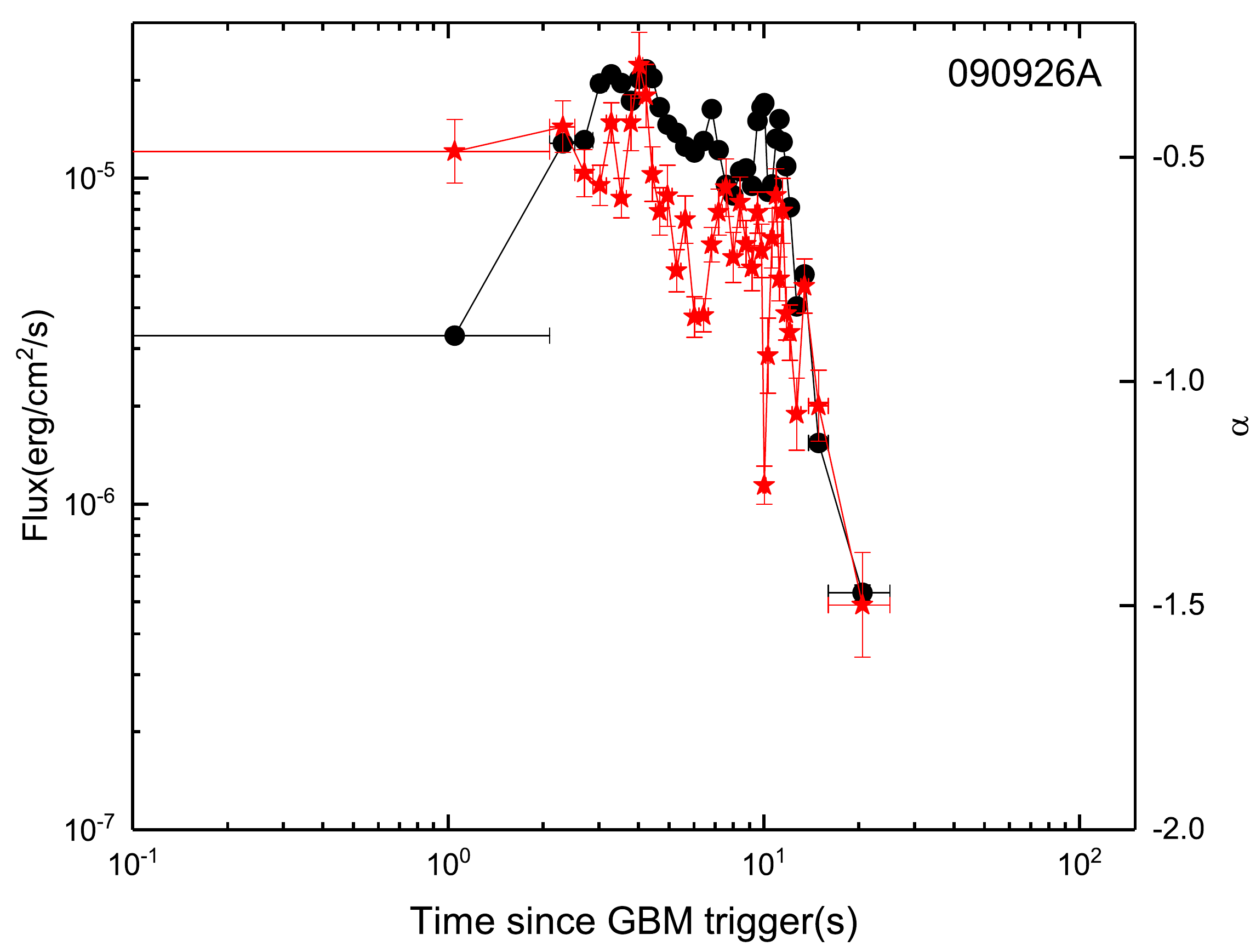}}
\resizebox{4cm}{!}{\includegraphics{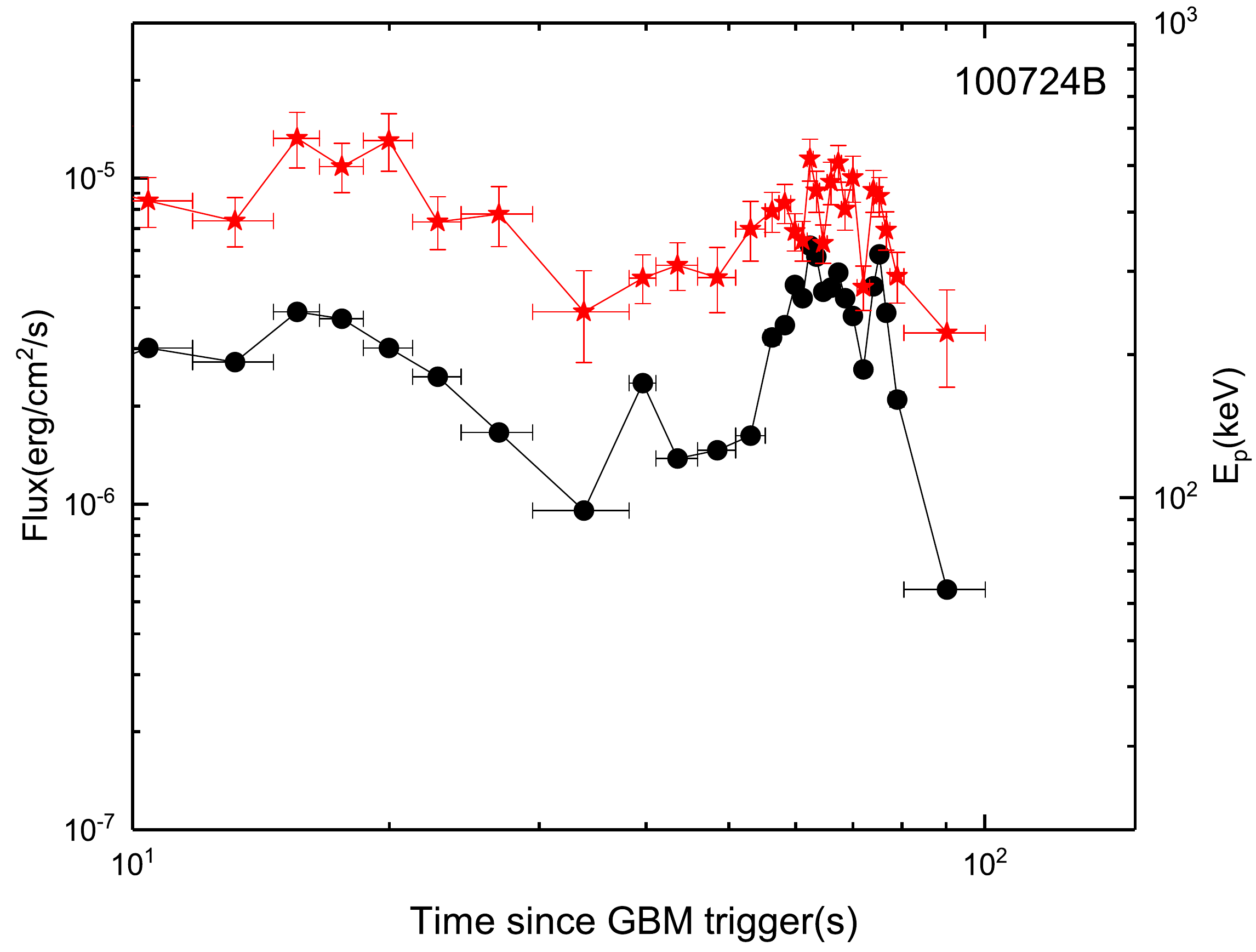}}
\resizebox{4cm}{!}{\includegraphics{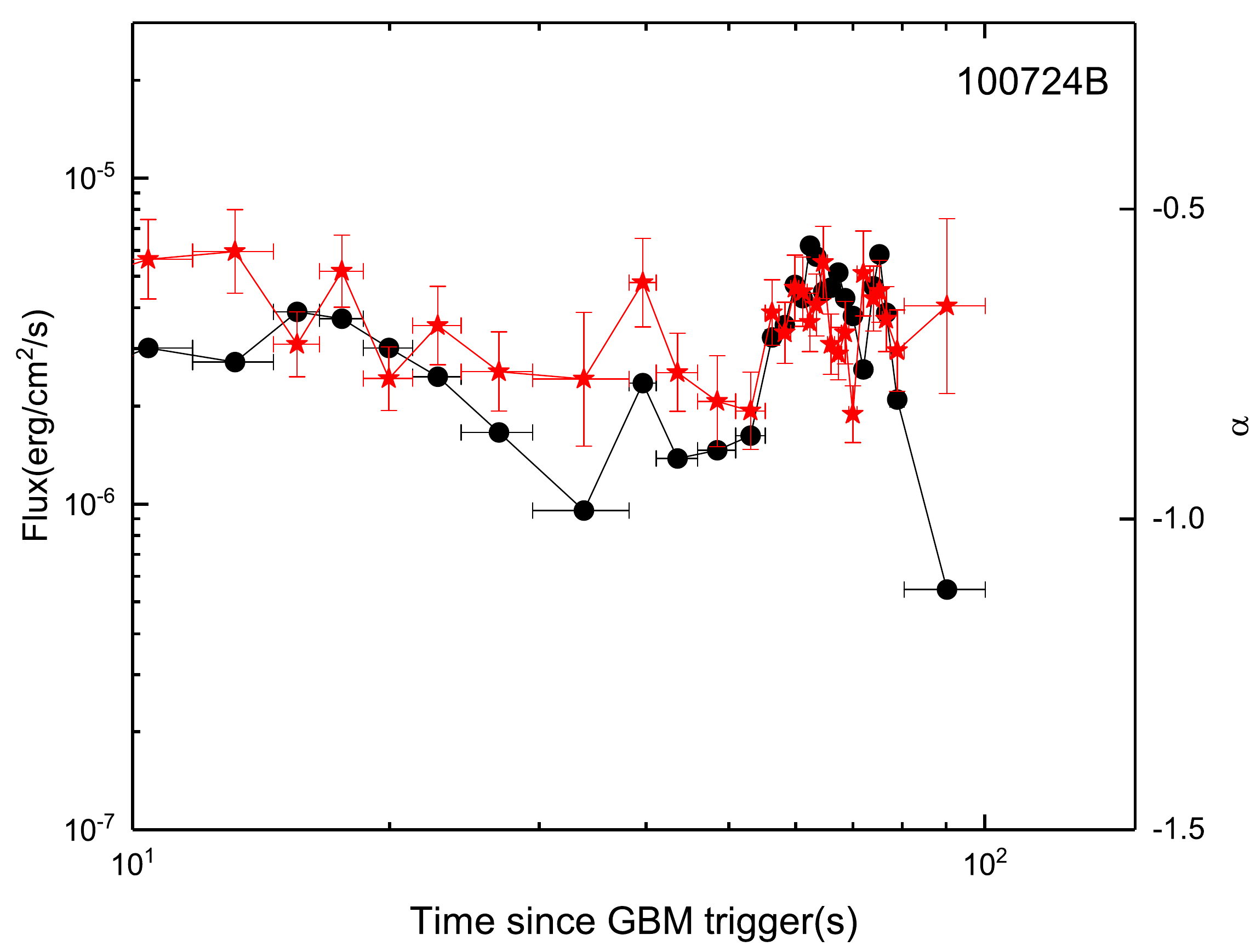}} 
\resizebox{4cm}{!}{\includegraphics{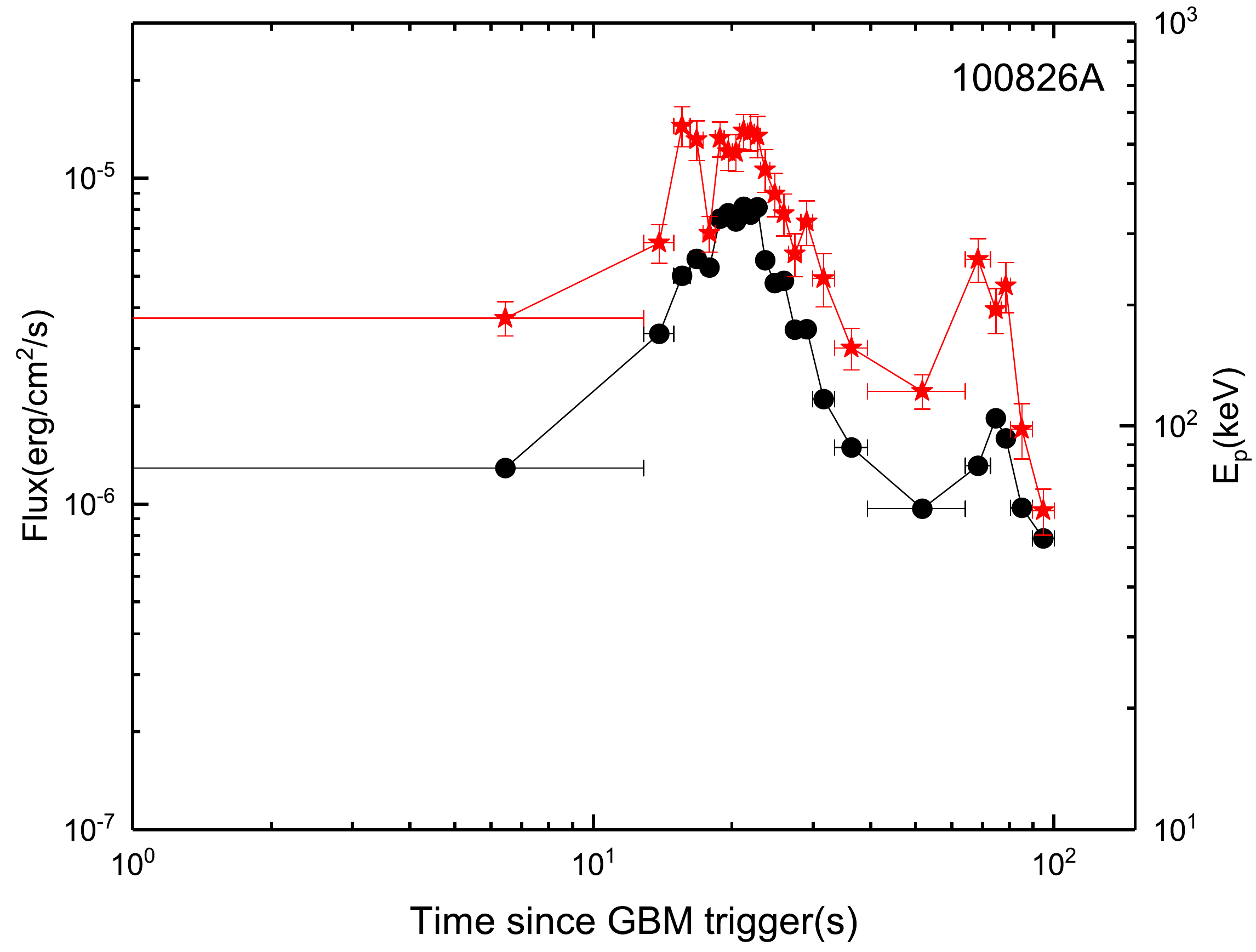}}
\resizebox{4cm}{!}{\includegraphics{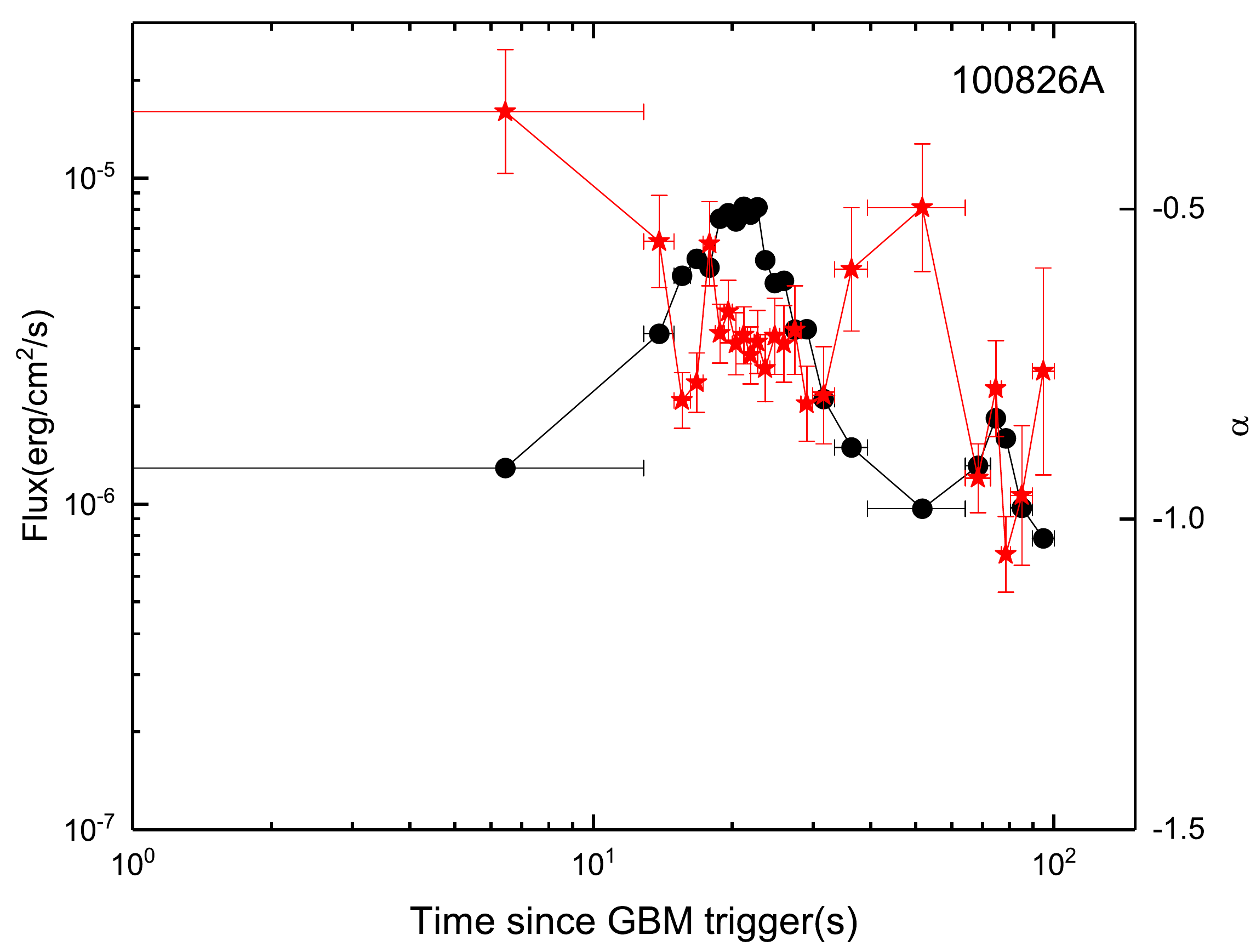}}
\resizebox{4cm}{!}{\includegraphics{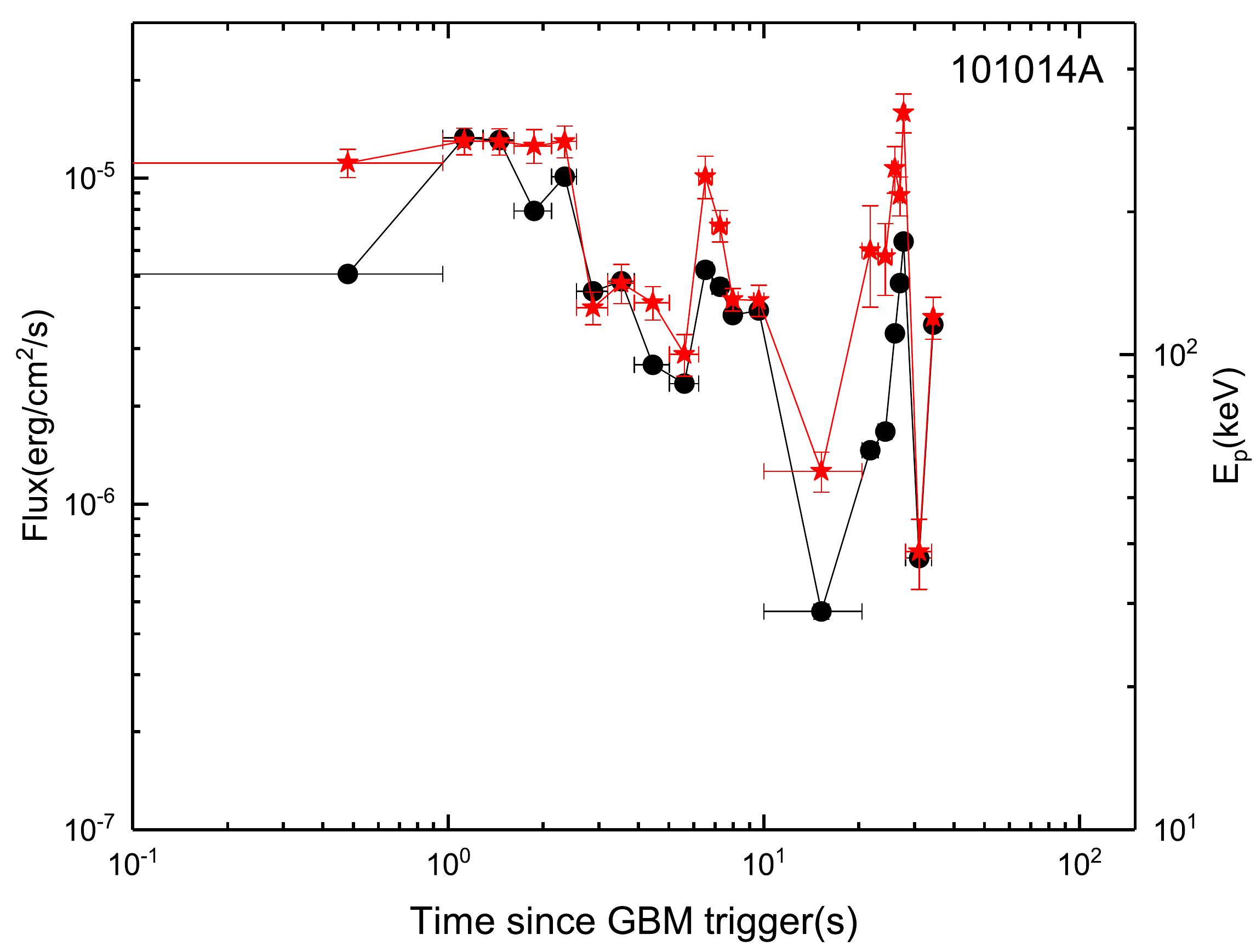}}
\resizebox{4cm}{!}{\includegraphics{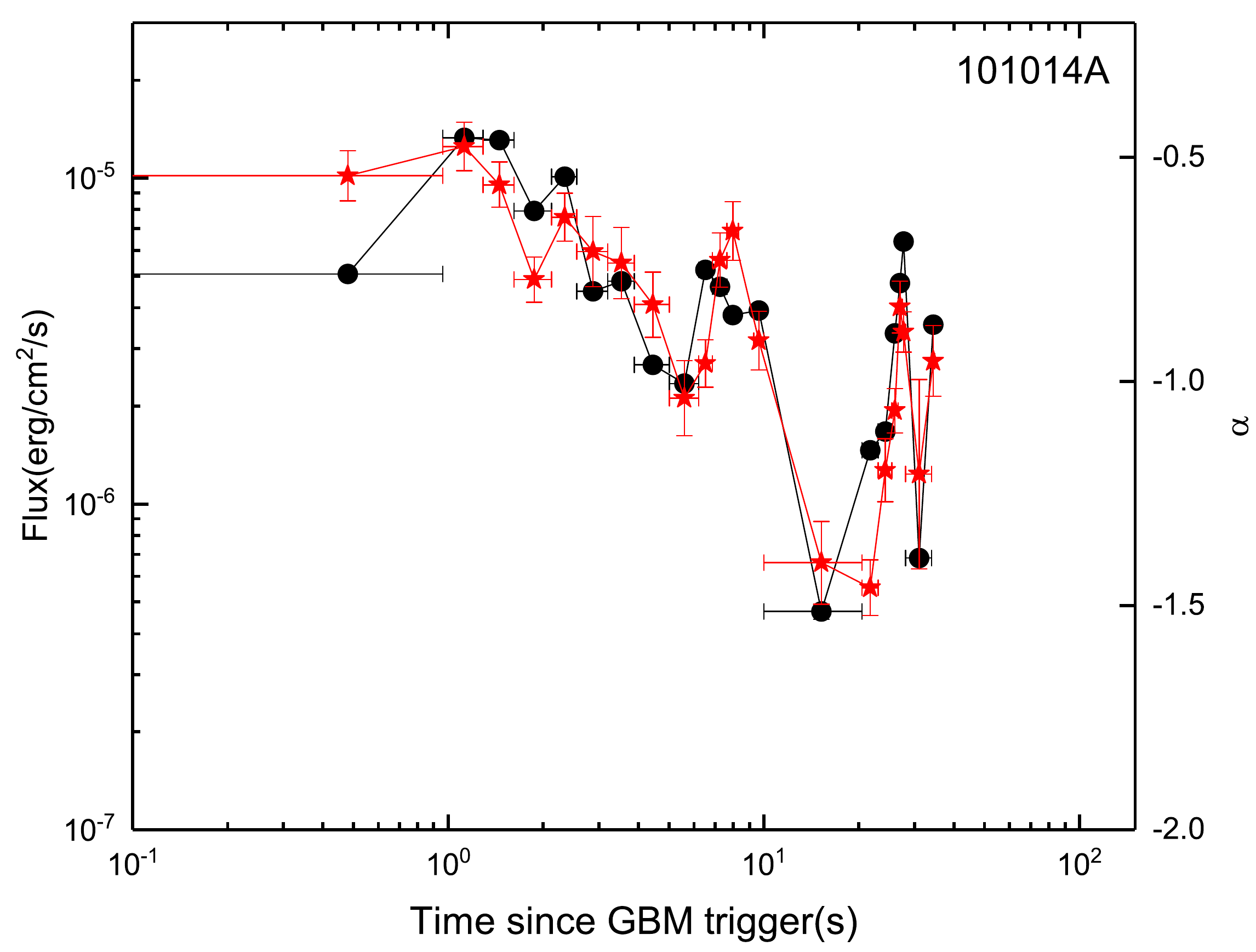}}
\resizebox{4cm}{!}{\includegraphics{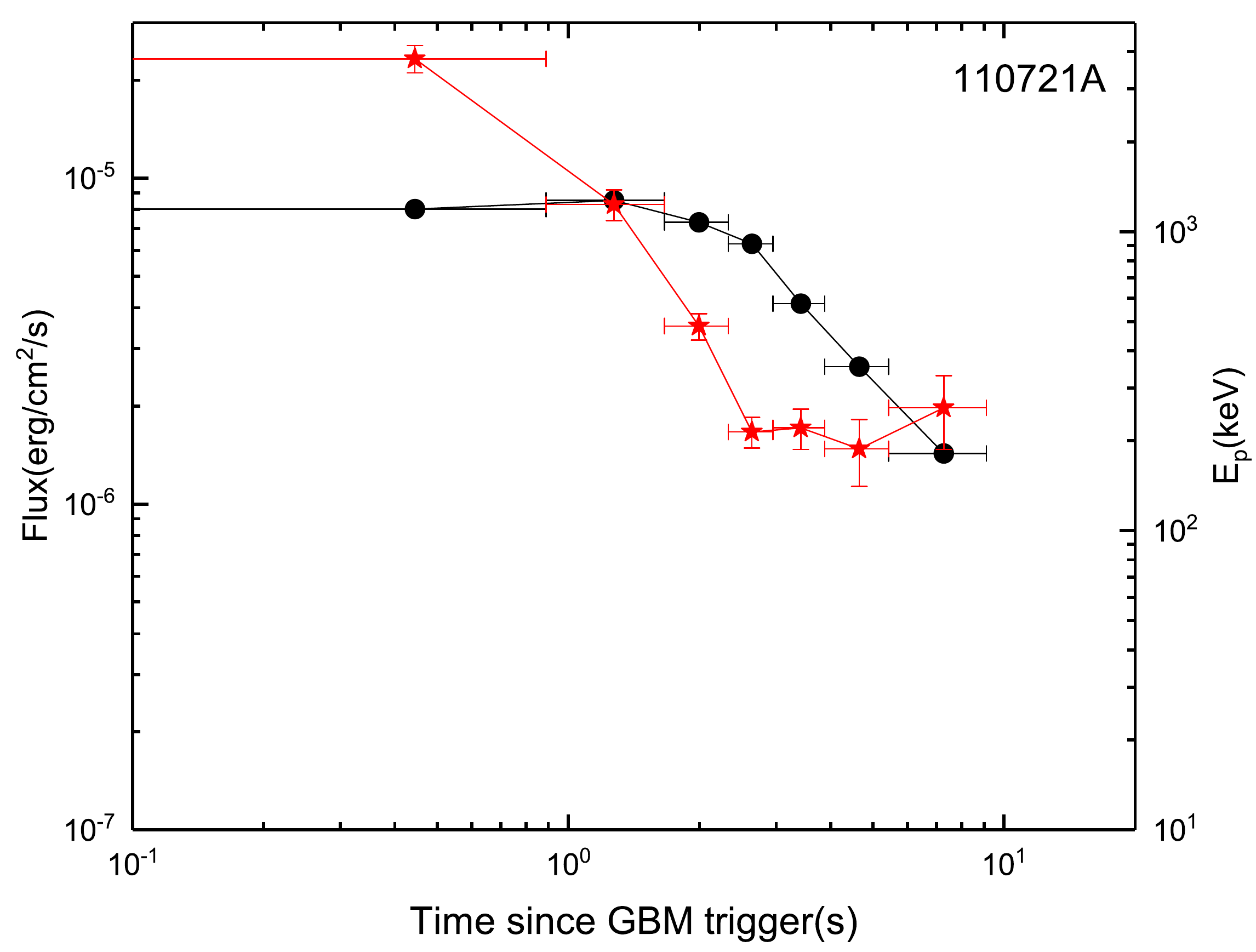}}
\resizebox{4cm}{!}{\includegraphics{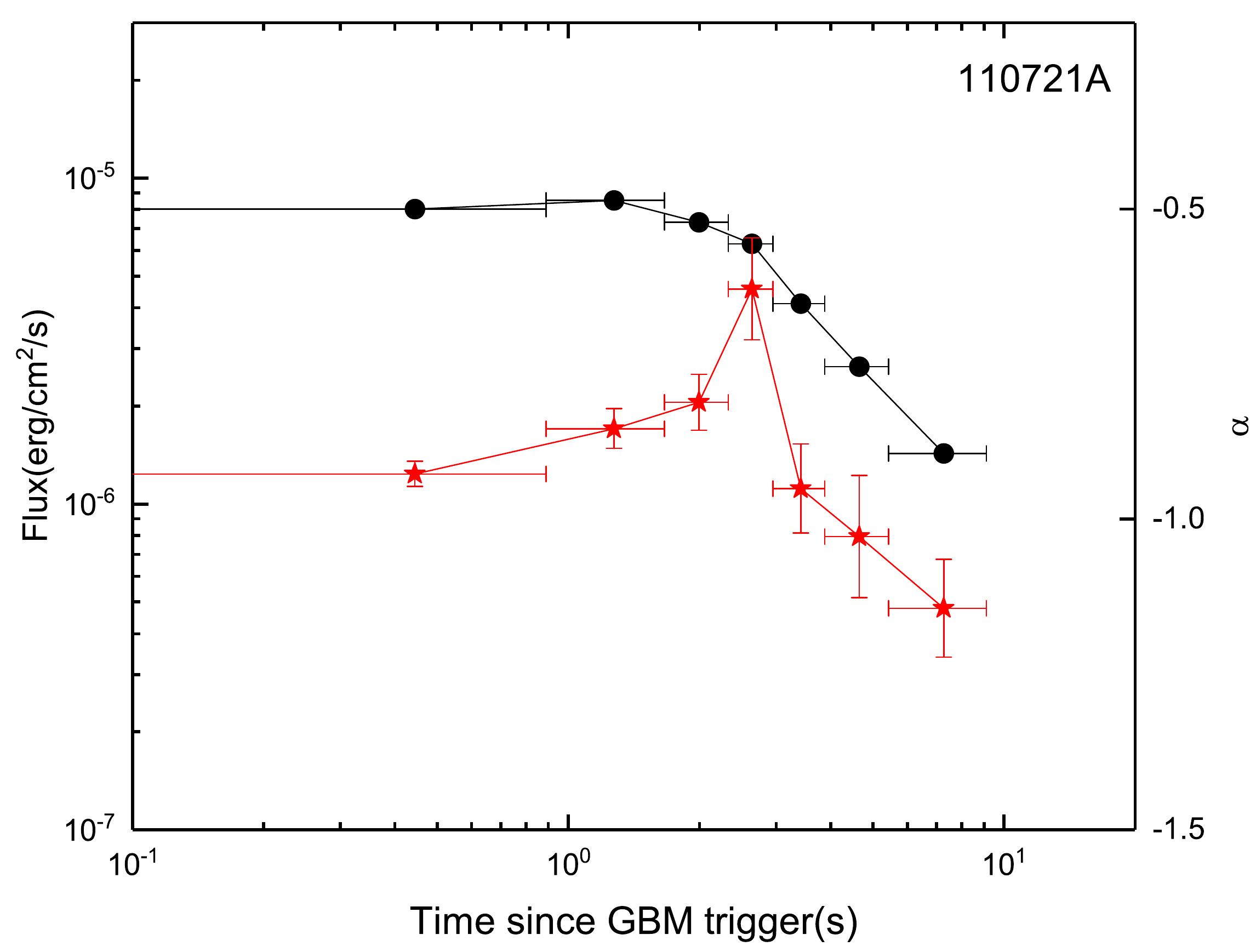}}
\resizebox{4cm}{!}{\includegraphics{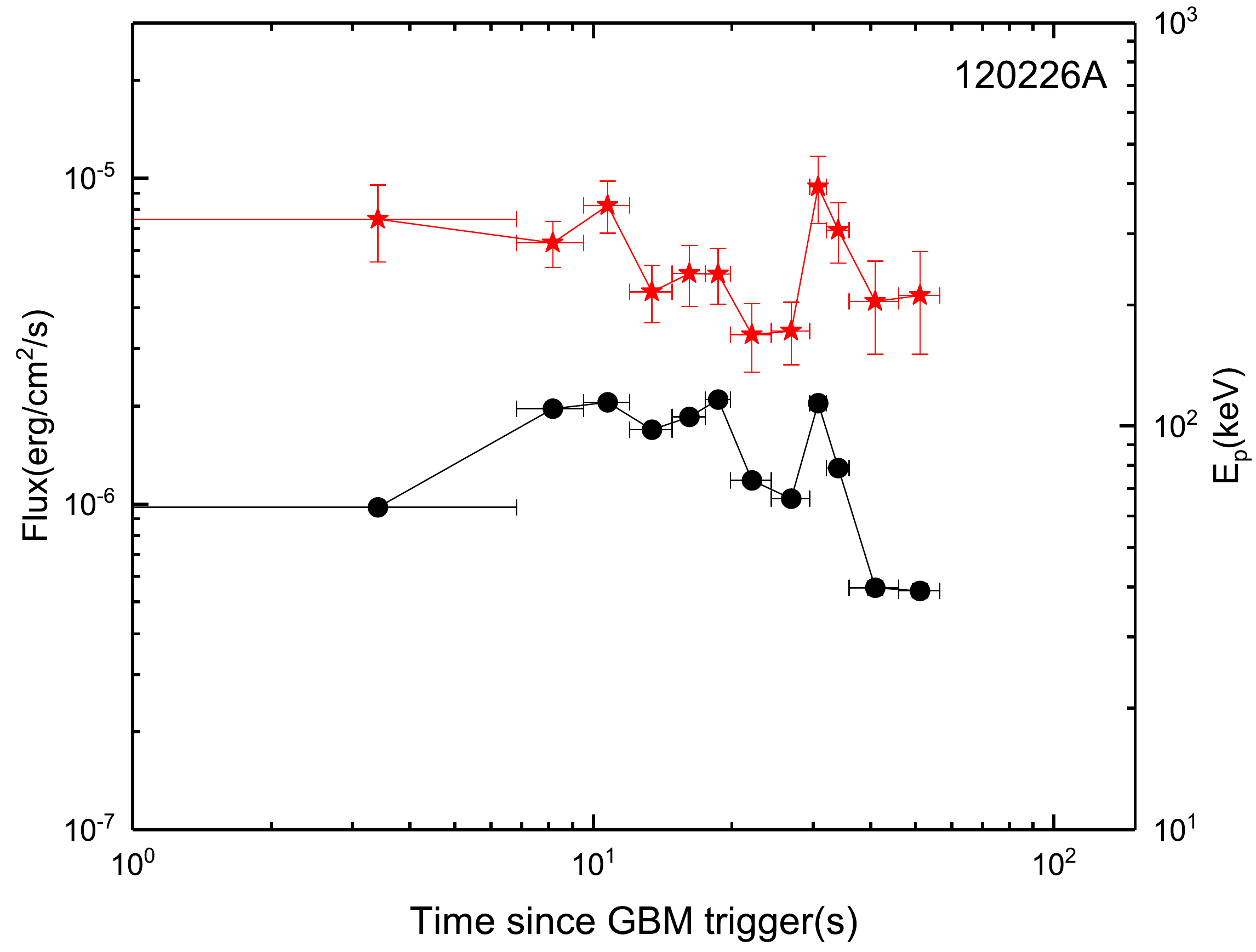}}
\resizebox{4cm}{!}{\includegraphics{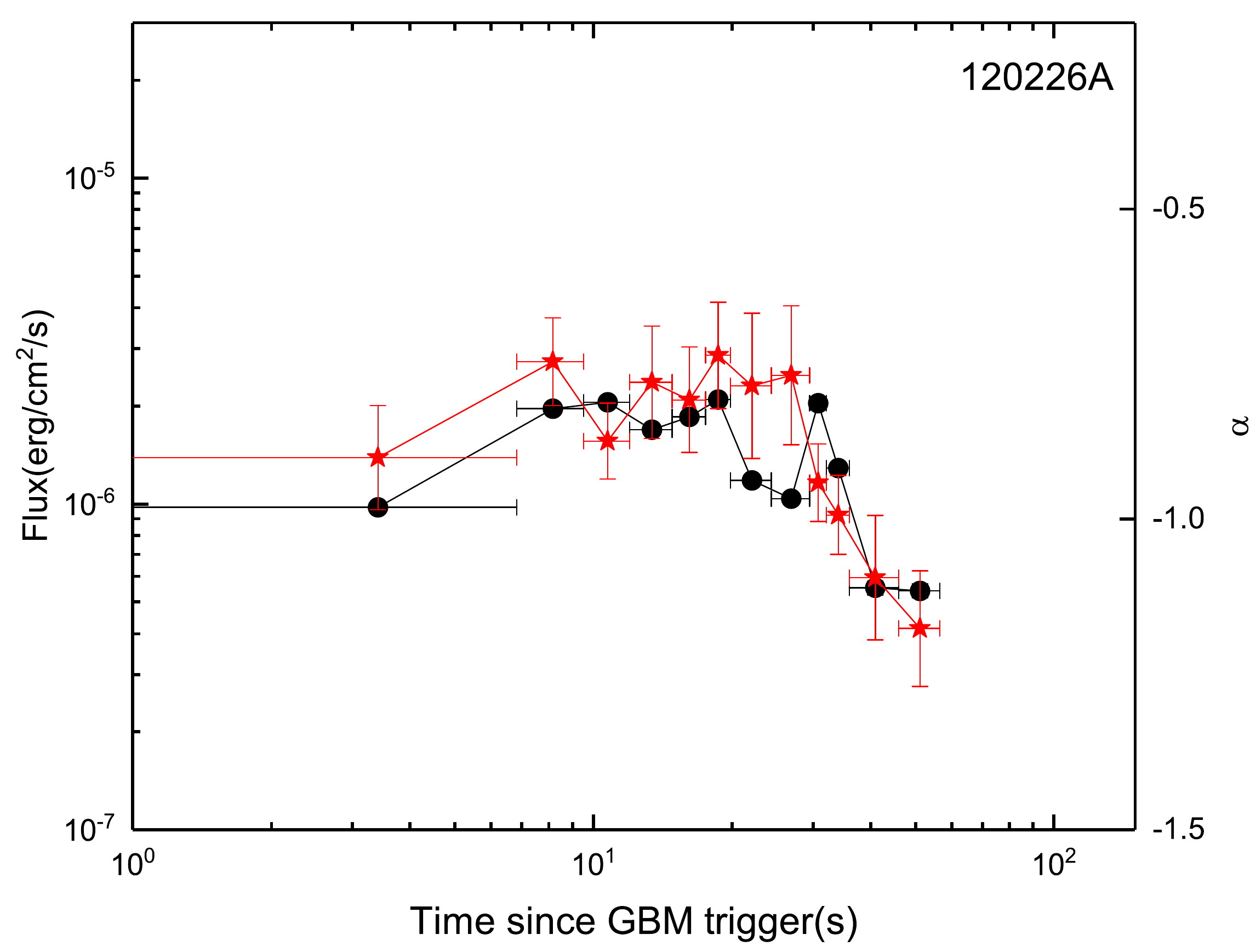}}
\resizebox{4cm}{!}{\includegraphics{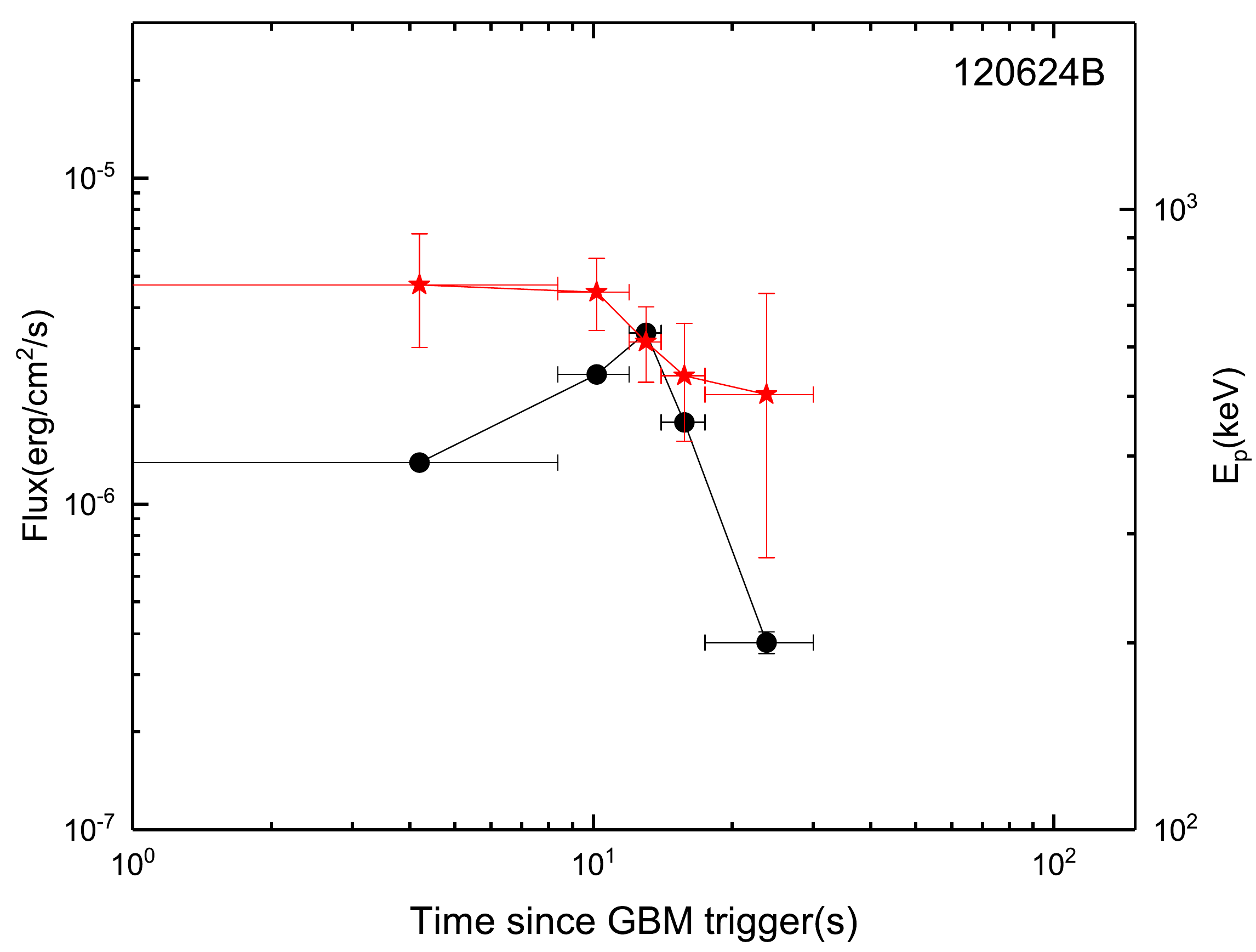}}
\resizebox{4cm}{!}{\includegraphics{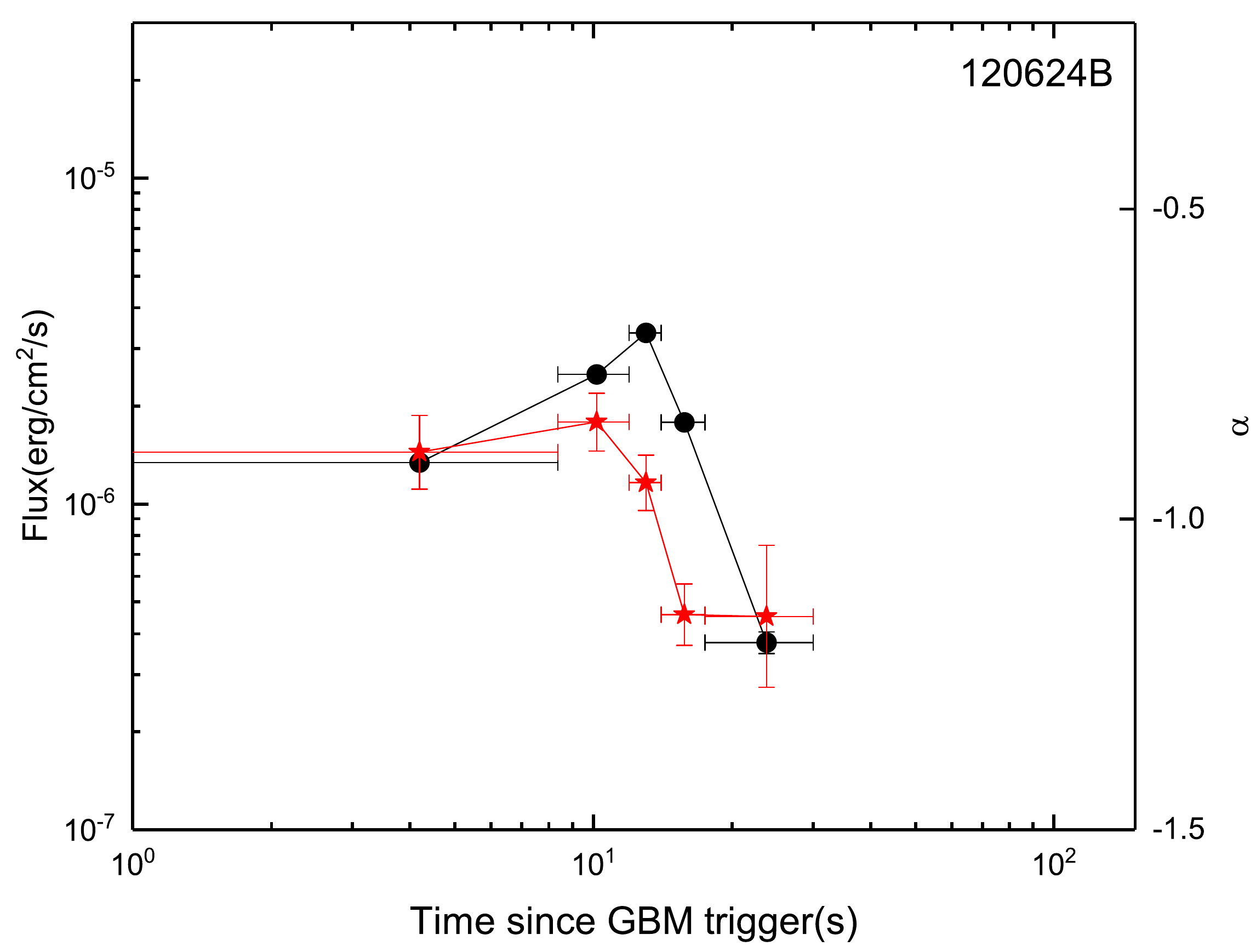}}
\resizebox{4cm}{!}{\includegraphics{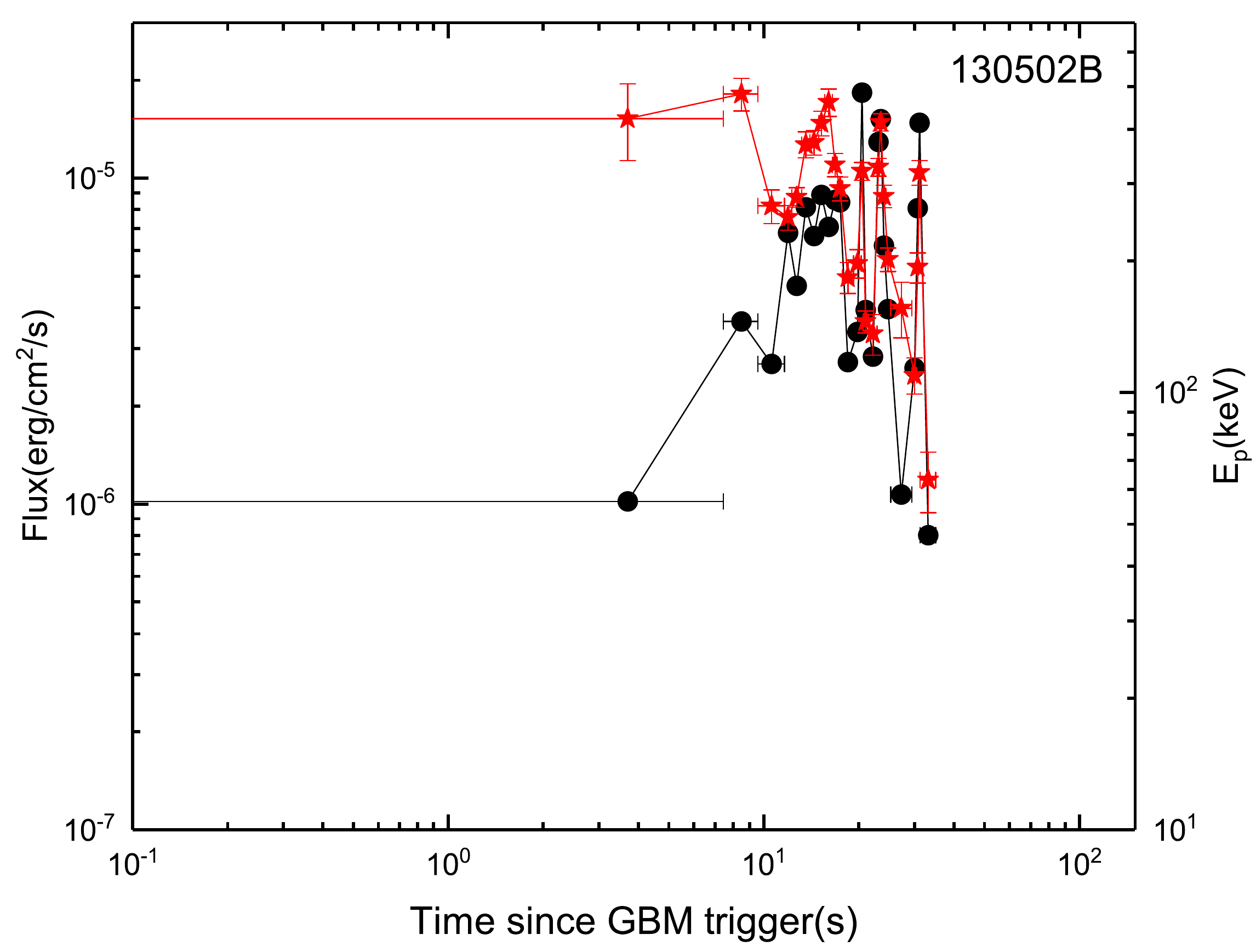}}
\resizebox{4cm}{!}{\includegraphics{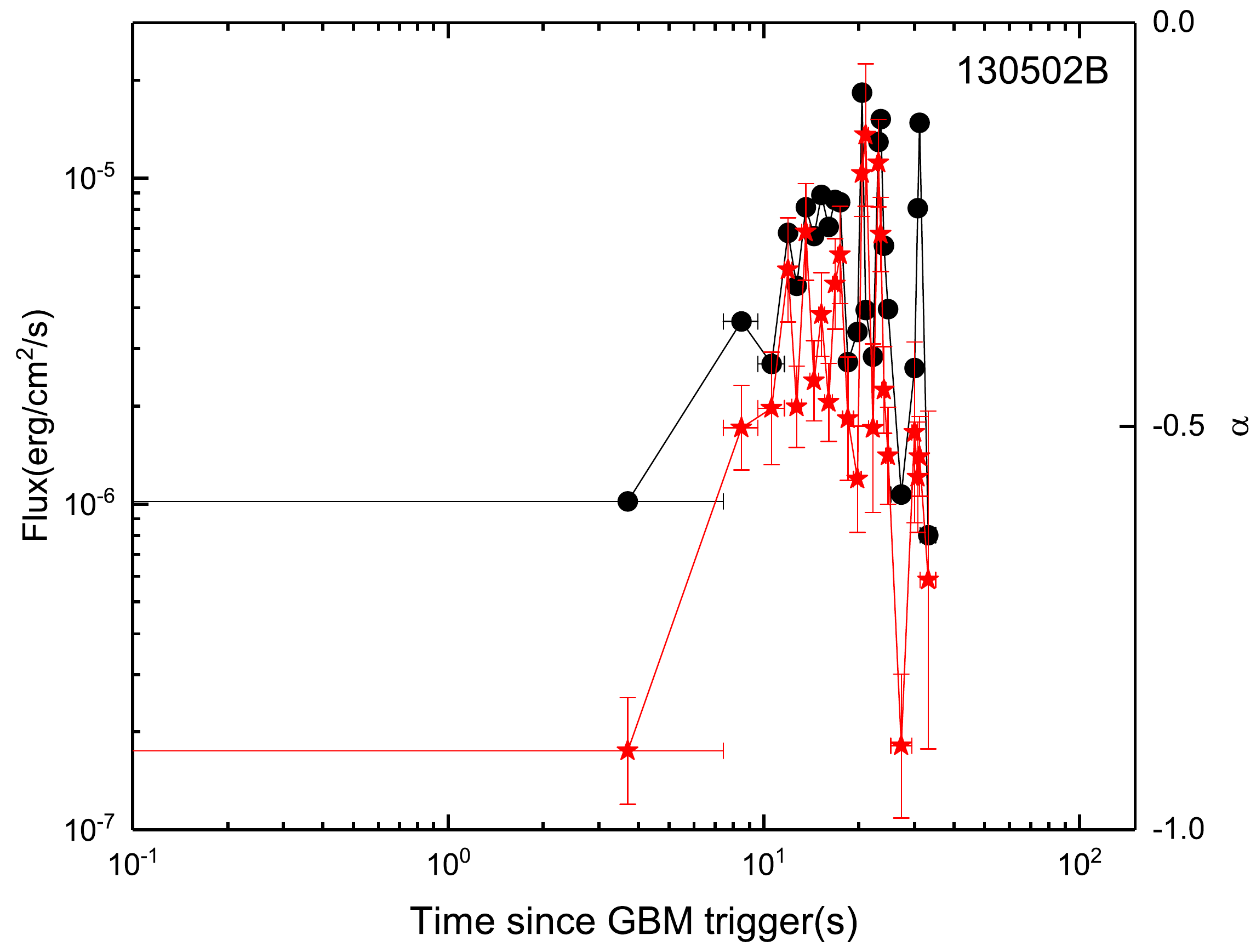}} 
\resizebox{4cm}{!}{\includegraphics{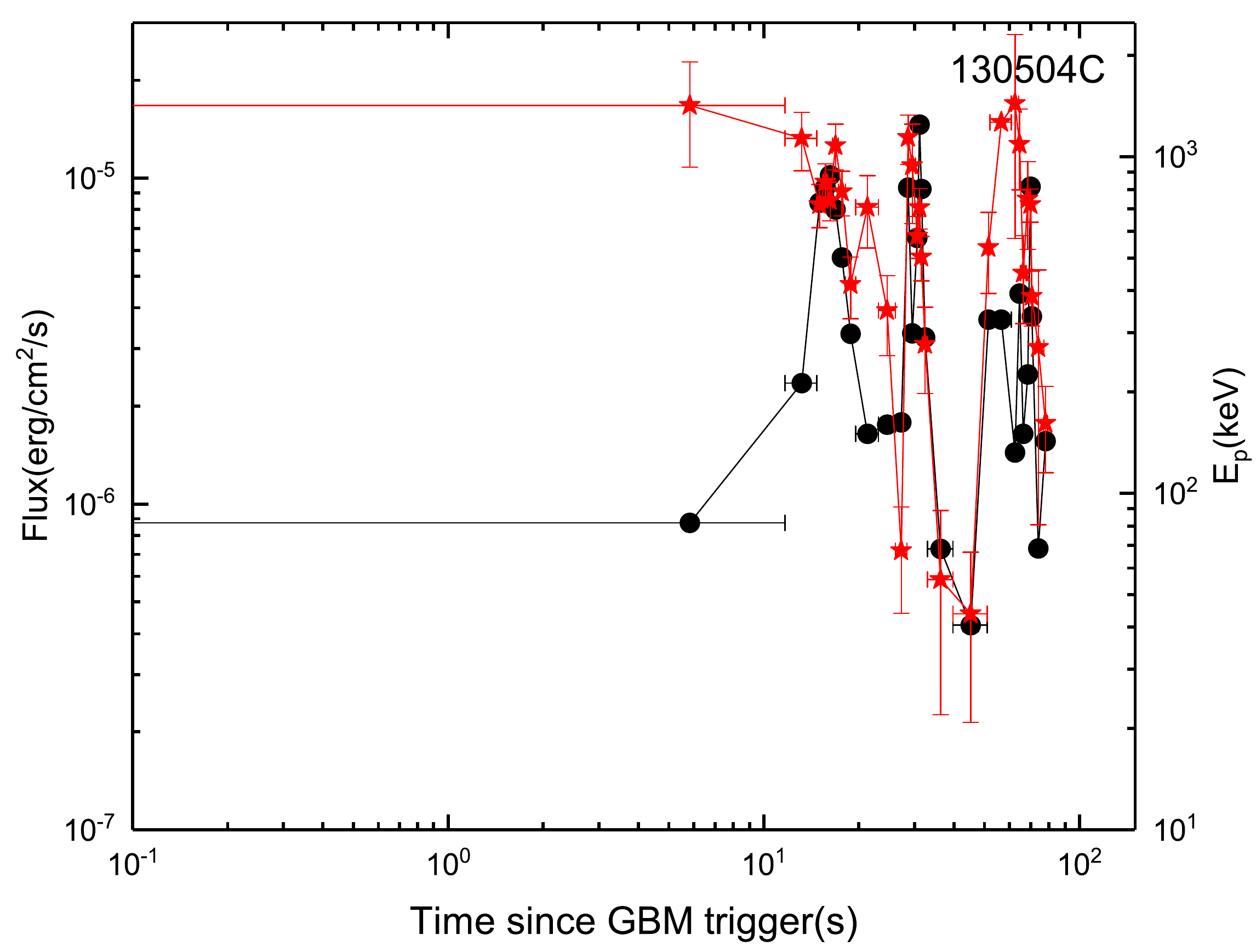}}
\resizebox{4cm}{!}{\includegraphics{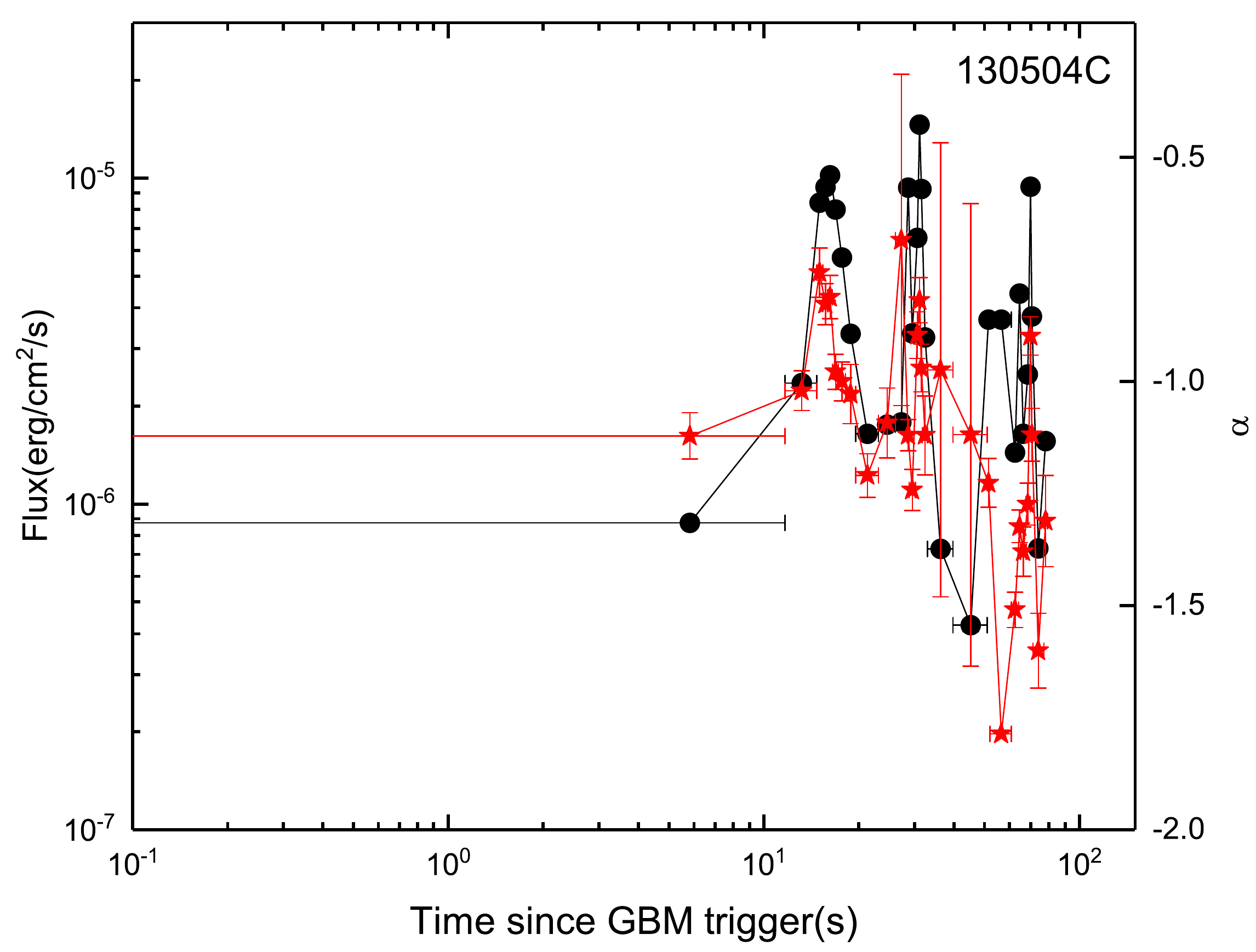}}    
\caption{Spectral evolutions. The temporal characteristics of energy flux for all bursts in our sample (the left-hand, y-axis), along with time evolutions of $E_{p}$ and $\alpha$, both are marked with red stars in the right-hand y-axis.\label{fig:spectral evolutions}}
\end{figure}

\addtocounter{figure}{-1}
\begin{figure}
\centering 
\resizebox{4cm}{!}{\includegraphics{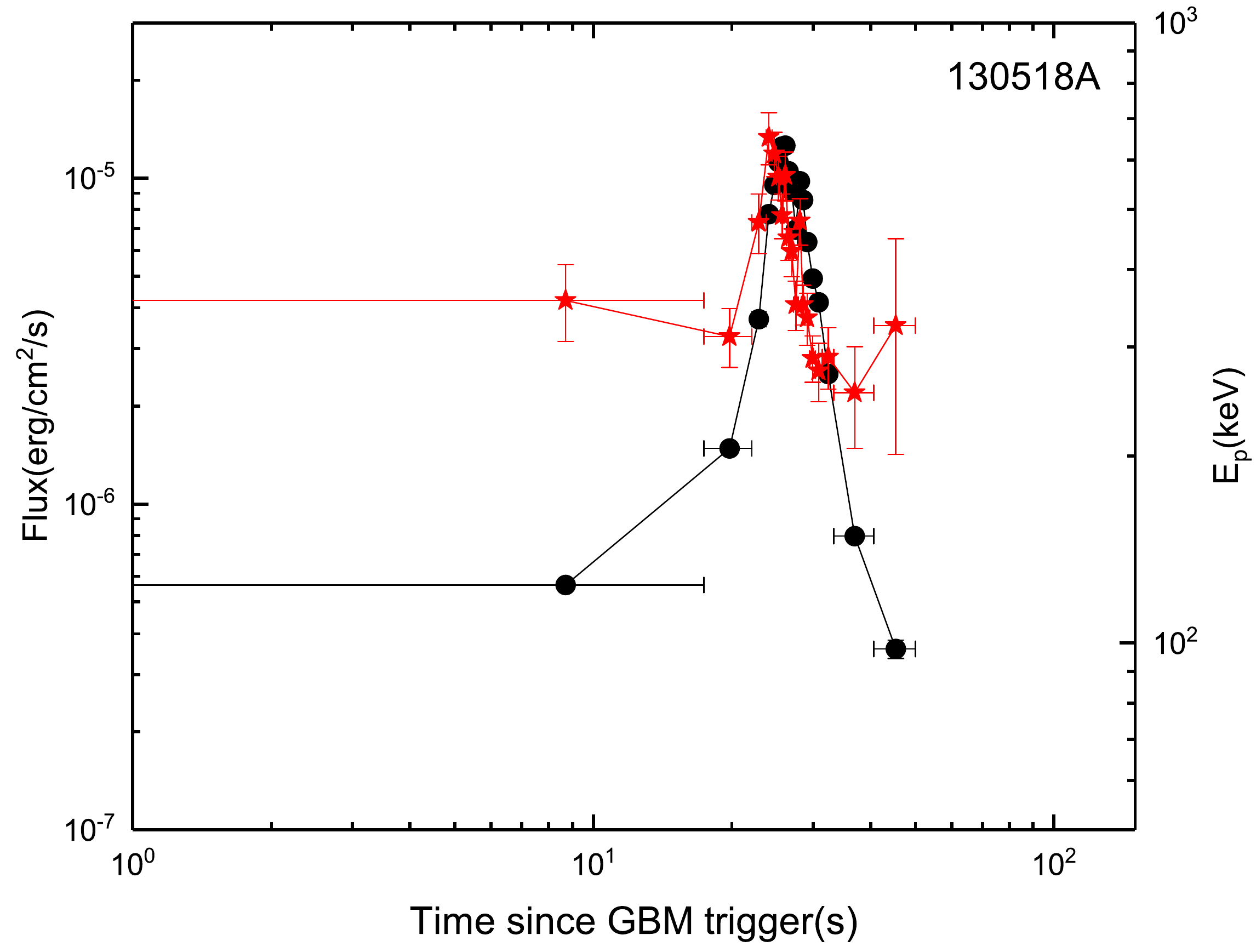}}
\resizebox{4cm}{!}{\includegraphics{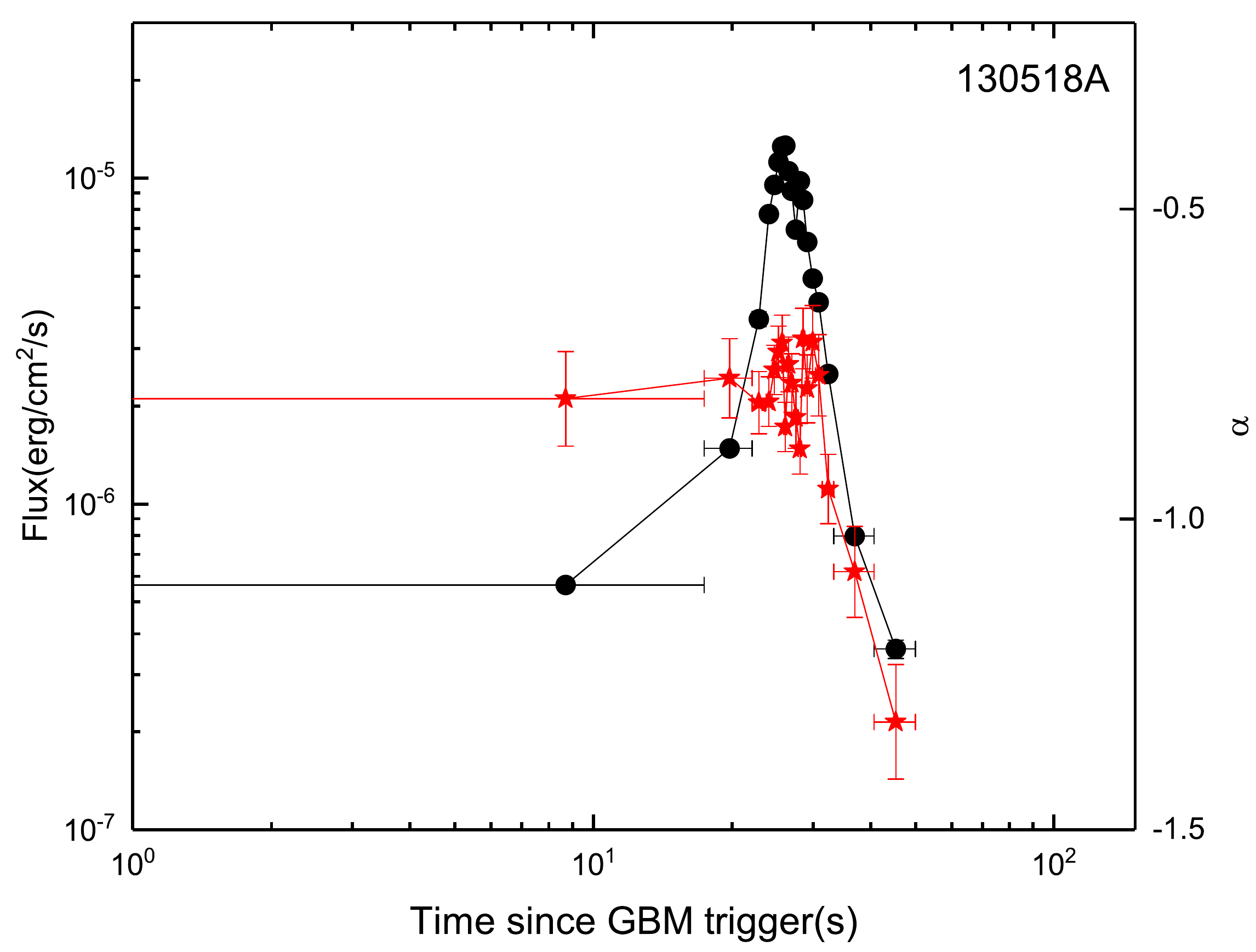}}
\resizebox{4cm}{!}{\includegraphics{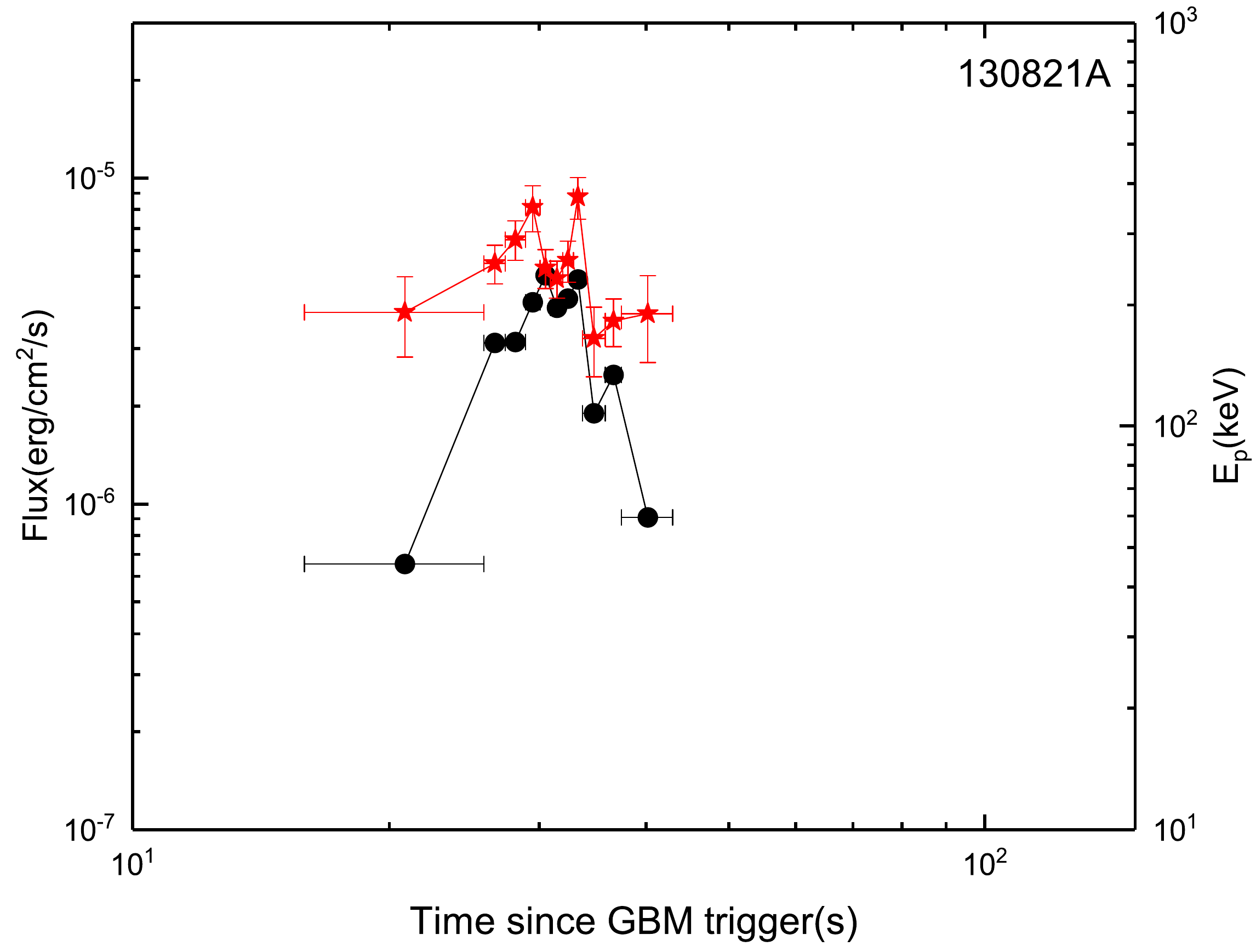}}
\resizebox{4cm}{!}{\includegraphics{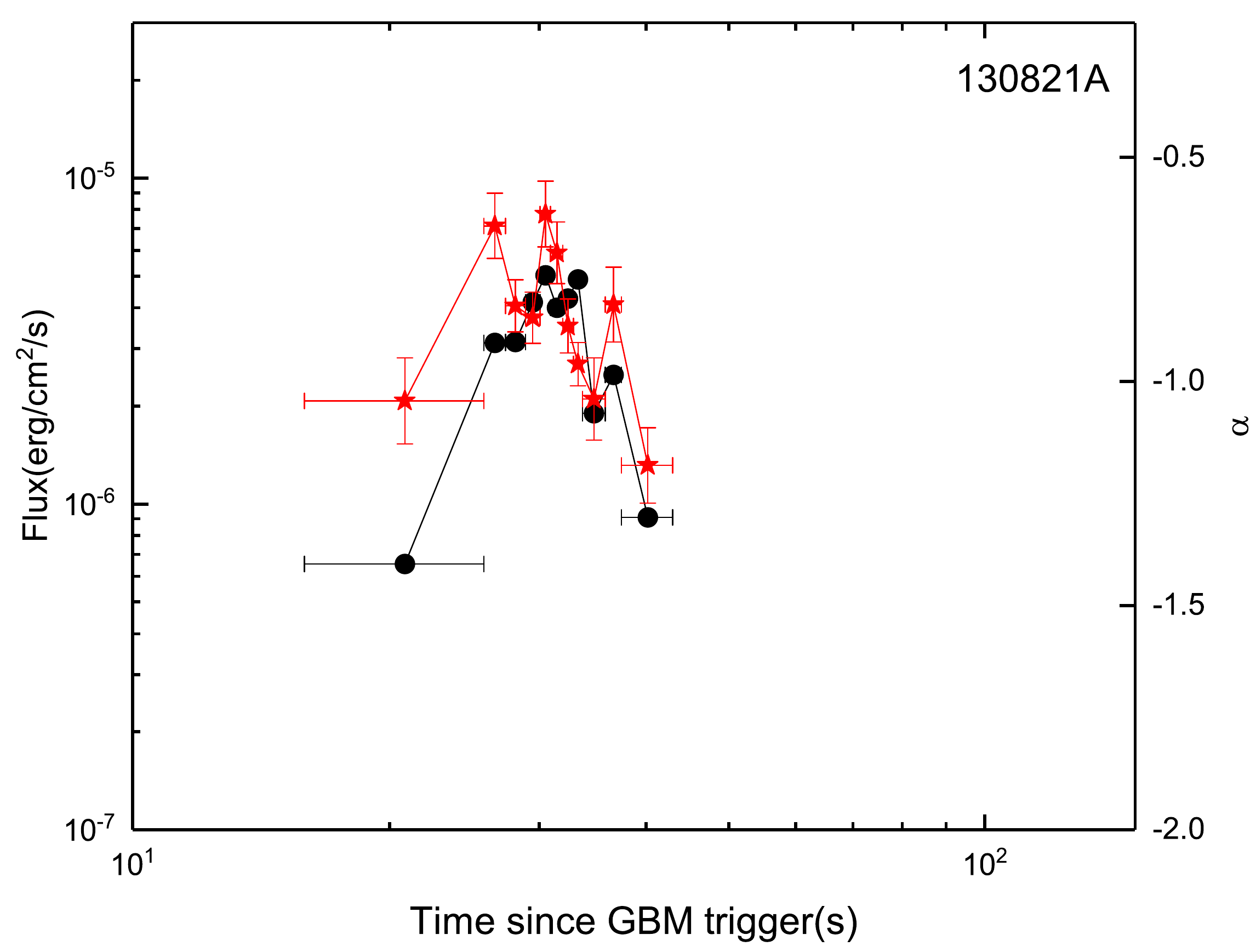}}   
\resizebox{4cm}{!}{\includegraphics{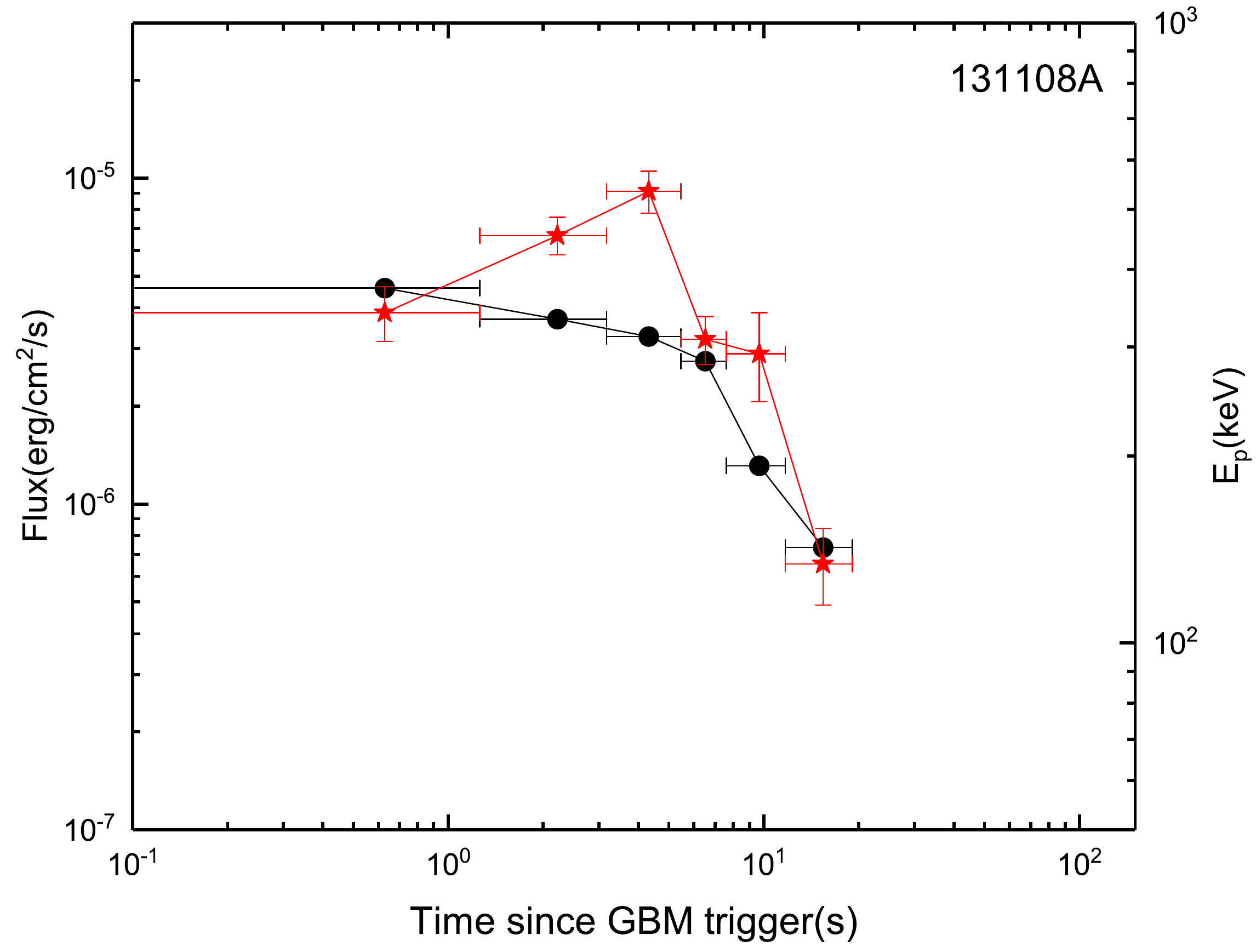}}
\resizebox{4cm}{!}{\includegraphics{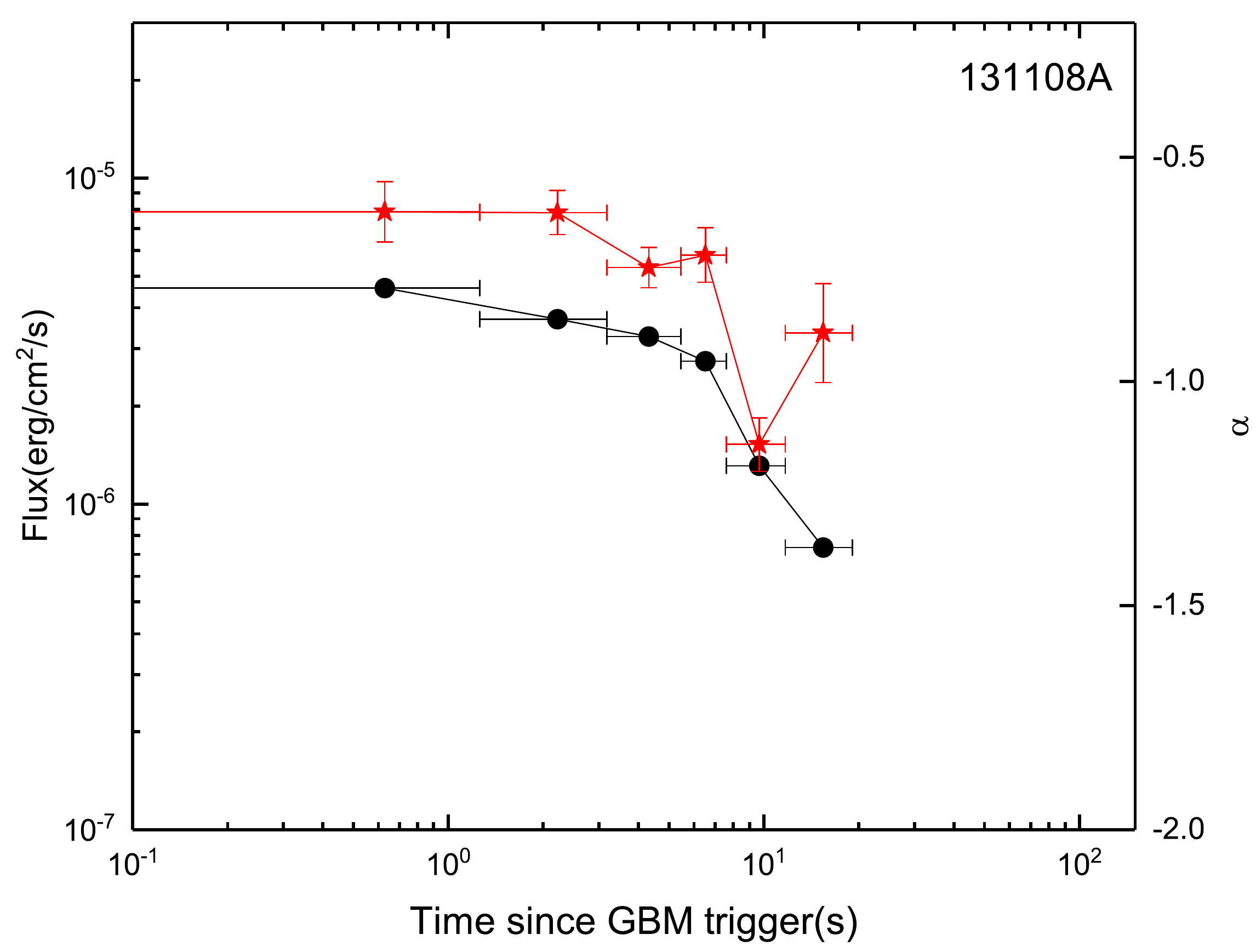}}
\resizebox{4cm}{!}{\includegraphics{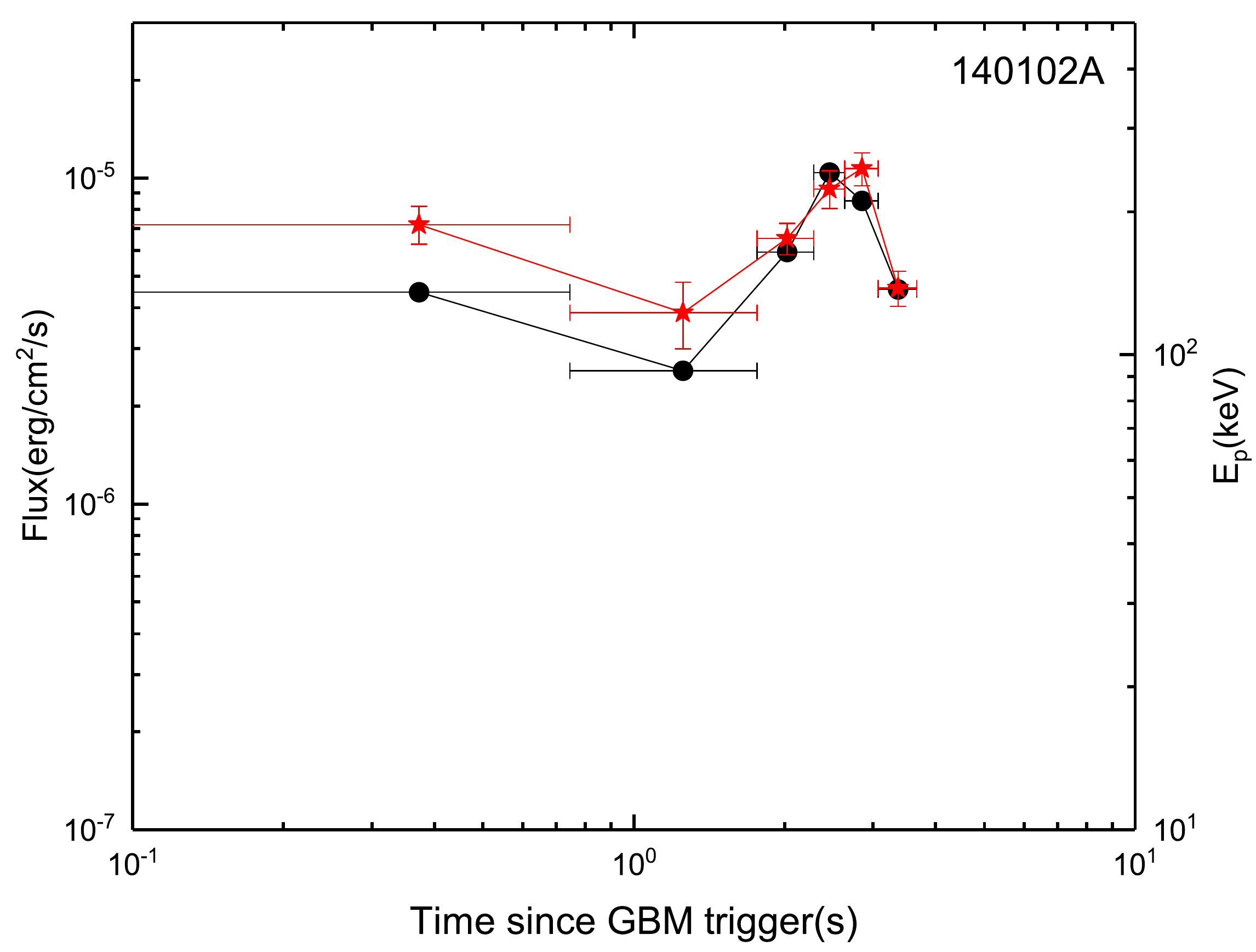}}
\resizebox{4cm}{!}{\includegraphics{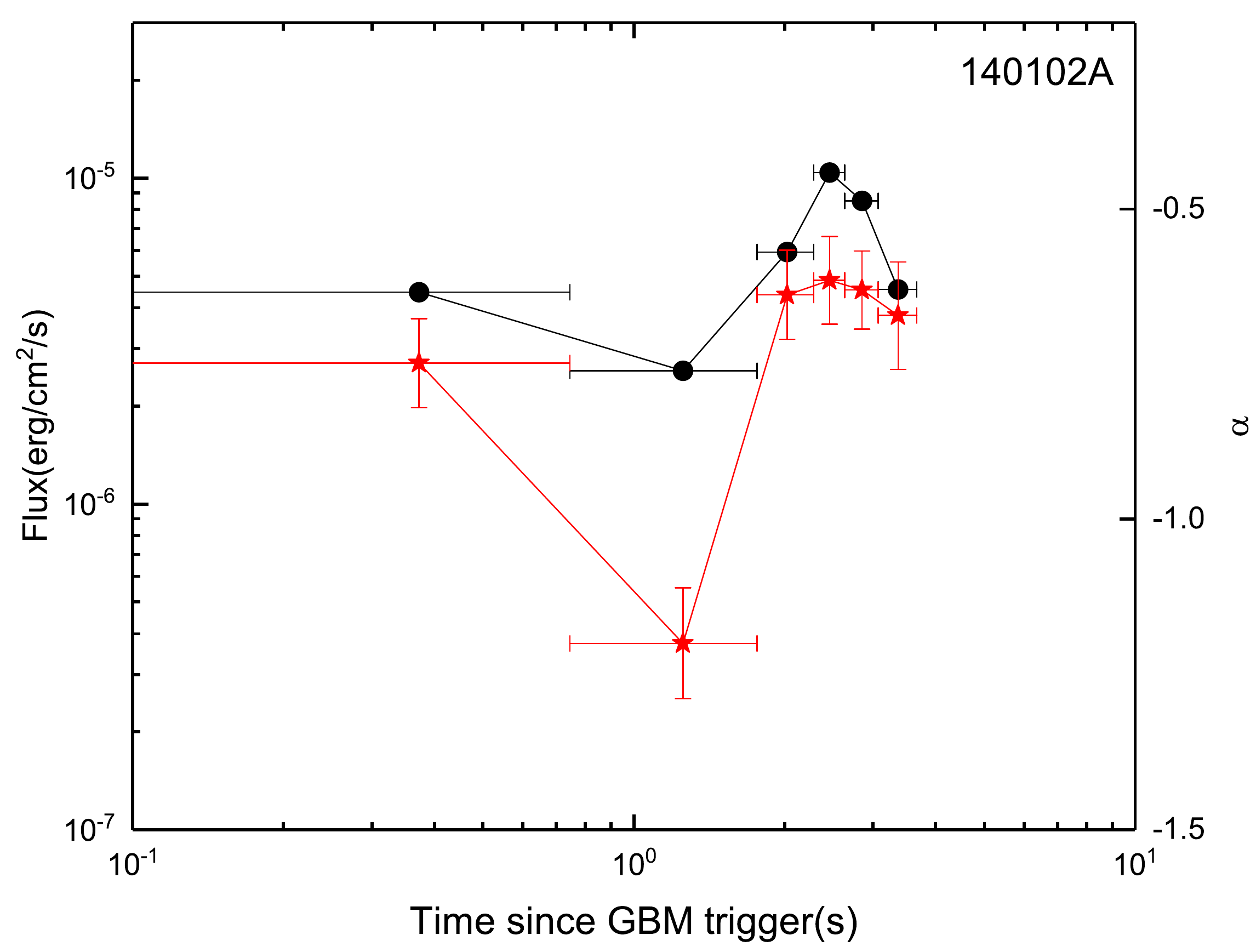}}
\resizebox{4cm}{!}{\includegraphics{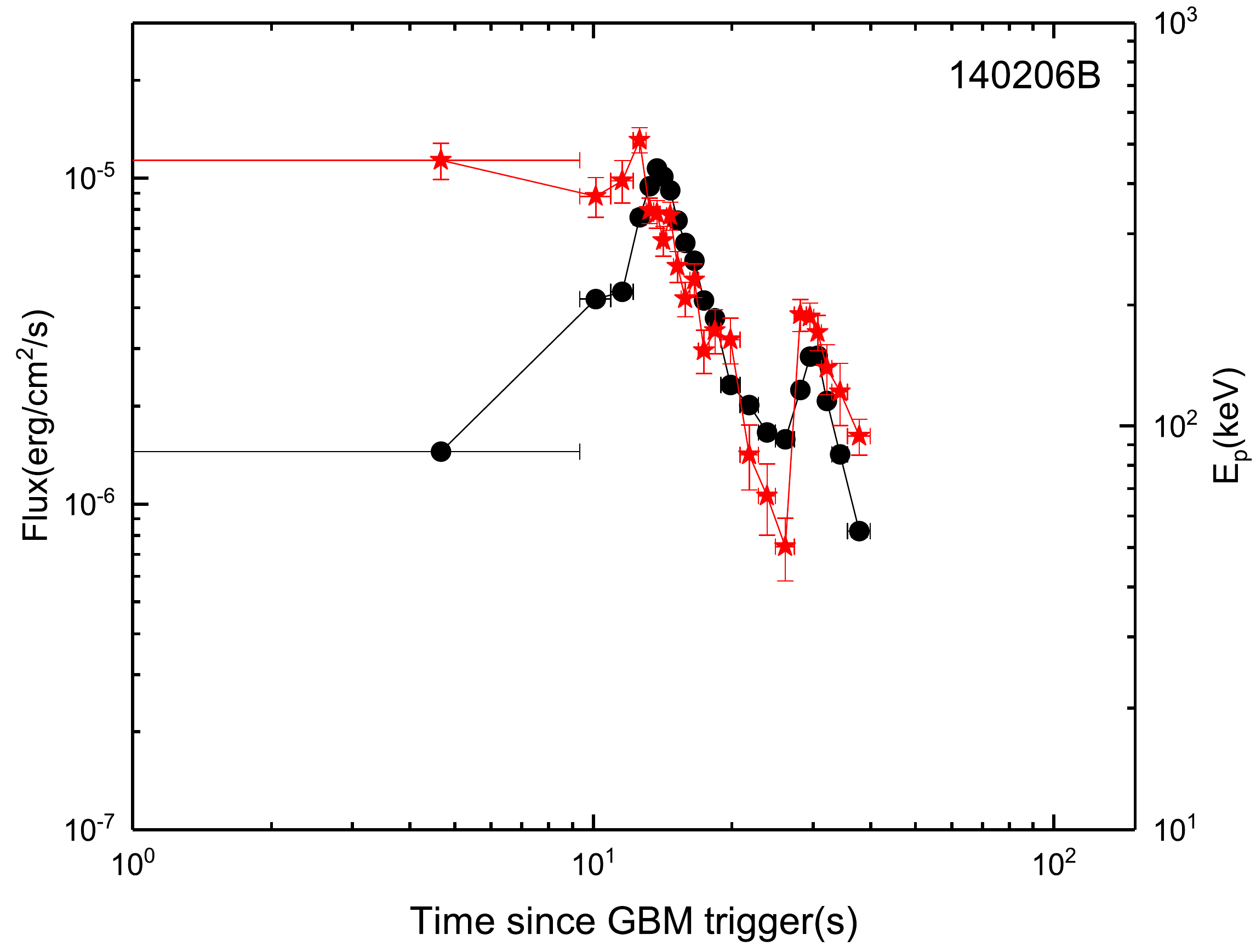}}
\resizebox{4cm}{!}{\includegraphics{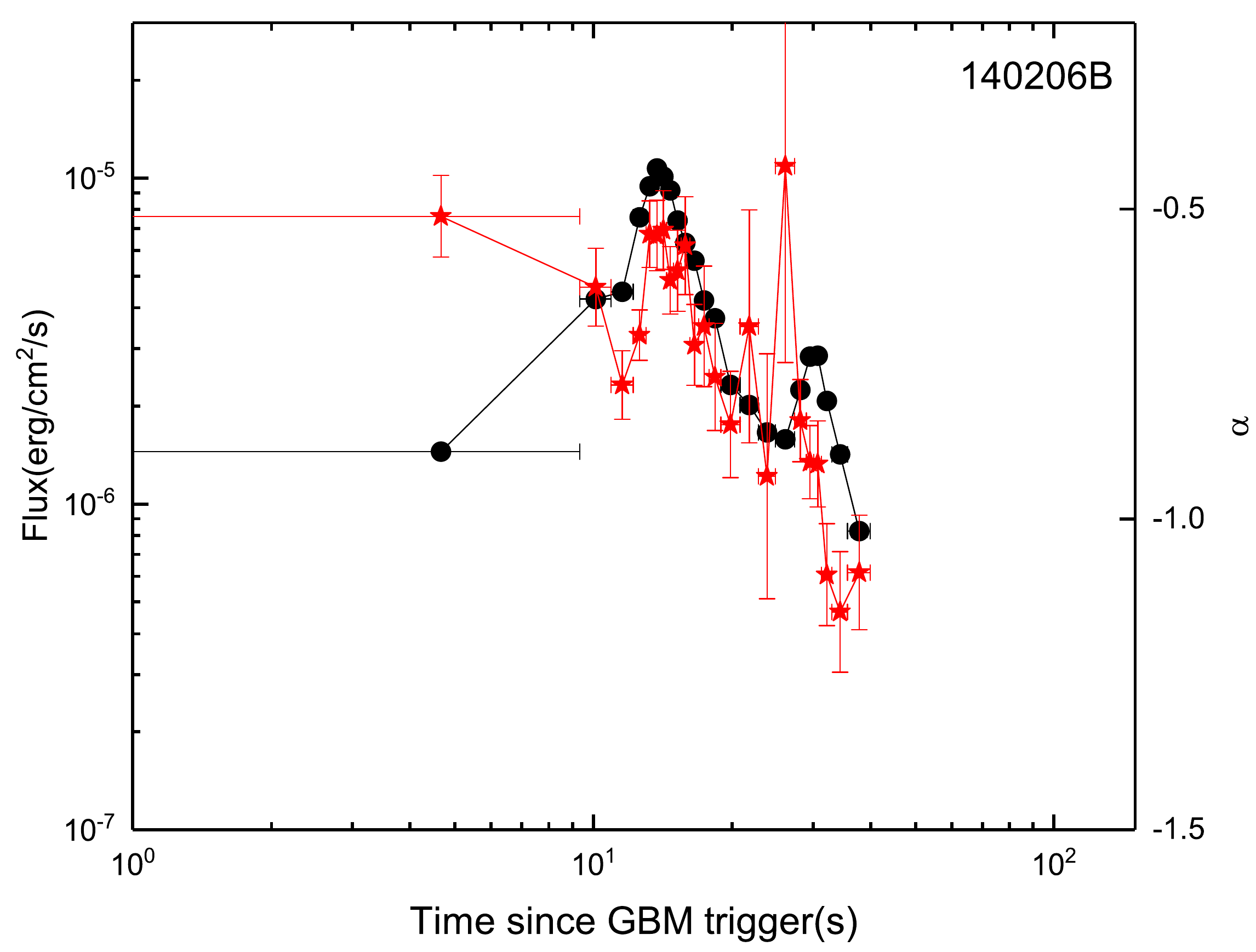}}
\resizebox{4cm}{!}{\includegraphics{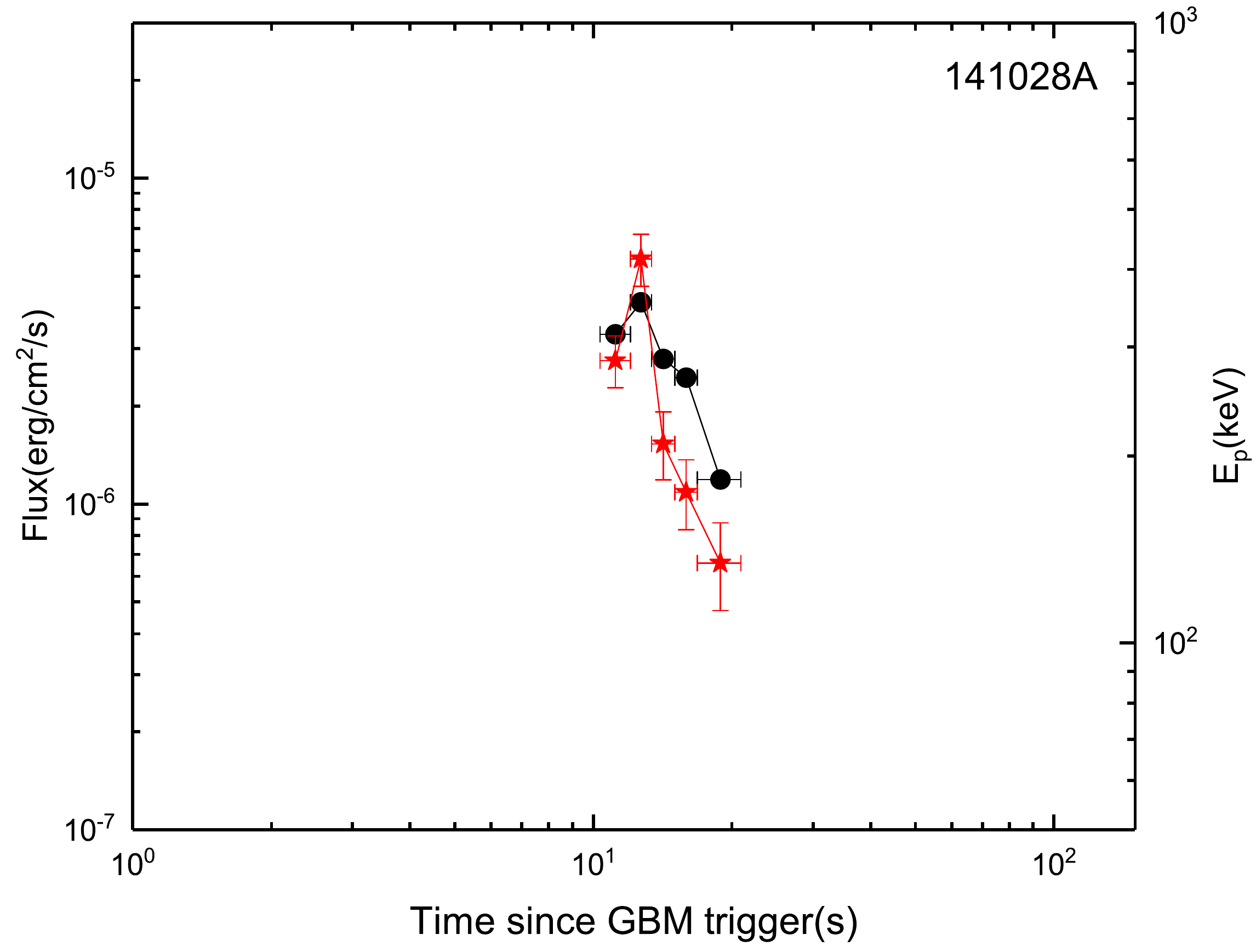}}
\resizebox{4cm}{!}{\includegraphics{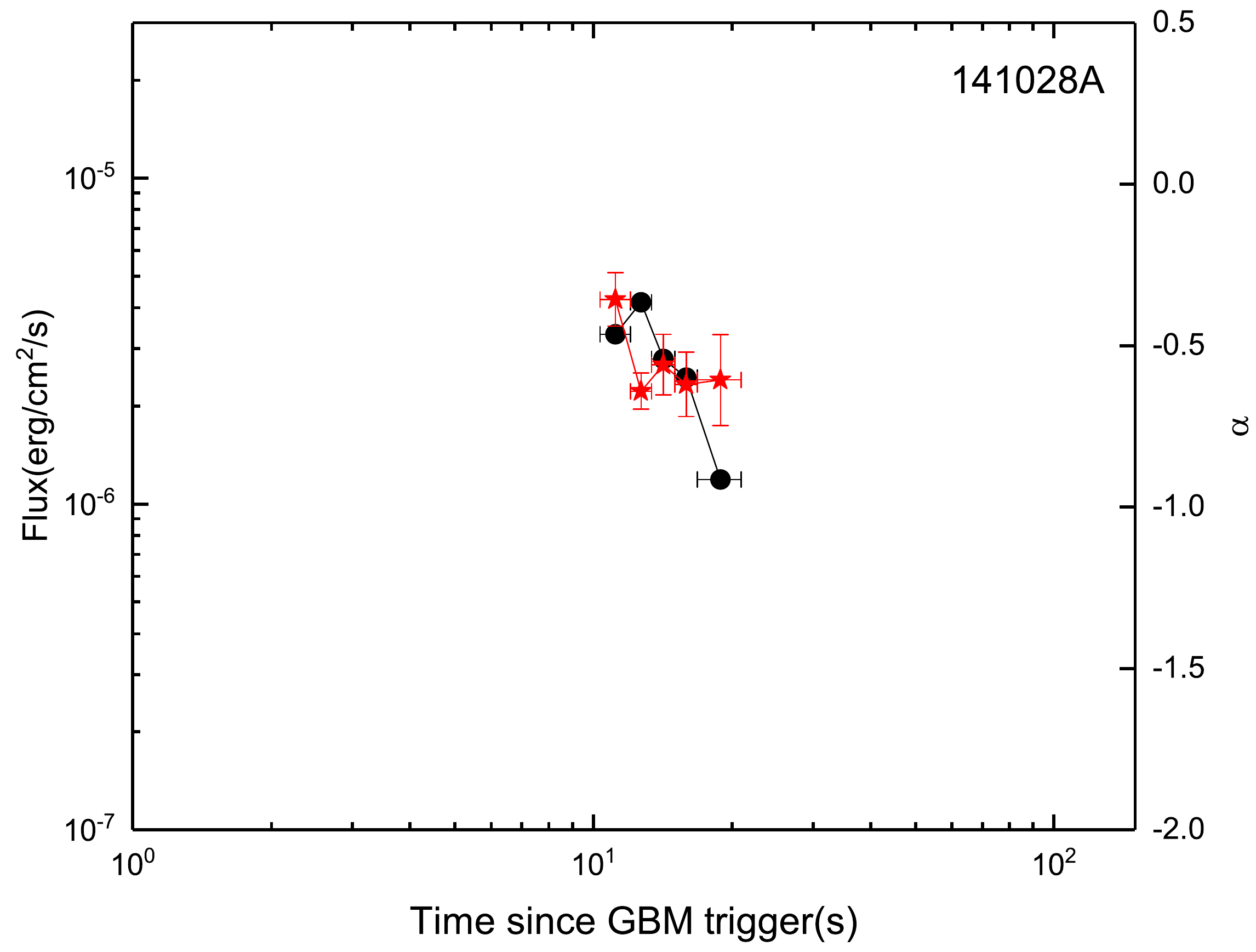}}
\resizebox{4cm}{!}{\includegraphics{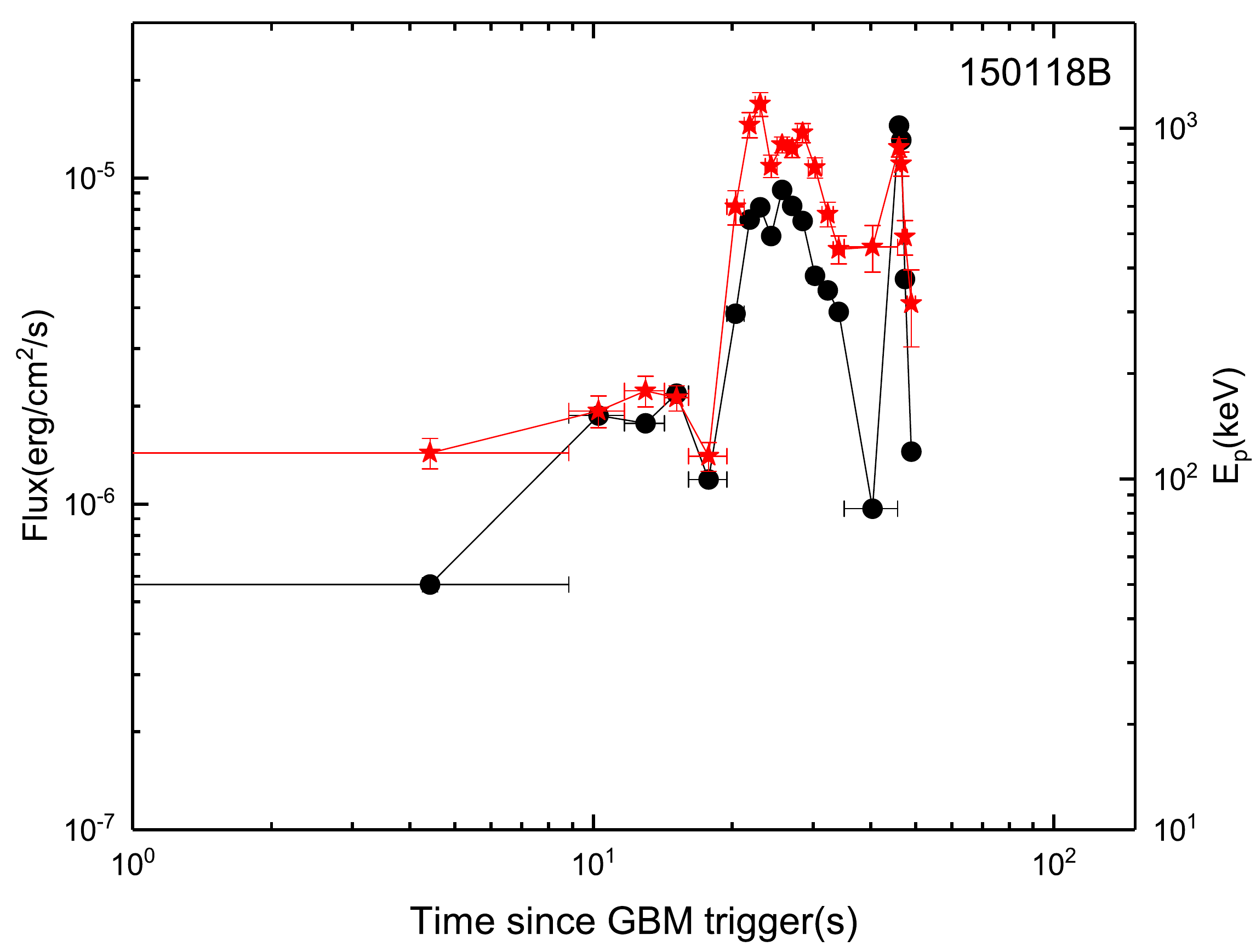}}
\resizebox{4cm}{!}{\includegraphics{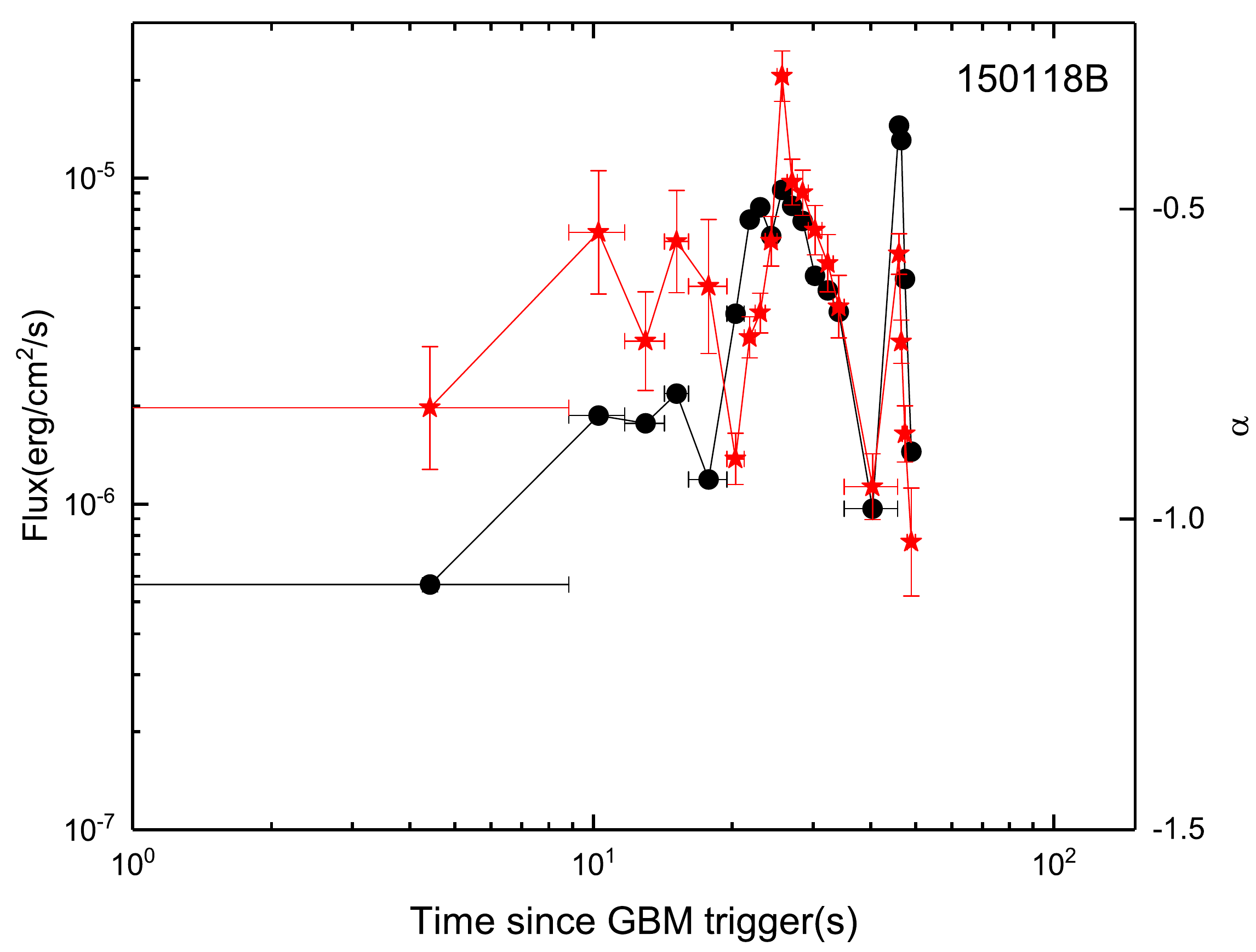}}
\resizebox{4cm}{!}{\includegraphics{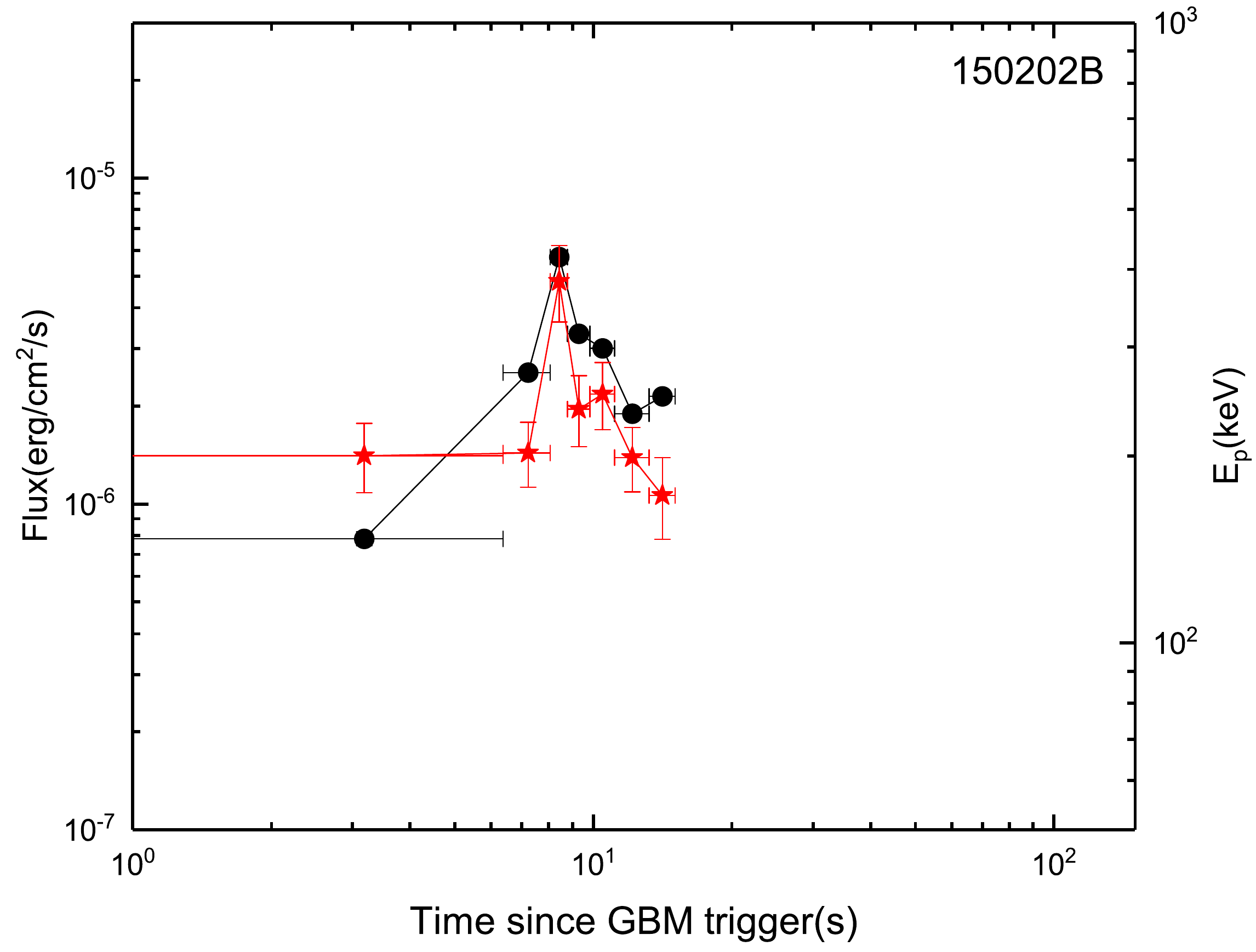}}
\resizebox{4cm}{!}{\includegraphics{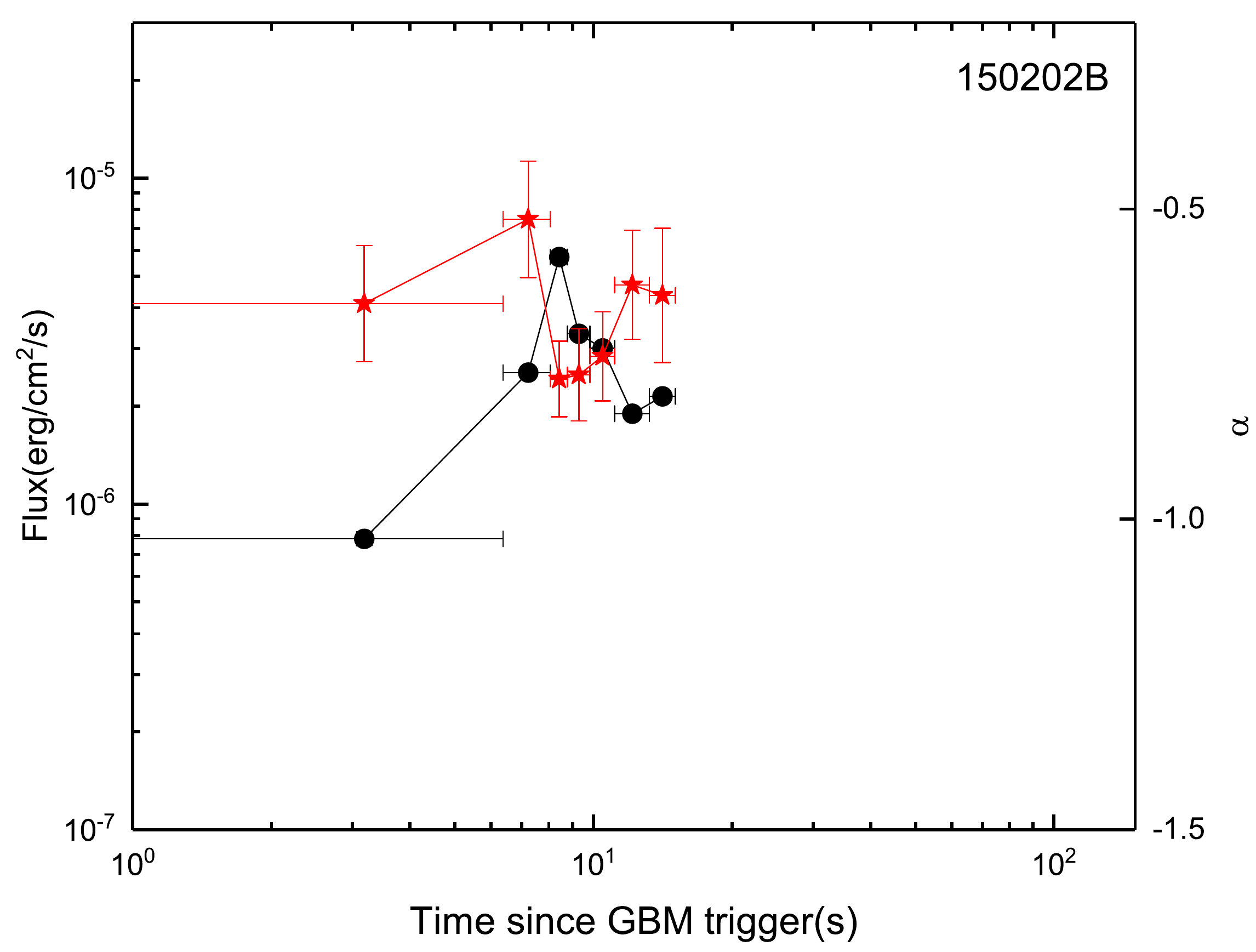}}
\resizebox{4cm}{!}{\includegraphics{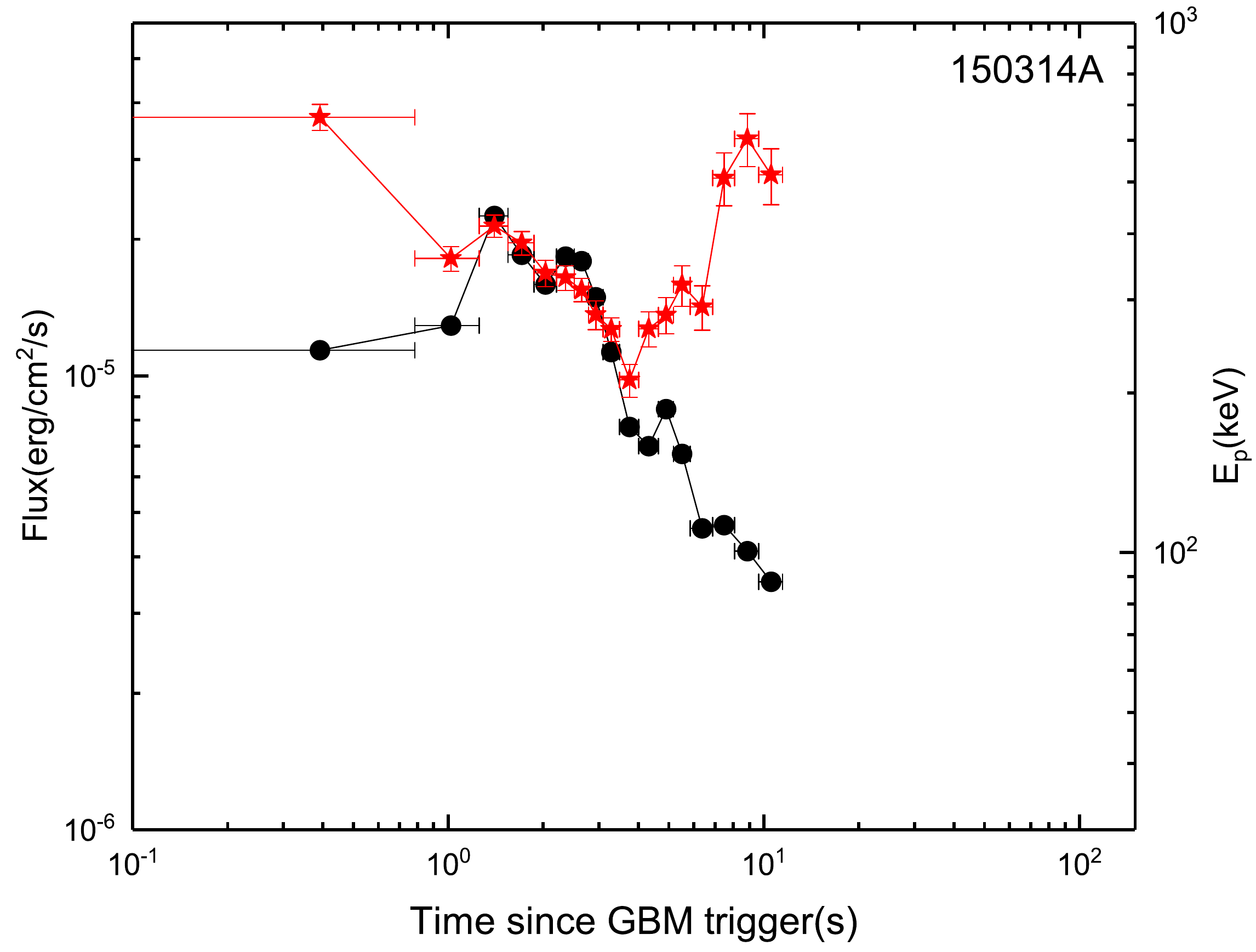}}
\resizebox{4cm}{!}{\includegraphics{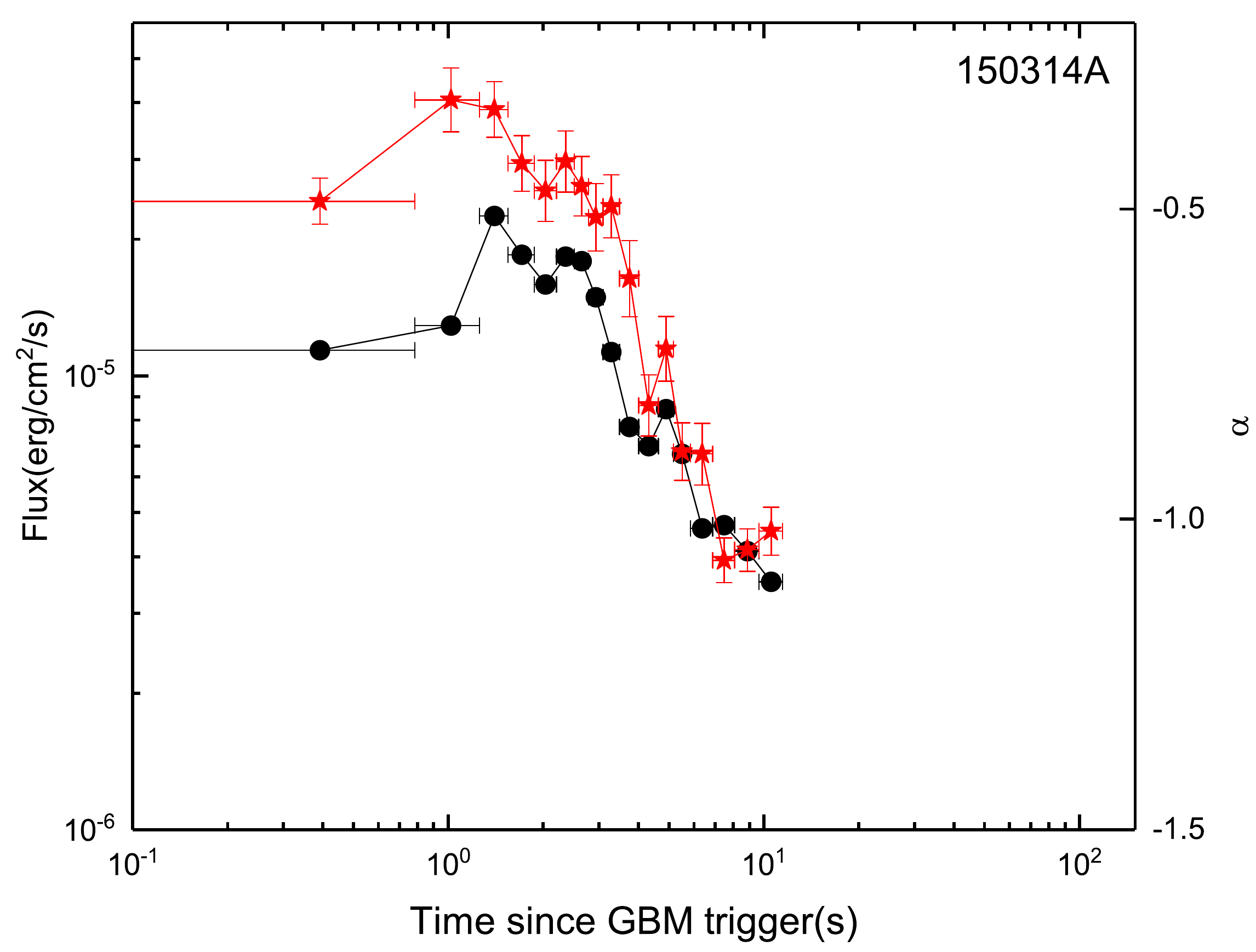}}
\resizebox{4cm}{!}{\includegraphics{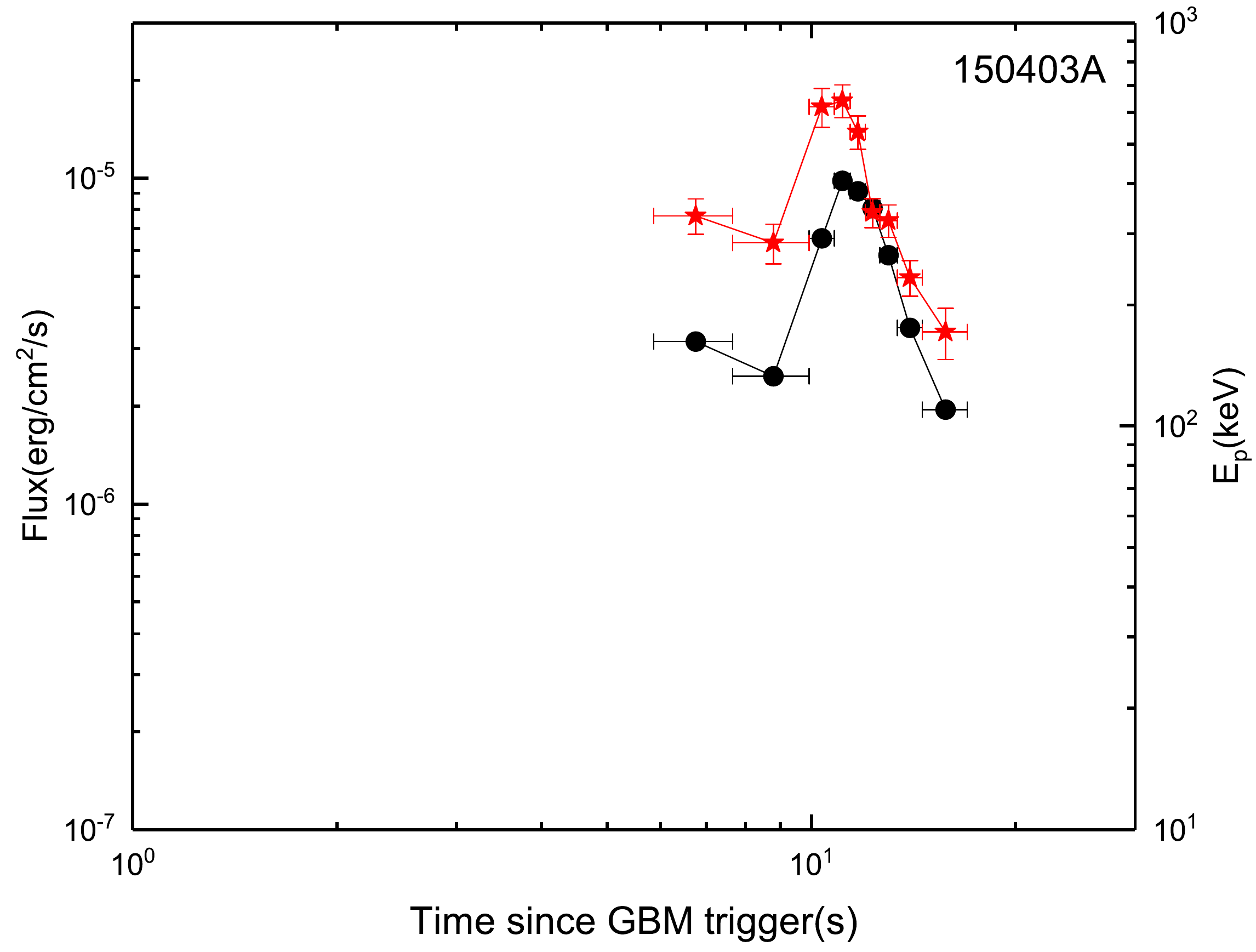}}
\resizebox{4cm}{!}{\includegraphics{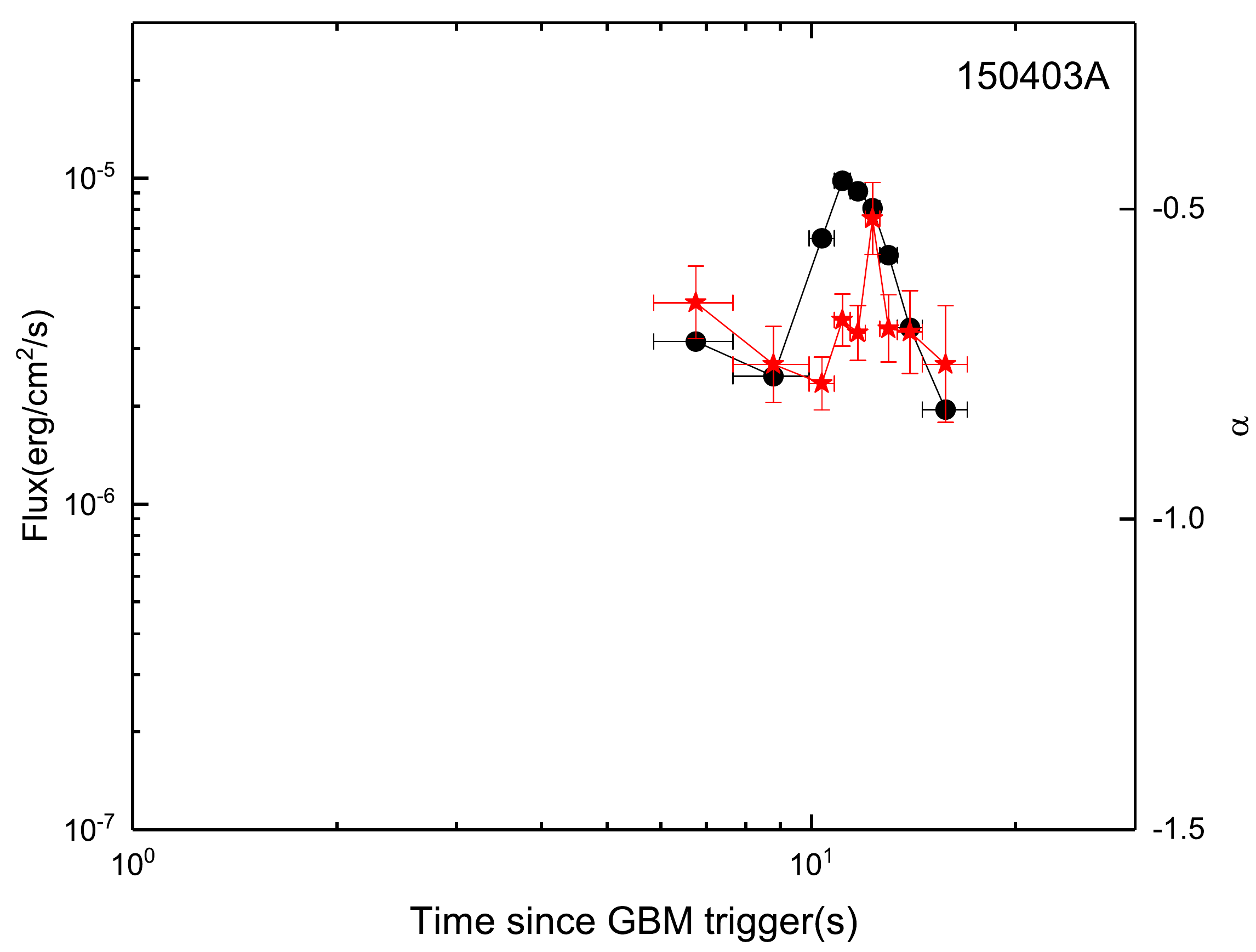}}
\resizebox{4cm}{!}{\includegraphics{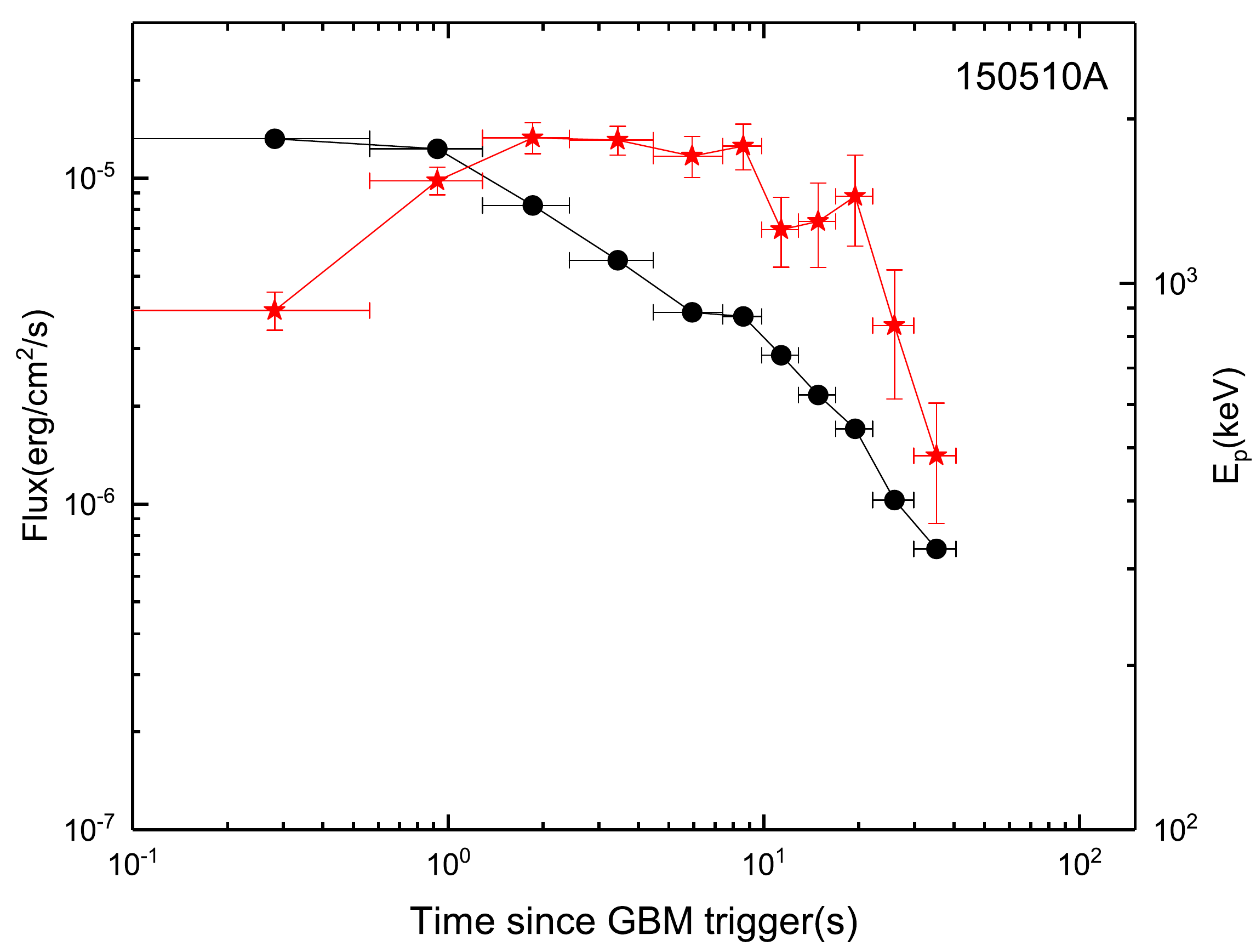}}
\resizebox{4cm}{!}{\includegraphics{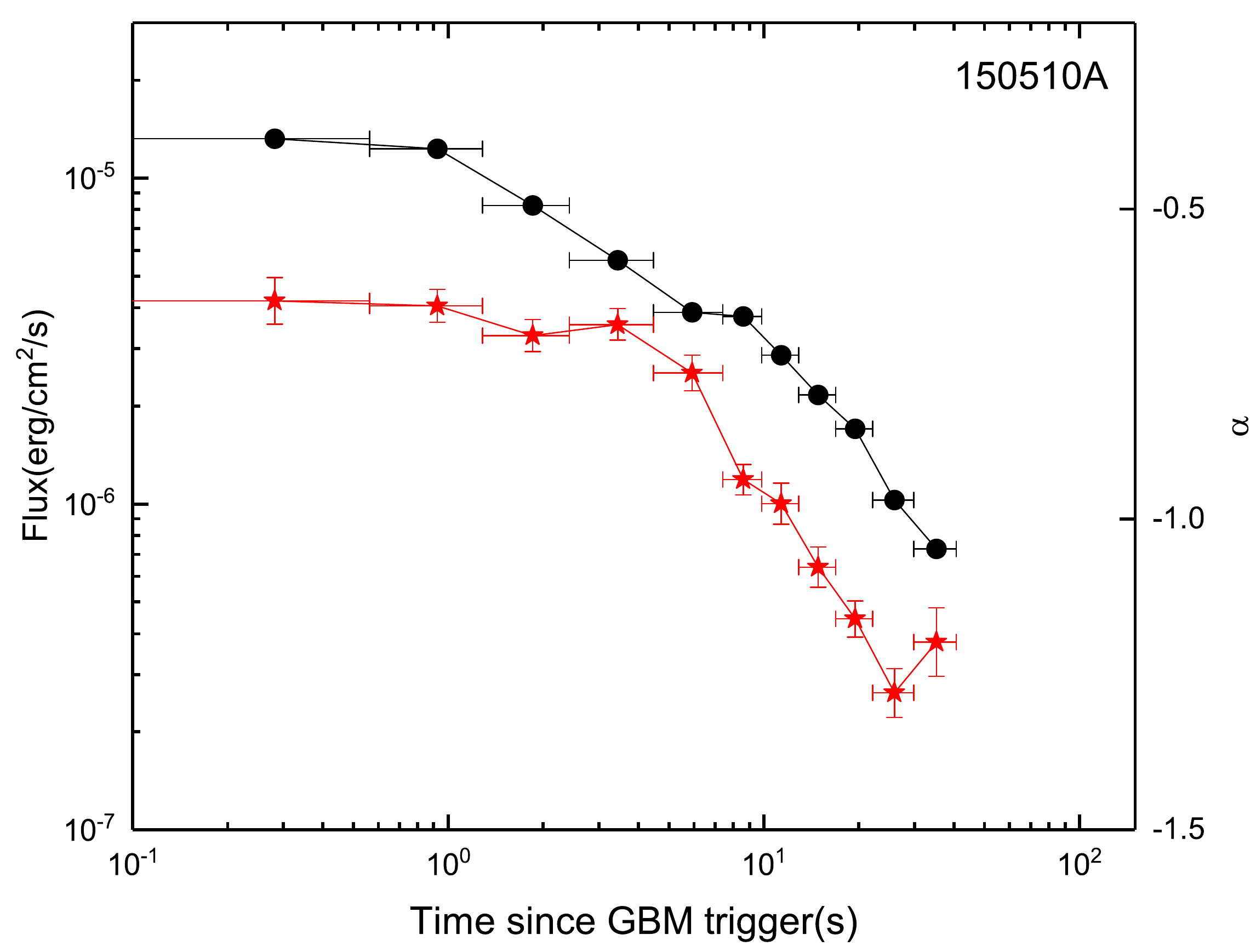}}
\resizebox{4cm}{!}{\includegraphics{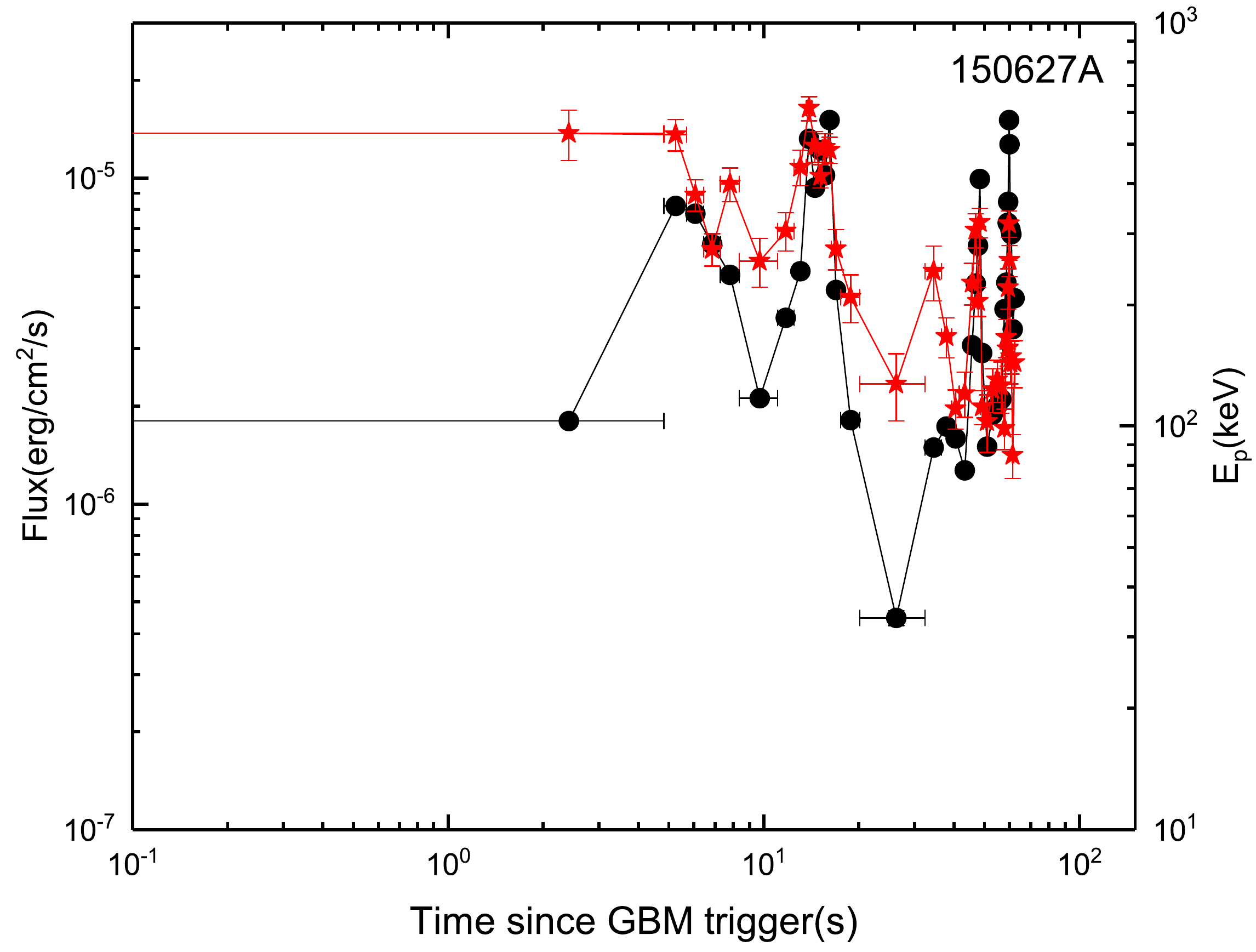}}
\resizebox{4cm}{!}{\includegraphics{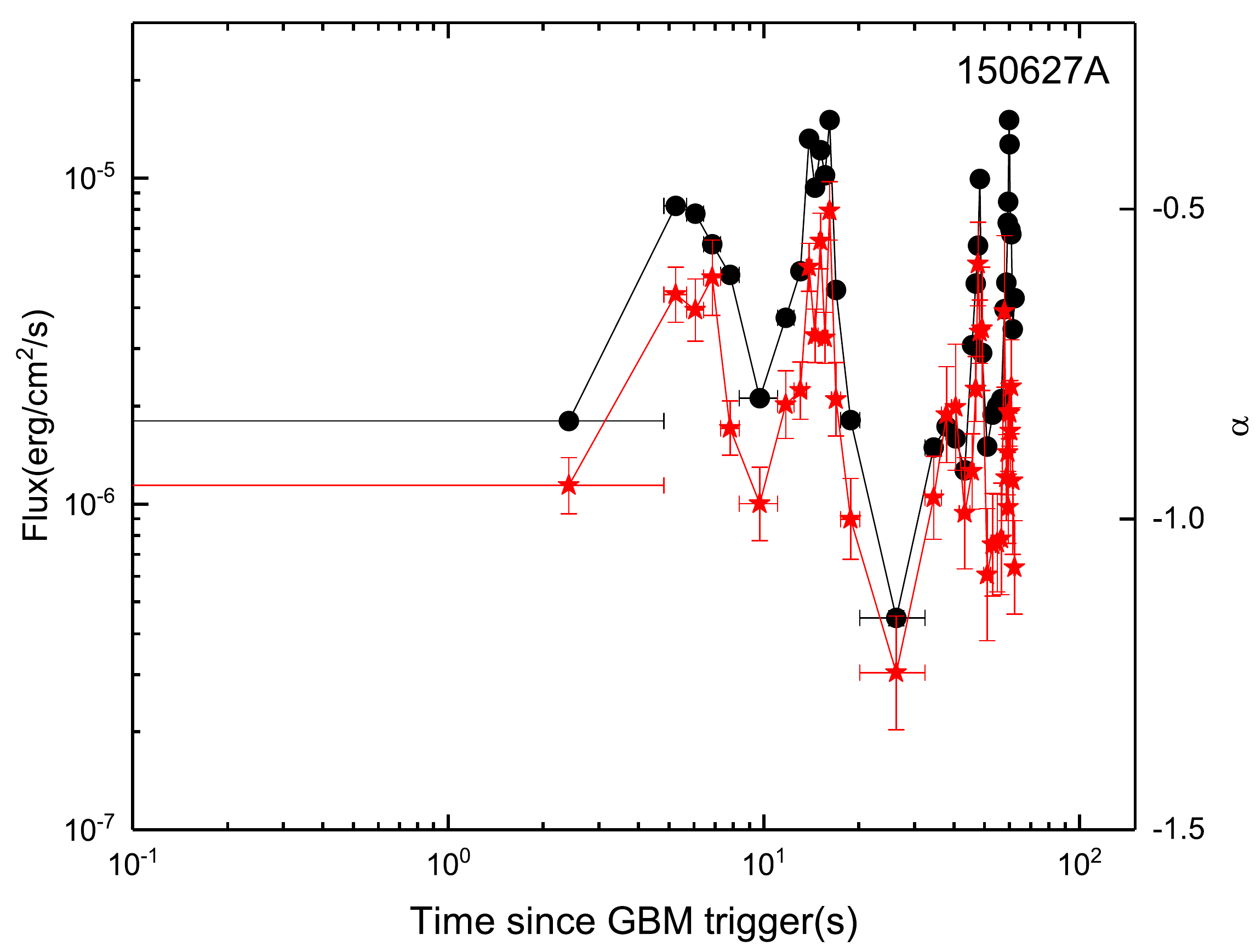}}
\resizebox{4cm}{!}{\includegraphics{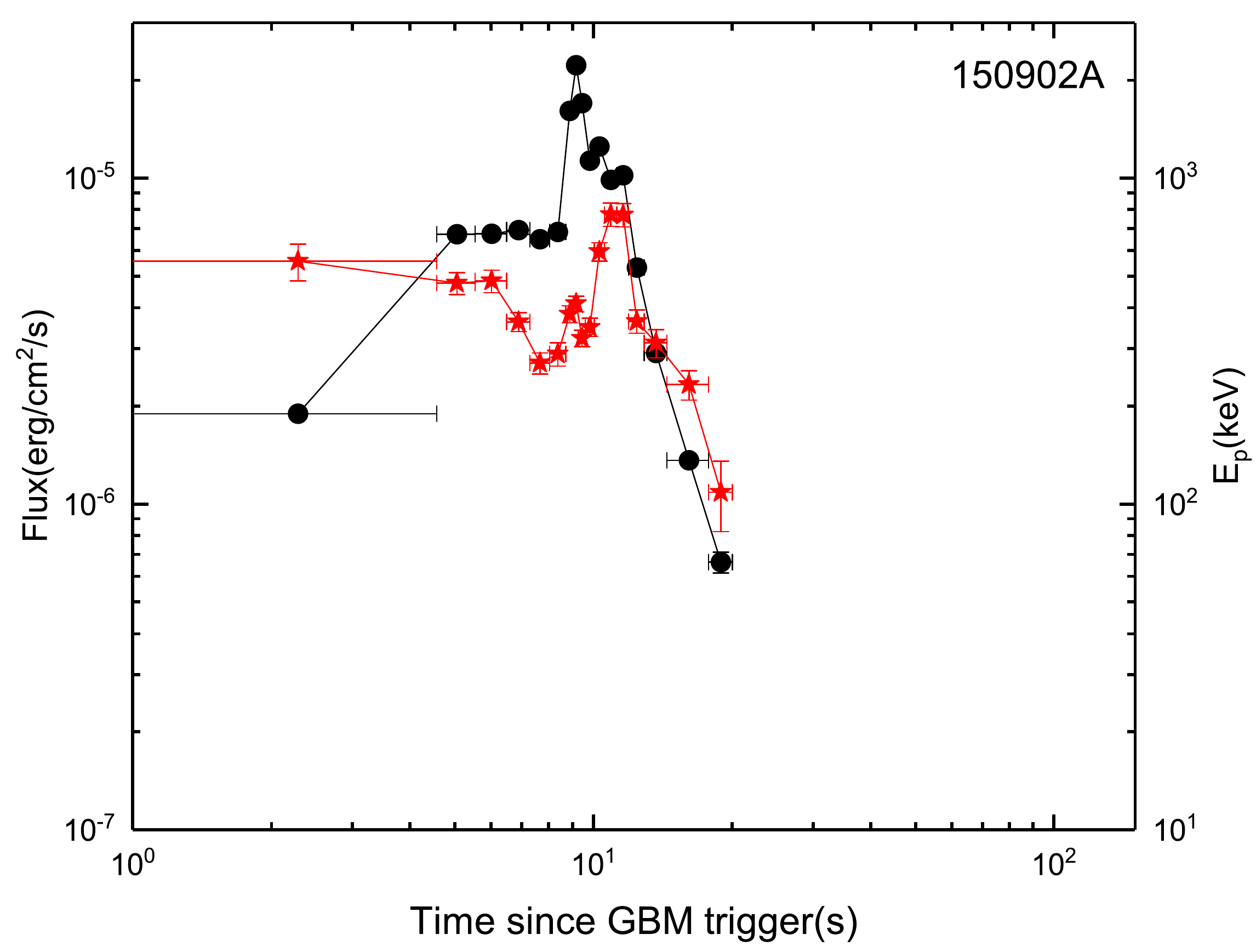}}
\resizebox{4cm}{!}{\includegraphics{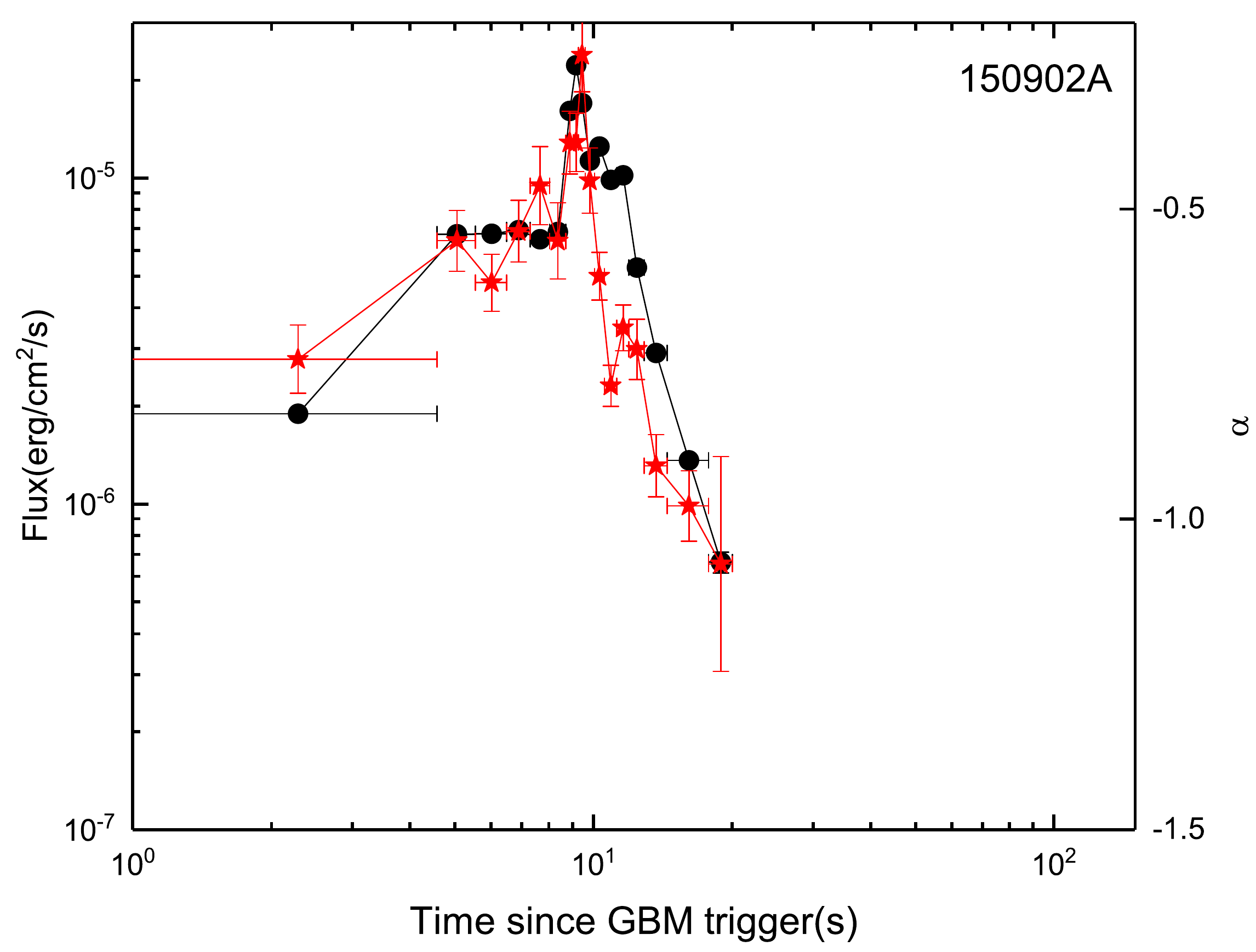}}
\resizebox{4cm}{!}{\includegraphics{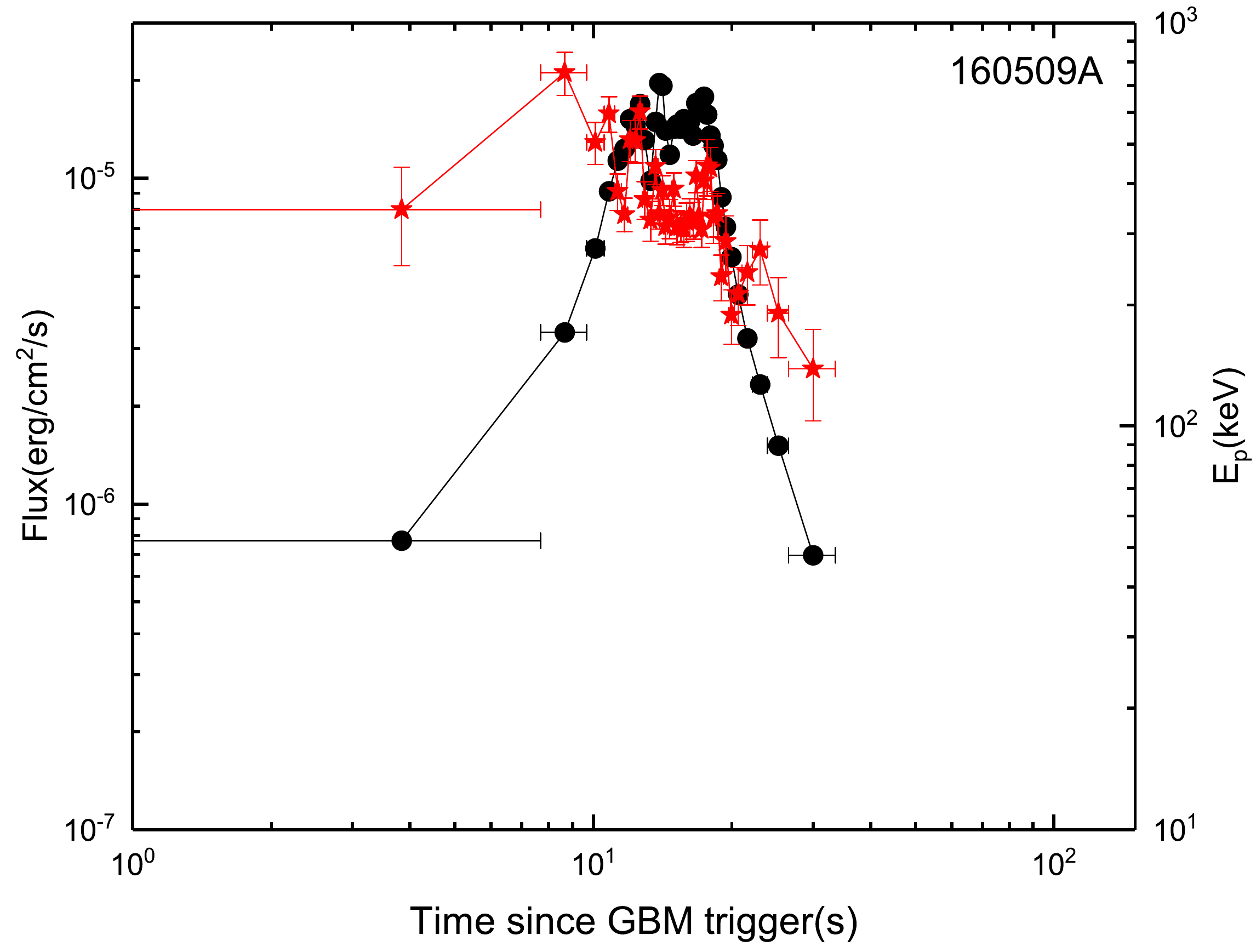}}
\resizebox{4cm}{!}{\includegraphics{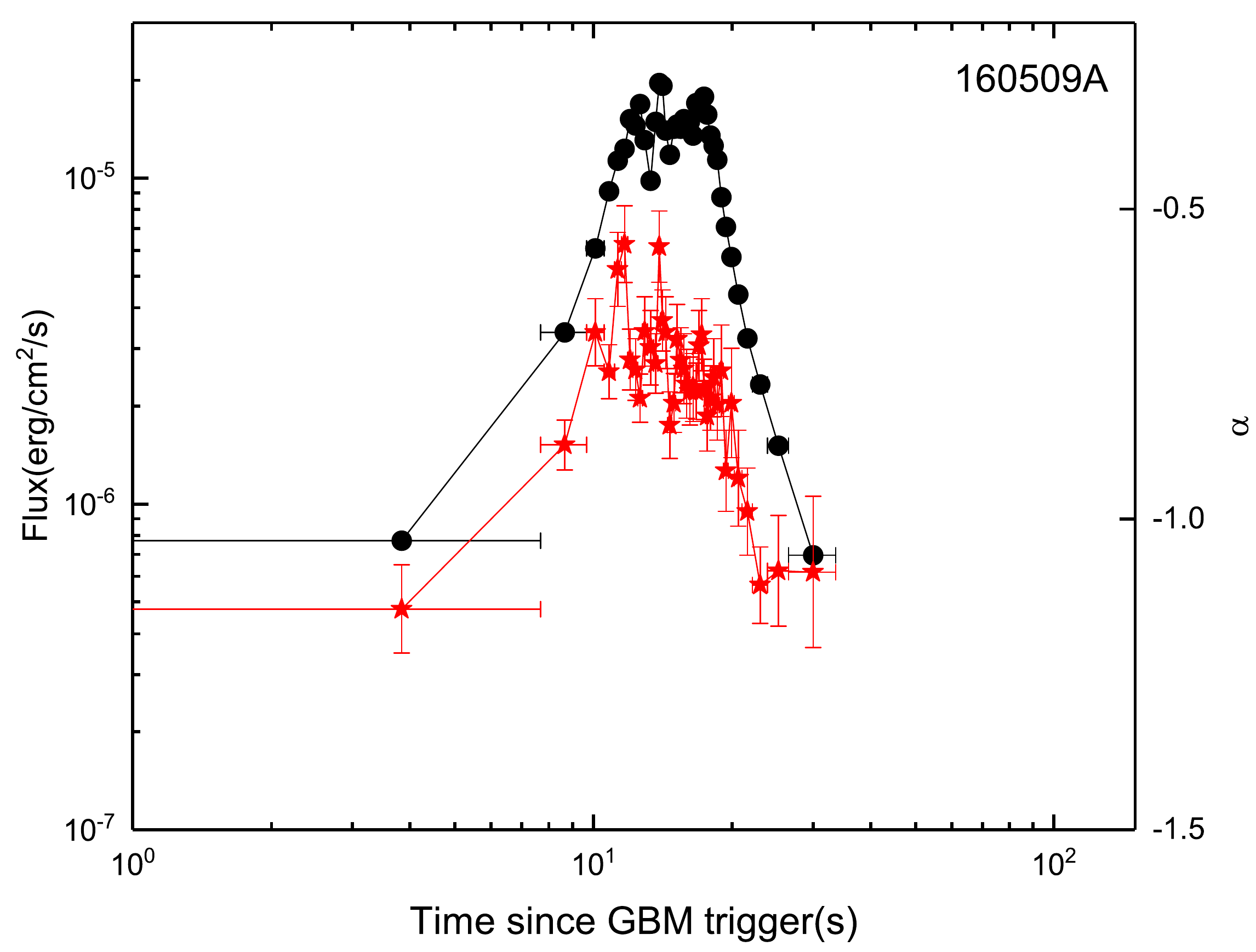}}
\caption{\it-continued}
\end{figure}

\addtocounter{figure}{-1}
\begin{figure}
\centering
\resizebox{4cm}{!}{\includegraphics{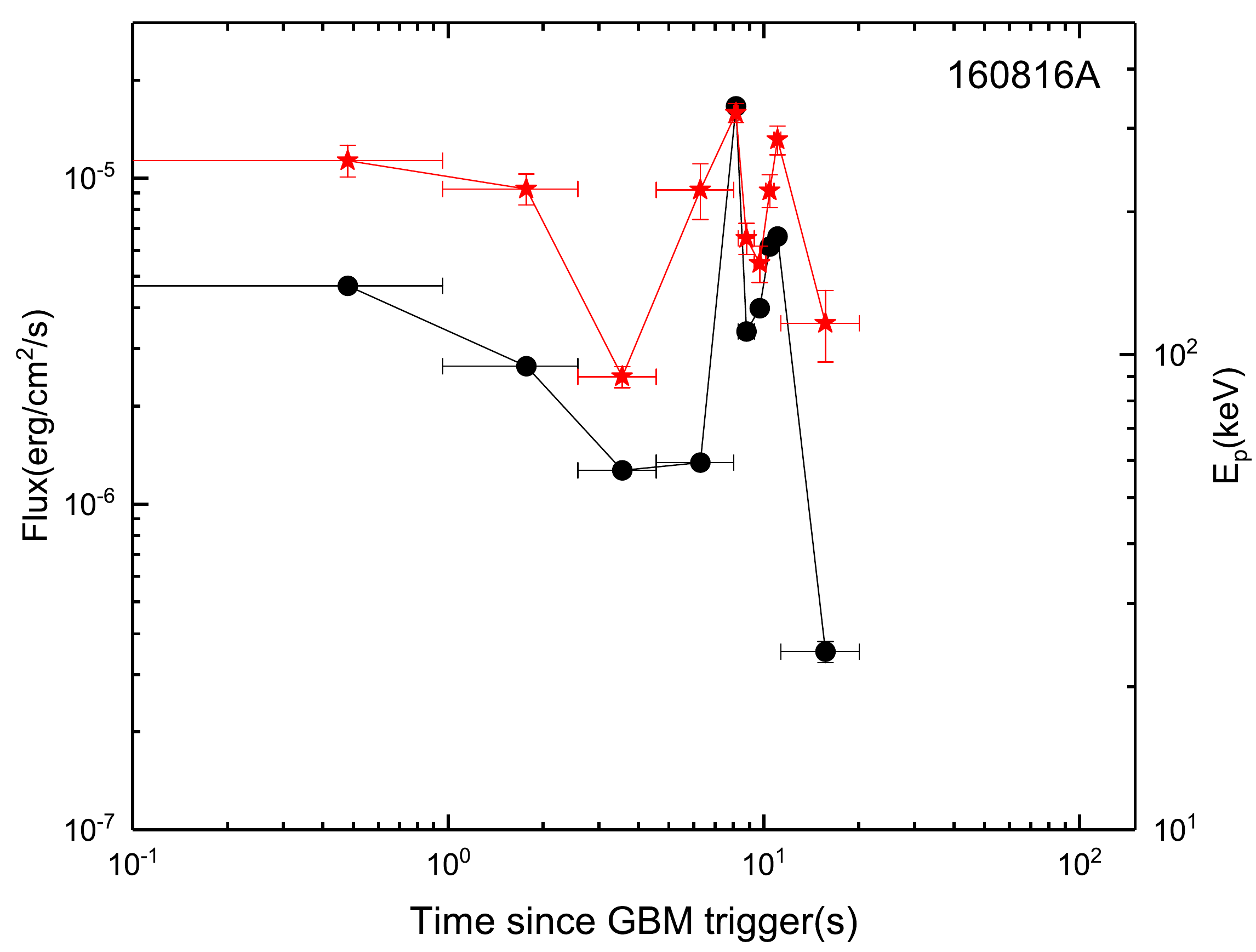}}
\resizebox{4cm}{!}{\includegraphics{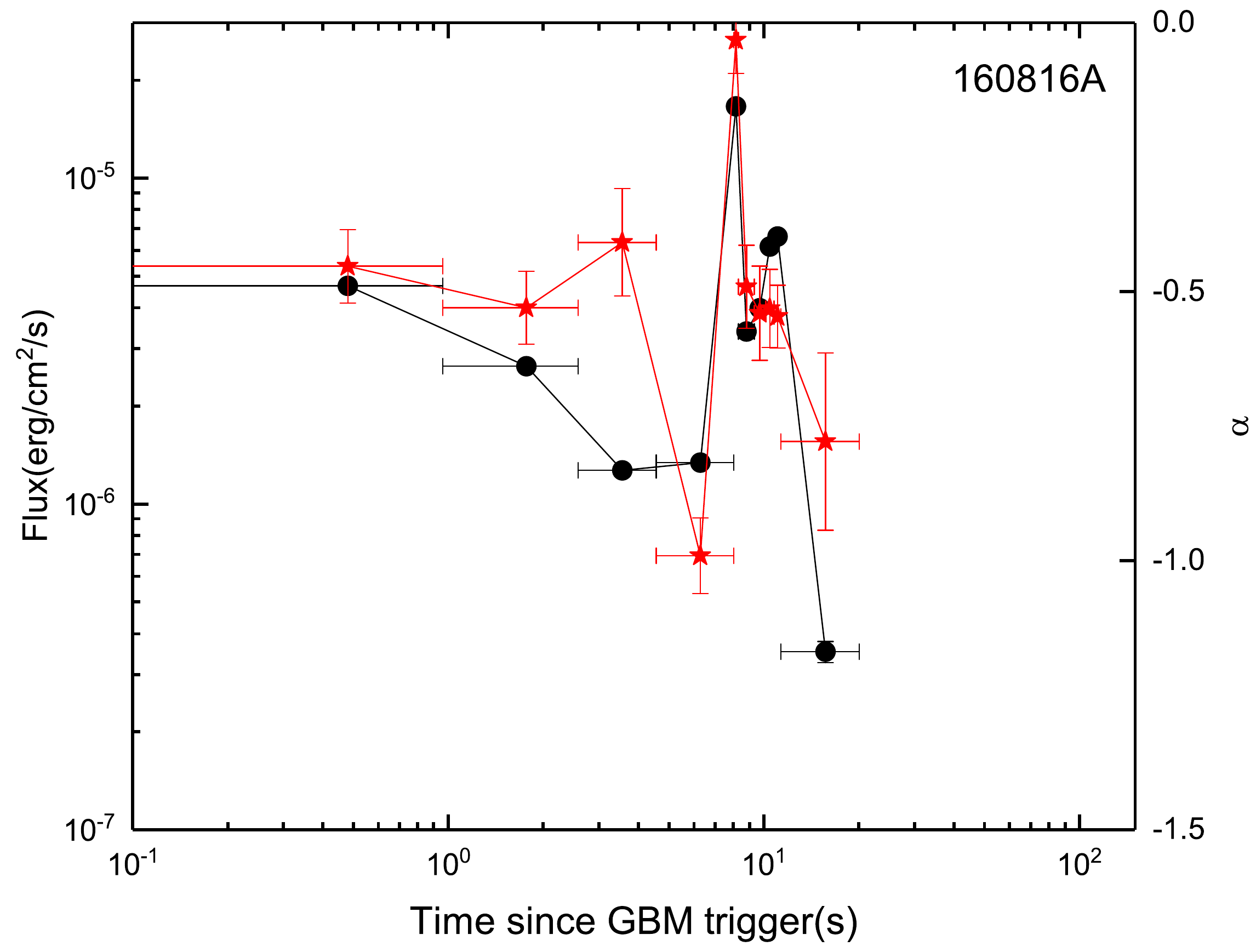}}
\resizebox{4cm}{!}{\includegraphics{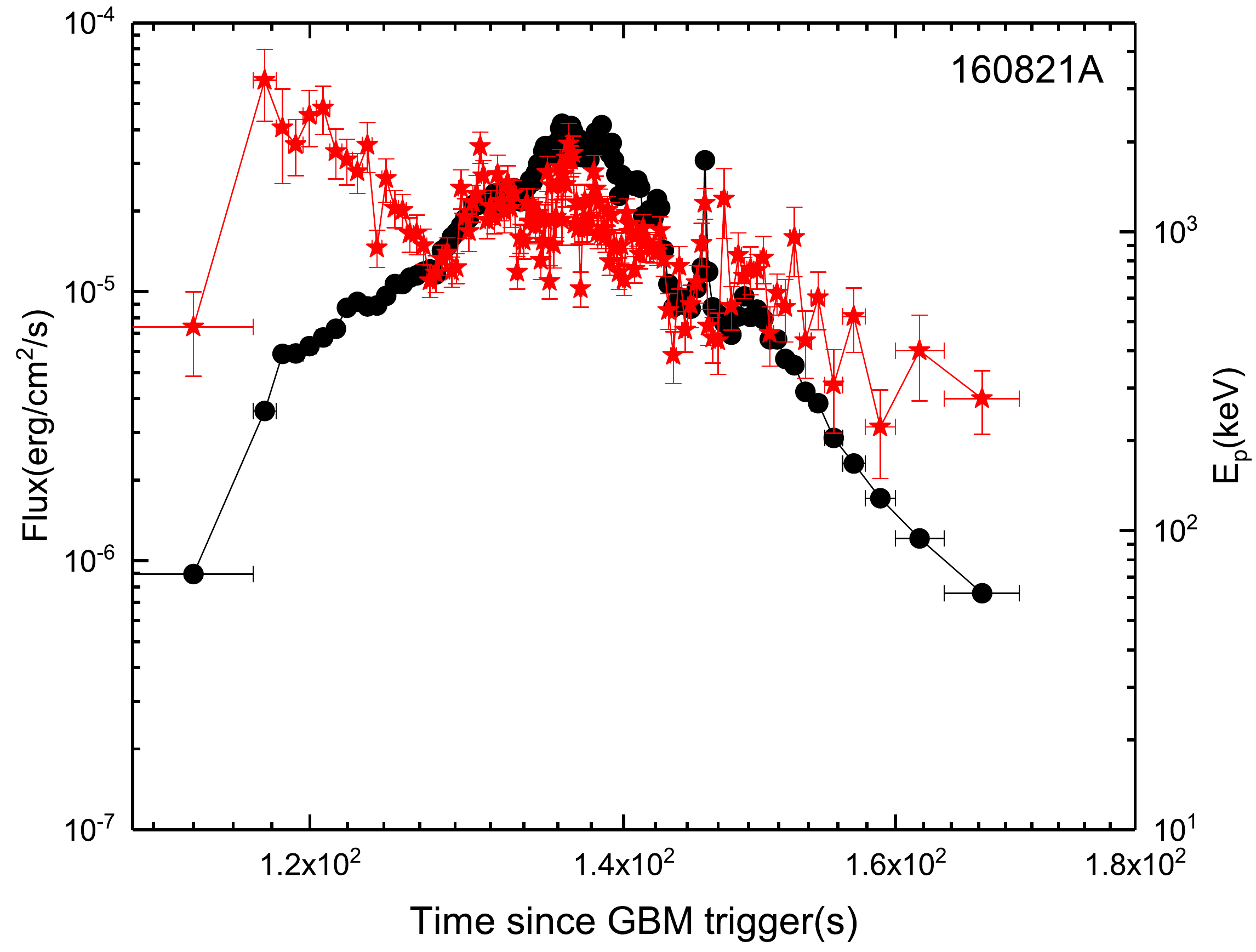}}
\resizebox{4cm}{!}{\includegraphics{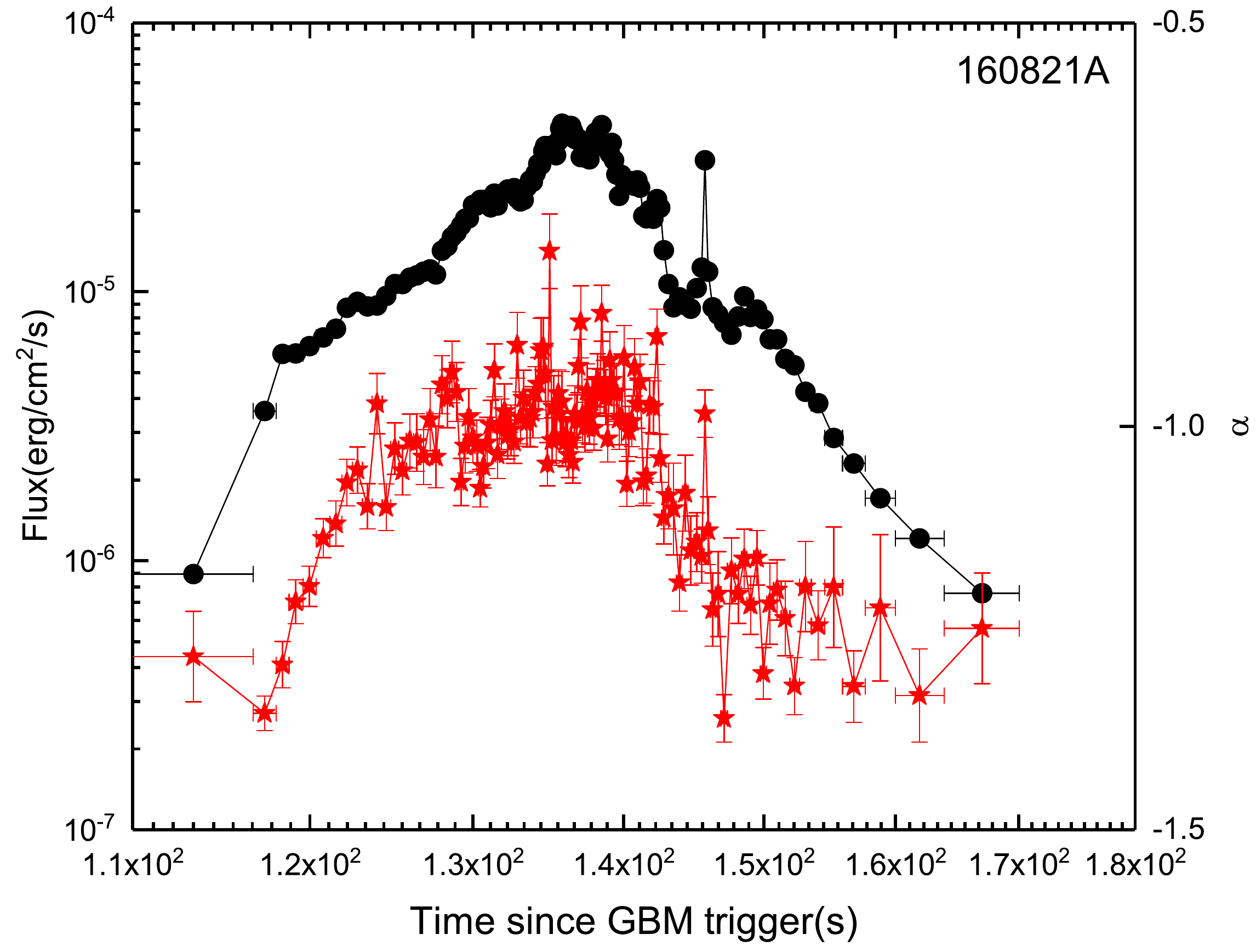}}
\resizebox{4cm}{!}{\includegraphics{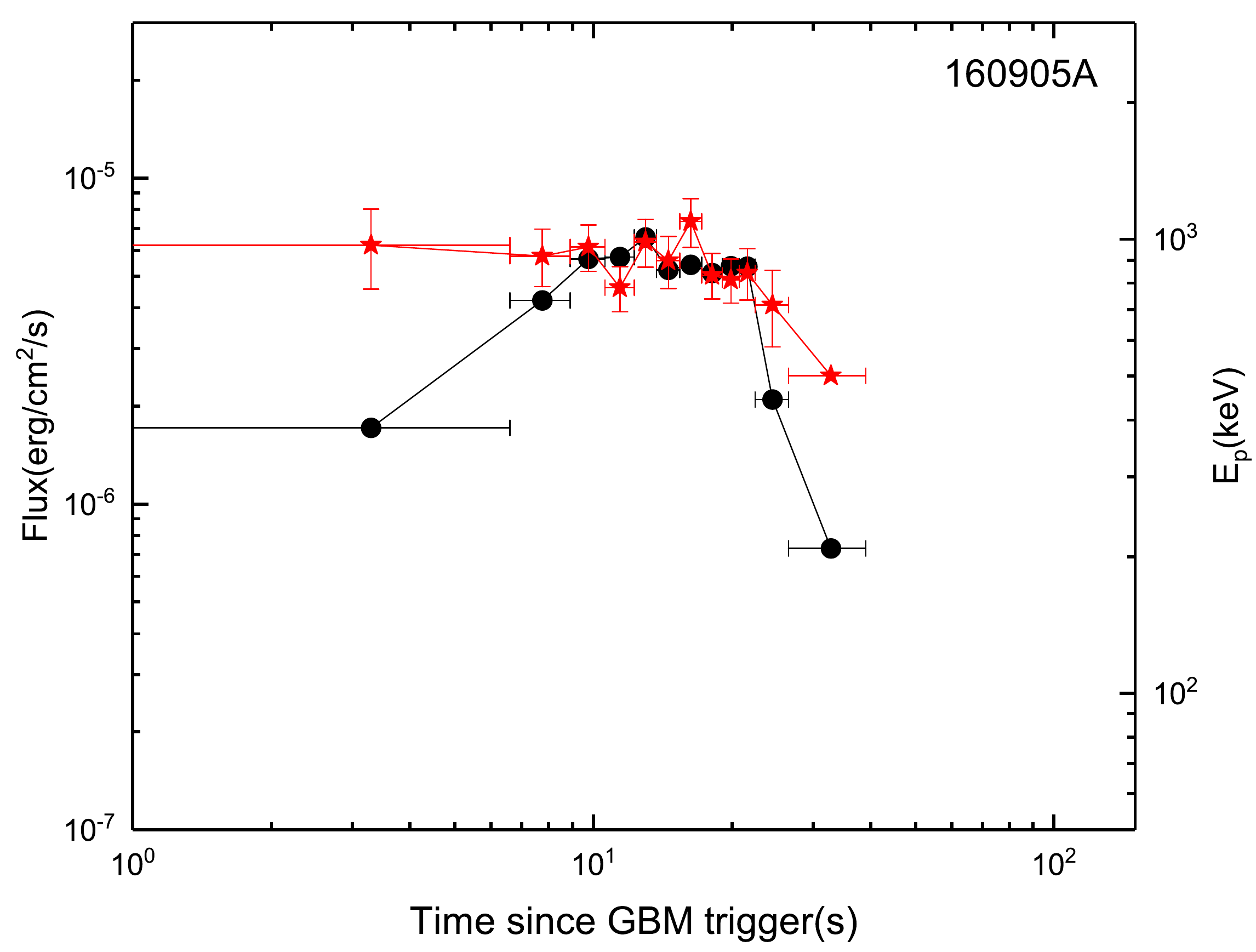}}
\resizebox{4cm}{!}{\includegraphics{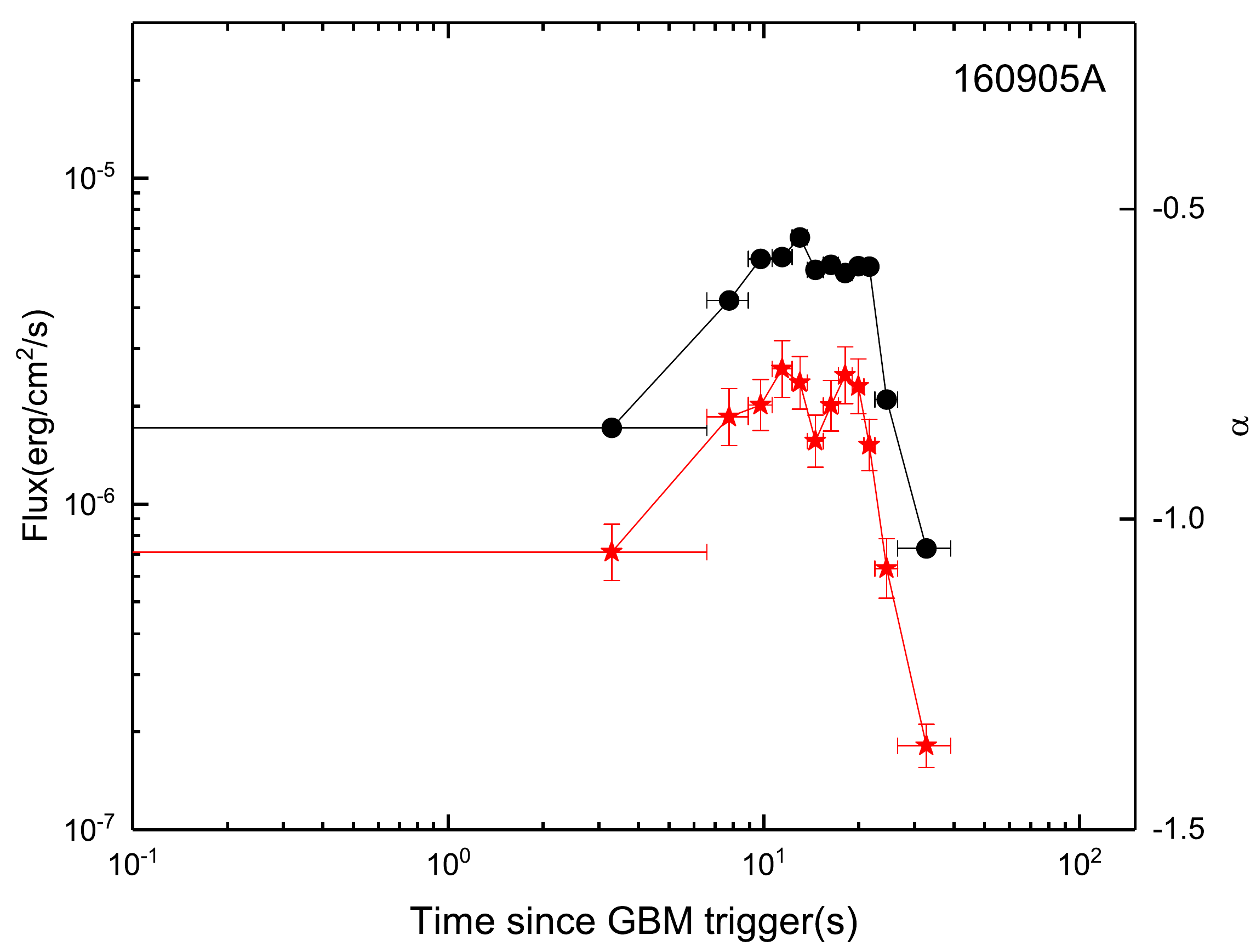}}
\resizebox{4cm}{!}{\includegraphics{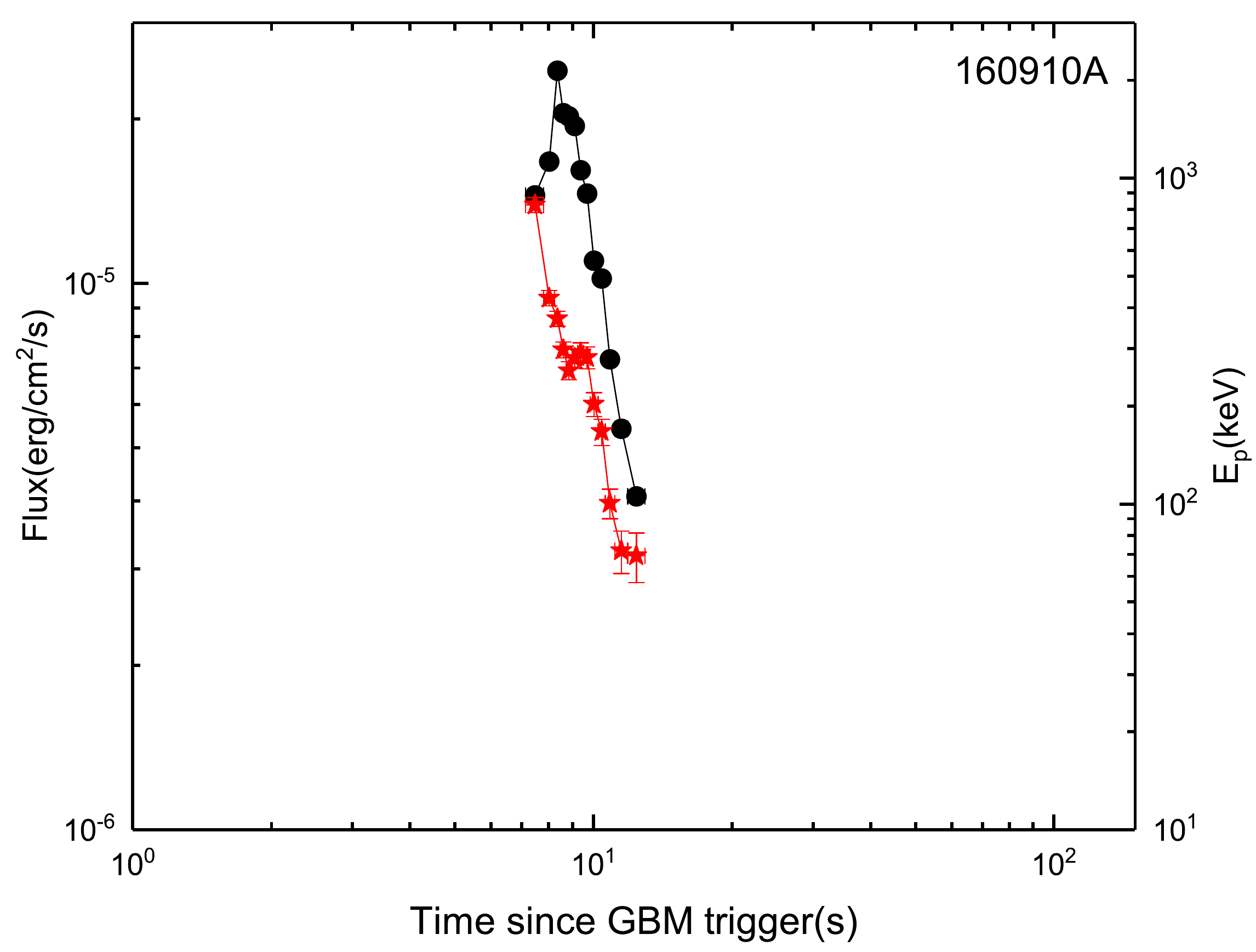}}
\resizebox{4cm}{!}{\includegraphics{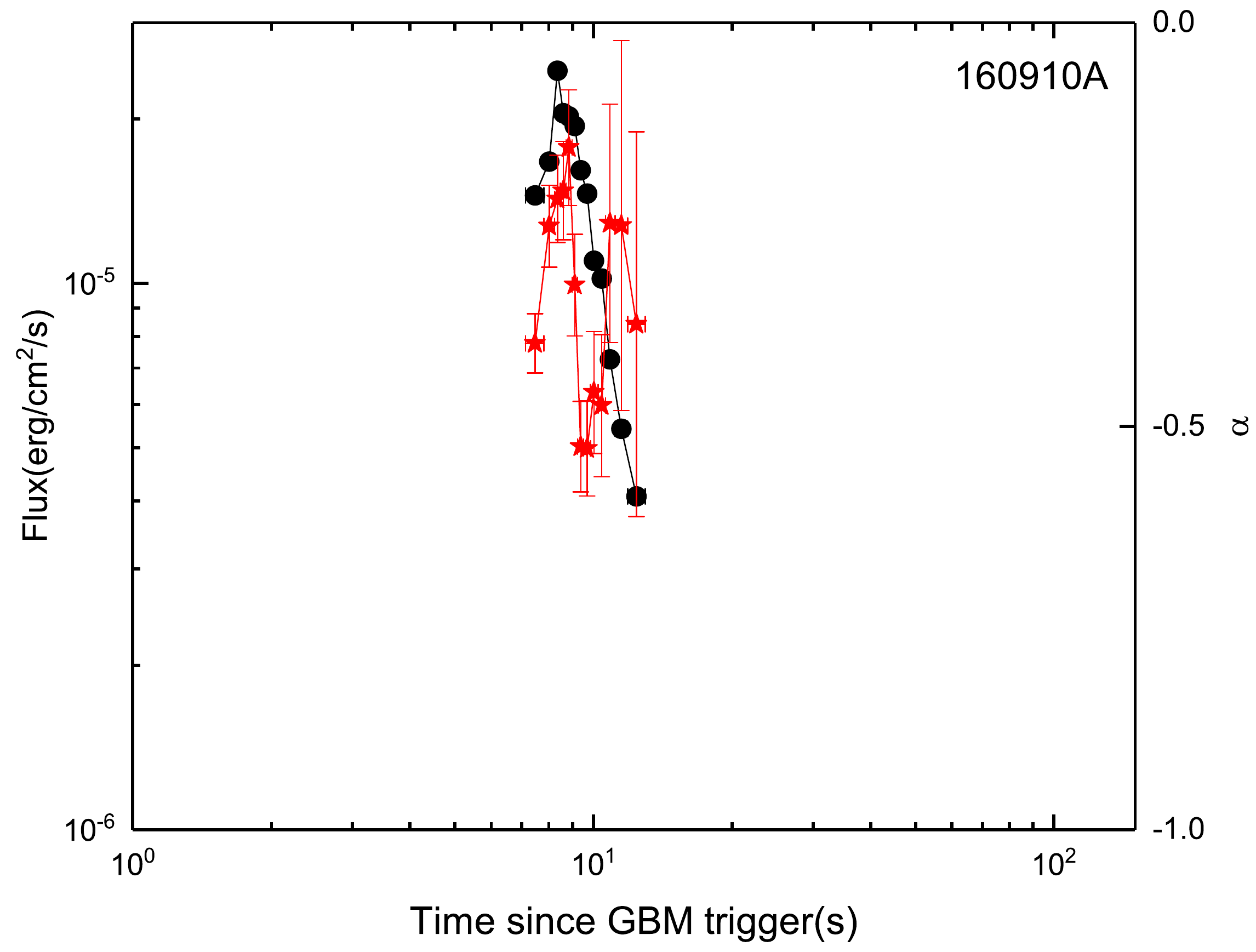}}
\resizebox{4cm}{!}{\includegraphics{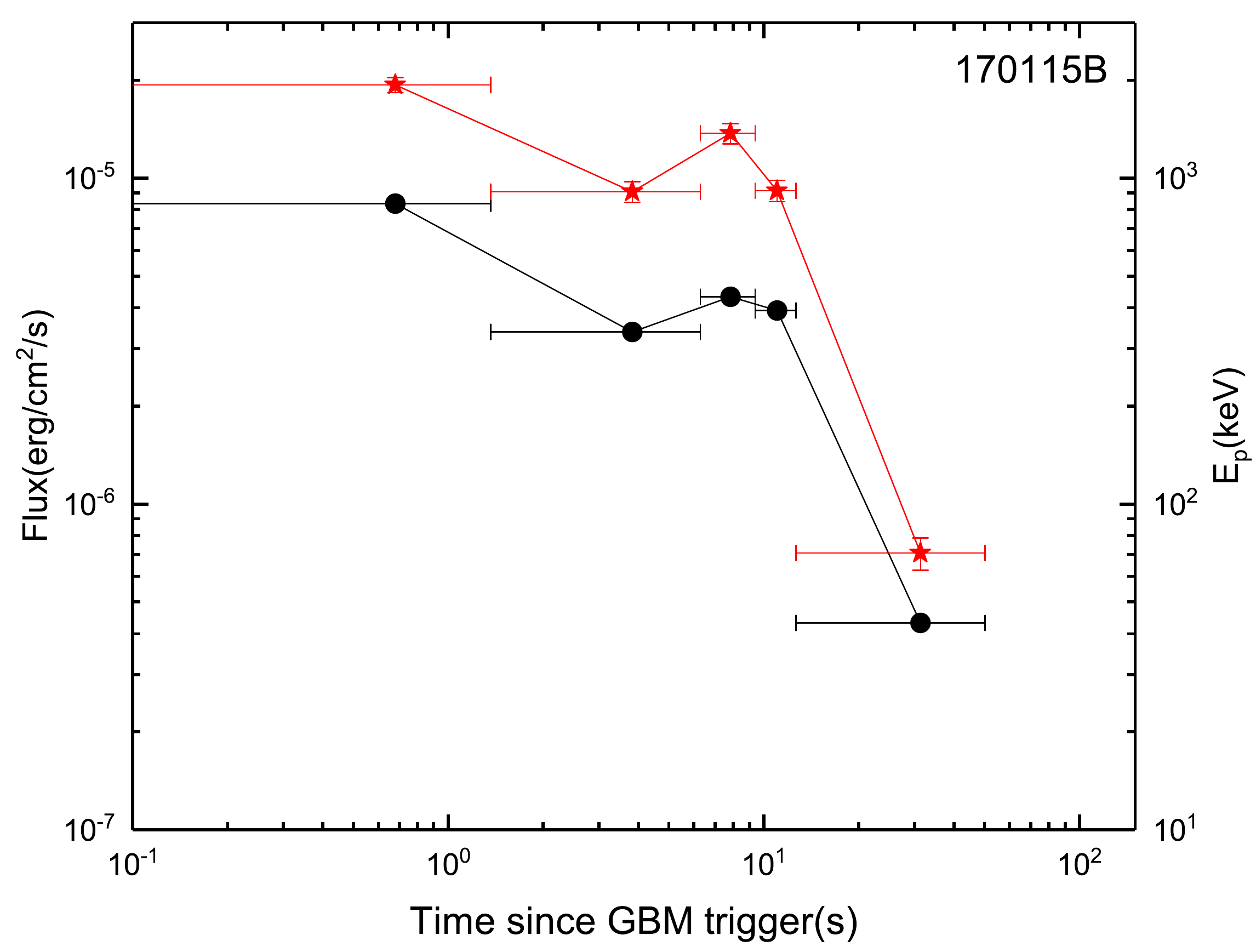}}
\resizebox{4cm}{!}{\includegraphics{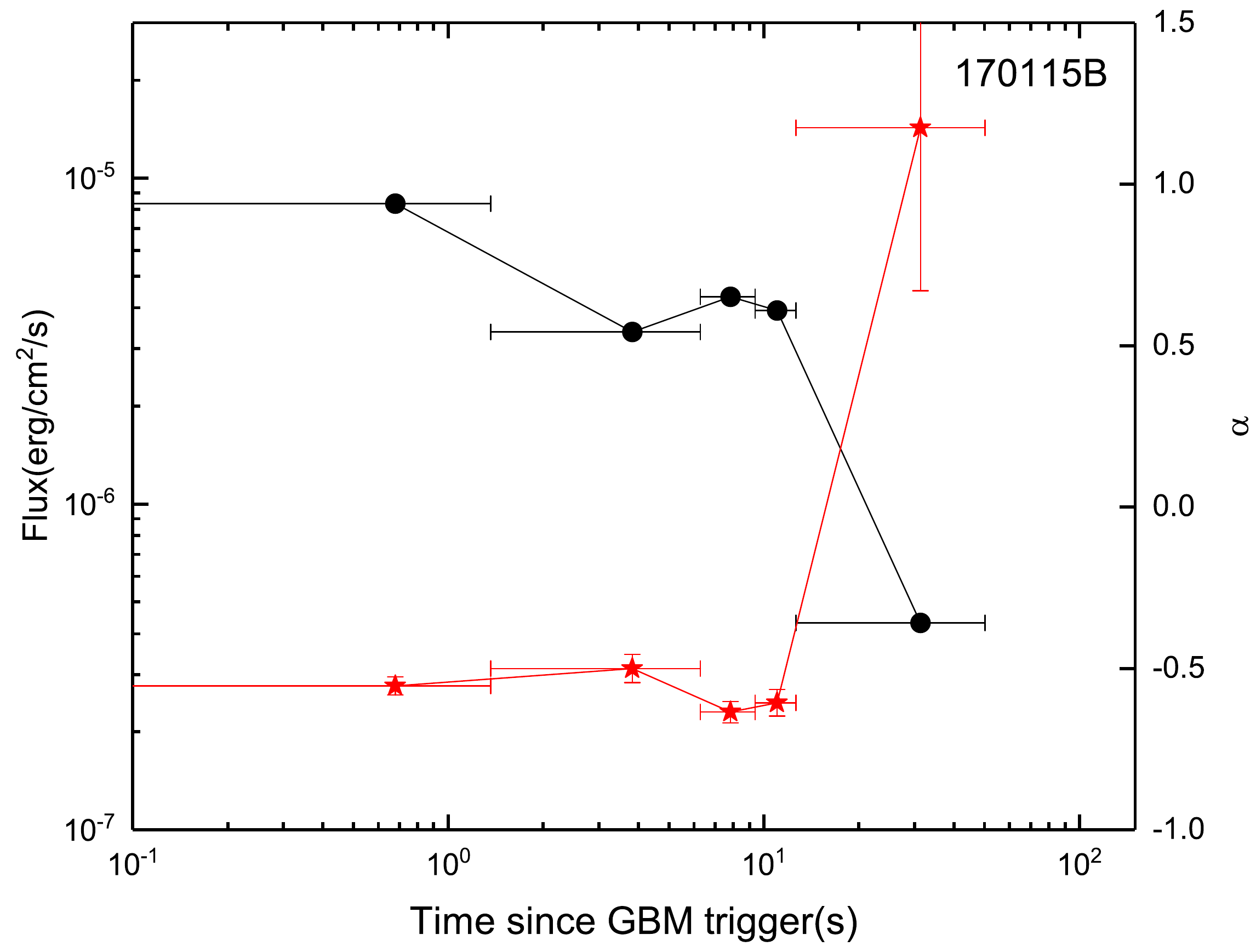}}
\resizebox{4cm}{!}{\includegraphics{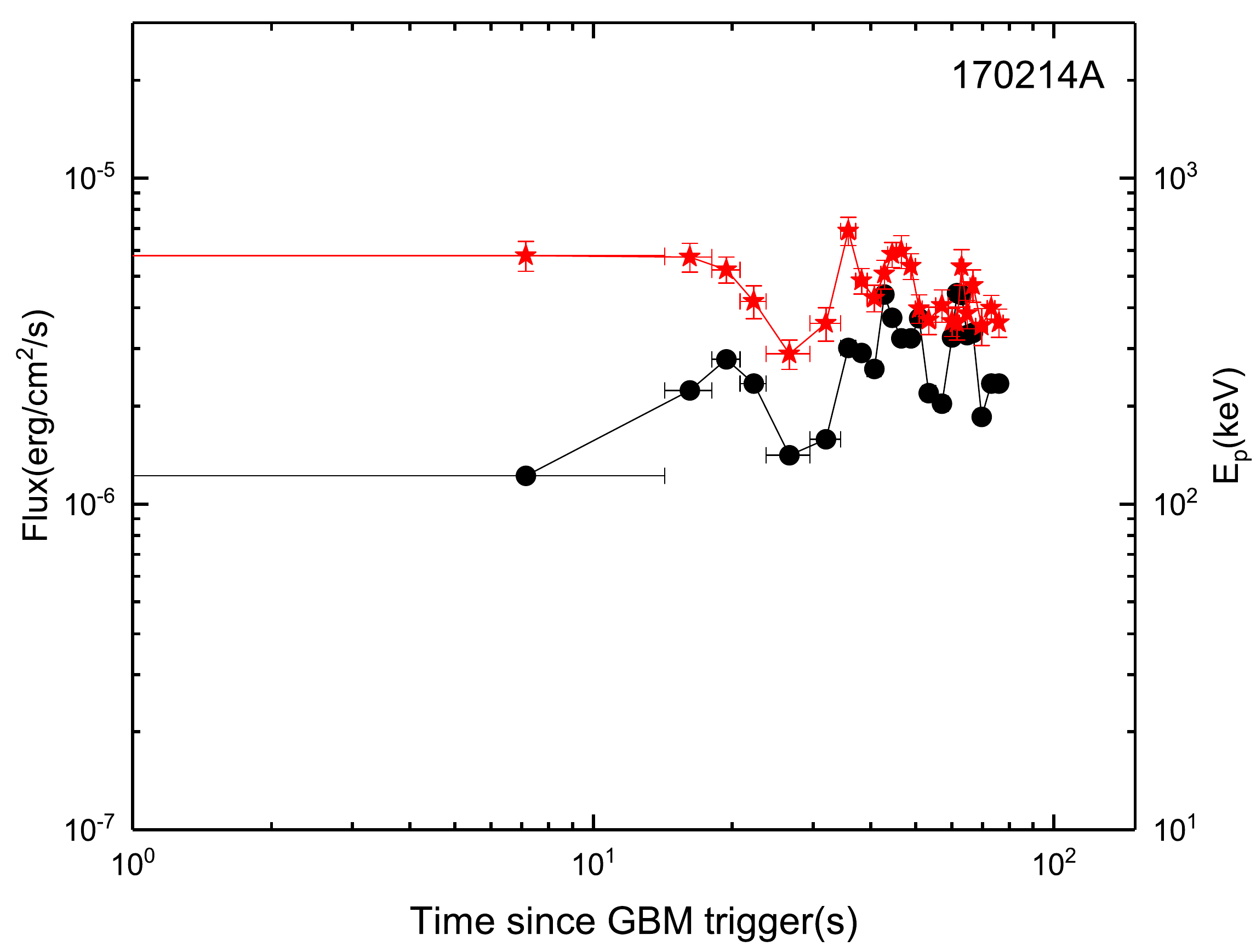}}
\resizebox{4cm}{!}{\includegraphics{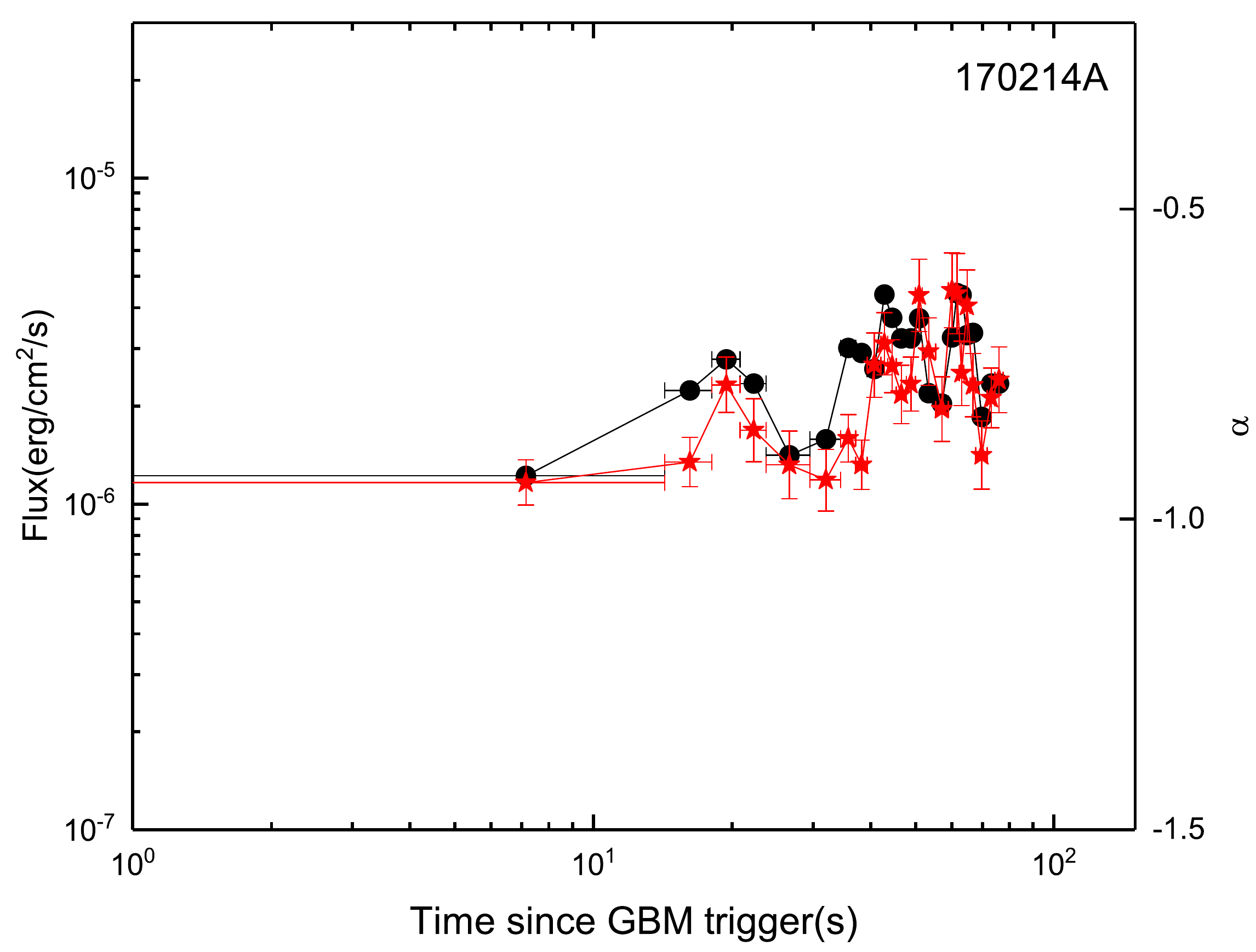}}
\resizebox{4cm}{!}{\includegraphics{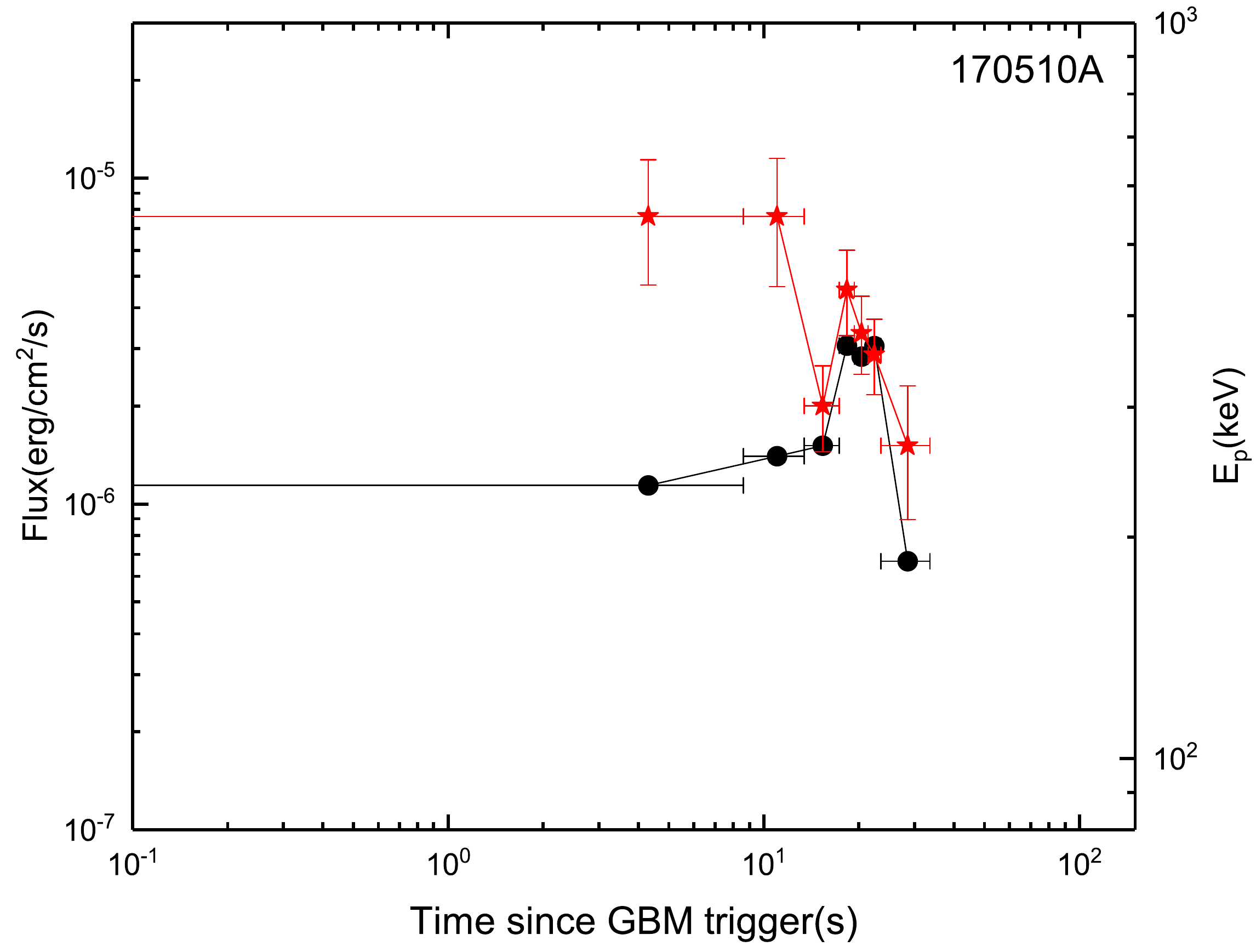}}
\resizebox{4cm}{!}{\includegraphics{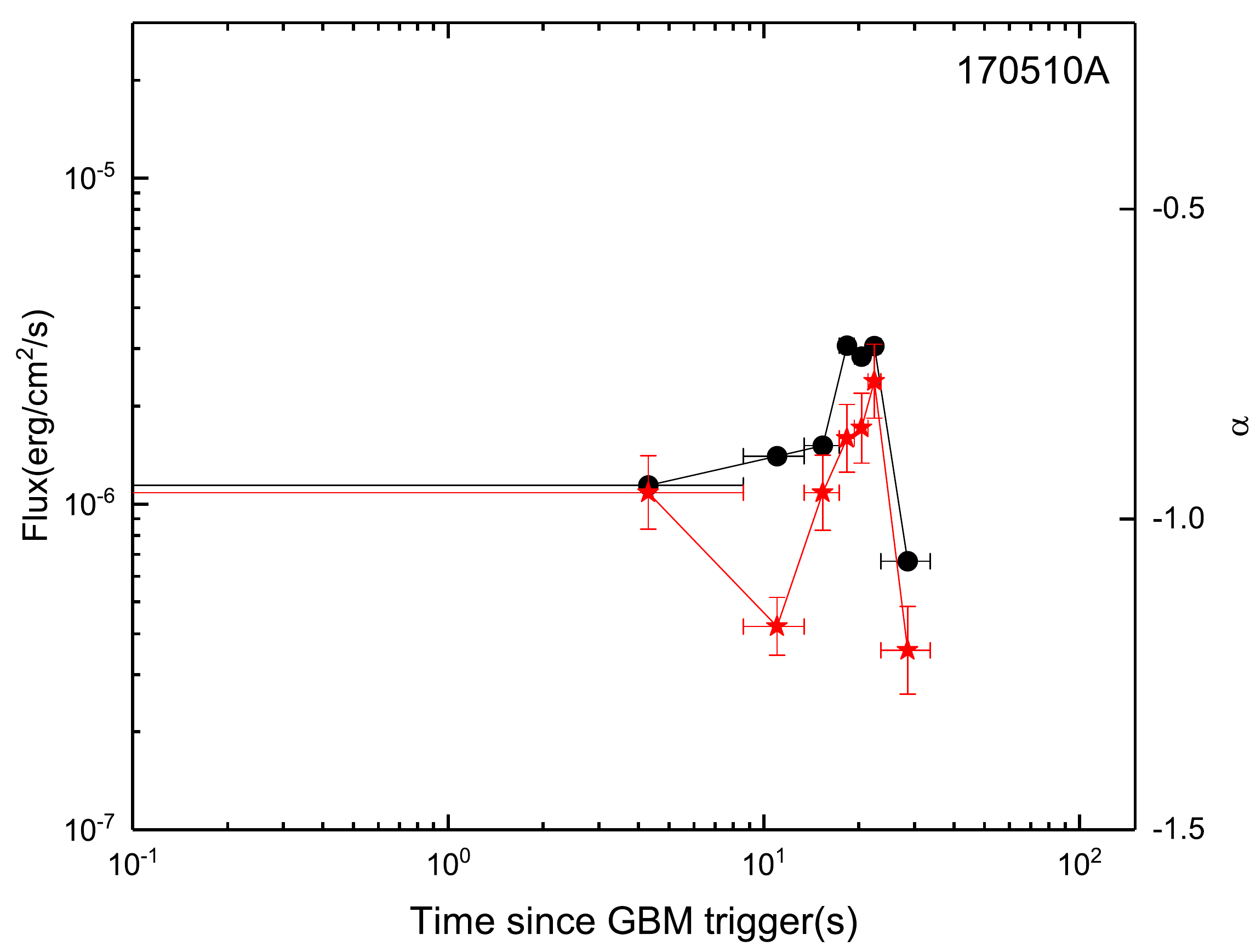}}
\resizebox{4cm}{!}{\includegraphics{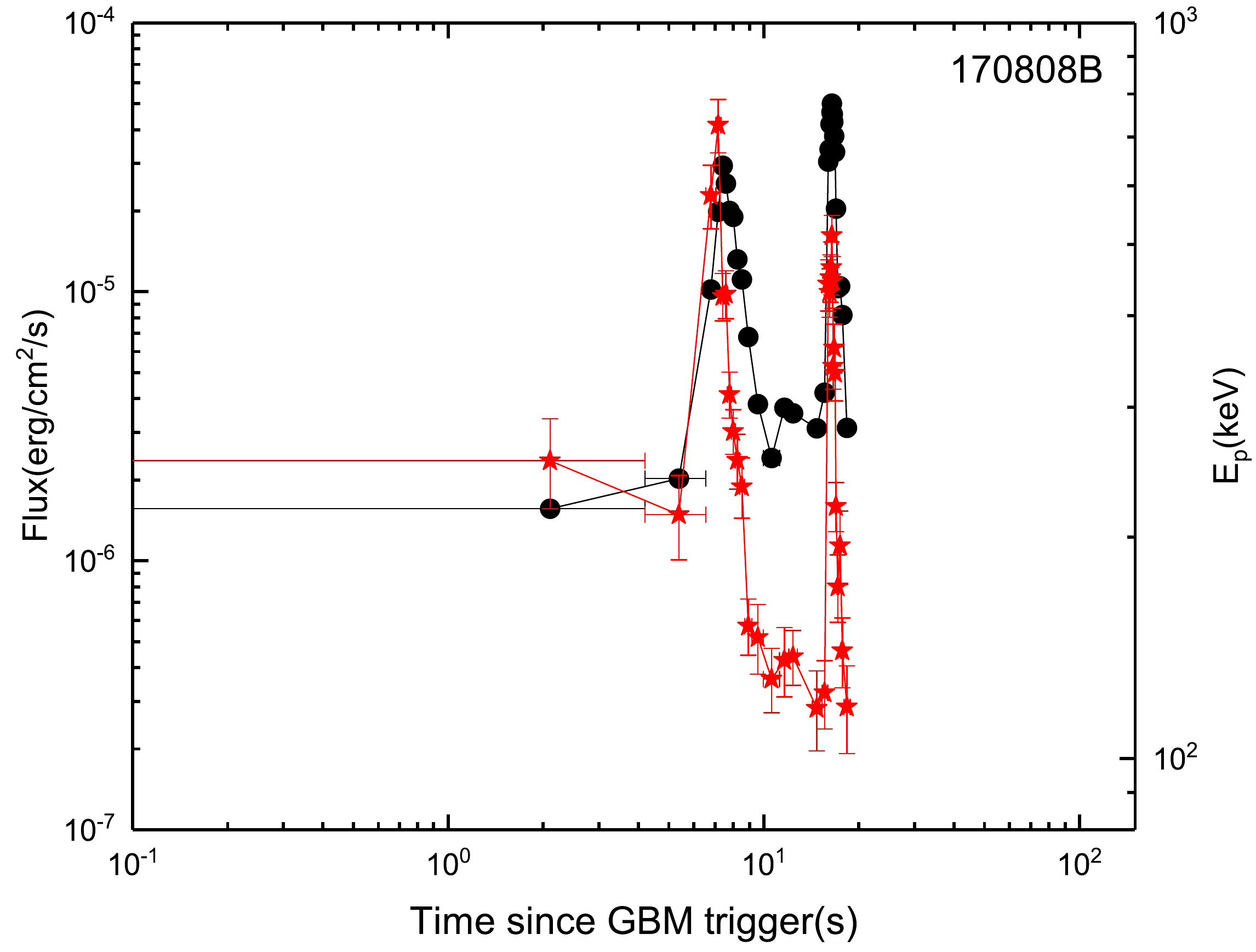}}
\resizebox{4cm}{!}{\includegraphics{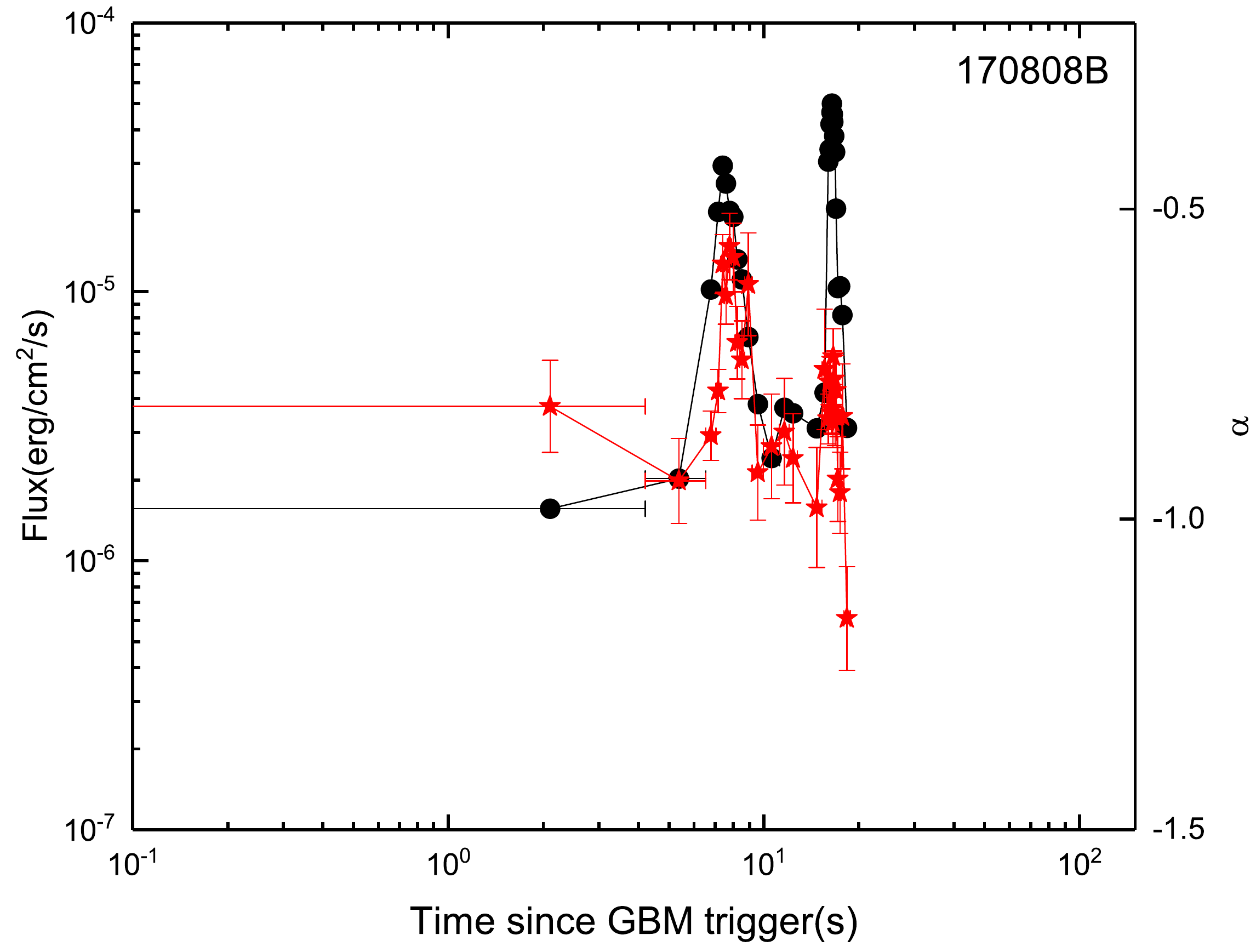}}
\resizebox{4cm}{!}{\includegraphics{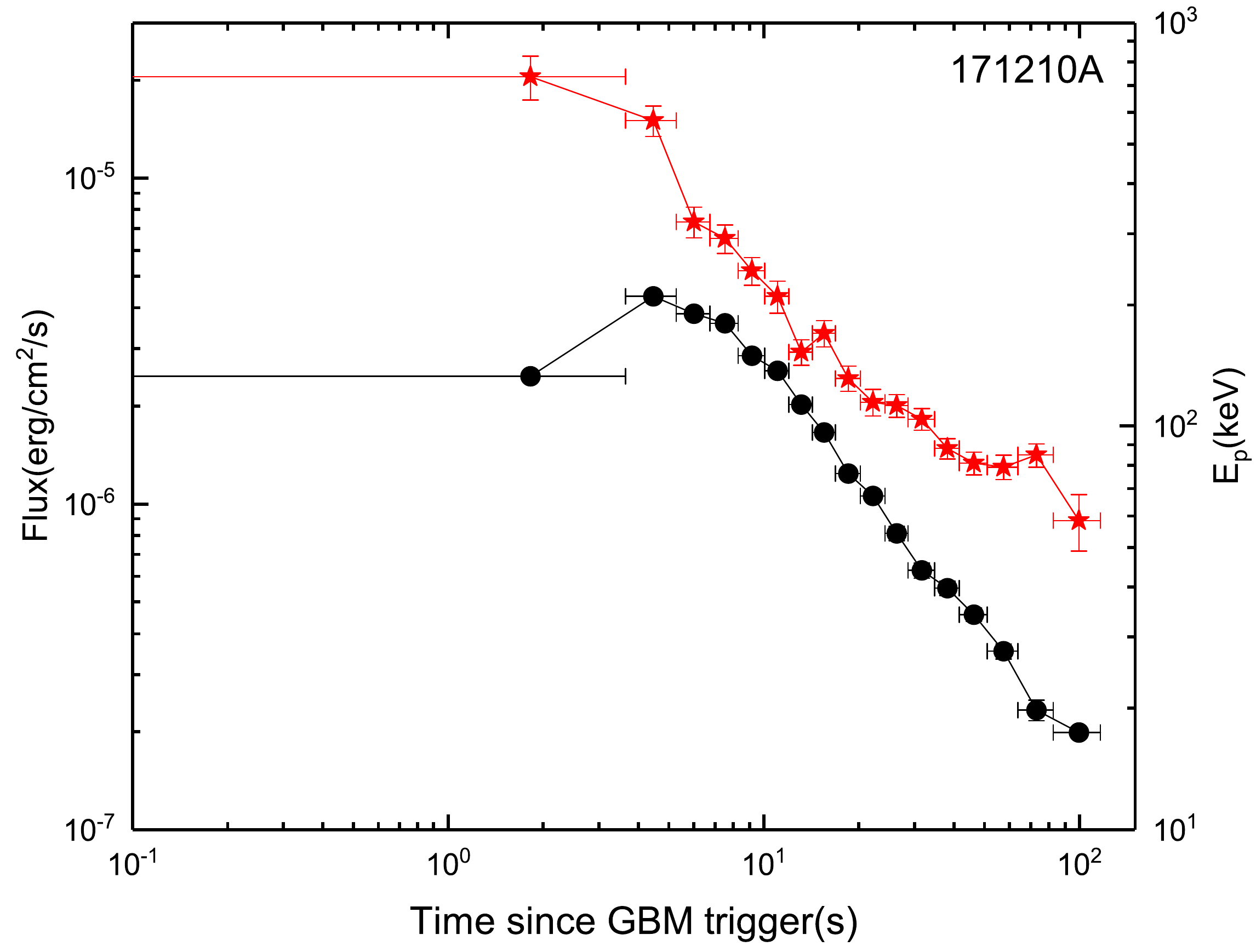}}
\resizebox{4cm}{!}{\includegraphics{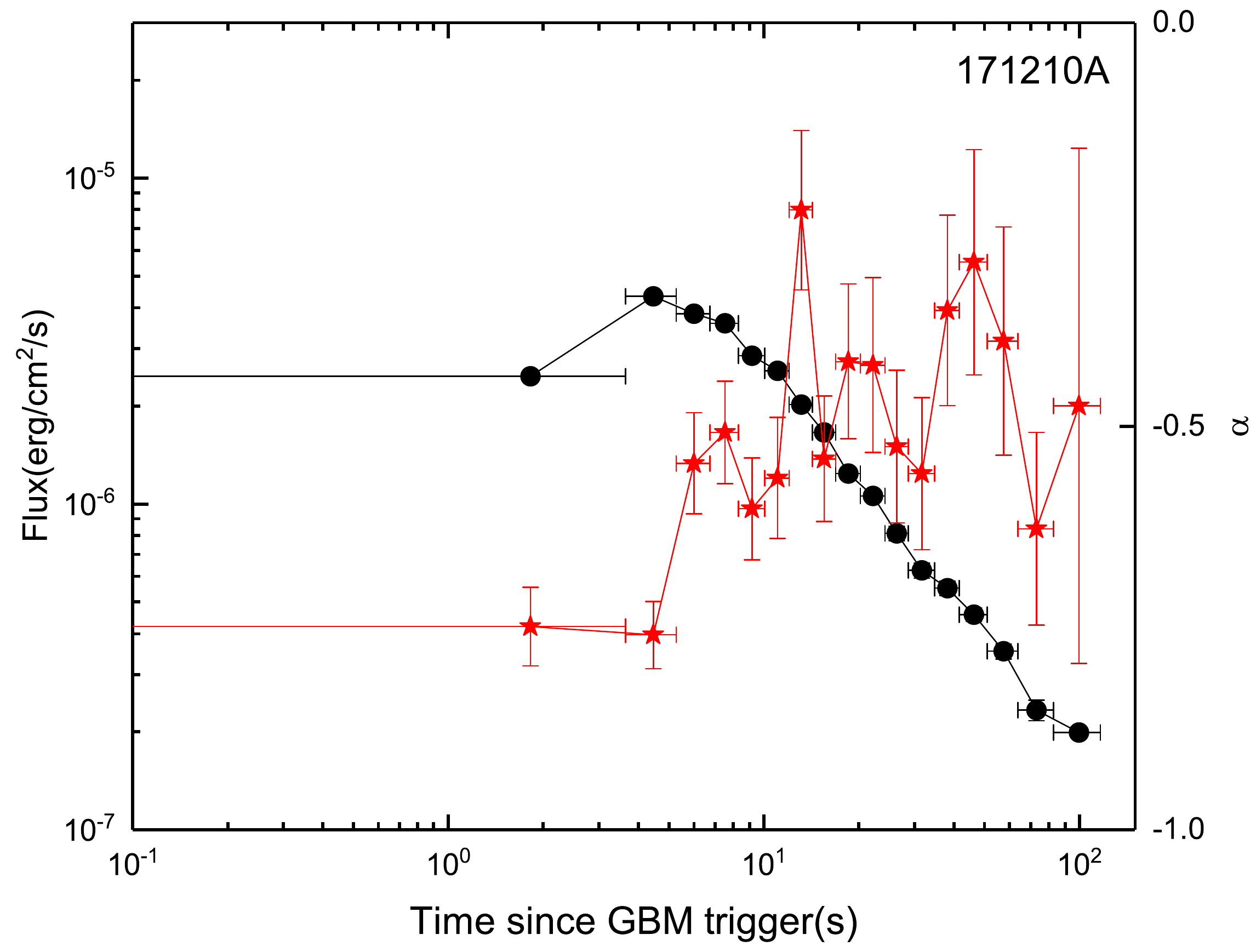}}
\resizebox{4cm}{!}{\includegraphics{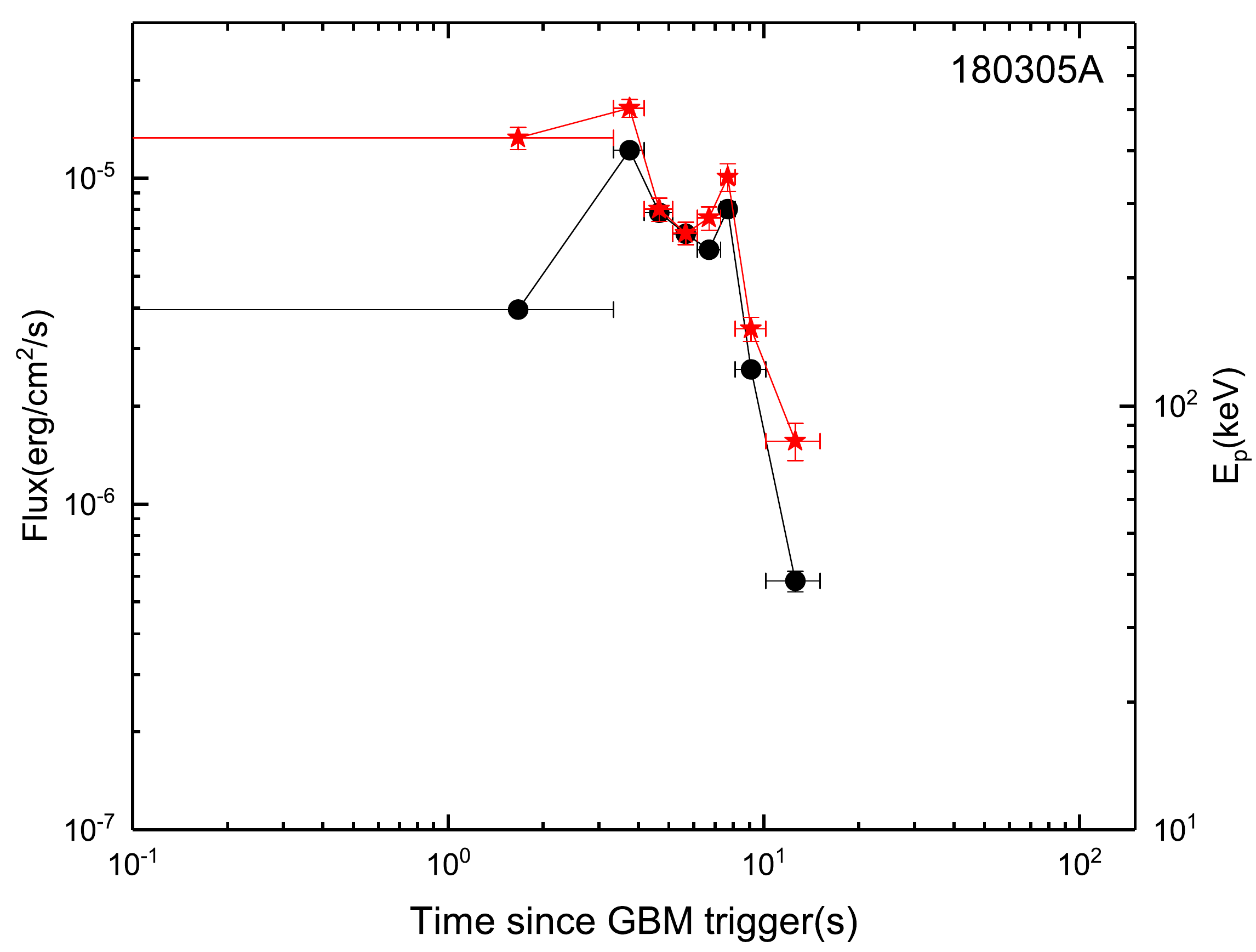}}
\resizebox{4cm}{!}{\includegraphics{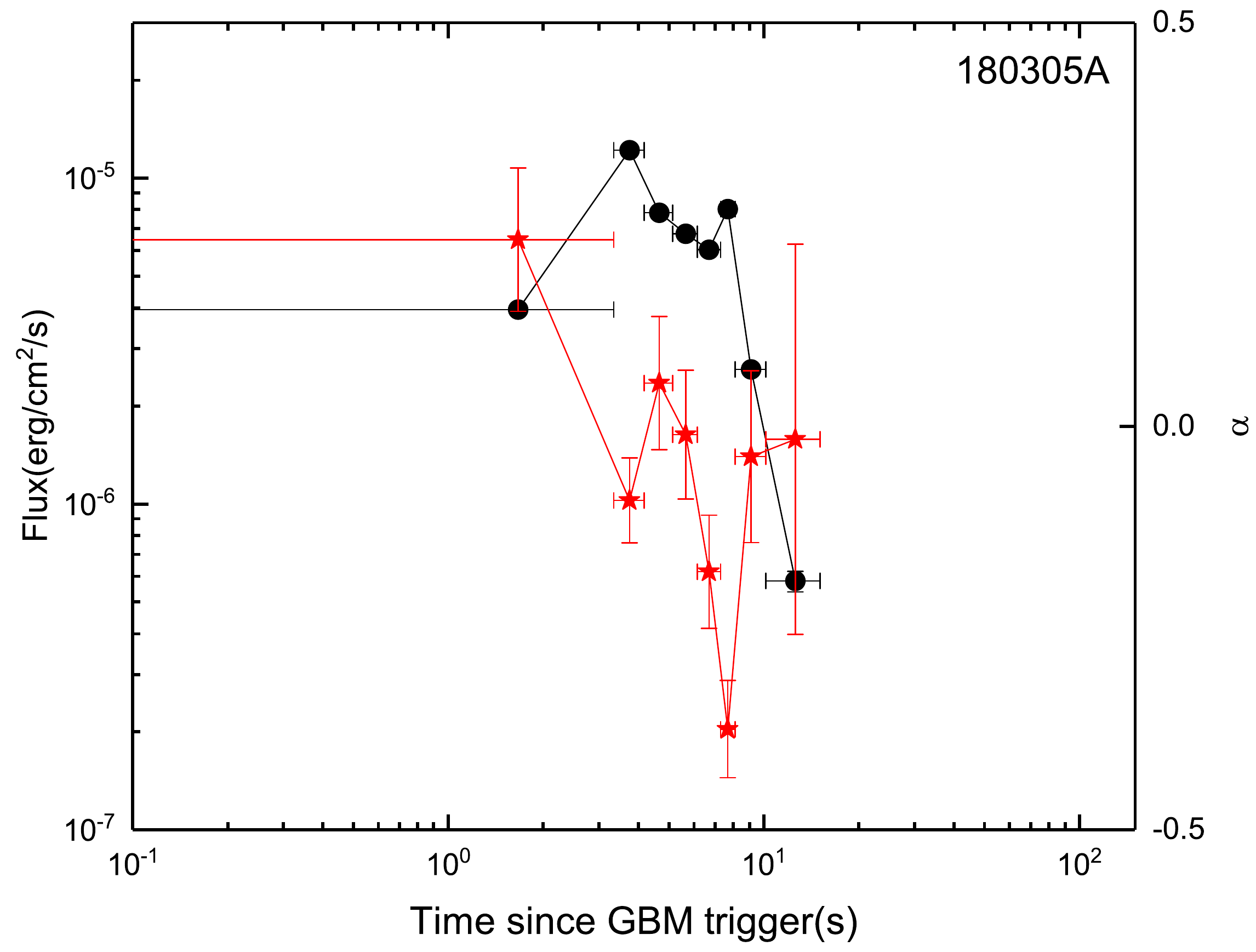}}
\caption{\it-continued}
\end{figure}

\begin{figure}
\gridline{\fig{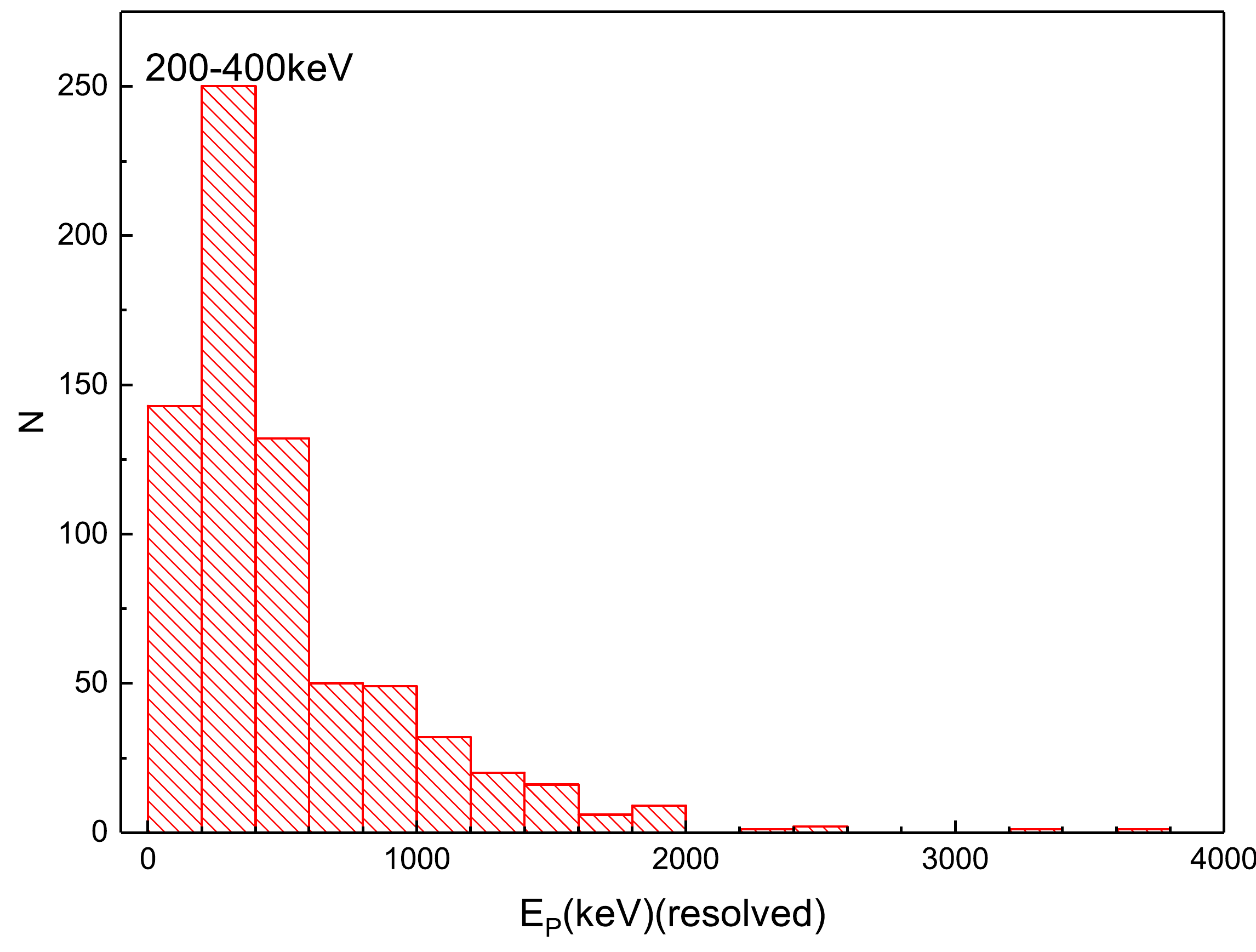}{0.5\textwidth}{}
          \fig{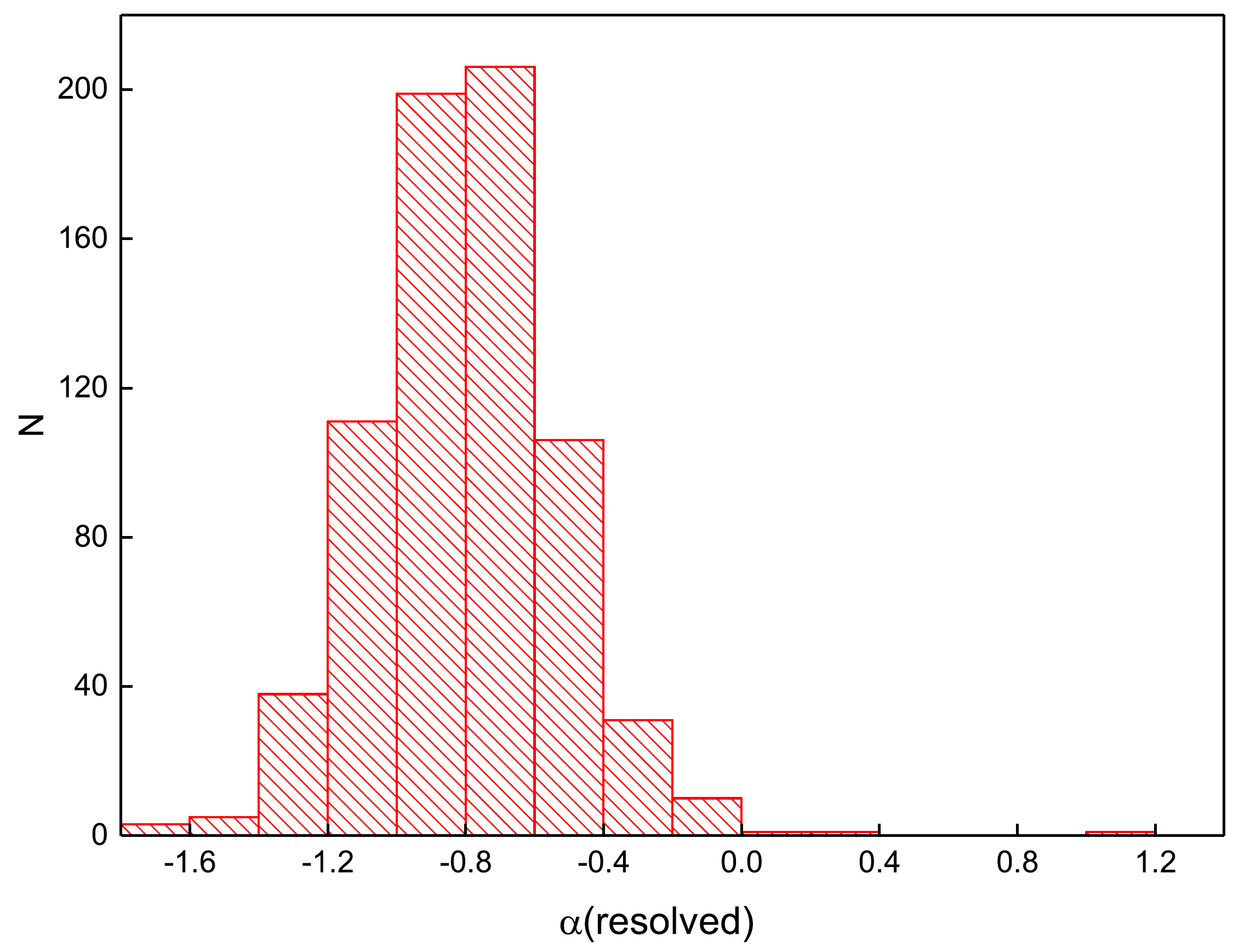}{0.5\textwidth}{}
          }
\caption{The histograms of $E_{p}$ and $\alpha$ in \edit1{\added{the}} detailed time-resolved spectra. The left panel is the histogram of $E_{p}$, the typical value of $E_{p}$ is from 200 to 400 keV. The right panel shows the histogram of $\alpha$, \edit1{\deleted{and}} the typical value is $\sim -0.8$. The typical value is consistent with the statistical study of \edit1{\added{a}} large sample in \edit1{\replaced{those}{the}} previous literatures both for $E_{p}$ and $\alpha$ in all \edit1{\replaced{411}{712}} spectra.\label{fig:resolved}}
\end{figure}

We \edit1{\deleted{will }}present the results of time-resolved spectral analysis and the evolution patterns of $E_{p}$ and $\alpha$ \edit1{\replaced{below}{in this section}}. The fitting results of the parameter correlations and the spectral evolutions of $E_{p}$ and $\alpha$ have been shown in Table \ref{tab:resolved_results}. Listed in this Table are the \edit1{\replaced{29}{36}} GRBs in our sample which satisfy our criteria in this study (Col.1), the detectors used (Col.2), the number of the time slice (Col.3), the Pearson's correlation coefficient r in the $E_{p}- F$ correlation (Col.4), the Pearson's correlation coefficient r in the $\alpha - F$ correlation (Col.5), the Pearson's correlation coefficient r in the $E_{p}- \alpha$ correlation (Col.6), the spectral evolution patterns of $E_{p}$ and $\alpha$ (Col.7), whether the values of $\alpha$ in \edit1{\added{the}} time-resolved spectral analysis are larger than the synchrotron limit ($-\frac{2}{3}$) or not (Col.8), \edit1{\replaced{what is the cluster of each burst (Col.9).}{the Pearson's correlation coefficient r in the $\alpha-F$ correlation obtained from the simulation (Col.9), the Pearson's correlation coefficient r in the $E_{p}-\alpha$ correlation obtained from the simulation (Col.10).}} \edit1{\deleted{And the Figure \ref{fig:spectral evolutions} represents the temporal characteristics of energy flux for all bursts in our sample (the left-hand, y-axis), along with time evolutions of the $E_{p}$ and $\alpha$, both are marked with red stars in the right-hand y-axis. That is to say,}}Figure \ref{fig:spectral evolutions} shows the spectral evolutions for \edit1{\replaced{all of the bursts in our sample.}{all the LLE bursts.}} The histograms of $E_{p}$ and $\alpha$ obtained by performing the detailed time-resolved spectral analysis have been shown in Figure \ref{fig:resolved}.

As described above, there are three types for the evolution patterns of peak energy $E_{p}$: (i) `hard-to-soft' trend; (ii) `flux-tracking' trend; (iii) `soft-to-hard' trend or chaotic evolutions. \edit1{\replaced{And the}{The}} recent study pointed out that the first two patterns are dominated\edit1{\deleted{ in $E_{p}$ evolution}}. A good fraction of GRBs follow `hard-to-soft' trend (about two-thirds), the rest should be the `flux-tracking' pattern (about one-third). While the low energy photon index $\alpha$ does not show \edit1{\added{a}} strong general trend compared with $E_{p}$ although it also evolves with time instead of remaining constant. All of these results can be contributed to the statistical study for the large sample of bursts in the previous literatures. \edit1{\deleted{However, maybe, }}\edit1{\replaced{our study will}{Our study may}} give birth to different and new progress in the field of the $Fermi$-LLE \edit1{\replaced{gamma ray}{gamma-ray}} bursts.

We investigate \edit1{\deleted{the }}Figure \ref{fig:spectral evolutions} in detail and \edit1{\replaced{identify them as six categories for the evolution patterns of $E_{p}$ and $\alpha$}{identify the evolution patterns of $E_{p}$ and $\alpha$ as six categories}}. In fact, \edit1{\replaced{there are four groups for}{five groups are enough to depict}} the evolution pattern of $E_{p}$, \edit1{\replaced{5}{6}} GRBs exhibit the `hard-to-soft' pattern; 2 GRBs undergo the transition from `soft-to-hard' to `hard-to-soft' (GRBs 131108A and 150510A); \edit1{\replaced{4}{5}} GRBs show the `intensity-tracking' (compared with flux); \edit1{\replaced{and 18}{22}} GRBs, a good fraction of those samples exhibit the `rough-tracking' (compared with flux) behavior\edit1{\added{; the other two, GRBs 150314A, 170510A, exhibit the chaotic evolutions}}. \edit1{\added{It is noticeable that, GRB 171210A, a special burst, shows the rough `flux-tracking' pattern with the superposition of `hard-to-soft' evolution.}} \edit1{\replaced{And it}{It}} is obvious that the `flux-tracking' pattern is very popular for most of the bursts, the total number include `intensity-tracking' and `rough-tracking' is \edit1{\replaced{22}{27}}, which means that \edit1{\replaced{the 75.9}{75}} percent of these bursts follow the `flux-tracking' pattern. For the evolution of $\alpha$, it consists of `hard-to-soft' pattern, `soft-to-hard' to `hard-to-soft' pattern, `intensity-tracking' pattern, `rough-tracking' pattern, `anti-tracking' pattern, `rough-tracking' combined with `hard-to-soft' pattern\edit1{\added{,}} and chaotic evolution pattern (all `-tracking' patterns based on the evolution of energy flux). \edit1{\replaced{2}{3}} GRBs exhibit the `hard-to-soft' pattern; 1 GRB undergoes the transition from `soft-to-hard' to `hard-to-soft' (GRB 110721A); 2 GRBs show `intensity-tracking' pattern; most of the bursts, \edit1{\replaced{20}{26}} GRBs exhibit `rough-tracking'; \edit1{\replaced{1 GRB exhibits}{3 GRBs exhibit}} the chaotic evolution; the rest \edit1{\replaced{three}{two}} GRBs, GRBs 150202B, 170115B\edit1{\deleted{ and 150510A }}, exhibit\edit1{\deleted{ the special behaviours, which the first two show}} the `anti-tracking' pattern\edit1{\added{.}} \edit1{\replaced{and the last one}{Similarly, we found that GRB 150510A}} shows the `rough-tracking' pattern combined with `hard-to-soft' pattern. All of these evolution patterns have been summarised in Table \ref{tab:resolved_results},\edit1{\deleted{and}} one can obtain the specific evolution pattern of $E_{p}$ and $\alpha$ for each burst from \edit1{\replaced{the}{this}} table.

In addition, from Figure \ref{fig:resolved} which has presented the histograms of $E_{p}$ and $\alpha$ obtained by performing the detailed time-resolved spectral analysis, the typical value is consistent with the statistical study of \edit1{\added{a}} large sample in \edit1{\replaced{those}{the}} previous literatures both for $E_{p}$ ($\sim$ 300 keV) and $\alpha$ ($\sim$ -0.8) in all \edit1{\replaced{411 spectra, but}{712 spectra. But}} such a value of $\alpha$ is inapplicable for some bursts such as GRBs \edit1{\replaced{080825C, 130305A, 141028A and 170115B}{080825C, 141028A, 170115B and 180305A}}, which the values of $\alpha$ for all slices are larger than the synchrotron limit (-$\frac{2}{3}$). Especially, GRB 170115B is different from the other three bursts because \edit1{\deleted{of the fact that }}the value of $\alpha$ ($\sim-0.8$) in \edit1{\added{the}} time-integrated spectrum is smaller than the synchrotron limit while the values in all \edit1{\added{the}} time-resolved spectra are larger than $-\frac{2}{3}$\edit1{\replaced{, however,}{. However,}} for the other three bursts, the value of $\alpha$ is larger than the limit both for \edit1{\added{the}} time-integrated spectrum and each time-resolved spectrum. \edit1{\replaced{And the evolution of it violate}{On the other hand, its evolution violates}} most of the bursts, which \edit1{\replaced{exhibit}{exhibits}} the `anti-tracking' \edit1{\replaced{behaviour}{behavior}} compared with energy flux, i.e., it is decreasing/increasing when the energy flux is increasing/decreasing. From Table \ref{tab:resolved_results}, one can \edit1{\added{also}} find that only \edit1{\replaced{7}{9}} GRBs can be classified as the kind that all of the values of $\alpha$ in the detailed time-resolved spectra \edit1{\replaced{don't}{do not}} exceed the synchrotron limit. \edit1{\replaced{And the values of $\alpha$ for the rest 18 GRBs}{The values of $\alpha$ for the rest of 23 GRBs}} in their detailed time-resolved spectra consist of the fraction that is larger than $-\frac{2}{3}$ and the fraction that does not exceed the synchrotron limit.

\subsection{Parameter Correlations} \label{subsec:subsec3.3}

\begin{figure}
\centering
\resizebox{4cm}{!}{\includegraphics{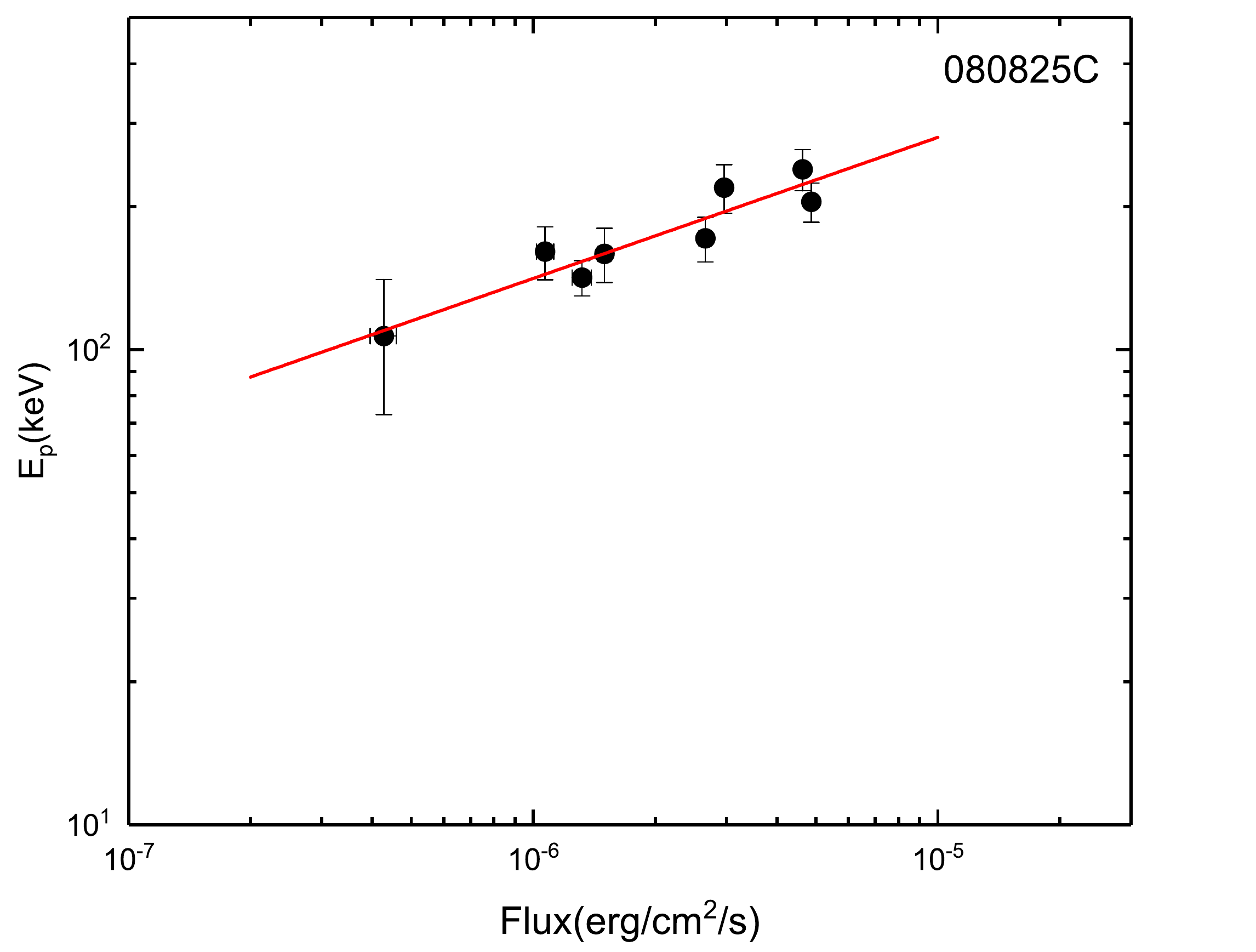}}
\resizebox{4cm}{!}{\includegraphics{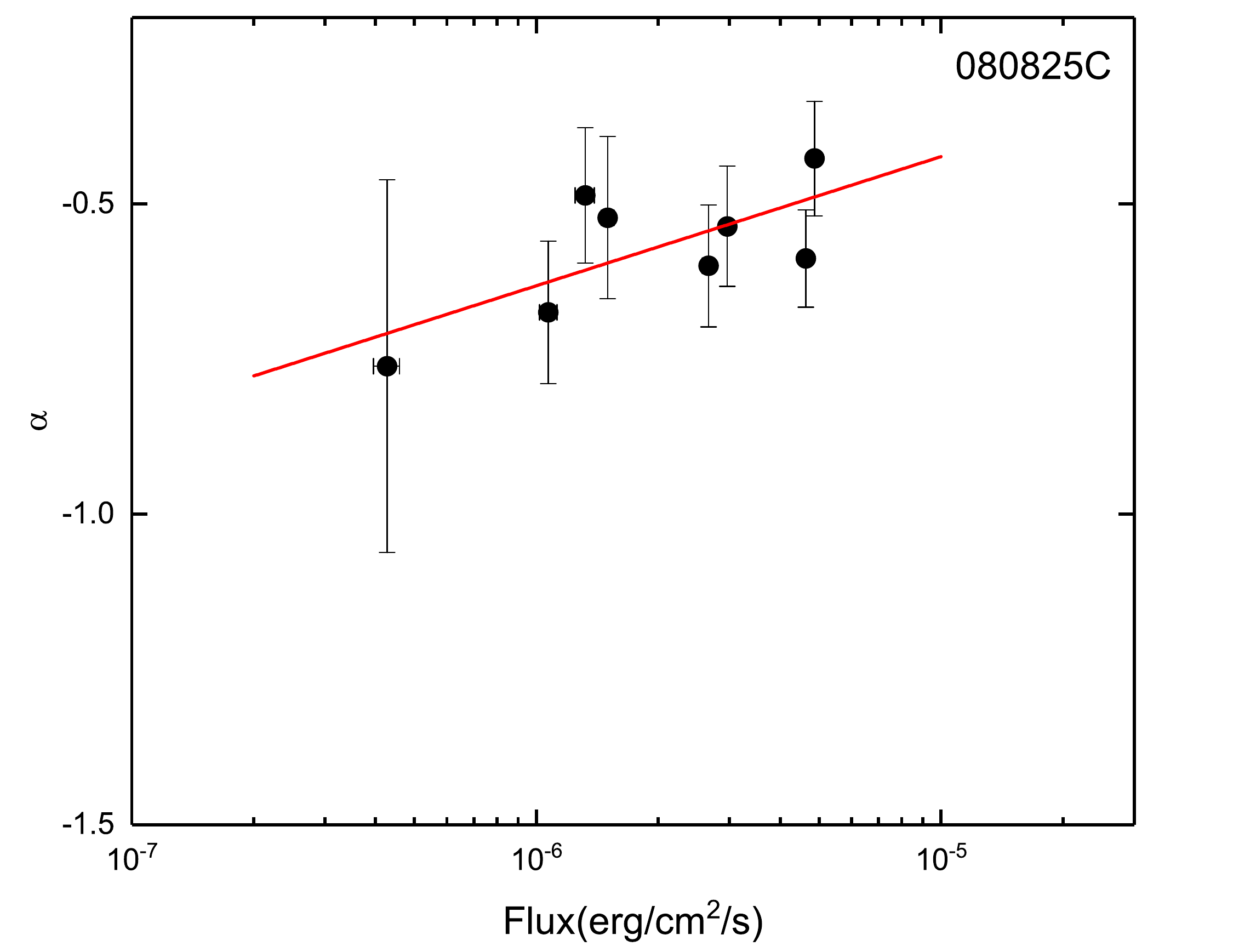}}
\resizebox{4cm}{!}{\includegraphics{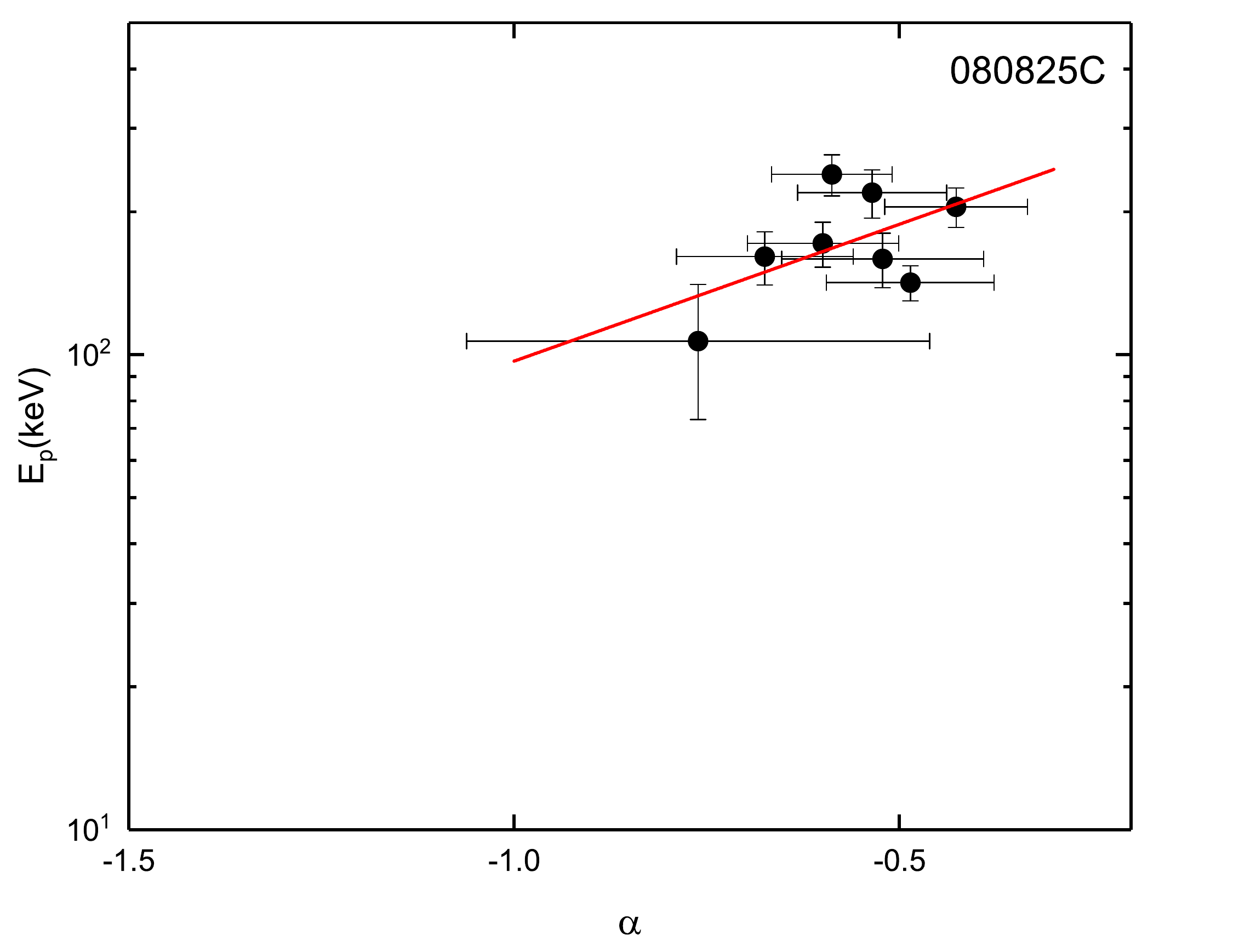}}
\resizebox{4cm}{!}{\includegraphics{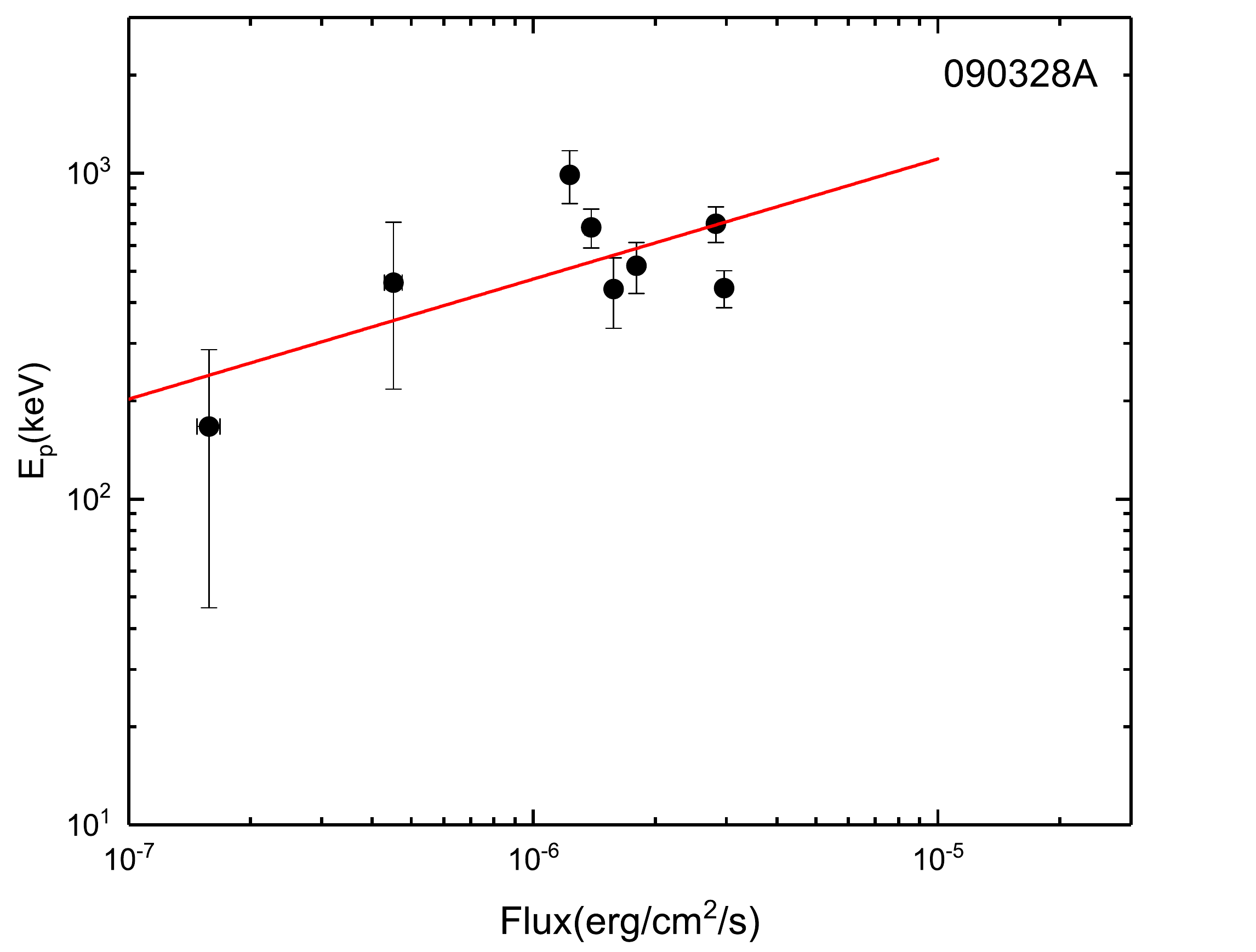}}
\resizebox{4cm}{!}{\includegraphics{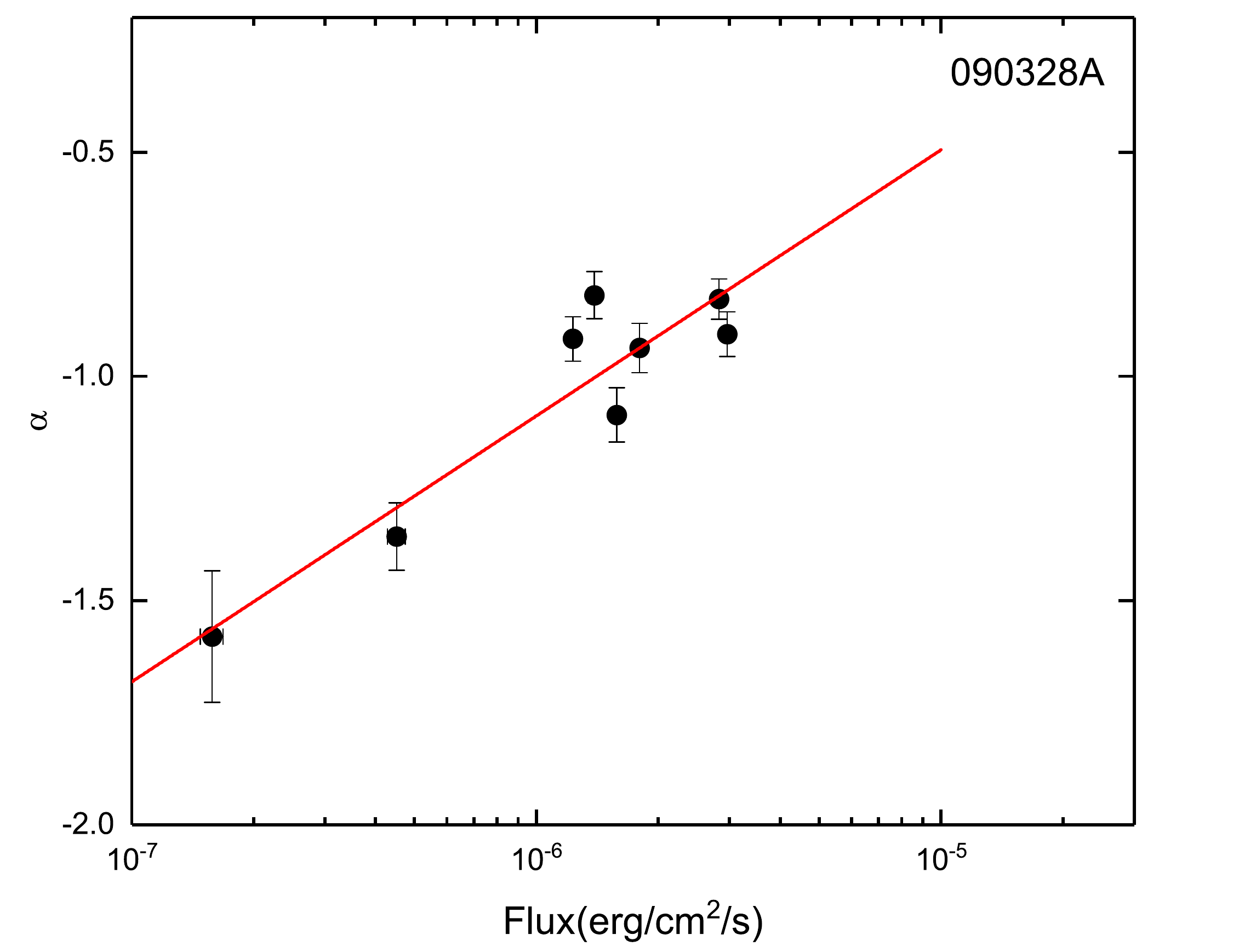}}
\resizebox{4cm}{!}{\includegraphics{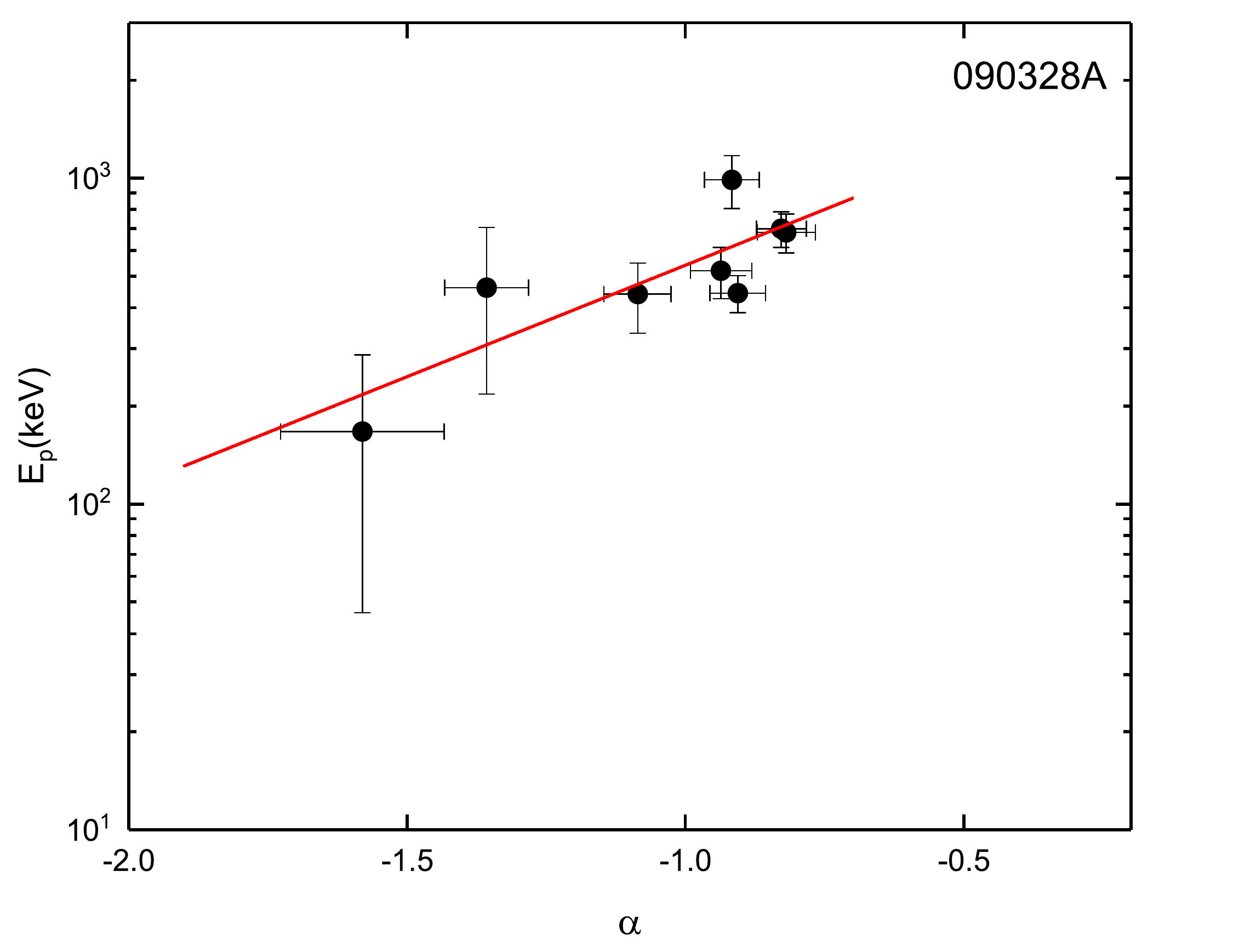}}
\resizebox{4cm}{!}{\includegraphics{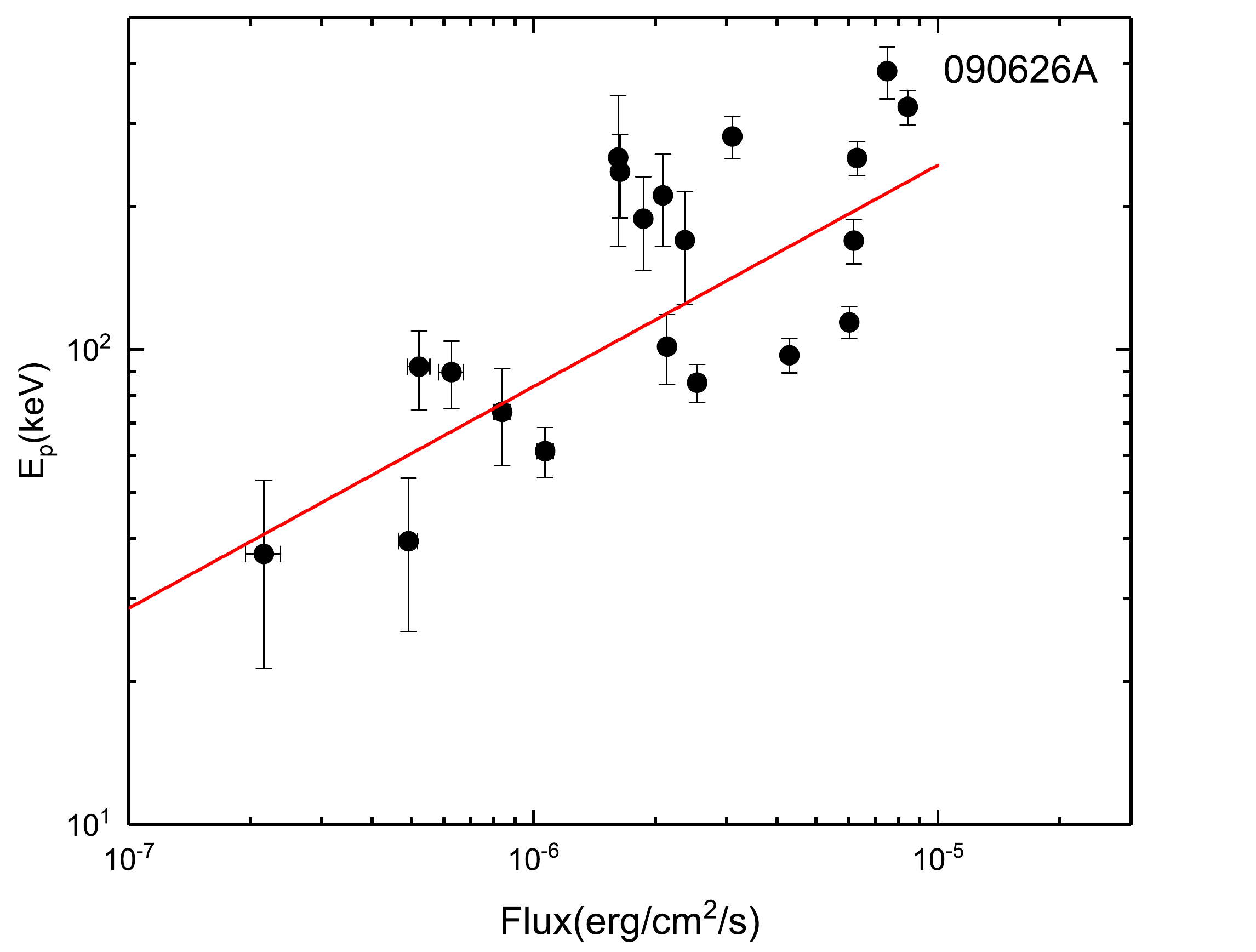}}
\resizebox{4cm}{!}{\includegraphics{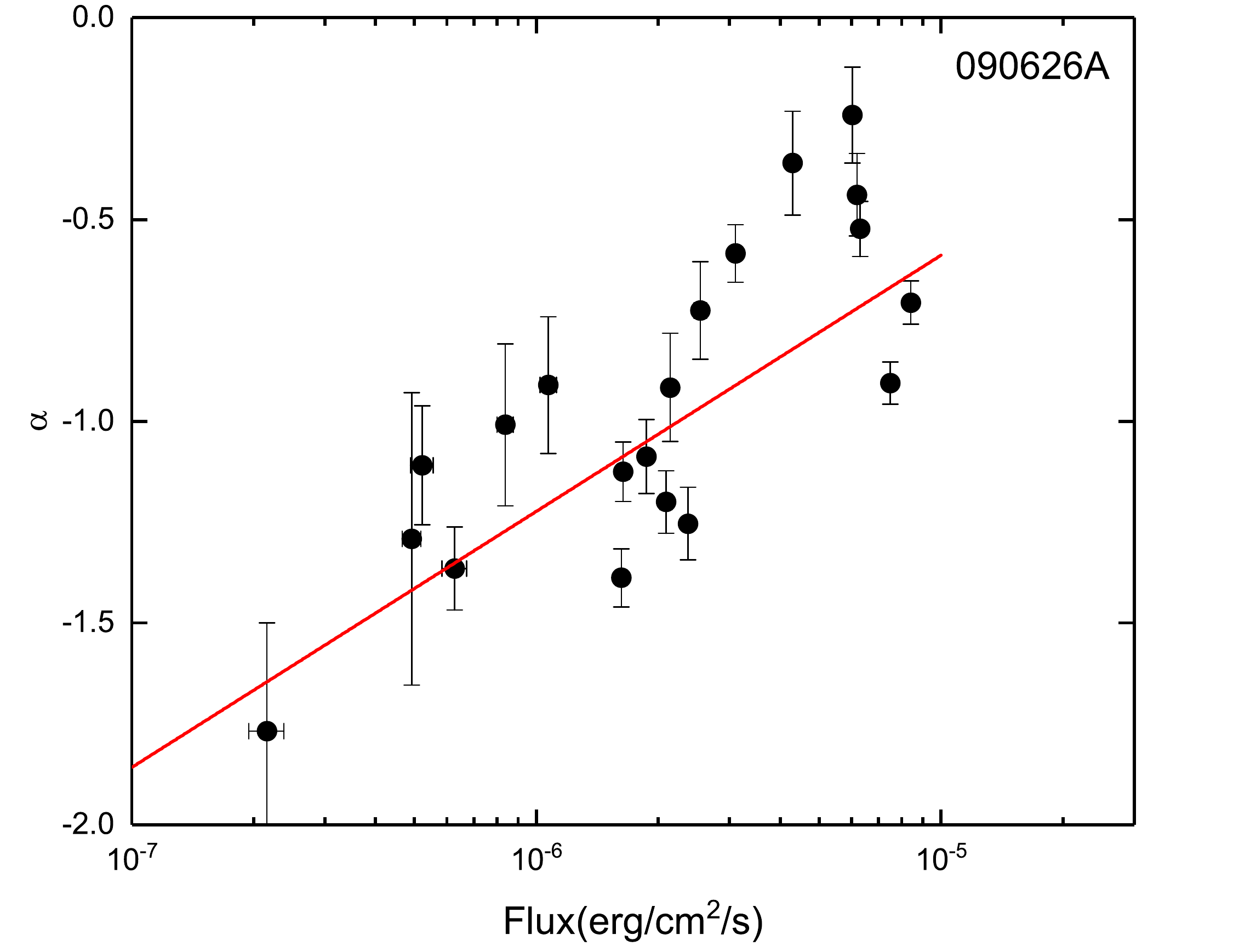}}
\resizebox{4cm}{!}{\includegraphics{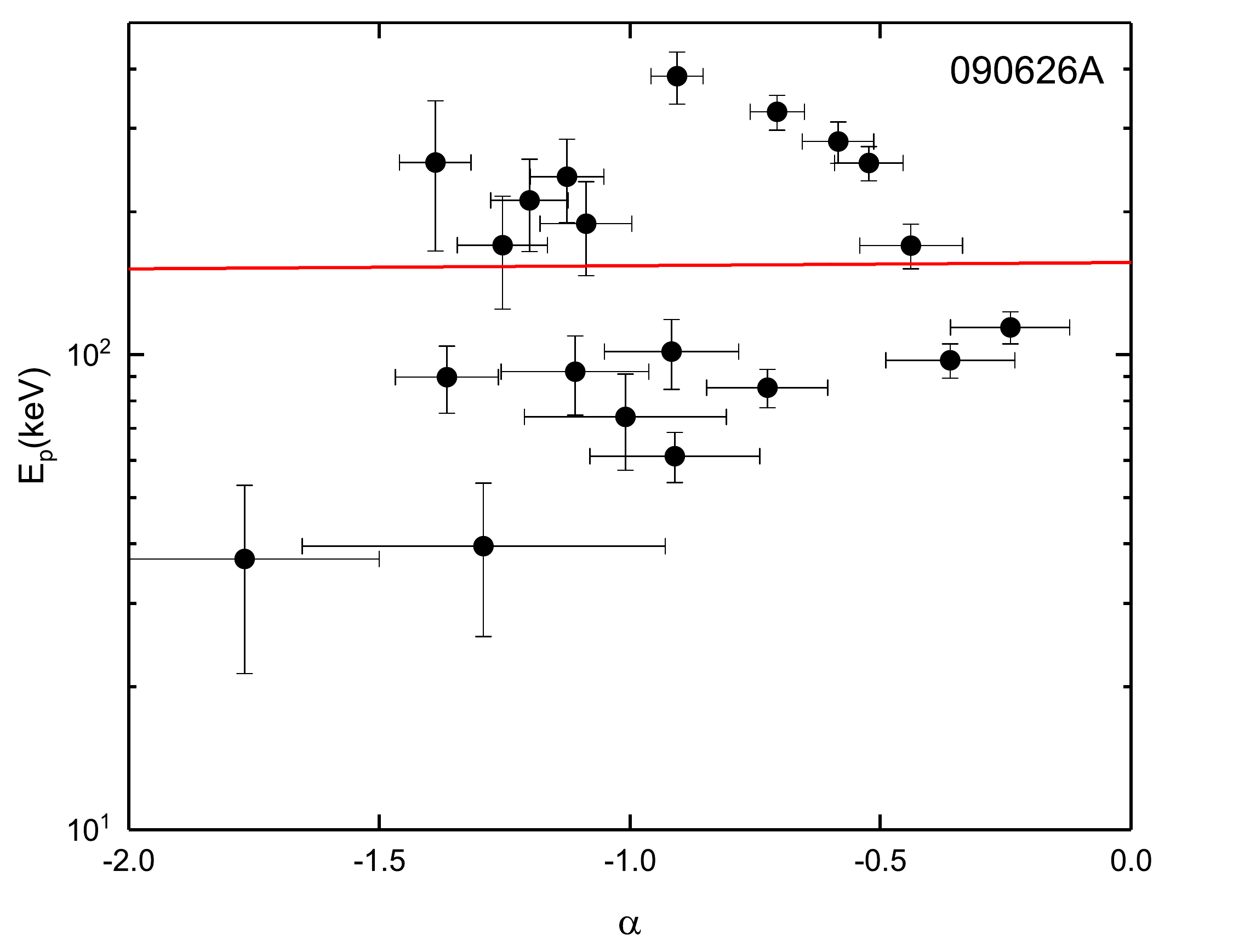}}
\resizebox{4cm}{!}{\includegraphics{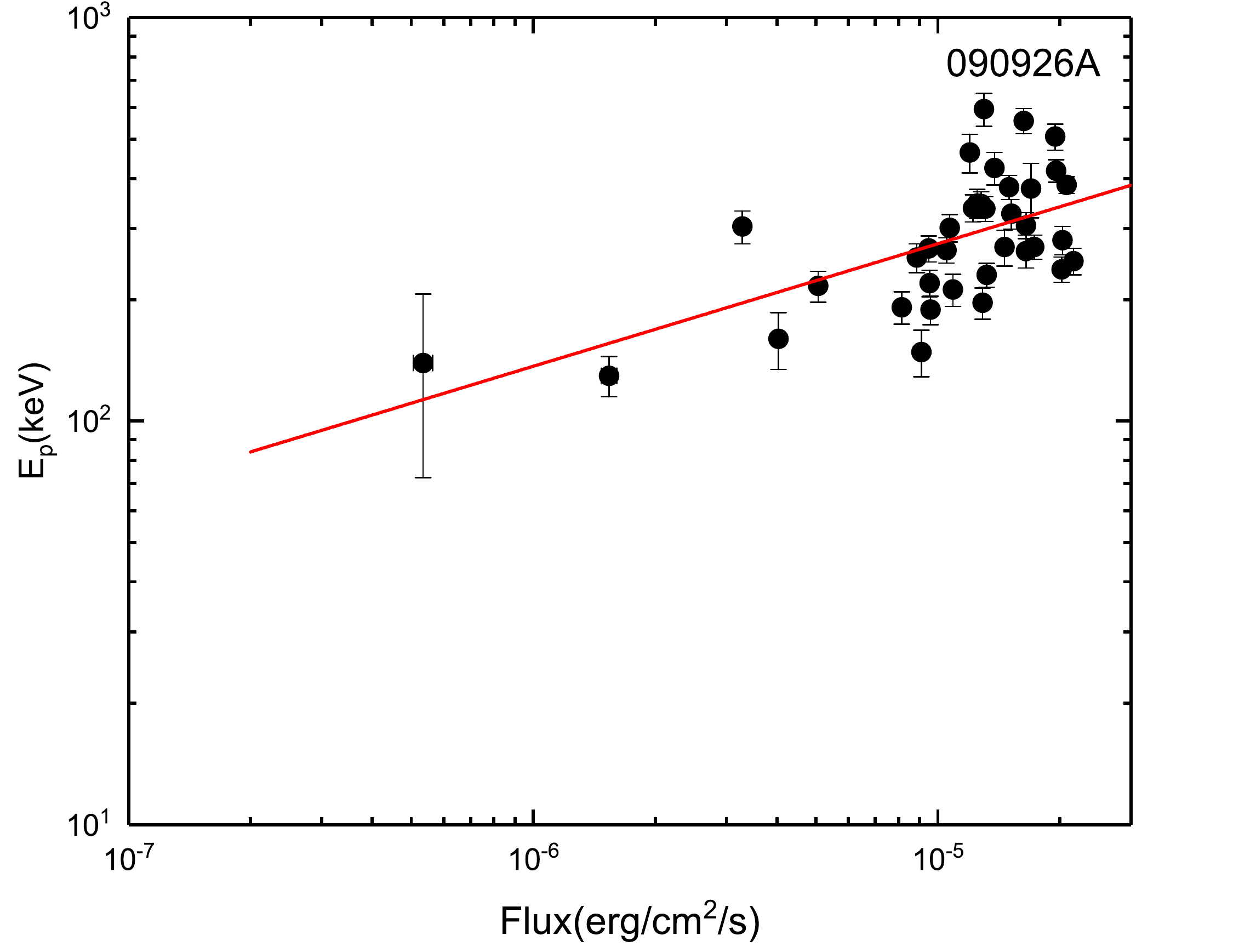}}
\resizebox{4cm}{!}{\includegraphics{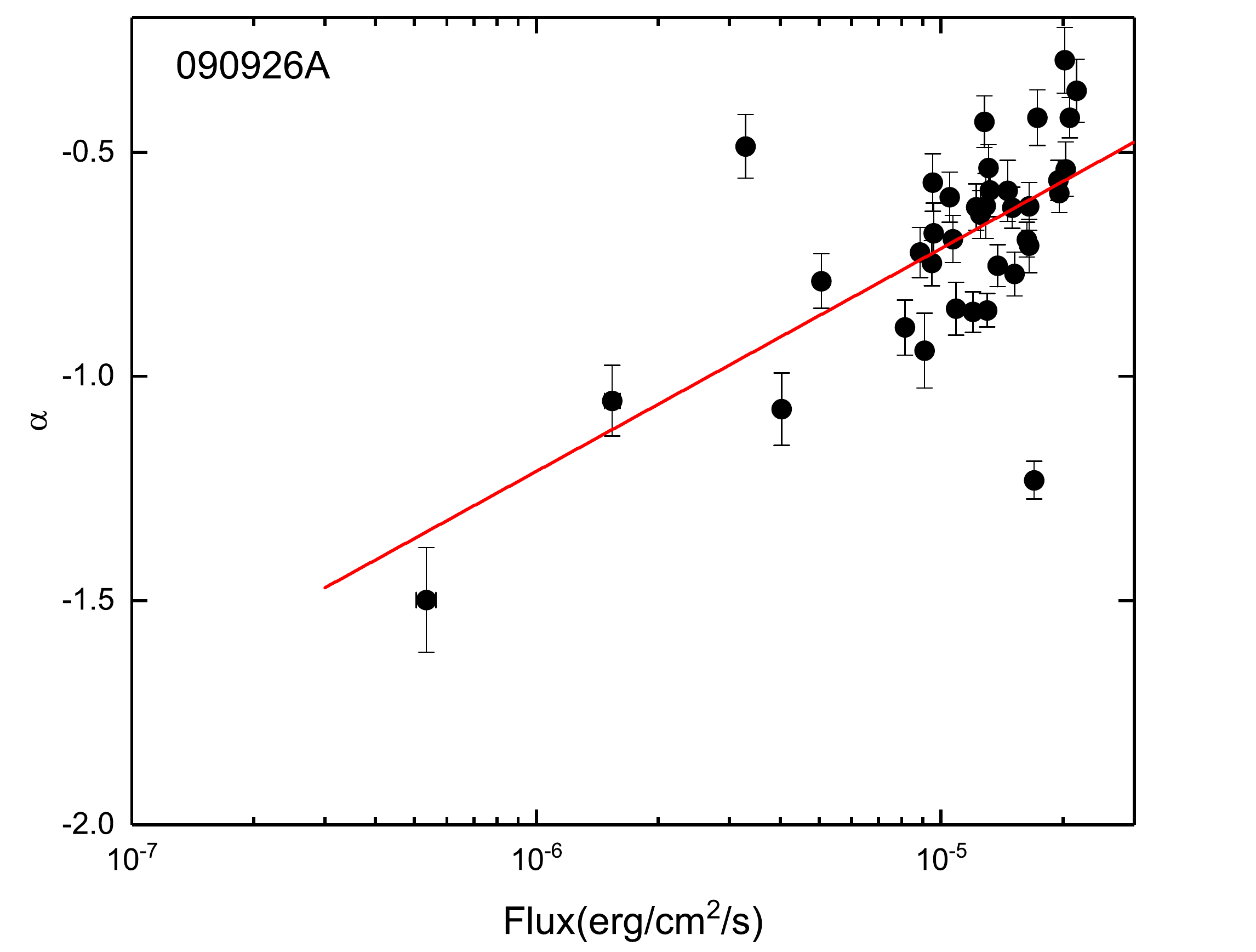}}
\resizebox{4cm}{!}{\includegraphics{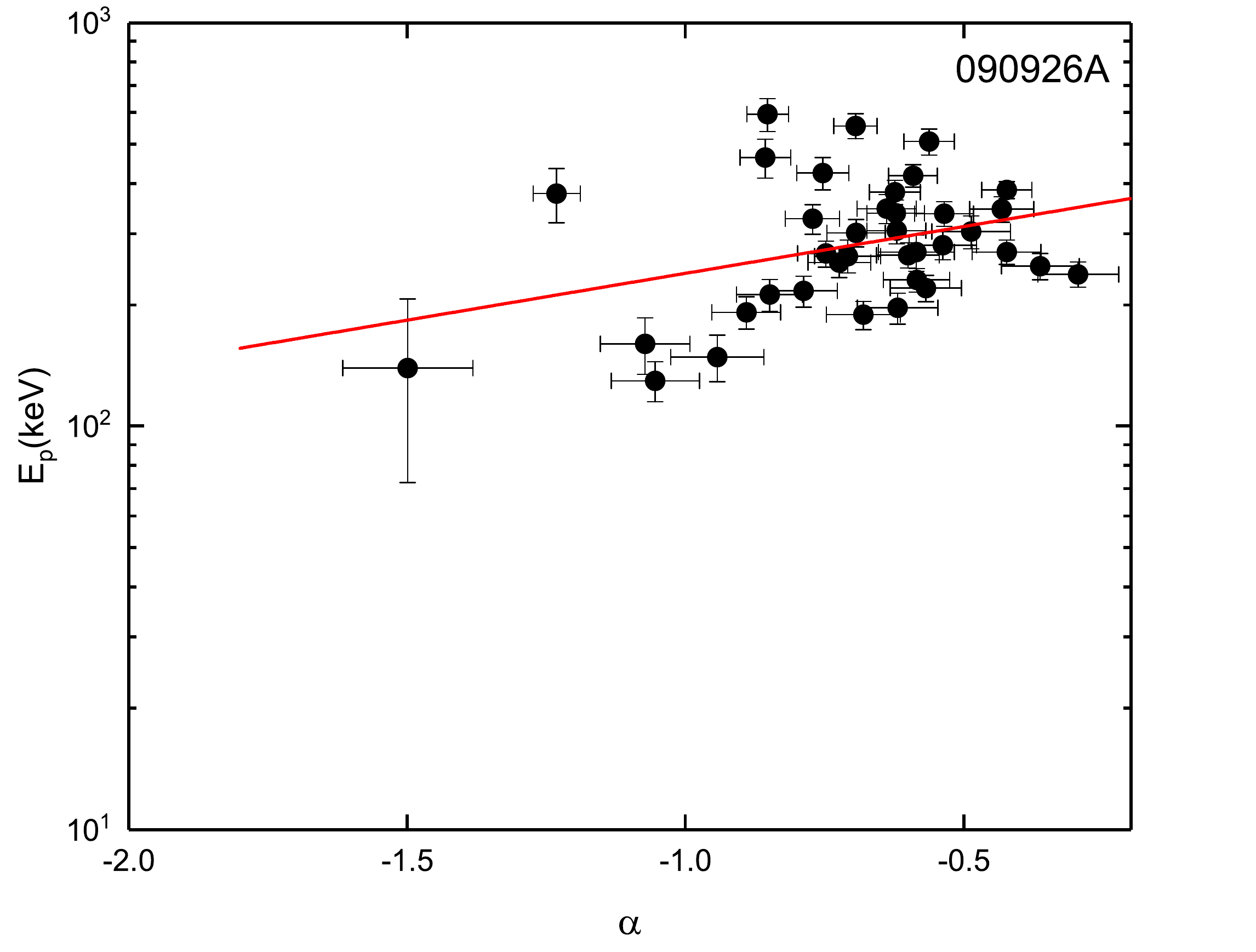}}
\resizebox{4cm}{!}{\includegraphics{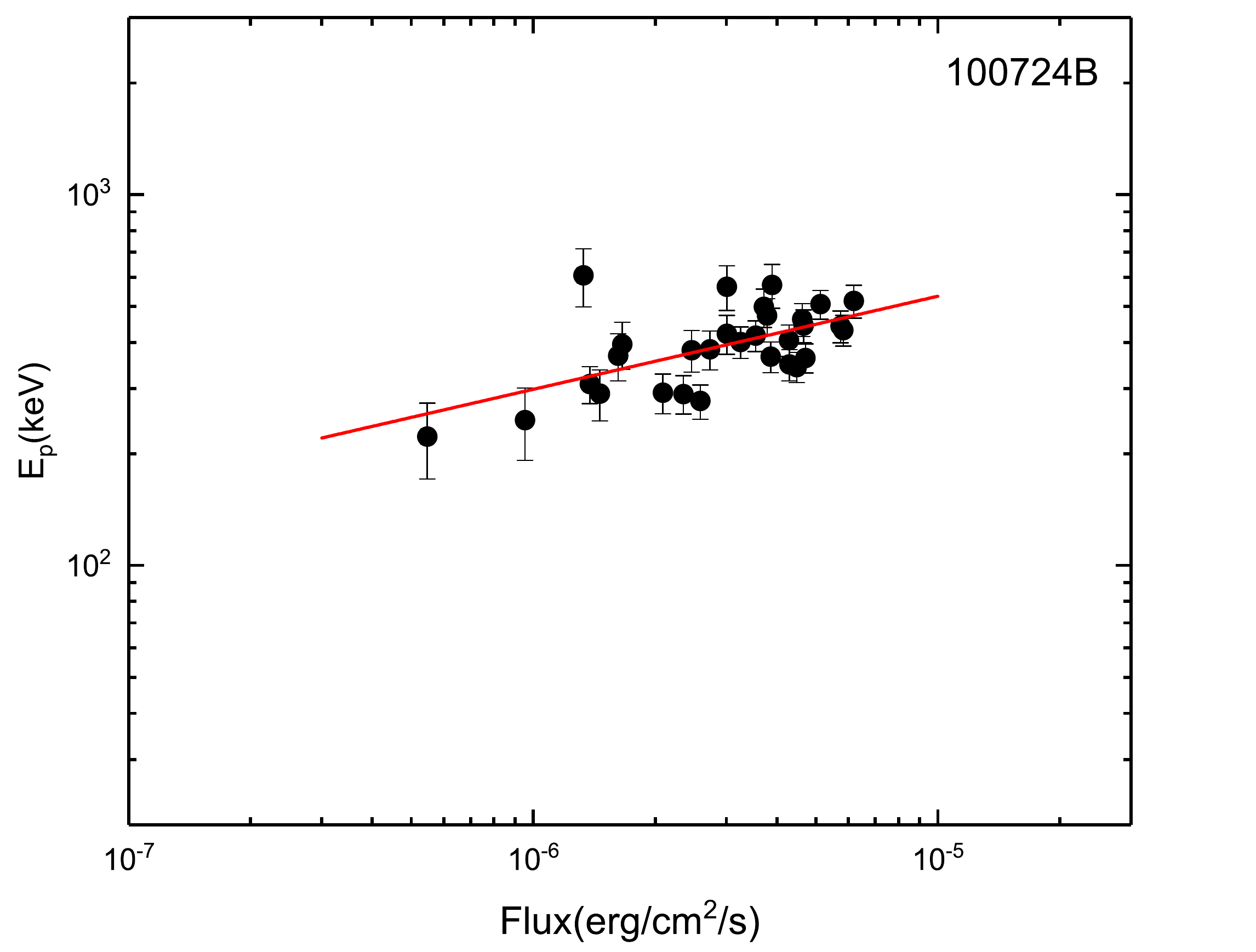}}
\resizebox{4cm}{!}{\includegraphics{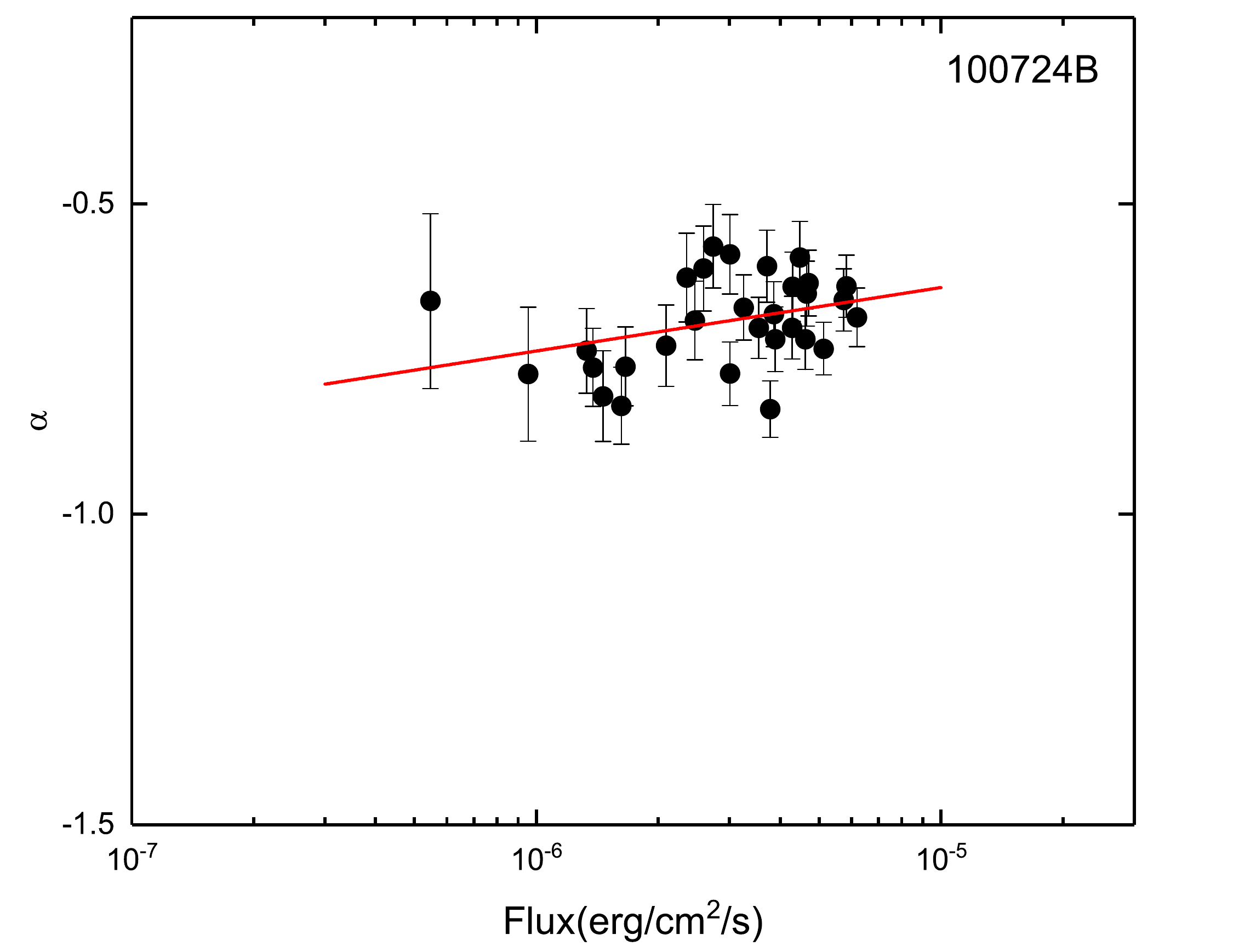}}
\resizebox{4cm}{!}{\includegraphics{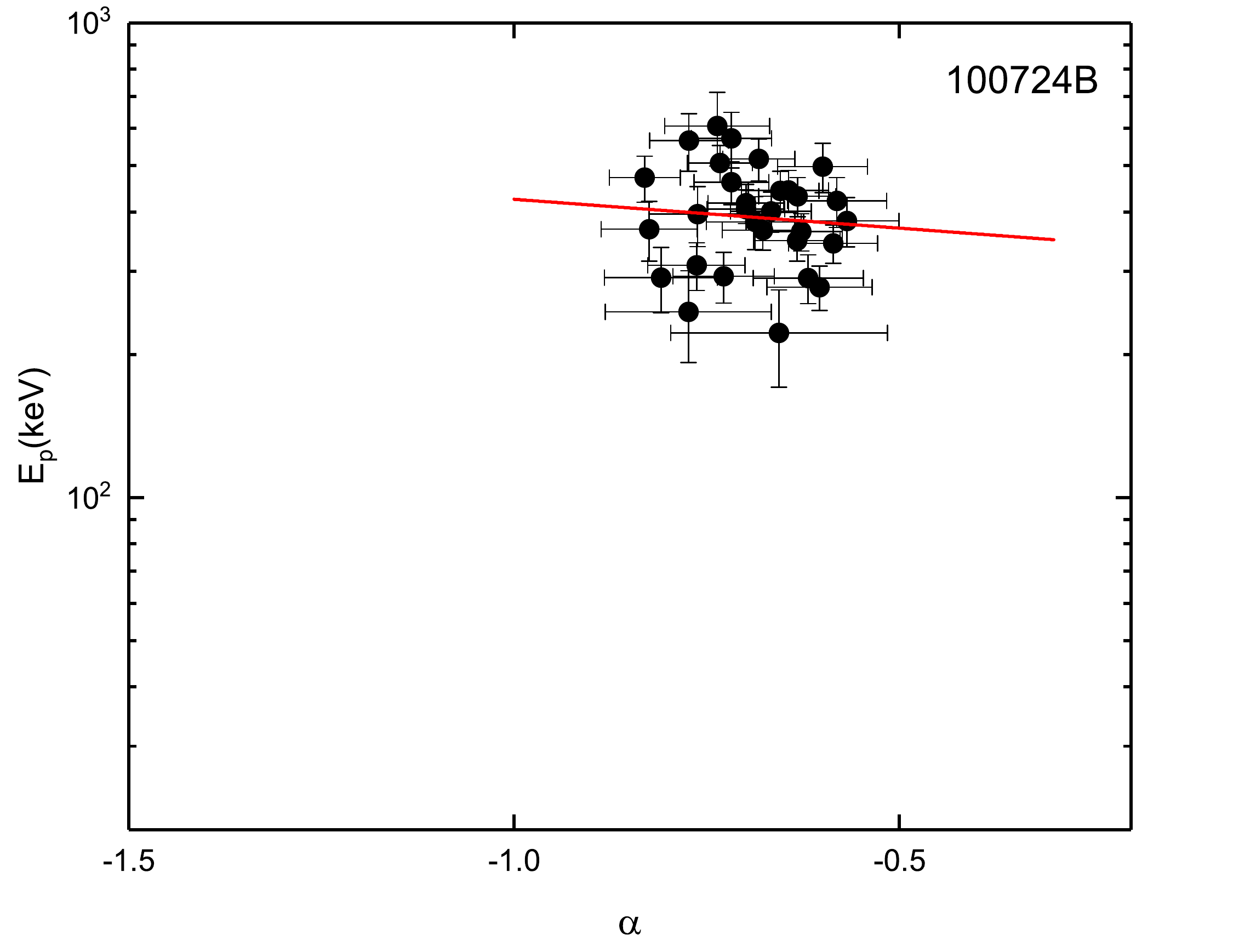}}
\resizebox{4cm}{!}{\includegraphics{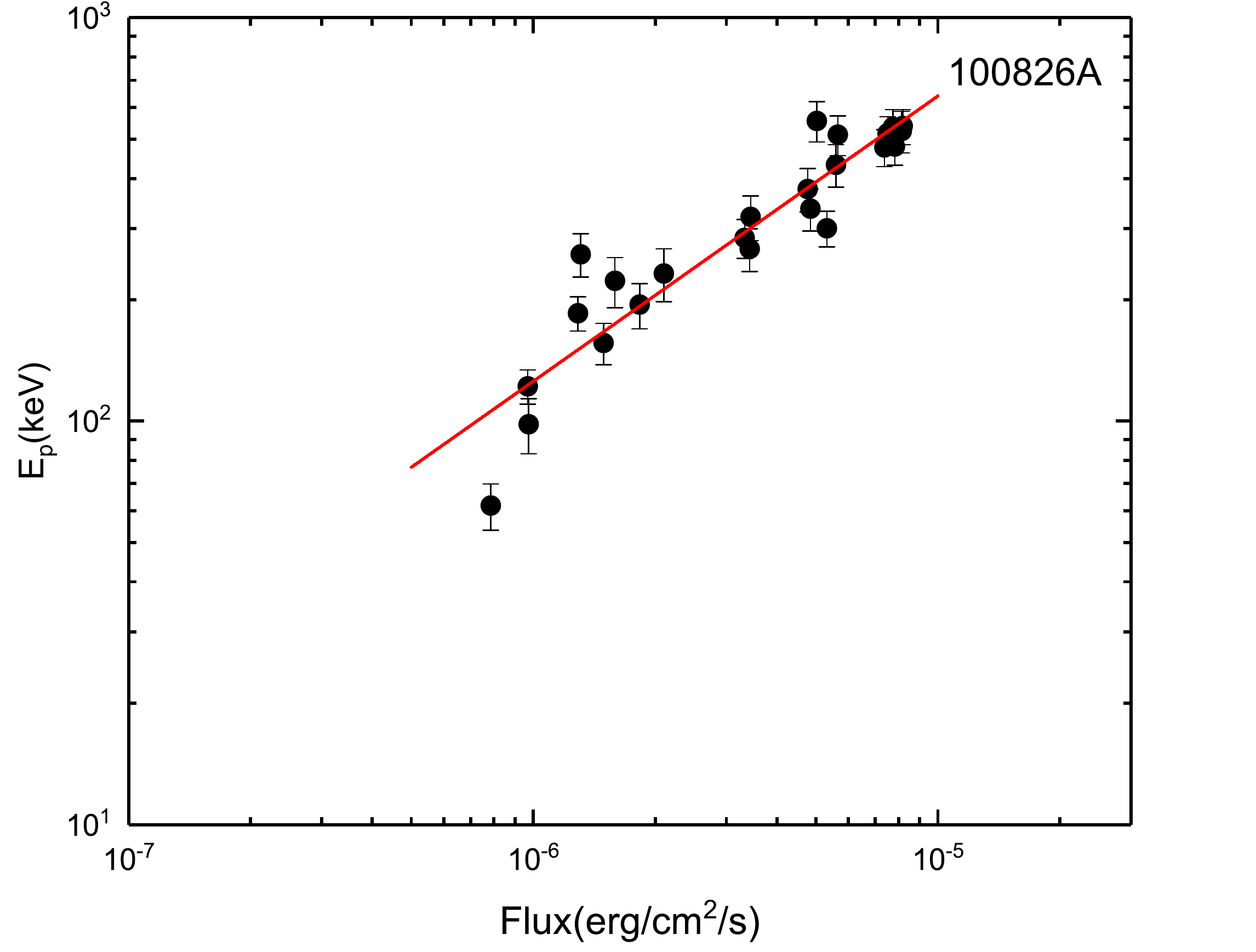}}
\resizebox{4cm}{!}{\includegraphics{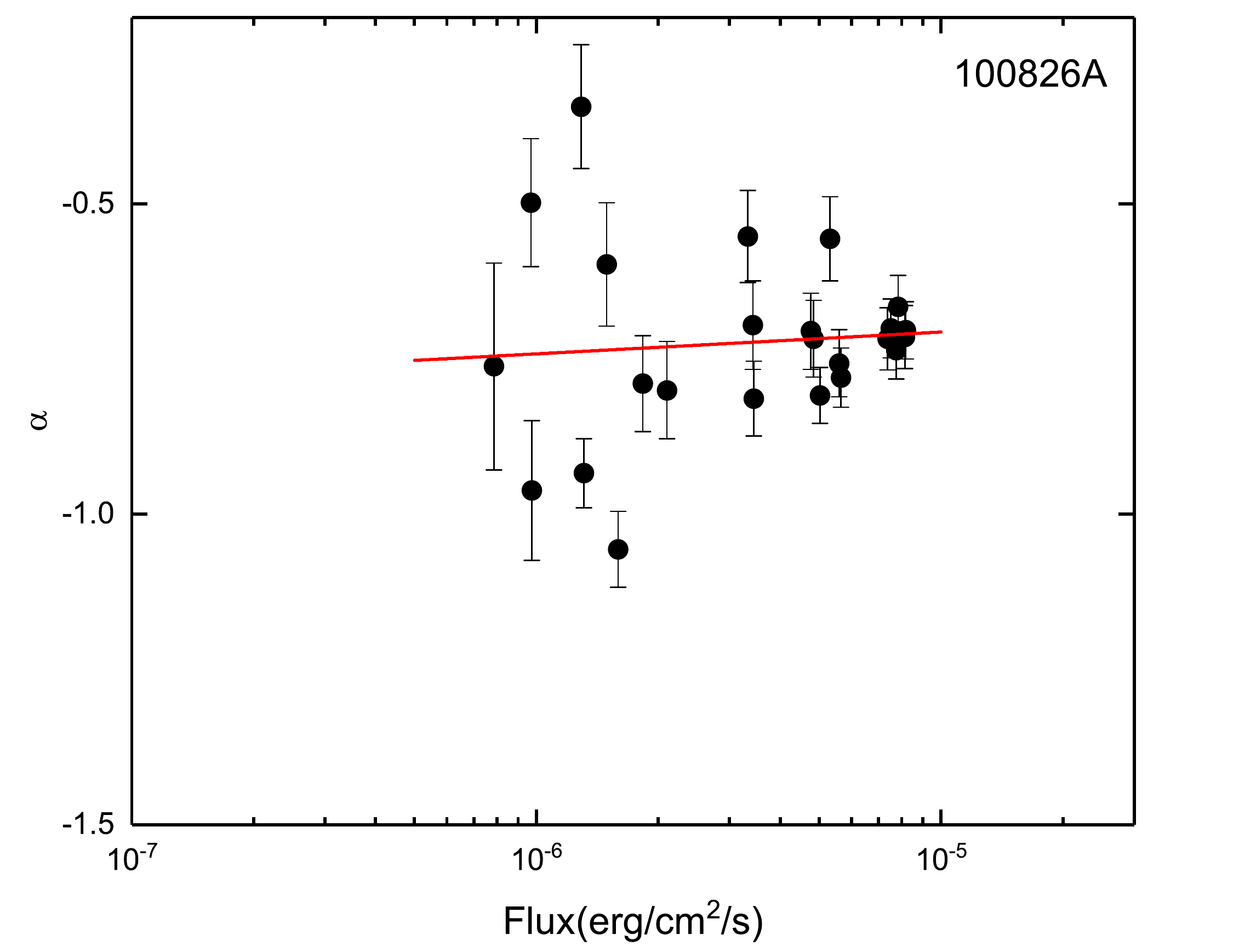}}
\resizebox{4cm}{!}{\includegraphics{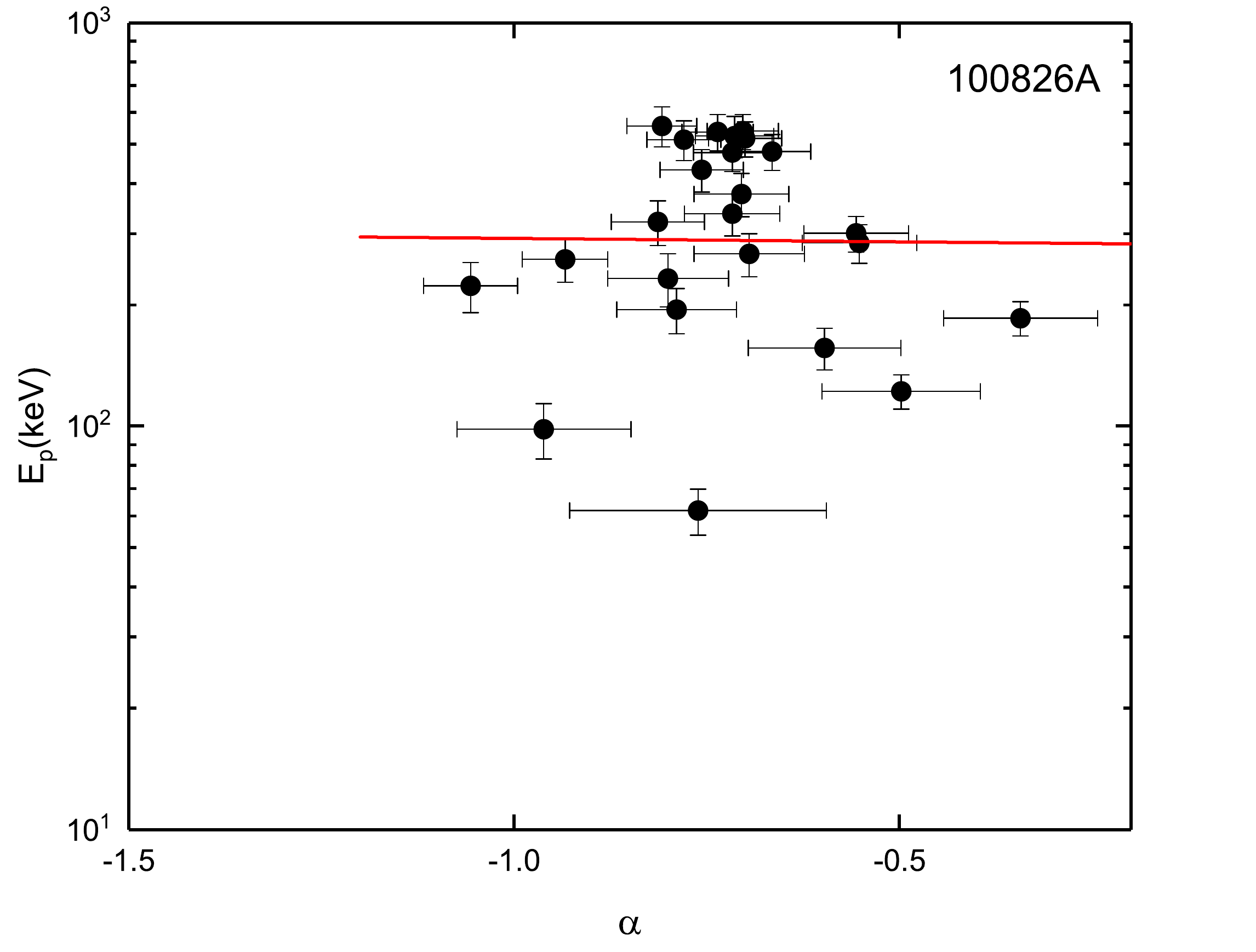}}
\resizebox{4cm}{!}{\includegraphics{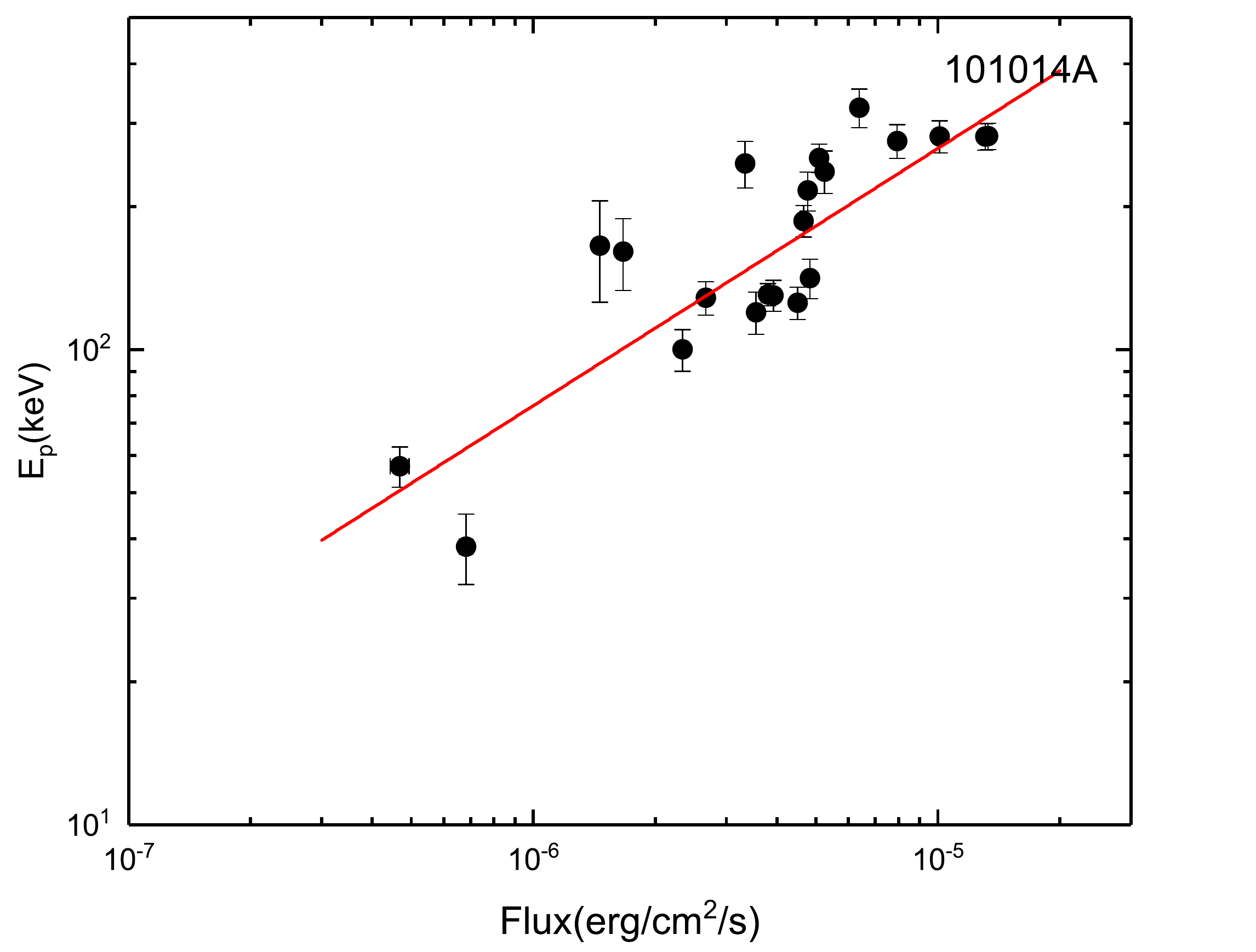}}
\resizebox{4cm}{!}{\includegraphics{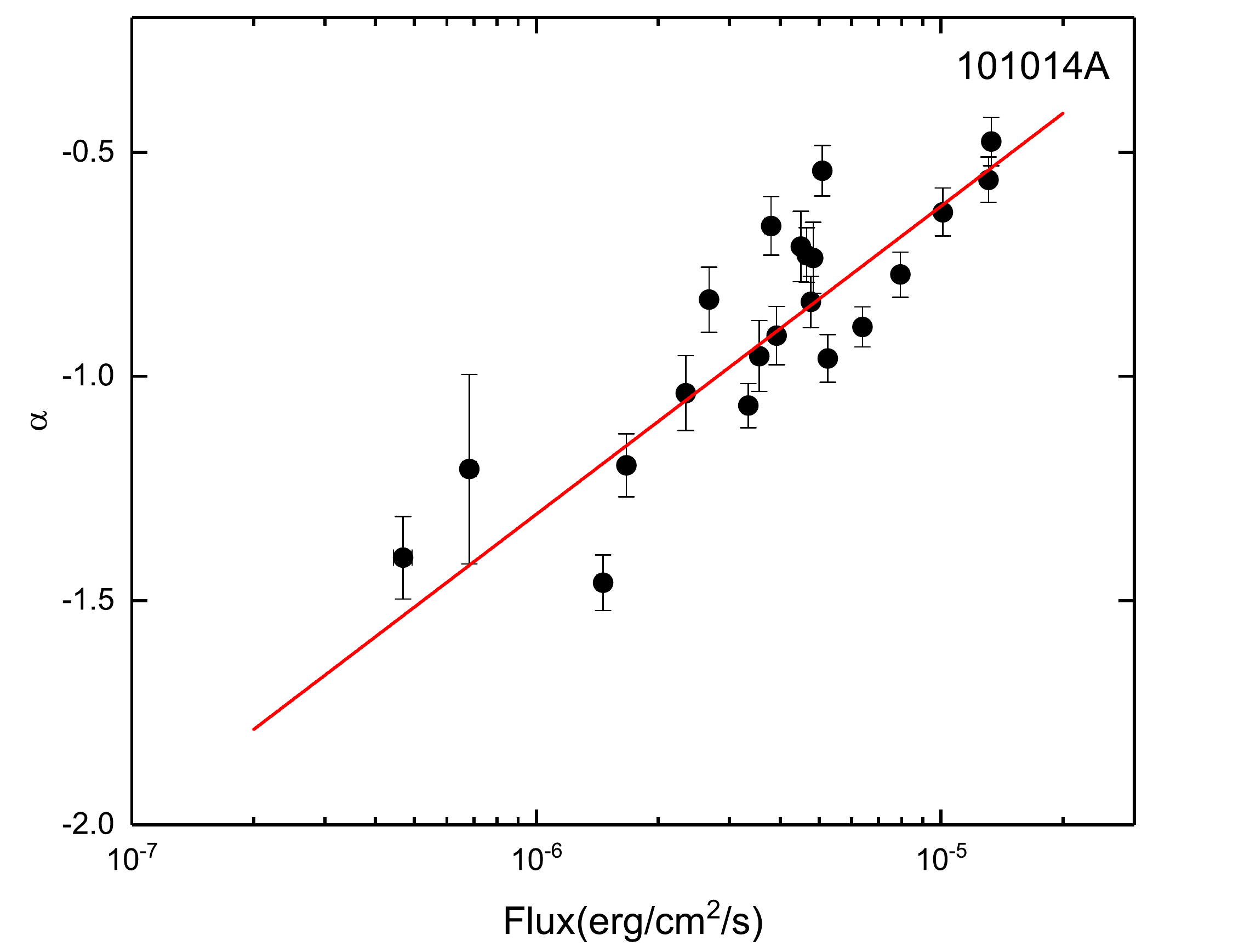}}
\resizebox{4cm}{!}{\includegraphics{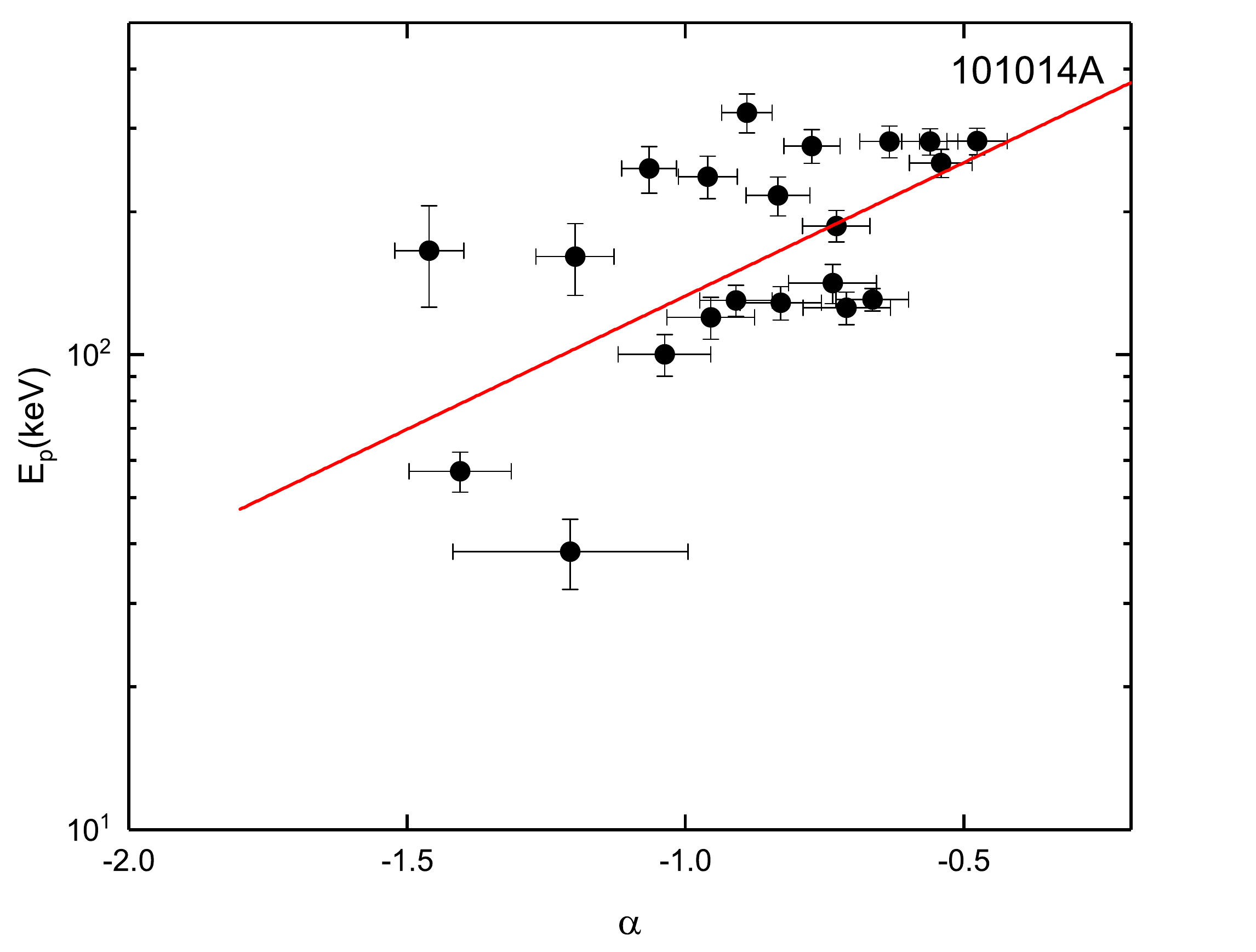}}
\resizebox{4cm}{!}{\includegraphics{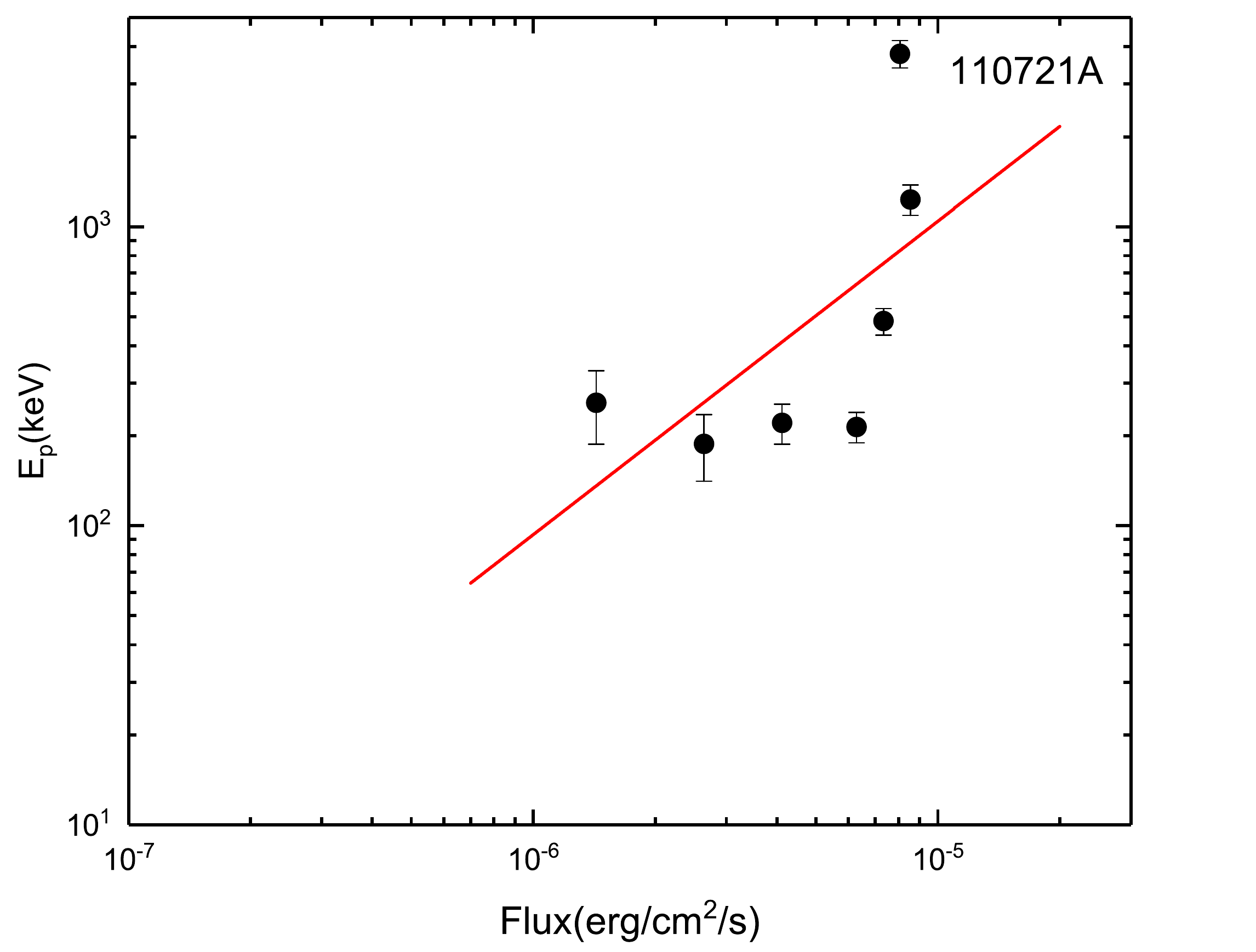}}
\resizebox{4cm}{!}{\includegraphics{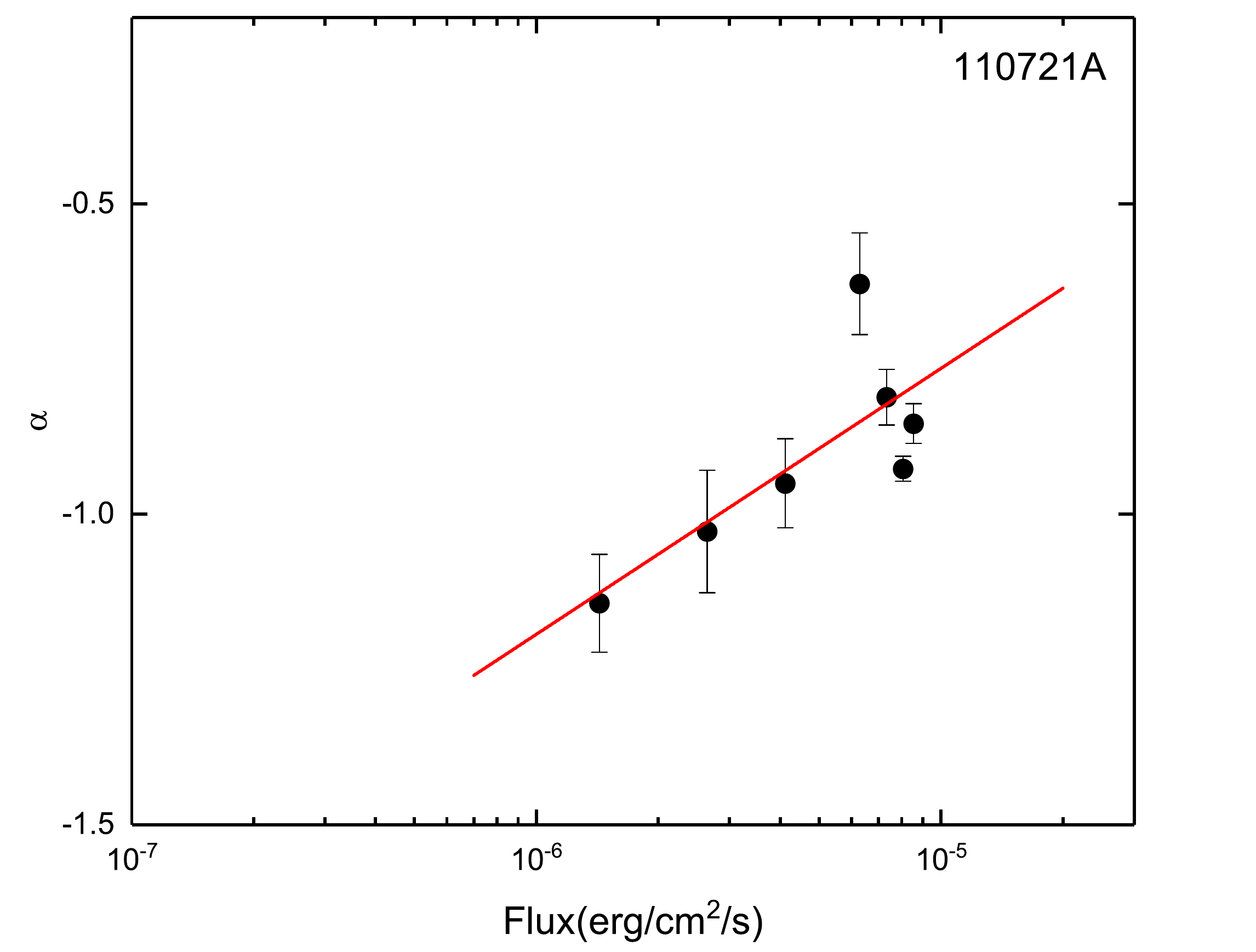}}
\resizebox{4cm}{!}{\includegraphics{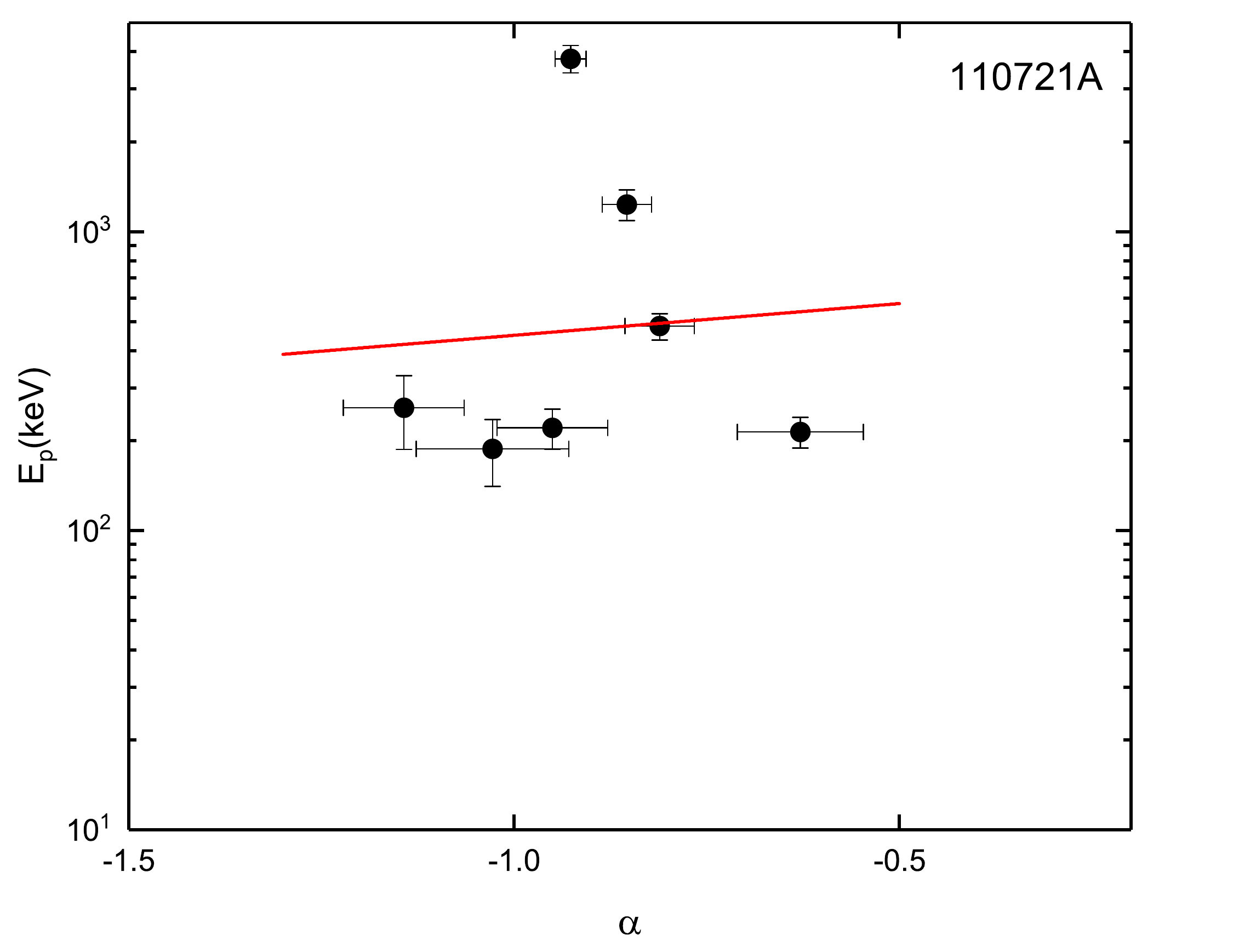}}
\caption{Parameter correlations. The correlations such as $E_{p}-F$, $\alpha-F$\edit1{\added{,}} and $E_{p}- \alpha$ obtained from the time-resolved spectra are shown for all of the bursts in our sample. \edit1{\replaced{The red solid line represent the best fit for them.}{The red solid line represents the best-linear-fitting result for each burst.}}\label{fig:parameter correlations}}
\end{figure}

\addtocounter{figure}{-1}
\begin{figure}
\centering
\resizebox{4cm}{!}{\includegraphics{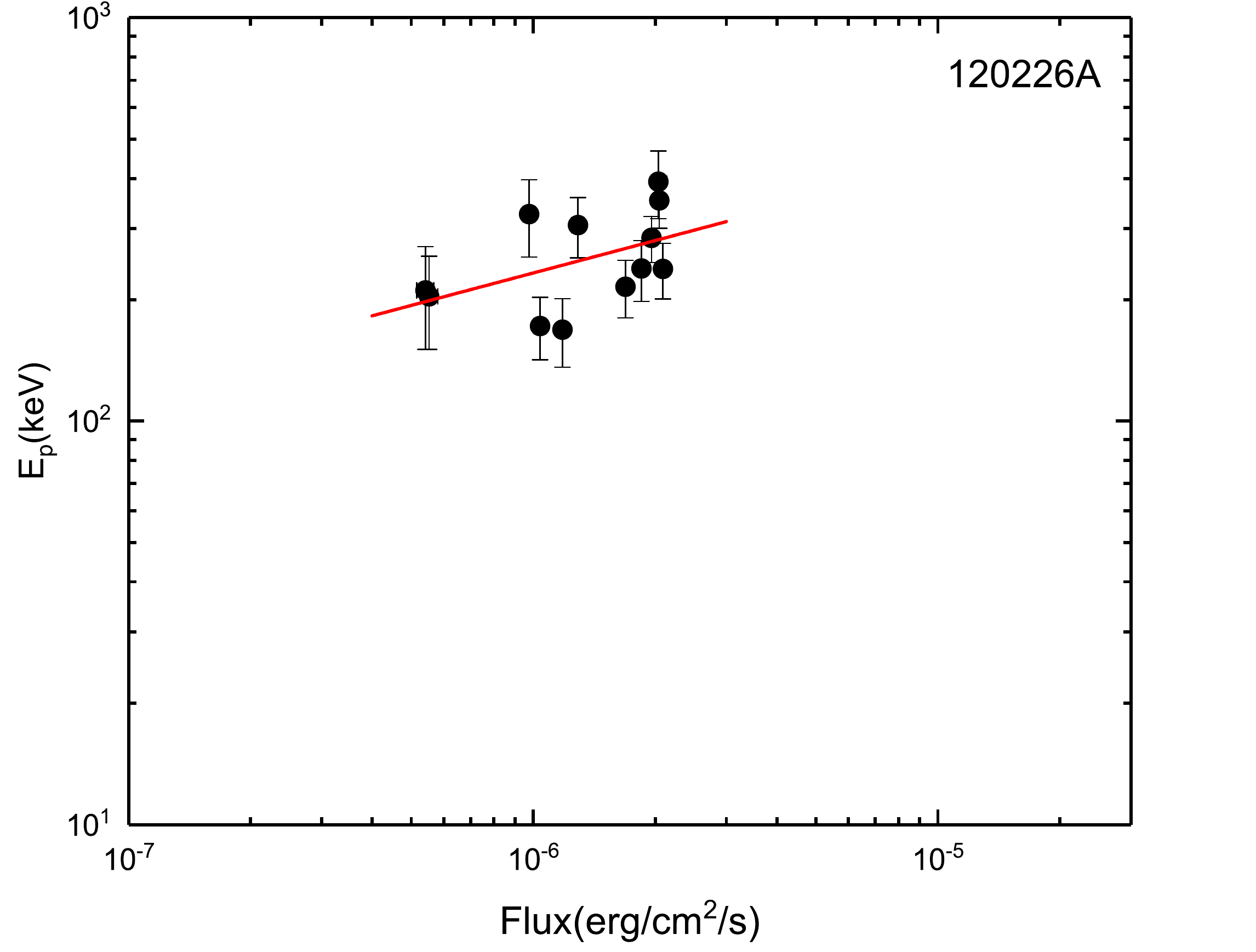}}
\resizebox{4cm}{!}{\includegraphics{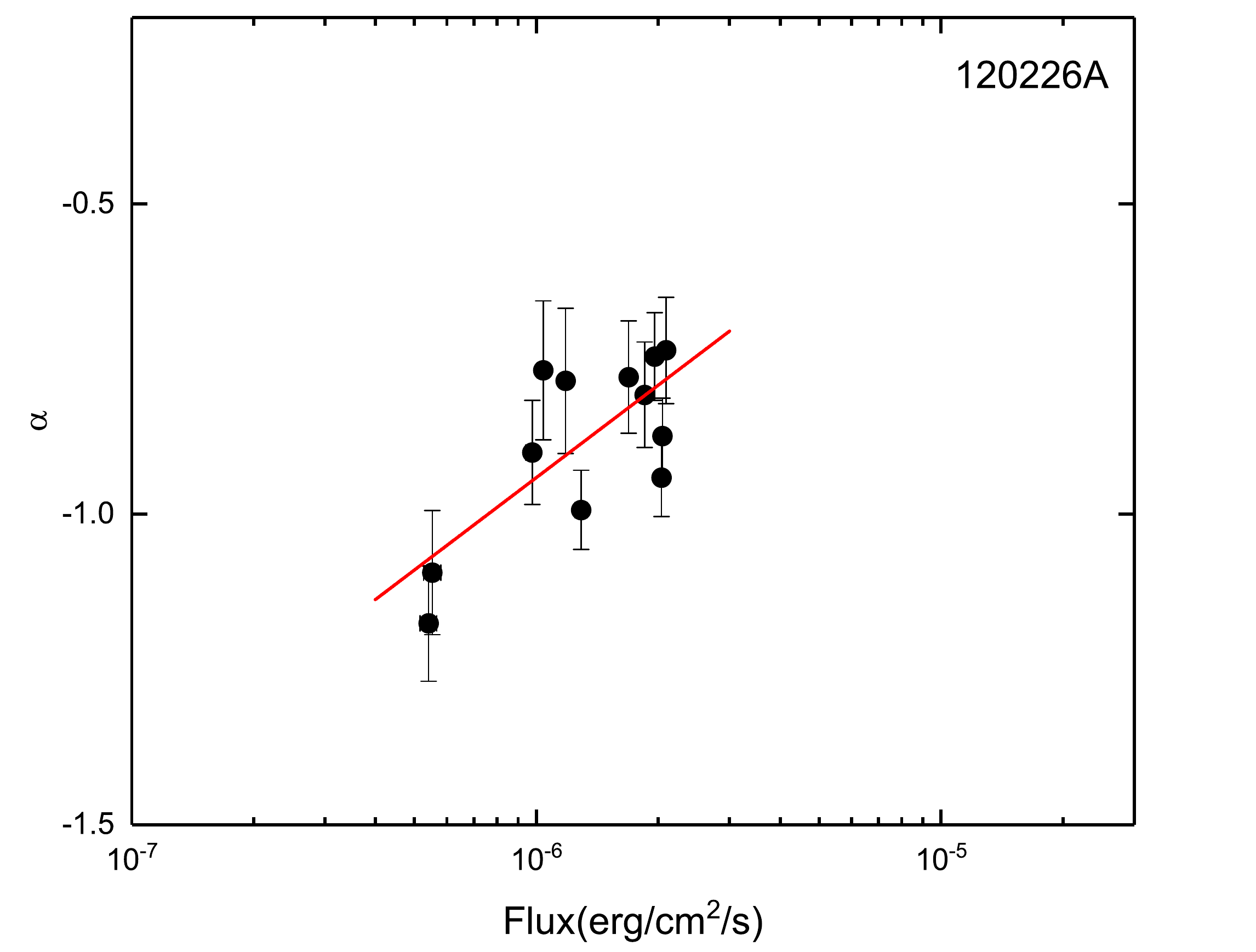}}
\resizebox{4cm}{!}{\includegraphics{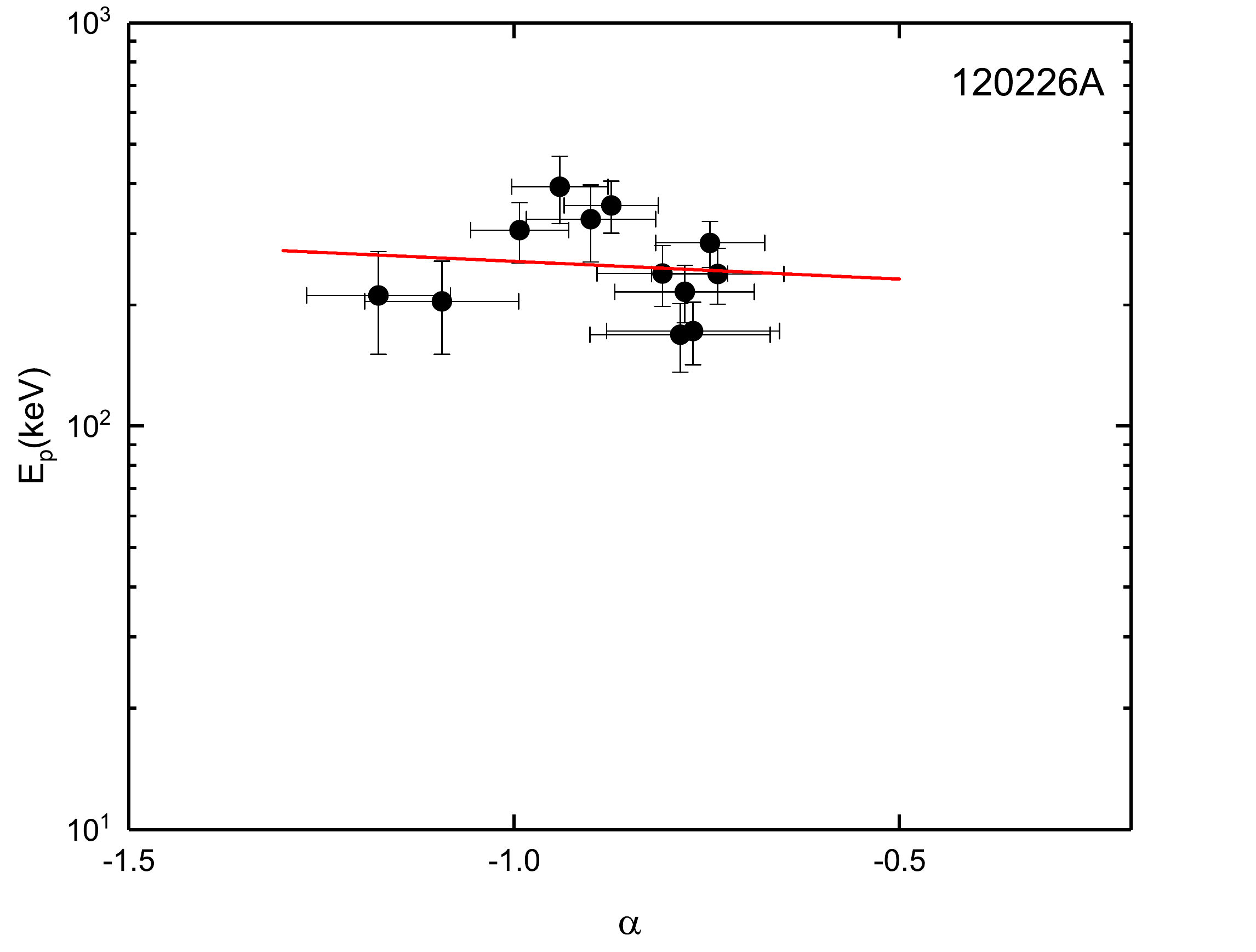}}
\resizebox{4cm}{!}{\includegraphics{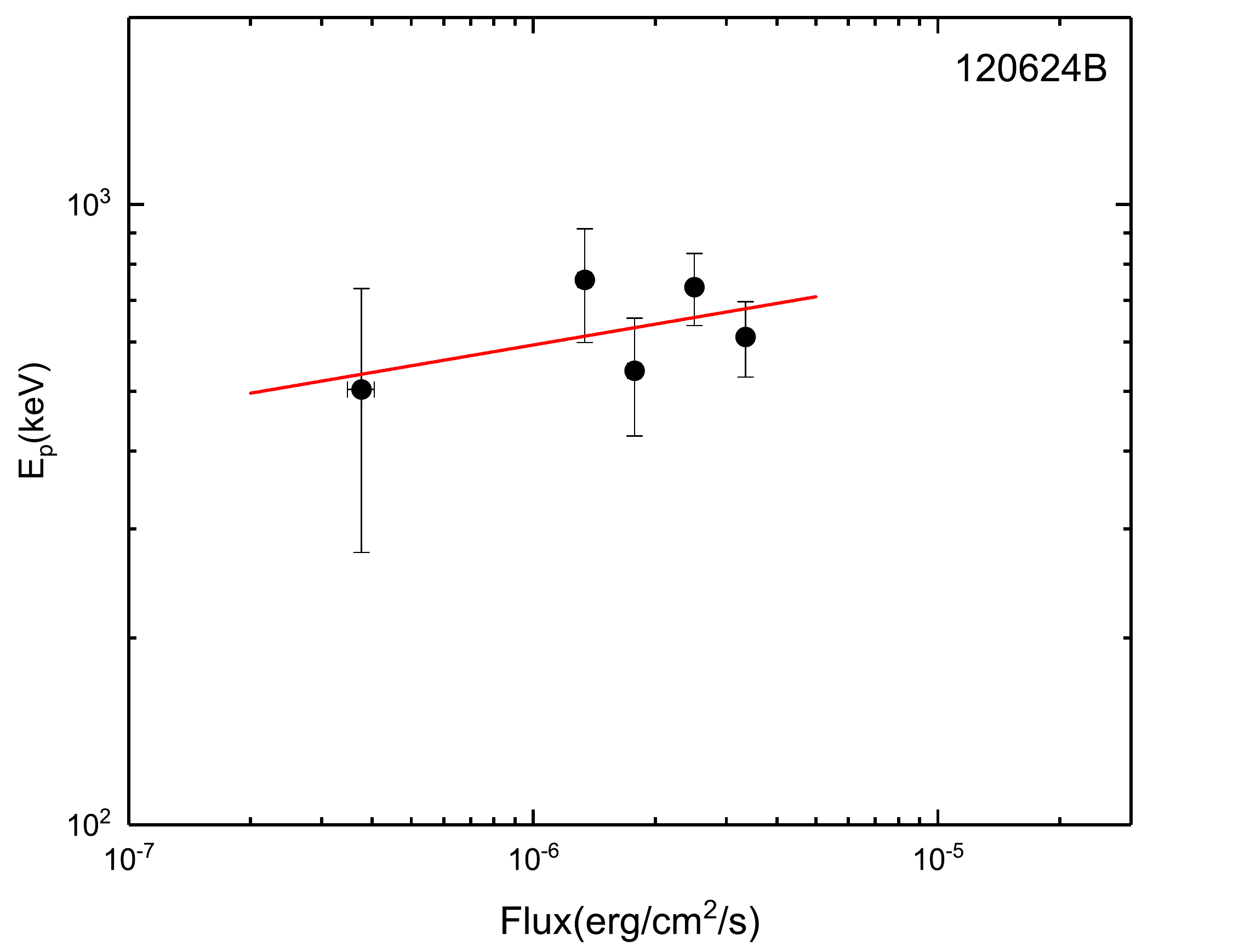}}
\resizebox{4cm}{!}{\includegraphics{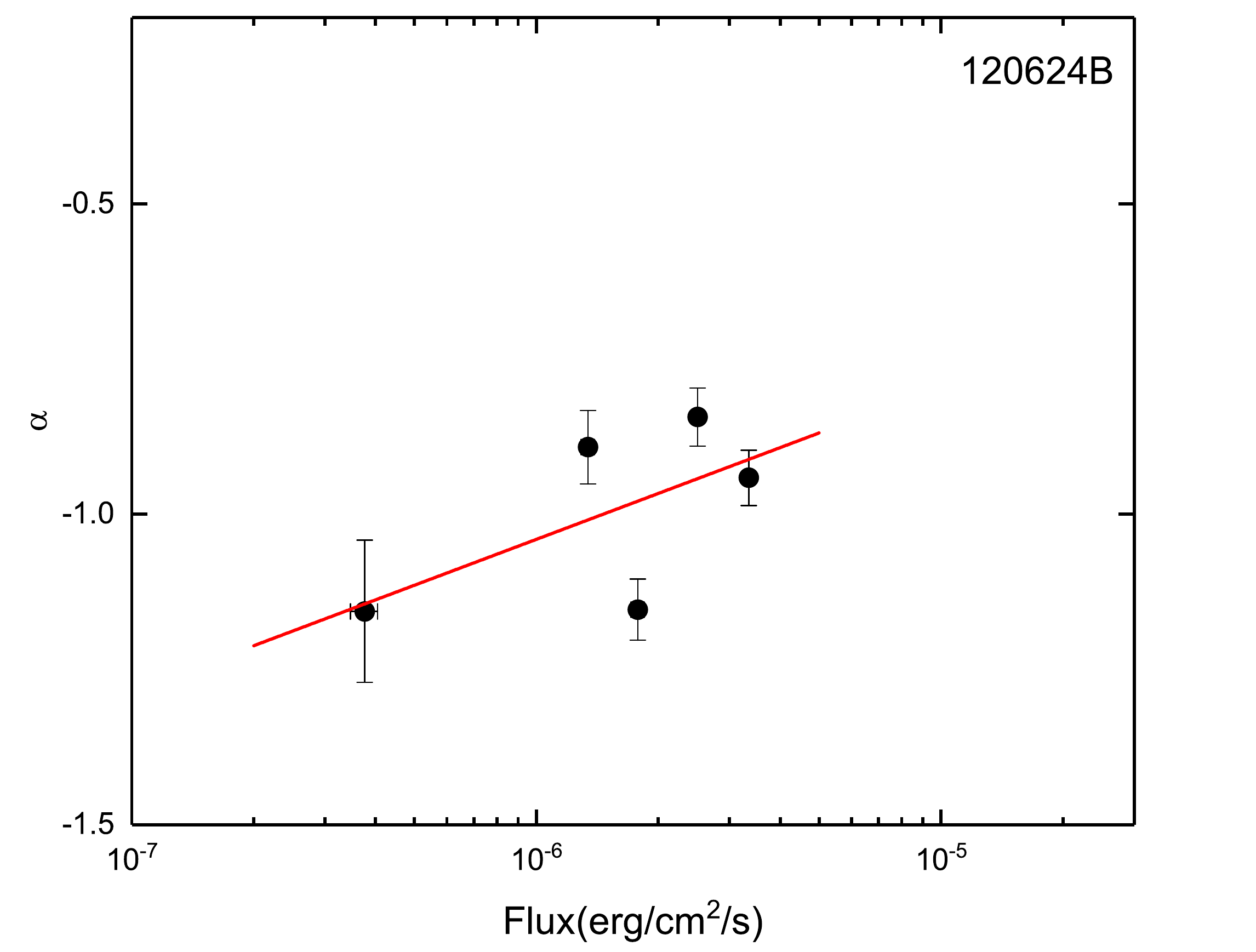}}
\resizebox{4cm}{!}{\includegraphics{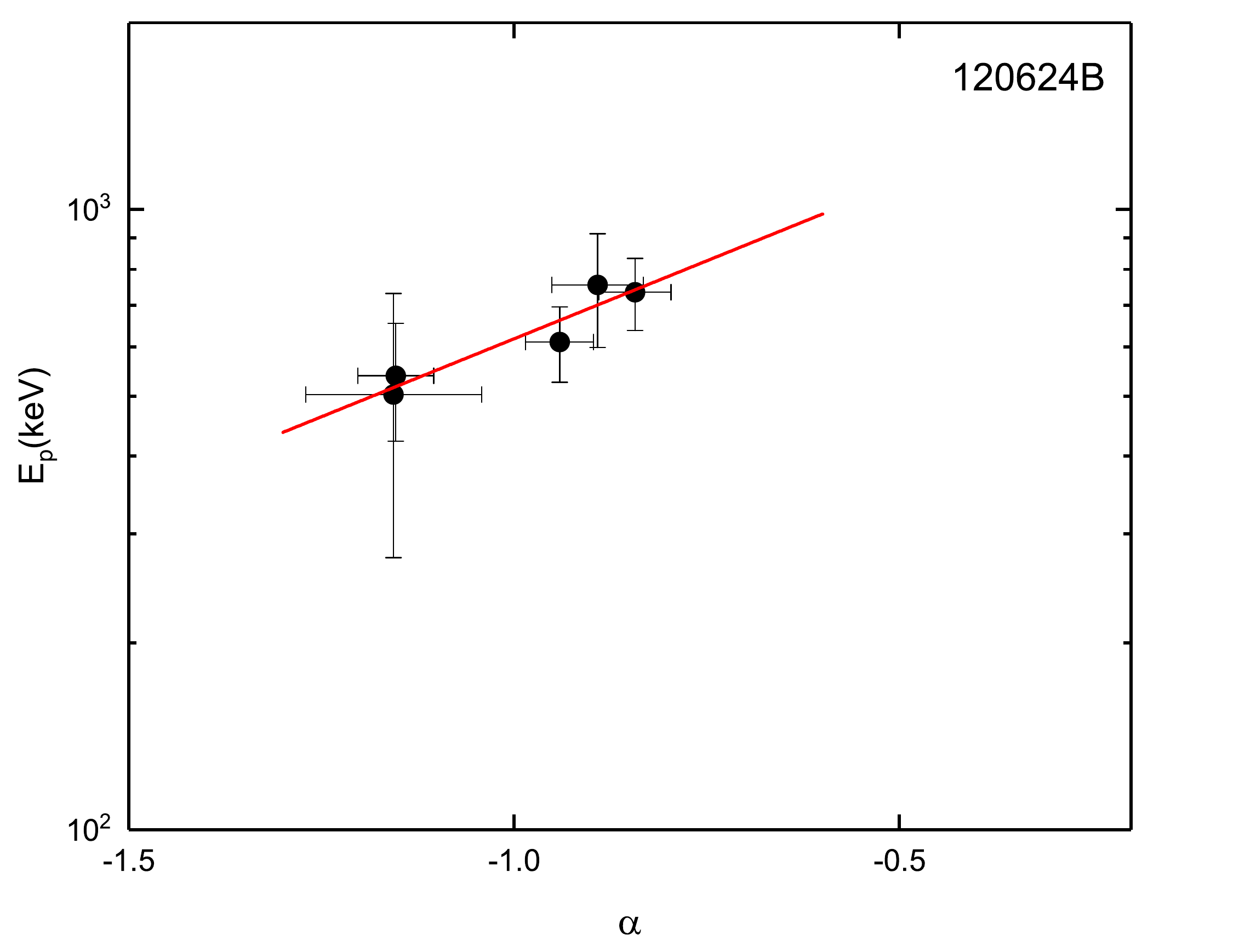}}
\resizebox{4cm}{!}{\includegraphics{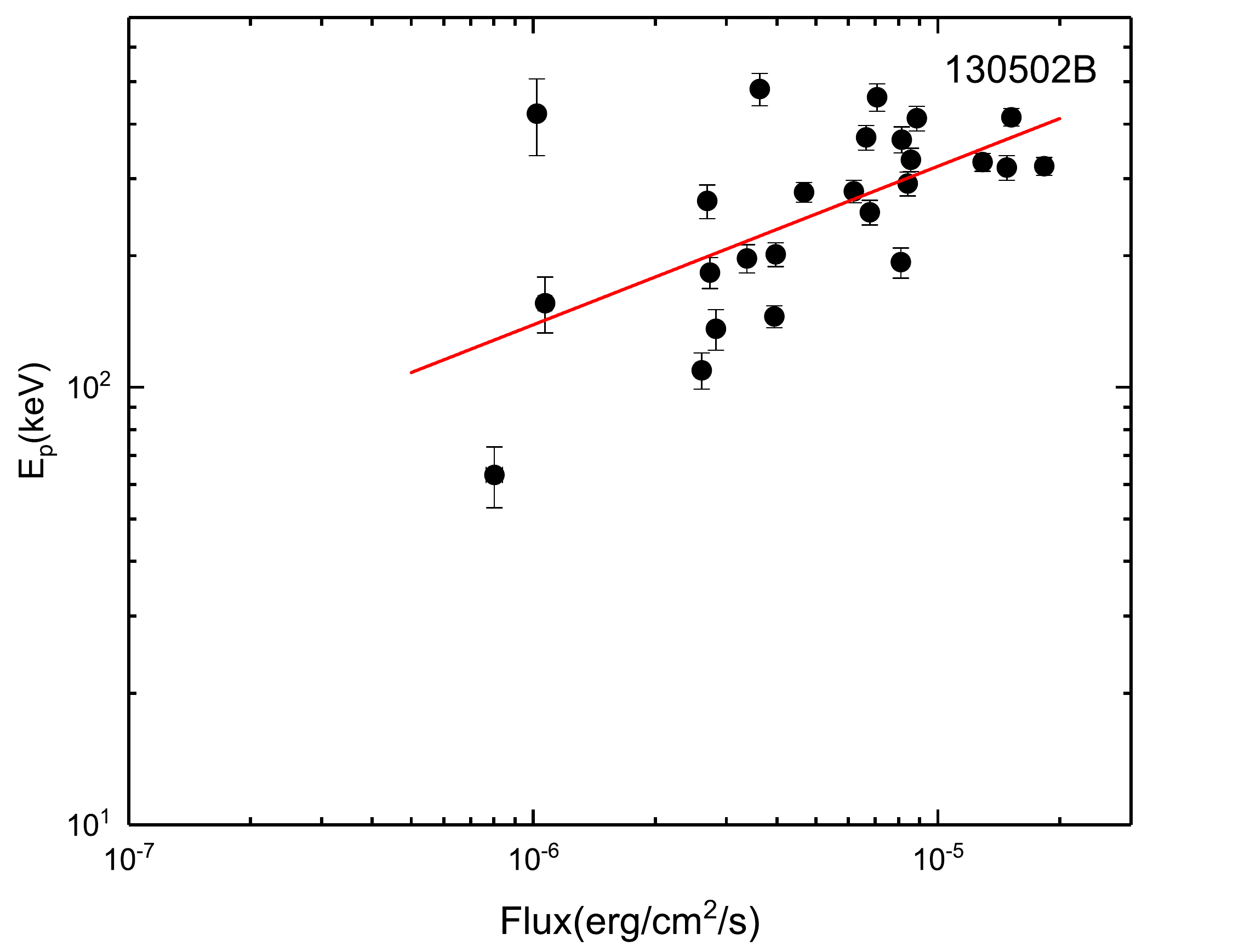}}
\resizebox{4cm}{!}{\includegraphics{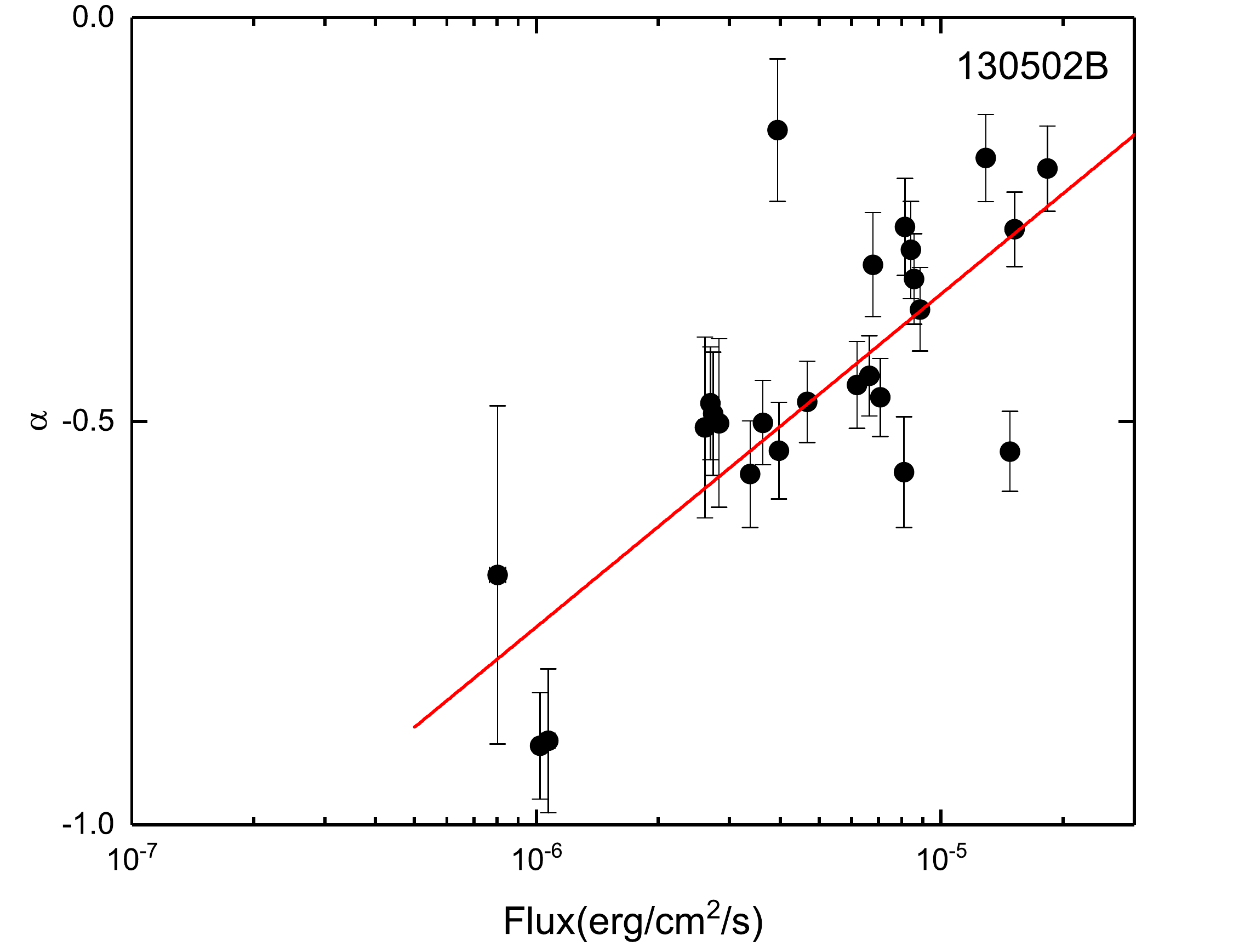}}
\resizebox{4cm}{!}{\includegraphics{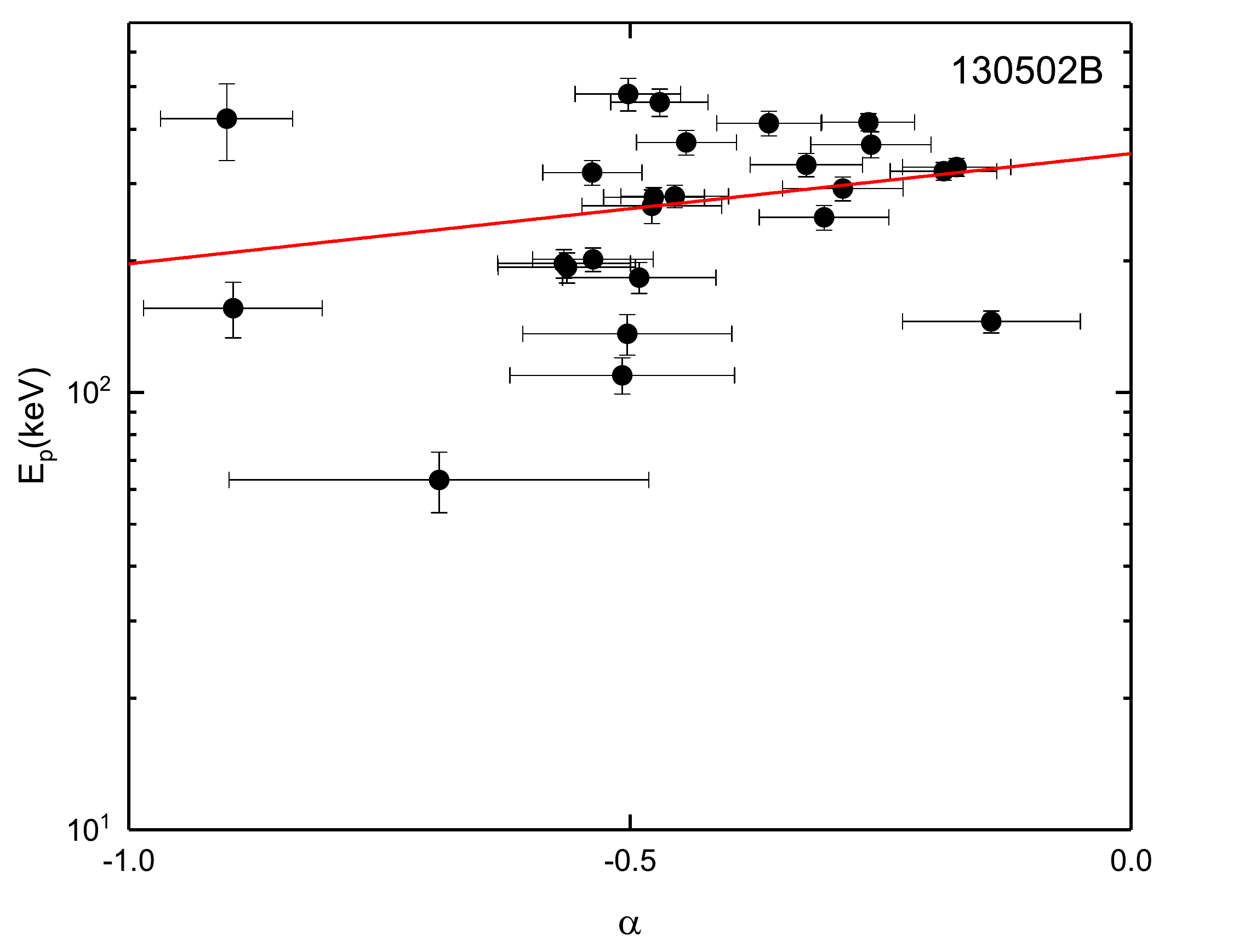}}
\resizebox{4cm}{!}{\includegraphics{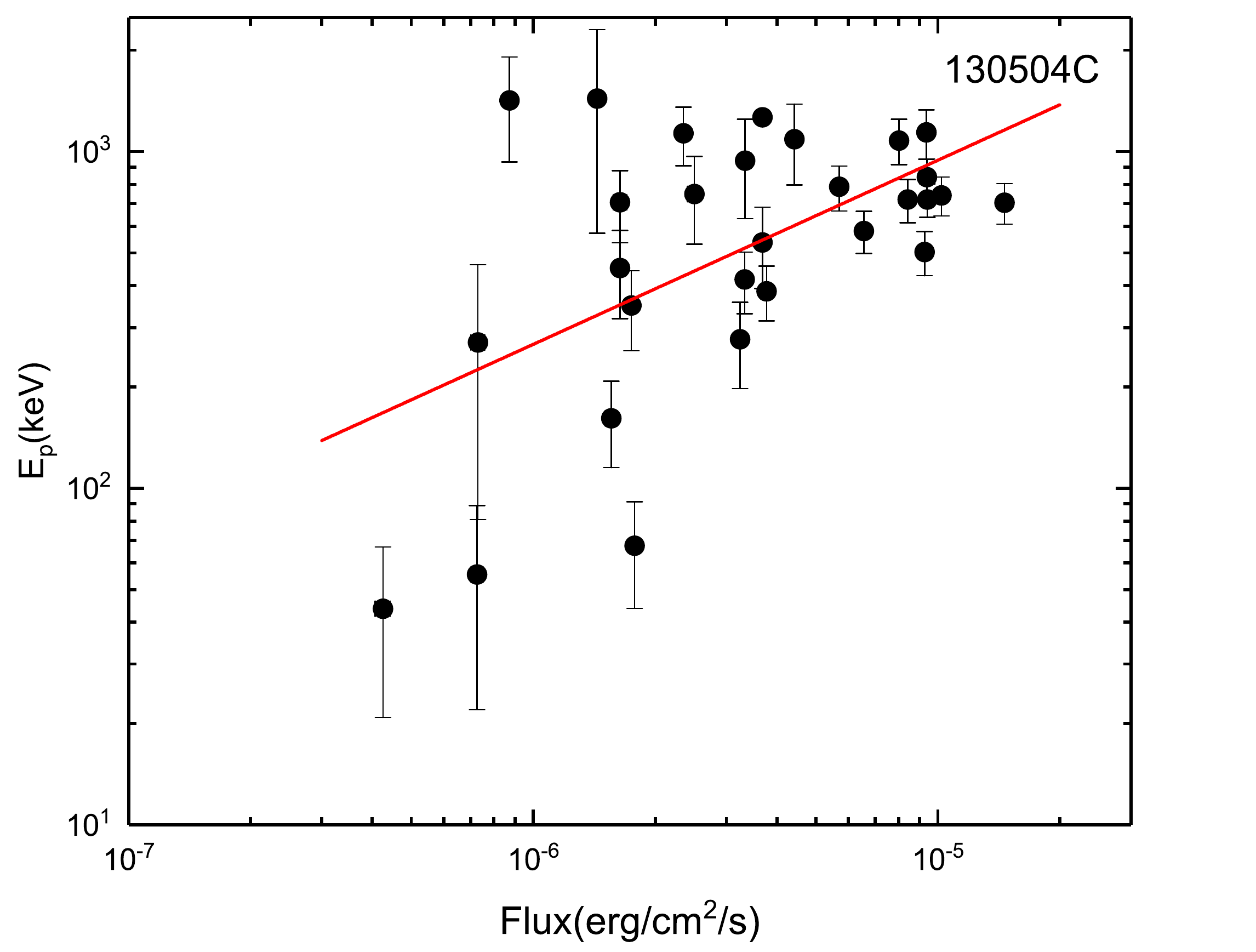}}
\resizebox{4cm}{!}{\includegraphics{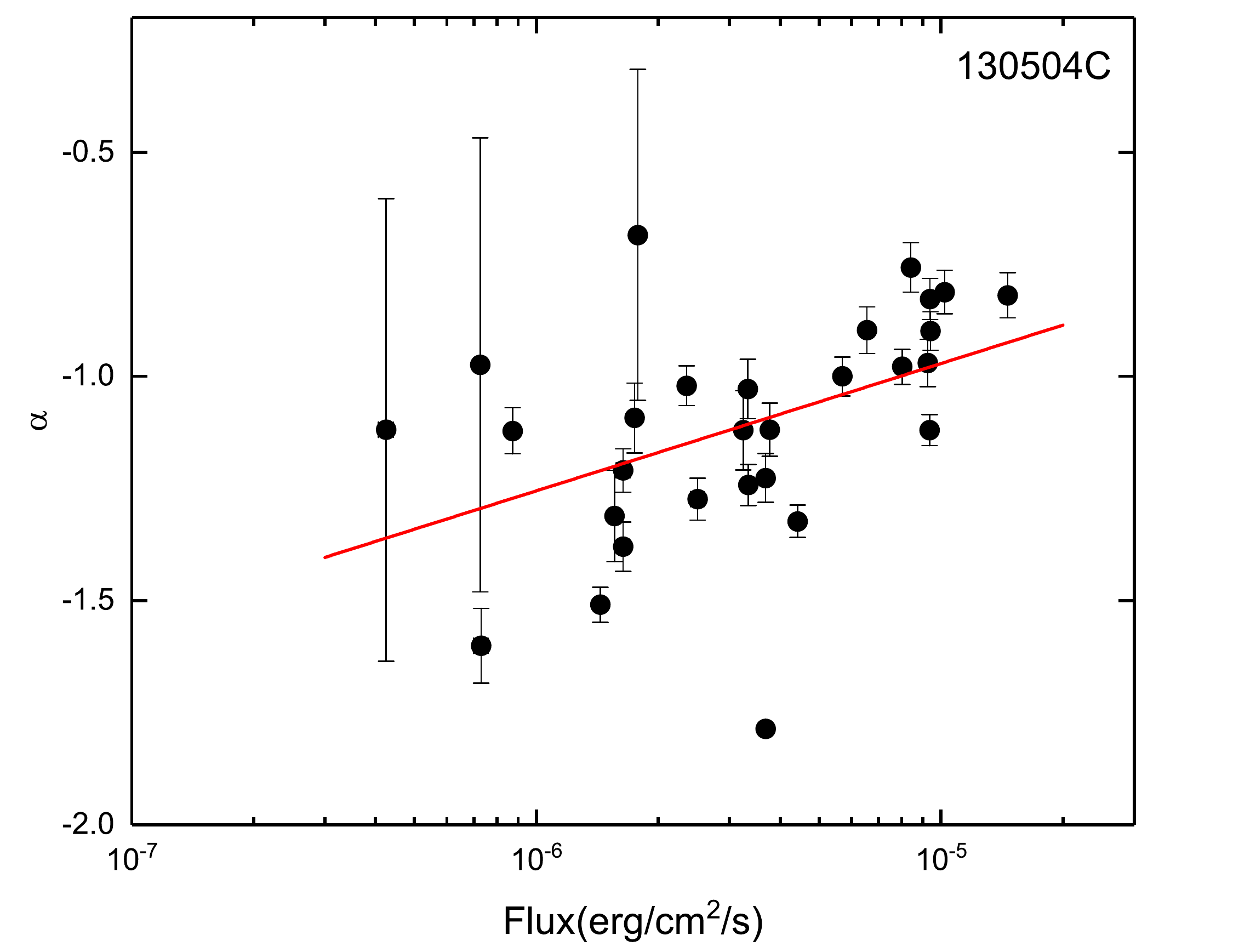}}
\resizebox{4cm}{!}{\includegraphics{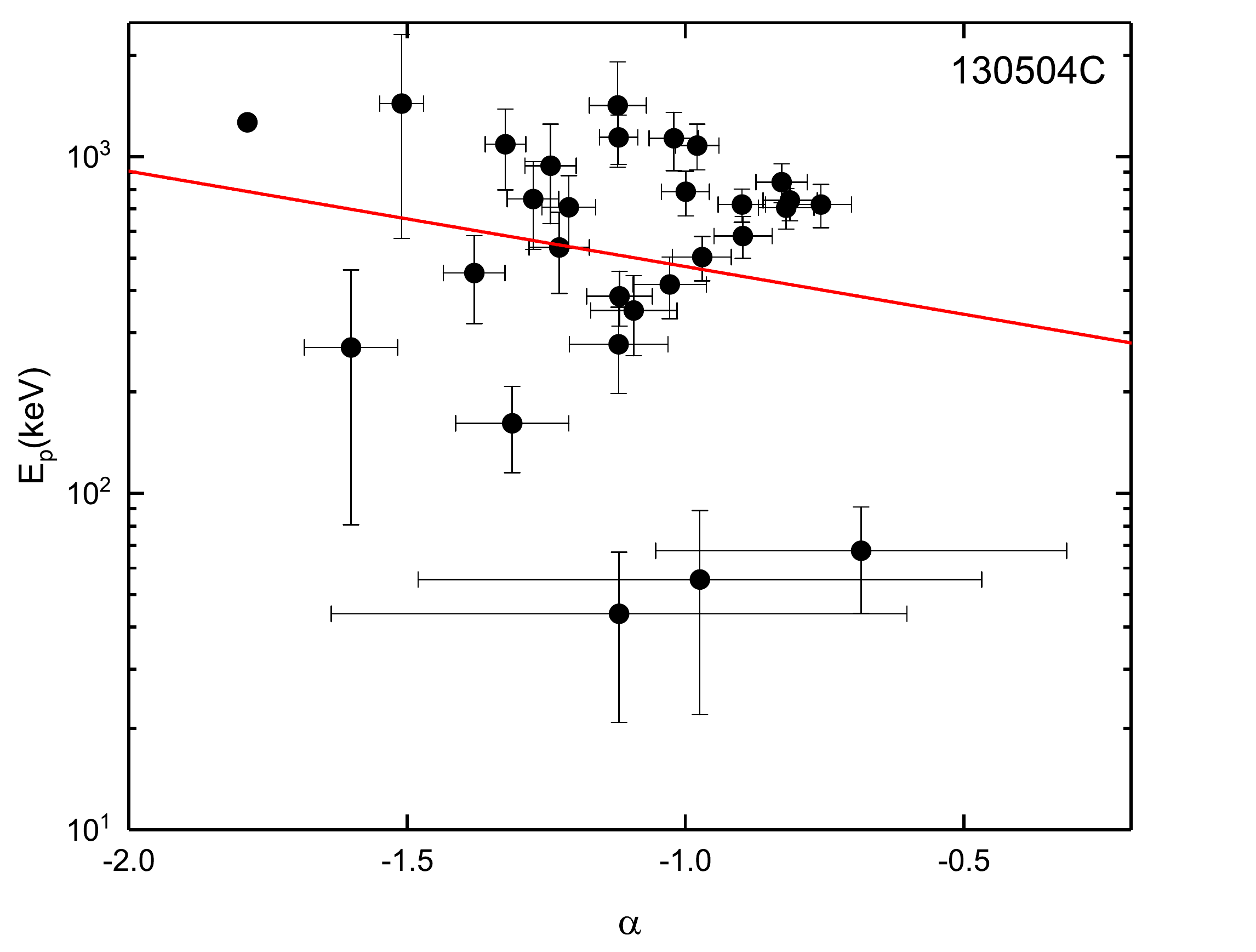}}
\resizebox{4cm}{!}{\includegraphics{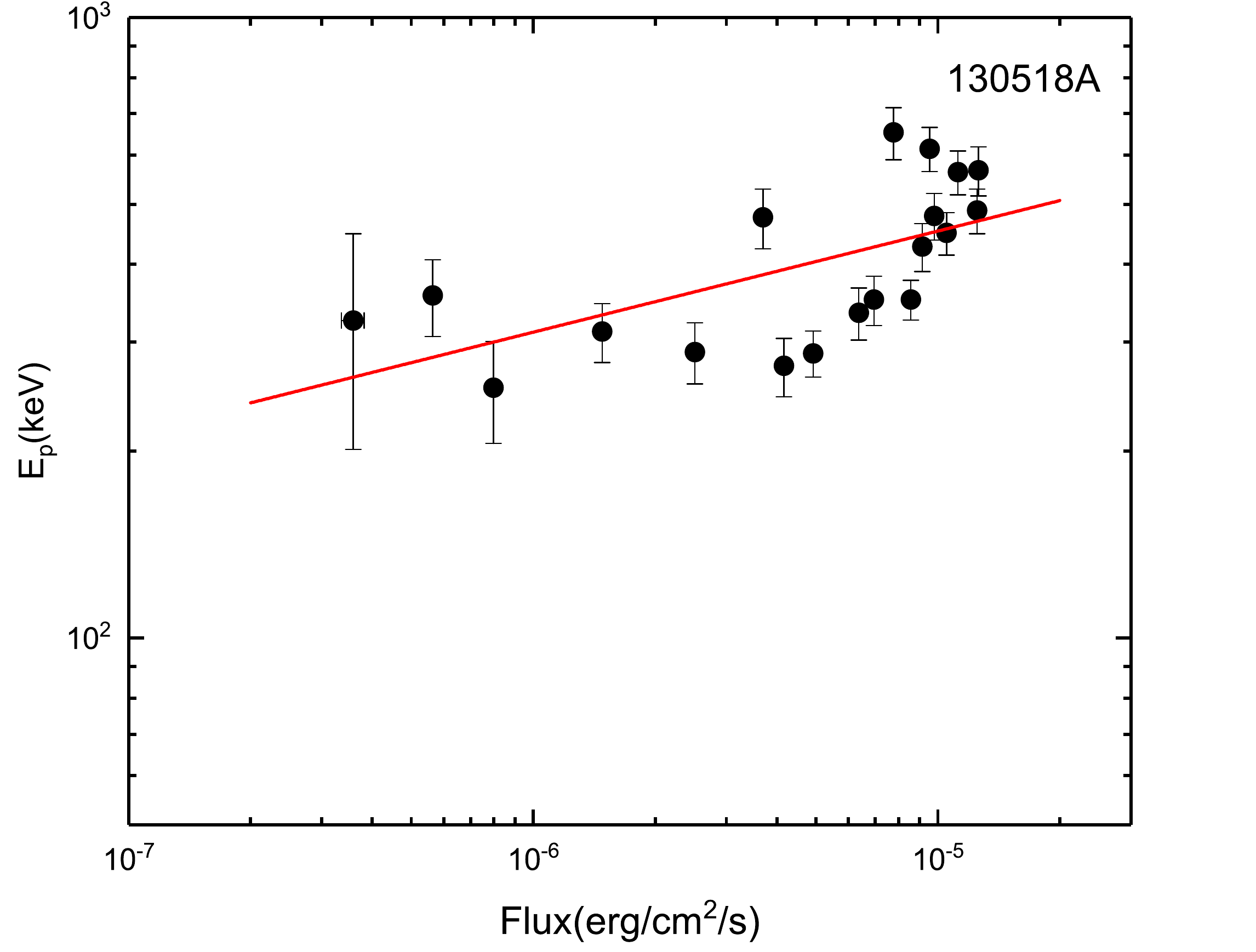}}
\resizebox{4cm}{!}{\includegraphics{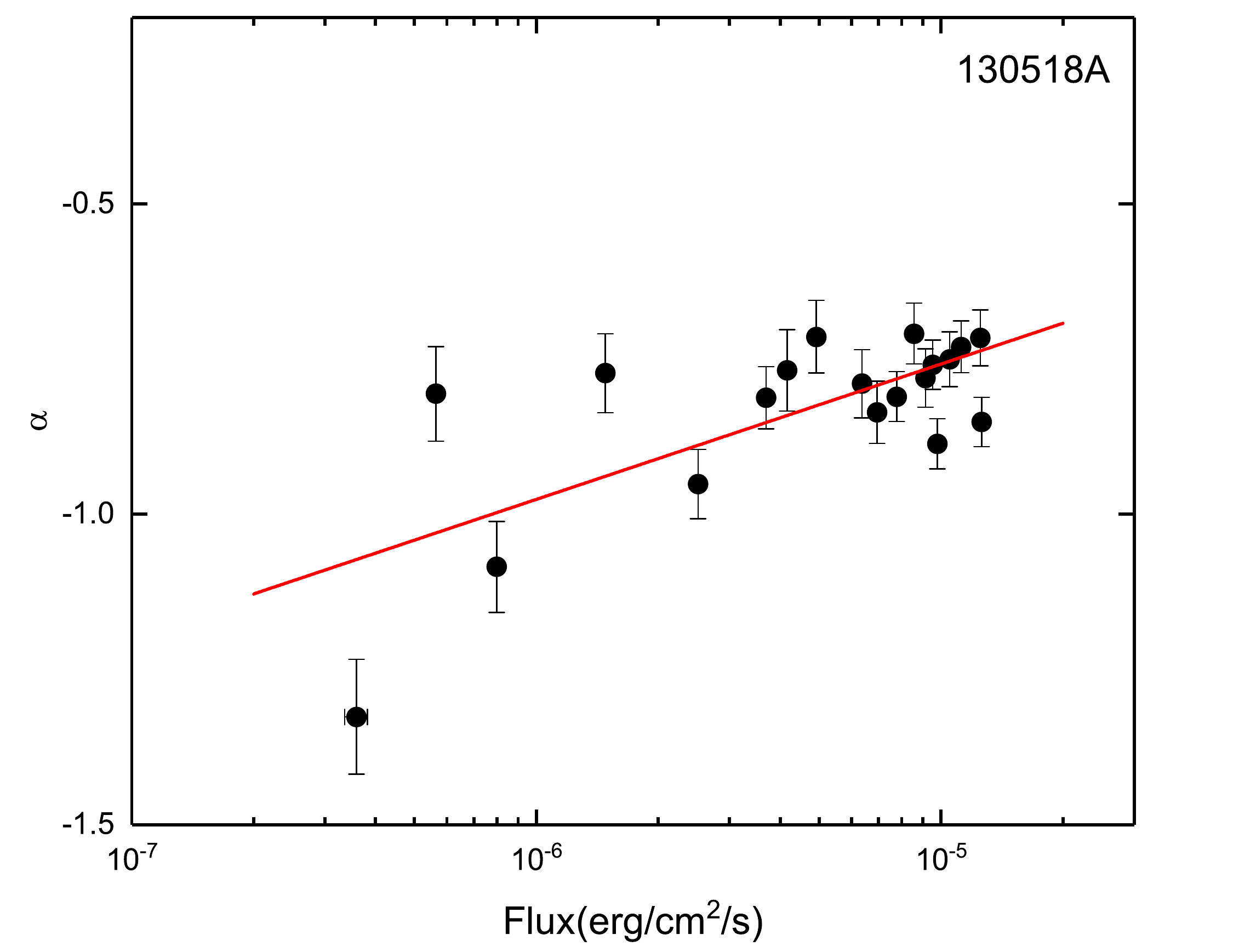}}
\resizebox{4cm}{!}{\includegraphics{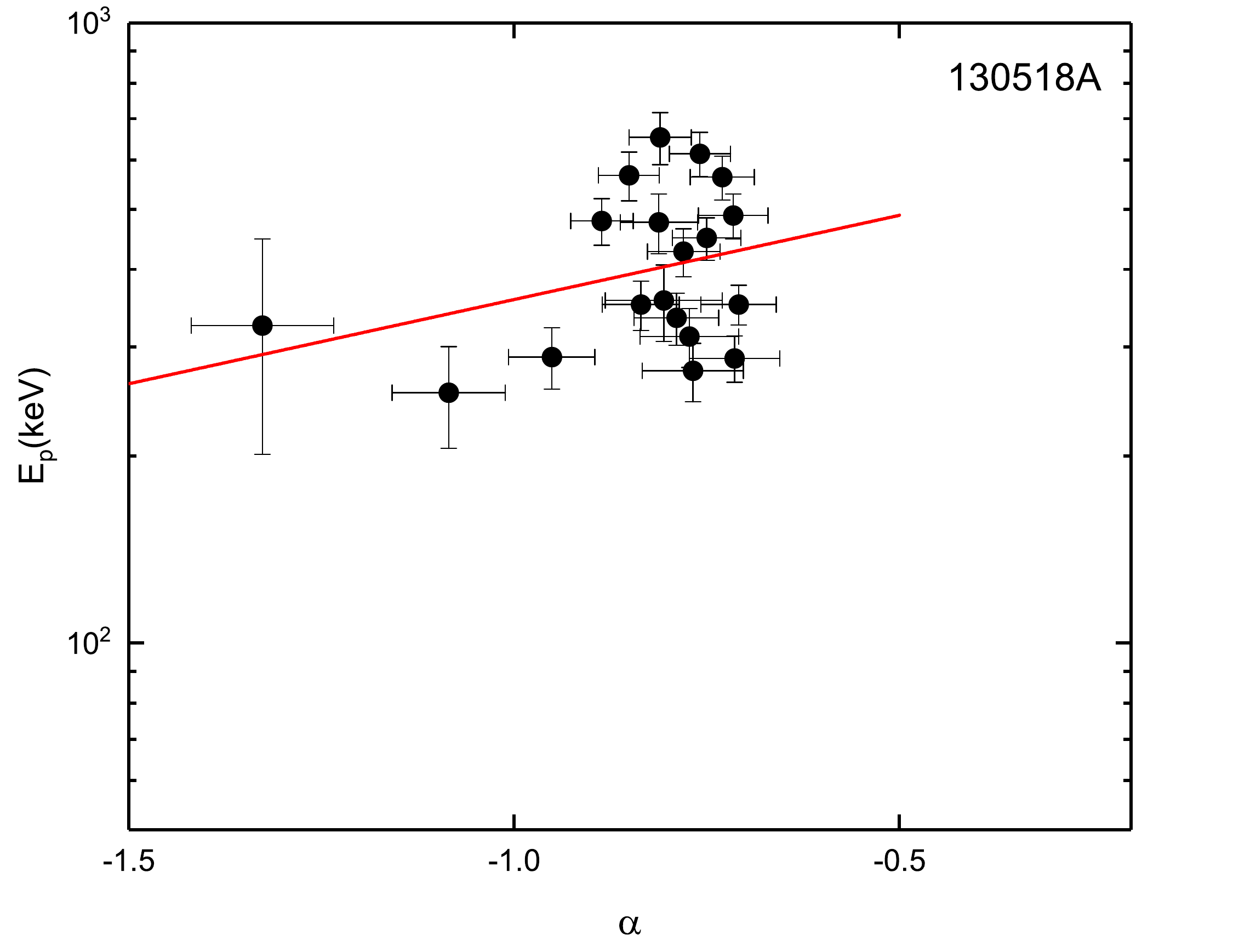}}
\resizebox{4cm}{!}{\includegraphics{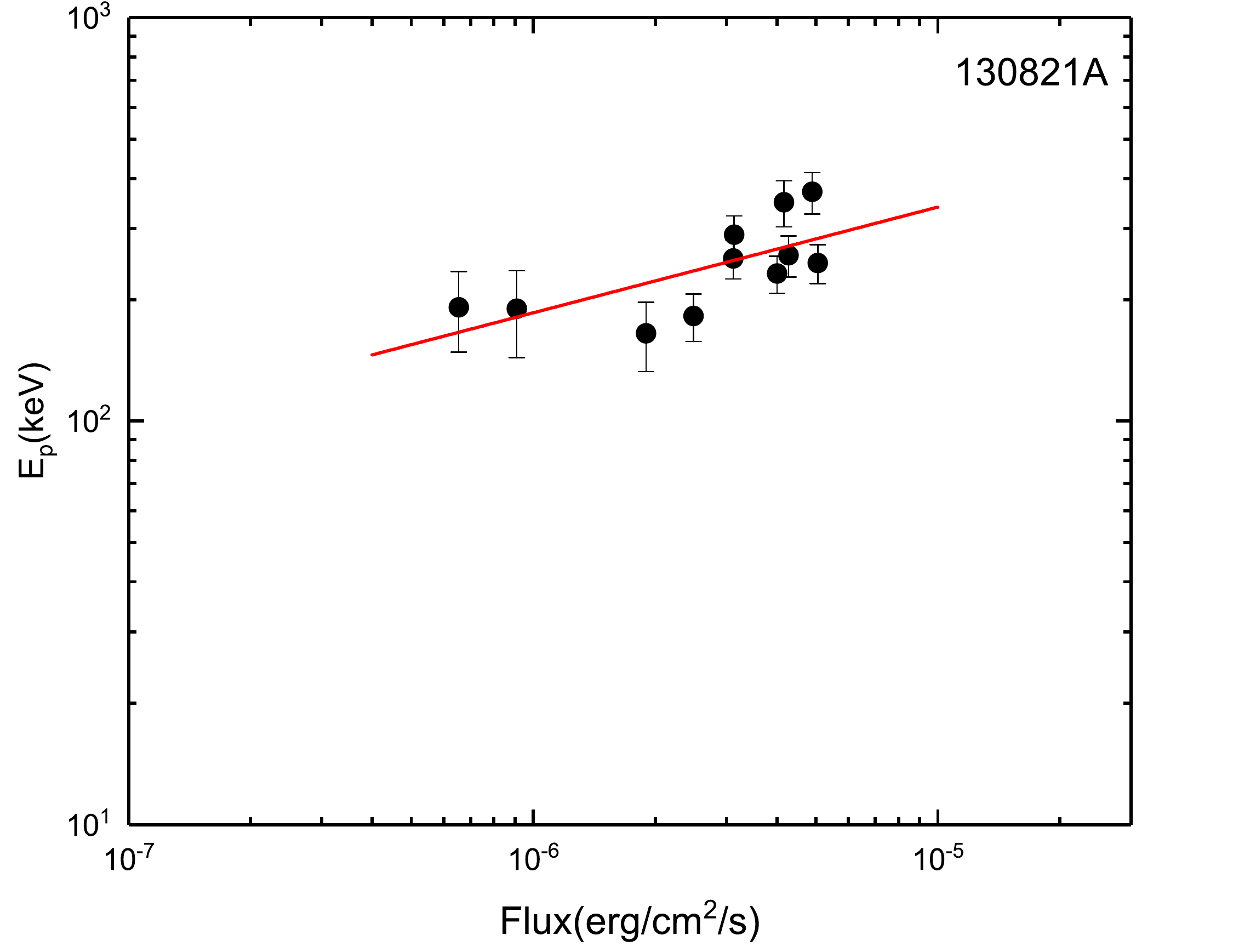}}
\resizebox{4cm}{!}{\includegraphics{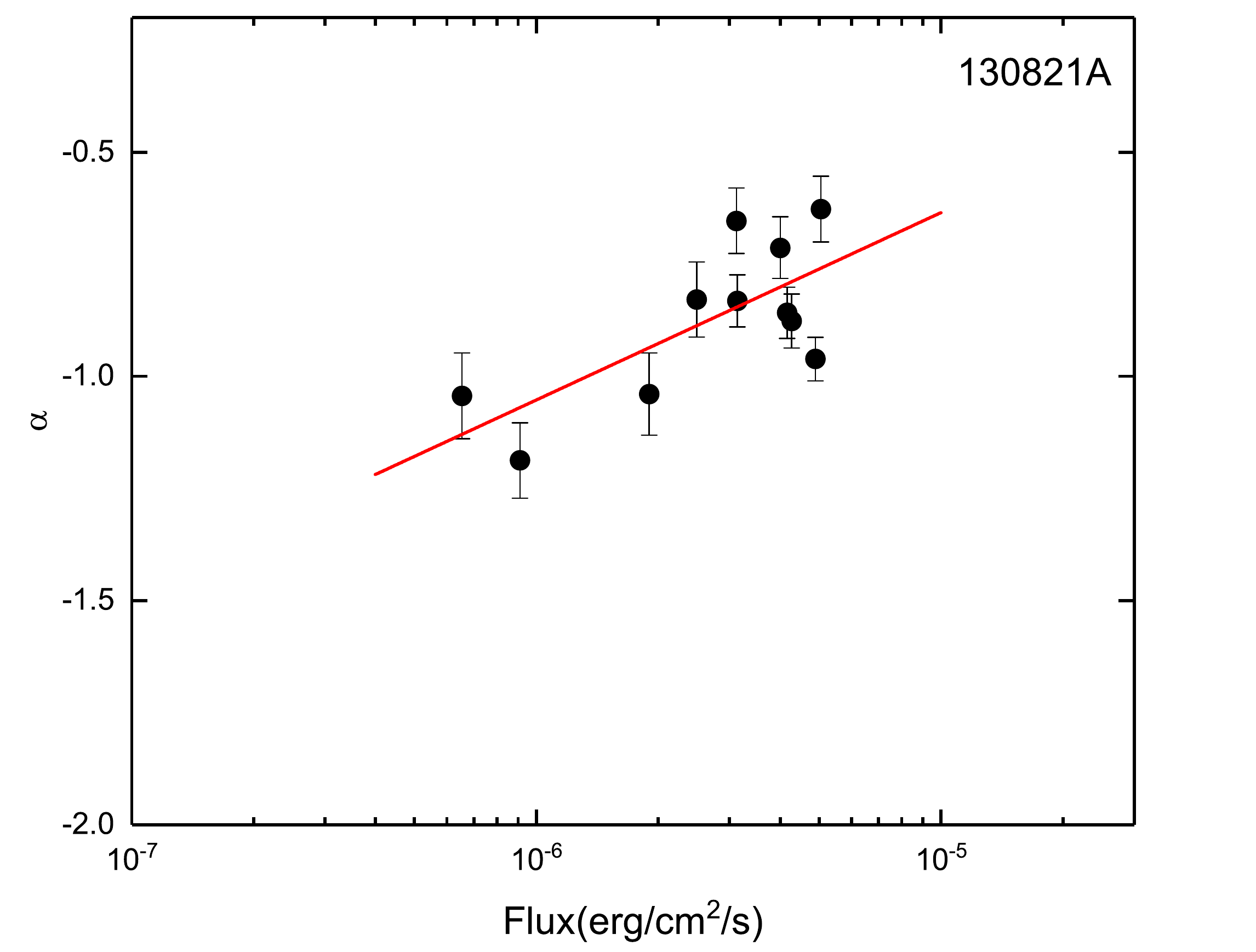}}
\resizebox{4cm}{!}{\includegraphics{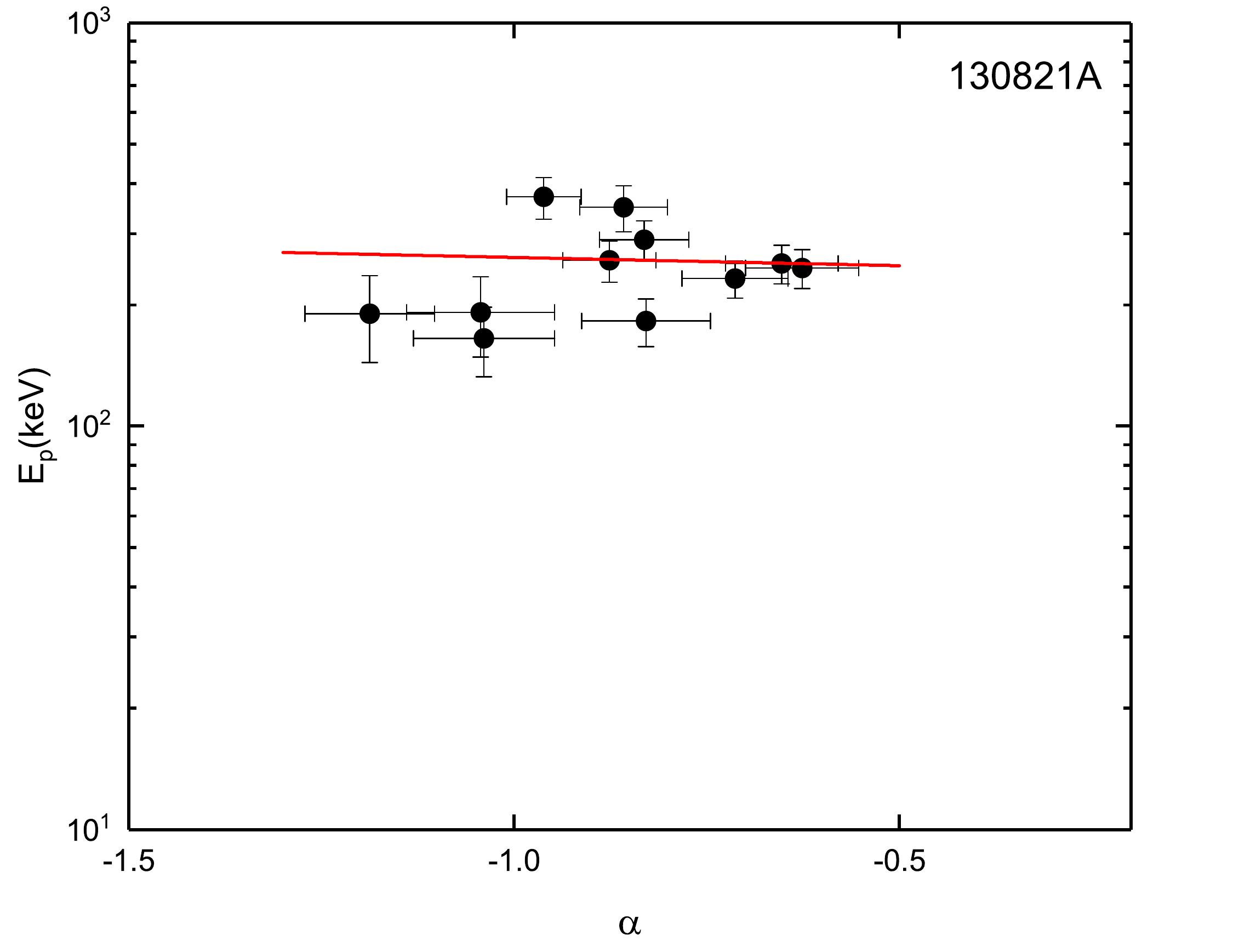}}
\resizebox{4cm}{!}{\includegraphics{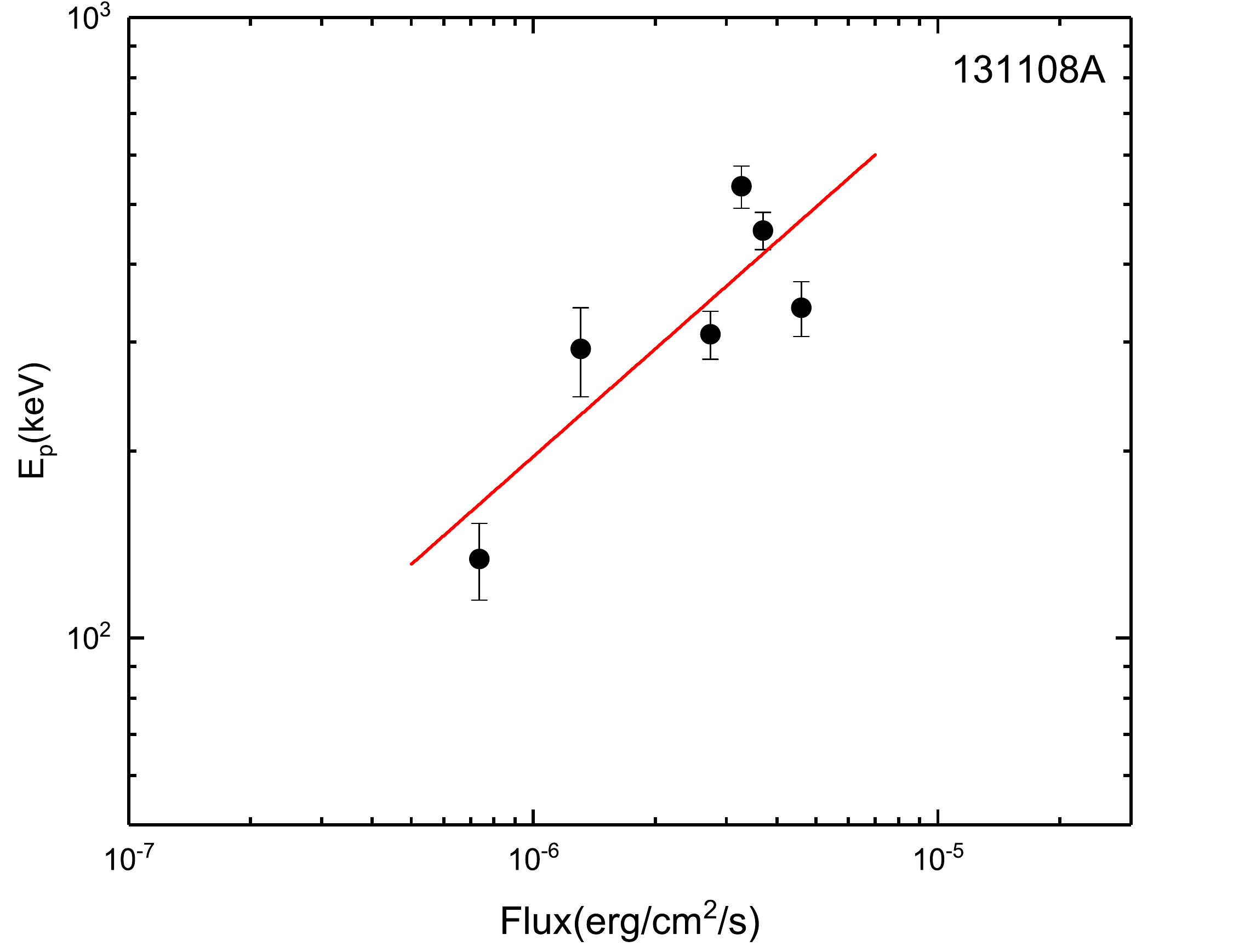}}
\resizebox{4cm}{!}{\includegraphics{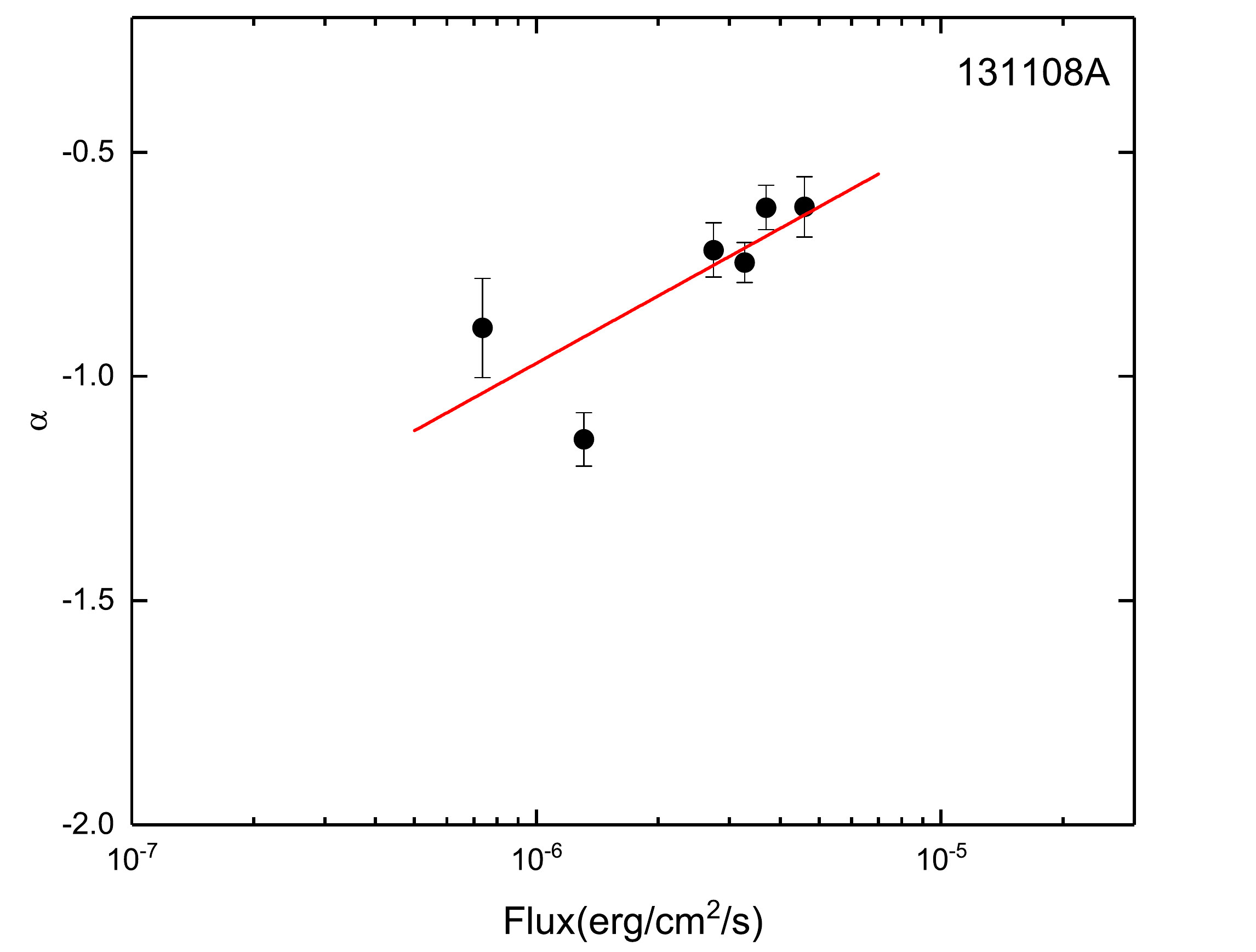}}
\resizebox{4cm}{!}{\includegraphics{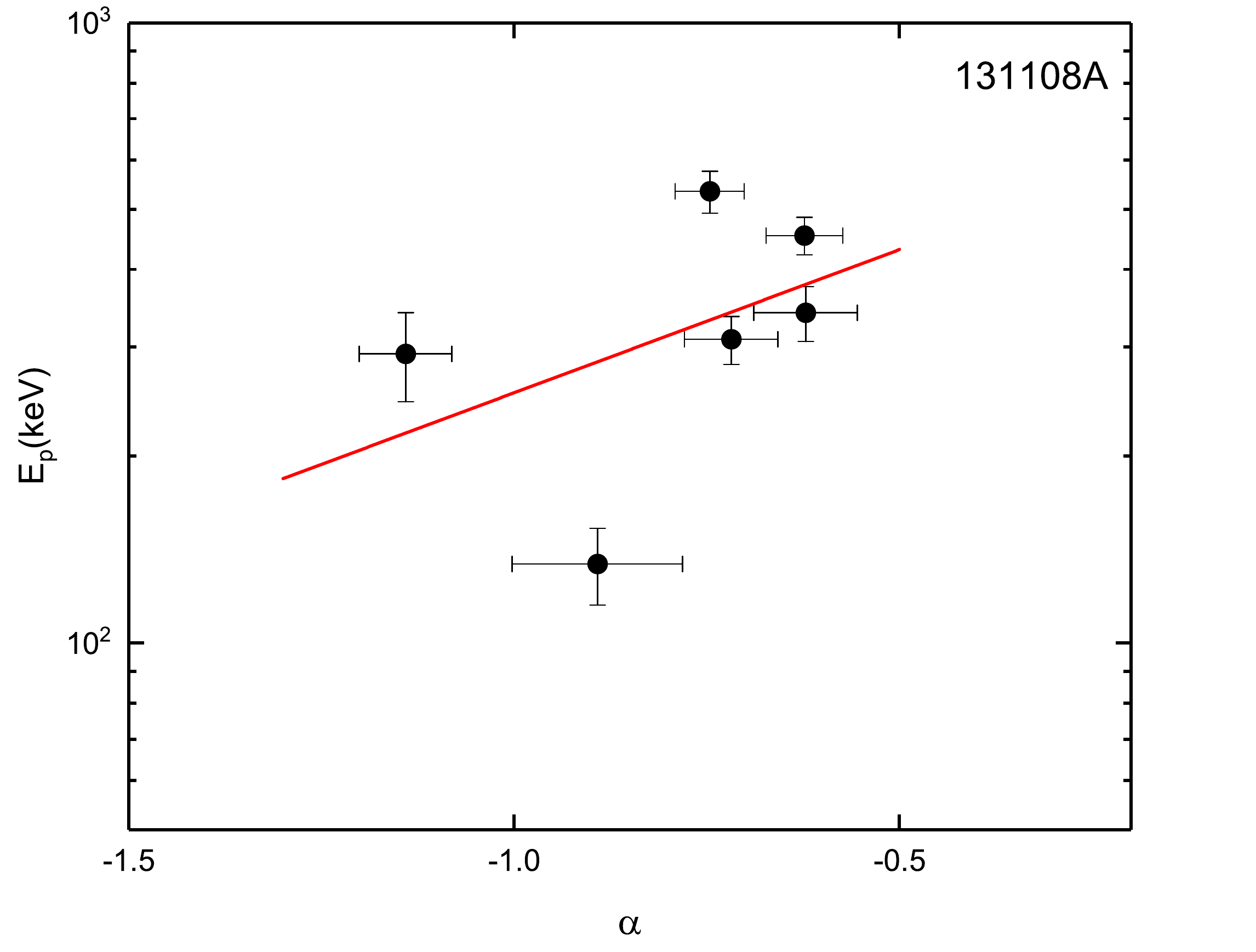}}
\resizebox{4cm}{!}{\includegraphics{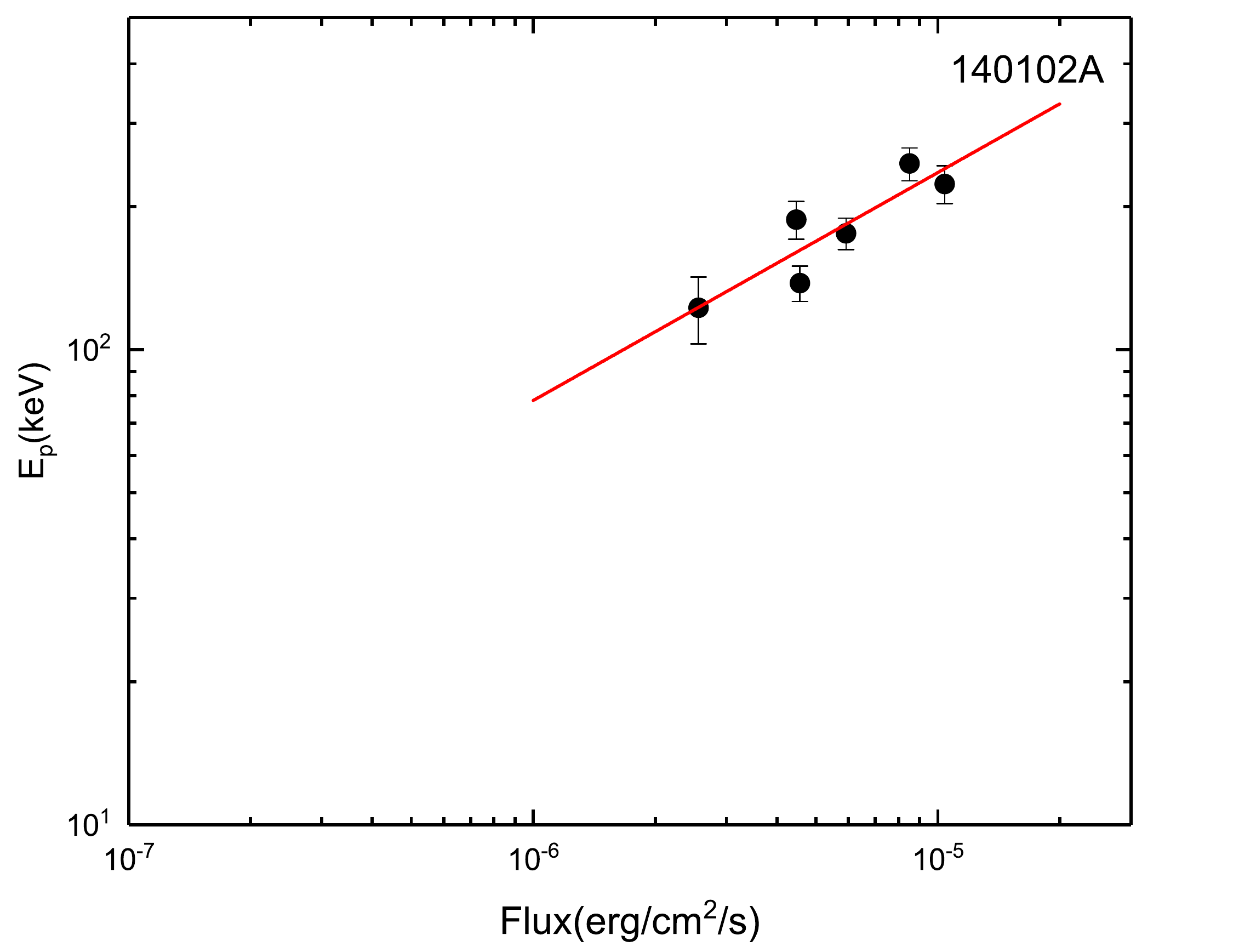}}
\resizebox{4cm}{!}{\includegraphics{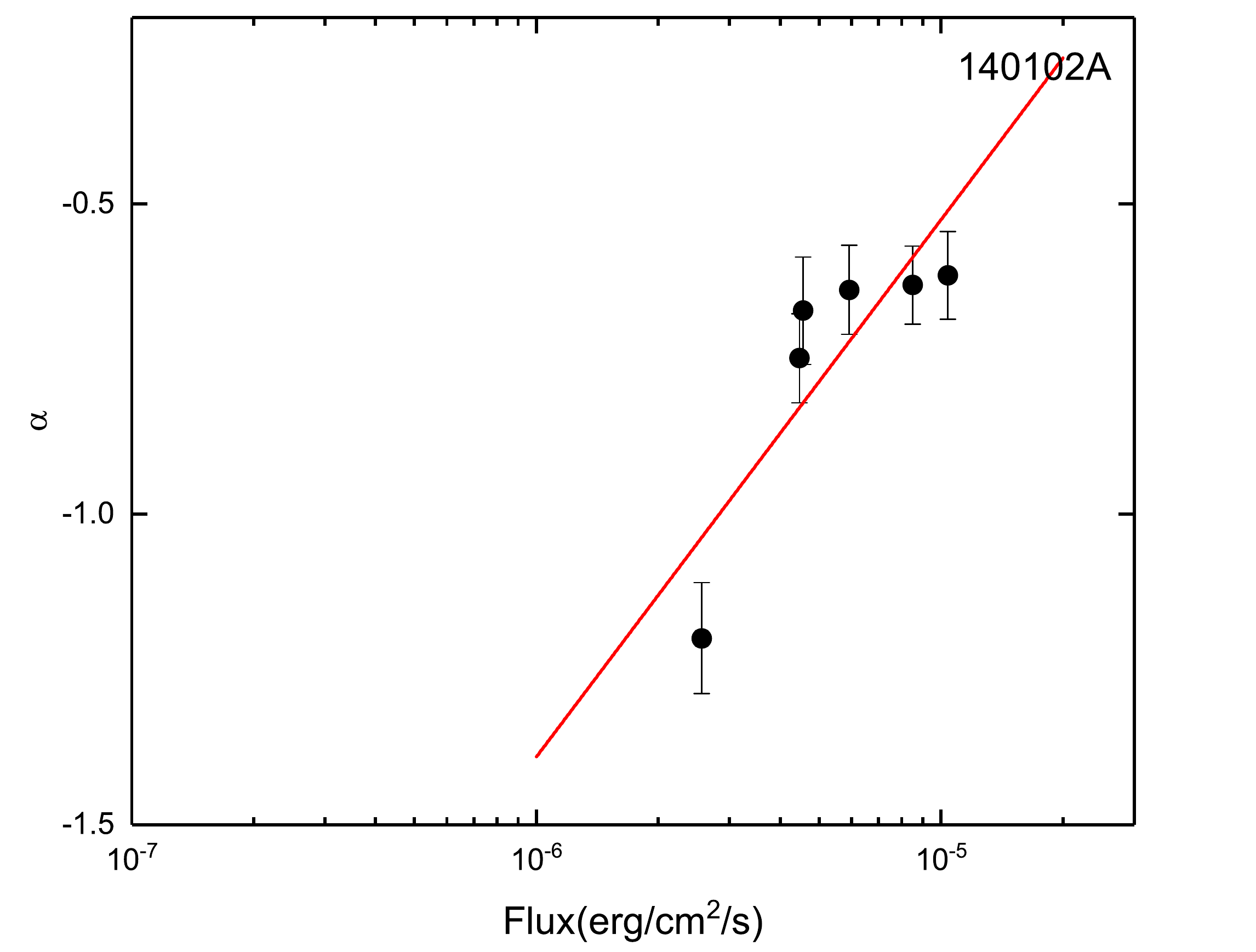}}
\resizebox{4cm}{!}{\includegraphics{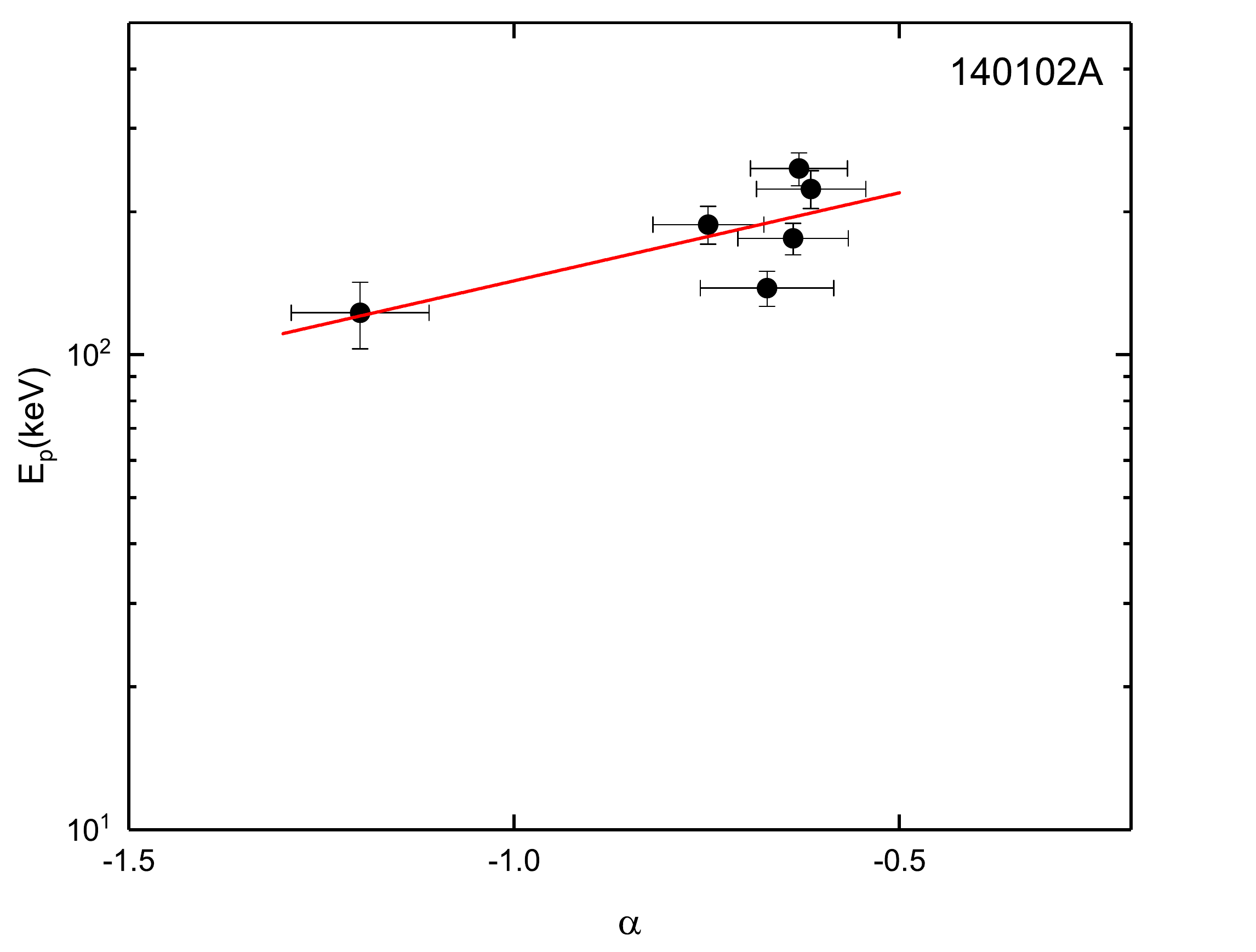}}
\resizebox{4cm}{!}{\includegraphics{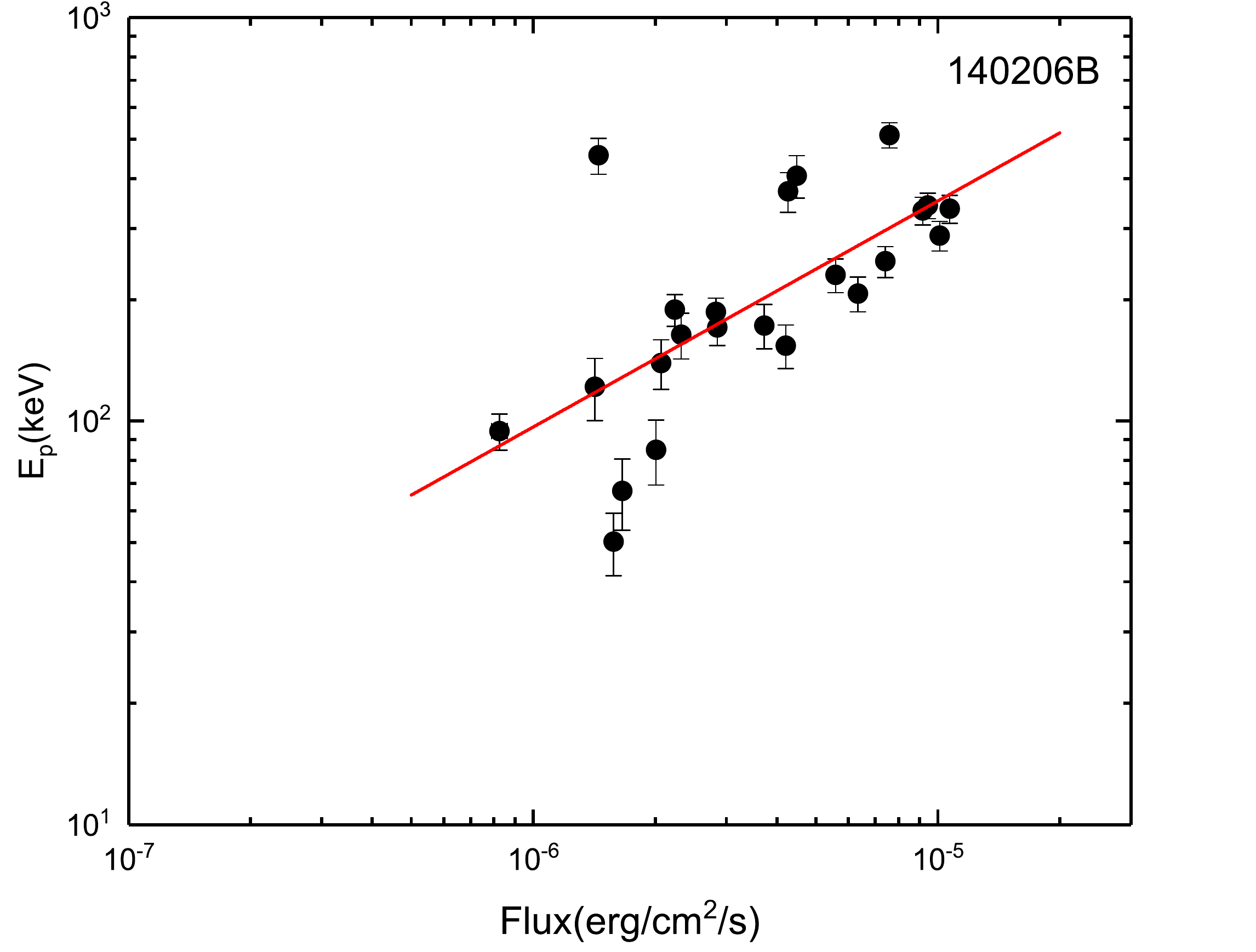}}
\resizebox{4cm}{!}{\includegraphics{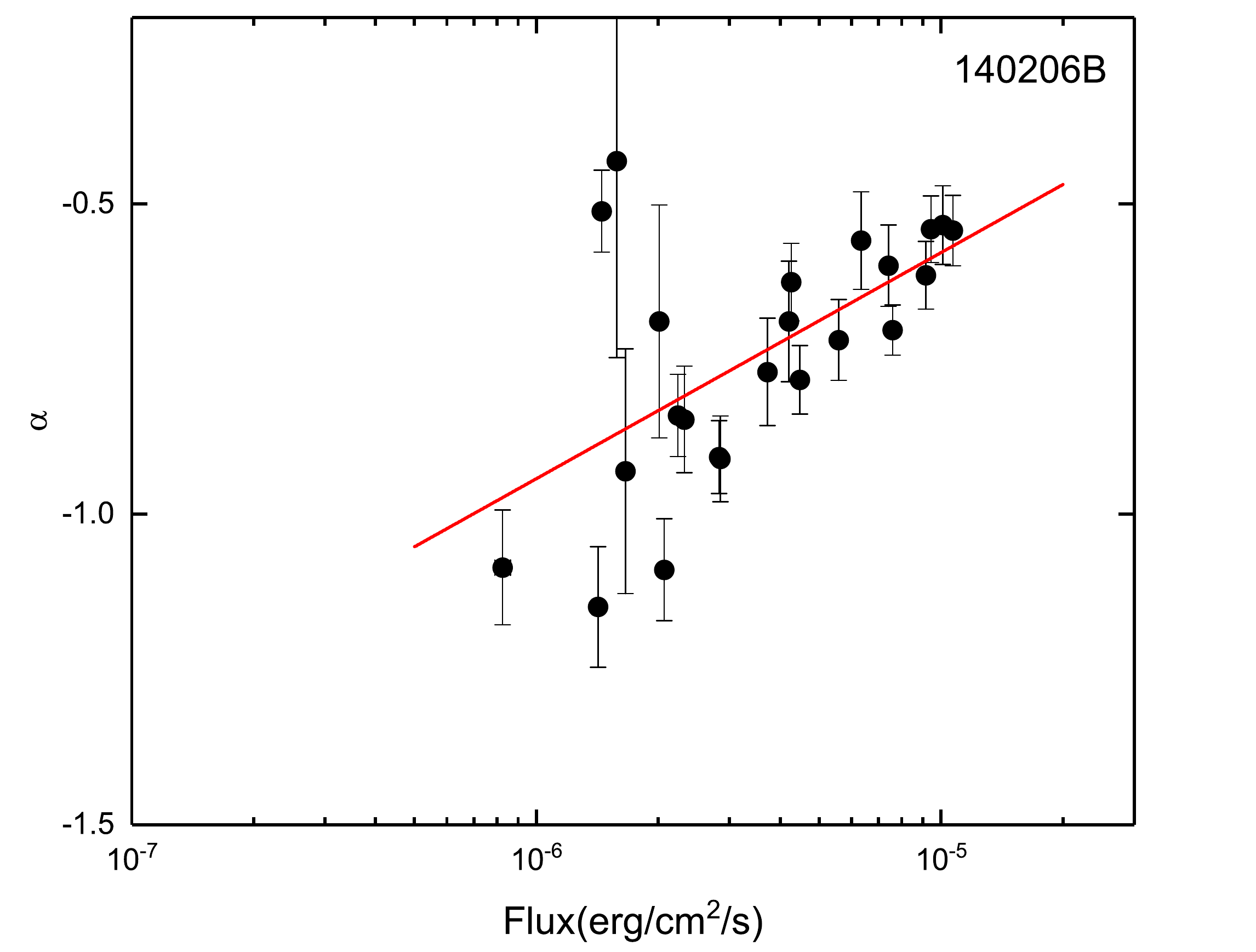}}
\resizebox{4cm}{!}{\includegraphics{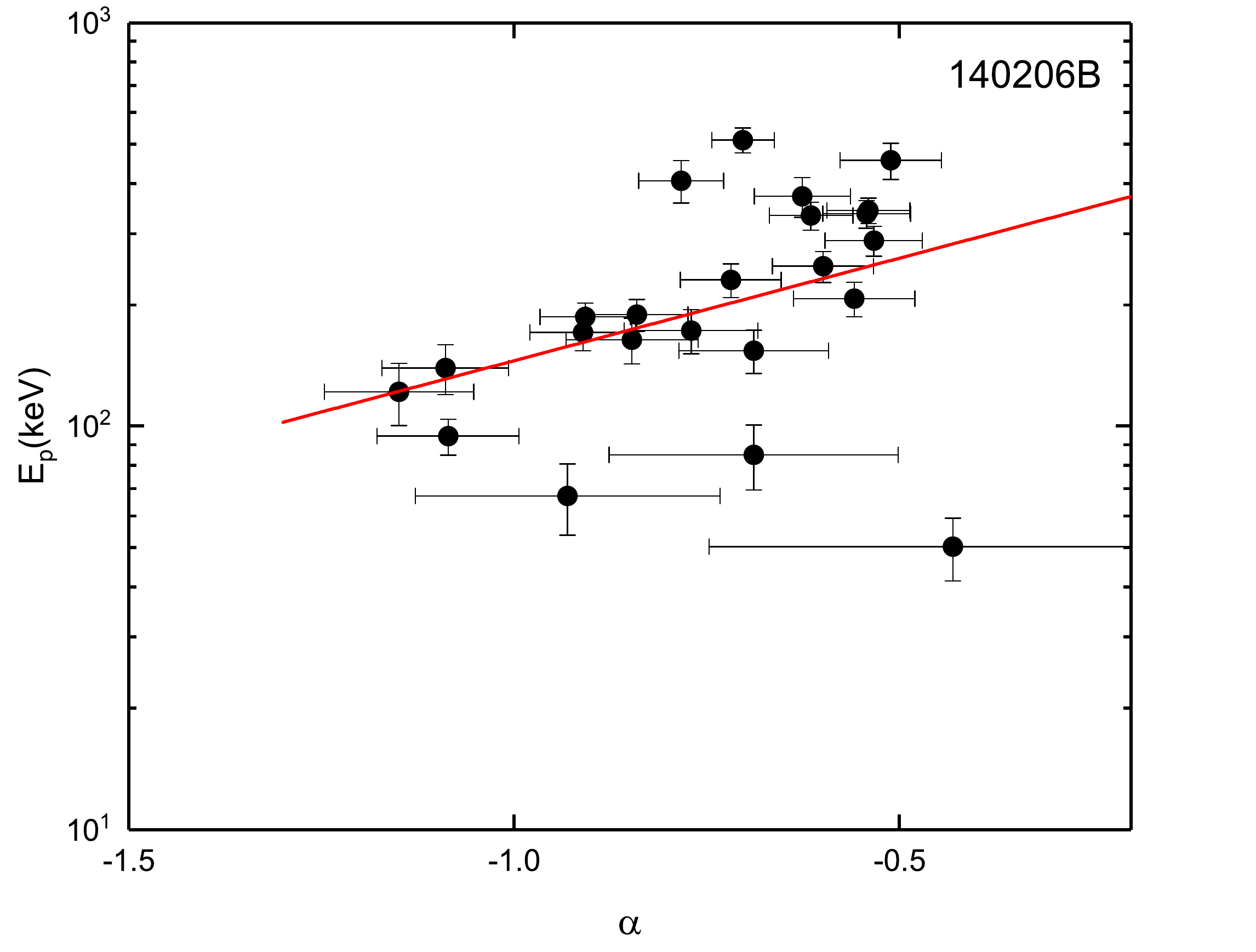}}
\resizebox{4cm}{!}{\includegraphics{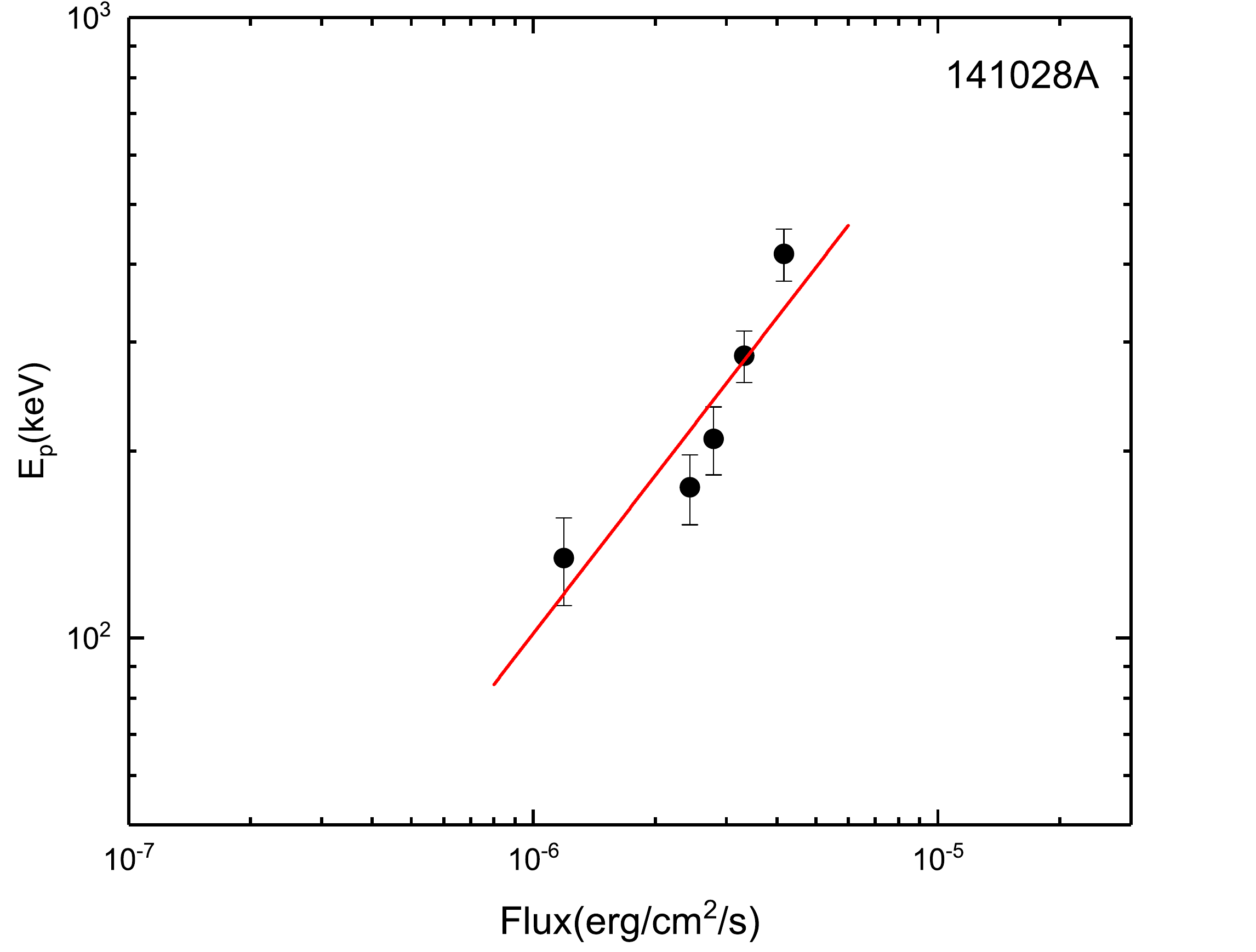}}
\caption{\it-continued}
\end{figure}

\addtocounter{figure}{-1}
\begin{figure}
\centering
\resizebox{4cm}{!}{\includegraphics{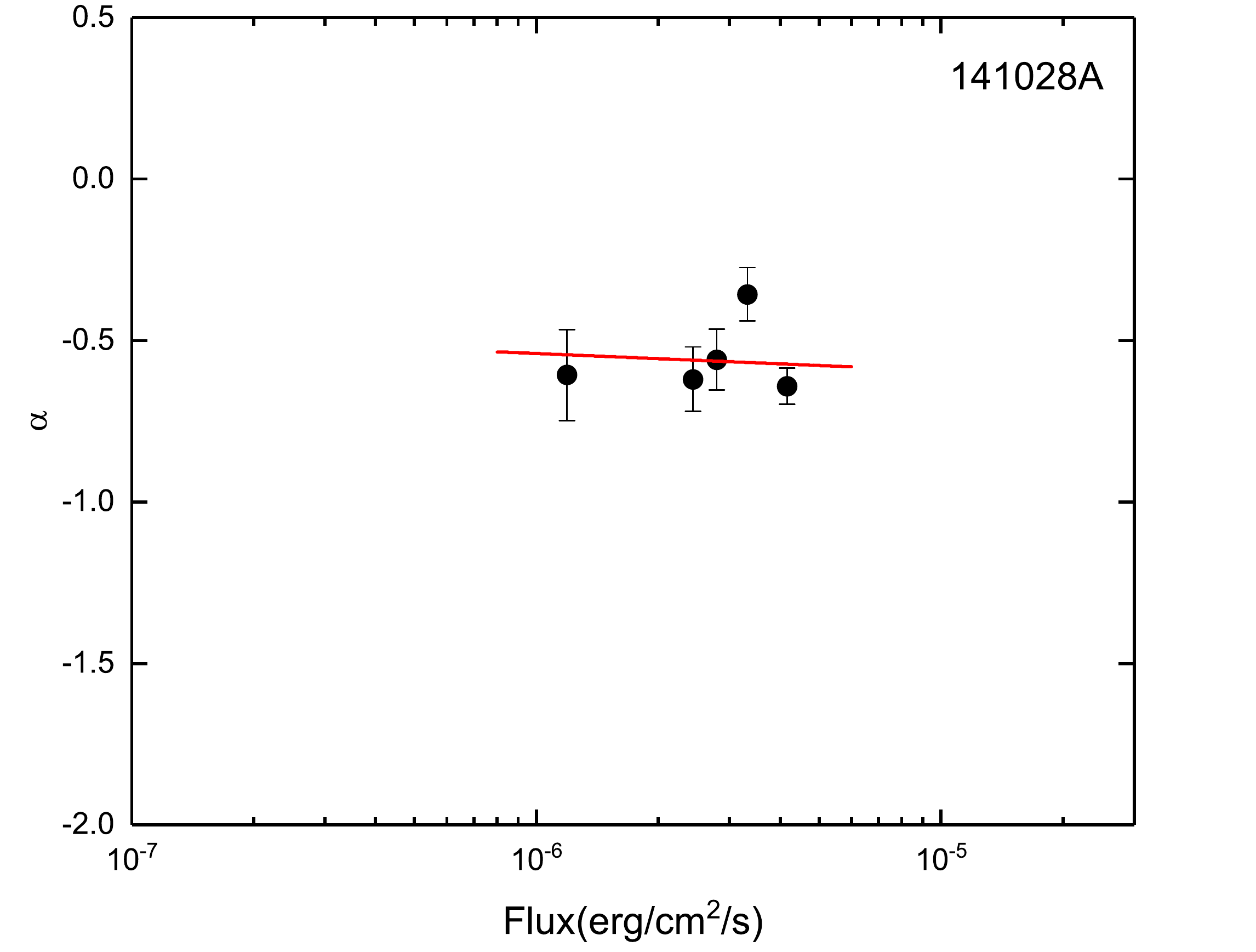}}
\resizebox{4cm}{!}{\includegraphics{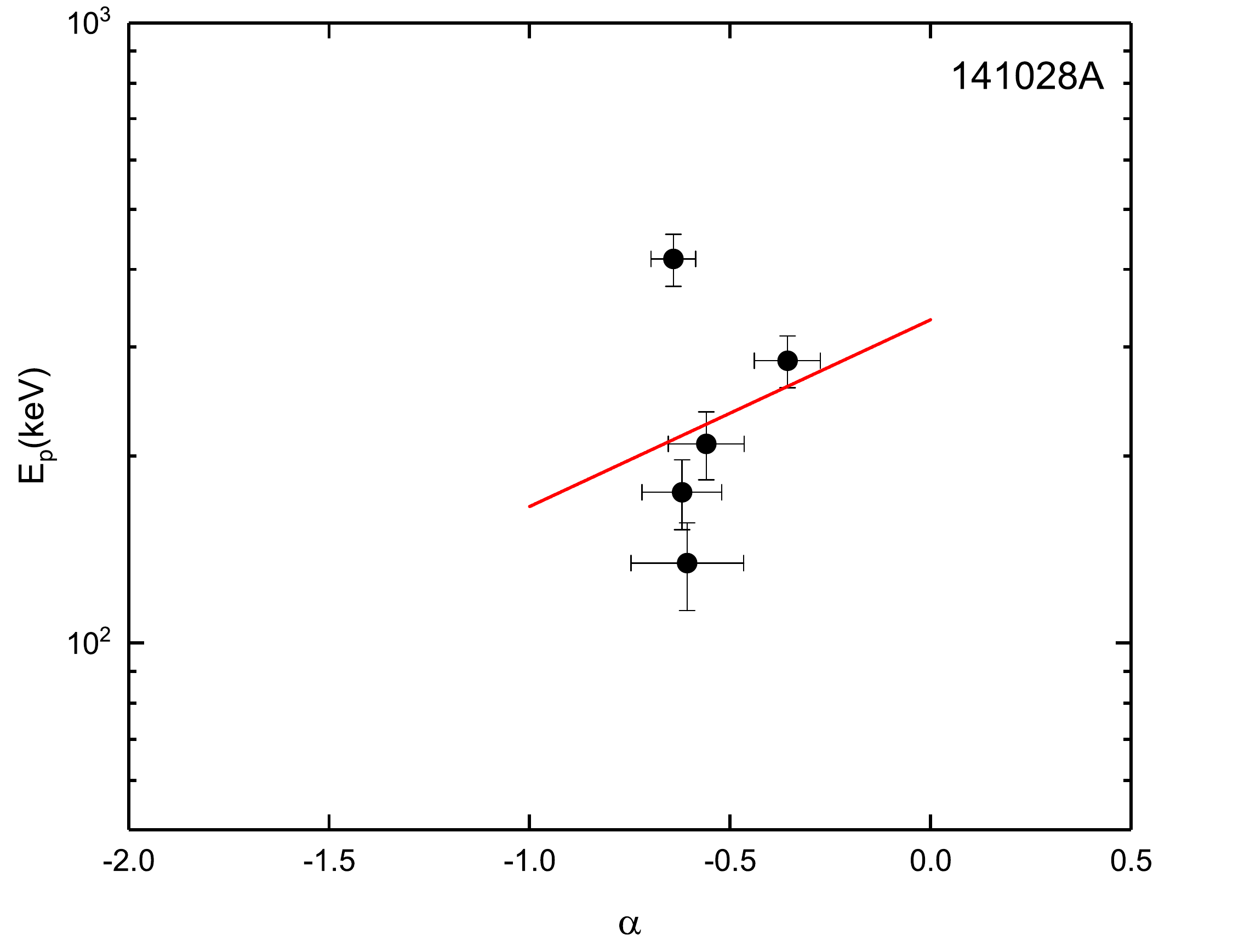}}
\resizebox{4cm}{!}{\includegraphics{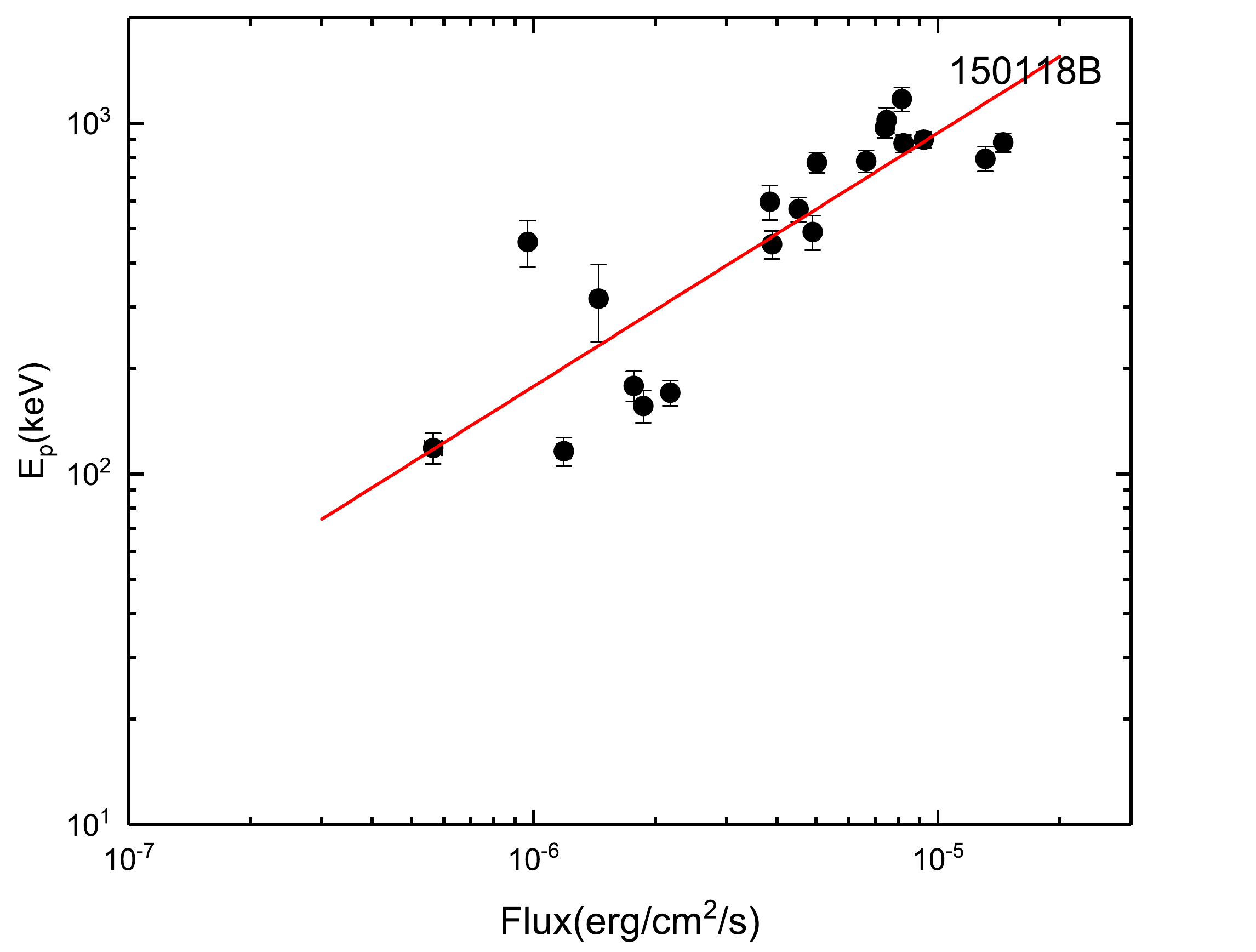}}
\resizebox{4cm}{!}{\includegraphics{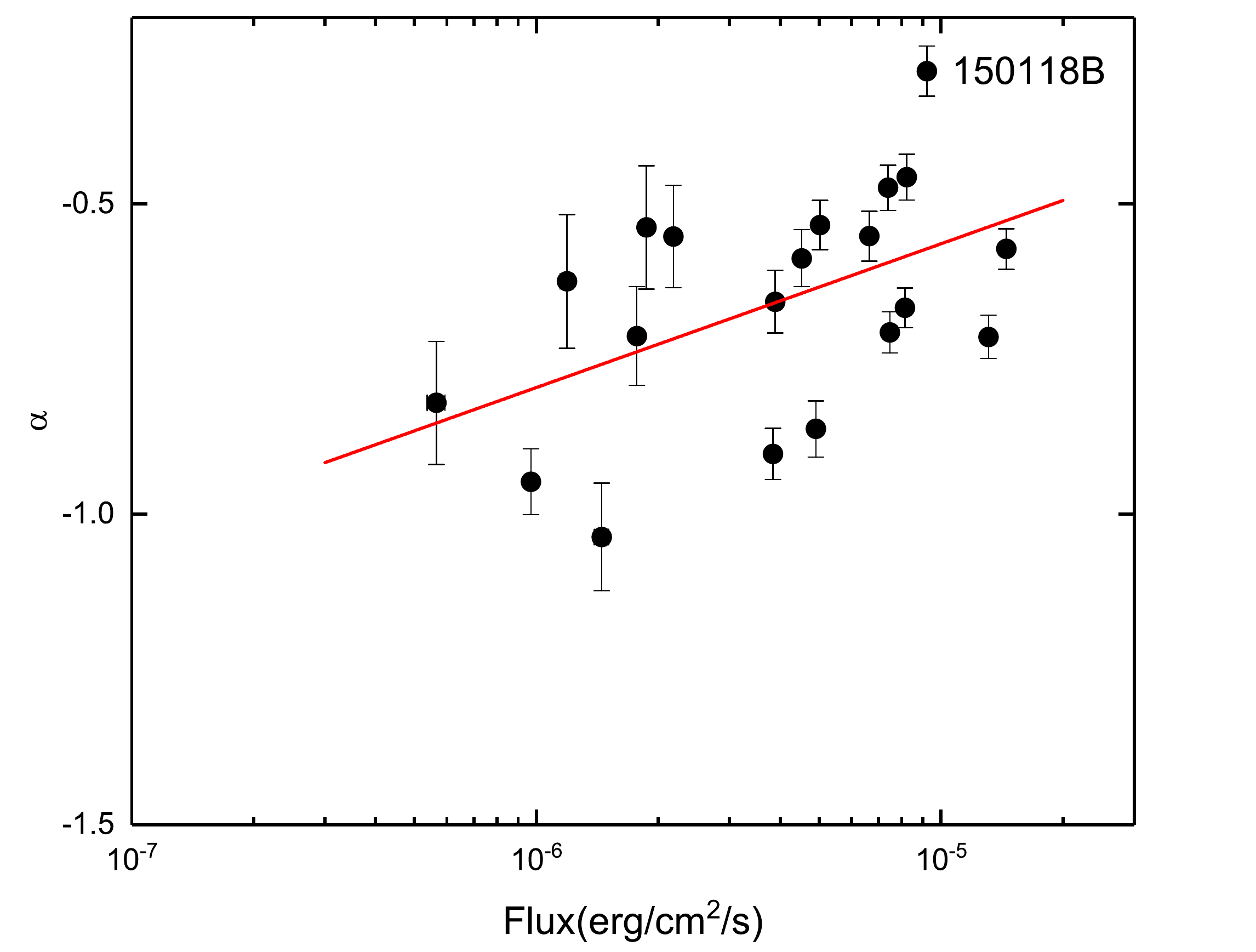}}
\resizebox{4cm}{!}{\includegraphics{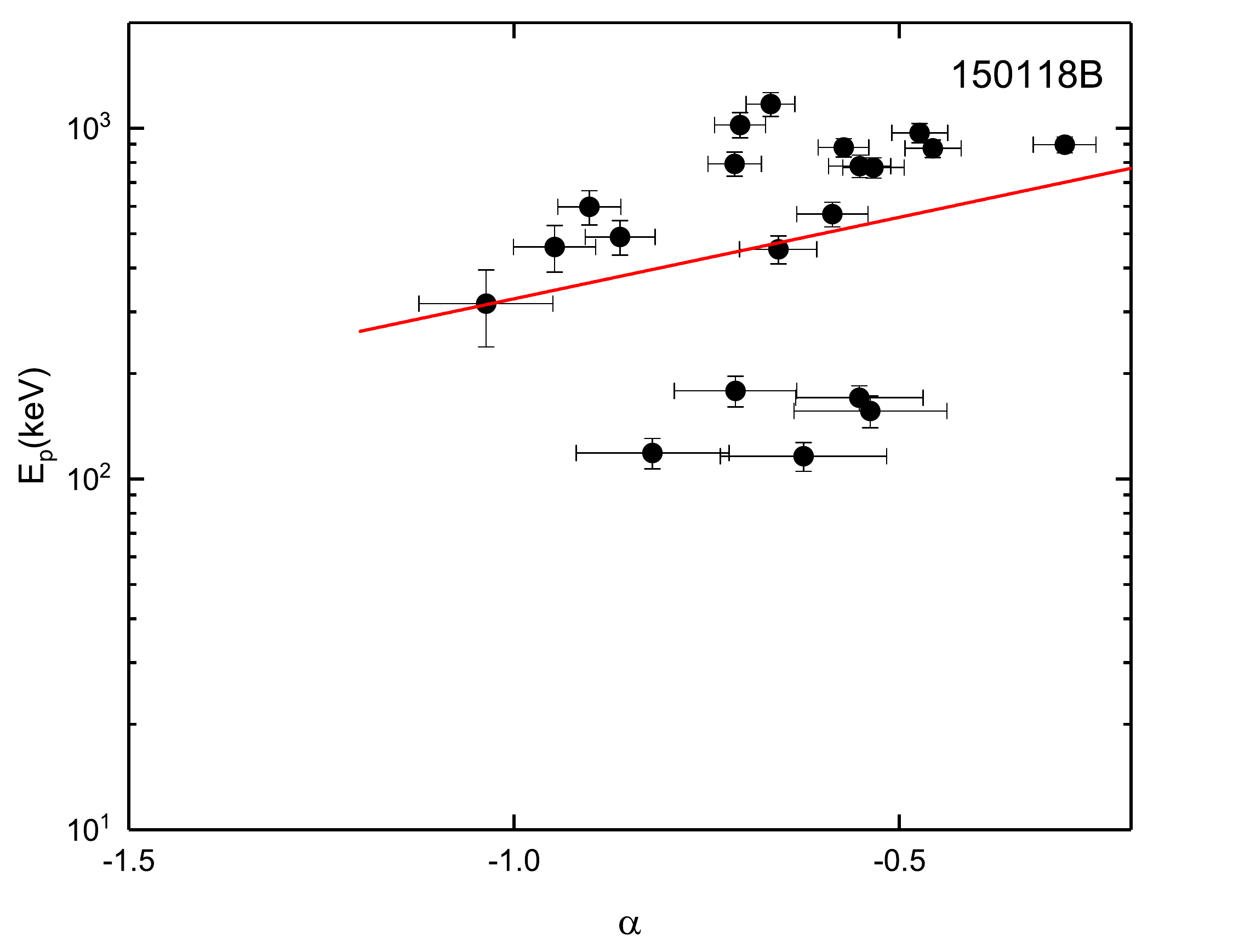}}
\resizebox{4cm}{!}{\includegraphics{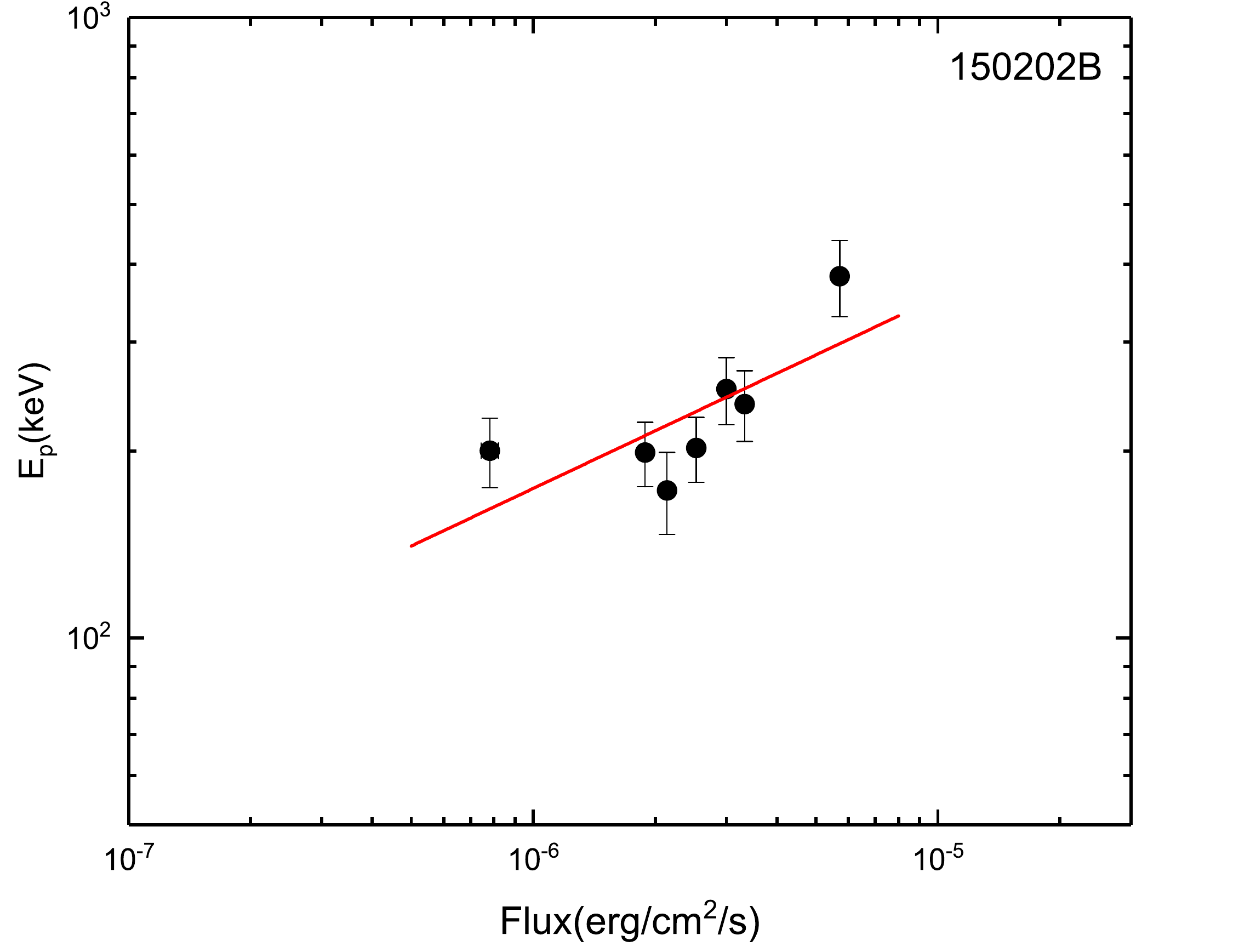}}
\resizebox{4cm}{!}{\includegraphics{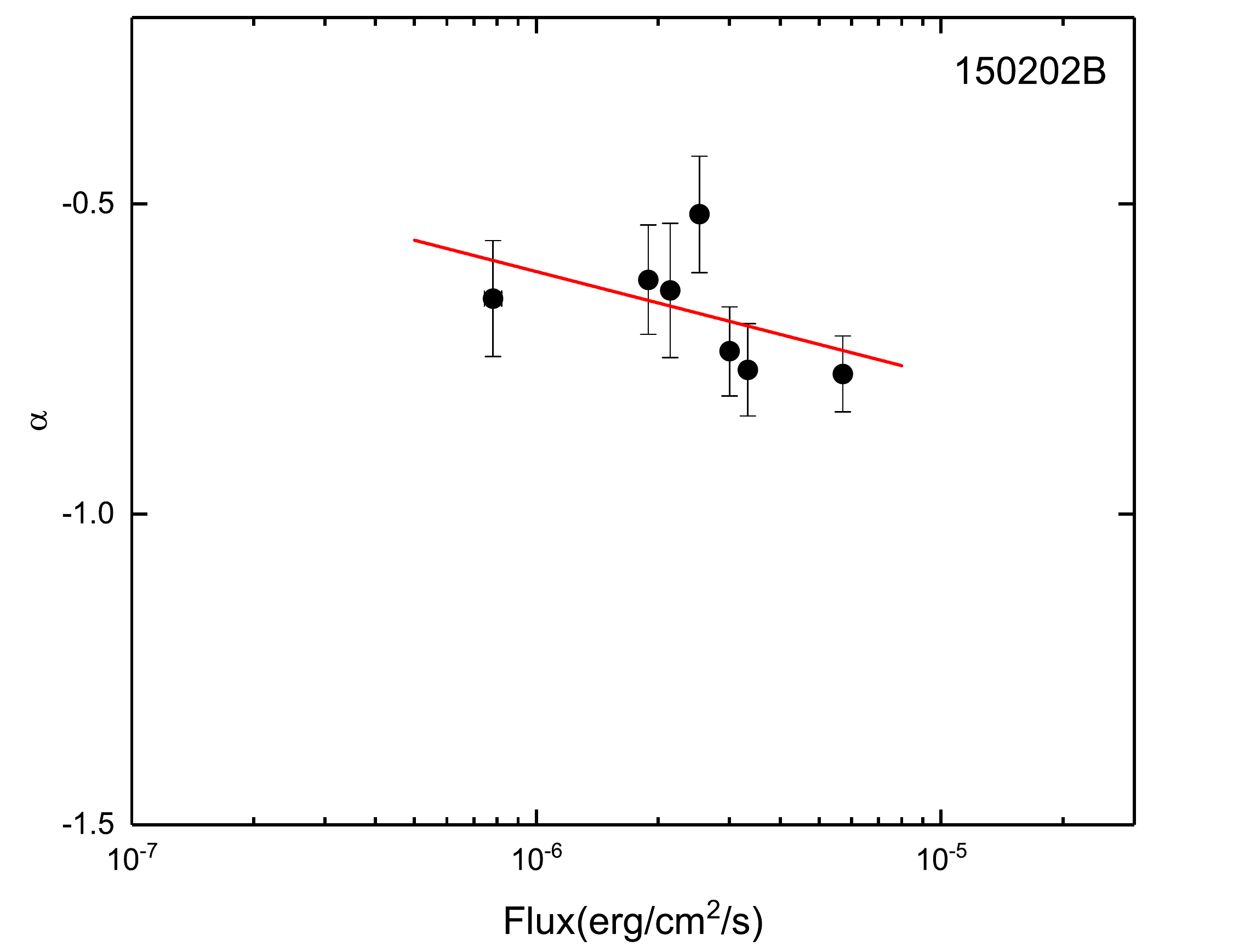}}
\resizebox{4cm}{!}{\includegraphics{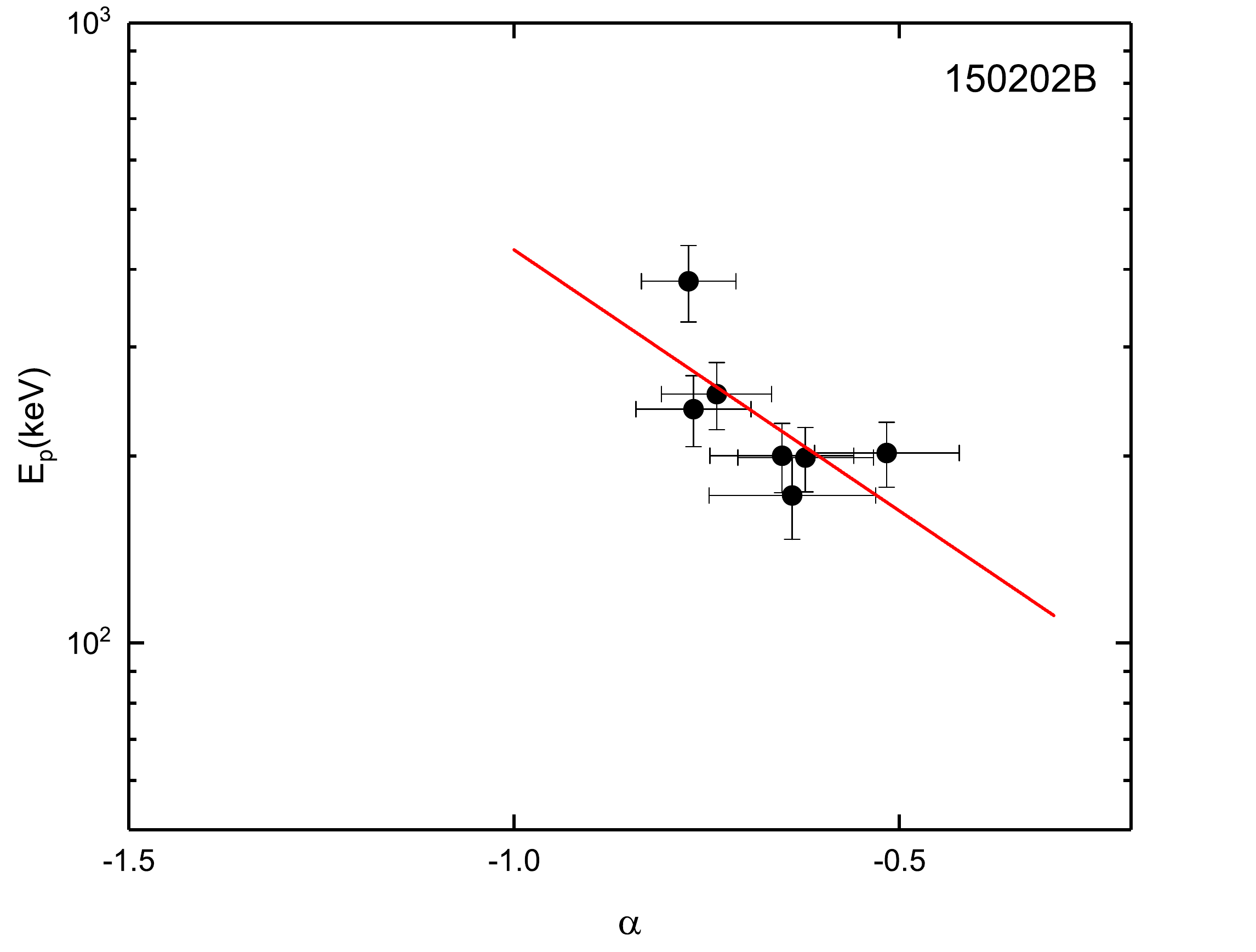}}
\resizebox{4cm}{!}{\includegraphics{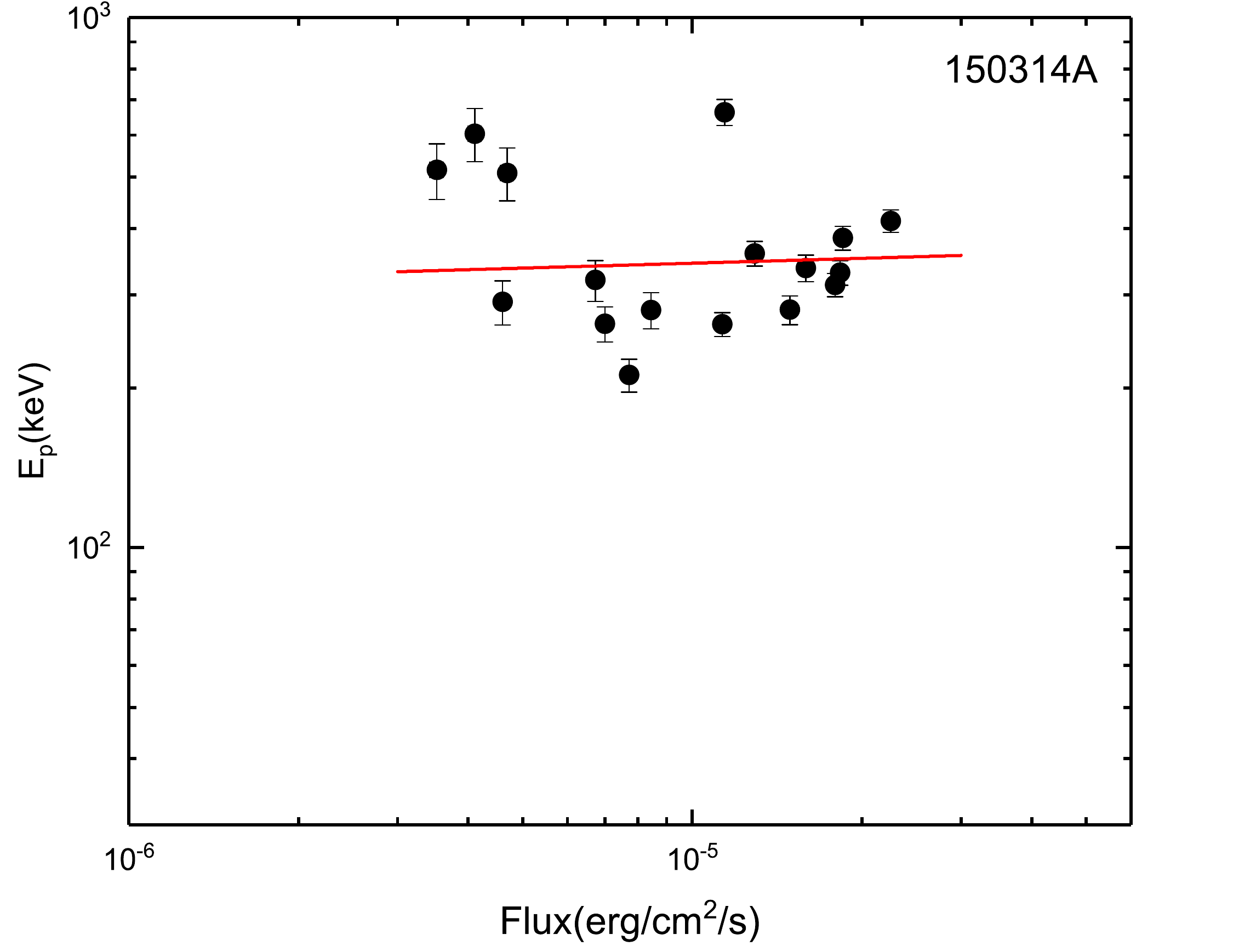}}
\resizebox{4cm}{!}{\includegraphics{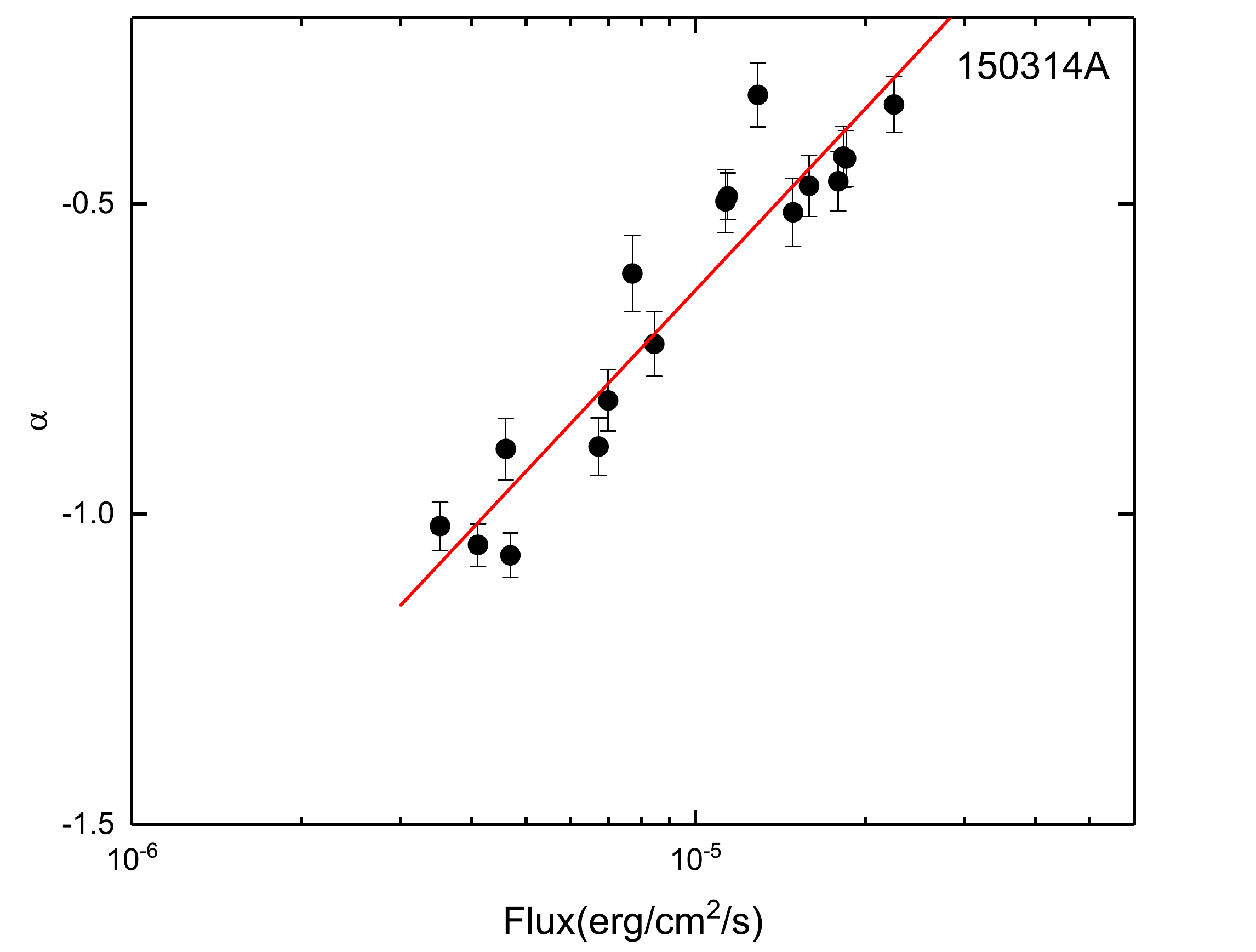}}
\resizebox{4cm}{!}{\includegraphics{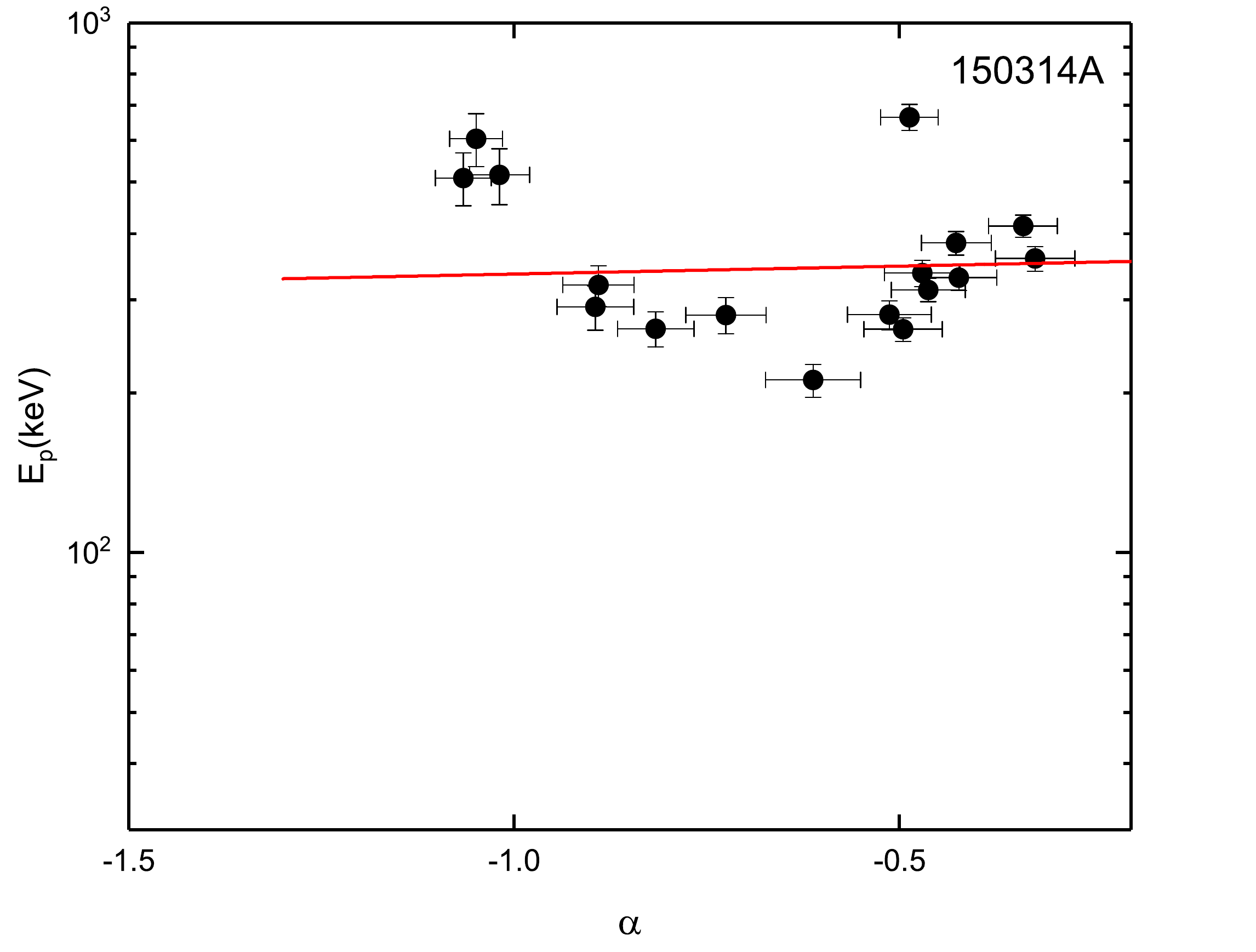}}
\resizebox{4cm}{!}{\includegraphics{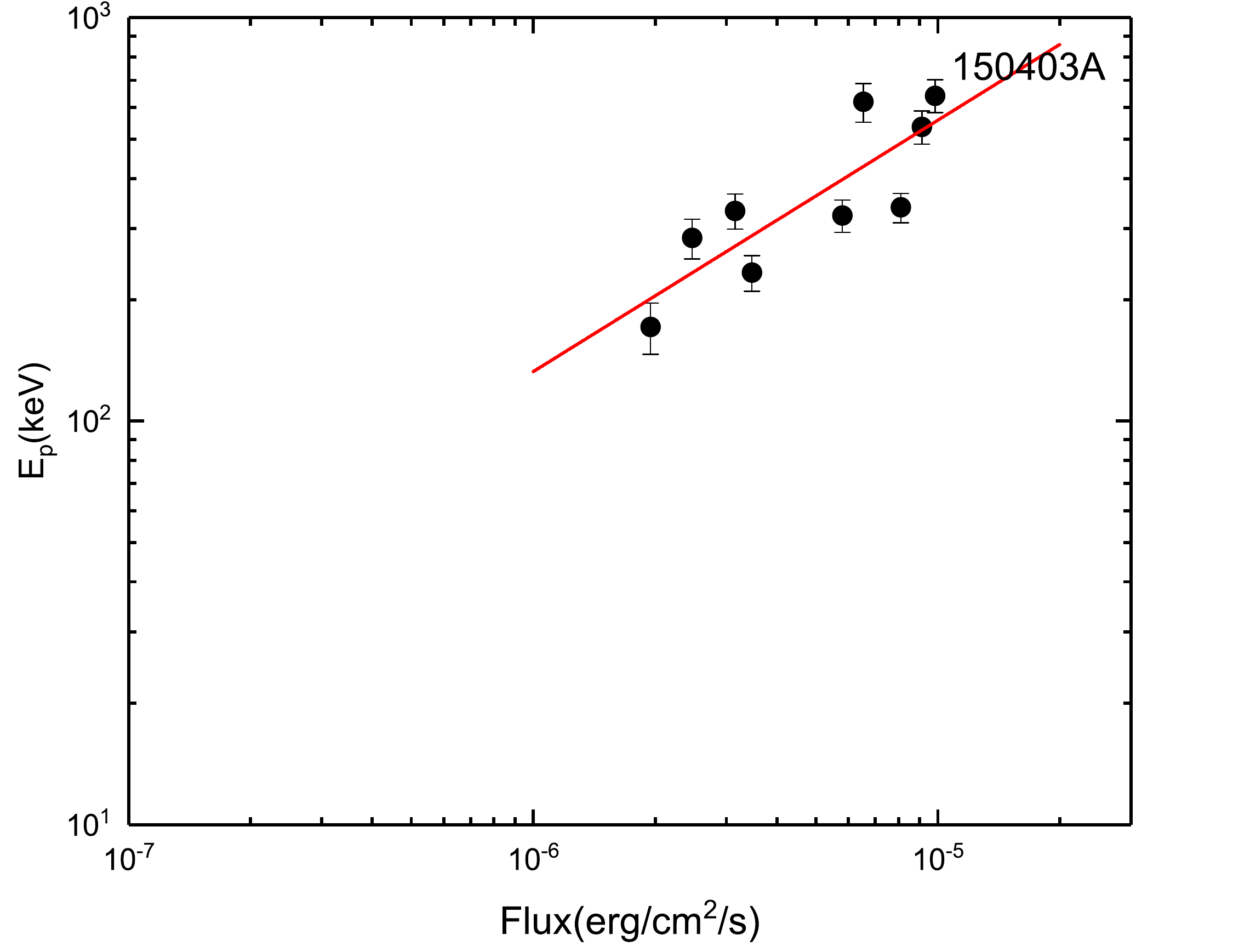}}
\resizebox{4cm}{!}{\includegraphics{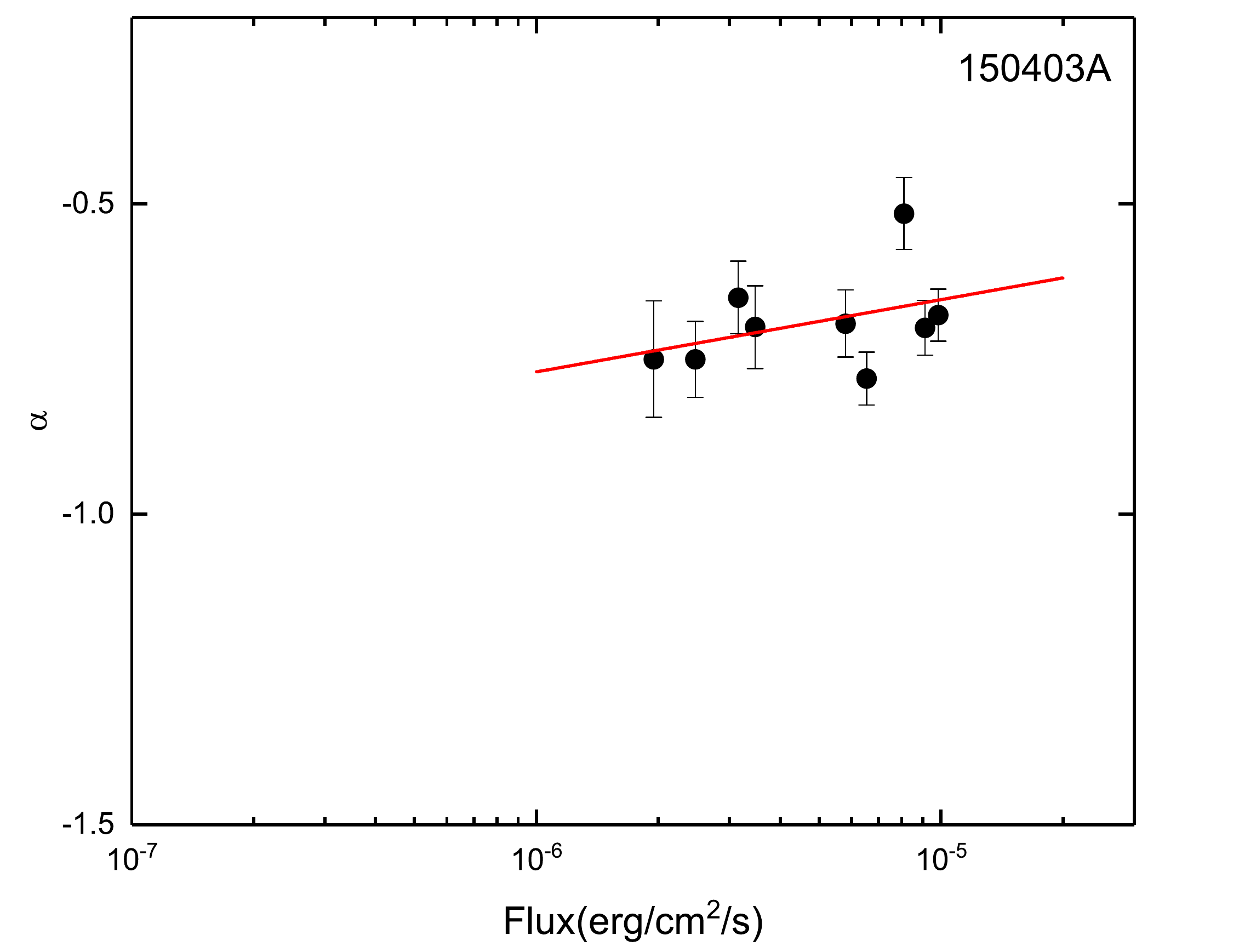}}
\resizebox{4cm}{!}{\includegraphics{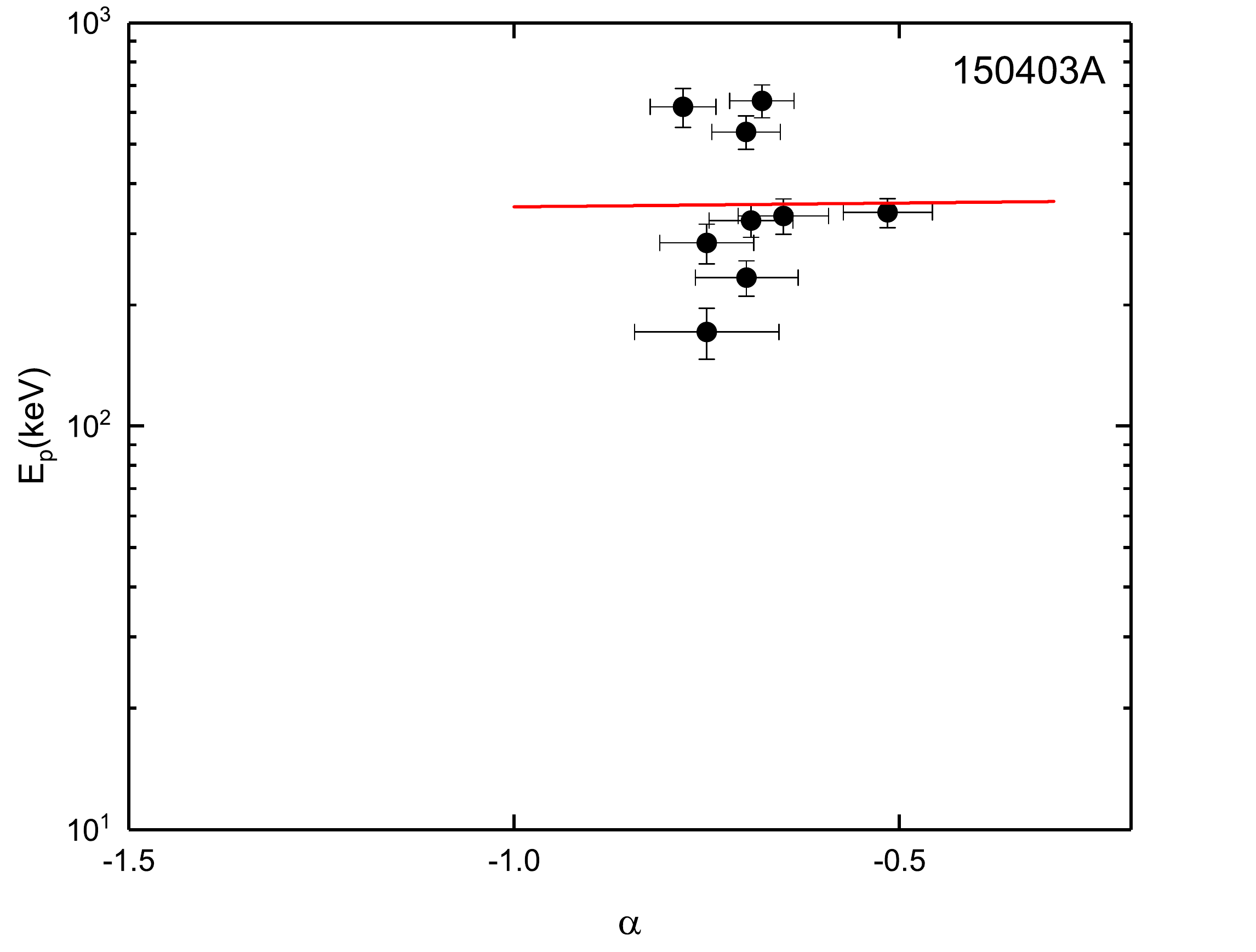}}
\resizebox{4cm}{!}{\includegraphics{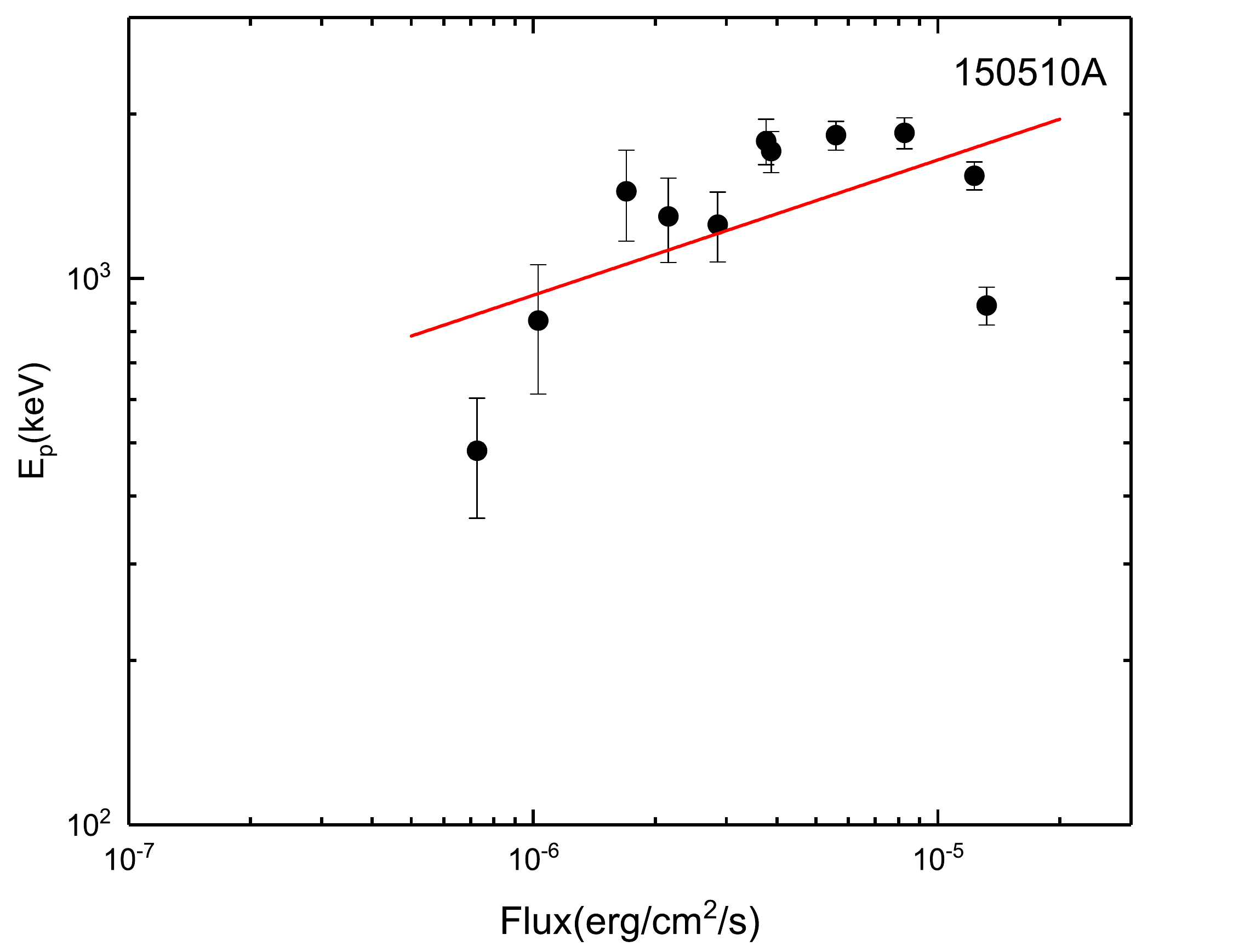}}
\resizebox{4cm}{!}{\includegraphics{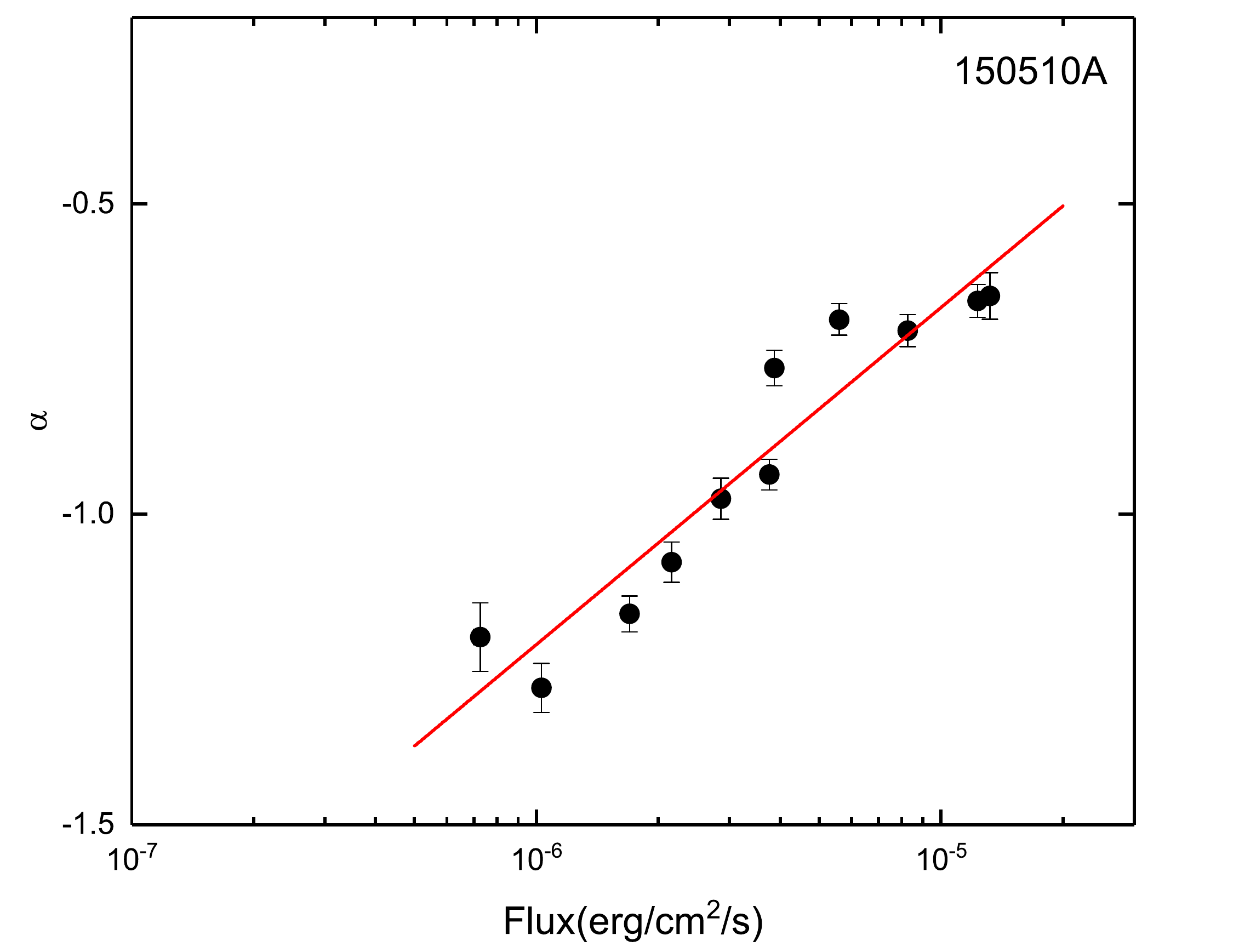}}
\resizebox{4cm}{!}{\includegraphics{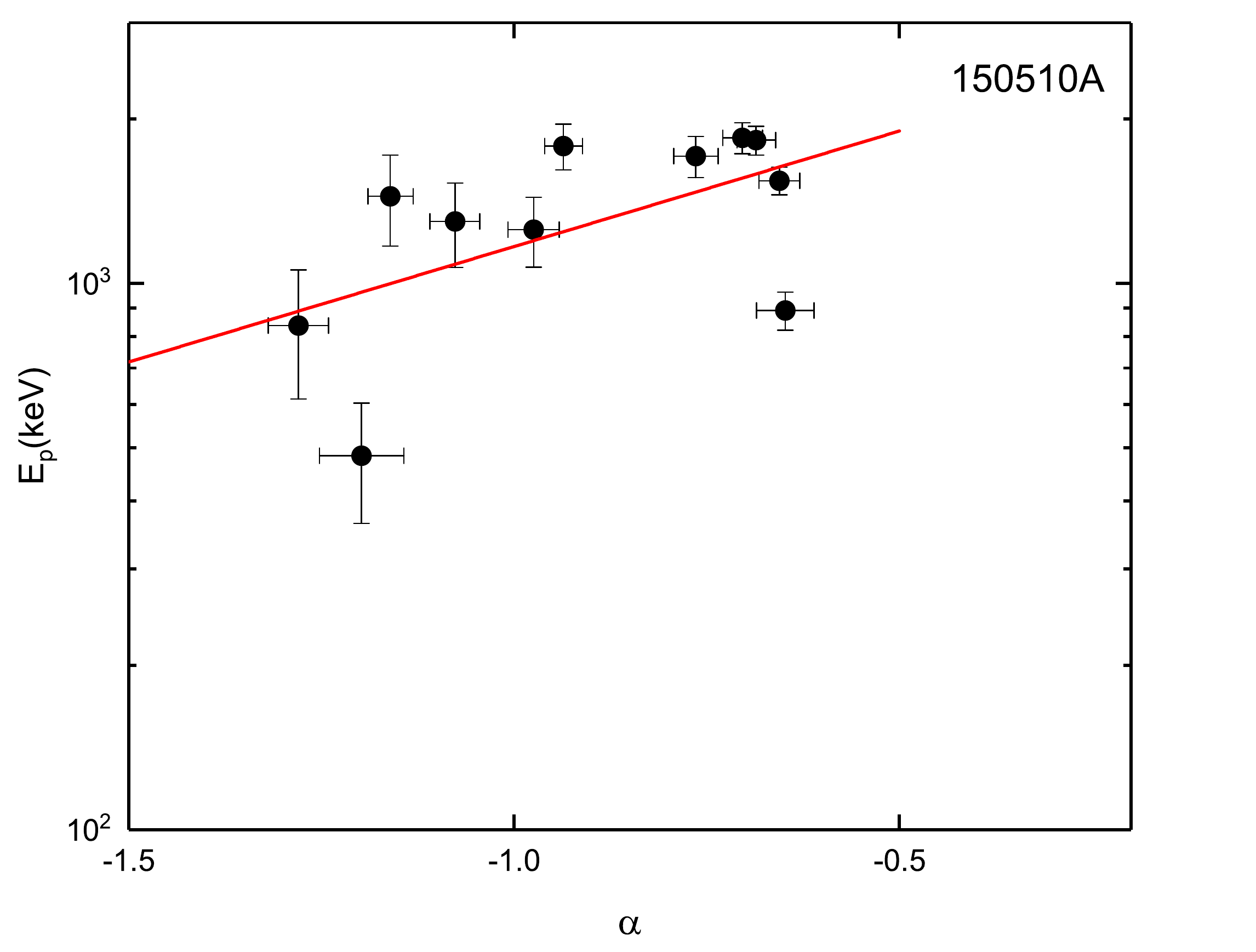}}
\resizebox{4cm}{!}{\includegraphics{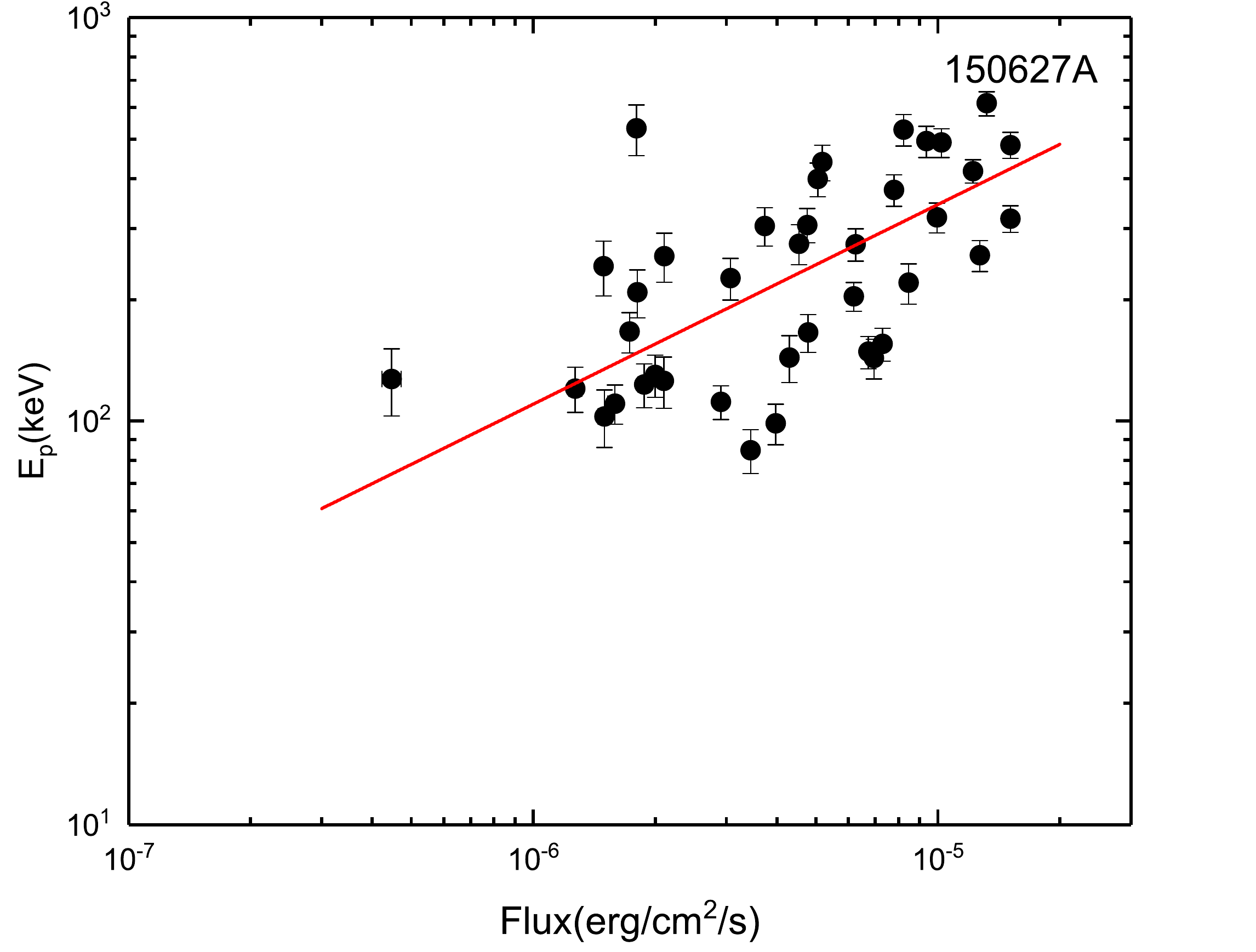}}
\resizebox{4cm}{!}{\includegraphics{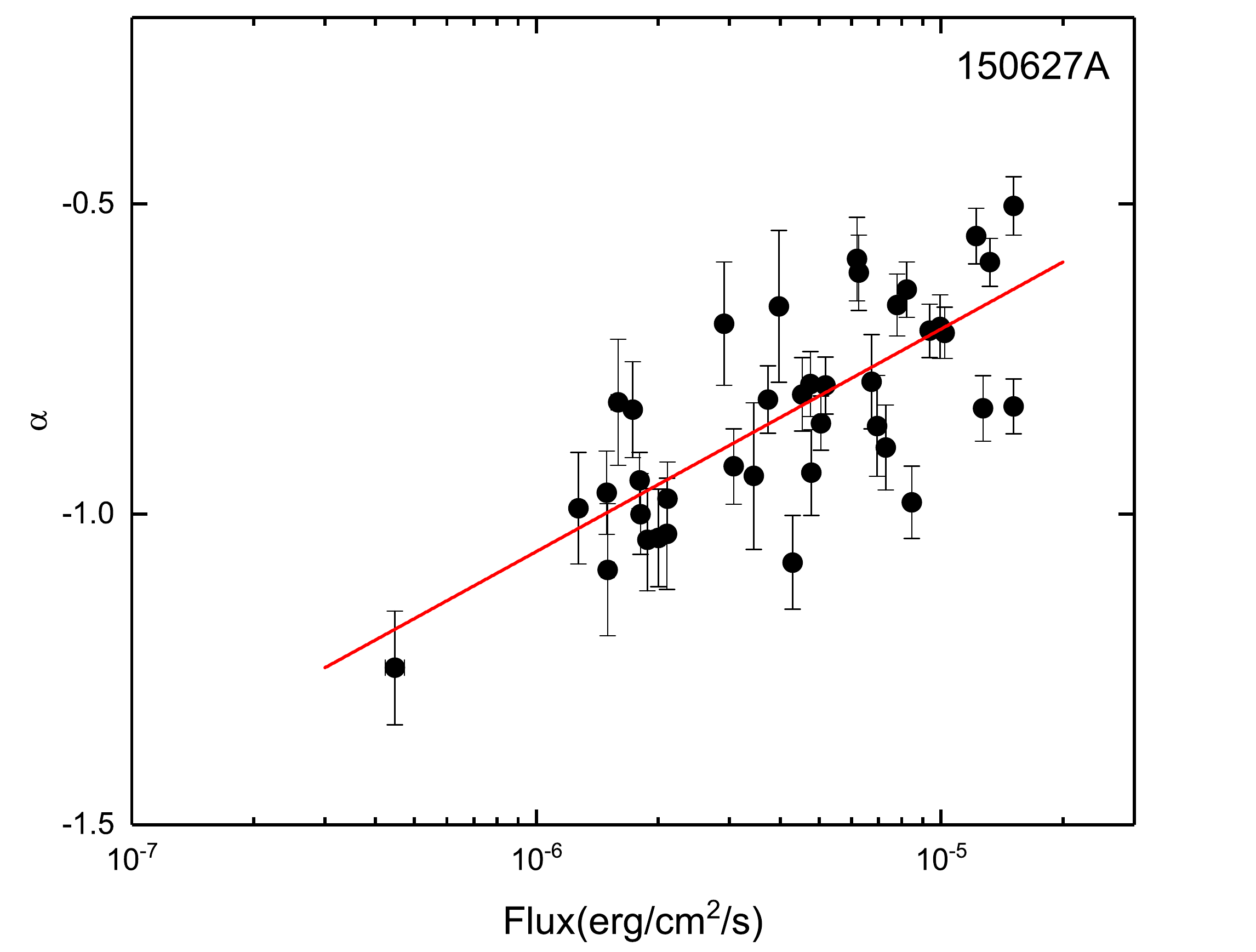}}
\resizebox{4cm}{!}{\includegraphics{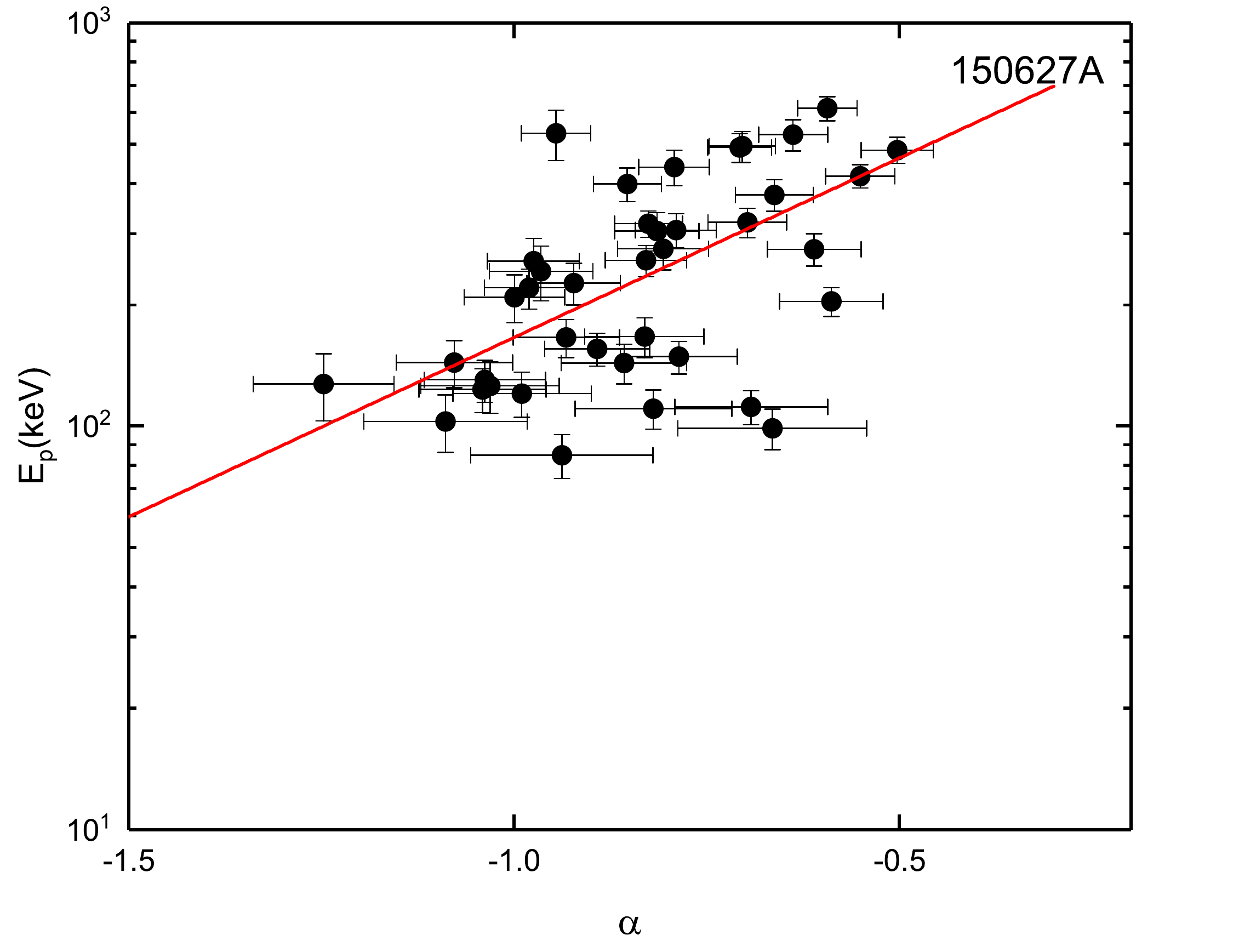}}
\resizebox{4cm}{!}{\includegraphics{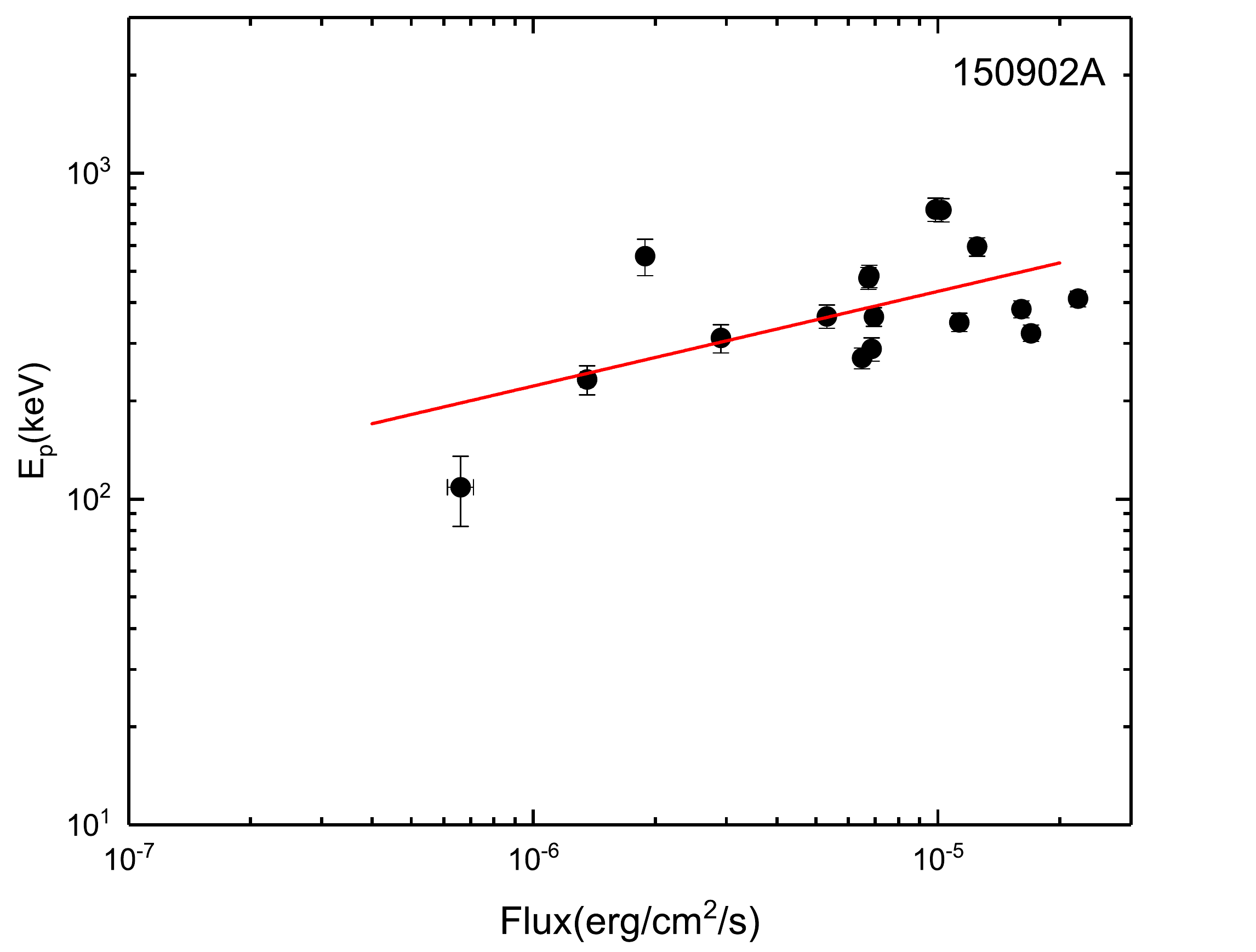}}
\resizebox{4cm}{!}{\includegraphics{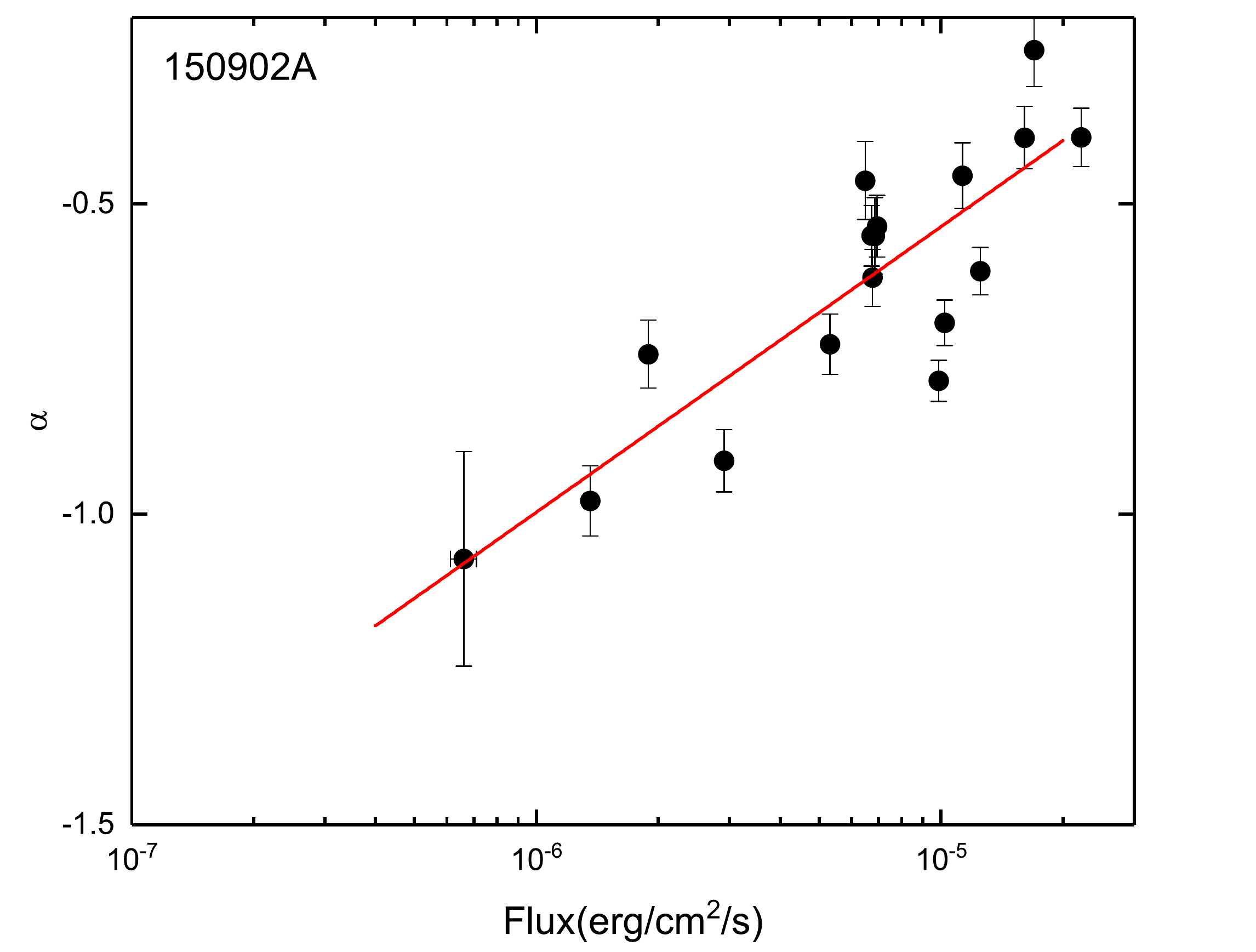}}
\resizebox{4cm}{!}{\includegraphics{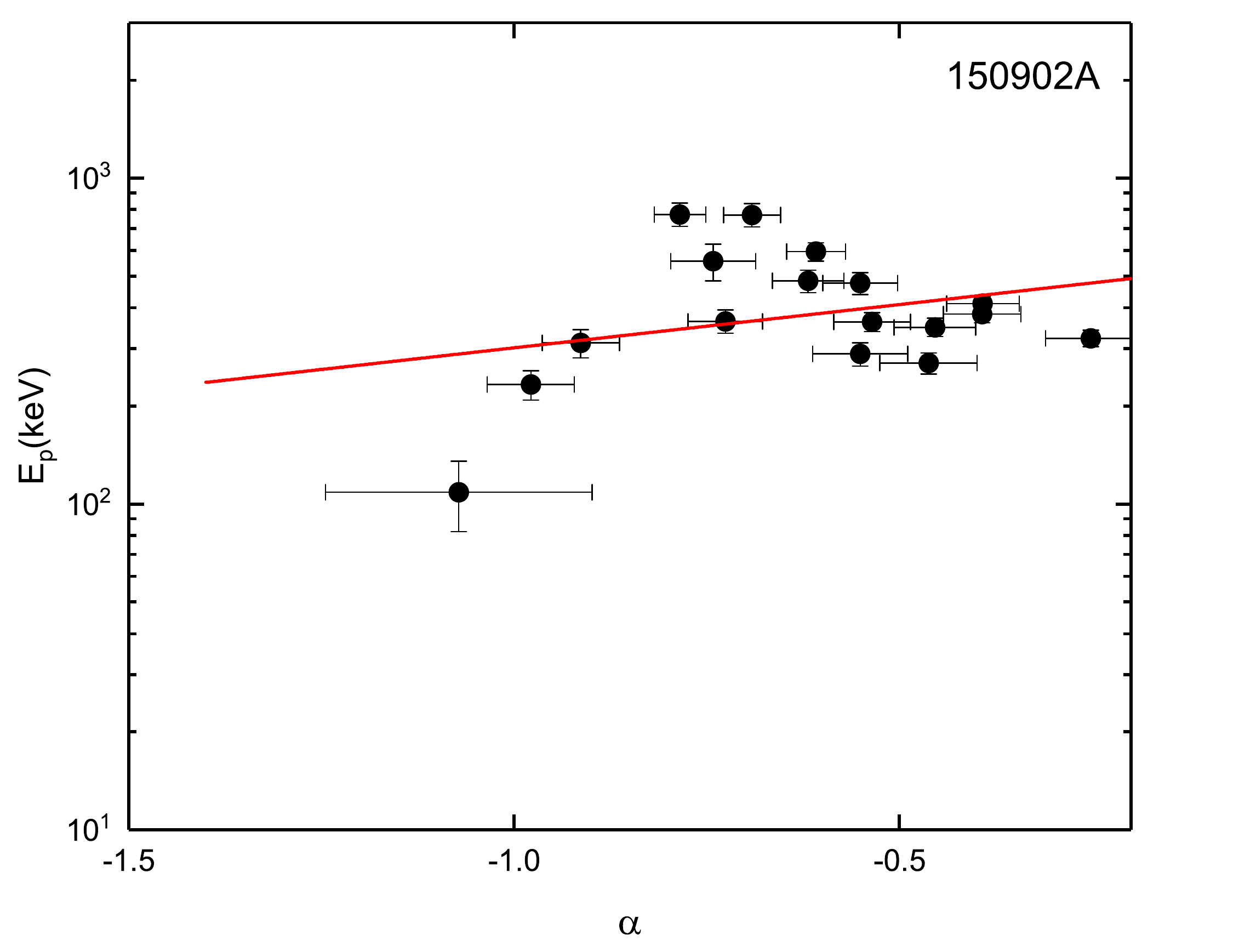}}
\resizebox{4cm}{!}{\includegraphics{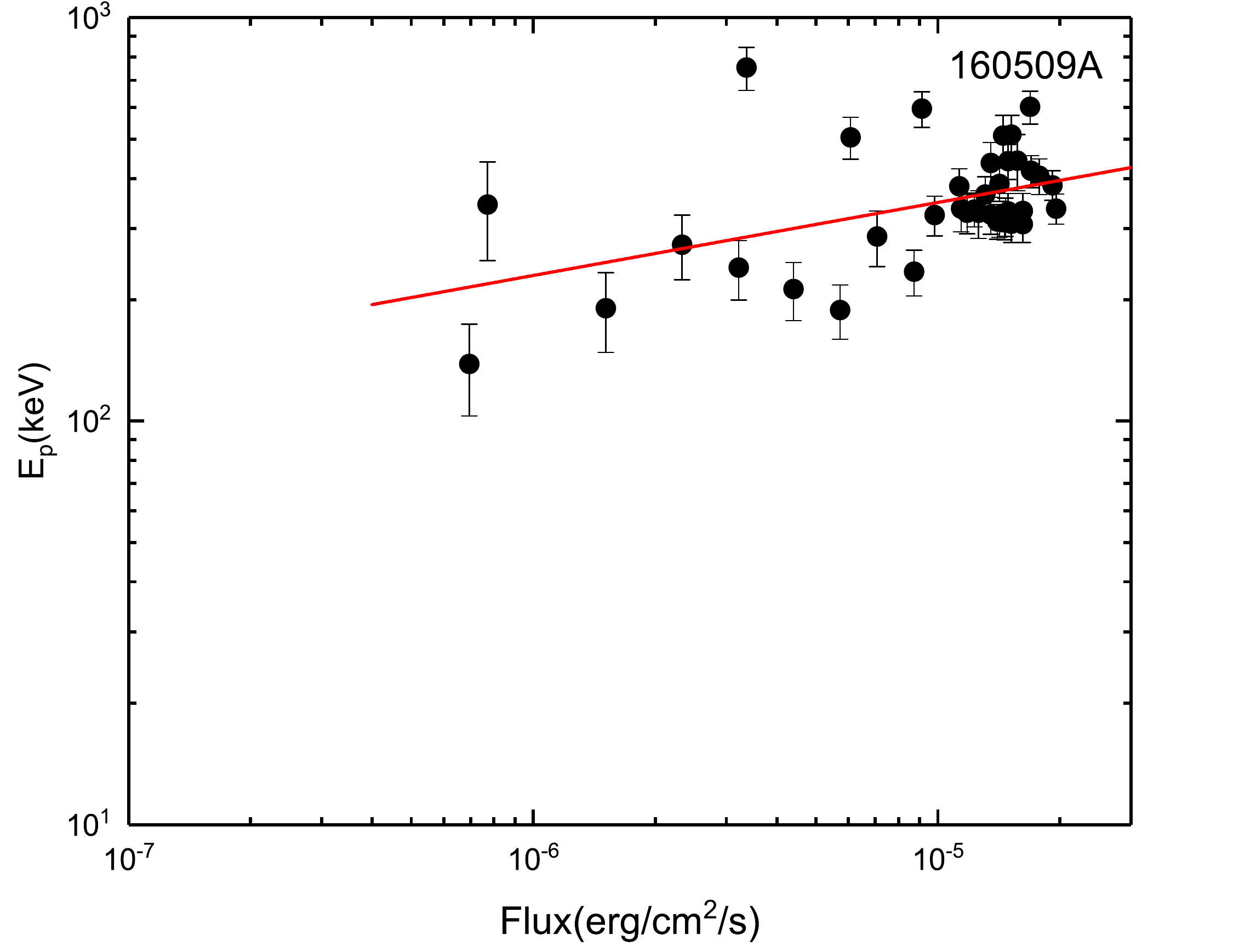}}
\resizebox{4cm}{!}{\includegraphics{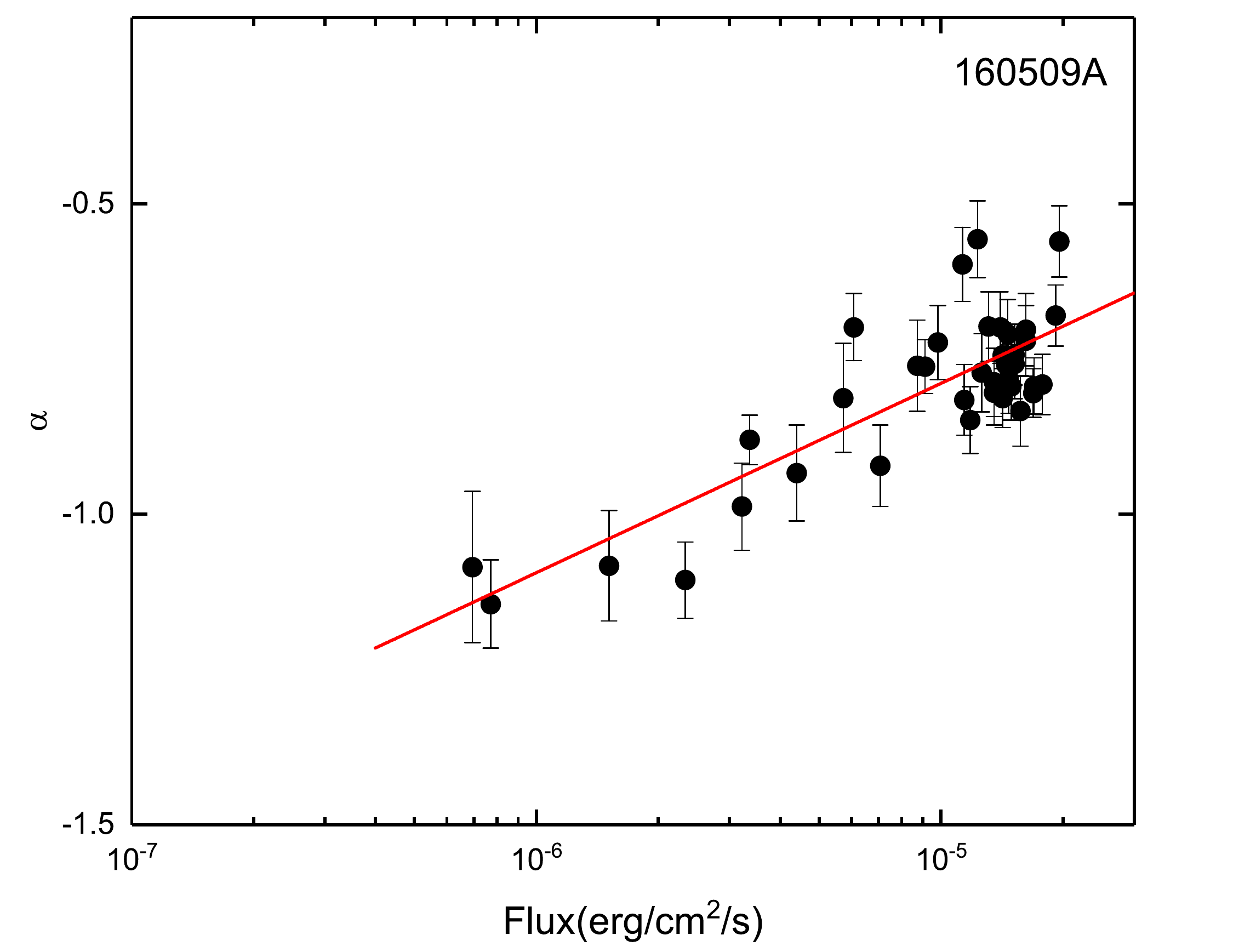}}
\resizebox{4cm}{!}{\includegraphics{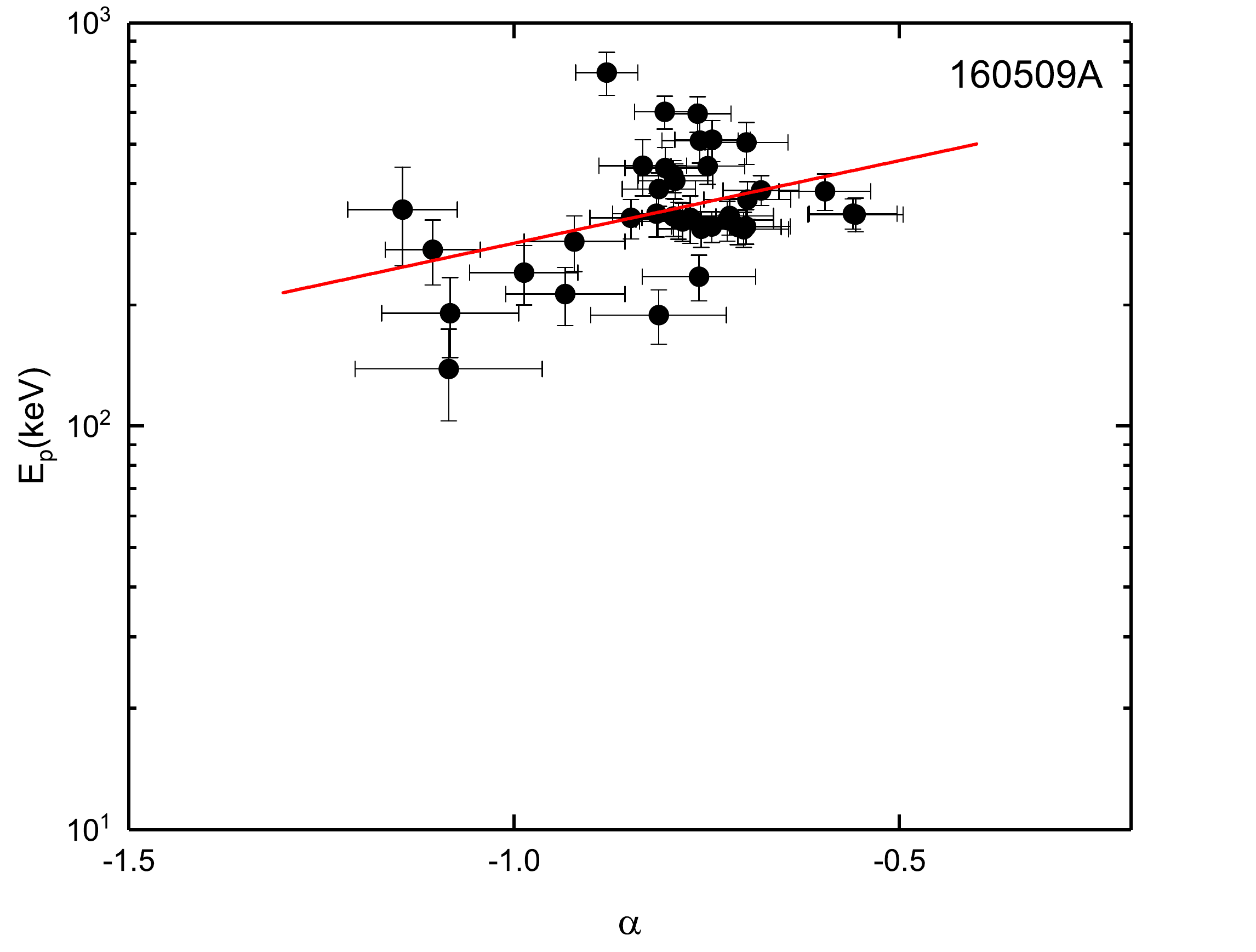}}
\resizebox{4cm}{!}{\includegraphics{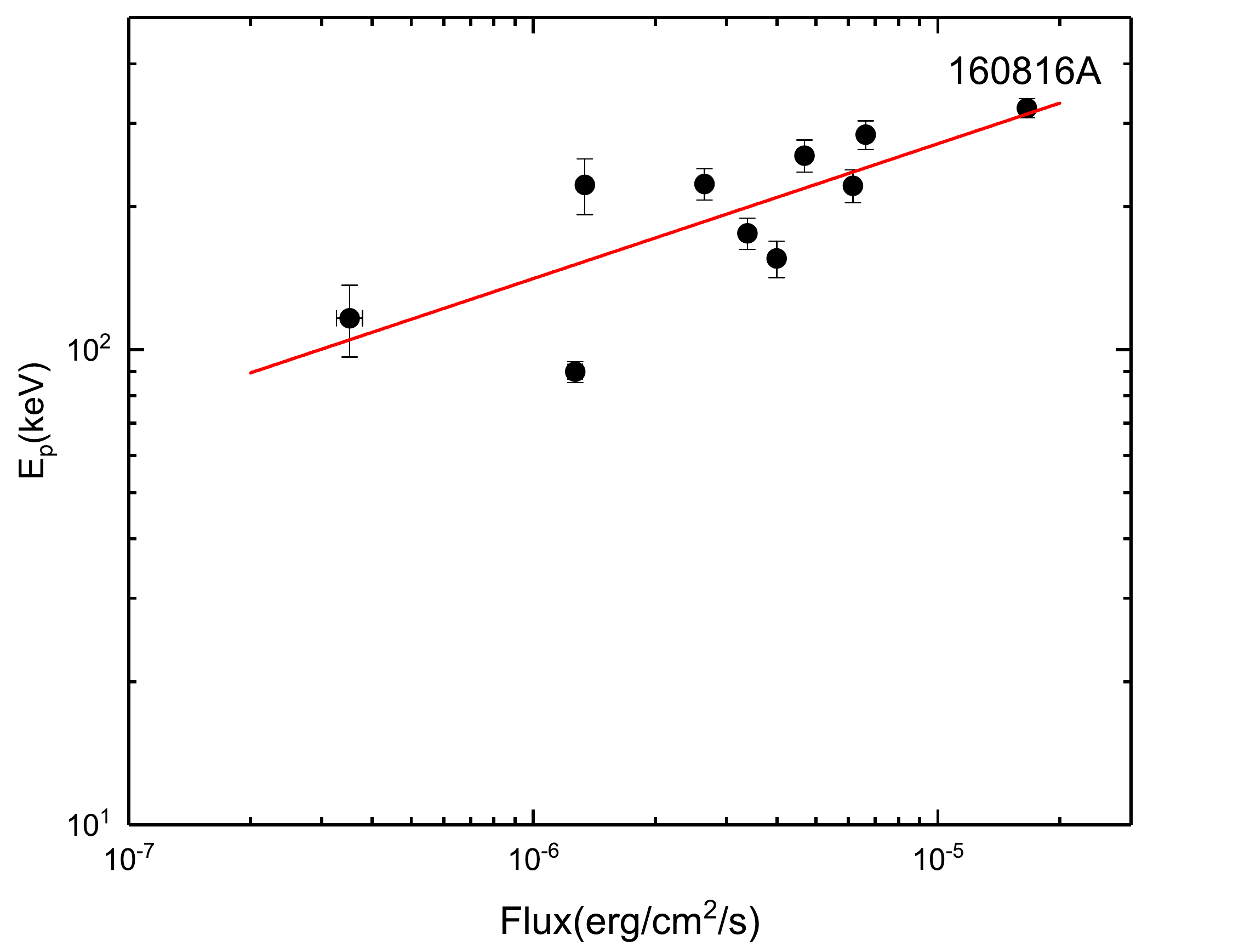}}
\resizebox{4cm}{!}{\includegraphics{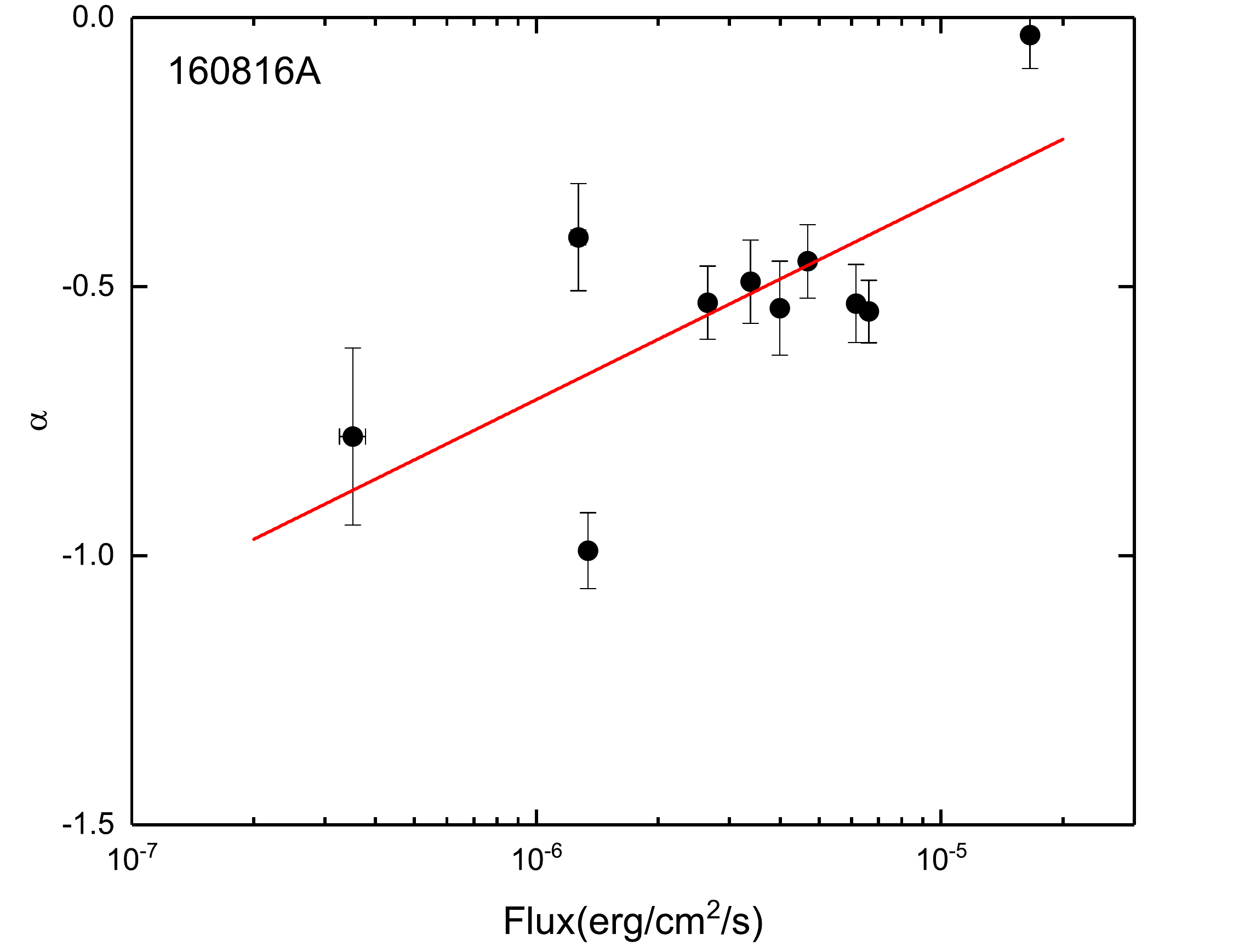}}
\caption{\it-continued}
\end{figure}

\addtocounter{figure}{-1}
\begin{figure} 
\centering
\resizebox{4cm}{!}{\includegraphics{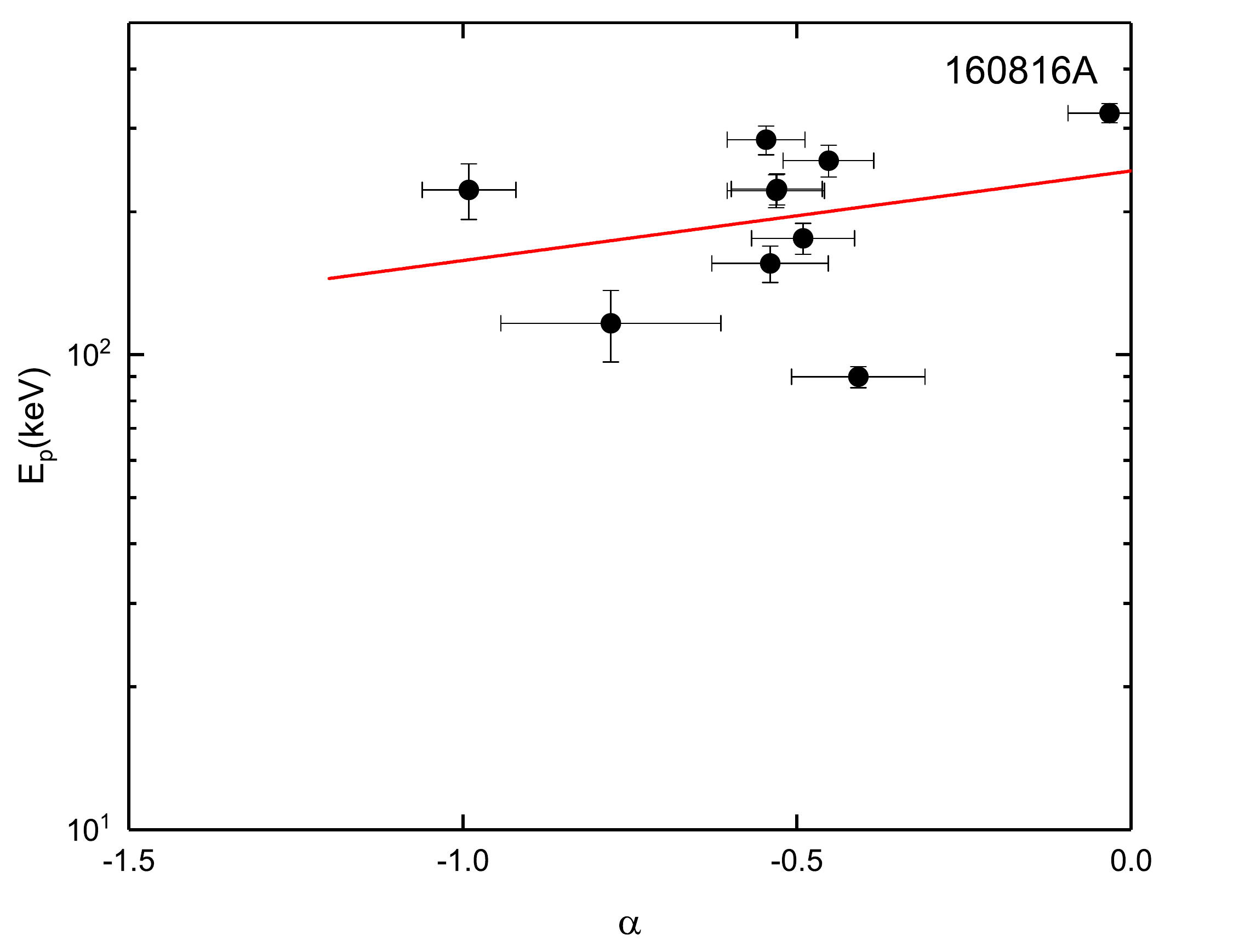}}
\resizebox{4cm}{!}{\includegraphics{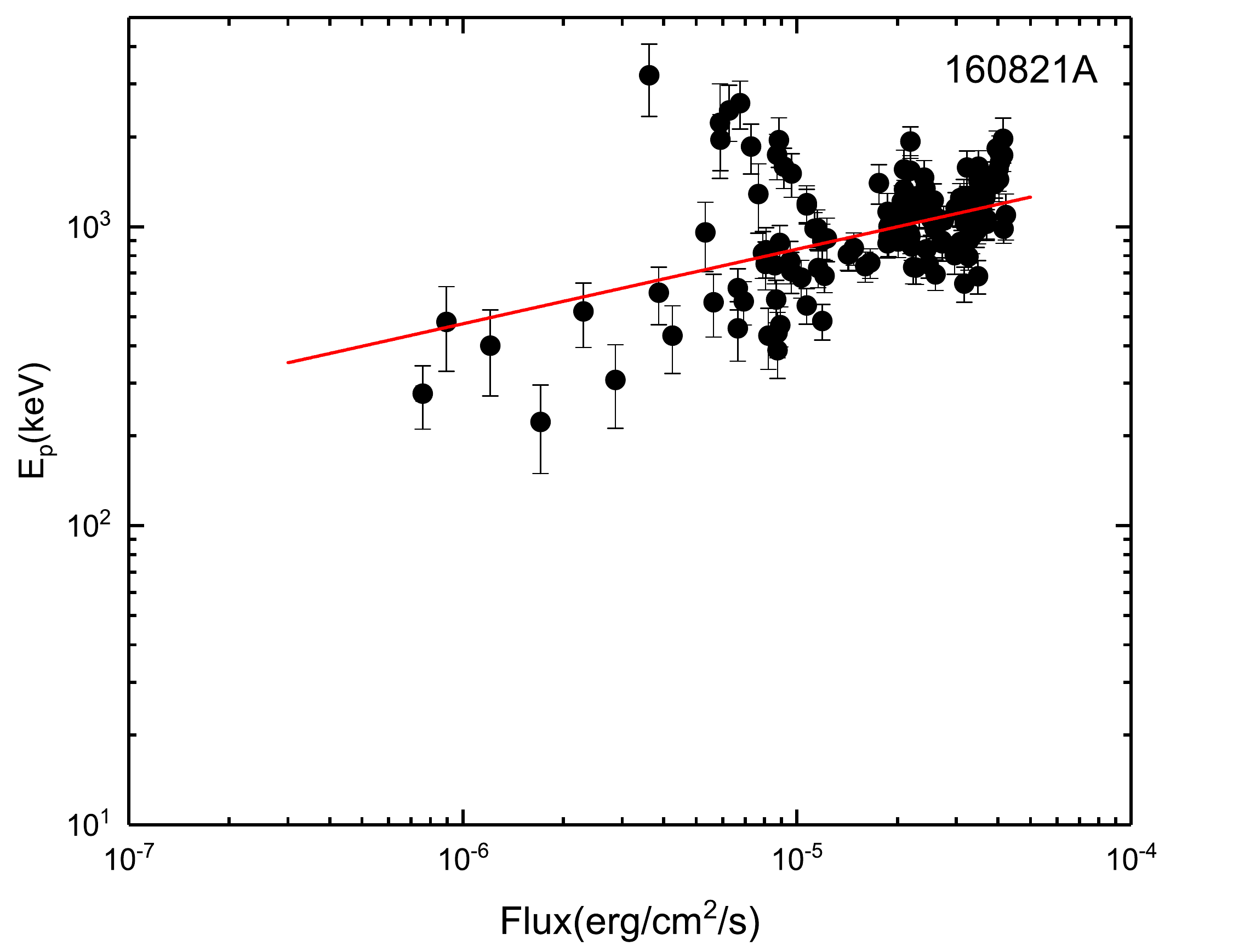}}
\resizebox{4cm}{!}{\includegraphics{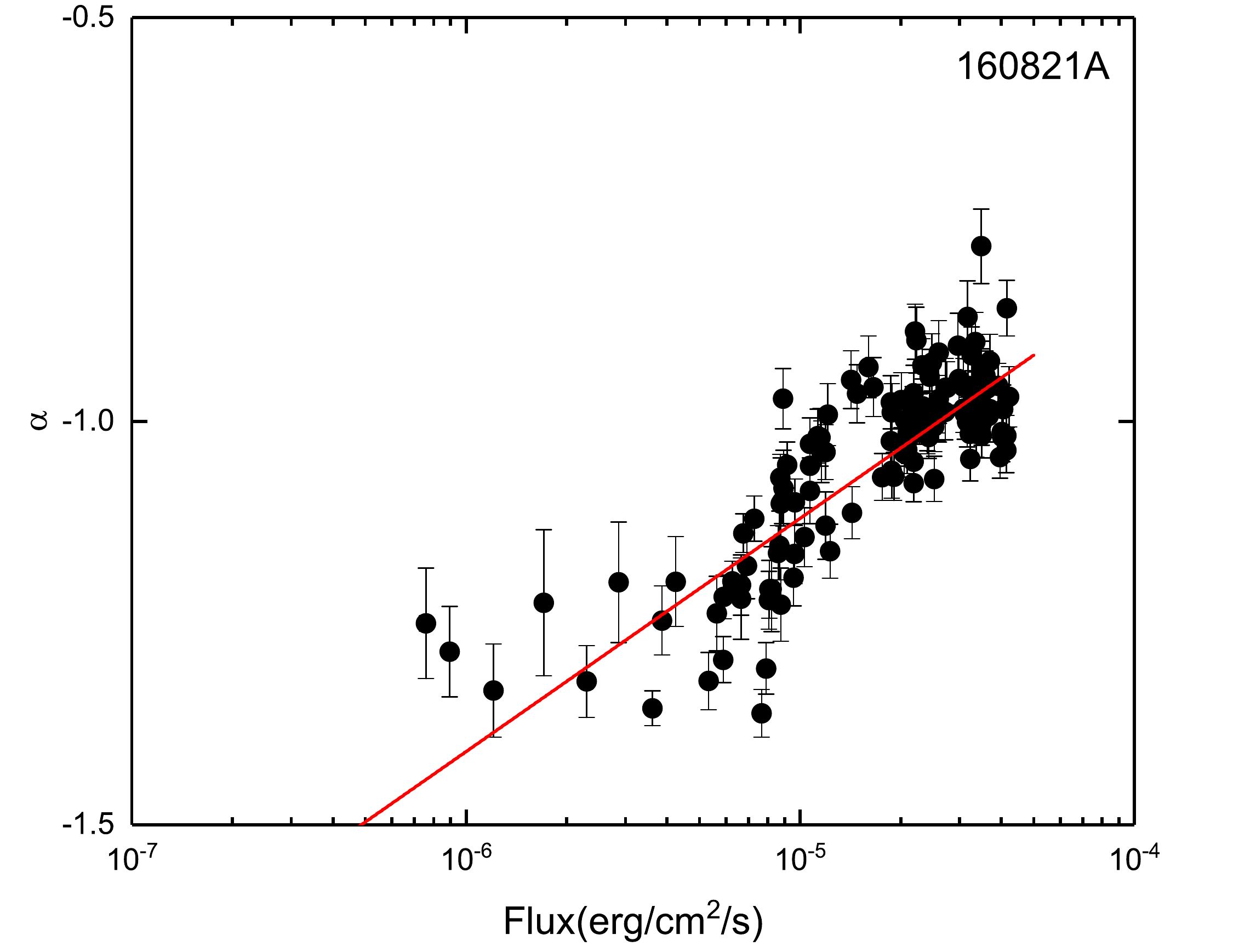}}
\resizebox{4cm}{!}{\includegraphics{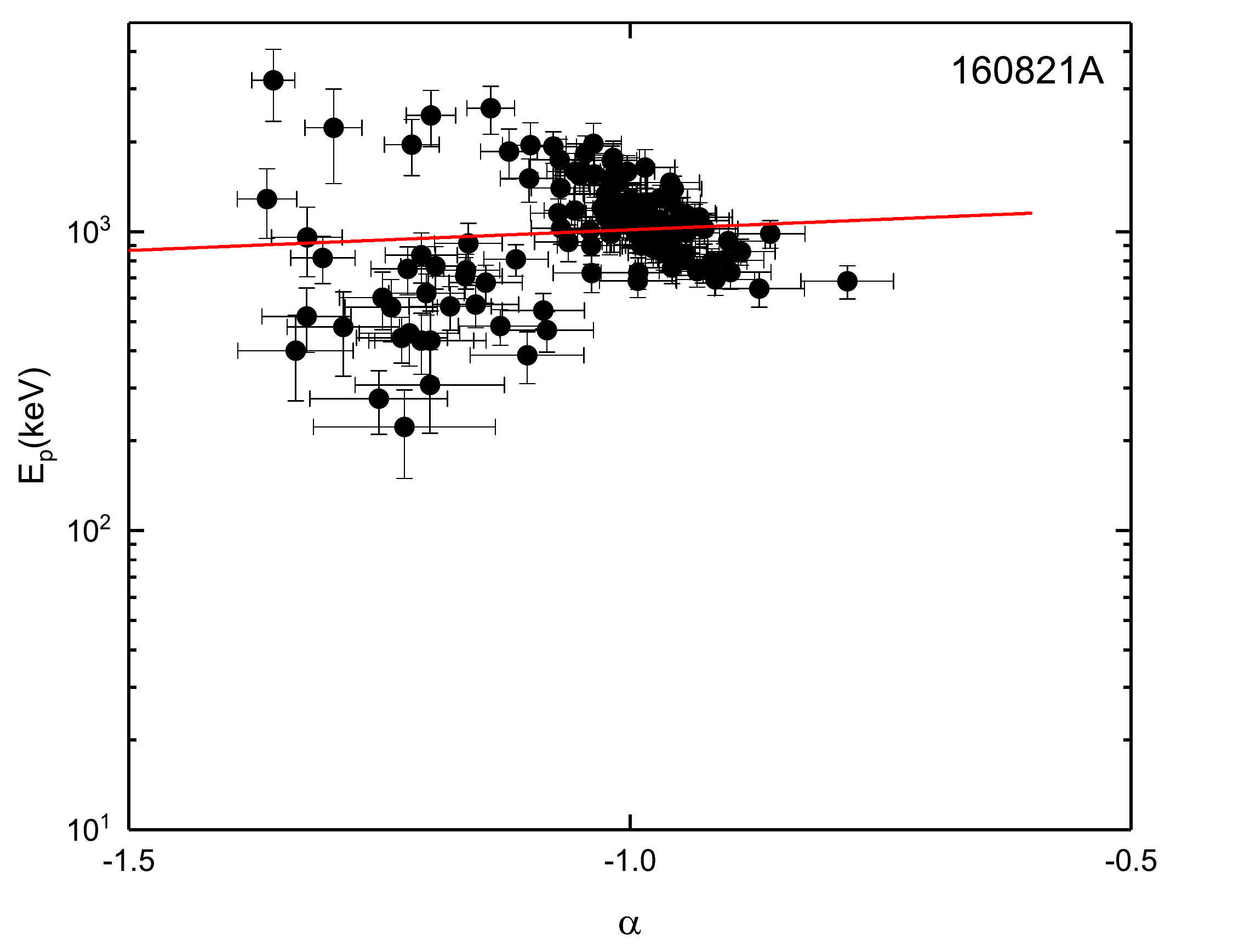}}
\resizebox{4cm}{!}{\includegraphics{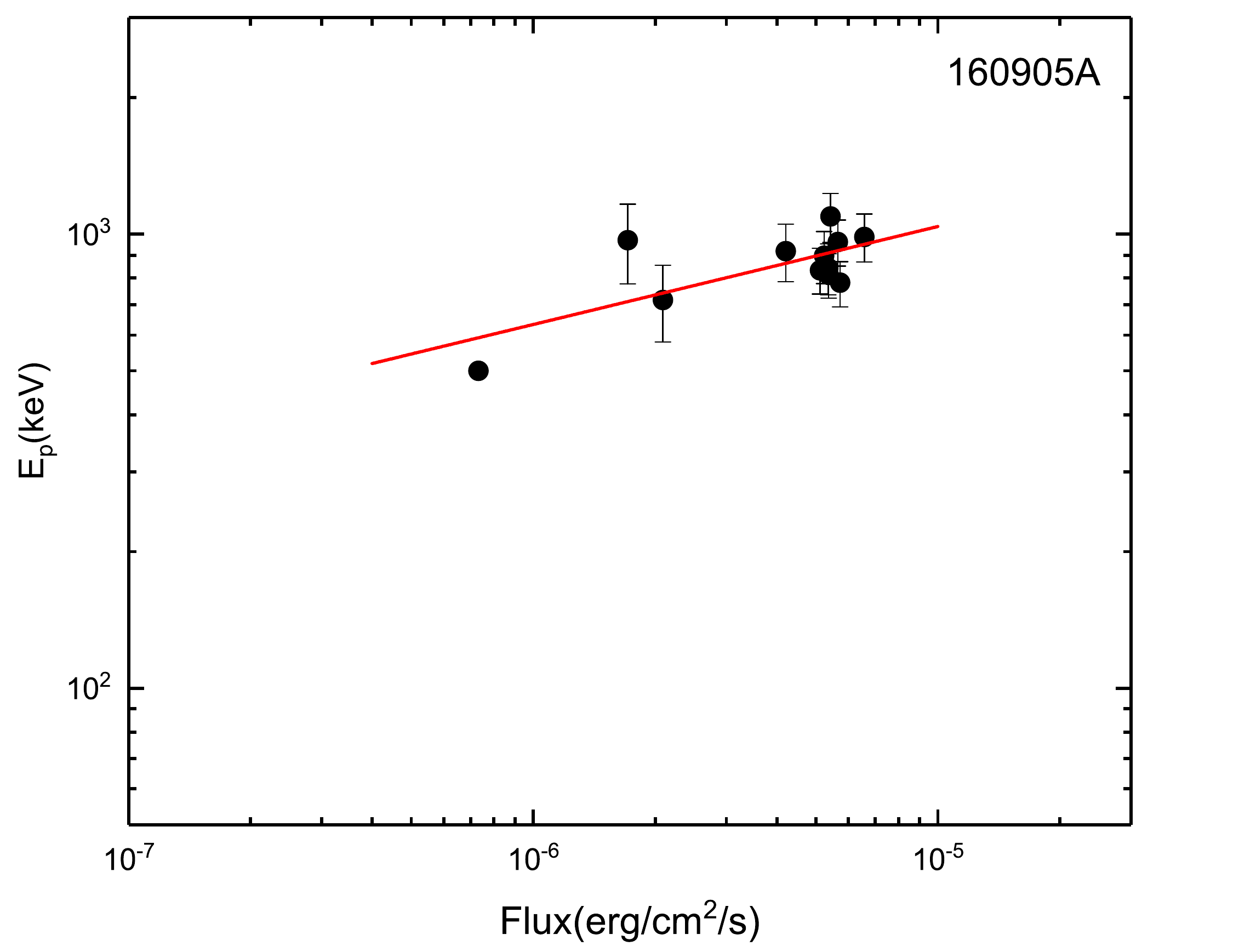}}
\resizebox{4cm}{!}{\includegraphics{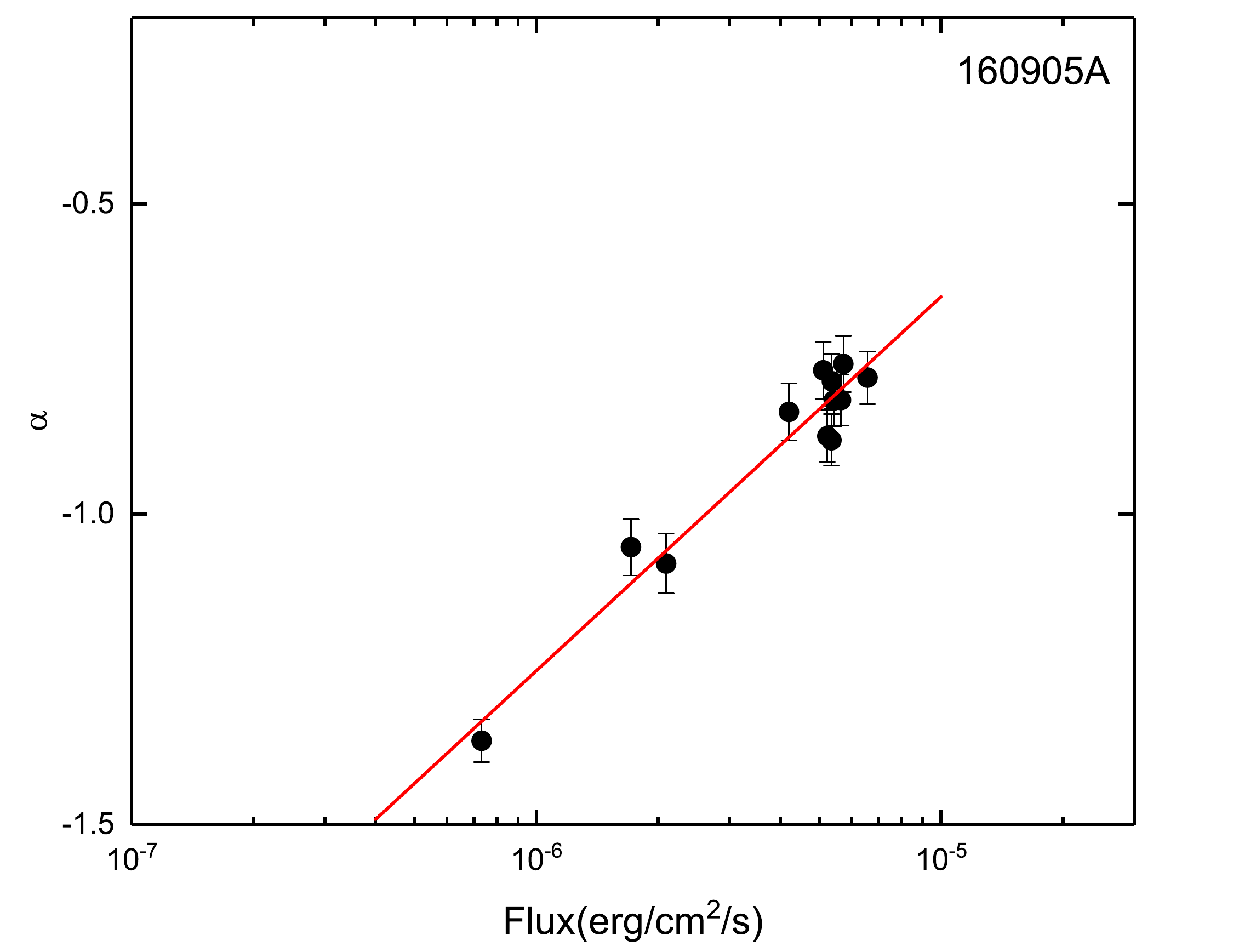}}
\resizebox{4cm}{!}{\includegraphics{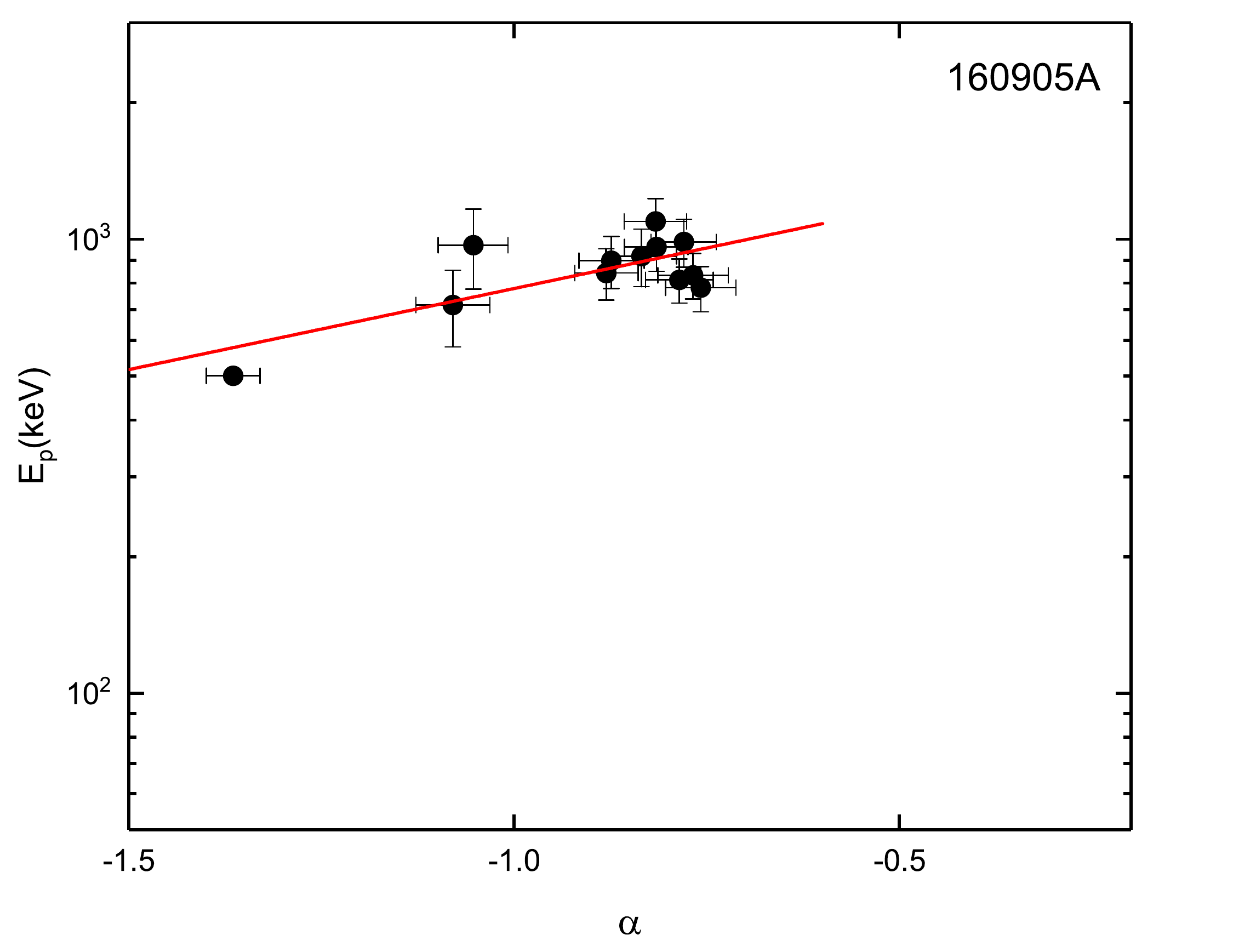}}
\resizebox{4cm}{!}{\includegraphics{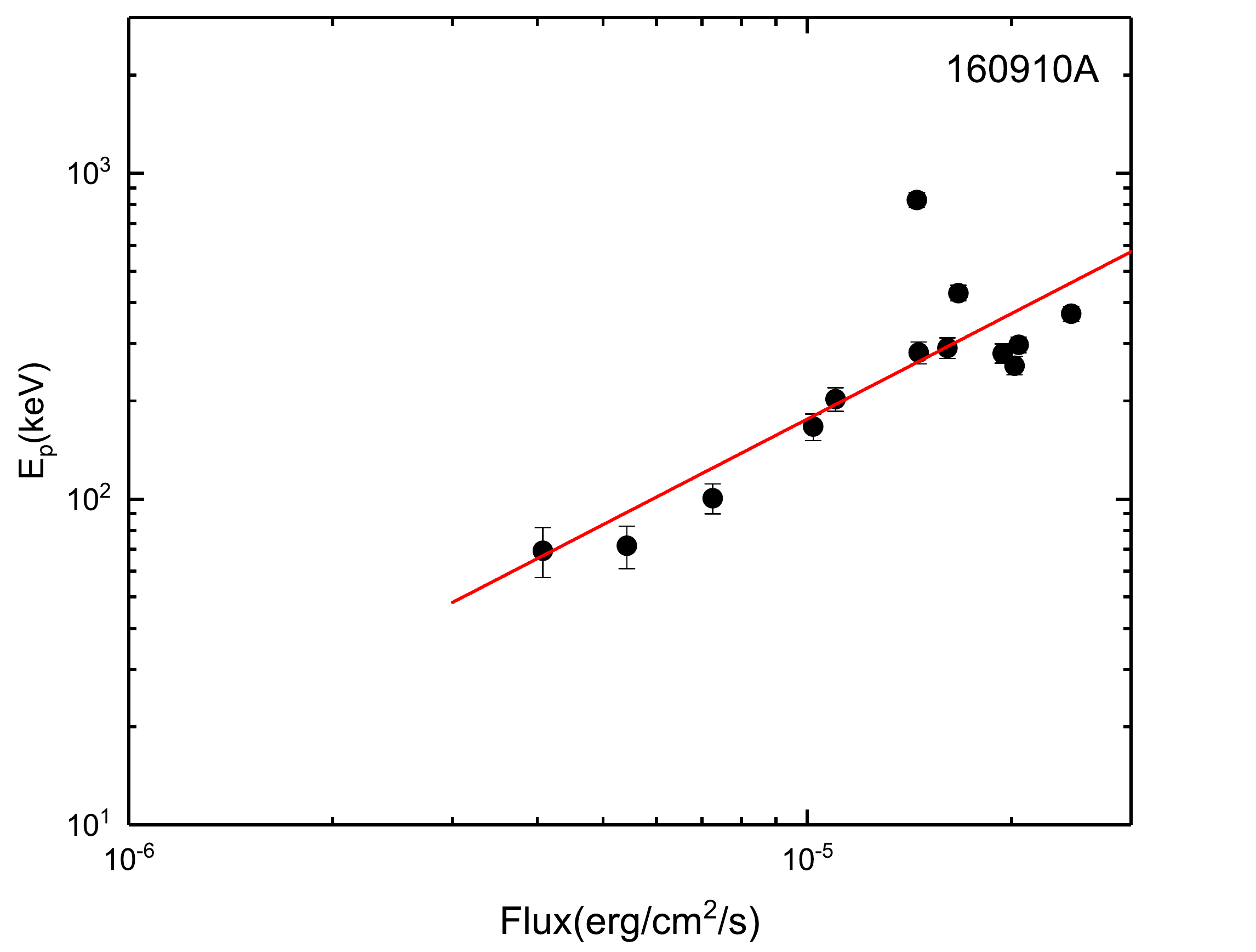}}
\resizebox{4cm}{!}{\includegraphics{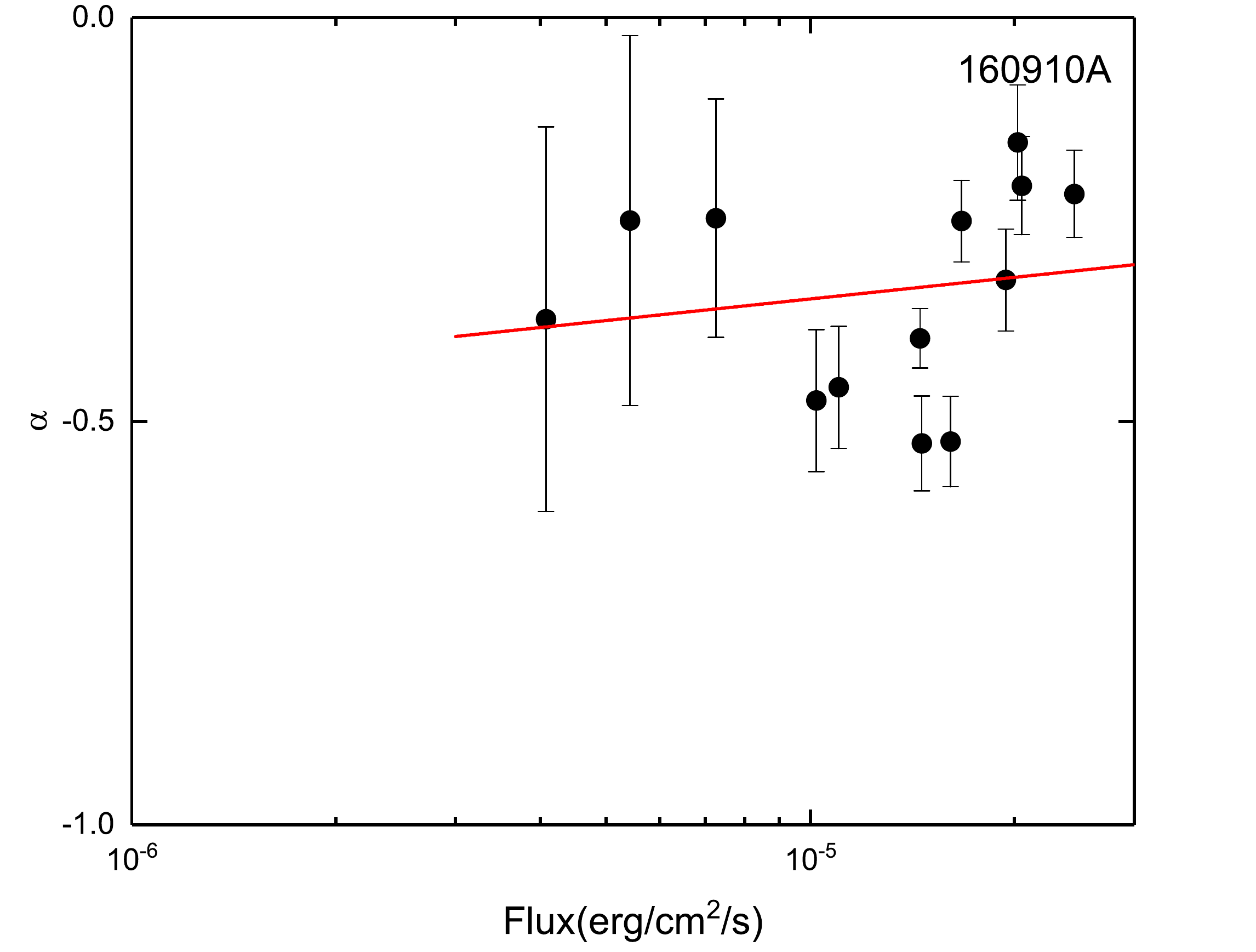}}
\resizebox{4cm}{!}{\includegraphics{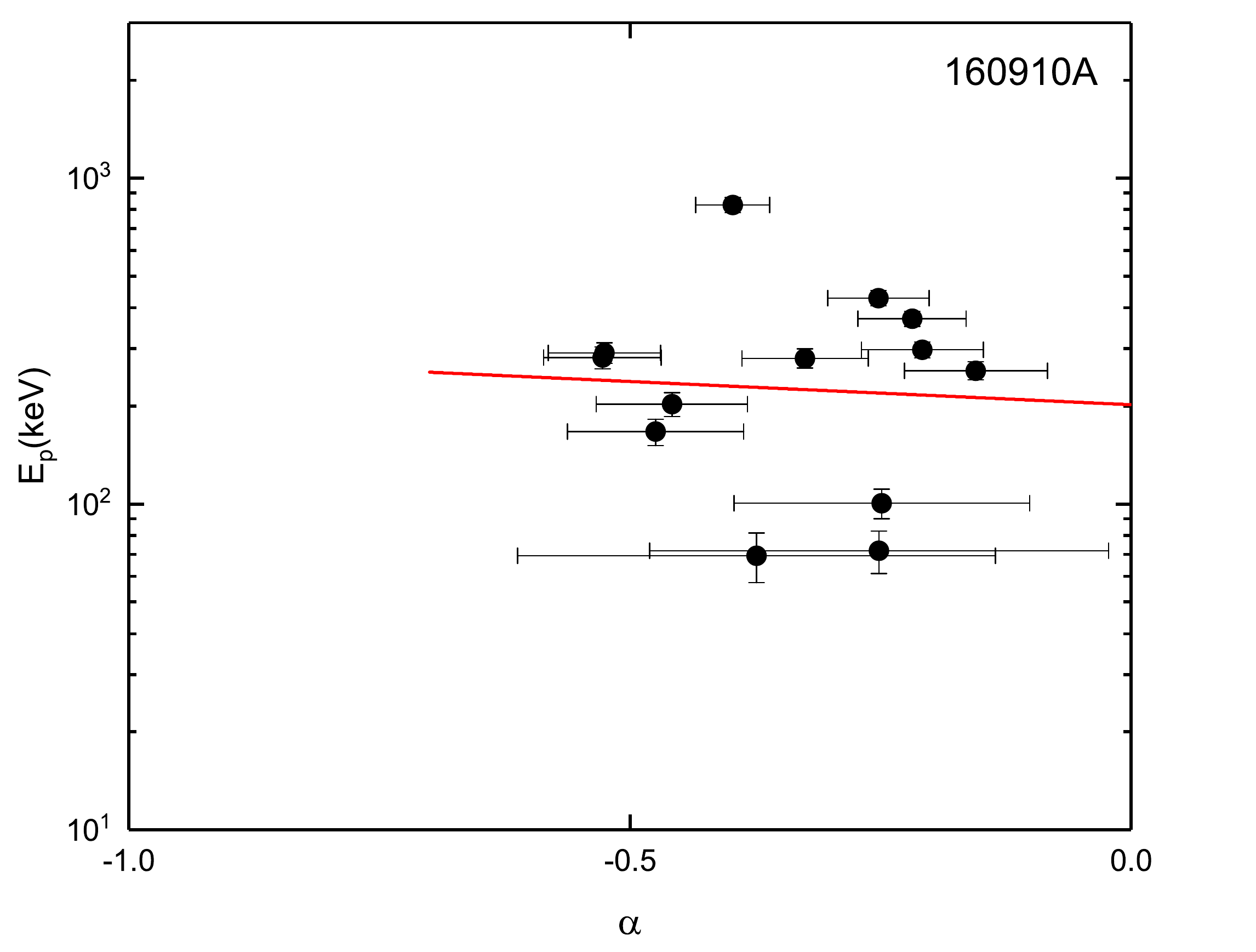}}
\resizebox{4cm}{!}{\includegraphics{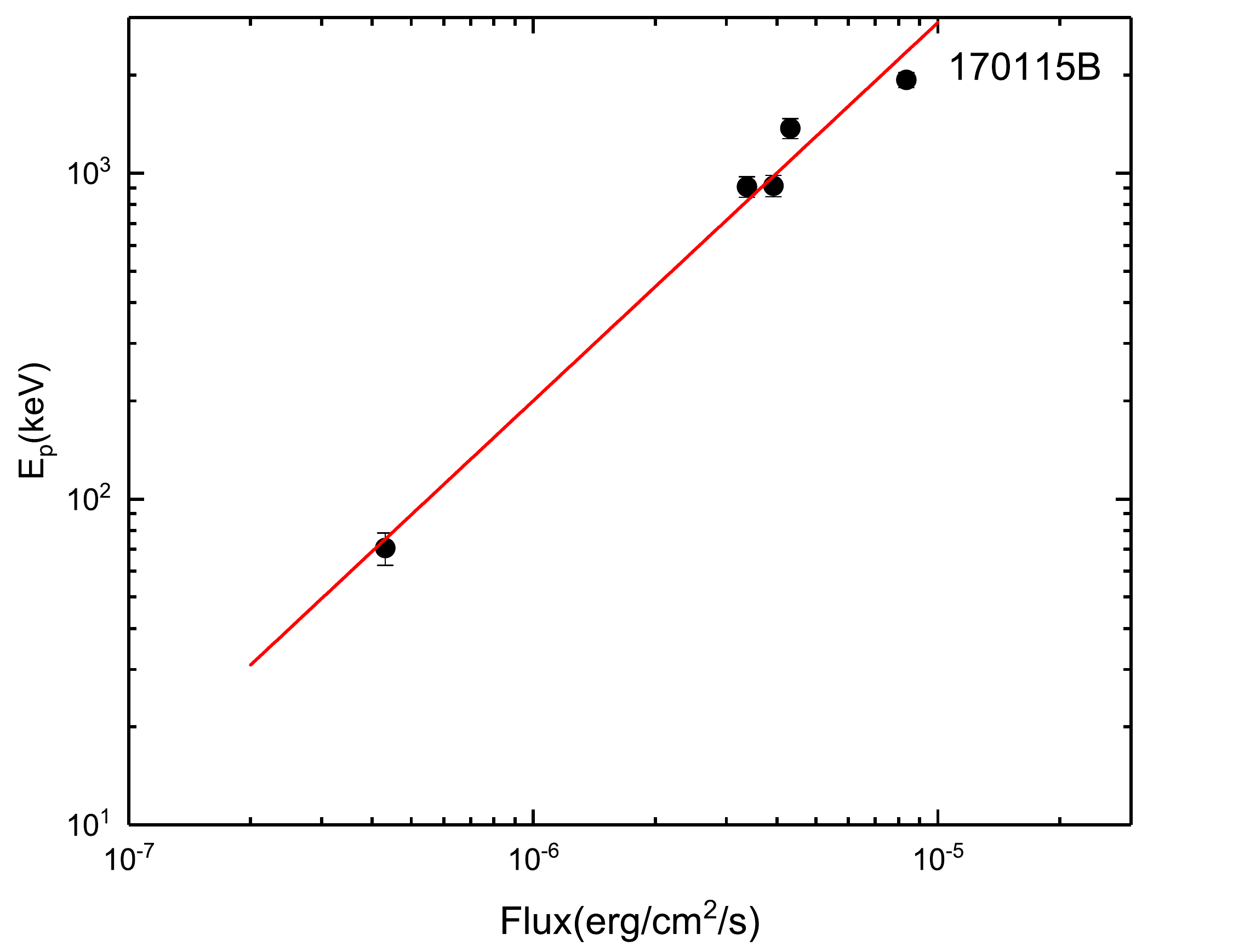}}
\resizebox{4cm}{!}{\includegraphics{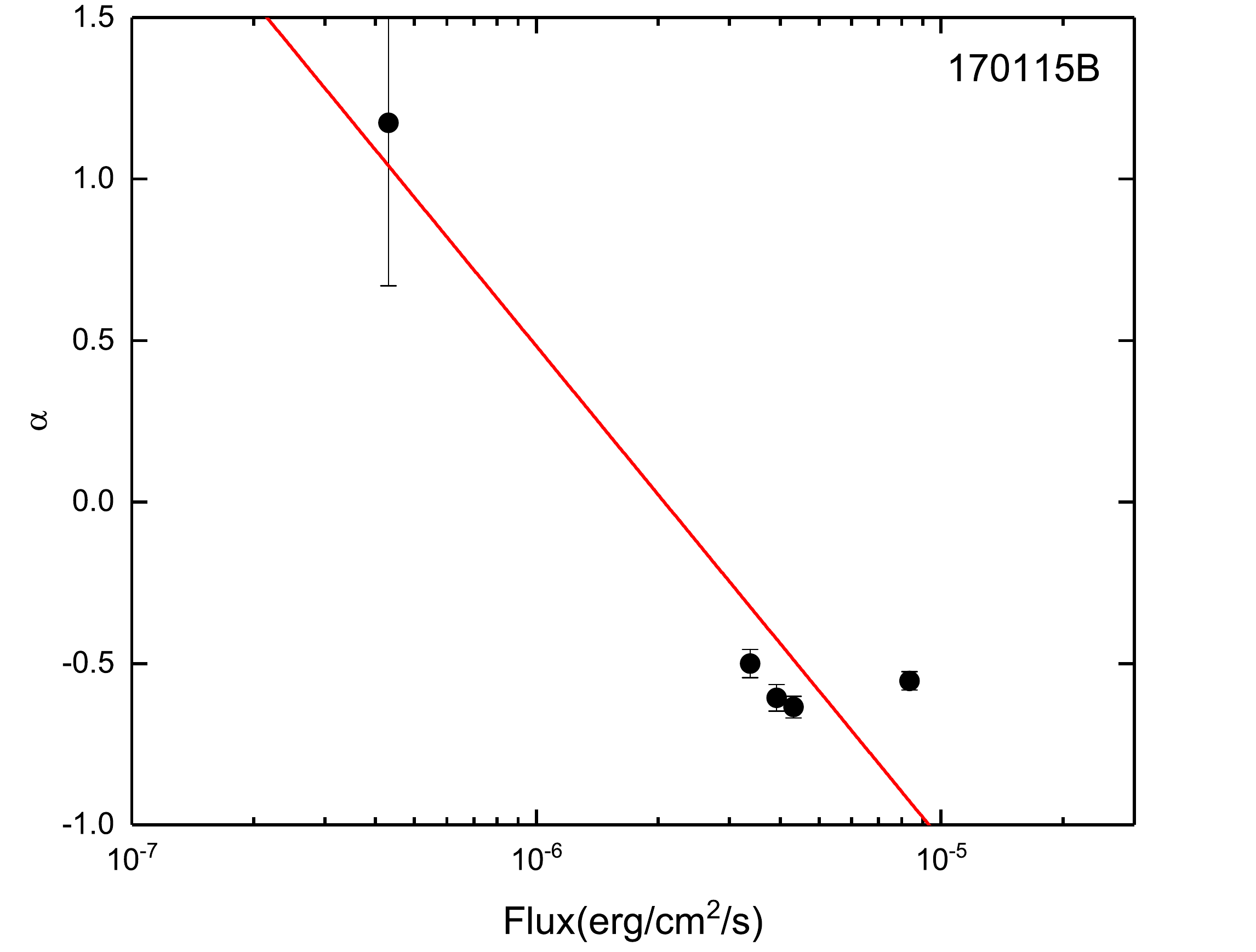}}
\resizebox{4cm}{!}{\includegraphics{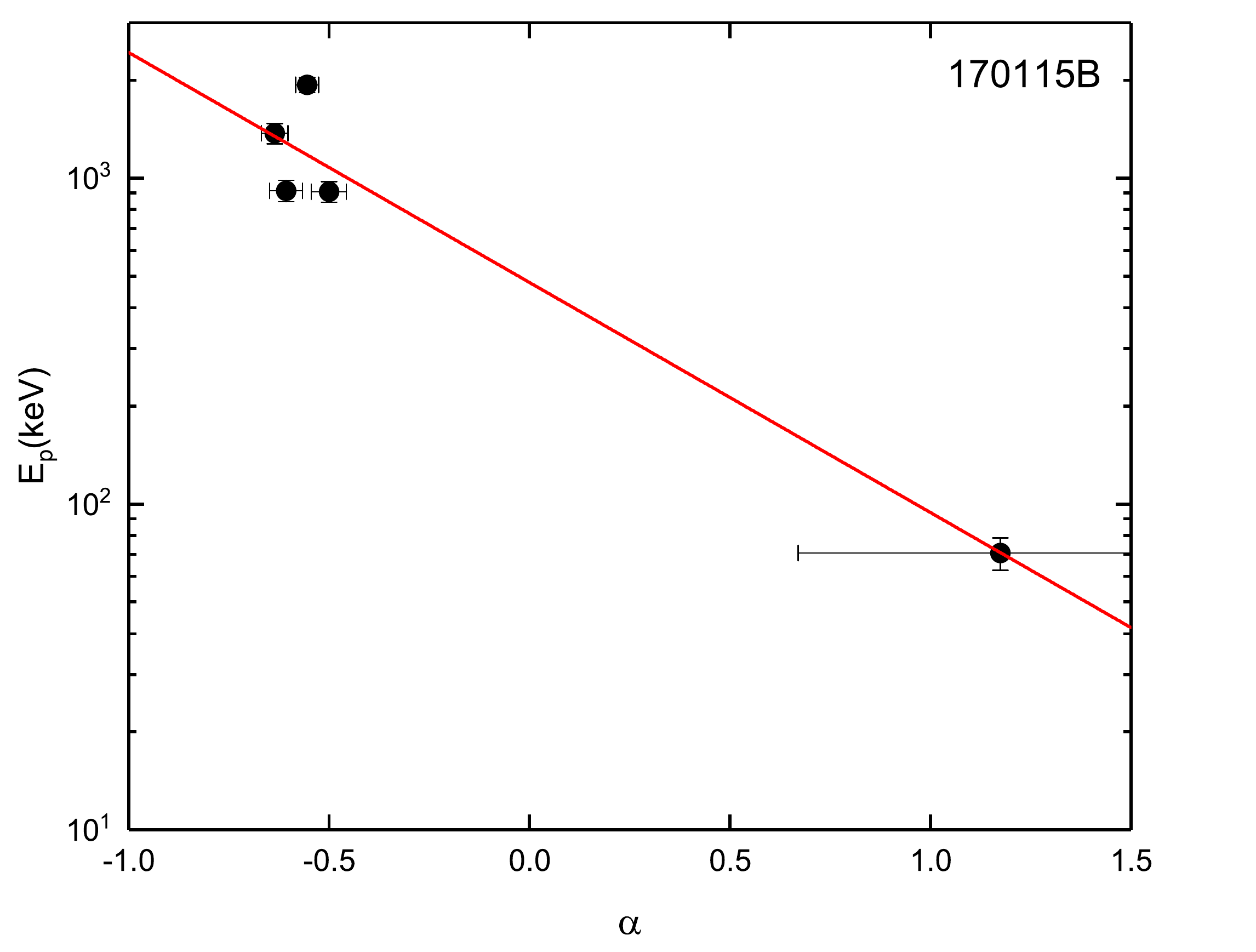}}
\resizebox{4cm}{!}{\includegraphics{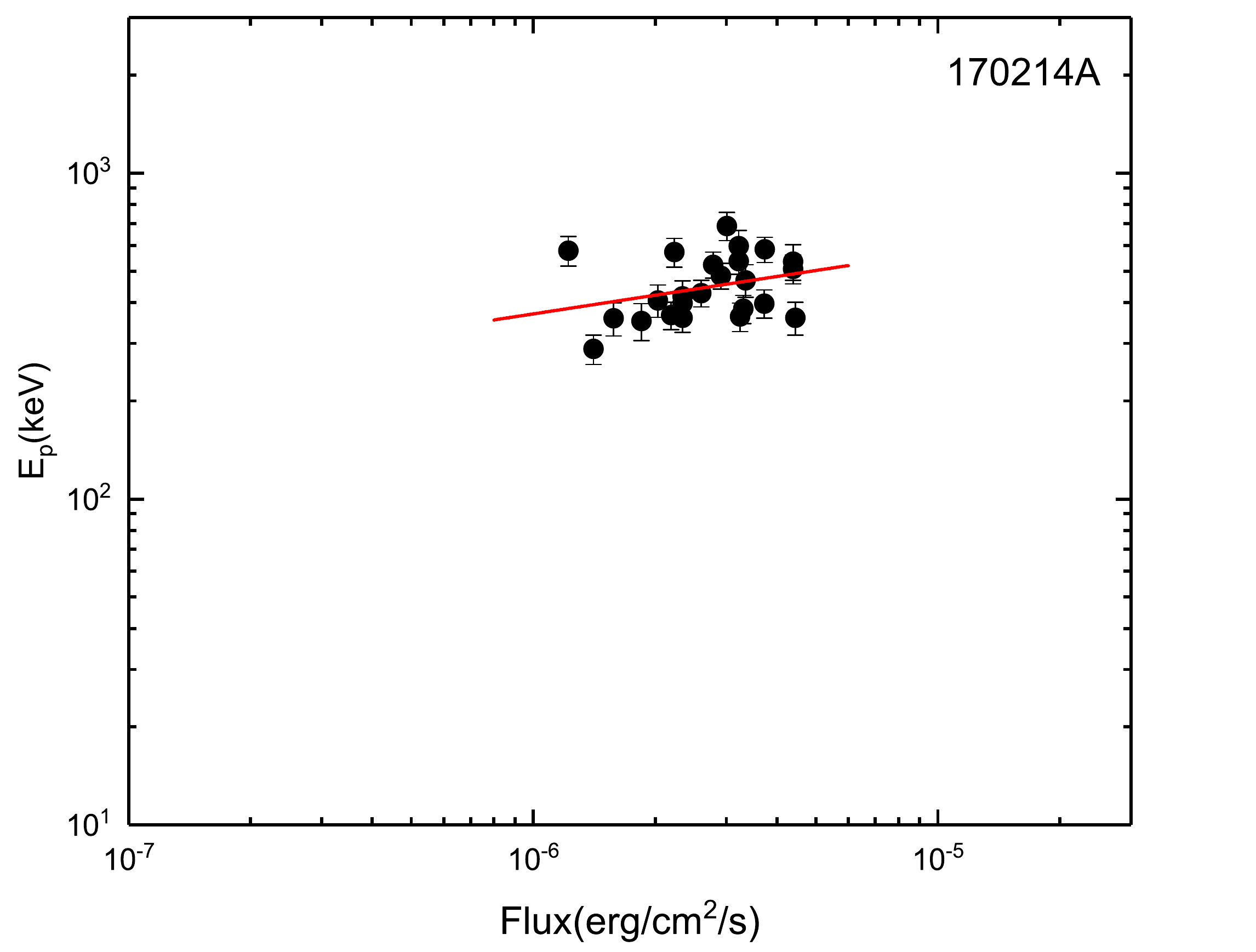}}
\resizebox{4cm}{!}{\includegraphics{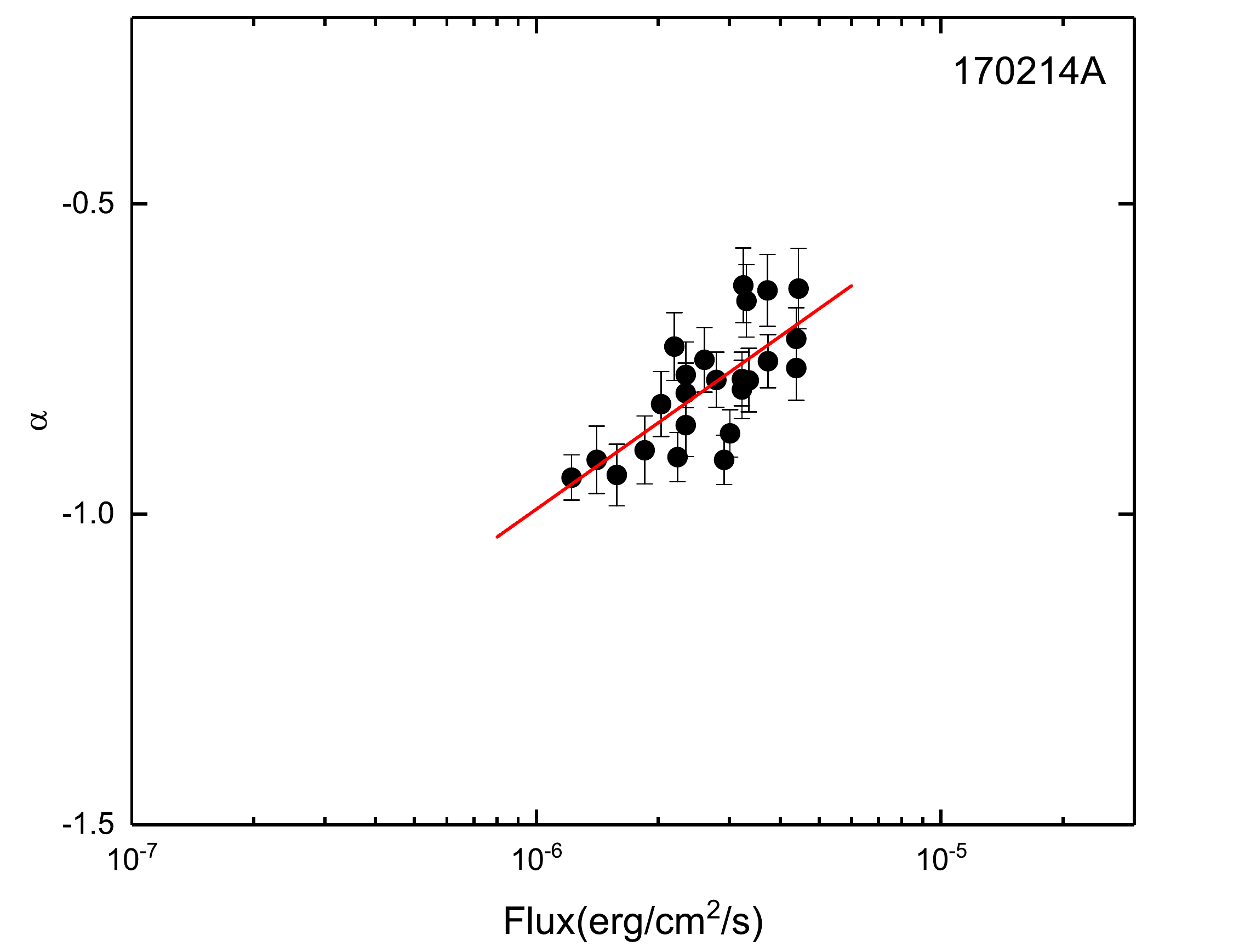}}
\resizebox{4cm}{!}{\includegraphics{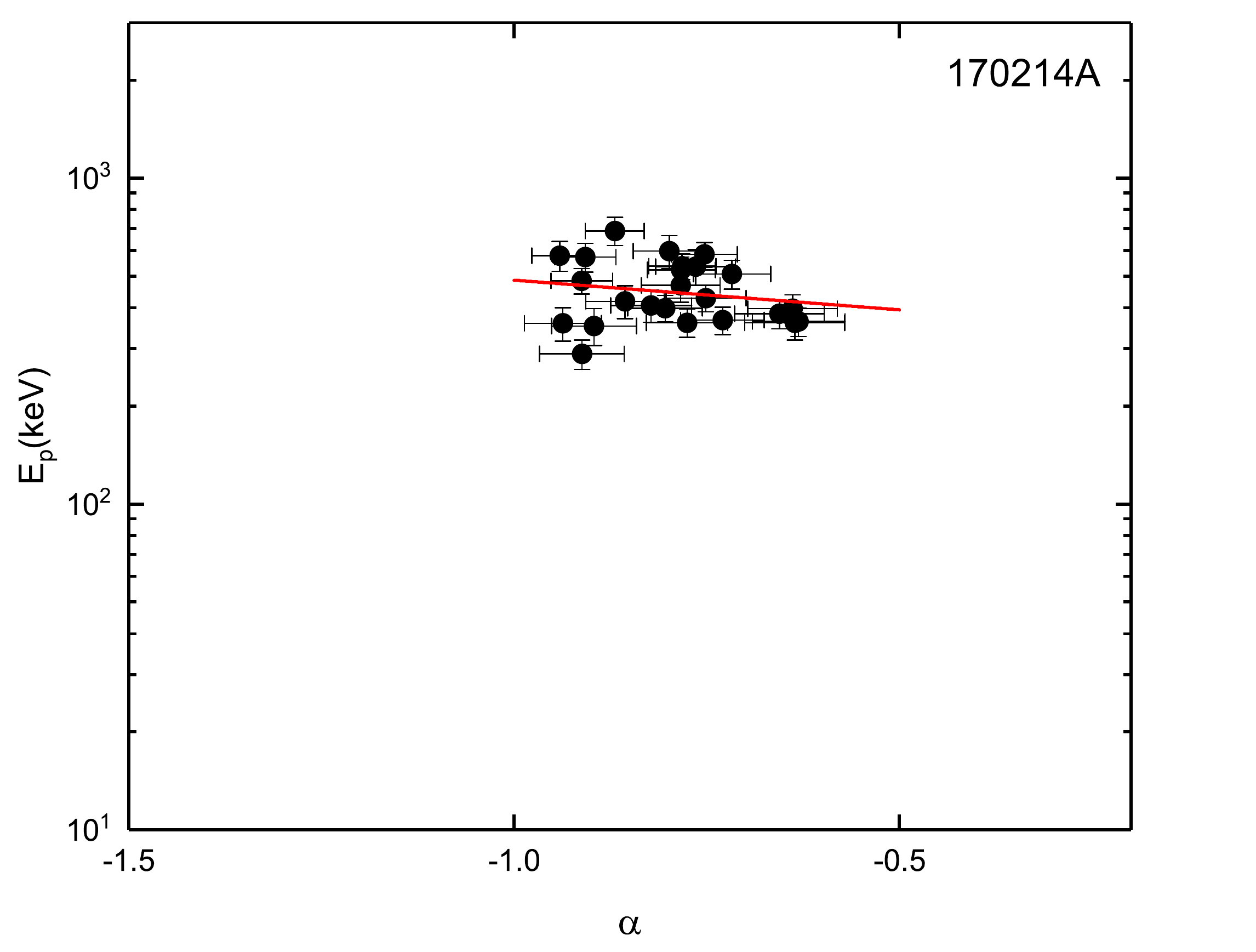}}

\resizebox{4cm}{!}{\includegraphics{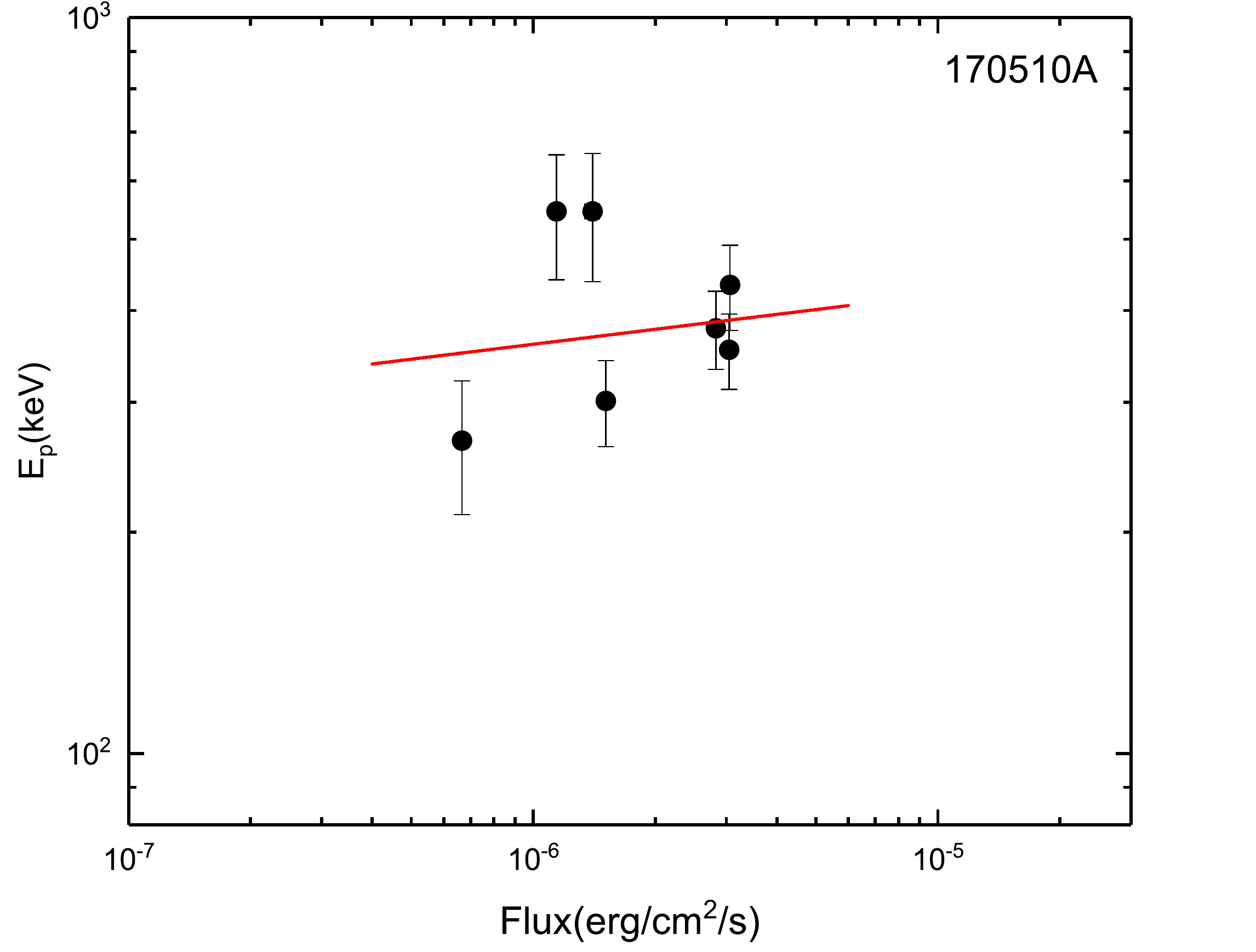}}
\resizebox{4cm}{!}{\includegraphics{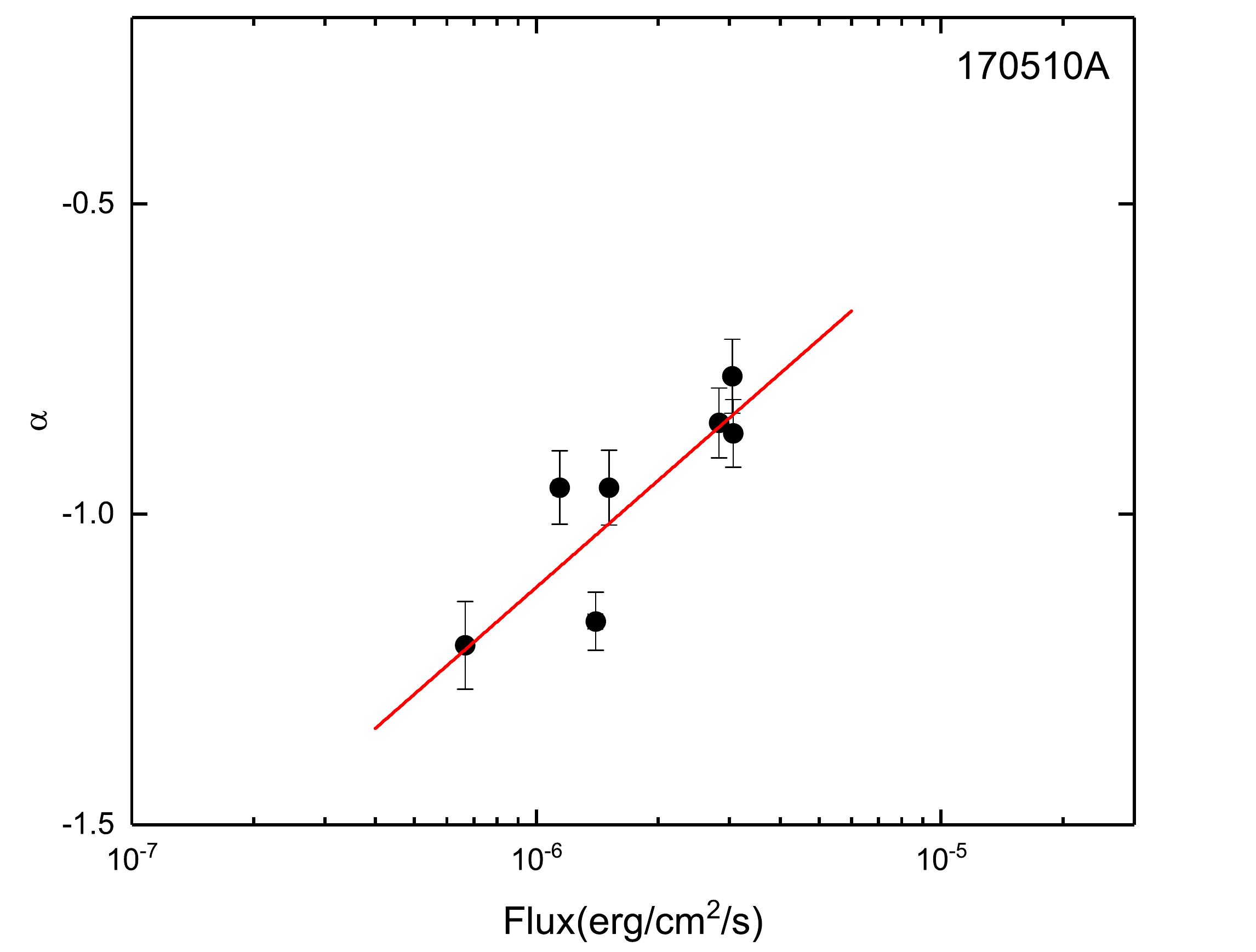}}
\resizebox{4cm}{!}{\includegraphics{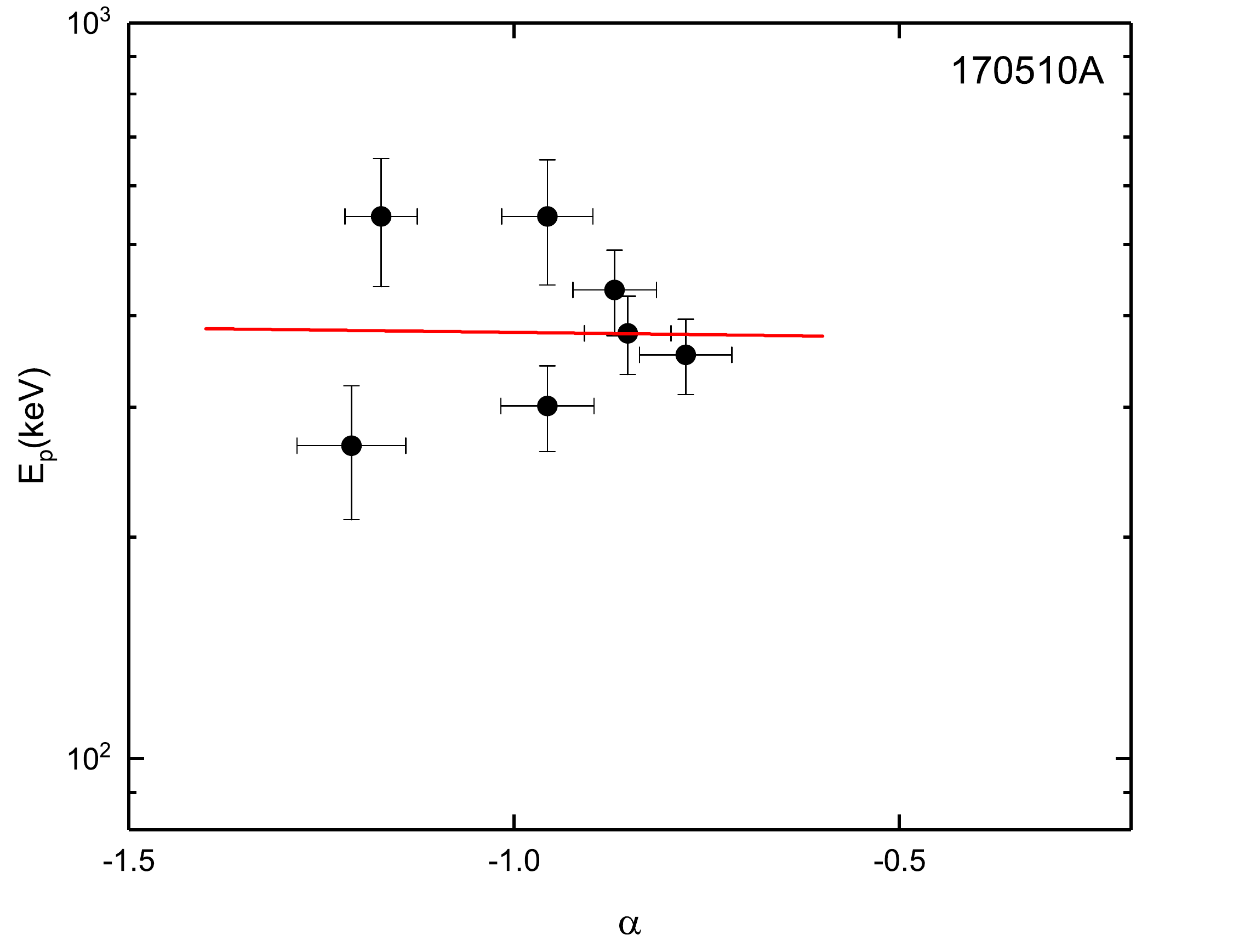}}
\resizebox{4cm}{!}{\includegraphics{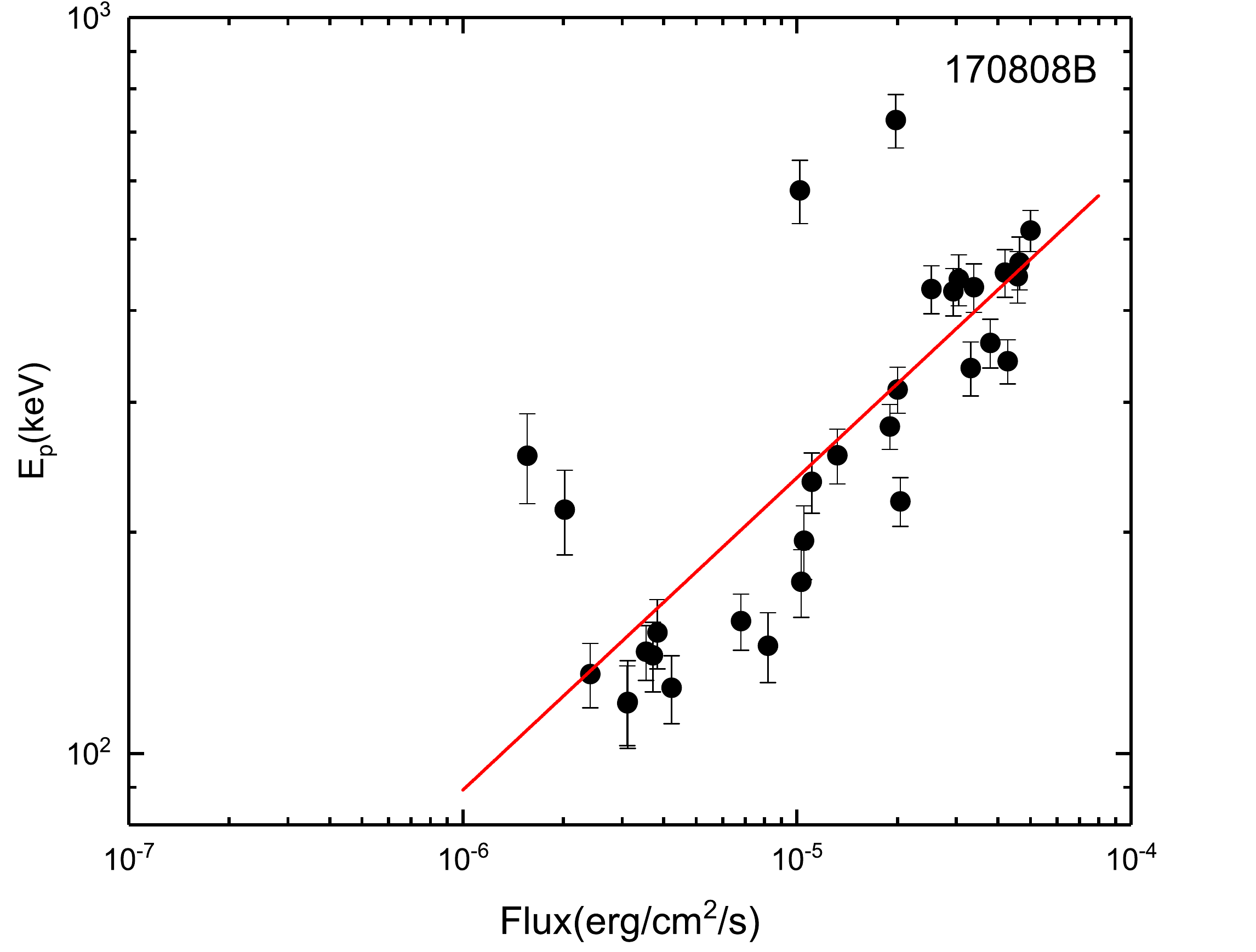}}
\resizebox{4cm}{!}{\includegraphics{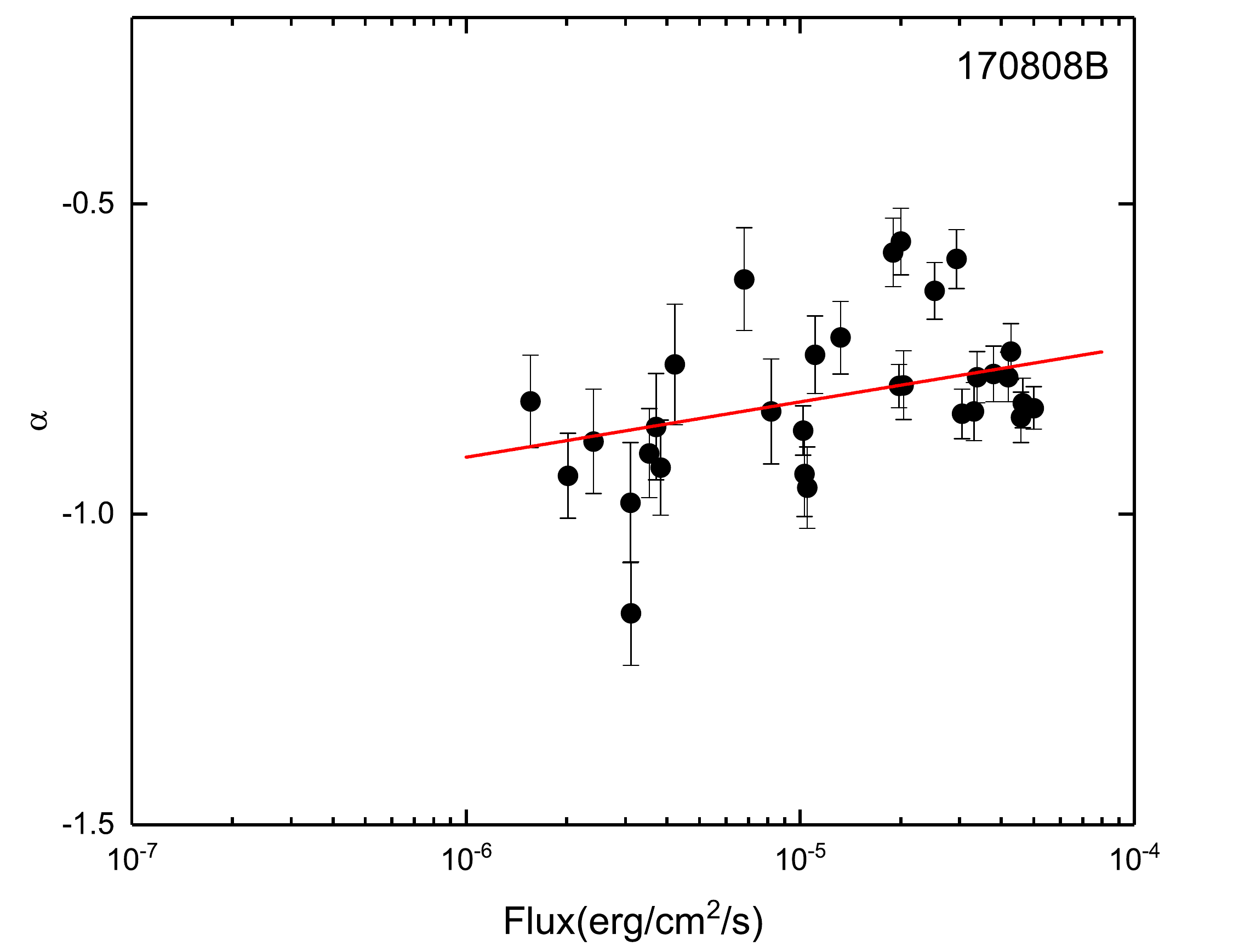}}
\resizebox{4cm}{!}{\includegraphics{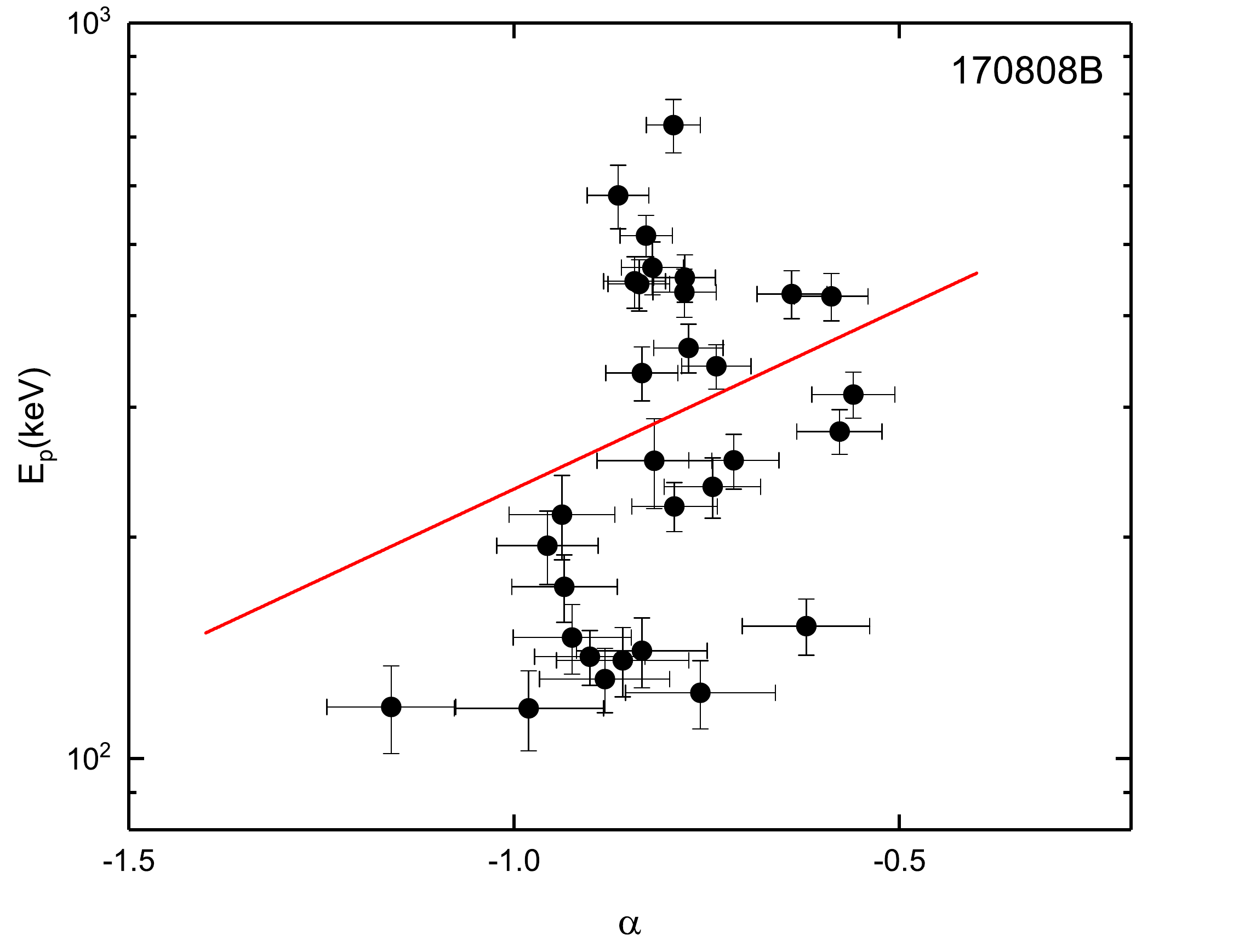}}
\resizebox{4cm}{!}{\includegraphics{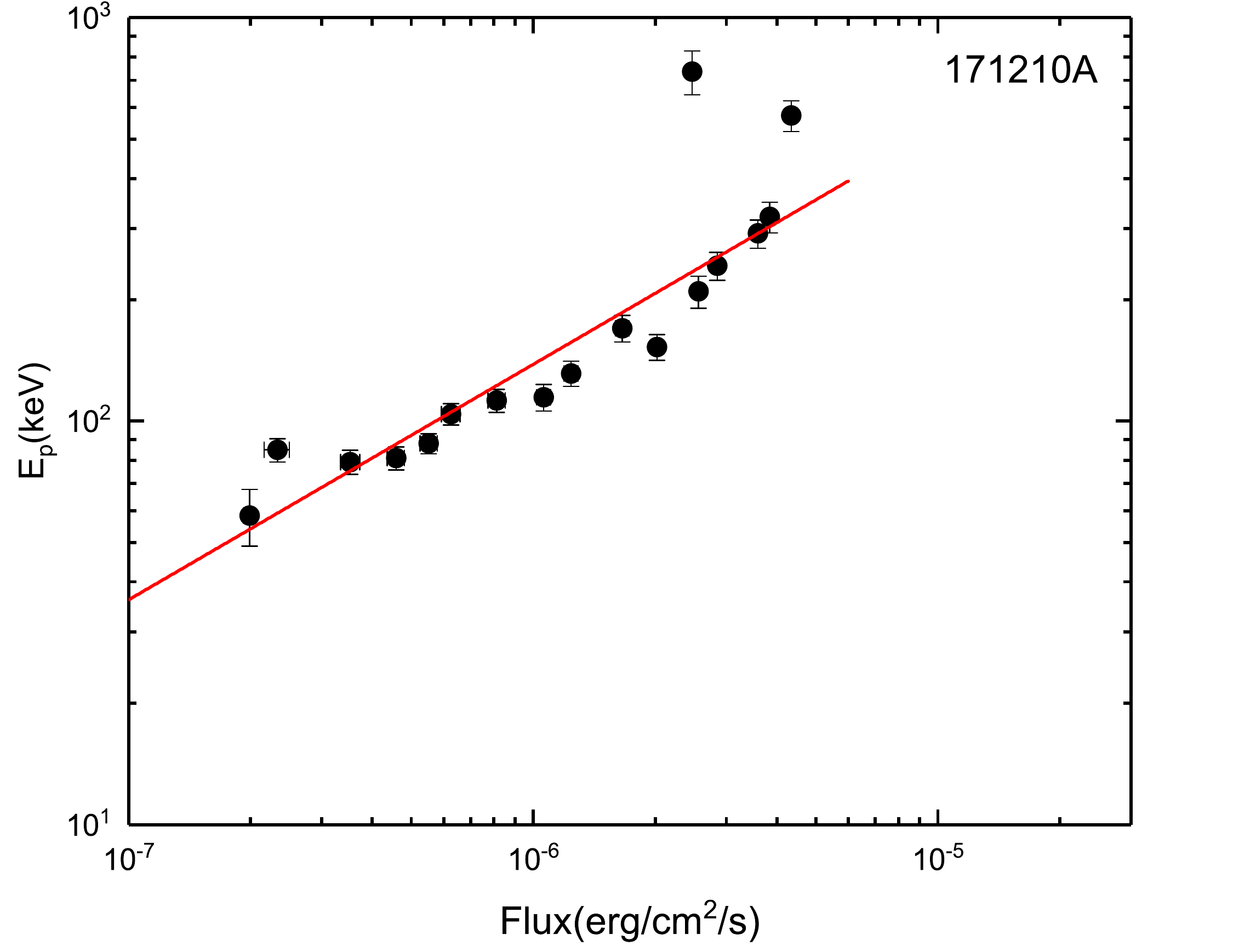}}
\resizebox{4cm}{!}{\includegraphics{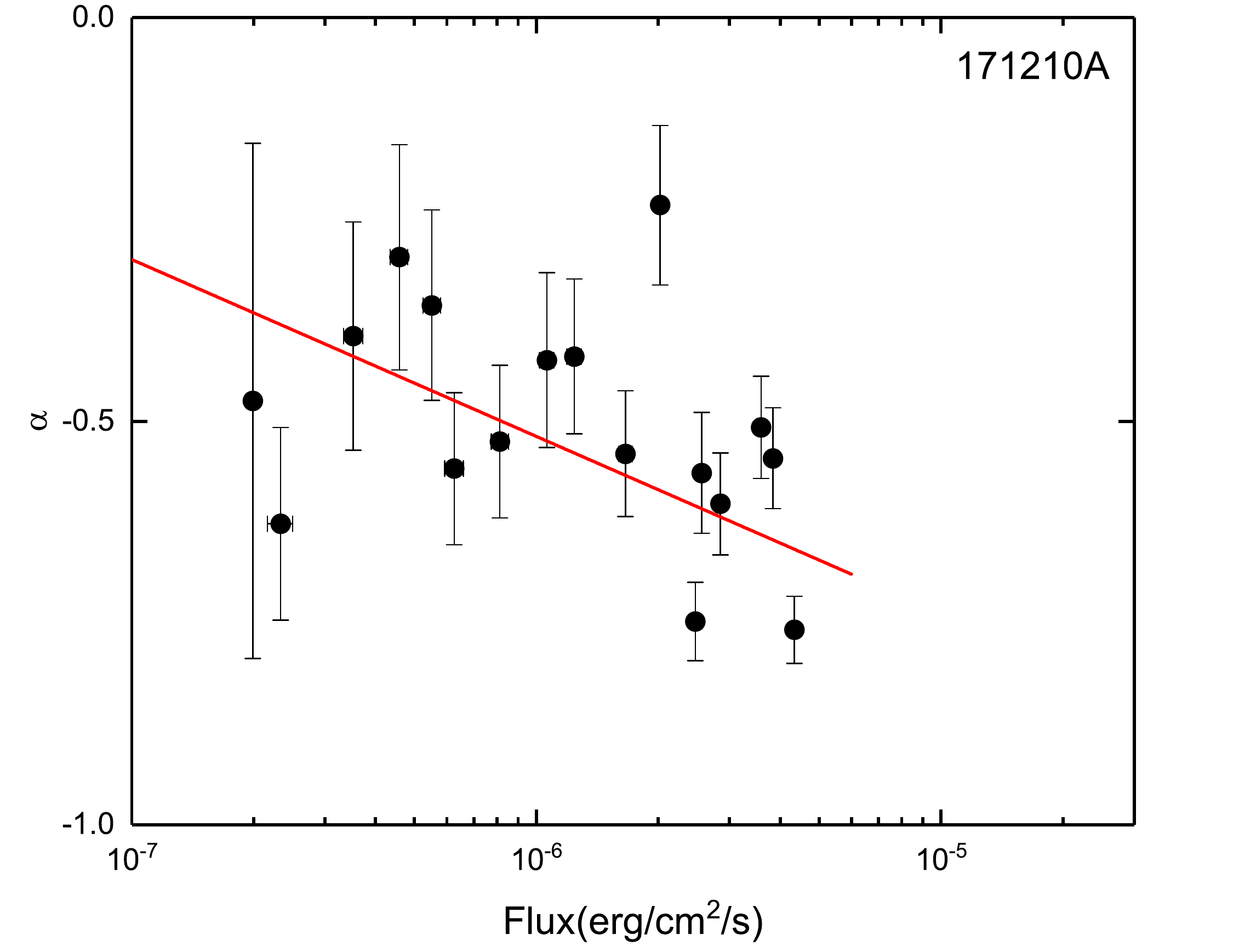}}
\resizebox{4cm}{!}{\includegraphics{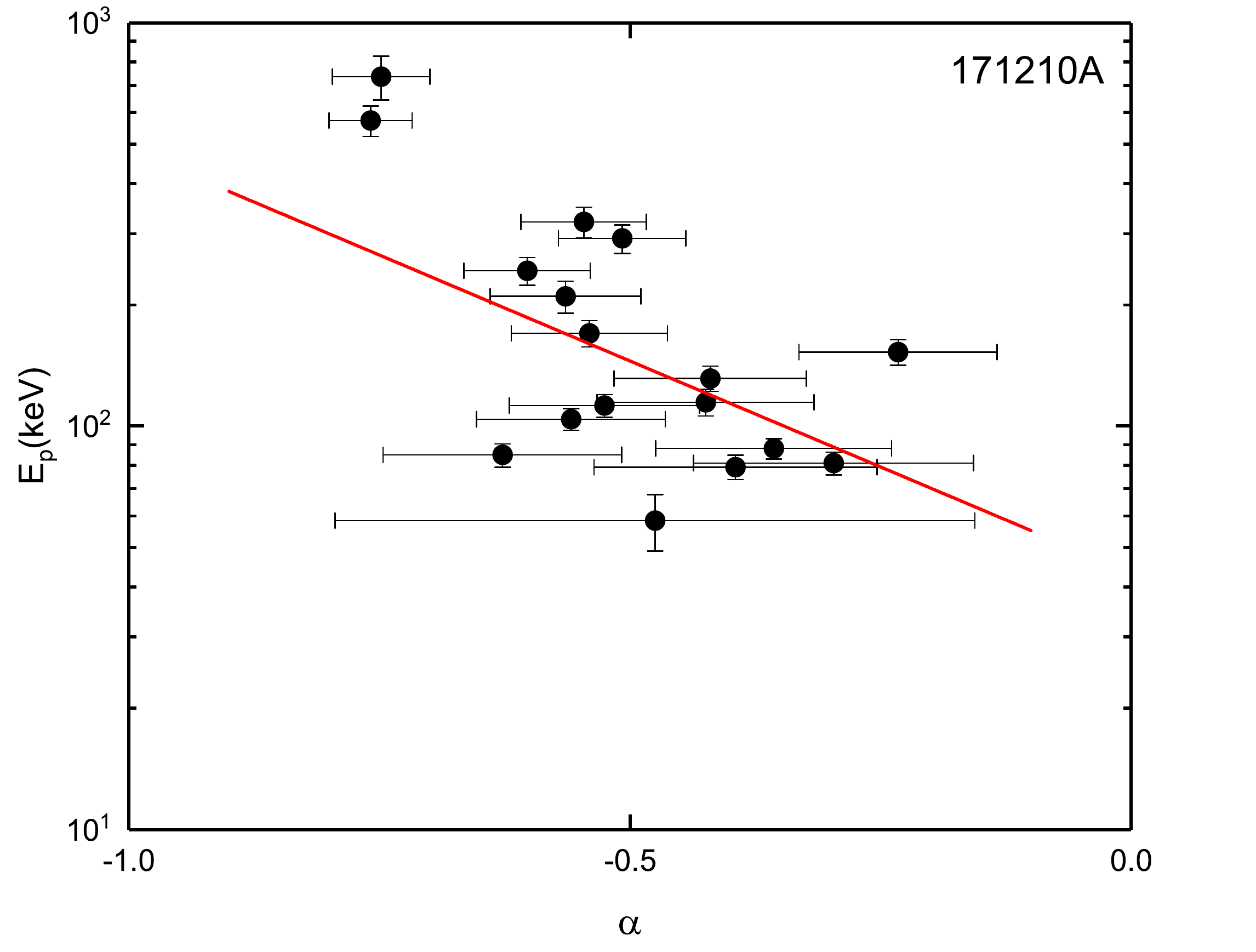}}
\resizebox{4cm}{!}{\includegraphics{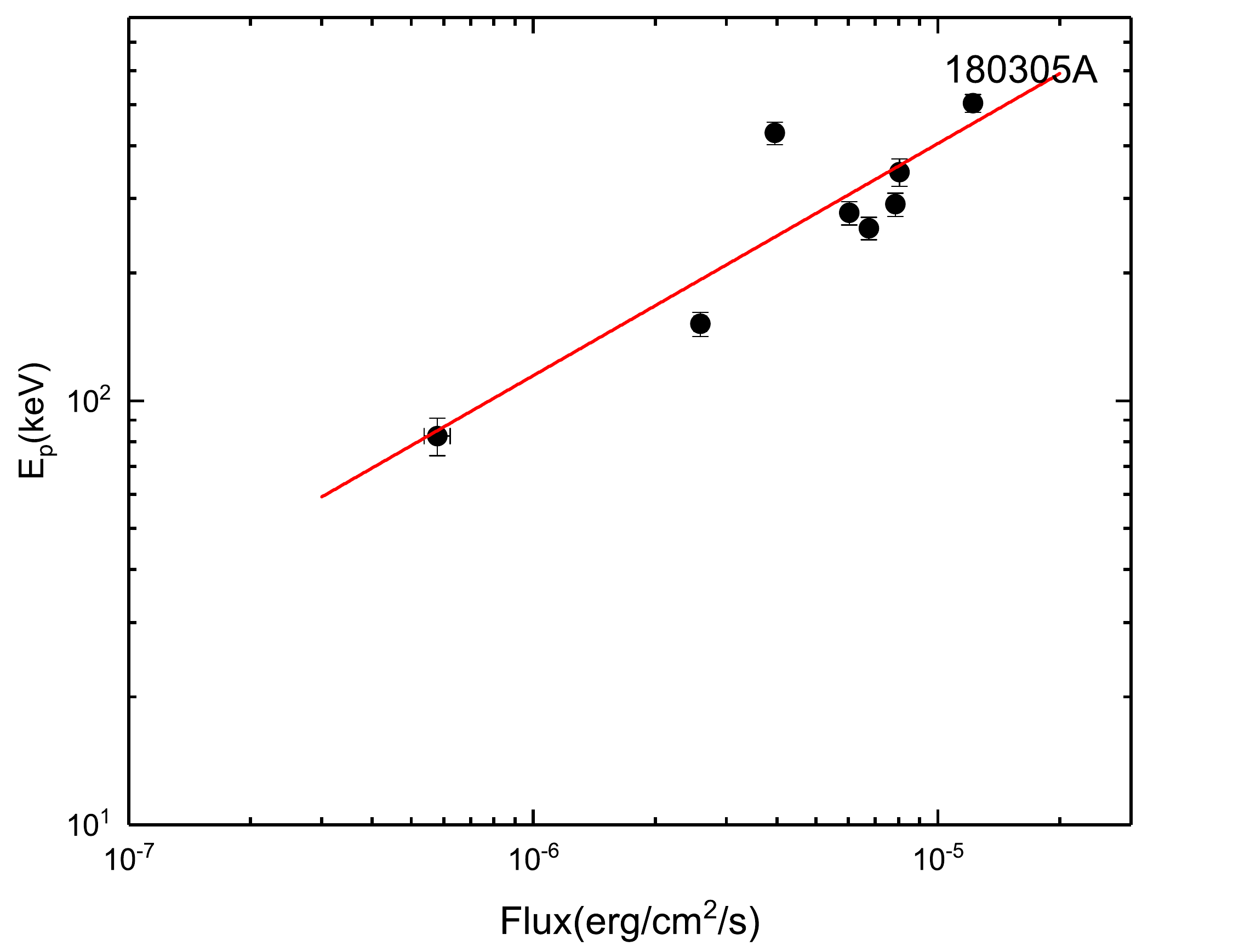}}
\resizebox{4cm}{!}{\includegraphics{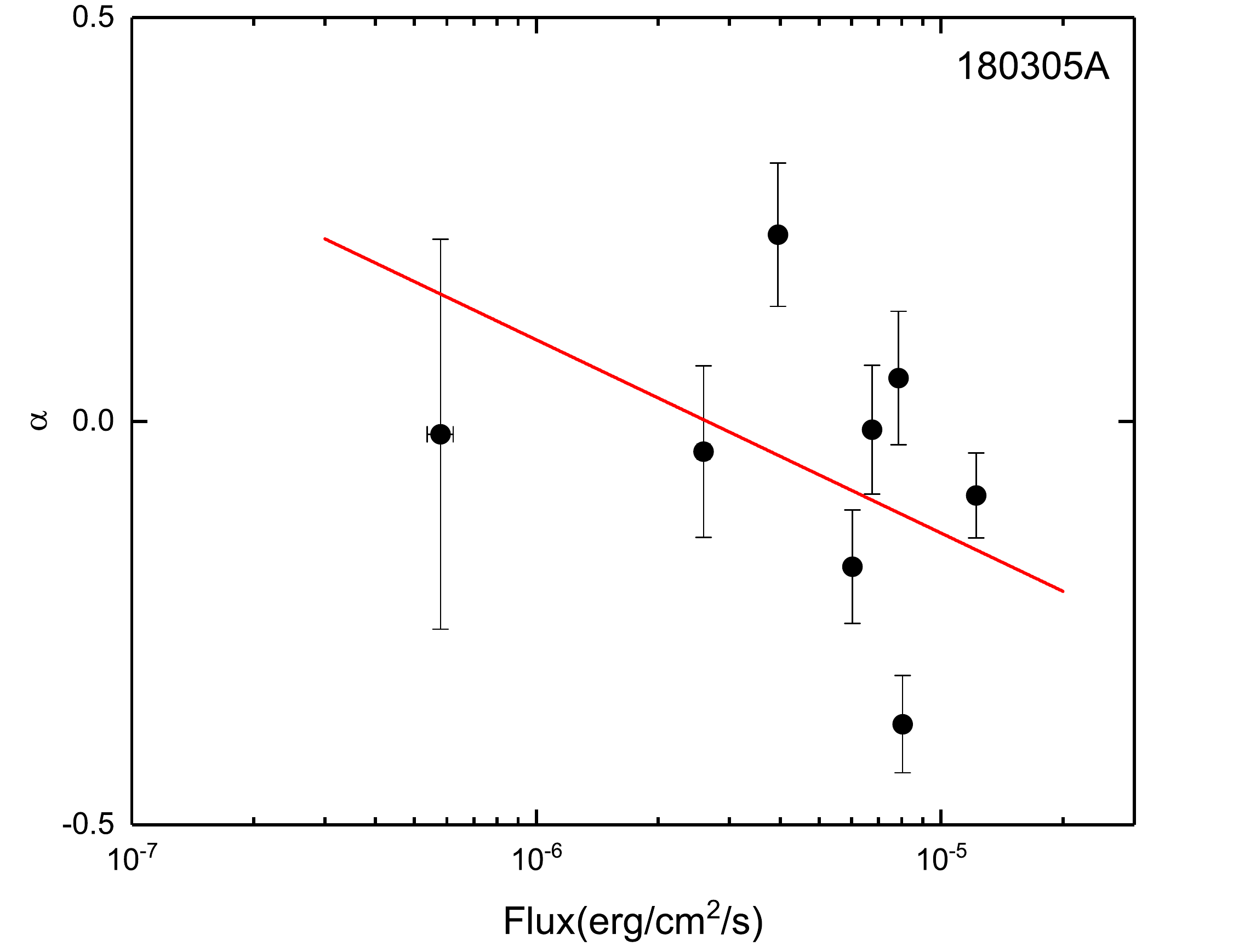}}
\resizebox{4cm}{!}{\includegraphics{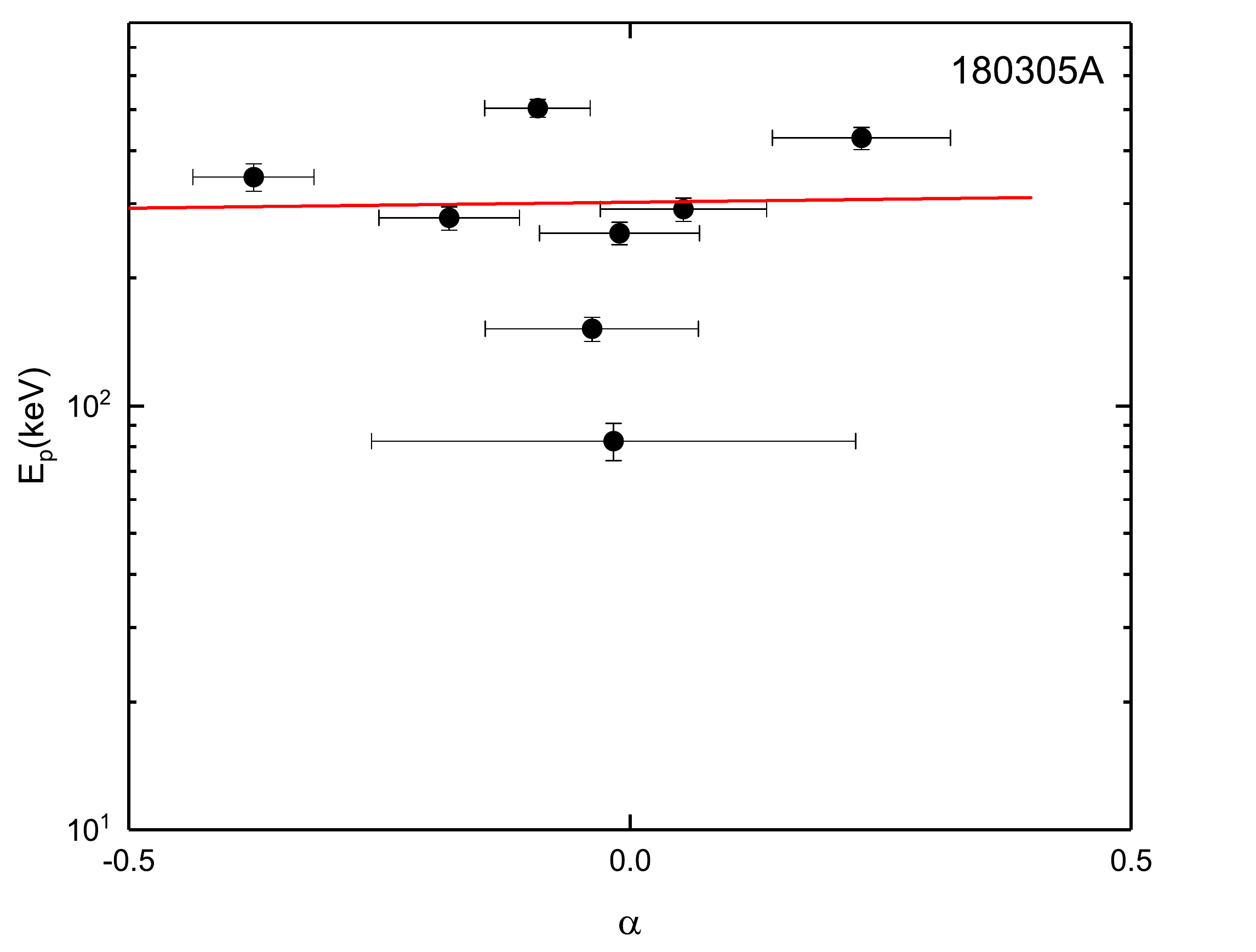}}
\caption{\it-continued}
\end{figure}

\begin{figure}[ht!]
\plotone{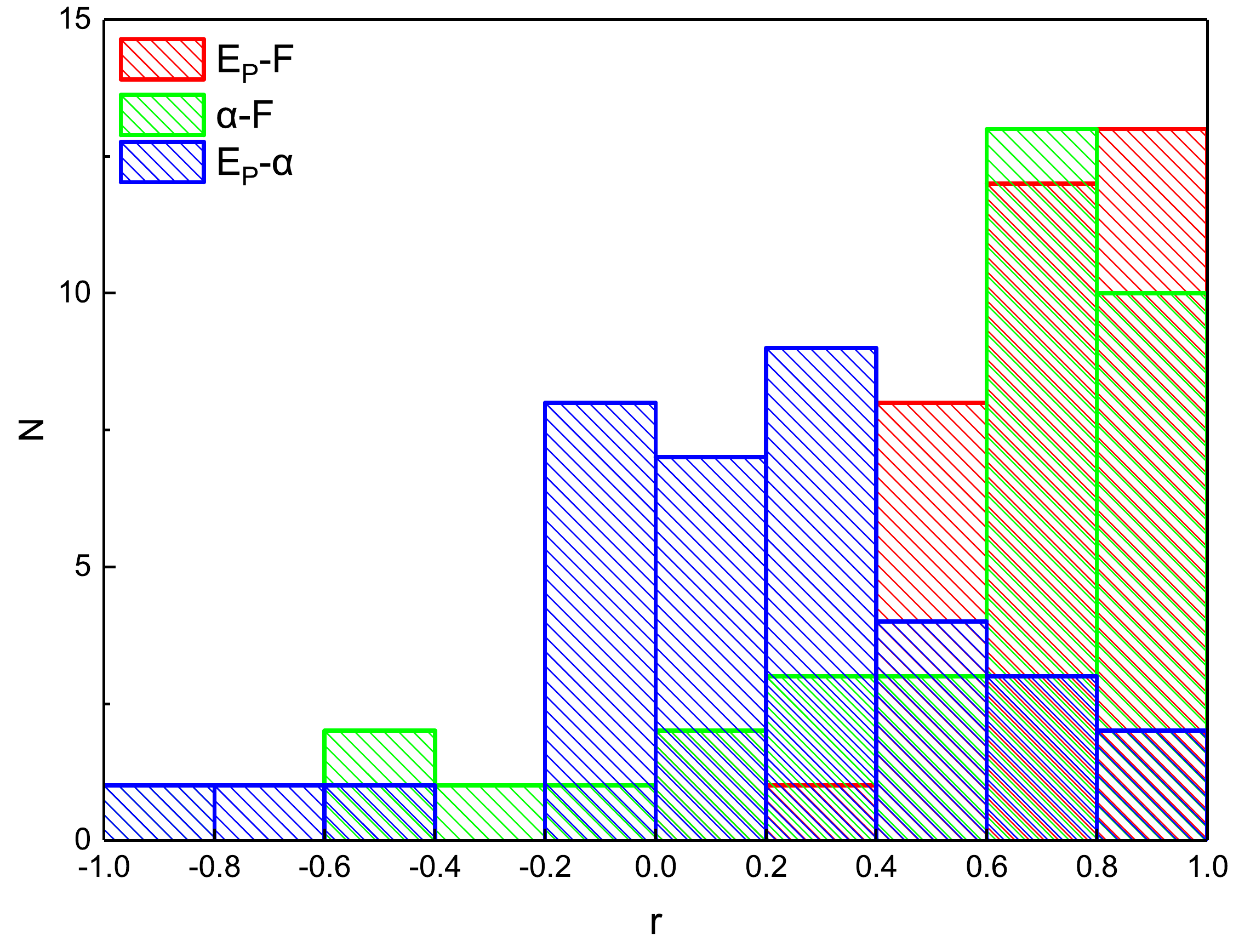}
\caption{The histograms of Pearson's correlation coefficient from the fitting results of parameter correlations such as $E_{p}-F$, $\alpha-F$\edit1{\added{,}} and $E_{p}- \alpha$. There is a strong monotonous positive correlation both for $E_{p}-F$ and $\alpha-F$ correlations in \edit1{\replaced{most bursts for our sample}{most of our bursts}}.\label{fig:statistical}}
\end{figure}

The parameter correlations may play an important role in revealing the nature of the prompt emission for gamma-ray bursts. \edit1{\replaced{Then}{In this section,}} the correlations such as $E_{p}-F$, $\alpha-F$\edit1{\added{,}} and $E_{p}- \alpha$ obtained from the time-resolved spectra are shown in Figure \ref{fig:parameter correlations} for all of the bursts in our sample. The fitting results of the parameter correlations (Pearson's correlation coefficient) have been shown in Table \ref{tab:resolved_results} (Col.4, Col.5, Col.6) as described in \ref{subsubsec:subsubsec3.2.2}. \edit1{\deleted{And the last figure,}}Figure \ref{fig:statistical}, the histograms of Pearson's correlation coefficient from the fitting results of parameter correlations such as $E_{p}-F$, $\alpha-F$\edit1{\added{,}} and $E_{p}- \alpha$ have been shown on it. 

In our analysis, we investigate \edit1{\deleted{the }}Figure \ref{fig:parameter correlations} in detail, then give the fitting results of the parameter correlations (Pearson's correlation \edit1{\replaced{coefficient}{coefficients}}) in Table \ref{tab:resolved_results}\edit1{\replaced{,}{.}} \edit1{\replaced{at the same time,}{Finally,}} the histograms of Pearson's correlation coefficient from the fitting results of all \edit1{\added{three}} parameter correlations \edit1{\replaced{was}{were}} presented in Figure \ref{fig:statistical}. Those previous analyses such as \citet{2001ApJ...548..770B}, \citet{2009MNRAS.393.1209F}, \citet{2010A&A...511A..43G}, \edit1{\added{and}} \citet{2019ApJ...886...20Y} have pointed out that, the $E_{p}- F$ relation \citep{1983Natur.306..451G}, i.e., the relation between the peak energy $E_{p}$ and energy flux $F$, exhibit three main types: (i) a non-monotonic relation (containing the positive and negative power-law segments while the break occurs at the peak flux); (ii) a monotonic relation which can be described by a single power-law; (iii) no clear trend. For all of our bursts, the most common \edit1{\replaced{behaviour}{behavior}} (in \edit1{\replaced{22}{25}} pulses) has a relation described by a single power-law which means that they have a strong positive relation. \edit1{\deleted{In fact, }}\edit1{\replaced{of}{Of}} these, \edit1{\replaced{11}{13}} GRBs have a very strong positive relation ($r\in(0.8,1.0)$, see Table \ref{tab:resolved_results} and Figure \ref{fig:statistical})\edit1{\replaced{ and}{,}} another \edit1{\replaced{11}{12}} GRBs have a strong positive relation ($r\in(0.6,0.8)$, also see Table \ref{tab:resolved_results} and Figure \ref{fig:statistical}). \edit1{\replaced{The rest 7 GRBs have the}{The rest of 11 GRBs have a}} positive correlation which is not strong or very strong, but the moderate correlation \edit1{\replaced{emerge in 6 GRBs}{emerged in 8 GRBs}}\edit1{\replaced{and the last one shows}{, the last three show}} \edit1{\replaced{the}{a}} weak correlation (\edit1{\replaced{GRB 170214A}{GRBs 150314A, 170214A, 170510A}}). In a word, \edit1{\replaced{in $75.9\%$}{$69.4\%$}} of these bursts show \edit1{\replaced{the}{a}} strong positive correlation and  \edit1{\replaced{in $24.1\%$}{$30.6\%$}} of \edit1{\replaced{the sample}{these bursts}} show \edit1{\replaced{the}{a}} weaker positive correlation compared with the former. However, these results are inconsistent with the study of 38 single pulses in \citet{2019ApJ...886...20Y}, which shows that 23 single pulses exhibit the non-monotonic relation and 13 pulses exhibit the monotonic relation (\edit1{\added{the}} two common \edit1{\replaced{behaviours}{behavior}} in their study).

Turning over to the $\alpha- F$ relation. The study of a large sample of single pulses in \citet{2019ApJ...886...20Y} shows a monotonic positive linear relation in the log-linear plots. In the study, the majority of the pulses show \edit1{\replaced{the}{a}} strong positive relation (28 pulses), 8 pulses have \edit1{\added{a}} very strong positive relation\edit1{\added{,}} and \edit1{\replaced{there are only 2 pulses which}{only 2 pulses}} have \edit1{\replaced{weak correlations}{a weak correlation}}. However, the results of our study present at least 6 types of monotonic linear relation in the log-linear plots. \edit1{\replaced{The most important type is that}{The strong positive correlation is most popular}}, \edit1{\replaced{there are 20 GRBs show the strong positive relations}{23 GRBs show this correlation}} ($r\in(0.6,1.0)$). \edit1{\replaced{And of these, 7 GRBs exhibit the very strong positive relations}{Of these, 10 GRBs exhibit a very strong positive correlation}} which means that the Pearson's correlation coefficient is larger than 0.8. \edit1{\replaced{2 GRBs show the moderate positive correlations}{Furthermore, 3 GRBs show a moderate positive correlation}} ($r\in(0.4,0.6)$). \edit1{\replaced{2 GRBs have weaker positive correlations}{3 GRBs have a weaker positive correlation}} ($r\in(0.2,0.4)$). \edit1{\replaced{There is no correlation between $\alpha$ and F for 3 GRBs.}{3 GRBs have no correlation between $\alpha$ and F.}} \edit1{\replaced{And the rest two GRBs}{The rest four GRBs}} are different from them in $\alpha-F$ correlation. \edit1{\replaced{One of them (GRB 150202B) shows the moderate negative correlation, while GRB 170115B shows the very strong negative correlation in this relation.}{Especially, GRB 170115B shows a very strong negative correlation in this relation.}}

Finally, the $E_{p}- \alpha$ correlation differs clearly from the first two relations. \edit1{\replaced{There are only}{Only}} 5 GRBs have \edit1{\added{a}} strong positive relation. Of these bursts, \edit1{\replaced{3}{2}} GRBs have a very strong positive relation, \edit1{\replaced{and 2}{3}} GRBs have a general strong positive relation. \edit1{\replaced{5}{Besides, 4}} GRBs have a moderate positive relation and \edit1{\replaced{8}{9}} GRBs have a weaker positive relation. \edit1{\replaced{There is no correlation between $E_{p}$ and $\alpha$ for \edit1{\replaced{9}{15}} GRBs}{15 GRBs have no correlation between $E_{p}$ and $\alpha$}}. Moreover, one can find that \edit1{\deleted{there are }}two bursts \edit1{\deleted{which }}have a strong negative correlation (GRB 150202B, 170115B). Especially, GRB 150202B has a general strong negative correlation while GRB 170115B has a very strong negative correlation with \edit1{\added{the}} value of $r=-0.97$. \edit1{\added{The last one (GRB 171210A) shows a moderate negative correlation.}}

It is noteworthy that there are two peculiar bursts, GRBs 150202B and 170115B, which have \edit1{\replaced{the}{an}} `anti-tracking' \edit1{\replaced{behaviours}{behavior}} compared with energy flux for \edit1{\added{the}} low energy photon index $\alpha$. The negative correlation exhibits both for their parameter correlations such as  $\alpha- F$ and $E_{p}- \alpha$ correlations. The Pearson's correlation coefficient of $\alpha-F$ is -0.48 for GRB 150202B, which means that \edit1{\replaced{it's}{it is}} a moderate negative correlation, and a strong negative correlation (r=-0.69) has been shown in $E_{p}-\alpha$ correlation for this burst. \edit1{\replaced{However, the}{Surprisingly, a}} very strong negative correlation has been exhibited both for $\alpha-F$ (r=-0.95) and $E_{p}-\alpha$ (r=-0.97) correlations for GRB 170115B. Additionally, the fact that the value of $\alpha$ in \edit1{\added{the}} time-integrated spectrum is smaller than the synchrotron limit while the values of $\alpha$ for all of the slices in \edit1{\added{the}} time-resolved spectra violate the limit for GRB 170115B can be found.

\subsection{Whether the Two Observed Strong Positive Correlations Are Intrinsic or Artificial} \label{subsec:simulation}

\begin{figure}
\centering
\resizebox{4cm}{!}{\includegraphics{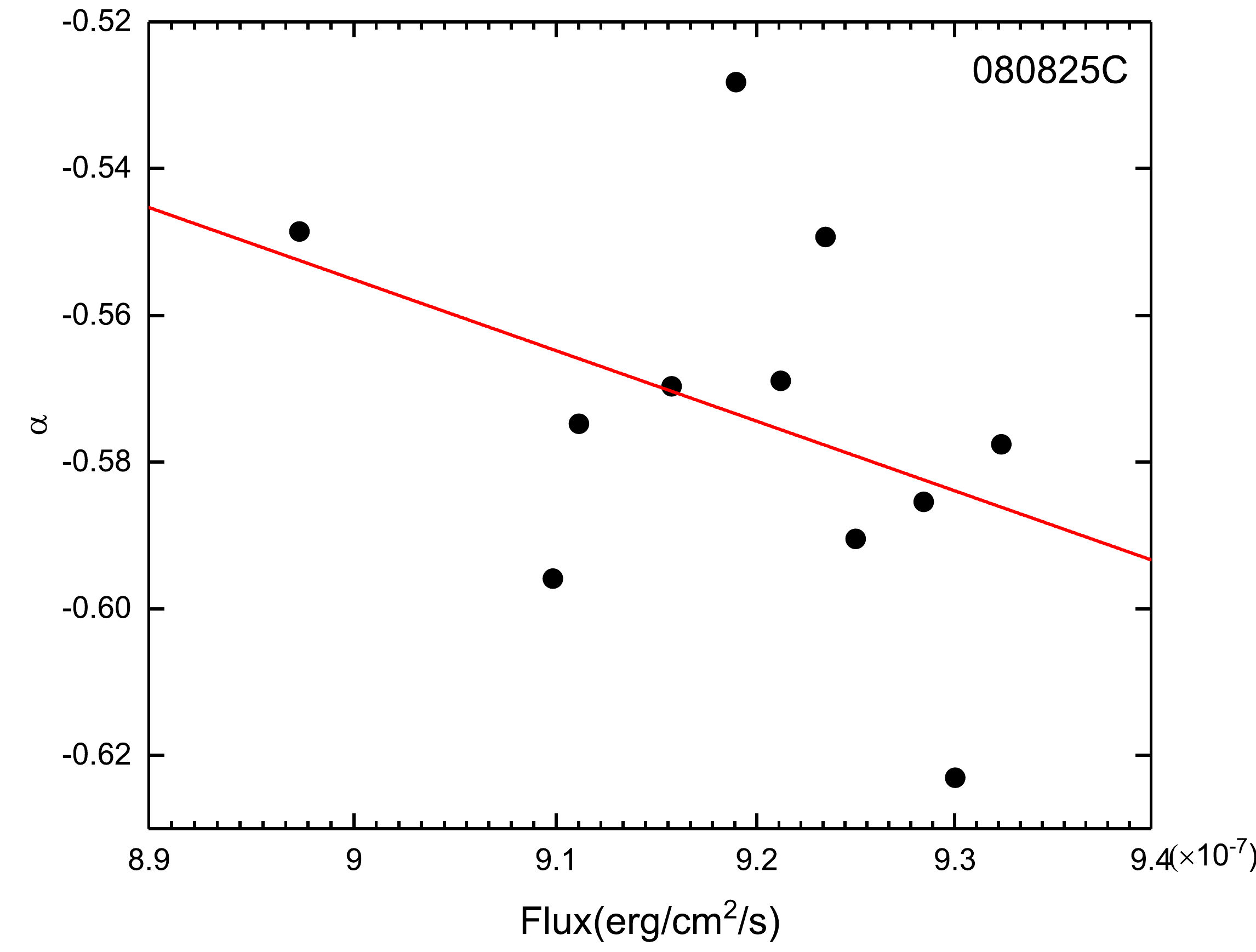}}
\resizebox{4cm}{!}{\includegraphics{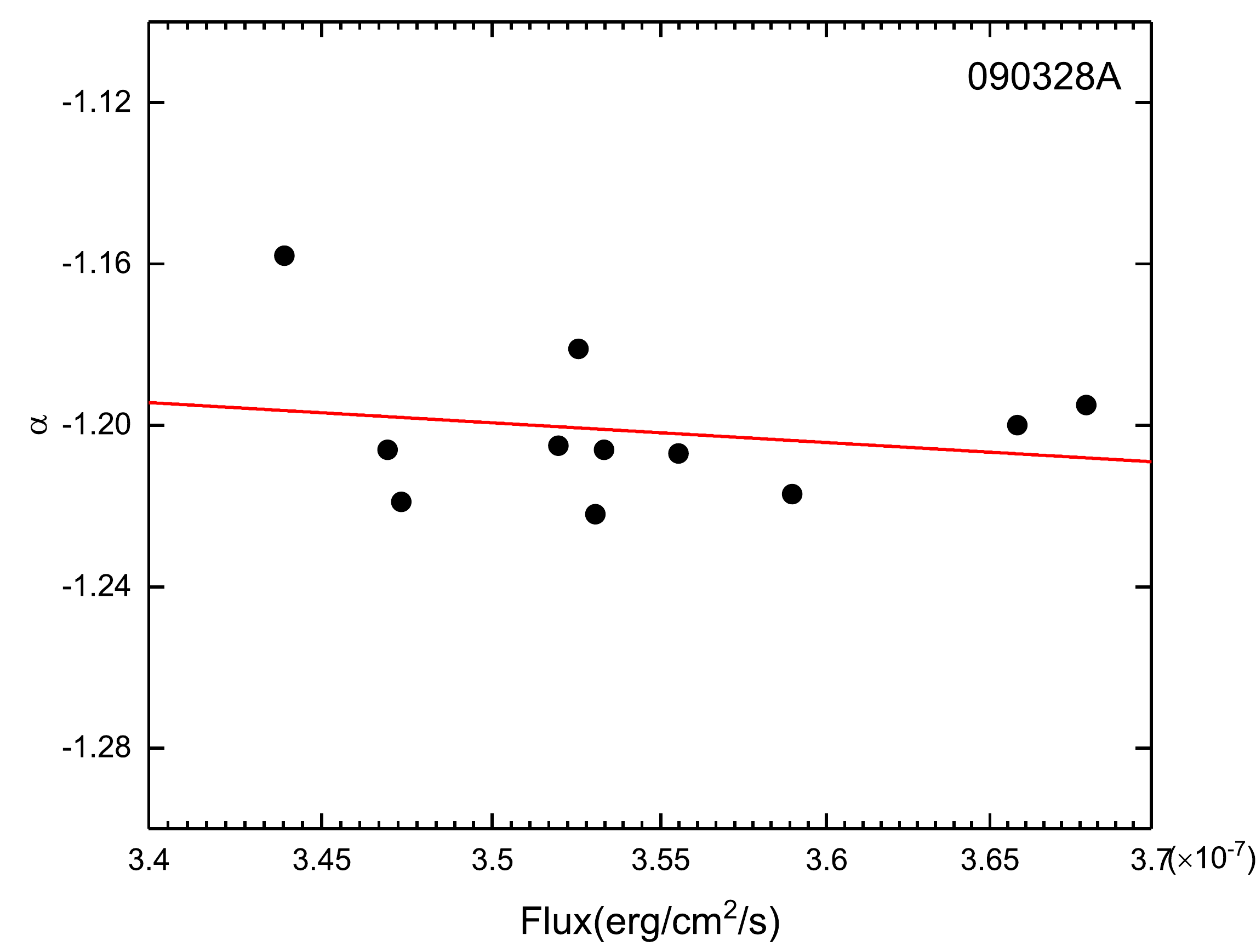}}
\resizebox{4cm}{!}{\includegraphics{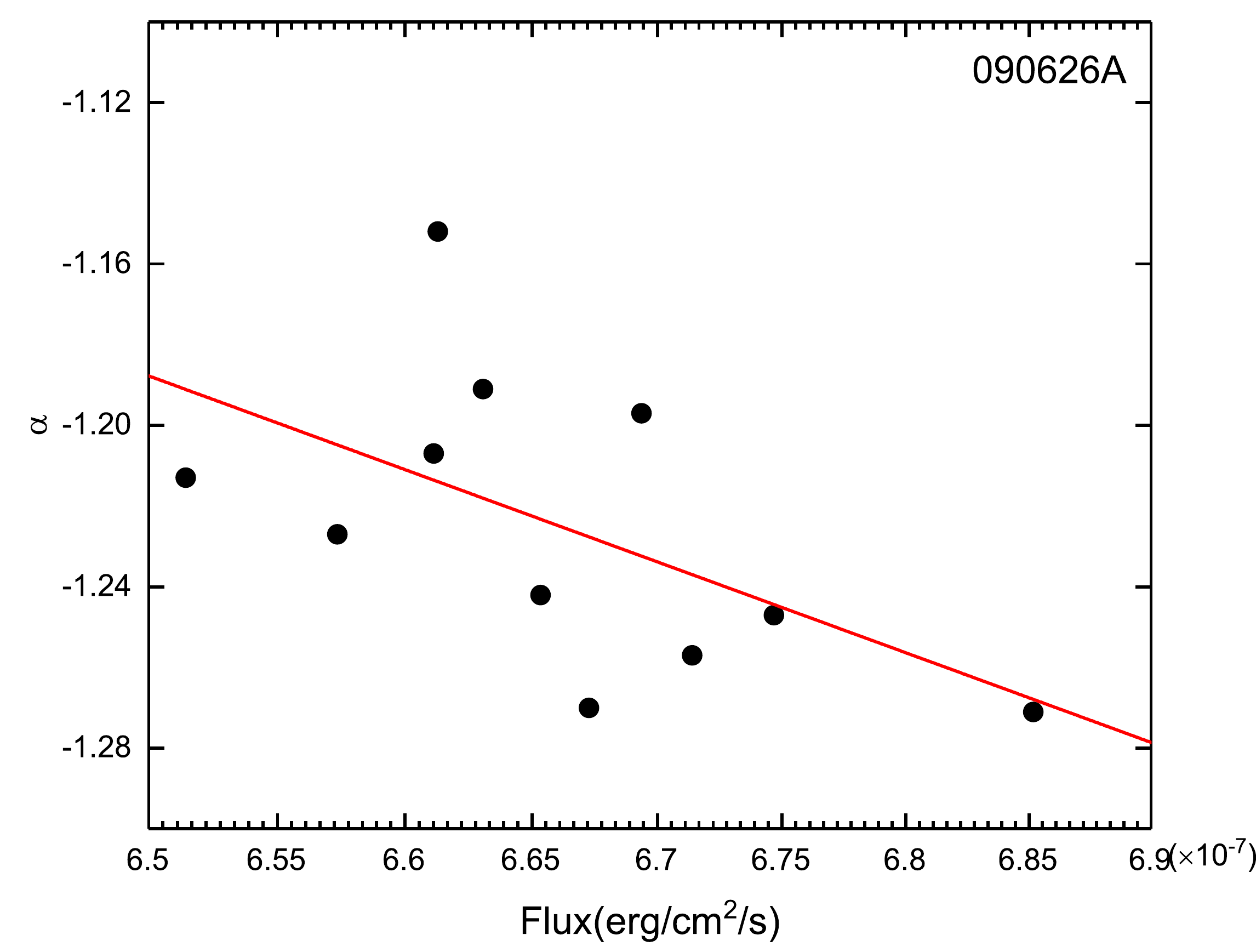}}
\resizebox{4cm}{!}{\includegraphics{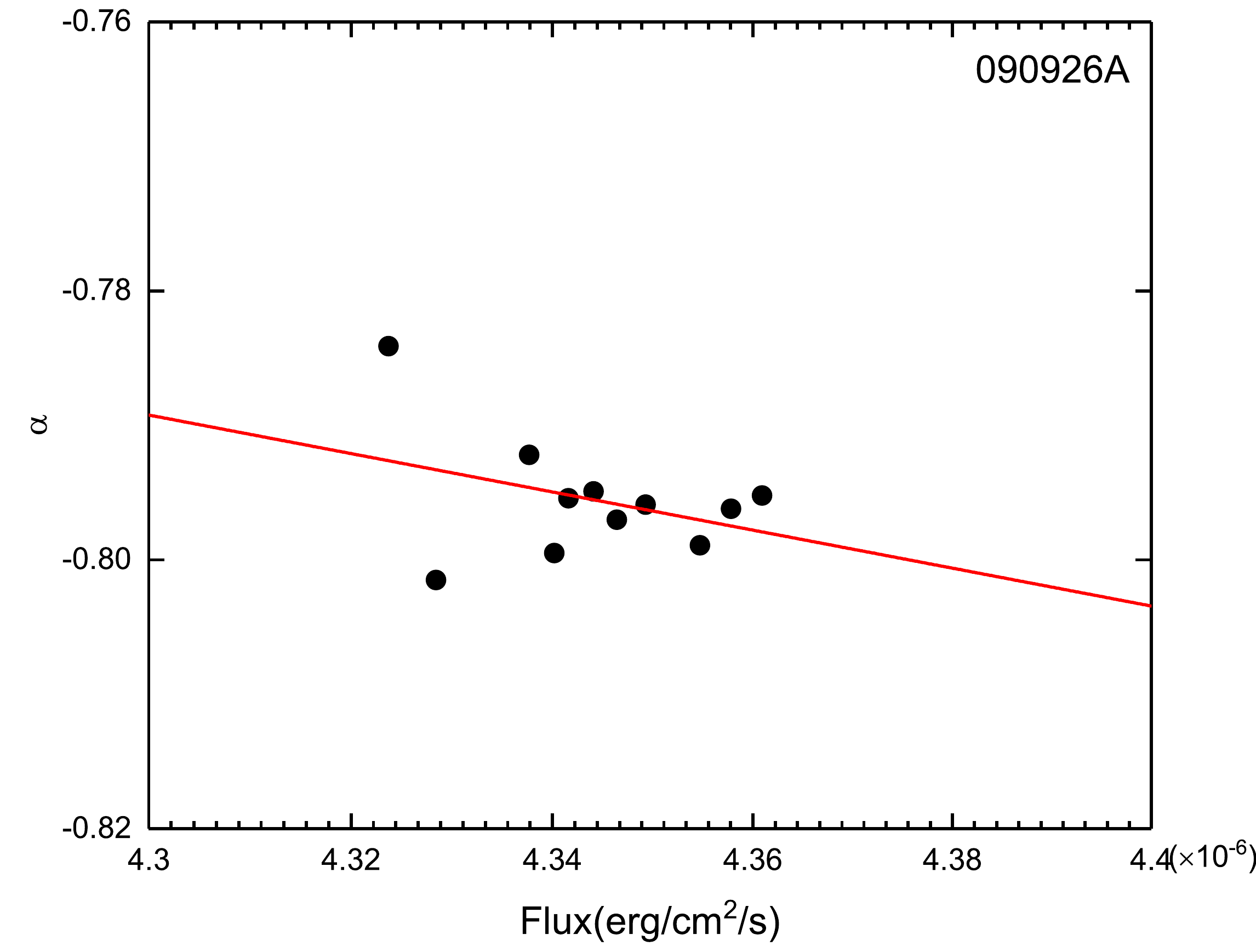}}
\resizebox{4cm}{!}{\includegraphics{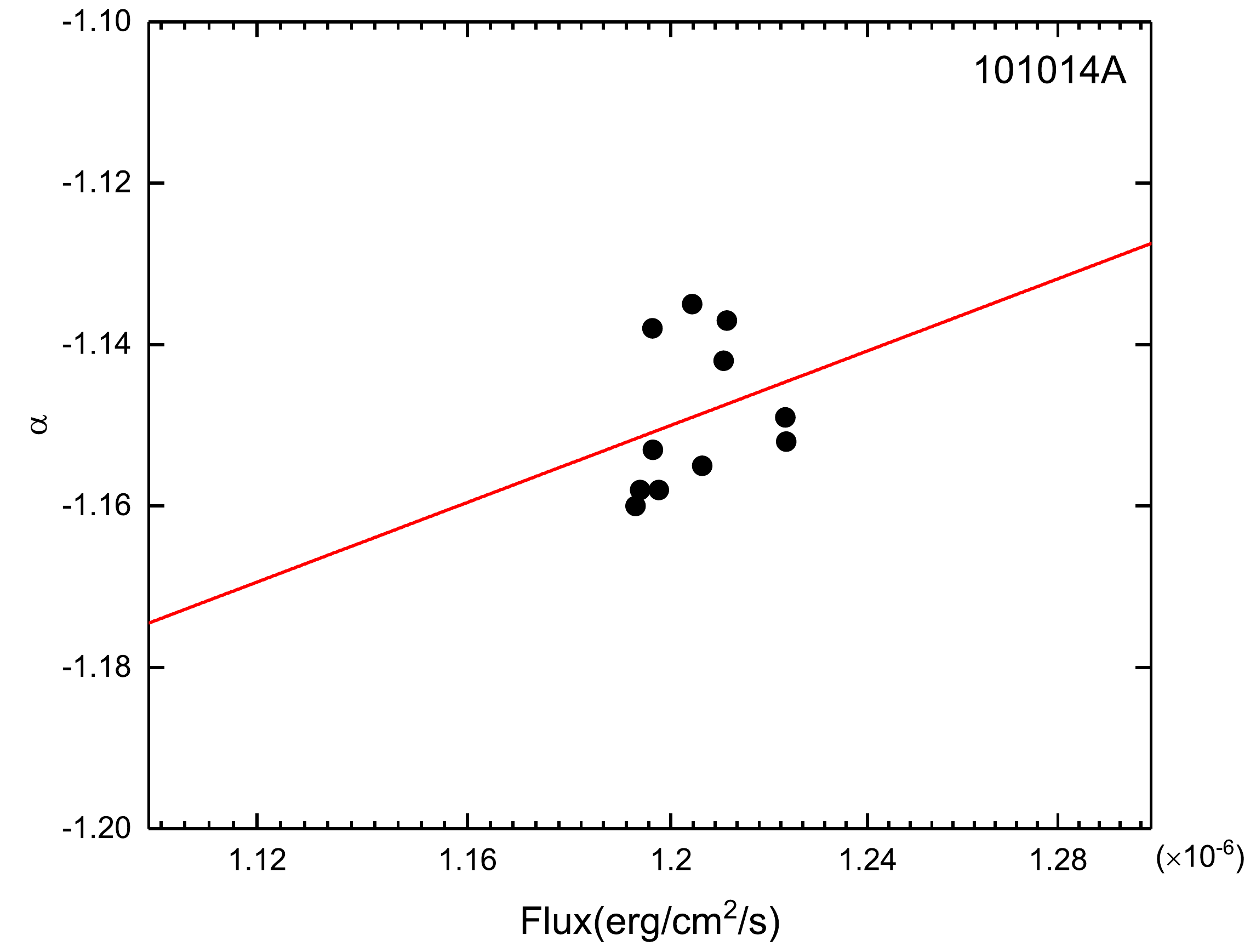}}
\resizebox{4cm}{!}{\includegraphics{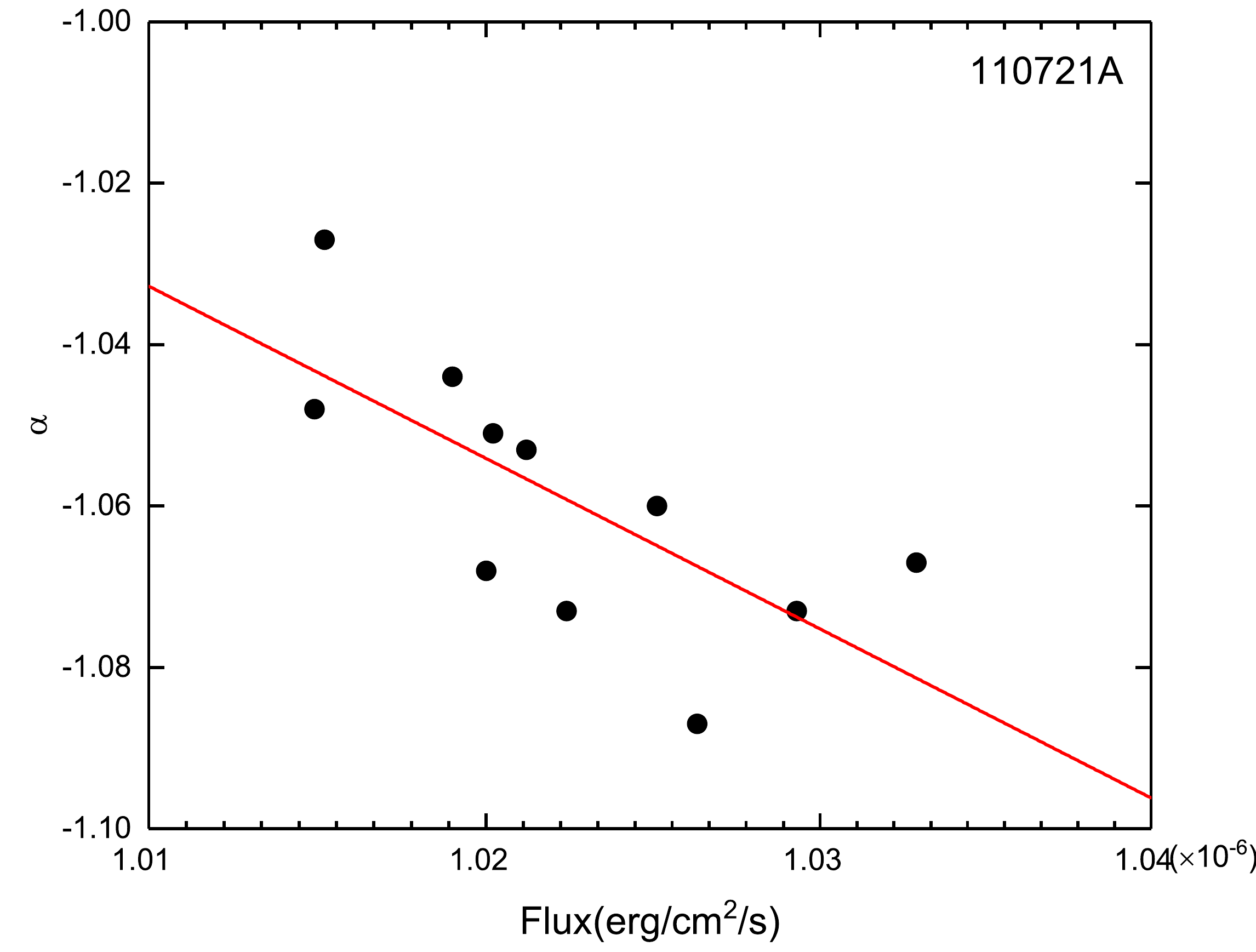}}
\resizebox{4cm}{!}{\includegraphics{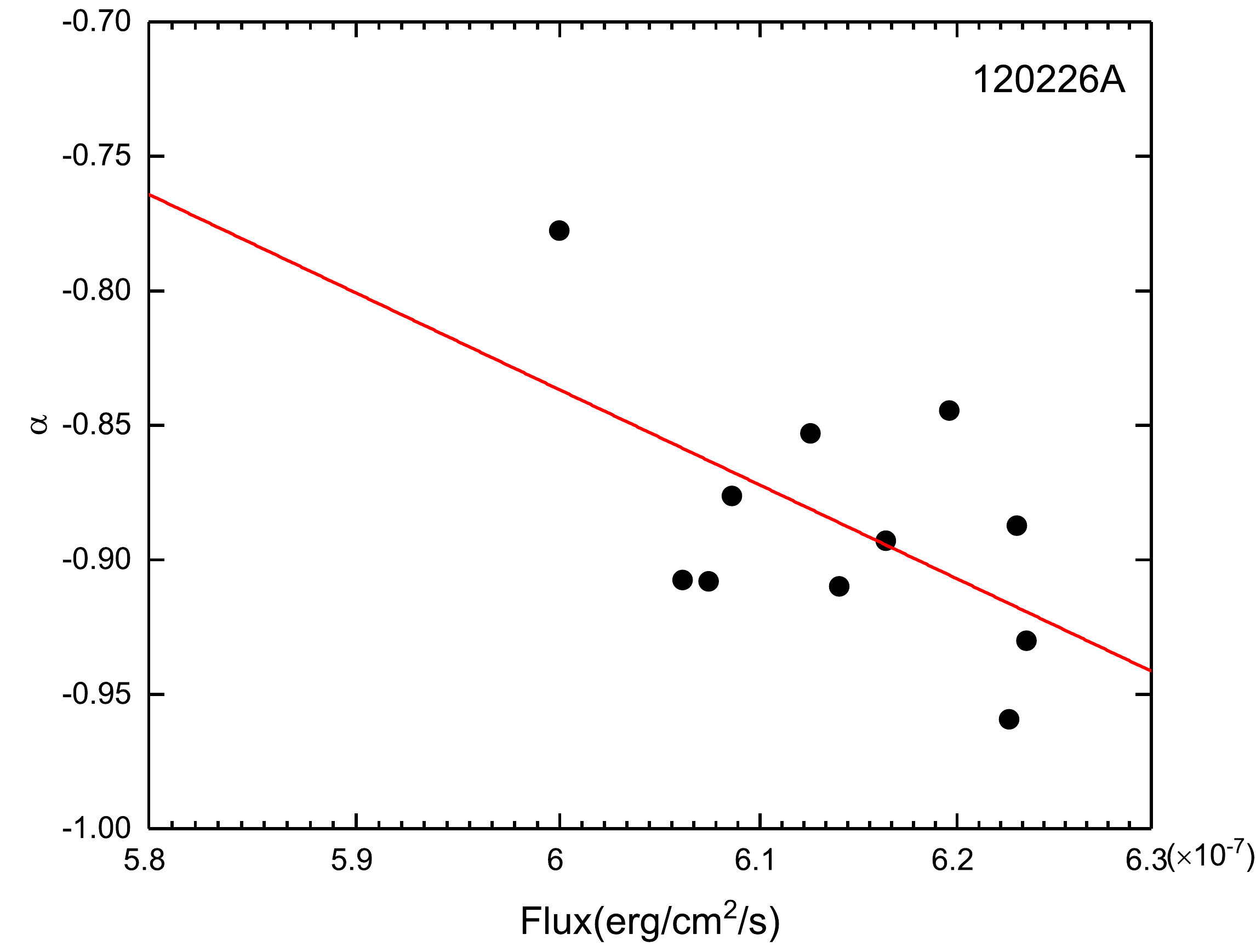}}
\resizebox{4cm}{!}{\includegraphics{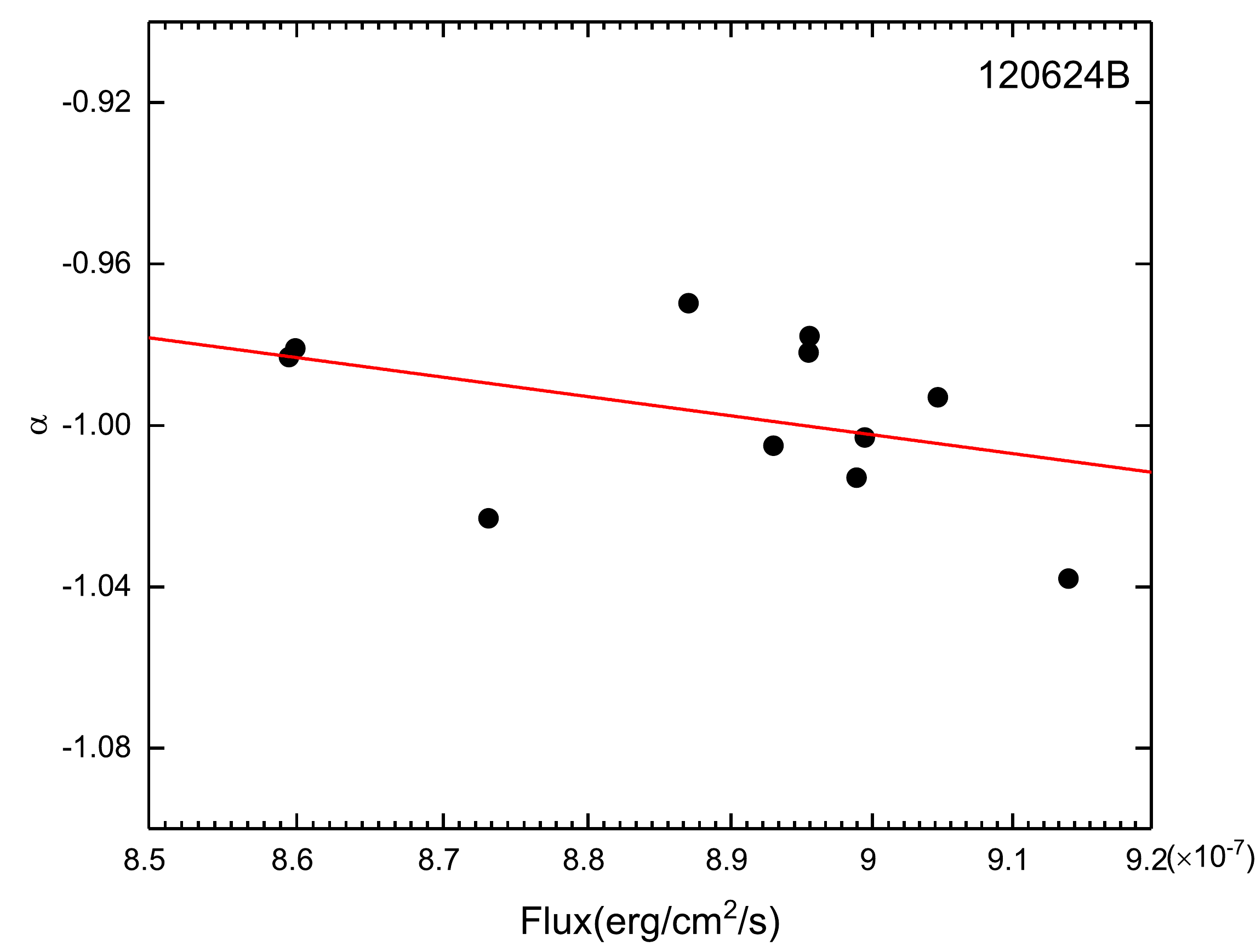}}
\resizebox{4cm}{!}{\includegraphics{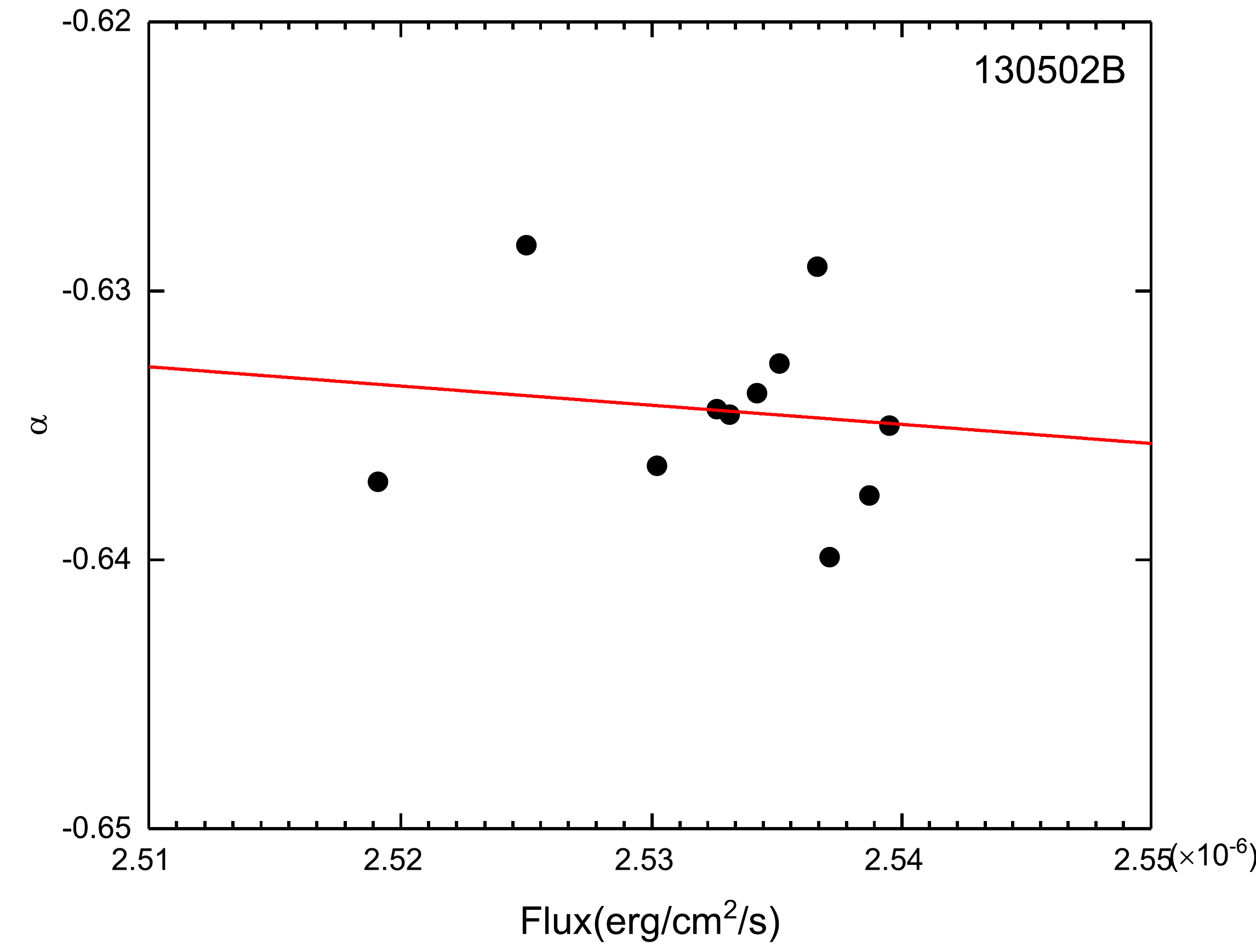}}
\resizebox{4cm}{!}{\includegraphics{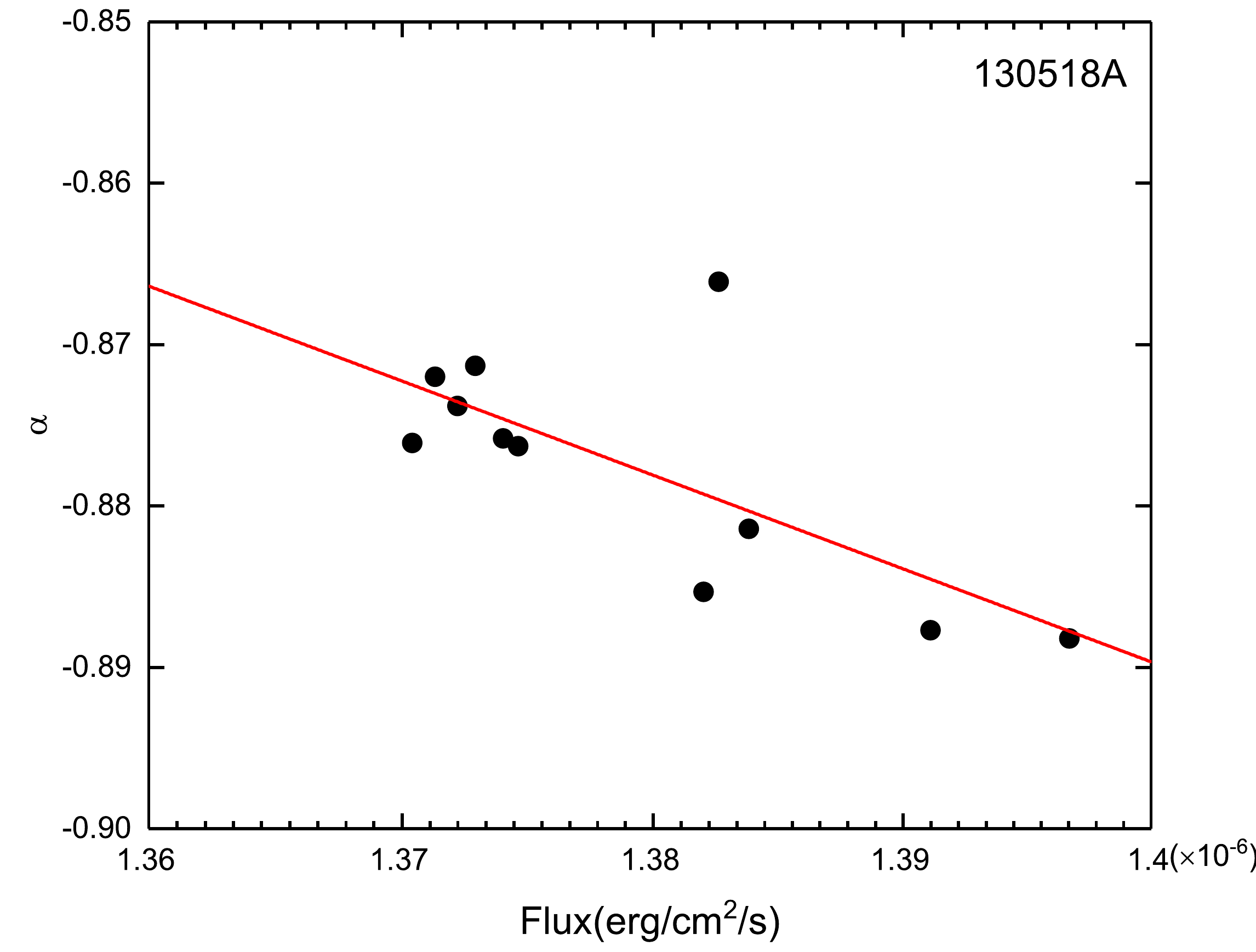}}
\resizebox{4cm}{!}{\includegraphics{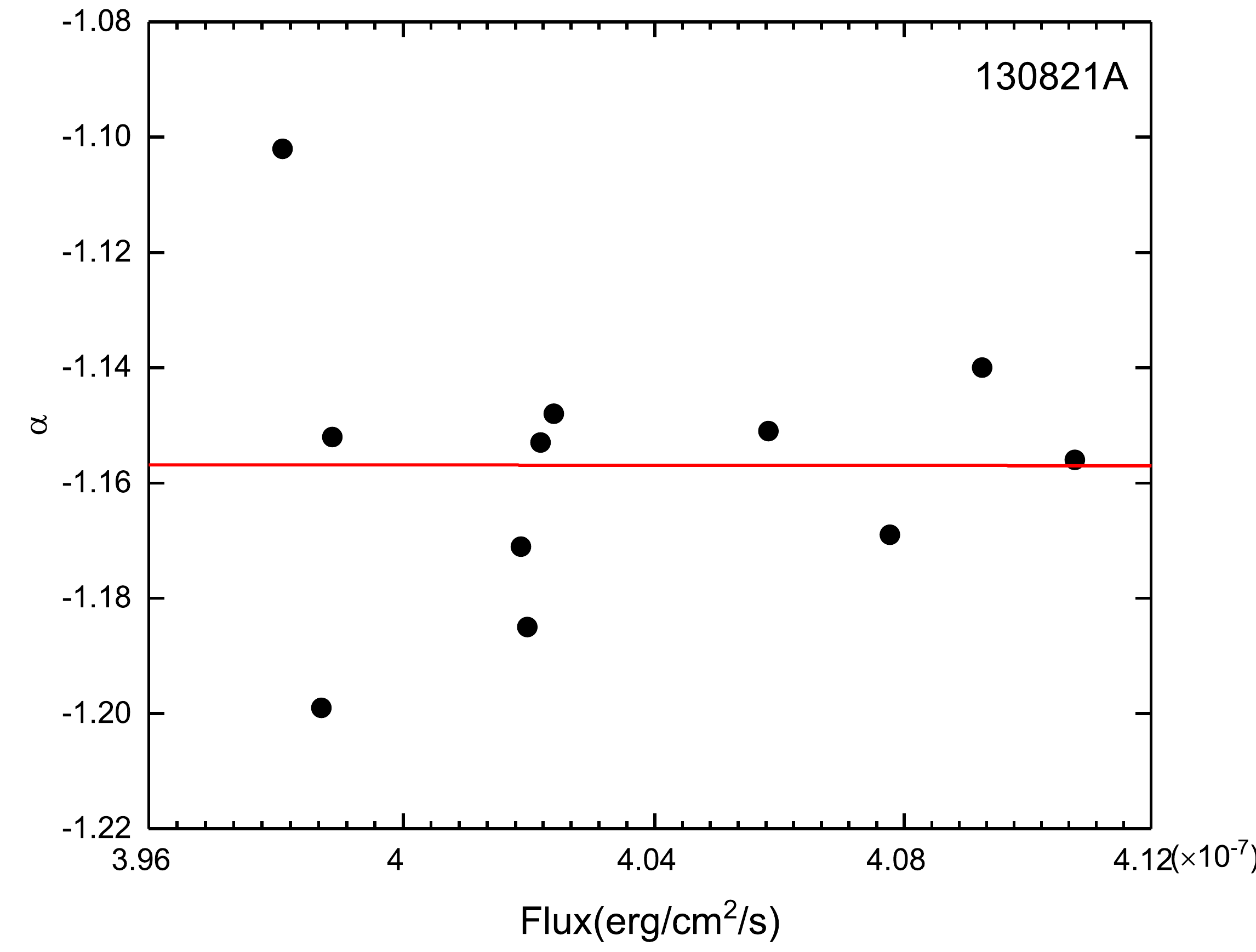}}
\resizebox{4cm}{!}{\includegraphics{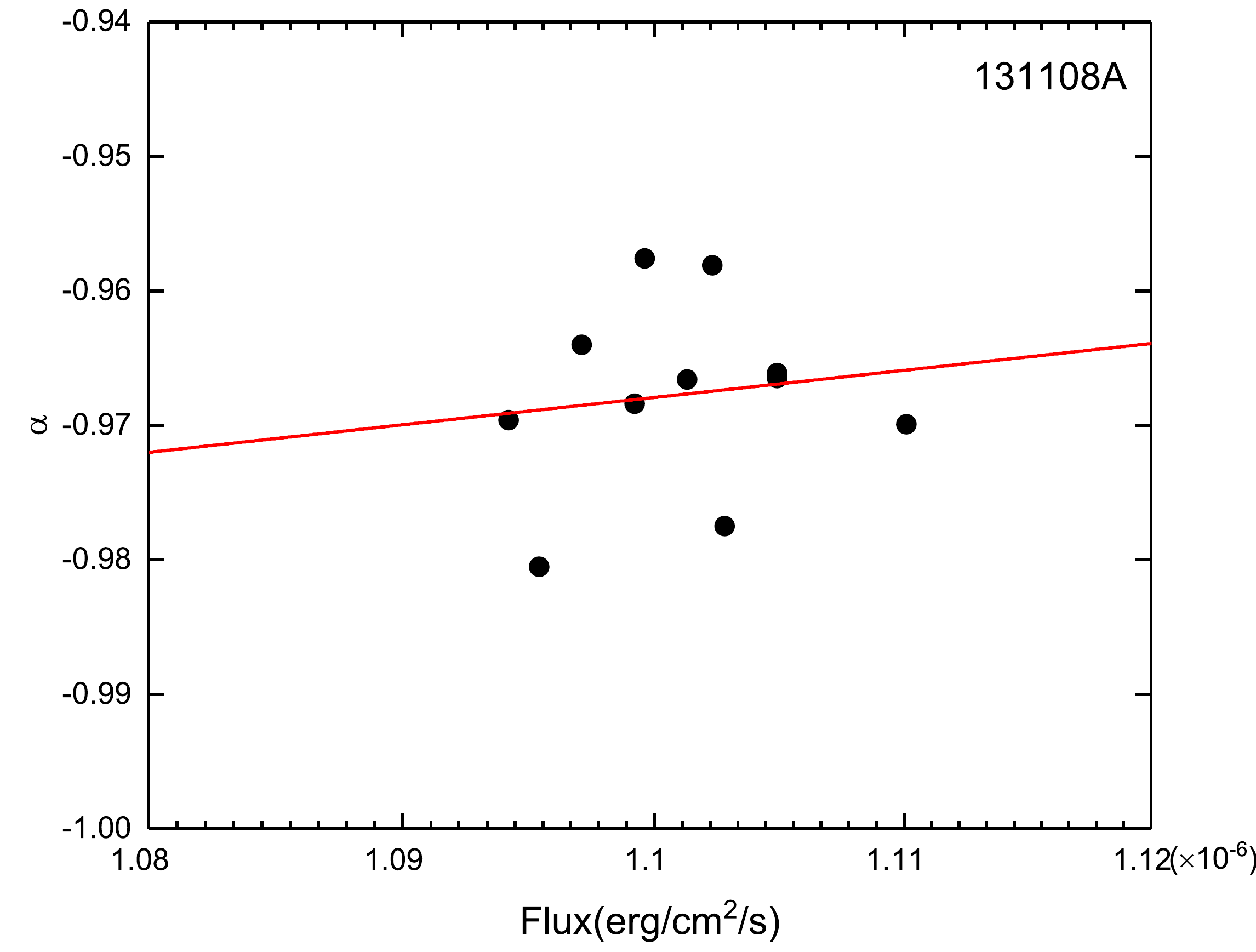}}
\resizebox{4cm}{!}{\includegraphics{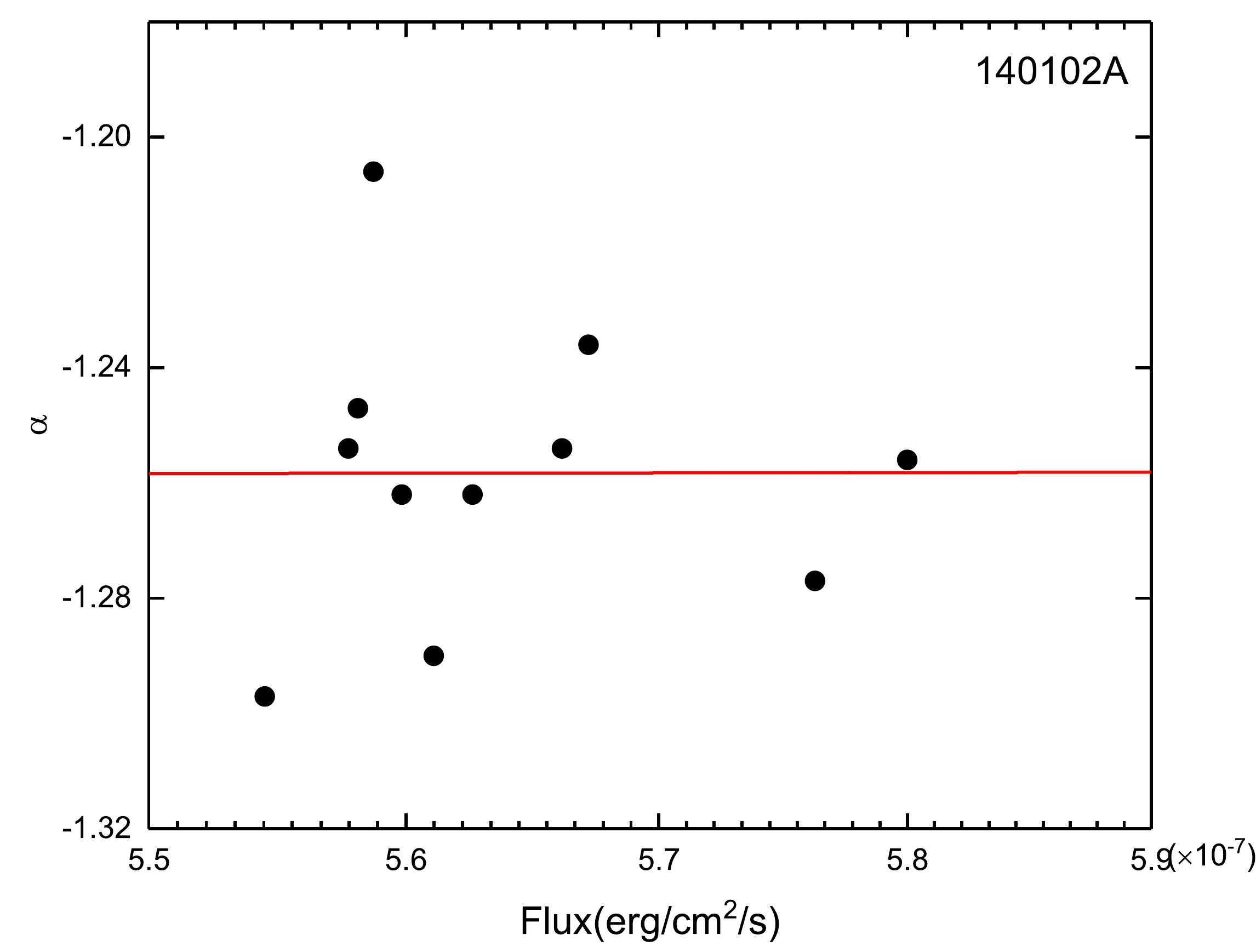}}
\resizebox{4cm}{!}{\includegraphics{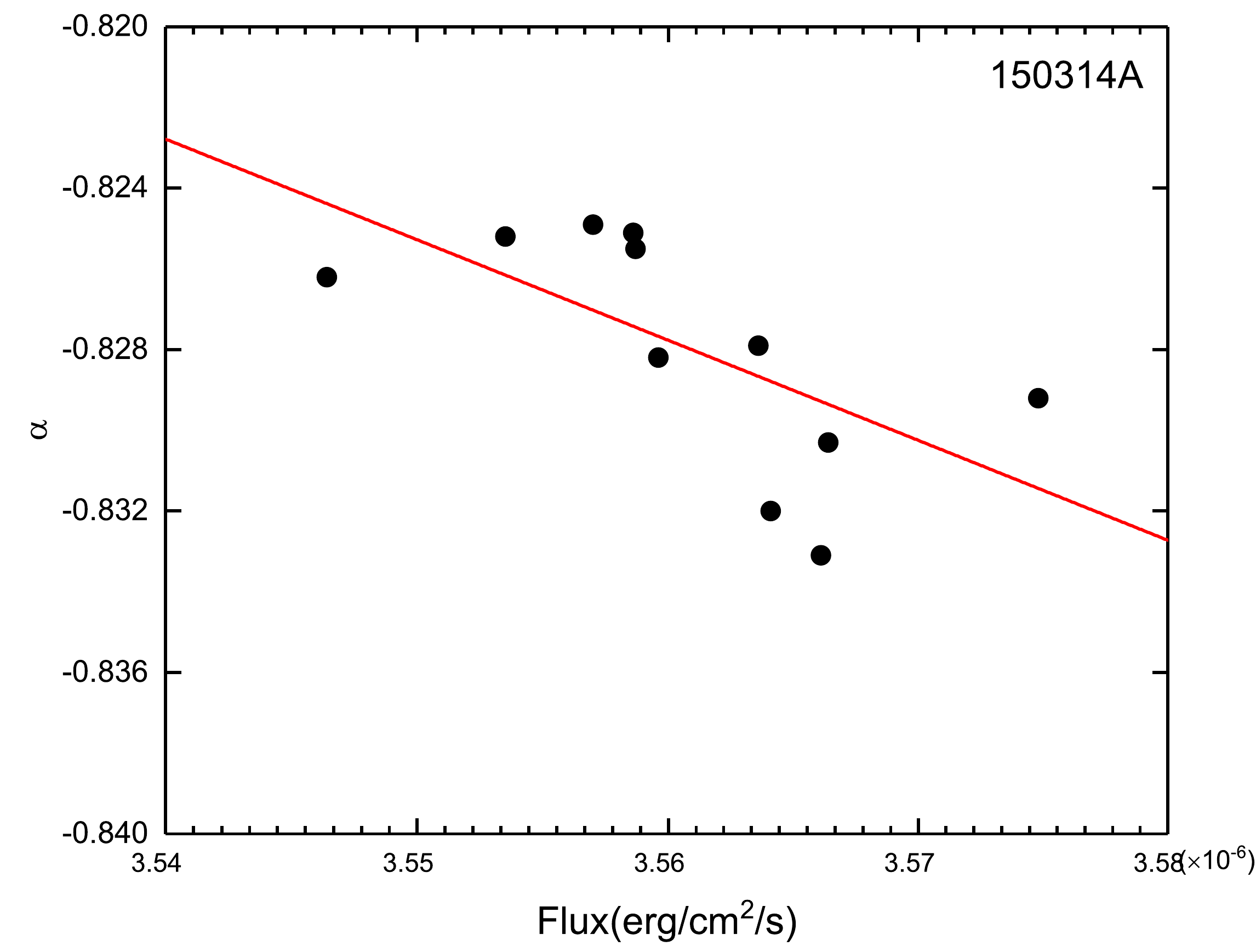}}
\resizebox{4cm}{!}{\includegraphics{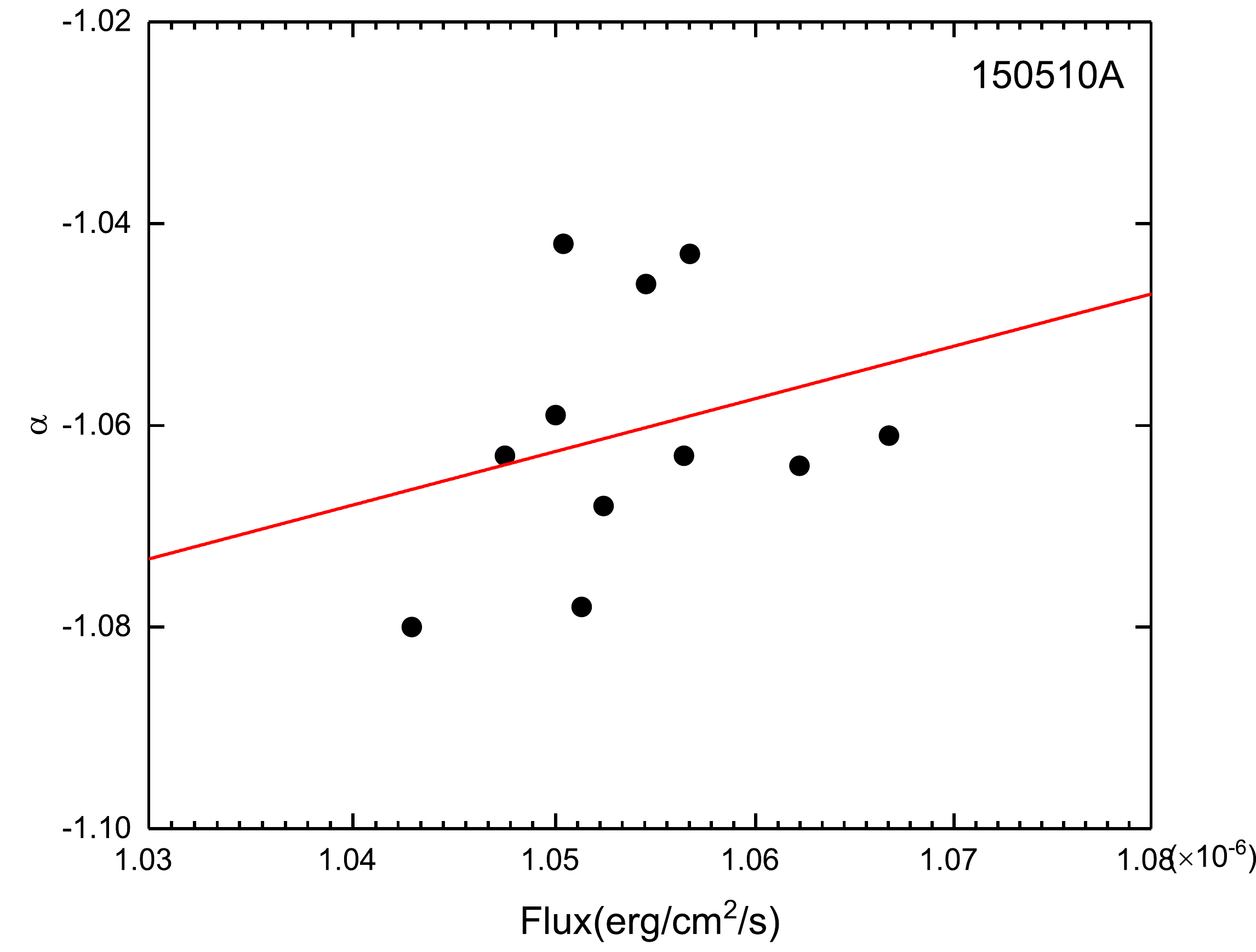}}
\resizebox{4cm}{!}{\includegraphics{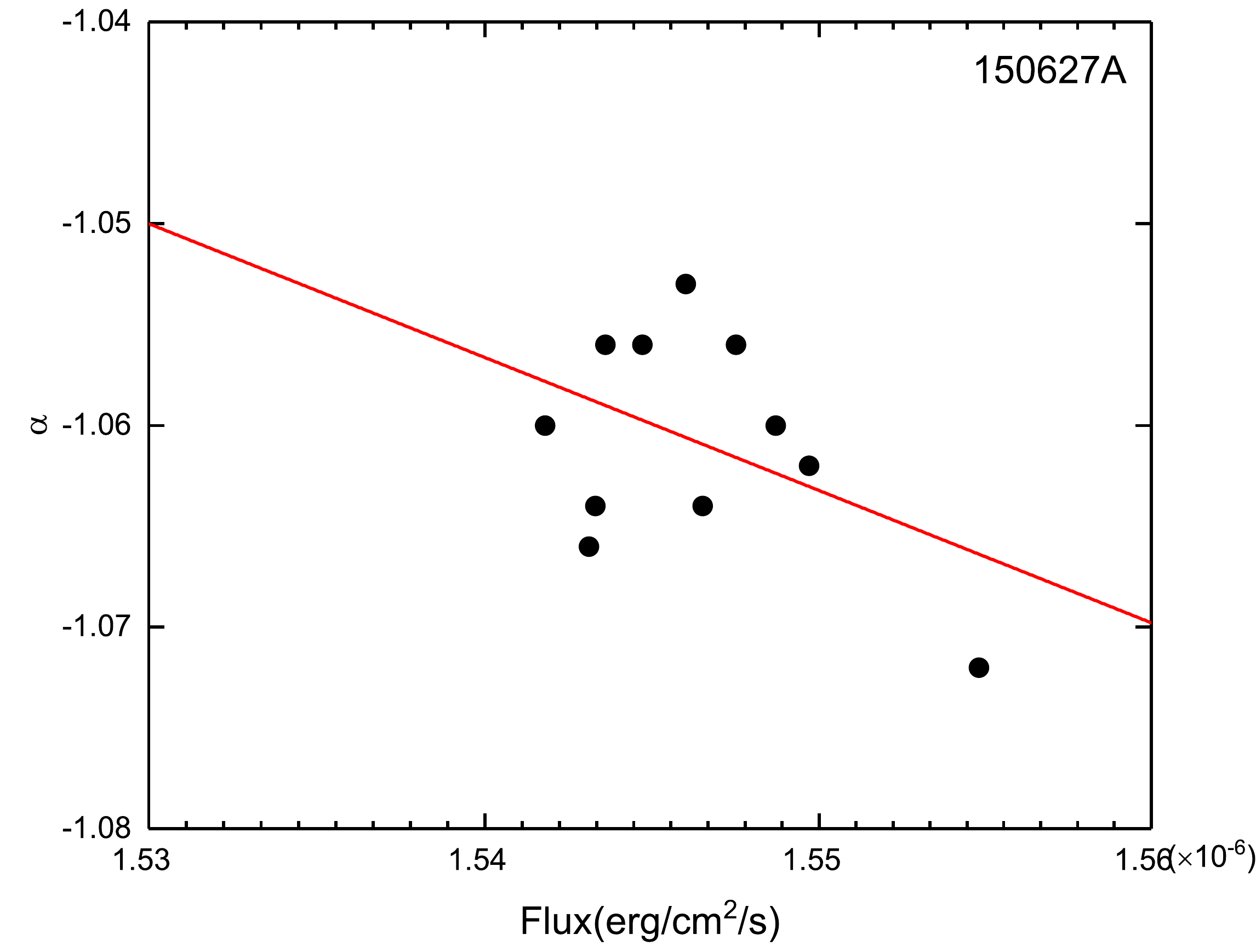}}
\resizebox{4cm}{!}{\includegraphics{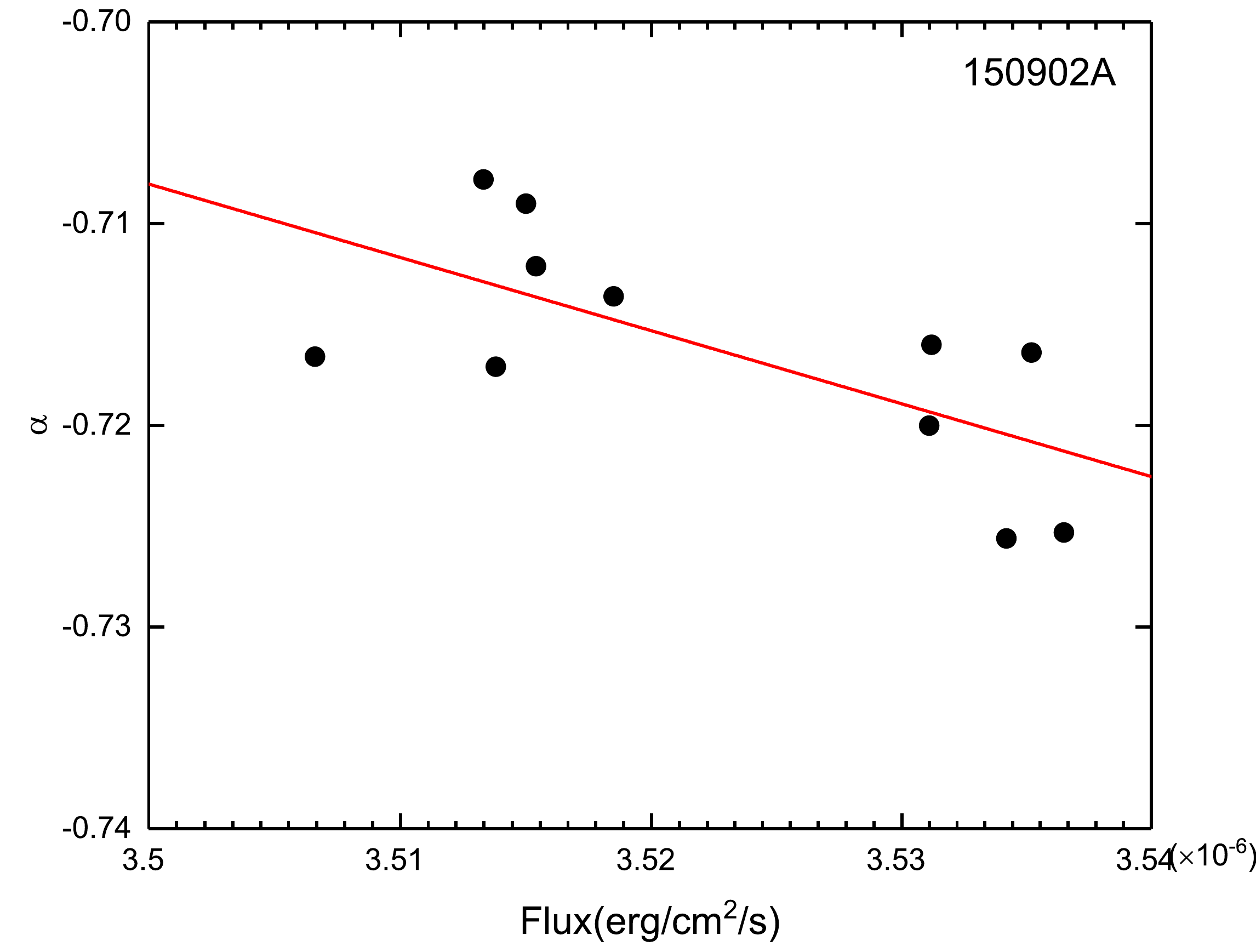}}
\resizebox{4cm}{!}{\includegraphics{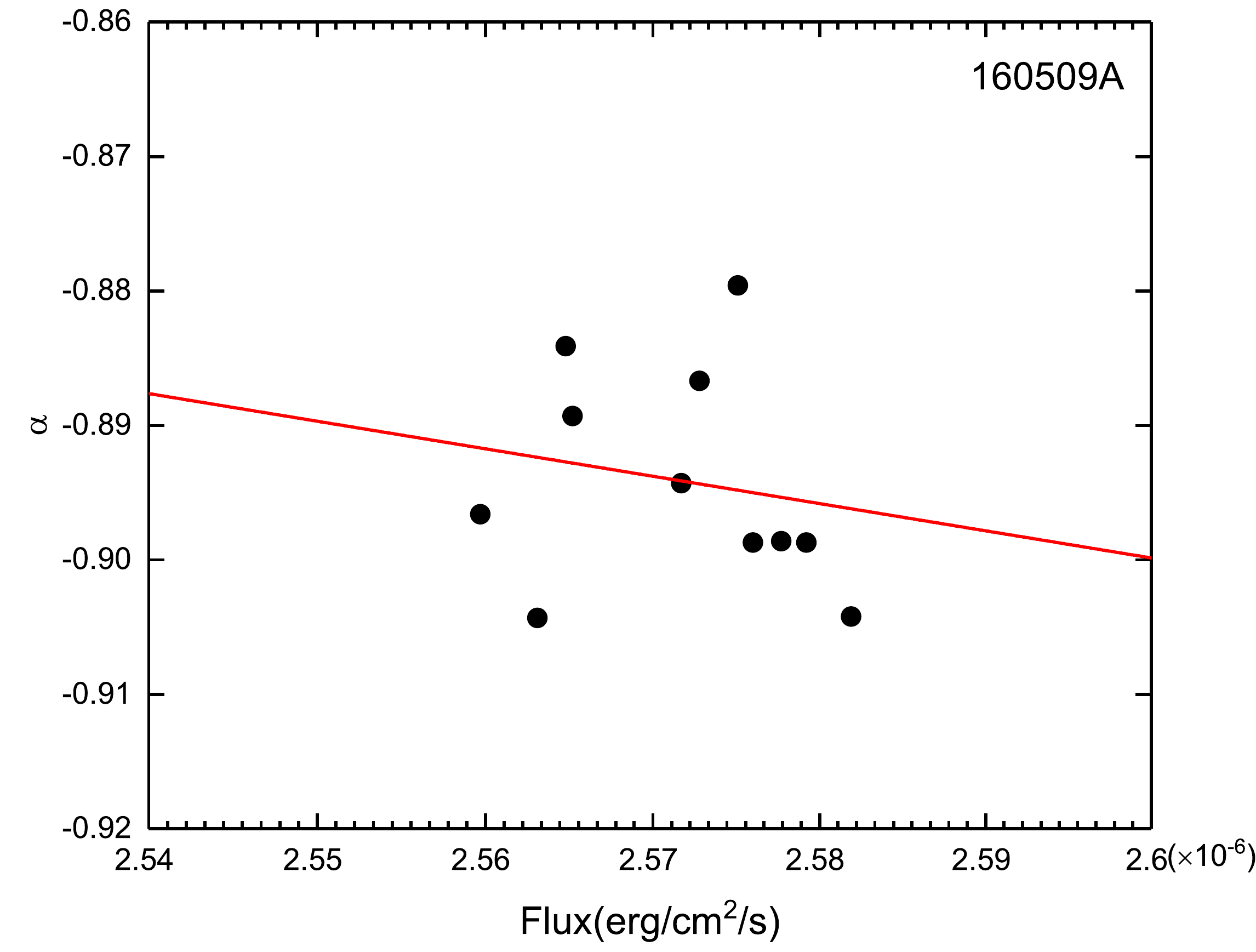}}
\resizebox{4cm}{!}{\includegraphics{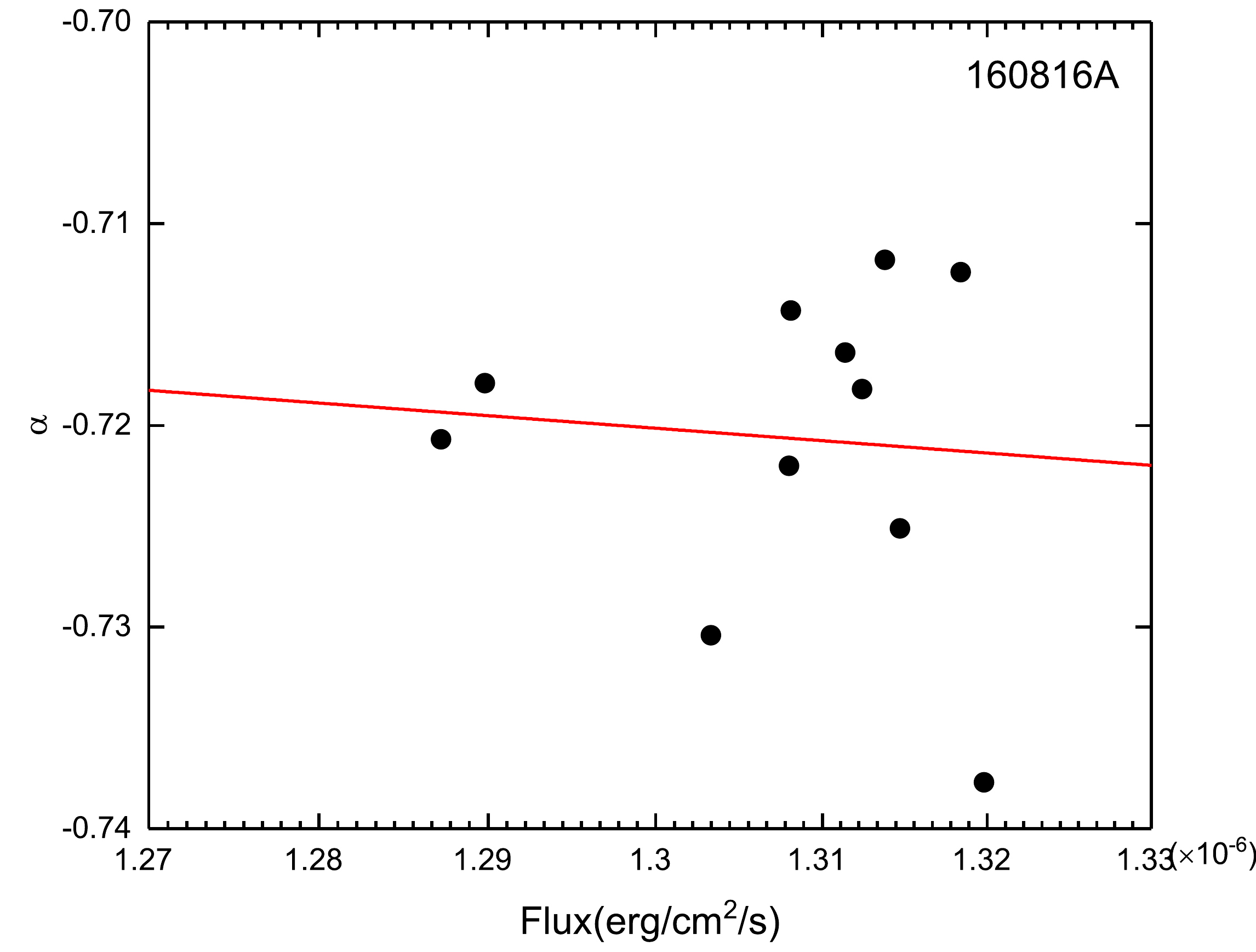}}
\resizebox{4cm}{!}{\includegraphics{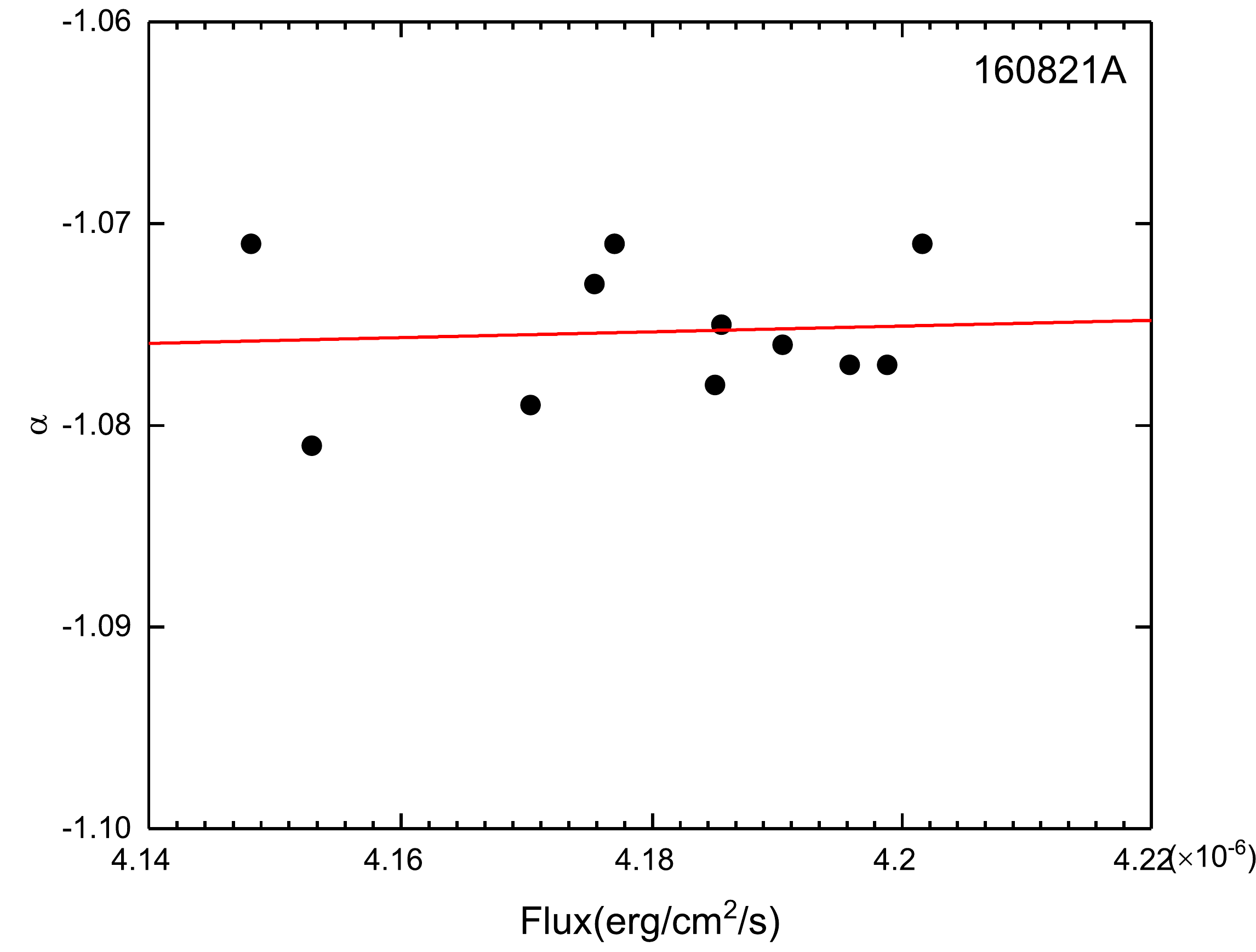}}
\resizebox{4cm}{!}{\includegraphics{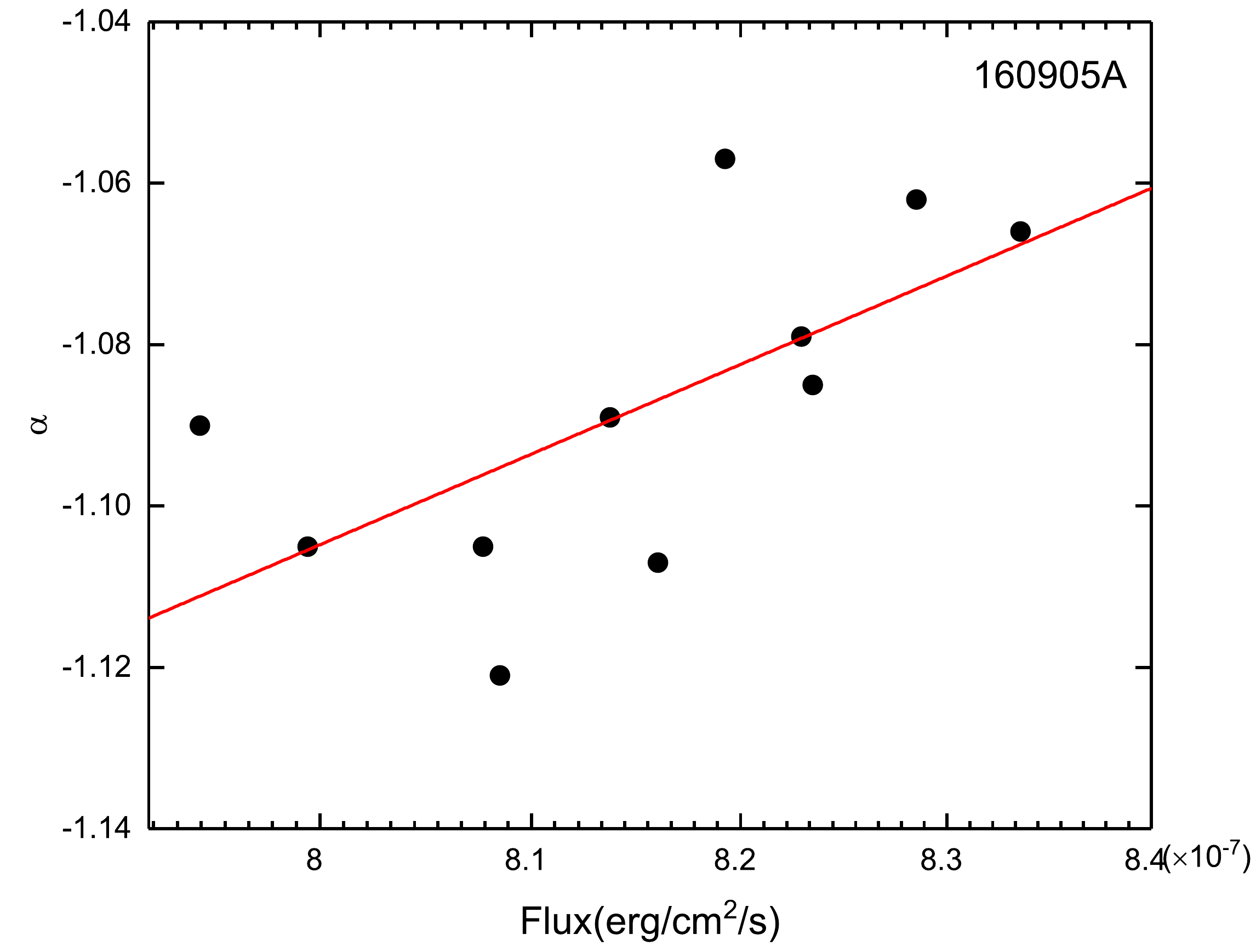}}
\resizebox{4cm}{!}{\includegraphics{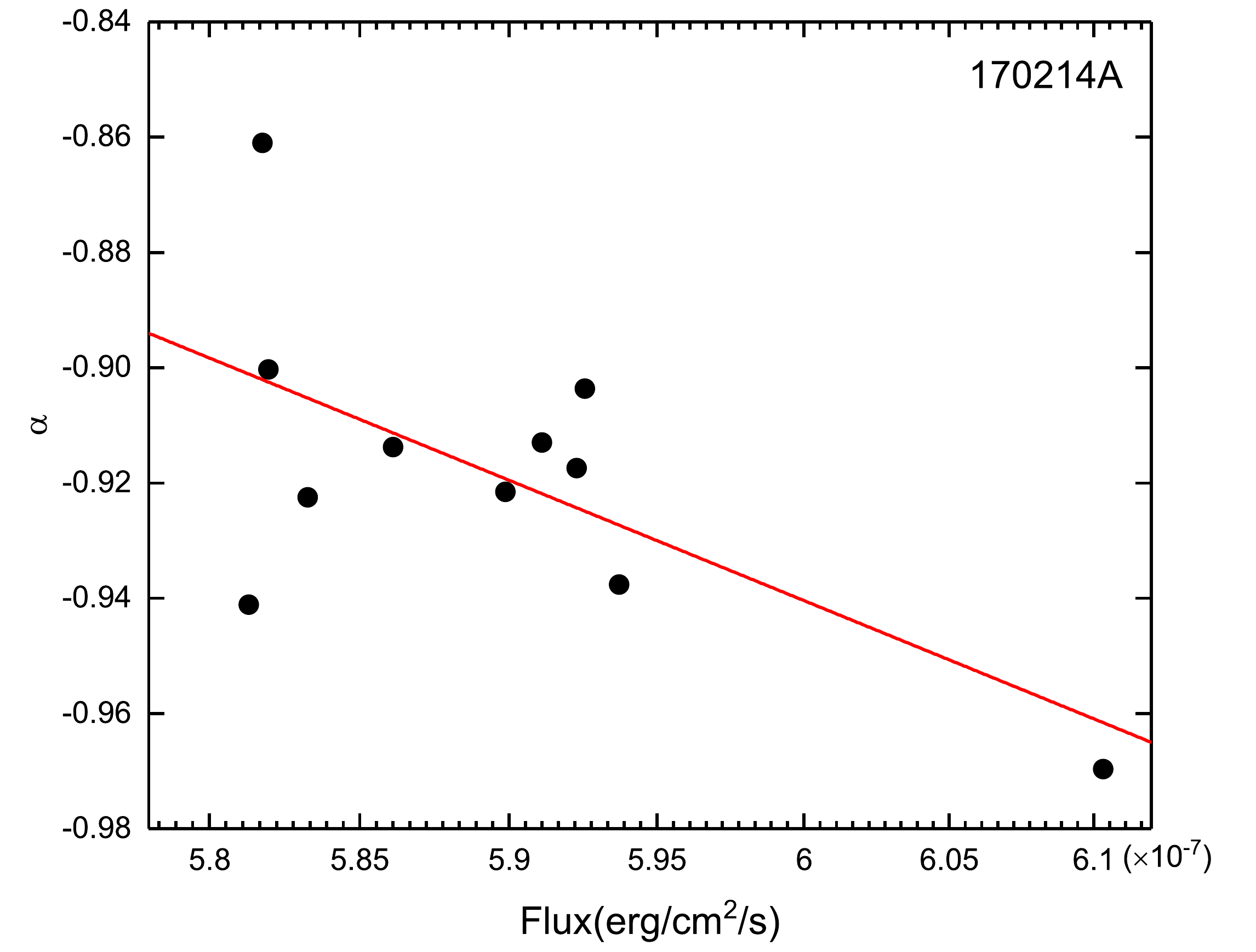}}
\resizebox{4cm}{!}{\includegraphics{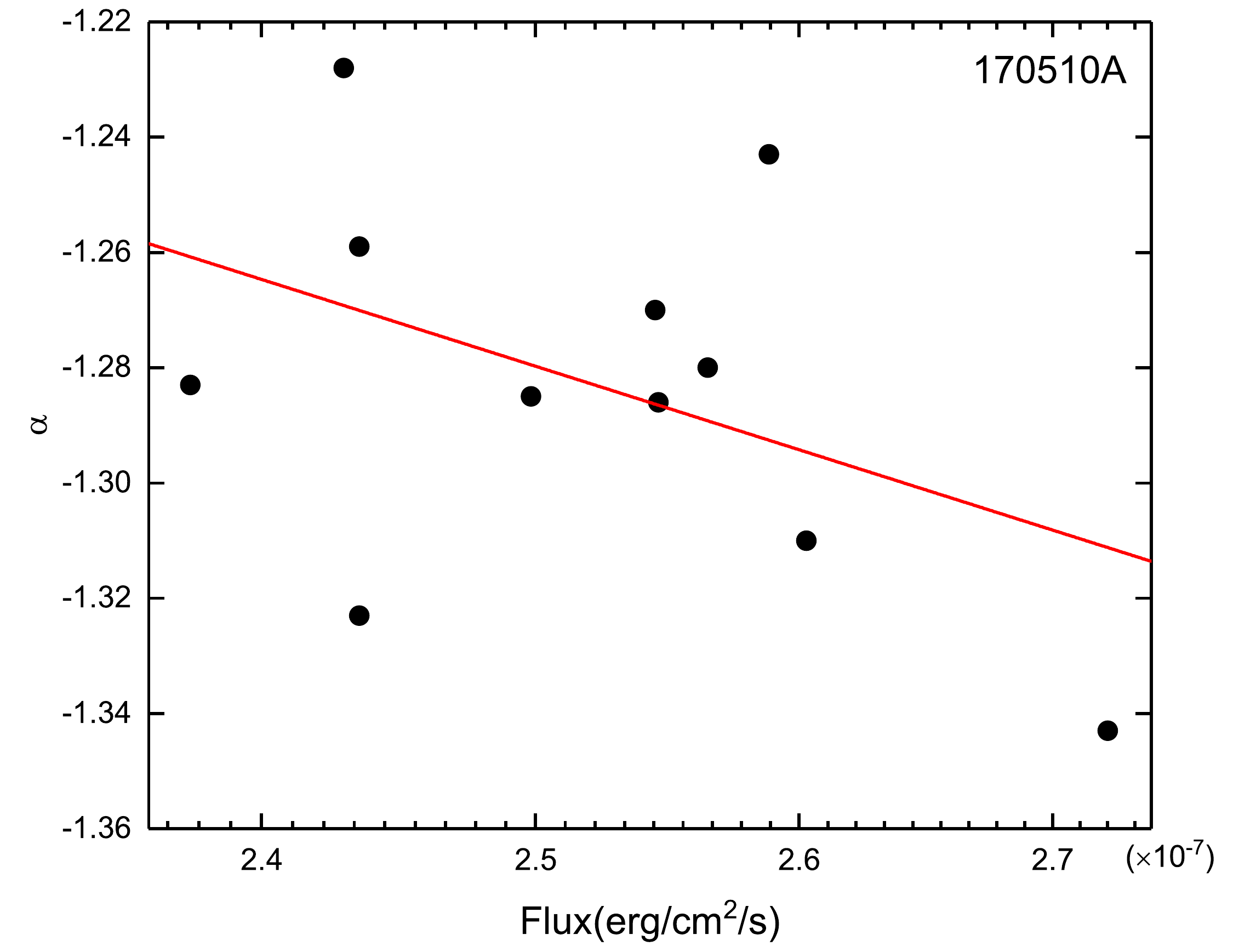}}
\caption{\edit1{\added{The $\alpha-F$ correlation from the simulation for 23 GRBs which exhibit a strong positive correlation in $\alpha-F$ correlation. The red solid line represents the best-linear-fitting result for each burst.}}\label{fig:simulation}}
\end{figure}

\begin{figure}
\centering
\resizebox{4cm}{!}{\includegraphics{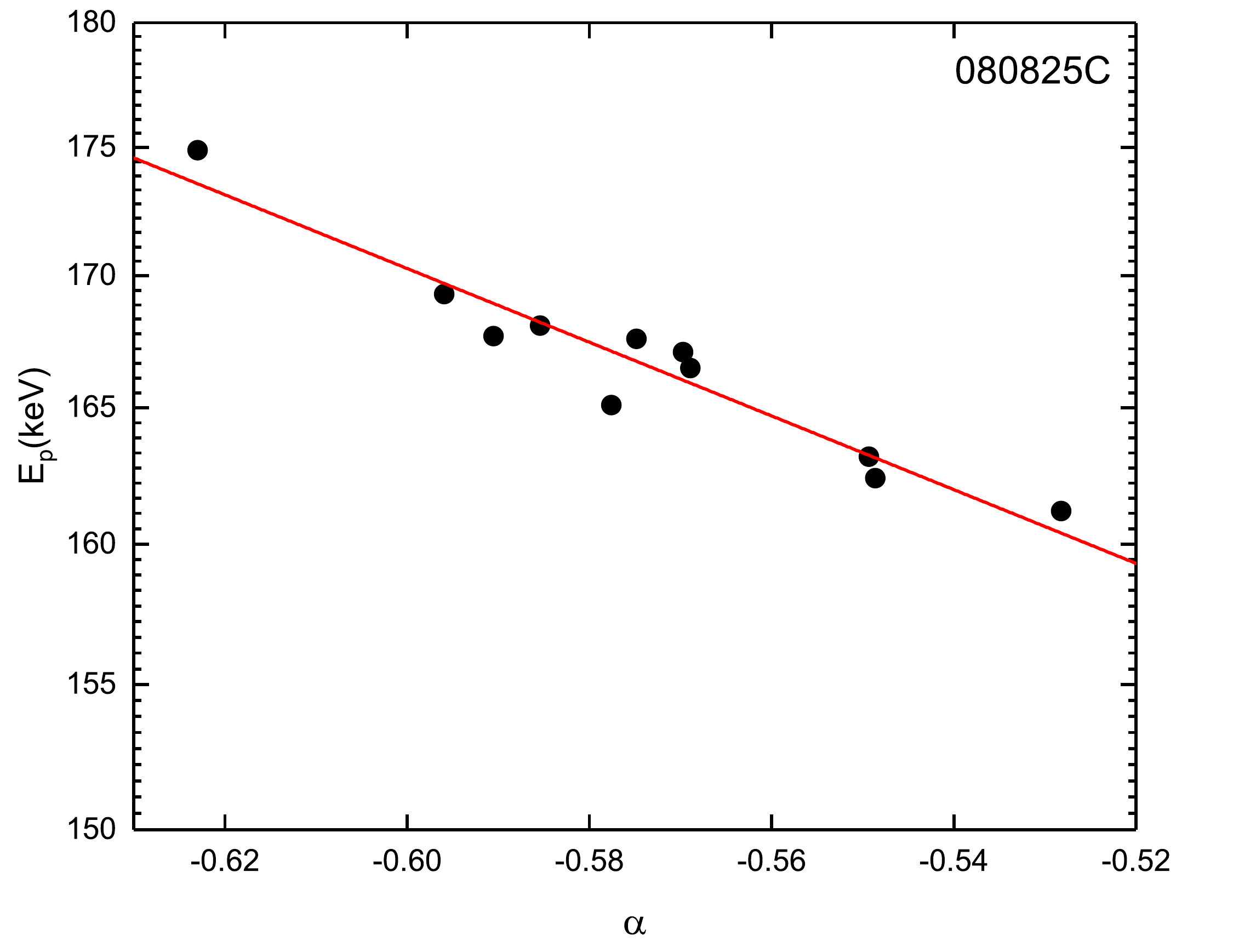}}
\resizebox{4cm}{!}{\includegraphics{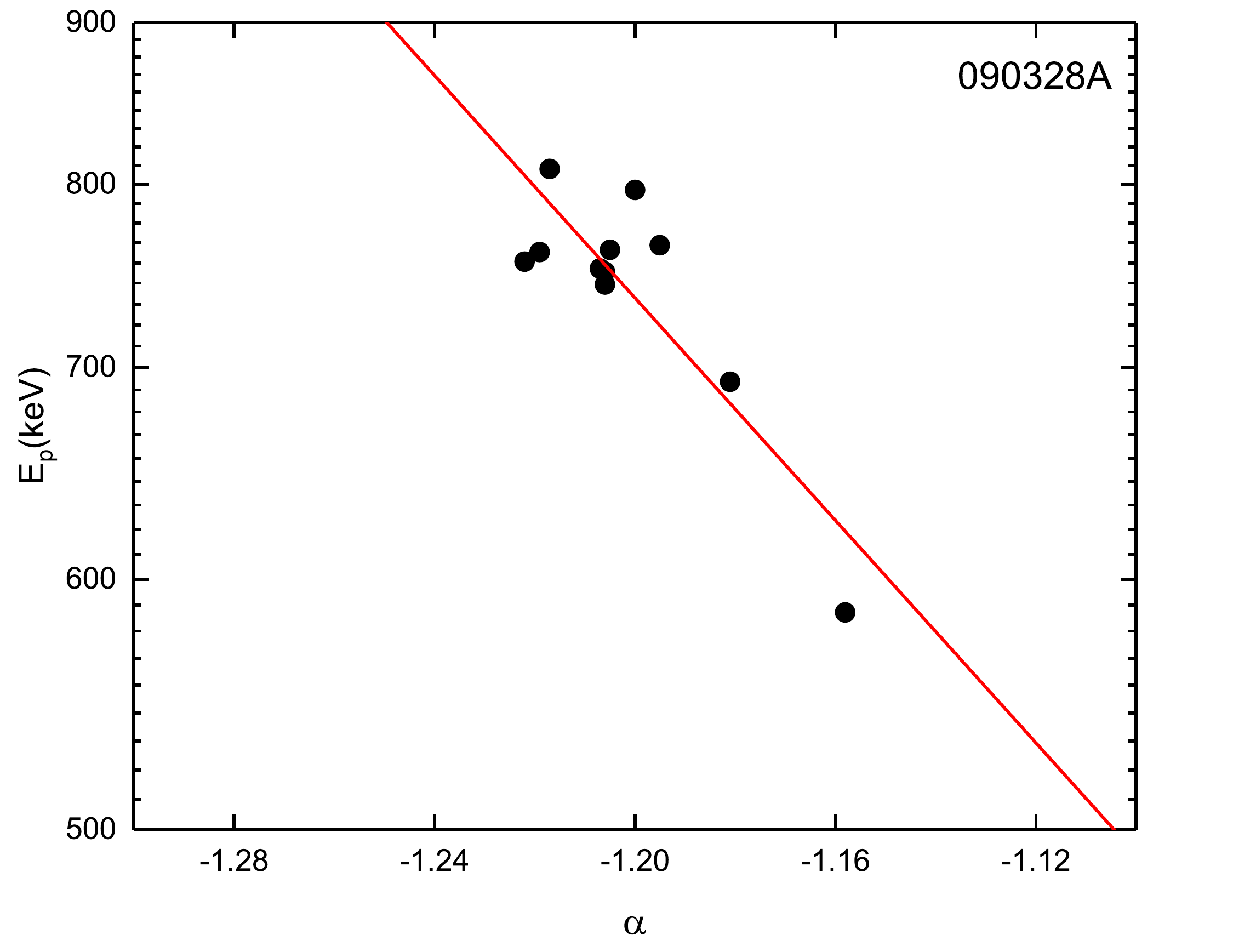}}
\resizebox{4cm}{!}{\includegraphics{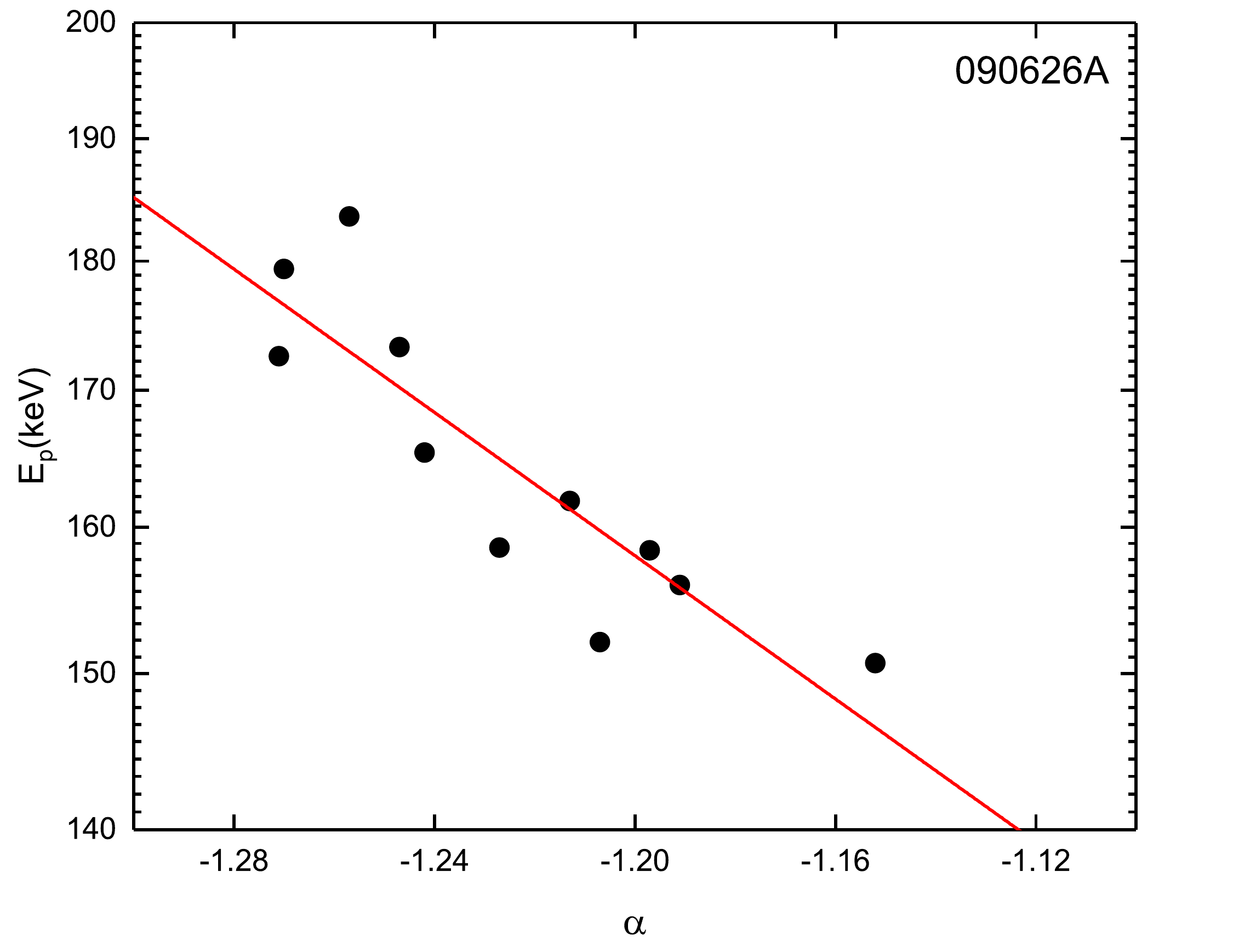}}
\resizebox{4cm}{!}{\includegraphics{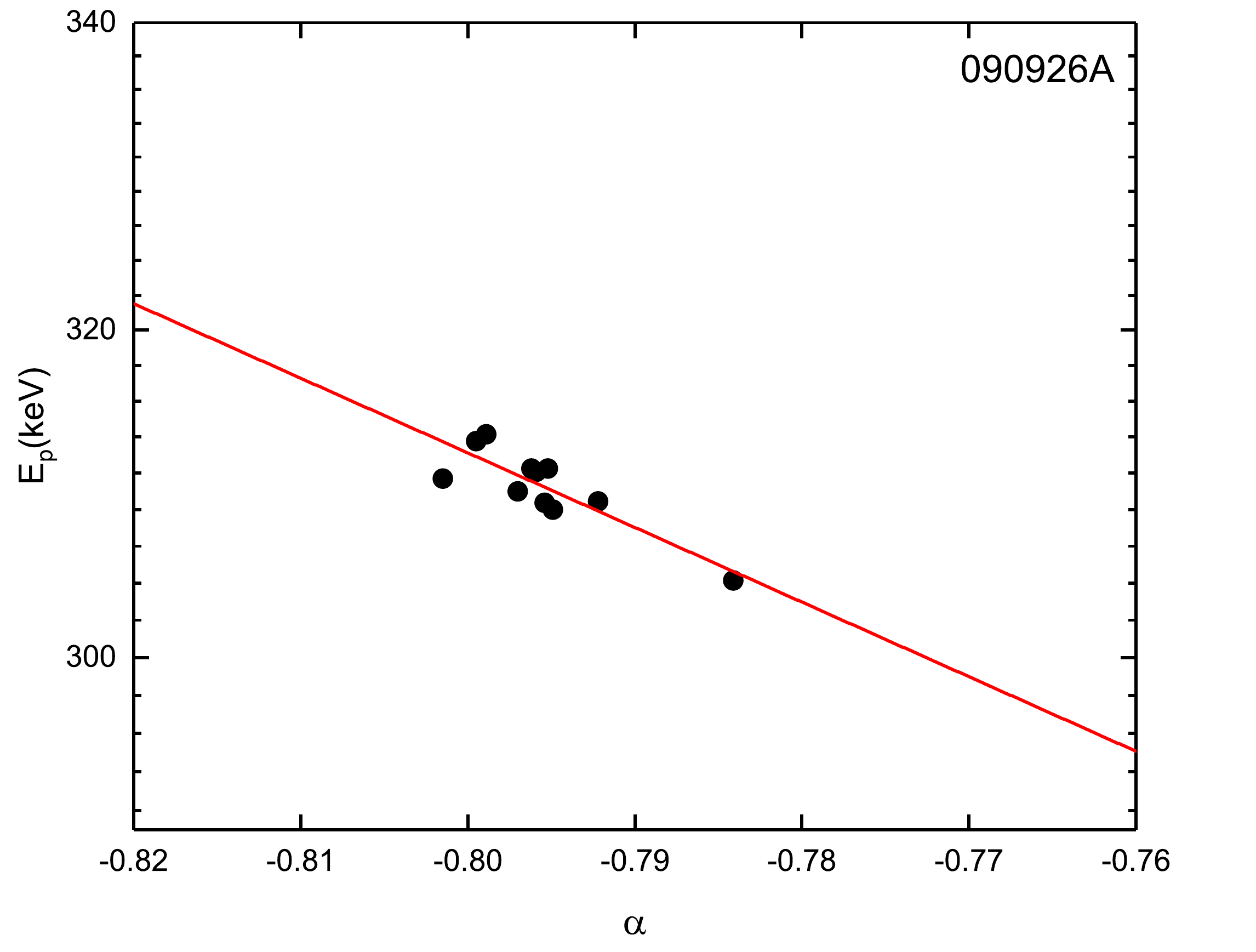}}
\resizebox{4cm}{!}{\includegraphics{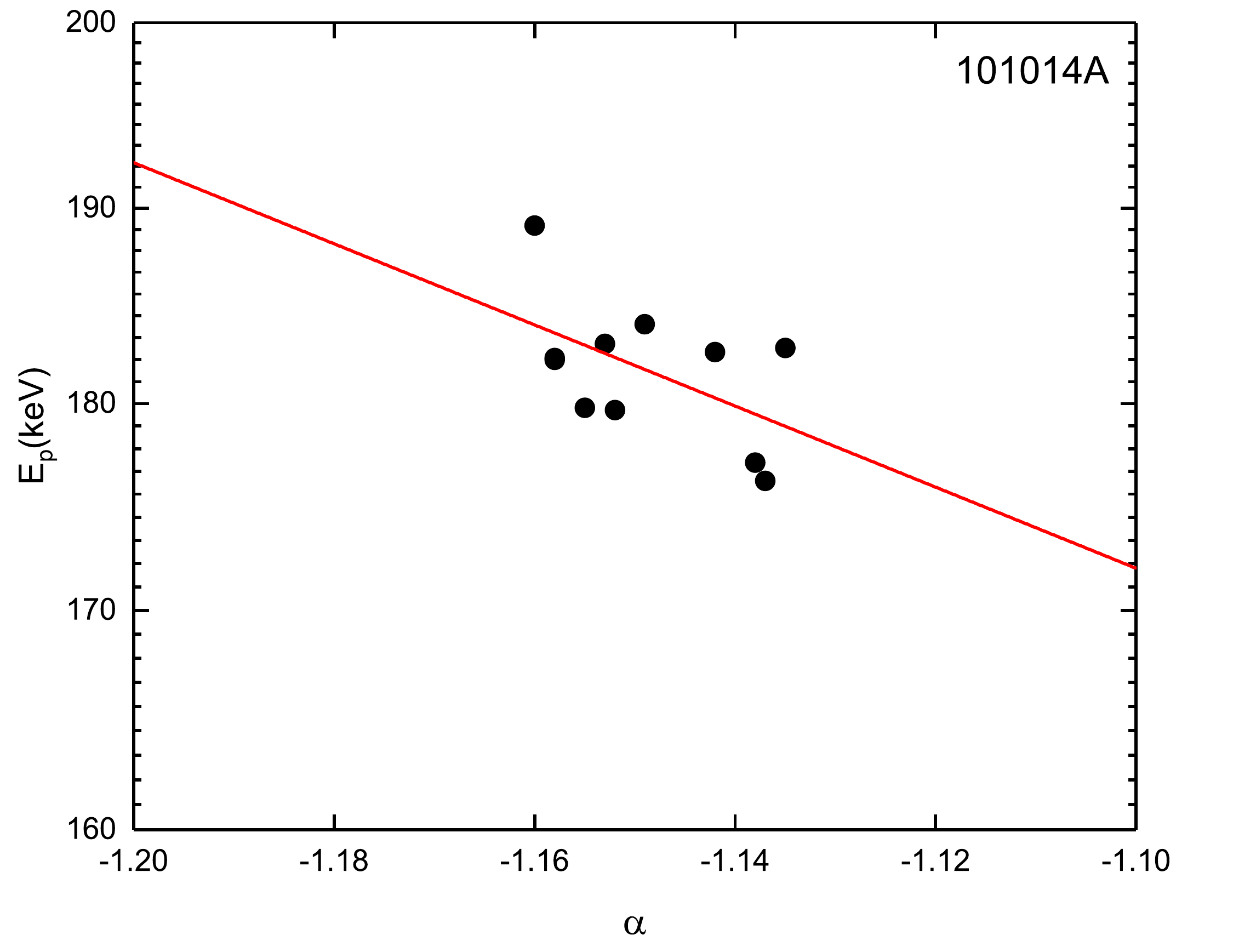}}
\resizebox{4cm}{!}{\includegraphics{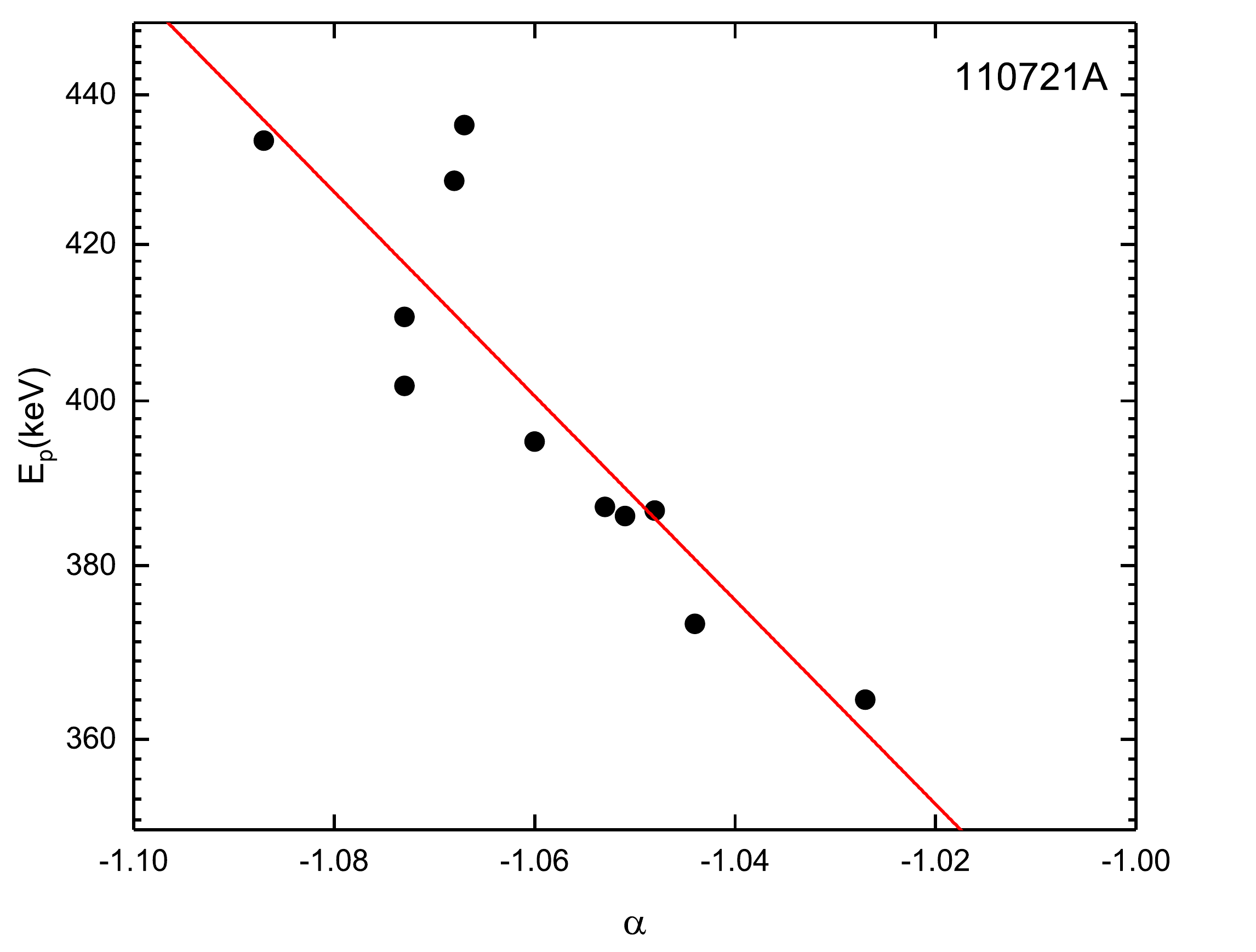}}
\resizebox{4cm}{!}{\includegraphics{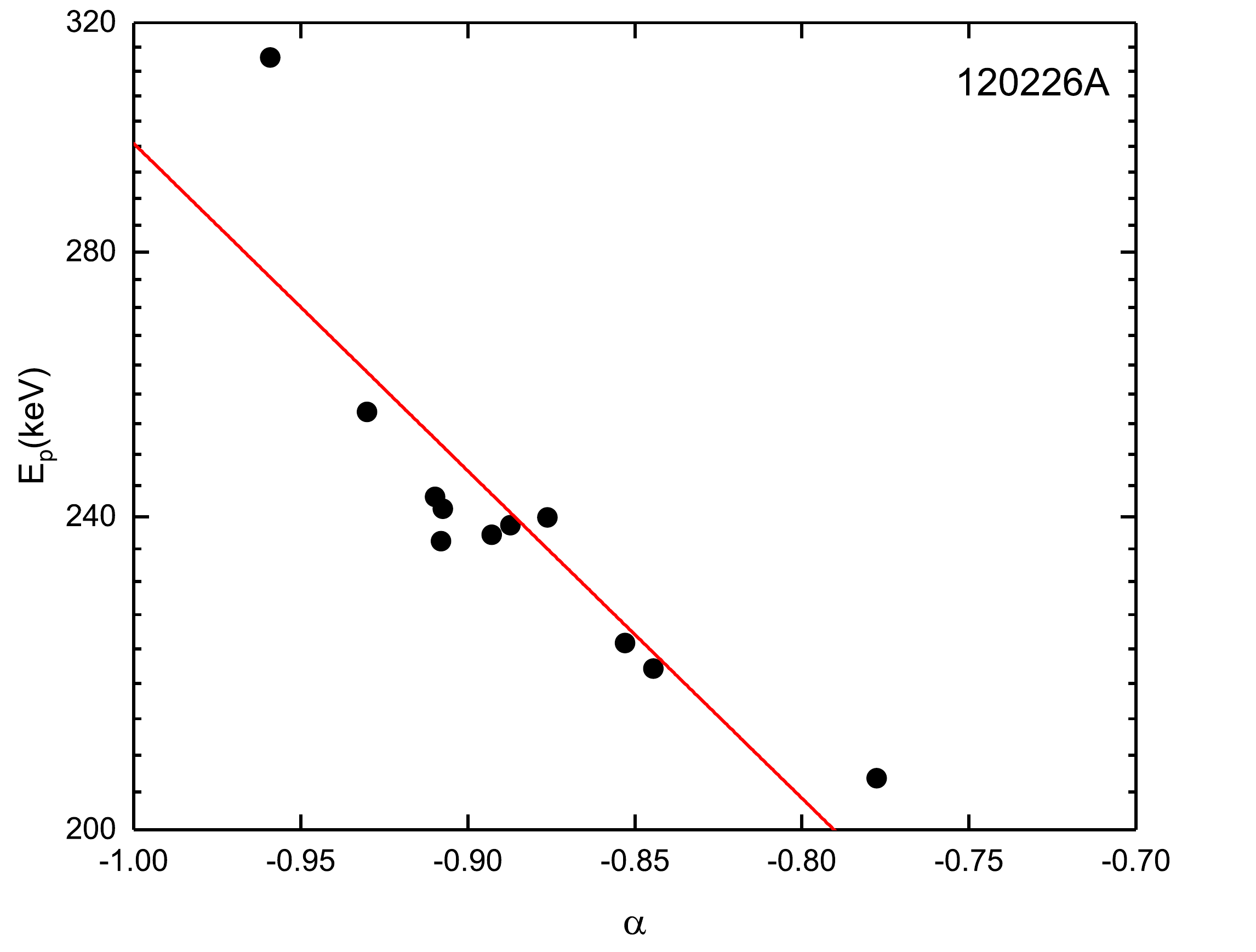}}
\resizebox{4cm}{!}{\includegraphics{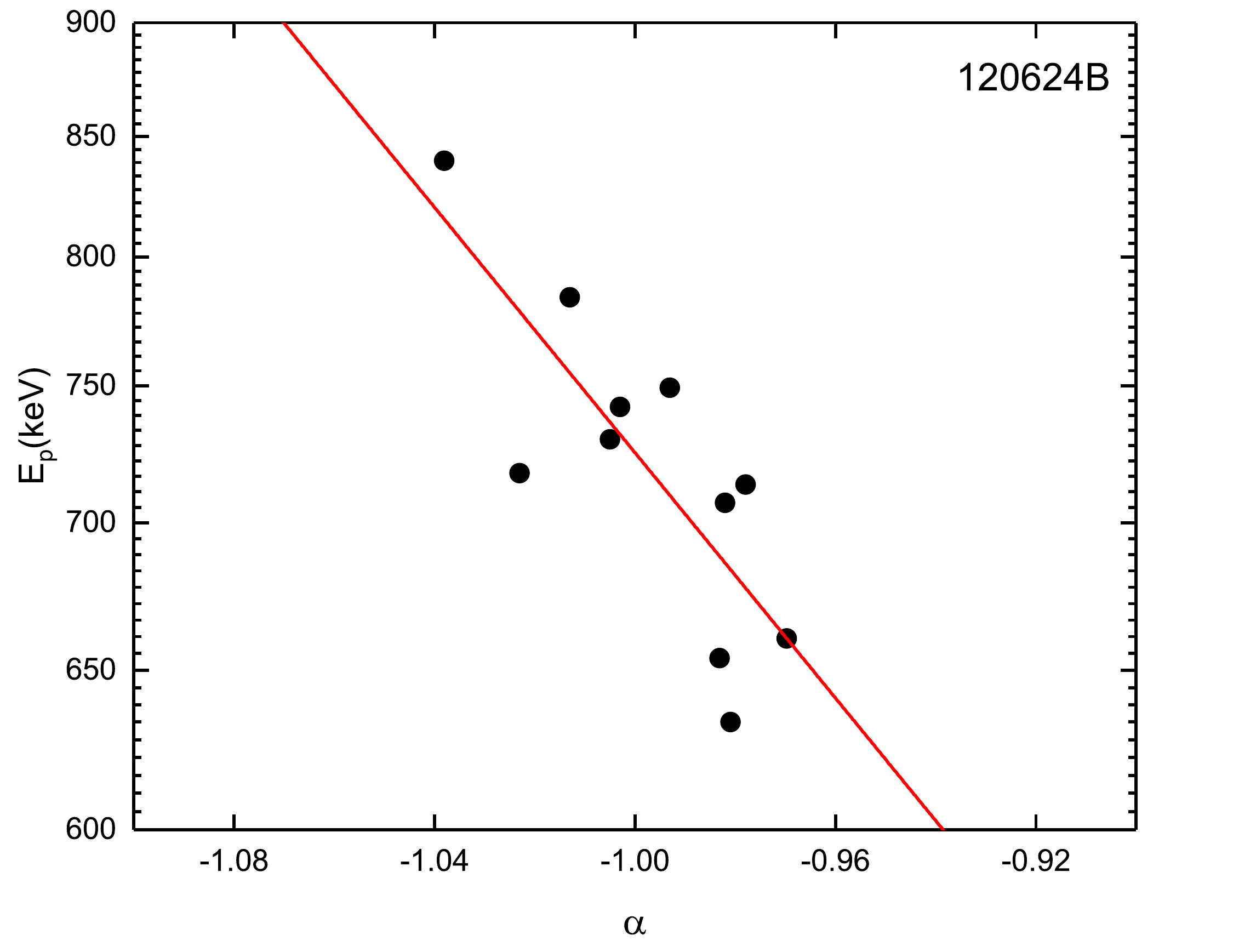}}
\resizebox{4cm}{!}{\includegraphics{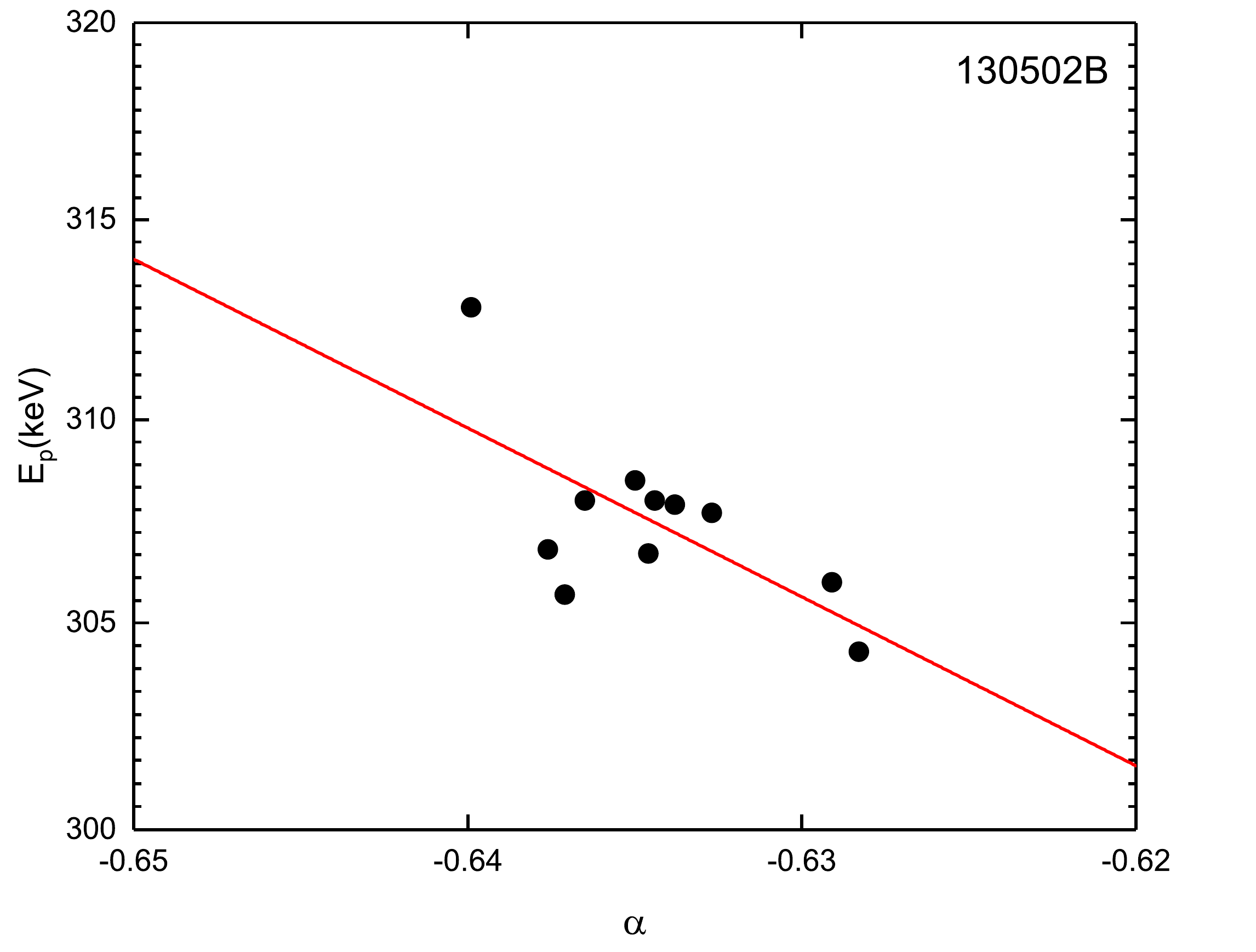}}
\resizebox{4cm}{!}{\includegraphics{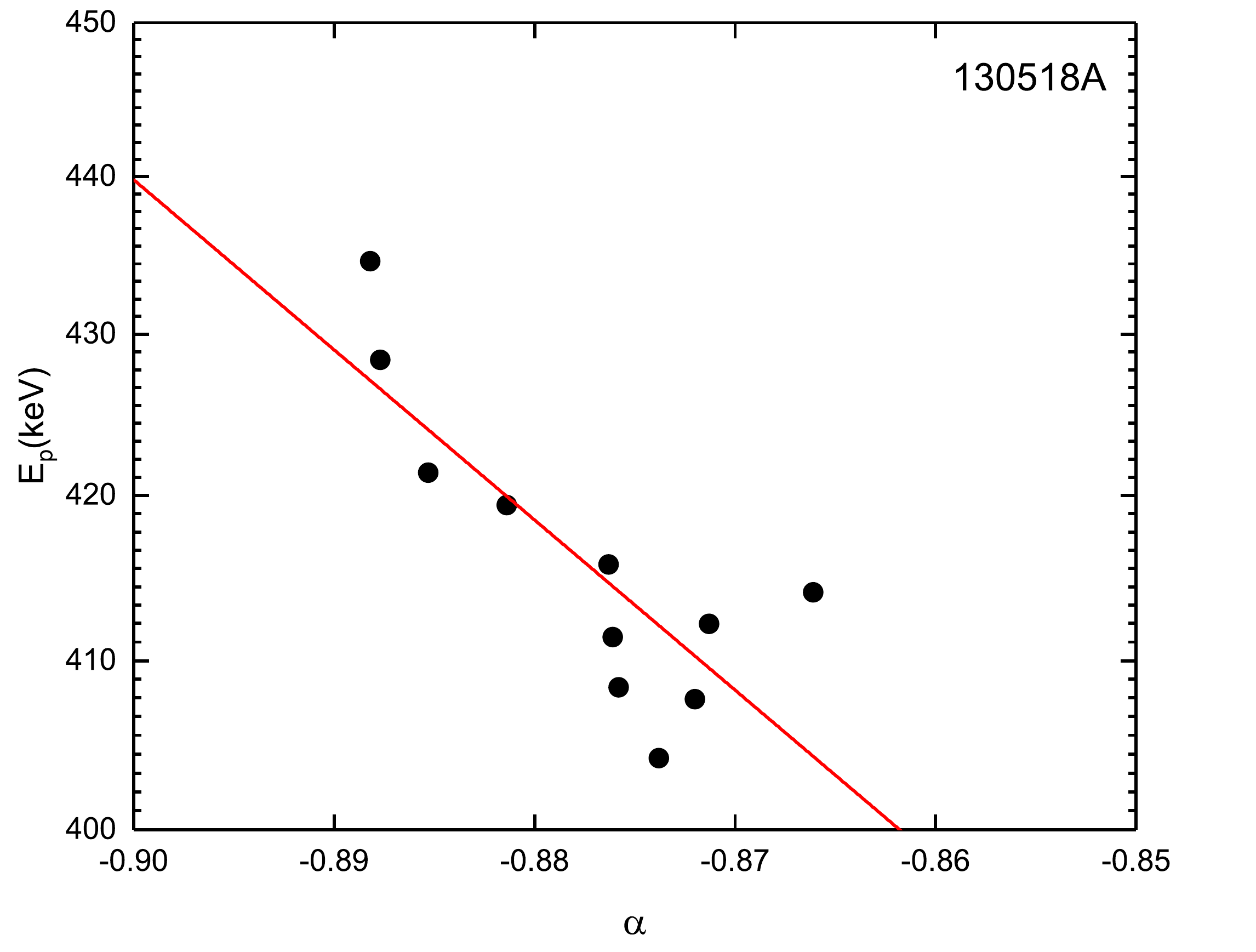}}
\resizebox{4cm}{!}{\includegraphics{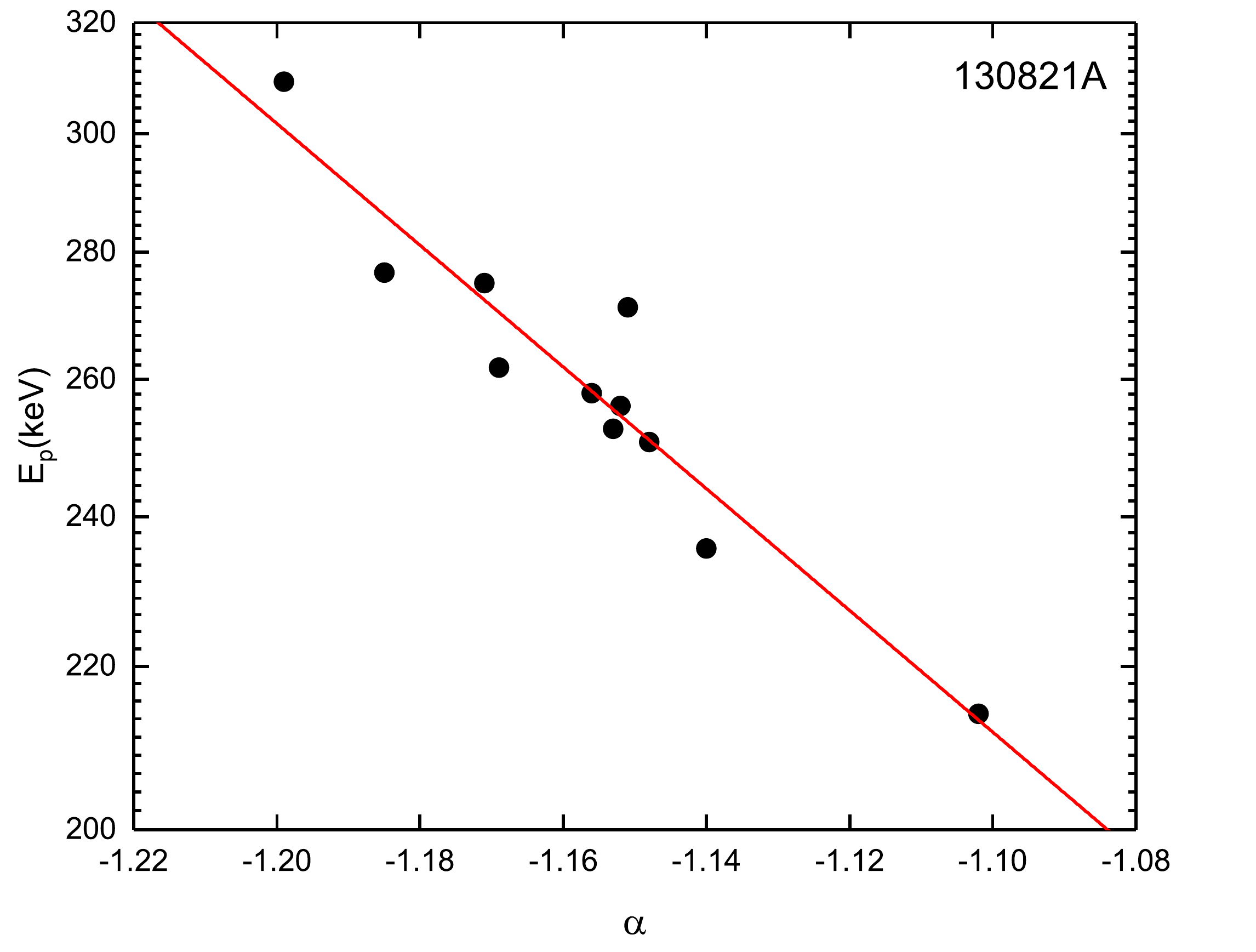}}
\resizebox{4cm}{!}{\includegraphics{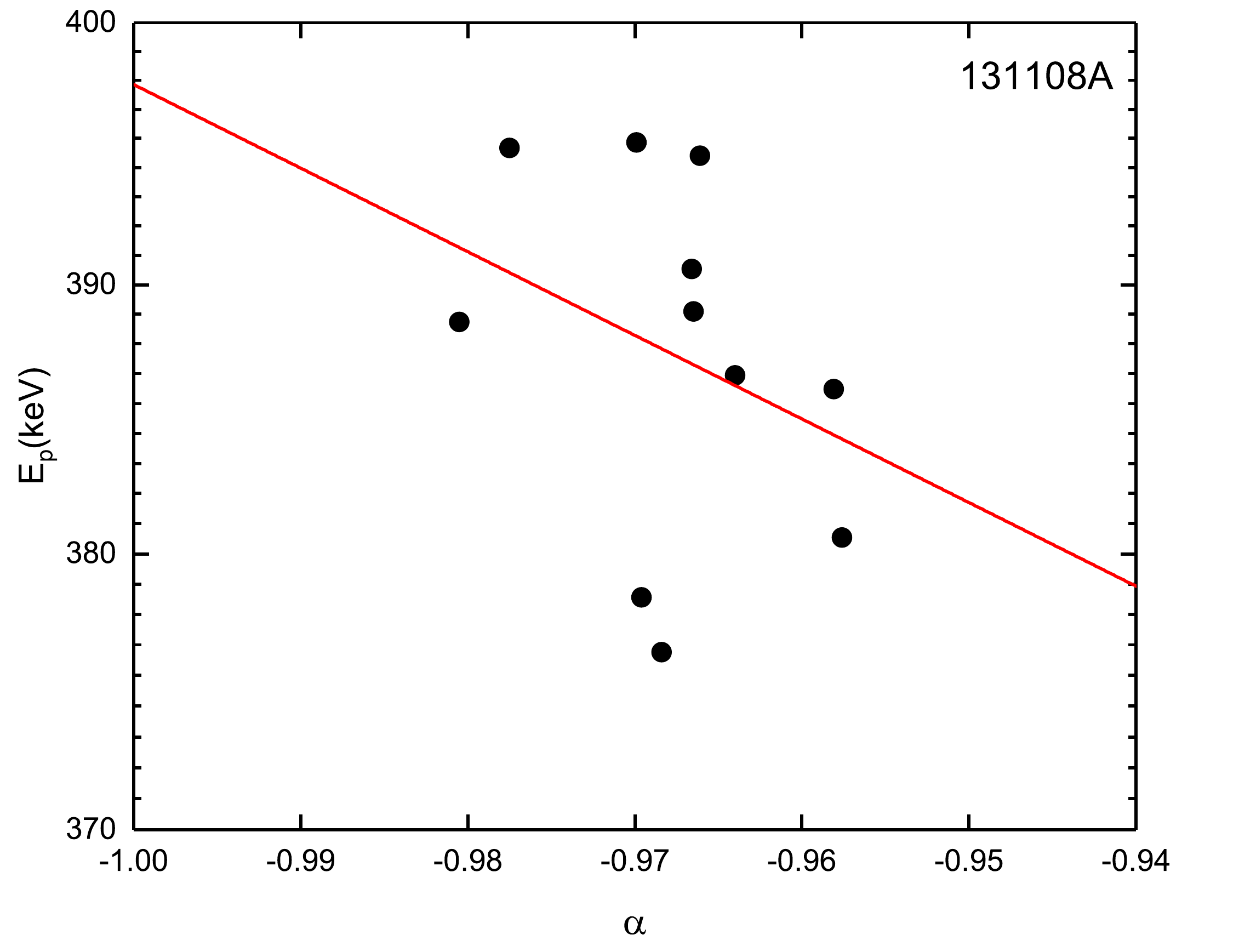}}
\resizebox{4cm}{!}{\includegraphics{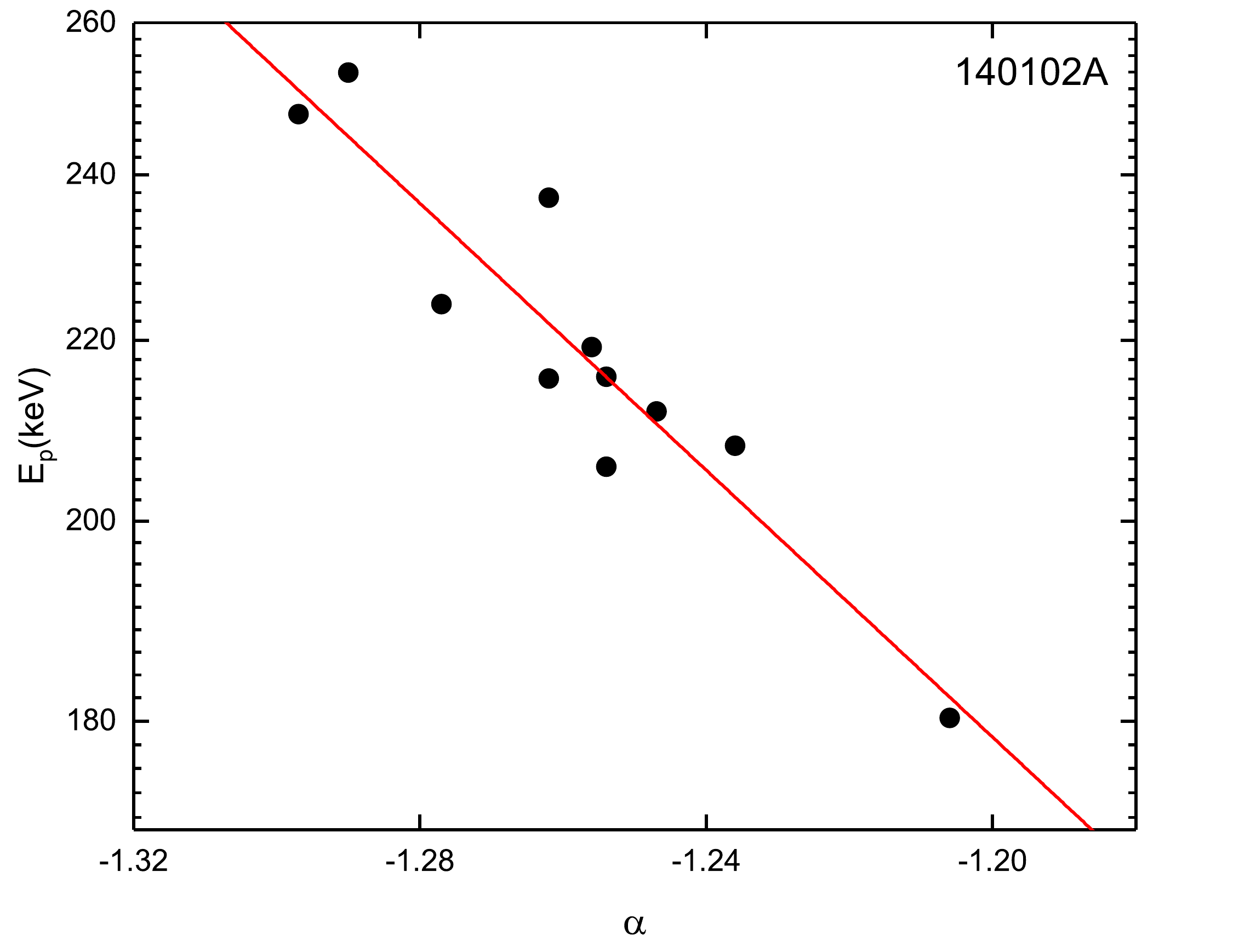}}
\resizebox{4cm}{!}{\includegraphics{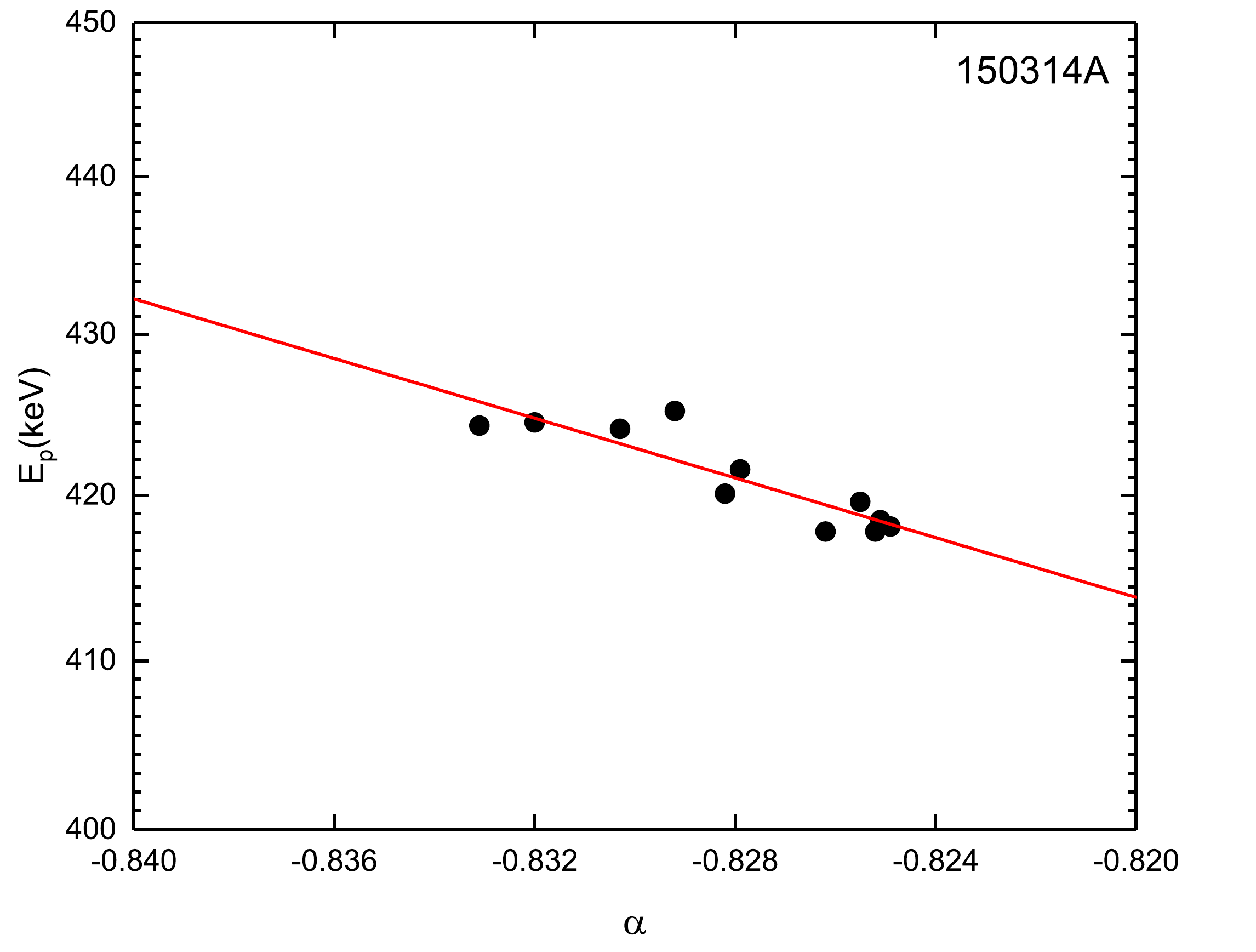}}
\resizebox{4cm}{!}{\includegraphics{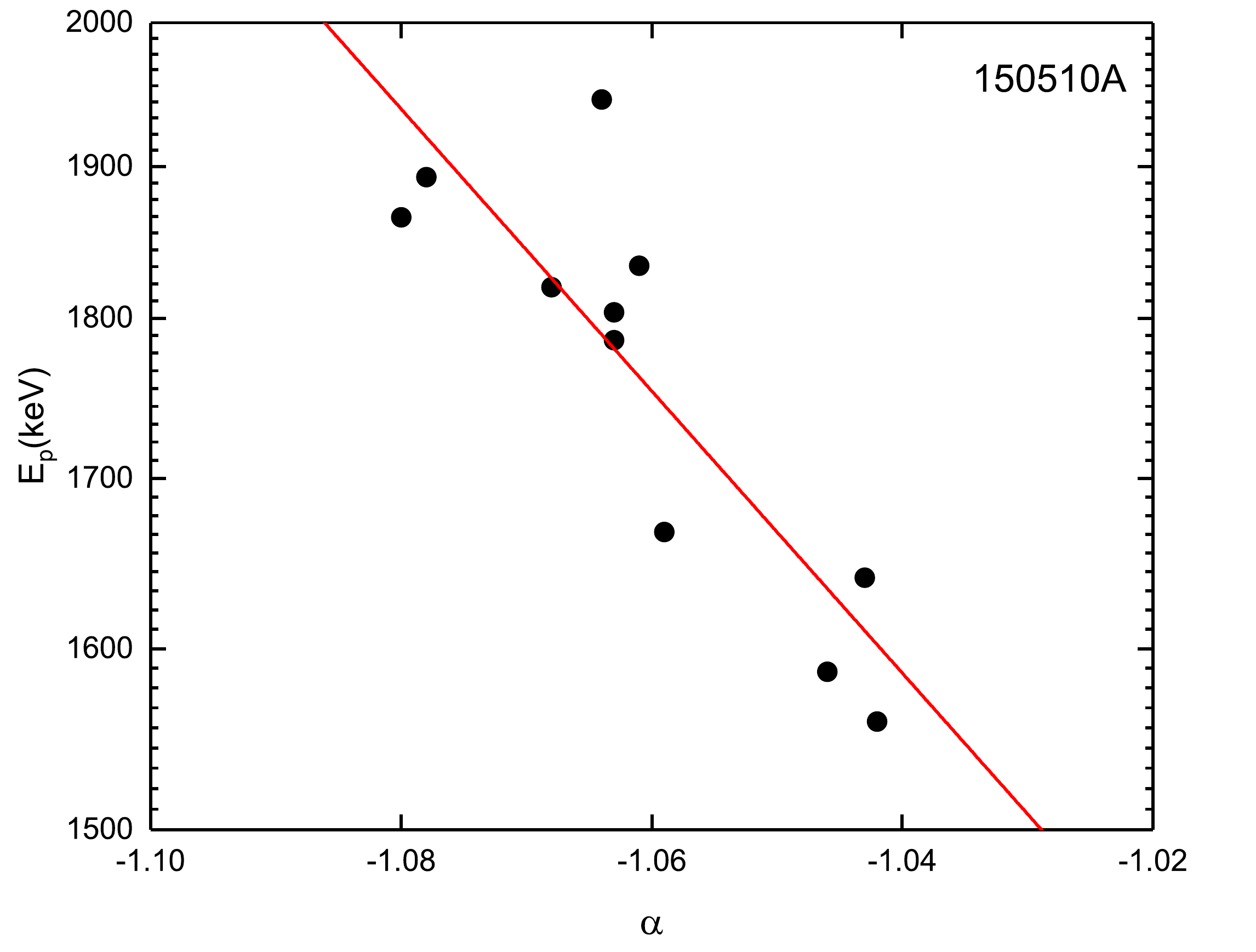}}
\resizebox{4cm}{!}{\includegraphics{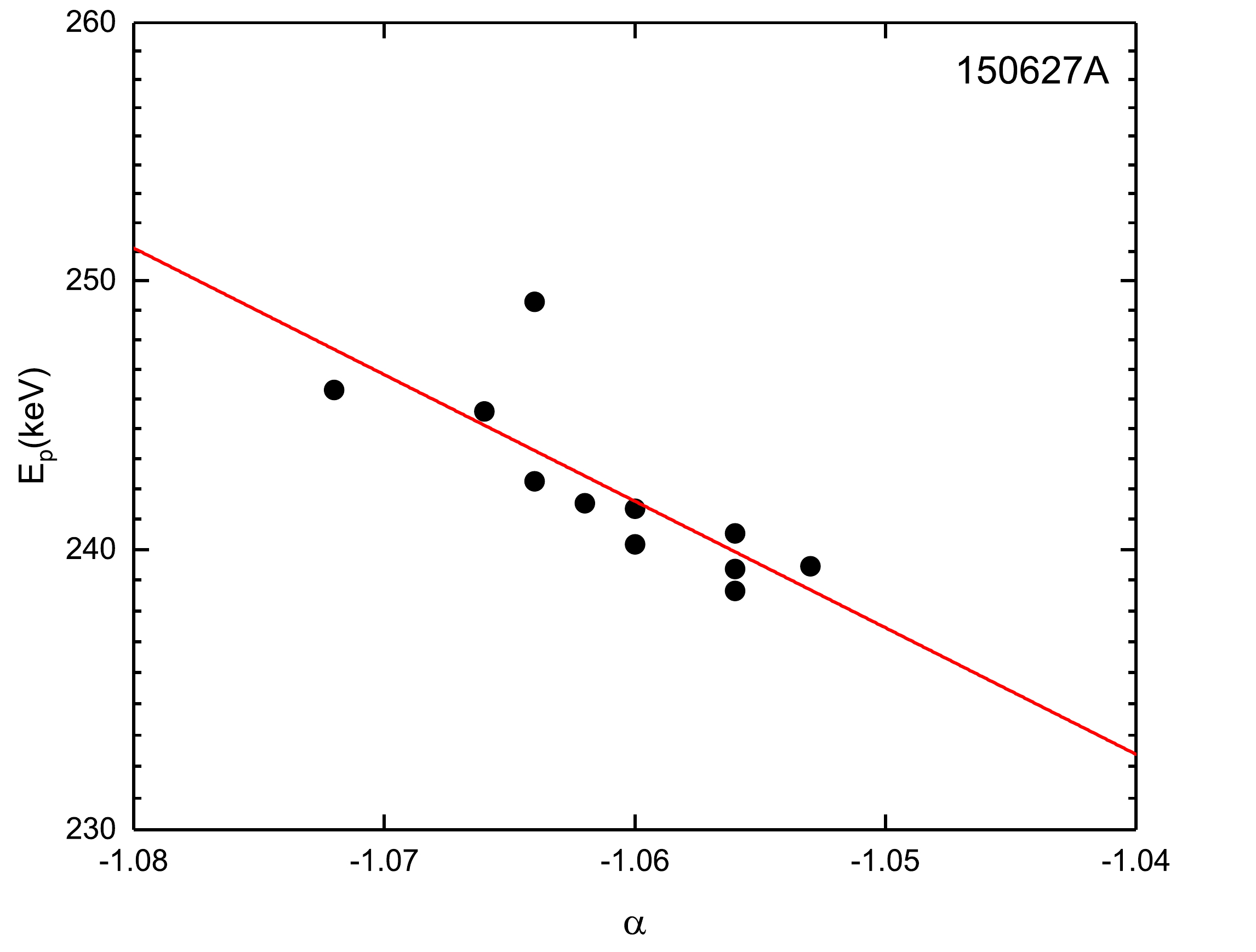}}
\resizebox{4cm}{!}{\includegraphics{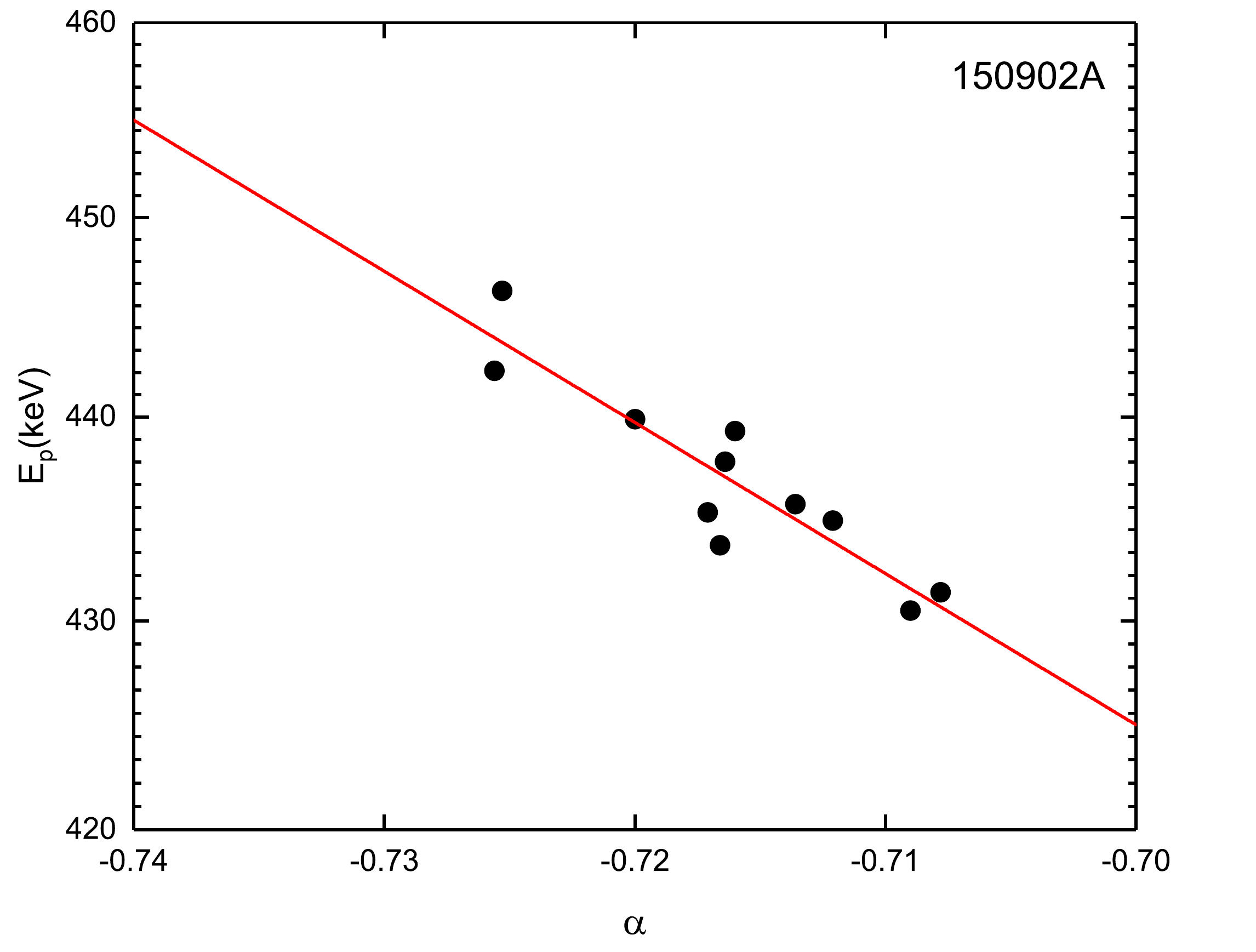}}
\resizebox{4cm}{!}{\includegraphics{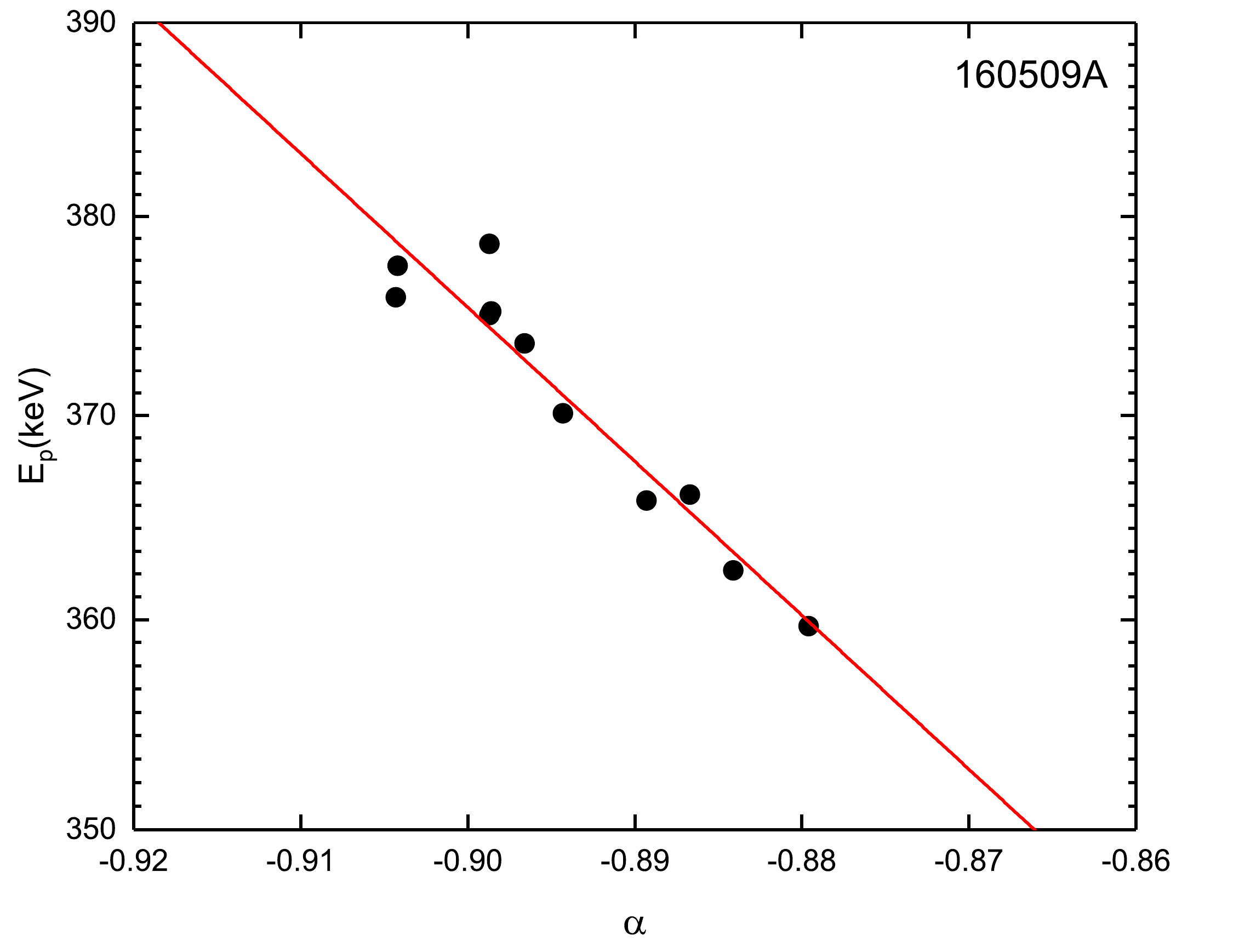}}
\resizebox{4cm}{!}{\includegraphics{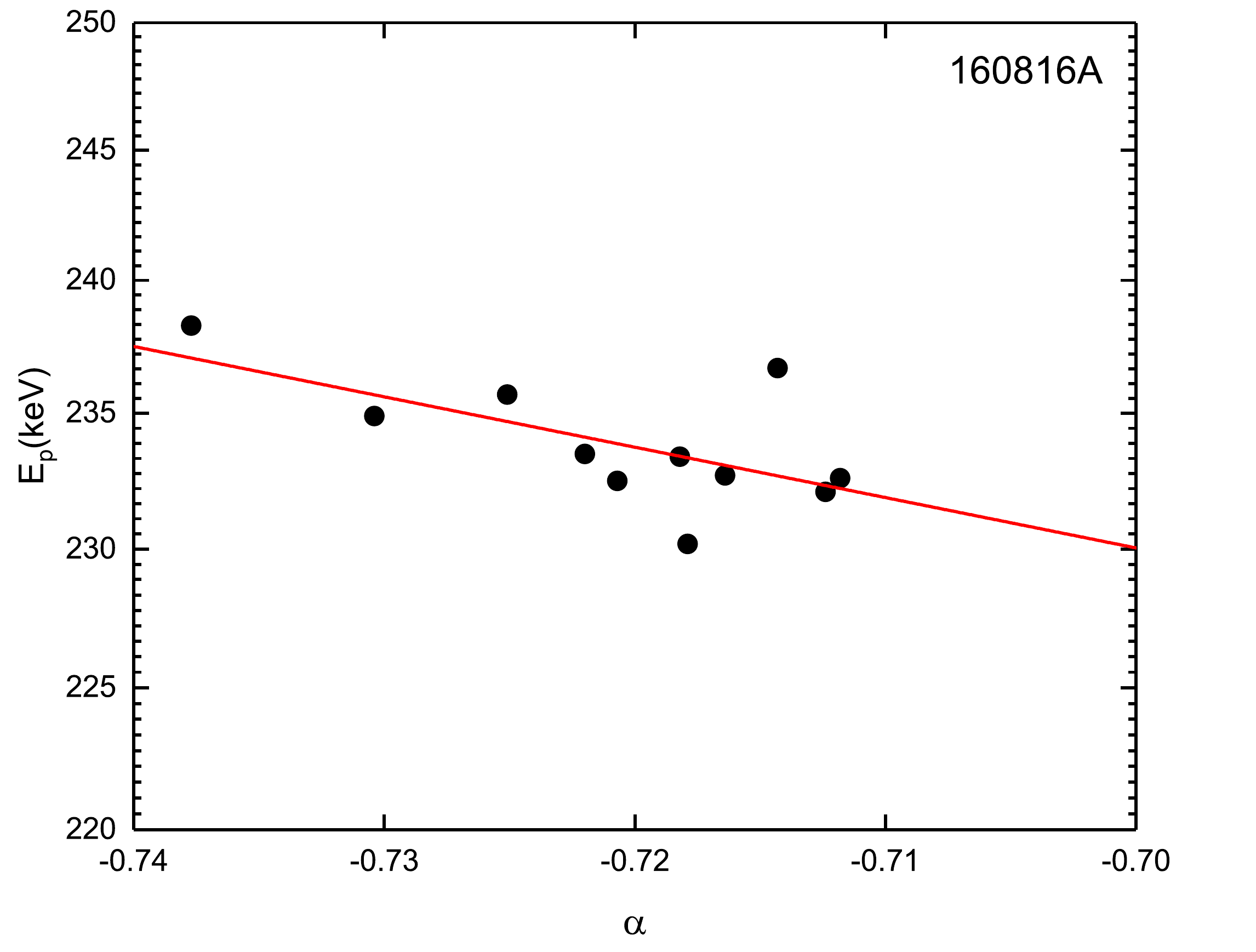}}
\resizebox{4cm}{!}{\includegraphics{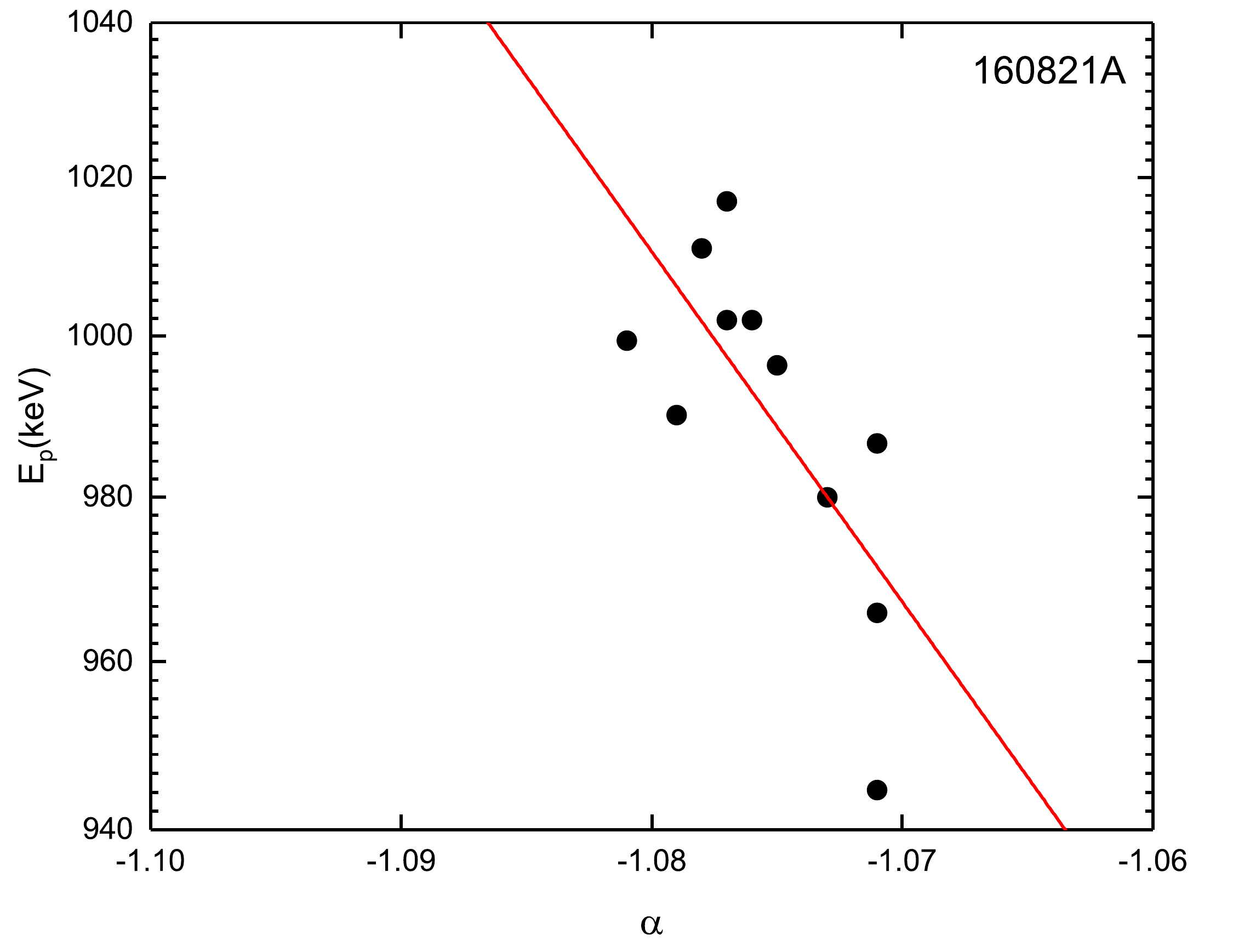}}
\resizebox{4cm}{!}{\includegraphics{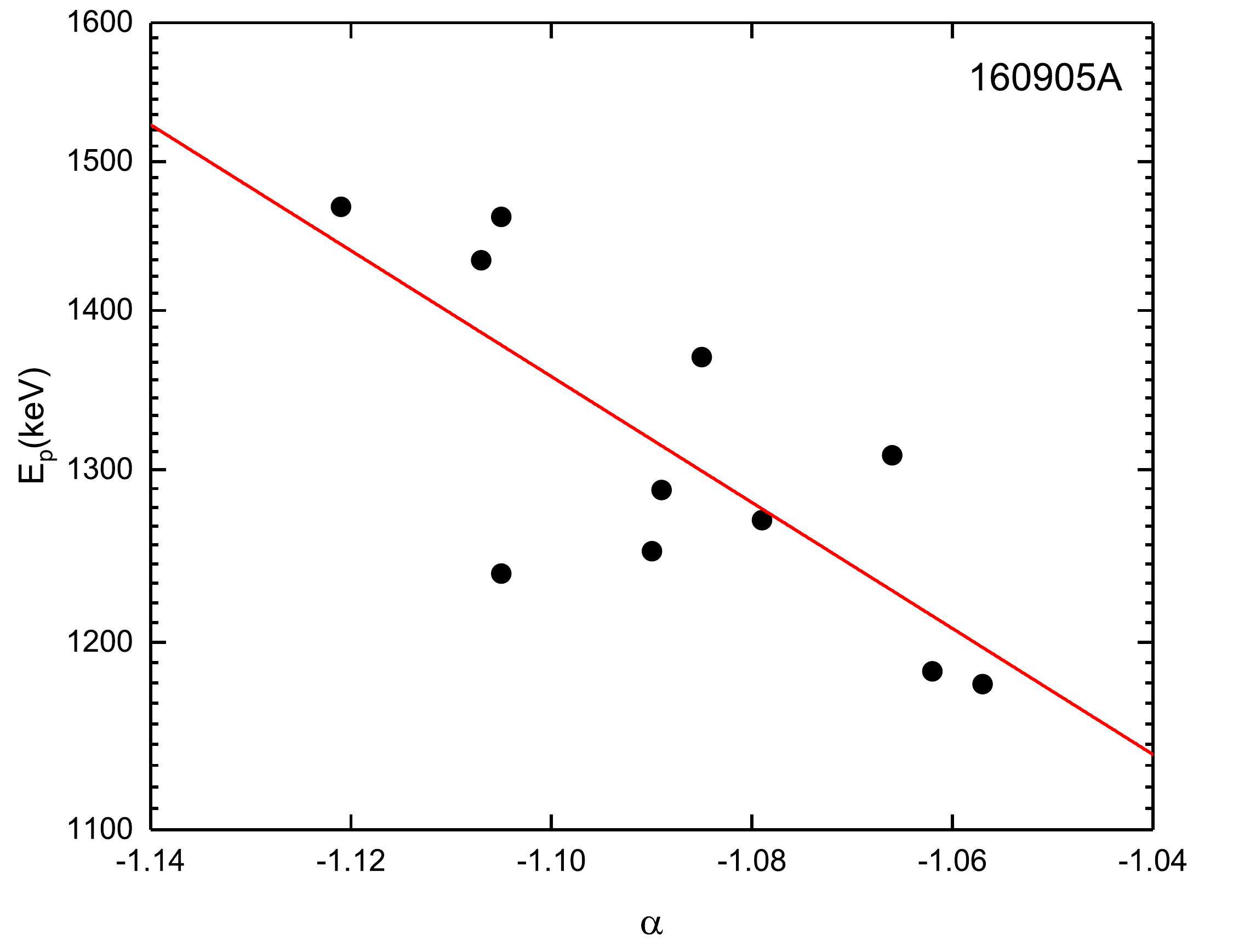}}
\resizebox{4cm}{!}{\includegraphics{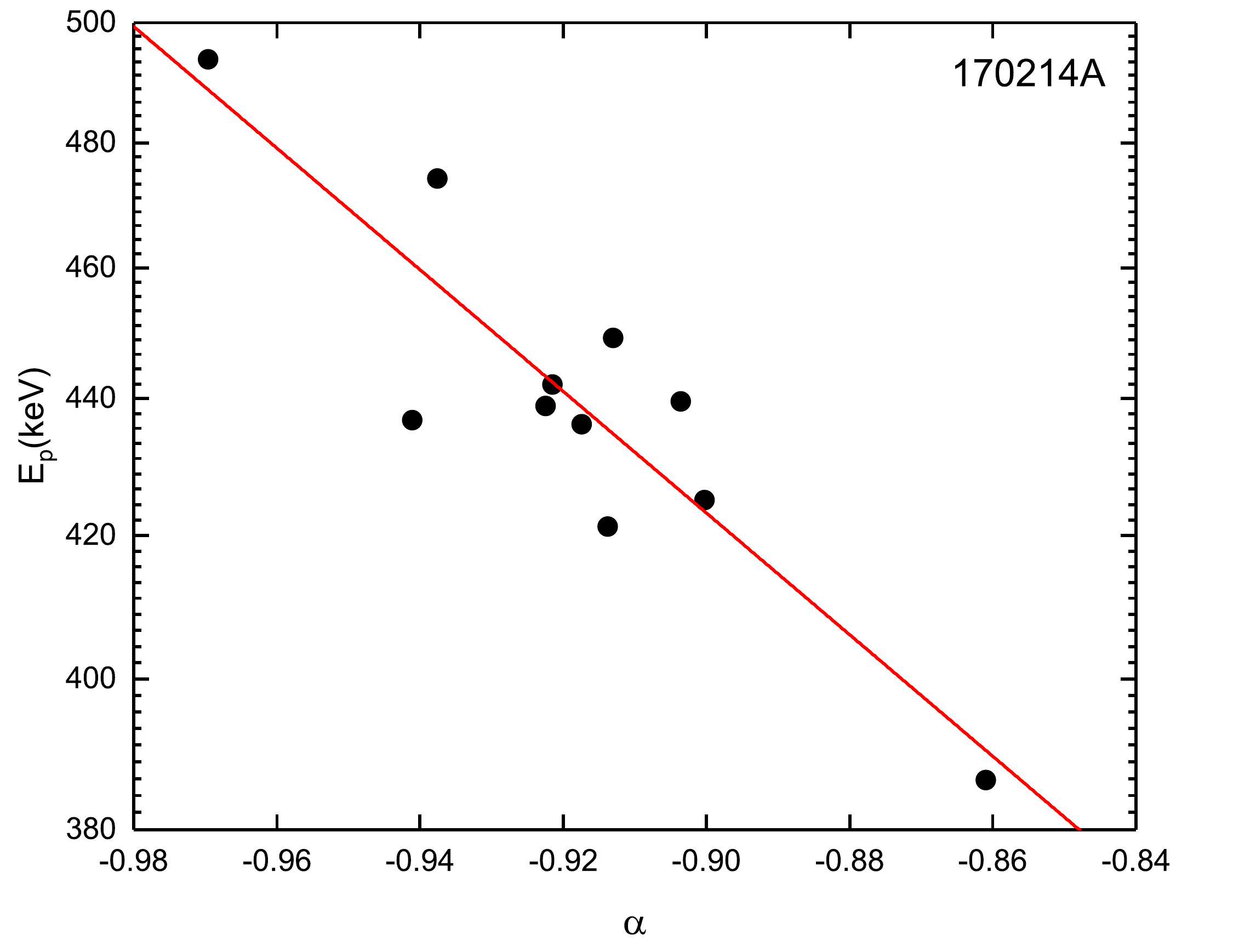}}
\resizebox{4cm}{!}{\includegraphics{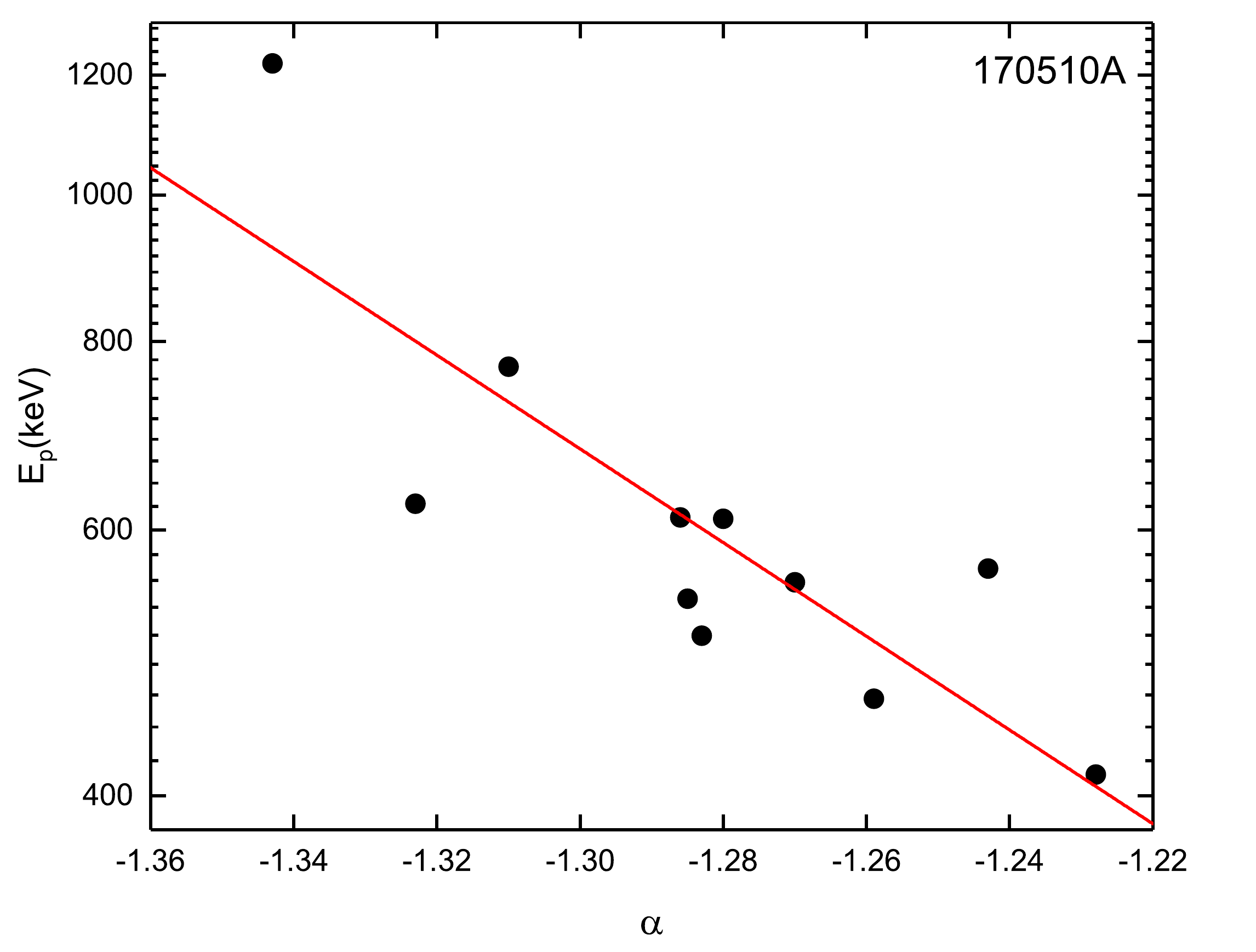}}
\caption{\edit1{\added{The $E_{p}-\alpha$ correlation from the simulation for 23 GRBs which exhibit a strong positive correlation in $\alpha-F$ correlation. The red solid line represents the best-linear-fitting result for each burst.}}\label{fig:simulation1}}
\end{figure}

\edit1{\added{
As said in Section \ref{subsec:subsec3.3}, we found that there are 23 GRBs show a strong positive correlation in $\alpha-F$ relation in our analysis. Also, five of these 23 GRBs have a strong positive correlation in $E_{p}-\alpha$. However, a physical mechanism (either synchrotron or photosphere emission) predicts a low-energy spectral index independent of the flux of the burst. On the other hand, \citet{2006ApJS..166..298K} pointed out that a strong anticorrelation was found between the peak energy $E_{p}$ and low energy spectral index $\alpha$ both for Band and COMP fits regardless of signal-to-noise ratio or the values of other parameters. In consideration of the differences between our results and the previous study, we performed a simulation to identify whether the two observed strong positive correlations are intrinsic or artificial.}}

\edit1{\added{
We performed the simulation analysis with the RMFIT package as a tool. We take the 23 GRBs which exhibit a strong positive correlation in $\alpha-F$ relation (5 GRBs also show a strong positive correlation in $E_{p}-\alpha$ relation among them) as a template to perform the simulations. The simulation procedure is as follows:
\begin{enumerate}
\item Extract the TTE data of the two brightest NaI and the corresponding BGO detectors of those GRBs (23 GRBs, see Figure \ref{fig:simulation} and Figure \ref{fig:simulation1}). We use the Band model with fixed input values of $E_{p}$, $\alpha$, $\beta$, and the normalization of the spectrum, which they are from the best Band-fitting parameters in the time-integrated spectrum for each burst, to produce an intrinsic spectrum.
\item Import the extracted data into RMFIT.
\item Perform a time-resolved spectral fitting analysis in different flux level (we changed the signal-to-noise ratio from 2 to 200, we used the values decreased by a step of a factor of 10 until the S/N was 2), and output the fitted parameters. 
\end{enumerate}}}

\edit1{\added{
Similarly, we show the two correlations $\alpha-F$ and $E_{p}-\alpha$ derived from the simulations in Figure \ref{fig:simulation} and Figure \ref{fig:simulation1}. In our simulations, only 1 GRB, GRB 160905A, shows a strong positive correlation (r=0.65) in $\alpha-F$ correlation. 21 GRBs show a strong anticorrelation except for the rest of 2 GRBs (GRBs 101014A, 131108A) in $E_{p}-\alpha$ correlation. Compared the simulated results with observed results (our fitting results), we think that the two observed strong positive correlations are artificial in our sample except for GRB 160905A in its $\alpha-F$ correlation.}}

\edit1{\added{As described in \citet{2002ApJ...565..182L}, a positive correlation between $E_{p}$ and $\alpha$ is expected due to the instrumental effect, even though the negative correlation is expected in the theory of gamma-ray bursts. If $E_{p}$ is close to the instrument's lower energy sensitivity limit, the low-energy spectral index $\alpha$ has not yet reached its asymptotic value, and $\alpha$ is softer than its  true value. In addition, because the spectrum with a low peak energy will exhibit most of its curvature near the low-energy edge of the instrument, smaller $E_{p}$ values will increase the uncertainty in the measurement of $\alpha$. Thus, we will observe the positive $E_{p}-\alpha$ correlation instead of the expected negative correlation in gamma-ray bursts. Combining with the `flux-tracking' pattern of $E_{p}$, on the other hand, it is naturally understandable that the positive $\alpha-F$ correlation will exhibit in the observation.}}

\section{conclusion and Discussion} \label{sec:sec4}

In this work, after performing the detailed time-resolved \edit1{\replaced{spectra}{spectral}} analysis of the bright gamma-ray bursts with the detection of $Fermi$-LLE in \edit1{\added{the}} prompt phase, \edit1{\replaced{we cluster these bursts and identified them as synchrotron origin or photosphere origin for their prompt emission phase.}{we presented all the spectra with the best Band-fitting results at peak flux for our bursts. To confirm whether our results are consistent with the $Fermi$ team results, we compared our results with the $Fermi$ GRB spectral catalog.}} Then we gave the evolution patterns \edit1{\replaced{for}{of the}} peak energy $E_{p}$ and low energy spectral index $\alpha$. \edit1{\added{Also, the statistical analysis for whether the low energy power-law indices $\alpha$ exceed the synchrotron limit were given.}} \edit1{\replaced{And}{Finally,}} the parameter correlations such as $E_{p}-F$, $\alpha-F$, and $E_{p}-\alpha$ \edit1{\replaced{also were}{were also}} presented in the analysis. \edit1{\added{To address whether the two observed correlations $\alpha-F$ and $E_{p}-\alpha$ are intrinsic or artificial, we have performed a simulation.}}

\edit1{\replaced{There are some interesting and new discoveries in our study,}{Meanwhile, some interesting phenomena were found in our $Fermi$-LLE bursts.}} such as:
\begin{enumerate}
\edit1{\deleted{\item
In light of those detailed analyses for all of the clusters, we think that there may be three categories for LLE-bursts, include photosphere origin (16 GRBs), synchrotron origin (3 GRBs) and photosphere origin with the superposition of synchrotron emission (10 GRBs). In a word, maybe, the photosphere component has been detected for about $90\%$ of these bursts in our study although there is no evidence to prove whether the thermal component has been detected for most of the bursts. Maybe, as said in \citet{2010MNRAS.407.1033B} and \citet{2010ApJ...725.1137L}, the Band spectral shapes can be produced by the photosphere.}}
\edit1{\added{\item
A single Band function is enough to perform the spectral fitting for every burst around their peak flux.}}
\item \edit1{\replaced{$79.3\%$}{$77.8\%$}} of the bursts have an $\alpha_{max}$ which is larger than the synchrotron limit ($-\frac{2}{3}$) in our\edit1{\deleted{ sample of}} bursts.
\item As we all know, the typical value of low energy photon index $\alpha$ is $\sim-1.0$ for the time-integrated spectrum, while the typical value of $\alpha$ in our sample is \edit1{\replaced{$\sim-0.8$}{$\sim-0.9$}}.
\item A good fraction of GRBs follow `hard-to-soft' trend (about two-thirds), and the rest should be the `flux-tracking' pattern (about one-third) in the previous literatures for $E_{p}$ evolution. However, it is obvious that the `flux-tracking' pattern is very popular for most of the bursts in our study include `intensity-tracking' (\edit1{\replaced{4}{5}} GRBs) and `rough-tracking' (\edit1{\replaced{18}{22}} GRBs)\edit1{\replaced{ and}{,}} the total number is \edit1{\replaced{22}{27}}, which means that \edit1{\replaced{$75.9\%$}{75\%}} of the bursts exhibit the `flux-tracking' pattern. \edit1{\replaced{And}{Additionally,}} the low energy photon index $\alpha$ does not show \edit1{\added{a}} strong general trend compared with $E_{p}$ although it also evolves with time instead of remaining constant in the previous literatures. While, \edit1{\replaced{22}{28}} GRBs exhibit `flux-tracking' pattern \edit1{\replaced{include}{which includes}} `intensity-tracking' (2 GRBs) and `rough-tracking' (\edit1{\replaced{20}{26}} GRBs) in our study. In a word, \edit1{\replaced{there are$75.9\%$}{$77.8\%$}} of \edit1{\replaced{the}{our}} bursts exhibit the `flux-tracking' pattern.
\item For the parameter correlations, from Section \ref{subsec:subsec3.3}, a majority of bursts exhibit \edit1{\replaced{the}{a}} strong (very strong) positive correlation (\edit1{\replaced{$75.9\%$}{$69.4\%$}}) between $E_{p}$ and $F$ (energy flux). \edit1{\replaced{And $70\%$}{$63.9\%$}} of \edit1{\replaced{the}{our}} bursts have \edit1{\replaced{the}{a}} strong (very strong) positive correlation between $\alpha$ and $F$. But there is no clear \edit1{\replaced{behaviour}{behavior}} in $E_{p}-\alpha$ correlation in our sample\edit1{\deleted{ of bursts}}. \edit1{\replaced{At last}{Finally}}, it is noteworthy that \edit1{\replaced{the}{a}} very strong negative correlation has been exhibited both for $\alpha-F$ and $E_{p}-\alpha$ correlations for GRB 170115B.
\edit1{\added{\item The two observed strong positive correlations ($\alpha-F$ and $E_{p}-\alpha$) are artificial in our sample except for GRB 160905A in its $\alpha-F$ correlation.}} 
\end{enumerate}

Over the last fifty years, the research in the field of gamma-ray bursts has made a lot of progress, but there are still some open questions \citep[e.g.,][]{2011CRPhy..12..206Z,2017SSRv..212..409D,2018pgrb.book.....Z,2019arXiv190202562P}. One of the questions is about the radiation mechanism in the prompt emission, which debated whether the GRB prompt emission is produced by the synchrotron radiation or the emission from the photosphere \citep{2014IJMPD..2330003V,2017IJMPD..2630018P}. However, a unified model has not been provided even though the physical models like the synchrotron model \citep{2016ApJ...816...72Z} and subphotospheric dissipation model \citep{2019MNRAS.485..474A} have been used to make the spectral fitting.

As we all know, the Band component in most observed gamma-ray burst spectra seems to be thought as synchrotron origin. Two possible cases should be considered: the first one is for the internal shock model \citep{1994ApJ...427..708P,1994ApJ...430L..93R}, which invokes a small radius. The second case invokes a large internal magnetic dissipation radius, so-called the Internal-Collision-induced MAgnetic Reconnection and Turbulence (ICMART) model \citep{2011ApJ...726...90Z}. For the internal shock model, the peak energy $E_{p}\propto L^{1/2}\gamma_{e,ch}^{2}R^{-1}(1+z)^{-1}$ can be derived from the synchrotron model in \citet{2002ApJ...581.1236Z}, where $L$ is the ``wind'' luminosity of the ejecta, $\gamma_{e,ch}$ is the typical electron Lorentz factor of the emission region, $R$ is the emission radius, and $z$ is the redshift of the burst. Then, the tracking behavior will emerge because of the natural relation of $E_{p}\propto L^{1/2}$. While a hard-to-soft evolution pattern of peak energy $E_{p}$ is predicted for the ICMART model \citep{2011ApJ...726...90Z,2014NatPh..10..351U}. On the other hand, \citet{2018ApJ...869..100U} also pointed out that the ``flux-tracking'' behavior could be reproduced within the ICMART model if other factors such as bulk acceleration are taken into account. Furthermore, \citet{2016ApJ...816...72Z} demonstrated that the synchrotron model can reproduce the $E_{p}$-tracking pattern through the data analysis for GRB 130606B. Therefore, the ``flux-tracking'' behavior of $E_{p}$ can be made with both these two synchrotron models. In a short, a hard-to-soft pattern and tracking behavior of $E_{p}$ can be reproduced successfully in the synchrotron model.

Meanwhile, the photosphere model can also produce an $E_{p}$-tracking pattern and a hard-to-soft pattern of $E_{p}$ successfully \citep{2014ApJ...785..112D,2019ApJ...882...26M}. But, this model predicts a hard-to-soft pattern of $\alpha$ instead of $\alpha$-tracking behavior. It is difficult to produce the observed $\alpha$-tracking behavior in this model. On one hand, the predicted $\alpha$ value ($\alpha \sim +0.4$) is much harder than the observed \citep{2014ApJ...785..112D}. The introduction of a special jet structure is necessary to reproduce a typical $\alpha \sim -1$ \citep{2013MNRAS.428.2430L}. On the other hand, this model invokes an even smaller emission radius than the internal shock model, so, the contrived conditions from the central engine are needed to reproduce the tracking pattern of $\alpha$. However, few bursts exhibit a hard-to-soft pattern in our sample. Besides, \citet{2019MNRAS.485..474A} used the physical subphotospheric model to fit the $Fermi$ data (include 6 LLE-bursts in our sample; GRBs 090926A,130518A, 141028A,150314A, 150403A, 160509A), only 171 out of 634 spectra are accepted (17 out of 135 spectra for the six LLE-bursts). As a result, we infer that the great majority of bursts in our sample are dominated by the synchrotron component even though the photosphere component is still not excluded in their prompt phases.

Additionally, the patterns of the peak energy $E_{p}$ evolution have close connections to the spectral lags \citep{2018ApJ...869..100U}. In general, the light curves at higher energies peak earlier than those at lower energies, named positive spectral lags. Reversely, the negative spectral lags, the higher-energy emission slightly lagging behind the lower-energy emission \citep{2016ApJ...825...97U}. The previous literature shows that only small fraction bursts show negative lags or no spectral lags \citep{1996ApJ...459..393N,2000ApJ...534..248N,2006ApJ...646..351L,2012MNRAS.419..614U}. \citet{2018ApJ...869..100U} studied and provided the connections between the patterns of the $E_{p}$ evolution and the types of spectral lags (positive or negative lags). According to \citet{2018ApJ...869..100U}, the positive spectral lags can occur if the peak energy exhibits a hard-to-soft evolution pattern, but the negative type can not occur. When the $E_{p}$ presents a flux-tracking behavior, both the positive and the negative types of spectral lags can occur. The clue to differentiate between the positive lags and the negative lags for $E_{p}$-tracking pattern comes from the peak location of the flux curve. The peak location of the flux curve slightly lags behind the peak of $E_{p}$ curve for the former, whereas there is no longer a visible lag between them for the latter \citep{2018ApJ...869..100U}. Assume that those bursts which exhibit a hard-to-soft pattern or flux-tracking pattern of peak energy $E_{p}$ occur spectral lags. Then, the positive type of spectral lags will occur at the six bursts which exhibit a hard-to-soft behavior of $E_{p}$ (GRBs 080825C, 090328A, 110721A, 120624B, 160910A, 171210A). The positive type of spectral lags will also occur at the 12 GRBs because of their peak location of flux curves slightly lags behind their peak of $E_{p}$ curves (GRBs 090926A, 100826A, 130502B, 130504C, 130518A, 140206B, 150118B, 150627A, 160509A, 160821A, 170214A, 170808B). The negative lags will occur at the rest of the bursts because there is no visible lag between the two peaks (GRBs 090626A, 100724B, 101014A, 120226A, 130821A, 140102A, 141028A, 150202B, 150403A, 160816A, 160905A, 170115B, 180305A).

\acknowledgments

We thank an anonymous referee for helpful suggestions. We also thank Lei-Ming Du, Zhao-Yang Peng, and Dao-Zhou Wang for their help. We acknowledge the use of the public data from the Fermi data archives.
This work is supported by the National Natural Science Foundation of China (grant No.11673006), the Guangxi Science Foundation (grant Nos. 2016GXNSFFA380006, 2017AD22006,  2018GXNSFDA281033), the One-Hundred-Talents Program of Guangxi colleges, and High level innovation team and outstanding scholar program in Guangxi colleges.



\end{document}